\documentclass[3p,preprint,sort&compress,square,comma,numbers,merge]{elsarticle}

\usepackage[greek,american]{babel}
\usepackage[utf8]{inputenc}
\usepackage{graphicx}
\usepackage{grffile}
\usepackage{ifpdf}
\ifpdf
  \usepackage{epstopdf}
\fi
\DeclareGraphicsExtensions{.png,.pdf,.jpg,.mps,.eps}
\graphicspath{{./tikz/},{./fig/},{./scripts/ideo/}}
\usepackage[dvipsnames,usenames]{color}
\usepackage{subfig}
\usepackage{fix2col}

\usepackage{amsmath,amssymb}
\usepackage{amsfonts}
\usepackage{dsfont}  %
\usepackage{mathrsfs}  %
\usepackage{esint} %
\usepackage{hepnames}
\usepackage{xpatch}

\makeatletter
\xpatchcmd\@HepConStyle
 {\edef\@upcode{\updefault}}
 {\ifdefined\shapedefault\edef\@upcode{\shapedefault}\else\edef\@upcode{\updefault}\fi}
 {}{}
\makeatother
\usepackage{import}
\usepackage{xspace}      %
\xspaceaddexceptions{]}
\usepackage{xpunctuate}  %
\usepackage{grffile}     %
\usepackage{mathtools}   %
\usepackage{braket}      %
\usepackage{oubraces}    %
\usepackage{nccmath}     %
\usepackage[autostyle=true,autopunct=true]{csquotes}  %
\usepackage[export]{adjustbox}  %
\usepackage{booktabs}    %
\usepackage{makecell}    %
\usepackage{array}%
\newcolumntype{$}{>{\global\let\currentrowstyle\relax}}
\newcolumntype{^}{>{\currentrowstyle}}
\newcommand{\rowstyle}[1]{\gdef\currentrowstyle{#1}#1\ignorespaces}
\RequirePackage{bigdelim}  %
\usepackage{multirow}
\usepackage{blindtext}
\usepackage{titletoc}  %
\usepackage{fancyhdr}  %
\usepackage{macros_small}
\usepackage[plainpages=false,colorlinks]{hyperref}
\usepackage[capitalise]{cleveref}  %
\crefformat{footnote}{#2\footnotemark[#1]#3}
\RequirePackage{bm}

\RequirePackage{slashed}

{\catcode`@=11
 \@ifundefined{extract@font}{
   \gdef\mathrm#1{{\rm#1}}  \gdef\mathit#1{{\it#1}}
   \message{- myunits: NFFS not present -- using OFFS}
 }{}

\protected\gdef\EE{\@ifnextchar*{\@EE}{\cdot\@EE*}}
\protected\gdef\@EE*#1{10^{#1}}

 \gdef\IN{\mathrm {I\kern-.18em N}}
 \gdef\IZ{\mathrm {Z\kern-.3em Z}}
 \gdef\IR{\mathrm {I\kern-.18em R}}
 \gdef\IC{\mathrm {\kern+.25em\rule[.05em]{.06em}{1.4ex}\kern-.31em C}}

 \gdef\lsim{\mathrel{\mathpalette\@versim<}}
 \gdef\gsim{\mathrel{\mathpalette\@versim>}}
 \gdef\@versim#1#2{\lower.5ex\vbox{\baselineskip\z@skip\lineskip-.01ex
   \ialign{$\m@th#1\hfil##\hfil$\crcr#2\crcr\sim\crcr}}}
}

\def\diffcb#1{\mbox{$\textrm{d}^{3}\!#1$}}

\def\diag{\mathop{\mathrm{diag}}} %
\def\disc{\mathop{\mathrm{disc}}} %
\renewcommand{\Re}{\mathop{\mathrm{Re}}} %
\renewcommand{\Im}{\mathop{\mathrm{Im}}} %
\def\bvec#1{\boldsymbol{\mathbf{#1}}}  %

\newcommand{\bracket}[2]{\mbox{$\left\langle{#1}\vphantom{#2}\right|\left.\!\!{#2}\vphantom{#1}\right\rangle$}}
\newcommand{\abs}[1]{\left| #1 \right|} %
\RequirePackage{upgreek} 

{\catcode`@=11
 \@ifundefined{extract@font}{
   \gdef\mathrm#1{{\rm#1}}  \gdef\mathit#1{{\it#1}}
   \message{- myunits: NFFS not present -- using OFFS}
 }{}

 \gdef\defUnit{\@ifnextchar[{\defUnit@}{\defUnit@[{},{}]}}
 \gdef\defUnit@[#1]#2#3{\bgroup\def\@tempa{#2}\def\@tempb{#3}%
   \def\@tempc{*}\defUnit@@#1;*,*;}
 \gdef\defUnit@@#1,#2;{\def\@tempd{#1}\ifx\@tempc\@tempd\egroup\else%
   \expandafter\xdef\csname #1\@tempa\endcsname{\noexpand\mathrm {#2\@tempb}}%
   \expandafter\defUnit@@ \fi}

 \defUnit[f,f;n,n;mu,\upmu;m,m;C,c;{},{};k,k]{m}{m}
 \defUnit{Lj}{lj}
 \defUnit[{},{};M,M]{Parsec}{ps}
 \defUnit{AstroE}{AE}
 \defUnit[f,f;n,n;mu,\upmu;m,m]{b}{b}
 \defUnit[m,m;{},{}]{rad}{rad}
 \defUnit[m,m;{},{}]{sr}{sr}
 \defUnit[f,f;p,p;n,n;mu,\upmu;m,m;{},{}]{s}{s}
 \defUnit{Min}{min}
 \defUnit{Hour}{h}
 \defUnit{Year}{y}
 \defUnit[{},{};k,k;M,M;G,G]{Hz}{Hz}
 \defUnit[n,n;mu,\upmu;m,m;k,k]{V}{\hskip-0.1em{}V}
 \defUnit[{},{};M,M;G,G;T,T]{V}{V}
 \defUnit[p,p;n,n;mu,\upmu;m,m;{},{};k,k]{A}{A}
 \defUnit[f,f;p,p;n,n;mu,\upmu;m,m;{},{}]{C}{C}
 \defUnit[m,m;{},{};k,k;M,M;G,G]{Ohm}{\Omega}
 \defUnit[m,m;{},{}]{Siemens}{S}
 \defUnit[p,p;n,n;mu,\upmu;m,m;{},{}]{F}{F}
 \defUnit[m,m;{},{}]{H}{H}
 \defUnit[n,n;mu,\upmu;m,m;{},{}]{T}{T}
 \defUnit[{},{};k,k]{Gauss}{Gauss}
 \defUnit{Weber}{Wb}
 \defUnit[{},{};k,k]{N}{N}
 \defUnit[mu,\upmu;m,m;{},{};k,k;M,M;G,G;T,T]{W}{W}
 \defUnit[mu,\upmu;m,m;{},{};k,k;M,M;G,G;T,T]{eV}{e\hskip-0.1em{}V}
 \defUnit[m,m;{},{};k,k;M,M;G,G;T,T]{J}{J}
 \defUnit[{},{};k,k]{Cal}{cal}
 \defUnit{Ry}{Ry}
 \defUnit[f,f;p,p;n,n;mu,\upmu;m,m;{},{};k,k]{g}{g}
 \defUnit[{},{};k,k;M,M;G,G]{t}{t}
 \defUnit{amu}{amu}
 \defUnit[m,m;{},{};k,k;M,M;G,G;T,T]{Bar}{bar}
 \defUnit[m,m;k,k;M,M;G,G;T,T]{bar}{bar}
 \defUnit[m,m;{},{};h,h;k,k;M,M;G,G;T,T]{Pa}{Pa}
 \defUnit[m,m;{},{};h,h;k,k;M,M;G,G;T,T]{atm}{atm}
 \defUnit{Poise}{P}
 \defUnit{Stokes}{St}
 \defUnit[m,m;{},{}]{K}{K}
 \defUnit{degC}{{^oC}}
 \defUnit{Candela}{cd}
 \defUnit{Lumen}{lm}
 \defUnit{Lux}{lx}
 \defUnit[m,m;{},{}]{Roentgen}{R}
 \defUnit[m,m;{},{}]{Rem}{rem}
 \defUnit[m,m;{},{}]{Sievert}{Sv}
 \defUnit[m,m;{},{}]{Gray}{Gy}
 \defUnit[m,m;{},{}]{Rad}{rd}
 \defUnit[{},{};k,k;M,M;G,G;T,T]{Bq}{Bq}
 \defUnit[m,m;{},{}]{Curie}{Ci}
 \defUnit[{},{};k,k]{mol}{mol}
 \defUnit{ppm}{ppm}
 \defUnit{ppb}{ppb}
 \defUnit{dB}{dB}
 \defUnit{au}{a.u.}
}

\newcommand*{\wrt}{{with respect to}\xspace}

\newcommand*{\eg}{e.g\xperiod}  %
\newcommand*{\Eg}{E.g\xperiod}
\newcommand*{\ie}{i.e\xperiod}  %
\newcommand*{\refCite}[1]{{Ref.~\cite{#1}}\xspace}
\newcommand*{\refsCite}[1]{{Refs.~\cite{#1}}\xspace}
\newcommand*{\RefCite}[1]{{Reference~\cite{#1}}\xspace}

\newcommand*{\one}{{(i)}\xspace}
\newcommand*{\two}{{(ii)}\xspace}
\newcommand*{\three}{{(iii)}\xspace}
\newcommand*{\four}{{(iv)}\xspace}

\newrobustcmd{\subfloatLabel}[1]{\subref{#1}}

\newcommand*{\orderOf}[1]{\ensuremath{\mathcal{O}(#1)}\xspace}

\def\app#1#2{%
  \mathrel{%
    \setbox0=\hbox{$#1\sim$}%
    \setbox2=\hbox{%
      \rlap{\hbox{$#1\propto$}}%
      \lower1.1\ht0\box0%
    }%
    \raise0.25\ht2\box2%
  }%
}
\newcommand*{\proptosim}{\mathpalette\app\relax}
\renewcommand*{\Re}{\operatorname{Re}}
\renewcommand*{\Im}{\operatorname{Im}}
\newcommand*{\dif}[1][]{\ifthenelse{\equal{#1}{}}{\mathop{}\!\mathrm{d}}{\mathop{}\!\mathrm{d}^{#1}}}
\newcommand*{\pdif}[1][]{\ifthenelse{\equal{#1}{}}{\mathop{}\!\partial}{\mathop{}\!\partial^{#1}}}
\newcommand*{\clebsch}[6]{\ensuremath{({#1},{#2};{#3},{#4}\,|\,{#5},{#6})}\xspace}
\newcommand*{\Underbrace}[2]{{\underbrace{#1}_{#2}}}
\newcommand*{\Overbrace}[2]{{\overbrace{#1}^{#2}}}

\DeclareMathOperator*{\argmax}{arg\,max}
\makeatletter
\DeclareRobustCommand\Widecheck[1]{{\mathpalette\@Widecheck{#1}}}
\def\@Widecheck#1#2{%
    \setbox\z@\hbox{\m@th$#1#2$}%
    \setbox\tw@\hbox{\m@th$#1%
       \widehat{%
          \vrule\@width\z@\@height\ht\z@
          \vrule\@height\z@\@width\wd\z@}$}%
    \dp\tw@-\ht\z@
    \@tempdima\ht\z@ \advance\@tempdima2\ht\tw@ \divide\@tempdima\thr@@
    \setbox\tw@\hbox{%
       \raise\@tempdima\hbox{\scalebox{1}[-1]{\lower\@tempdima\box
\tw@}}}%
    {\ooalign{\box\tw@ \cr \box\z@}}}
\makeatother
\newcommand*{\widecheck}[1]{\makebox[0pt]{$\phantom{#1}\Widecheck{\phantom{#1}}$}#1}

\newcommand*{\tMin}{\ensuremath{{\abs{t}_\text{min}}}\xspace}
\newcommand*{\acc}{\ensuremath{{\epsilon}}\xspace}  %

\newcommand*{\yGJ}{\ensuremath{{y_{\text{GJ}}}}\xspace}
\newcommand*{\zGJ}{\ensuremath{{z_{\text{GJ}}}}\xspace}

\newcommand*{\zHF}{\ensuremath{{z_{\text{HF}}}}\xspace}

\newcommand*{\yGJv}{\ensuremath{{\hat{y}_{\text{GJ}}}}\xspace}

\newcommand*{\yHFv}{\ensuremath{{\hat{y}_{\text{HF}}}}\xspace}
\newcommand*{\zHFv}{\ensuremath{{\hat{z}_{\text{HF}}}}\xspace}
\newcommand*{\anglesGJ}{\ensuremath{{\Omega_{\text{GJ}}}}\xspace}
\newcommand*{\anglesHF}{\ensuremath{{\Omega_{\text{HF}}}}\xspace}
\newcommand*{\thetaGJ}{\ensuremath{{\vartheta_{\text{GJ}}}}\xspace}
\newcommand*{\thetaHF}{\ensuremath{{\vartheta_{\text{HF}}}}\xspace}
\newcommand*{\cosThetaGJ}{\ensuremath{{\cos \thetaGJ}}\xspace}
\newcommand*{\cosThetaHF}{\ensuremath{{\cos \thetaHF}}\xspace}
\newcommand*{\phiGJ}{\ensuremath{{\phi_{\text{GJ}}}}\xspace}
\newcommand*{\phiHF}{\ensuremath{{\phi_{\text{HF}}}}\xspace}
\newcommand*{\IG}{\ensuremath{I^G}\xspace}
\newcommand*{\JP}{\ensuremath{J^P}\xspace}
\newcommand*{\JPC}{\ensuremath{J^{PC}}\xspace}
\newcommand*{\Mrefl}{\ensuremath{M^\refl}\xspace}
\newcommand*{\IGJPC}{\ensuremath{\IG\,\JPC}\xspace}
\newcommand*{\JPCMrefl}{\ensuremath{\JPC\,\Mrefl}\xspace}
\newcommand*{\JPMrefl}{\ensuremath{\JP\,\Mrefl}\xspace}

\newcommand*{\refl}{\ensuremath{\varepsilon}\xspace}  %
\newcommand*{\reflOp}{\ensuremath{\hat{\Pi}_y}\xspace}  %
\newcommand*{\natur}{\ensuremath{\eta}\xspace}  %
\newcommand*{\naturEx}{\ensuremath{\eta_\text{ex}}\xspace}  %
\newcommand*{\intens}{N}
\DeclareMathOperator{\relintens}{Relint}
\DeclareMathOperator{\ovl}{Ovl}
\DeclareMathOperator{\coh}{Coh}
\newcommand*{\phase}{\Delta \phi}
\newcommand{\tpl}{\mathfrak{t}} %
\newcommand*{\Paqpr}{\ensuremath{\Paq^\prime}\xspace}
\newcommand*{\qqbar}{\ensuremath{\Pq\Paq}\xspace}
\newcommand*{\qqbarpr}{\ensuremath{\Pq\Paqpr}\xspace}
\newcommand{\pbar}{\Pap}

\newcommand{\Xminus}{\relax\ifmmode\mathrm{X^-}%
                \else$\mathrm{X^-}$\fi}%
\newcommand{\X}{\relax\ifmmode\mathrm{X}%
                \else$\mathrm{X}$\fi}%
\newcommand*{\etaOrPr}{\ensuremath{\eta^{(}{'}\protect\vphantom{\eta}^{)}}\xspace}
\newcommand*{\Reg}{\ensuremath{\mathbb{R}}\xspace}
\newcommand*{\Pom}{\ensuremath{\mathbb{P}}\xspace}

\DeclareRobustCommand{\Pch}{\HepParticle{h}{\Pqc}{}\xspace}

\DeclareRobustCommand{\Pbh}{\HepParticle{h}{\Pqb}{}\xspace}
\DeclareRobustCommand{\Pbgh}{\HepParticle{\eta}{\Pqb}{}\xspace}
\DeclareRobustCommand{\Pbgc}{\HepParticle{\chi}{\Pqb}{}\xspace}
\usepackage{fp}
\FPeval{\plotHeightRatioOneDTwoD}{550/511}           %
\FPeval{\widthOneD}{1/(1+\plotHeightRatioOneDTwoD)}  %
\newlength{\totalPlotWidth}
\newlength{\onePlotWidth}
\newlength{\twoPlotWidth}
\newlength{\twoPlotWidthTwoD}
\FPeval{\widthOneD}{1/(2+\plotHeightRatioOneDTwoD)}  %
\newlength{\threePlotWidth}
\newlength{\threePlotWidthTwoD}
\newcommand*{\eqPunctSpacing}{\;}

\journal{Progress in Particle and Nuclear Physics}

\DeclareMathVersion{normal2} %

\begin{document}

\selectlanguage{american}

\clearpage{}%
\begin{frontmatter}

  \title{Light-Meson Spectroscopy with COMPASS}

  \author[bonn]{B.\ Ketzer}
  \ead{bernhard.ketzer@uni-bonn.de}

  \author[tum]{B.\ Grube}
  \ead{bgrube@tum.de}

  \author[ihep,tum]{D.\ Ryabchikov}
  \ead{dmitry.ryabchikov@ihep.ru}

  \address[bonn]{Universit\"at Bonn, Helmholtz-Institut f\"ur Strahlen- und Kernphysik, 53115 Bonn, Germany}
  \address[tum]{Technische Universit\"at M\"unchen, Physik-Department, 85748 Garching, Germany}
  \address[ihep]{State Scientific Center Institute for High Energy Physics of National Research Center \enquote{Kurchatov Institute}, 142281 Protvino, Russia}

  \date{\today}

  \begin{abstract}

    Despite decades of research, we still lack a detailed quantitative
    understanding of the way quantum chromodynamics (QCD) generates
    the spectrum of hadrons. Precise experimental studies of
    the hadron excitation spectrum and the dynamics of hadrons help to
    improve models and to test effective theories and lattice
    QCD simulations. In addition, QCD seems to allow hadrons beyond the three-quark and quark--antiquark
    configurations of the constituent-quark model. These so-called
    exotic hadrons contain additional constituent (anti)quarks
    or excited gluonic fields that contribute to the quantum numbers
    of the hadron.  Hadron spectroscopy is currently one of the most
    active fields of research in hadron physics.  The COMPASS
    experiment at the CERN SPS is studying the
    excitation spectrum of light mesons, which are composed of up,
    down, and strange quarks.  The excited mesons are produced via the strong
    interaction, \ie by Pomeron exchange, by scattering a
    \SI{190}{\GeVc} pion beam off proton or nuclear targets.  On heavy
    nuclear targets, in addition the electromagnetic interaction
    contributes in the form of quasi-real photon exchange at very low
    four-momentum transfer squared.  COMPASS has performed the most
    comprehensive analyses to date of isovector resonances decaying
    into \etaPi, \etaPrPi, or \threePi final states.  In this review,
    we give a general and pedagogical introduction into scattering
    theory and the employed partial-wave analysis techniques. We also
    describe novel methods developed for the high-precision COMPASS
    data.  The COMPASS results are summarized and compared to previous
    measurements.  In addition, we discuss possible signals for exotic
    mesons and conclude that COMPASS data provide solid evidence for
    the existence of the manifestly exotic \PpiOne[1600], which has
    quantum numbers forbidden for a quark-model state, and of the
    \PaOne[1420], which does not fit into the quark-model spectrum.
    By isolating the contributions from quasi-real photon exchange,
    COMPASS has measured the radiative widths of the \PaTwo and, for
    the first time, that of the \PpiTwo and has tested predictions of
    chiral perturbation theory for the process
    $\pi^- + \gamma \to \threePi$.

  \end{abstract}


\end{frontmatter}

\begin{center}
  (Published in \emph{Progress in Particle and Nuclear Physics 113 (2020) 103755})
 \end{center}
\thispagestyle{empty}

\tableofcontents
\clearpage{}%
\clearpage{}%
\section{Introduction}
\label{sec:intro}

The understanding of the fundamental building blocks of matter has
been a long-standing quest of mankind.  According to the Greek
philosopher Empedocles (around 500~B.C.), the universe is composed of
Air, Fire, Water, and Earth. It was Democritus (460 to 371~B.C.) who
had the vision of fundamental indivisible
constituents~\cite{taylor:1999}:
\begin{quote}
  By convention [there is] sweet and by convention [there is] bitter,\\
  by convention [there is] hot, by convention [there is] cold, by
  convention [there is] color;\\ but in reality [there are only]
  atoms\footnote{From Ancient Greek \textgreek{ἄτομος} meaning indivisible.} and
  void.
  \begin{flushright}
    In:
    \begin{minipage}[t]{0.4\linewidth}
      {\small\sc Sextus Empiricus}\\
      {\small\emph{Against the Mathematicians VII.135.}}
    \end{minipage}
  \end{flushright}
\end{quote}

It took more than 2000~years before in the 19th century Mendeleev and
Meyer noticed patterns in the chemical properties of the chemical
elements, indicating an underlying symmetry and hinting at atoms as
their basic constituents.
This strategy was later to be repeated to find even more fundamental
building blocks of matter.
In the 20th century, scattering experiments with momentum transfers
large enough to probe the internal structure of the colliding objects
subsequently revealed the substructure of atoms, which are systems of
negatively charged electrons and a positively charged nucleus bound
together by the electromagnetic force, and that of atomic nuclei,
which are composed of protons and neutrons held together by the
nuclear or strong force.  In addition to protons and neutrons, a large
number of other strongly interacting particles, both with half-integer
spin (called baryons\footnote{The term baryon was coined by
  A.~Pais~\cite{Pais:1953} and was derived from the Greek word
  \textquote{\textgreek{βαρύς}} for heavy, because at that time most
  known particles that were considered elementary had lower masses
  than the baryons.}) and integer spin (called
mesons\footnote{Originally, C.~D.~Anderson and S.~H.~Neddermeyer
  proposed the term \textquote{mesotron}~\cite{Anderson:1938} that was
  derived from the Greek word \textquote{\textgreek{μέσος}} for
  intermediate for particles with masses between that of the electron
  and the proton.  The term was later changed to
  meson~\cite{Bhabha1939}.}), was observed in experiments with cosmic
rays and later at particle accelerators.  The lightest member of the
meson family is the pion, which was discovered by Powell in 1947. The
kaon, the lightest particle containing strange quarks, was observed by
Rochester and Butler in 1947.  Many heavier mesons and baryons
followed, suggesting that pions, protons and neutrons (\ie nucleons)
were merely the lightest members of a large family of strongly
interacting particles.  Most of the heavier species of mesons and
baryons are extremely unstable and decay after a very short time of
the order of $\EE*{-24}\,\s$.  Today, more than 200~mesons and more
than 100~baryons have been identified~\cite{Tanabashi:2018zz}.

As for the chemical elements, such a proliferation of particles called
for a systematic order in the zoo of hadrons.  Sorting the then-known
baryons and mesons by their quantum numbers strong isospin and
strangeness, which are known to be conserved in strong
interactions,\footnote{The strong isospin characterizes the
  approximate symmetry of the strong interaction between proton and
  neutrons, or \Pqu~and \Pqd~quarks, which is explicitly broken by the
  small mass difference between~\Pqu and~\Pqd.  In the rest of the
  paper, the term \enquote{isospin} always refers to the strong
  isospin.}  hinted towards a further substructure of these strongly
interacting particles.  In the 1960s, Gell-Mann and Zweig suggested
that hadrons could be made of more elementary objects called quarks.
Three types or flavors of quarks, up~(\Pqu), down~(\Pqd),
strange~(\Pqs), carrying different charge, isospin, and strangeness
were required to explain the quantum numbers of the observed hadrons.
What started out merely as a mathematical tool, soon turned out to
become reality, when electron-scattering experiments at SLAC began to
show that protons did indeed contain point-like constituents, called
partons by Feynman.  In the 1970s, experiments colliding high-energy
electrons with their antiparticles, positrons, revealed the existence
of heavier types of quarks, carrying new quantum numbers called charm
and bottomness (or beauty).  In 1999, the sixth and heaviest quark,
called top, was observed at Fermilab.

According to our present-day understanding, matter is composed of
fundamental point-like fermions, \ie particles with spin $\hbar/2$,
which are summarized in \cref{tab:fermions}.
While the quarks experience the strong force, the leptons do not. The
stable matter around us is entirely composed of fermions of the first
generation, \ie electrons and up and down quarks. The heavier fermions
can be produced in high-energy particle collisions at accelerators or
in cosmic-ray interactions.  While the masses of the charged leptons
can be measured directly, the masses of the quarks need to be
determined indirectly, since they are confined inside hadrons and
never appear as isolated particles (see below).  The quark masses
given in \cref{tab:fermions} are so-called current masses, which are
the mass values of the bare quarks appearing in the QCD
Lagrangian~\cite{pdg_quark_masses:2018}.  They are determined by
comparing measured hadron properties with calculations, \eg using
Lattice QCD (see \cref{sec:pheno.lattice}) or chiral perturbation
theory (see \cref{sec:theory.chiPT}) for the case of light quarks.  In
order to render the physical quantities calculated from the theory
finite, a renormalization scheme needs to be applied, which requires
the introduction of a scale parameter~$\mu$.  The most commonly
applied scheme in QCD perturbation theory is the
$\overline{\text{MS}}$ scheme and a typical scale is $\mu = 2\,\GeV$.

\begin{table}[tbp]
  \centering
  \renewcommand{\arraystretch}{1.2}
  \caption{Fundamental fermions with quantum numbers $Q$~(charge),
    $I$~(strong isospin), $I_3$~($z$-component of strong isospin),
    $\mathsf{S}$~(strangeness), $\mathsf{C}$~(charm),
    $\mathsf{B}$~(beauty or bottomness), $\mathsf{T}$~(topness),
    $Y$~(strong hypercharge).  Except for~$I$, which follows the
    standard quantum mechanics rules for addition of angular momenta,
    all other quantum numbers are additive.  The masses given for the
    quarks are so-called current masses and are determined in the
    $\overline{\text{MS}}$ renormalization scheme at a scale of
    $2\,\GeV$~\cite{pdg_quark_masses:2018}.  The mass values for
    electron and muon are rounded.  For the precise values of the
    lepton masses and their uncertainties
    see~\cite{Tanabashi:2018zz}.}
  \label{tab:fermions}
  \begin{tabular}{llrrrrrrrrr}
    \toprule
    Fermion & Mass & $Q$ & $I$ & $I_3$ & $\mathsf{S}$ & $\mathsf{C}$ &
    $\mathsf{B}$ & $\mathsf{T}$ & $Y$
    & Generation \\
            & [$\MeV/c^2$] & [$e$] &&&&&& \\
    \midrule
    \Pem & $0.511$ & $-1$ & 0 & 0 & 0 & 0 & 0 & 0 & 0 & I \\
    $\nu_e$ &      & 0    & 0 & 0 & 0 & 0 & 0 & 0 & 0 & \\
    \Pgmm & $105.658$ & $-1$ & 0 & 0 & 0 & 0 & 0 & 0 & 0 & II \\
    $\nu_{\Pgm}$ &      & 0    & 0 & 0 & 0 & 0 & 0 & 0 & 0 & \\
    $\tau^-$ & $1776.86$ & $-1$ & 0 & 0 & 0 & 0 & 0 & 0 & 0 & III \\
    $\nu_{\Pgt}$ &      & 0    & 0 & 0 & 0 & 0 & 0 & 0 & 0 & \\
    \midrule
    $\Pqu$     & $2.16^{+0.49}_{-0.26}$ & $+\frac{2}{3}$ & $\frac{1}{2}$ & $+\frac{1}{2}$
                                     &0&0&0&0& $\frac{1}{3}$ & I \\
    $\Pqd$     & $4.67^{+0.48}_{-0.17}$ & $-\frac{1}{3}$ & $\frac{1}{2}$ & $-\frac{1}{2}$
                                     &0&0&0&0& $\frac{1}{3}$ & \\
    $\Pqc$     & $(1.27\pm 0.02)\EE{3}$ & $+\frac{2}{3}$ &0& 0 &0&1&0&0&0 & II \\
    $\Pqs$     & $93^{+11}_{-5}$ & $-\frac{1}{3}$ &0& 0 &$-1$&0&0&0&$-\frac{2}{3}$ & \\
    $\Pqt$     & $(172.9\pm0.4)\EE{3}$ & $+\frac{2}{3}$ &0& 0
                                     &0&0&0&1&0 & III \\
    $\Pqb$     & $(4.18^{+0.03}_{-0.02})\EE{3}$ & $-\frac{1}{3}$ &0& 0
                                     &0&0&$-1$&0&0 & \\
    \bottomrule
  \end{tabular}
\end{table}

In the standard model of particle physics, the lepton number, \ie the
number of leptons minus the number of antileptons,
$N_l - N_{\overline{l}}$, is a strictly conserved quantity.
Similarly, the baryon number~$\mathcal{B}$ is strictly conserved.
Quarks are assigned $\mathcal{B} = +1/3$, antiquarks
$\mathcal{B} = -1/3$, while leptons have $\mathcal{B} = 0$.  By
convention, the non-zero flavor quantum number of a quark ($I_3$,
$\mathsf{S}$, $\mathsf{C}$, $\mathsf{B}$, or $\mathsf{T}$) carries the
same sign as its charge.\footnote{We use sans serif fonts for the
  flavor quantum numbers~$\mathsf{S}$, $\mathsf{C}$, $\mathsf{B}$,
  and~$\mathsf{T}$ in order to distinguish them from quantities
  introduced later, \eg the charge-conjugation parity~$C$.}  The
generalized Gell-Mann--Nishijima formula relates the charge (in units
of the elementary charge~$e$) to the other additive quantum numbers of
the quarks,
\begin{equation}
  \label{eq:intro.gmn}
  Q = I_3 + \frac{\mathcal{B} + \mathsf{S} + \mathsf{C} + \mathsf{B} +
    \mathsf{T}}{2}\eqPunctSpacing.
\end{equation}
Flavor quantum numbers are conserved in strong interactions, which is
reflected in the conservation of the strong
hypercharge~$Y$,\footnote{There are different definitions of the
  hypercharge, \eg
  $Y = \mathcal{B} + \mathsf{S} \pm \mathsf{C} + \mathsf{B} \pm
  \mathsf{T}$, such that all quarks have fractional~$Y$, which causes
  the hypercharge of all hadrons to be integer.  Our definition
 follows~\cite{pdg_quark_model:2018}, which defines~$Y$ in
  SU(3)$_{\text{flavor}}$ from the generator~$\lambda_8$, \ie
  $Y = 1 / \sqrt{3} \lambda_8$.  The corresponding definition for SU(4)$_{\text{flavor}}$ centers the weight
  diagrams around zero also for
  $\mathsf{C} \neq 0$.} defined as
\begin{equation}
  \label{eq:intro.Y}
  Y = \mathcal{B} + \mathsf{S} - \frac{\mathsf{C}}{3} +
  \frac{\mathsf{B}}{3} - \frac{\mathsf{T}}{3}\eqPunctSpacing.
\end{equation}

In quantum field theory, the interactions between fermions are
described by the exchange of virtual particles of rest mass~$m$,
so-called gauge bosons. The exchanged particle is assigned a
4-momentum~$q$, which is the difference between the 4-momenta of the
particles entering and leaving the interaction vertex.
Since in general $q^2 \neq m^2$ for virtual particles, they are
commonly referred to as \textquote{not being on their mass shell},
which means that they do not obey the relativistic relation between
energy~$E$, rest mass~$m$, and three-momentum~$\bvec{p}$,\footnote{Unless
  explicitly stated, we use natural units, \ie $\hbar = c = 1$.}
\begin{equation}
  \label{eq:energy--momentum}
  E^2 = m^2 + \bvec{p}^2\eqPunctSpacing.
\end{equation}
The
virtual particles couple to a conserved property generally called
charge.  The electromagnetic interaction between particles carrying
electric charge, for example, is described in Quantum Electrodynamics
(QED) by the exchange of virtual photons, massless electrically
neutral spin-1 particles.  The range of the force is given by the
Compton wavelength of the virtual particle:
\begin{equation}
  \label{eq:compton_wavelength}
  \lambda = \frac{1}{m}\eqPunctSpacing.
\end{equation}
Yukawa postulated that the short-range interaction between nucleons is
due to the exchange of massive integer-spin particles, which he
identified as pions.

In Quantum Chromodynamics (QCD), the field theory of strong
interaction, the interaction between quarks is mediated by the
exchange of massless spin-1 particles called gluons, which couple to
the strong charge or color.  There are three different strong charges,
or colors, termed red, green, and blue.  In contrast to photons,
however, the gluons also carry charge, \ie color, themselves and hence
can also couple to themselves.\footnote{This is sometimes referred to
  as Yang--Mills theory, a gauge theory with non-Abelian symmetry.}  It
is this self-interaction of the gluons which generates a wealth of new
phenomena for strongly interacting particles, the two most important
being confinement and asymptotic freedom of quarks, which will be
explained in the following.  At large momentum transfers,
corresponding to distance scales much smaller than the size of a
nucleon of about $1\,\fm$, the quarks inside hadrons behave
essentially as free particles.  This asymptotic freedom allows the
interaction to be treated perturbatively with a small coupling
parameter~$\alpha_{\text{s}}$ as series expansion parameter.  It is in
this regime, where predictions from QCD have been verified to very
high precision. For larger distance scales approaching
\orderOf{1\,\fm}, the cloud of virtual gluons and \qqbar~pairs around
quarks modifies their effective strong charge, similarly to the
screening of the electric charge by the cloud of virtual photons
fluctuating into $e^+e^-$~pairs.  Due to the self-interaction of
gluons, however, there is an anti-screening of the strong charge,
leading to an increased coupling strength at larger distances.  This
can be modeled by a potential between stationary quarks that consists
of a Coulomb-like term for small distances~$r$ and increases linearly
for larger distances, \ie
\begin{equation}
  \label{eq:intro.potential}
  V(r) = -\frac{4}{3}\, \frac{\alpha_{\text{s}}}{r} + k\, r\eqPunctSpacing,
\end{equation}
resembling a string with a constant string tension~$k$ (see
\cref{sec:pheno.trajectories}).  With increasing distance, the energy
density stored in the string becomes high enough to create a new
quark--antiquark pair from the vacuum.  As a consequence, free quarks
are not observed in nature.\footnote{By the same argument, there are
  also no free gluons.  As the quarks, they are confined into hadrons.
  Hence the range of the strong interaction is limited to
  \orderOf{1\,\fm} although the gluons are massless, \ie
  \cref{eq:compton_wavelength} does not apply.}  It is one of the
major questions of modern physics, how this phenomenon of confinement
emerges from the underlying theory of QCD.

In order to account for confinement, hadrons are postulated to be
color-neutral objects, or, more precisely, color-singlets, which are
invariant under rotations in color space.  The simplest possible
configurations to obtain color-neutral objects are combinations of
three quarks of different color each, or of a quark and an antiquark
of a given color and the corresponding anti-color (or complementary
color).  Experiments on deep-inelastic scattering of leptons (\Pem,
$\Pgm$, or $\Pgn$) off protons and neutrons established that the nucleon
is a complicated dynamic system composed of three valence quarks,
which account for the quantum numbers of the nucleon, and an infinite
sea of virtual quarks, antiquarks, and gluons.  Almost $50\%$ of the
total momentum of the nucleon is carried by the gluons and only about
$30\%$ of the spin~$1/2$ of the nucleon is carried by the
quarks~\cite{Adolph:2015saz}.  While the hard scattering processes
between the lepton and the quark can be precisely calculated in QCD,
the parton distribution functions (PDF), that describe the momentum
and spin degrees of freedom of quarks and gluons inside the nucleon,
cannot be calculated from first principles.  The large effective
coupling at distance scales \orderOf{1\,\fm} corresponding to hadrons
makes it impossible to solve QCD in this regime by a series expansion
in the effective coupling parameter.  While a wealth of experimental
data and phenomenological models on PDFs is available for nucleons,
very little is known about the quark and gluon structure of the
lightest mesons.

Similarly, the excitation spectrum of strongly interacting particles
is still far from being understood, despite a 50-year long history of
world-wide experimental and theoretical efforts.  Various models and
effective theories have been developed to describe composite systems
of quarks and gluons and their complex excitation spectrum.  Despite
its conceptual simplicity, neglecting the afore-mentioned dynamic
internal structure, the constituent-quark model, which is discussed in
more detail in \cref{sec:pheno.qm}, can explain many properties of
observed hadrons.  Based on the concept of flavor symmetry, mesons are
described in the quark model as bound states of an effective
quark--antiquark system, while baryons are composed of three effective
quarks.
QCD, however, allows for a much larger variety of color-singlet bound
systems, including multi-quark states, molecule-like systems, or,
owing to the fact that also gluons carry the charge of the strong
interaction, systems where gluons explicitly contribute to the quantum
numbers of the hadronic state.  It is one of the most important
questions in present-day hadronic physics whether such states exist or
not.

Recently, precise ab-initio calculations of the properties of
ground-state baryons and mesons were performed by solving QCD on a
discrete space--time lattice (lattice QCD) (see
\eg~\refCite{Kronfeld:2012uk} for a review).  First qualitative
results have been obtained for excited states.  It is still a long
way, however, until lattice QCD will yield precision calculations for
excited states decaying into multi-particle final states.

On the experimental side, a plethora of excited hadronic states has
been observed at proton and electron accelerators.  Only in the last
15~years with the advent of high-intensity particle beams, modern
particle detectors, and readout technology, it became possible to
collect sufficiently large data samples both in the light and
heavy-quark sectors to achieve the statistical precision to observe
first signs of states not fitting into the \qqbar or $qqq$ model of
mesons and baryons.  At the same time, many states predicted by the
quark model are still unobserved or need confirmation.  The COMPASS
experiment is one of the key-players in the light-quark sector.  This
review will summarize the current status of results from this
experiment and put them into perspective \wrt the ultimate goal of
explaining the structure of hadrons in terms of quarks and gluons, the
fundamental building blocks of Quantum Chromodynamics.

The paper is structured as follows: After an introduction into the
strong interaction and its embedding in the Standard Model of particle
physics in \cref{sec:intro}, we review the phenomenology of mesons in
\cref{sec:pheno}. Starting from the quark model and the concept of
flavor symmetry, we discuss the expected multiplet structure of mesons
and the excitation spectrum of light mesons and confront it with
experimental results.  The experimentally observed approximate linear
relation (\textquote{Regge trajectories}) between spin and
mass-squared of mesons is shown to arise from a simple string model.
Finally, we define the notion of exotic mesons, and show how lattice
QCD may contribute to resolving some of the open
questions. \Cref{sec:theory} gives a general and pedagogical overview
over some of the theoretical concepts employed to analyze scattering
experiments using the $S$-matrix approach. We discuss the kinematics
for general two-body reactions and introduce the partial-wave
expansion for spinless particles. Resonances are introduced both in
the Breit--Wigner approximation and as poles of the analytic $S$-matrix
in the complex plane. The consequences of unitarity of the $S$-matrix
are discussed. Models which satisfy unitarity of the partial-wave
amplitudes, like the $K$-matrix or the $N$-over-$D$ method are briefly
introduced. The consequences of unitarity on production reactions and
final-state interactions are discussed. We continue by explaining some
of the important concepts underlying Regge theory, which lead to an
asymptotic description of scattering processes at high energies. The
section ends with a brief discussion of chiral perturbation theory,
which is the low-energy limit of QCD.  \Cref{sec:reactions} describes
the dominant reaction mechanisms in the COMPASS kinematic domain, \ie\
at high energies of the incoming particle beam, before \cref{sec:exp}
introduces the COMPASS experimental apparatus used for hadron
spectroscopy.  Based on the more general account in \cref{sec:theory},
\cref{sec:pwa} provides a detailed derivation of the partial-wave
analysis (PWA) formalism used to decompose the measured data into
partial-wave amplitudes with a given spin-parity assignment. Special
attention is paid to new developments which are expected to reduce the
model-dependence of the analysis, like the freed-isobar PWA
(\cref{sec:pwa_cells:freed_isobar}), or help to better constrain the
fit by fulfilling some of the fundamental properties of the $S$-matrix
(\cref{sec:pwa.unitary_model}). In \cref{sec:results} we finally
present the most important results from the analysis of diffractive
dissociation reactions, while in \cref{sec:results_3pic_primakoff} we
discuss the results from Primakoff reactions. The review closes with
conclusions and an outlook in \cref{sec:conclusions_outlook}. In
several appendices we provide more detailed mathematical formulas,
which are useful for PWA practitioners.
\clearpage{}%
\clearpage{}%
\section{Phenomenology of Mesons}
\label{sec:pheno}

In the following section, we review some phenomenological aspects of
mesons. The quark model by Gell-Mann and Zweig explains the multiplet
structure of observed mesons when classified by total spin, isospin
and strong hypercharge, by identifying them as composite objects of a
quark and an antiquark (see \cref{sec:pheno.qm.quantum-numbers}). The
excitation spectrum can be calculated in the quark model by
parameterizing the quark--antiquark interaction by effective
Hamiltonians, which include a Coulomb-like attractive potential that
dominates at small separations, and a linearly increasing potential
that dominates for larger distances.  We confront a recent
relativistic quark-model calculation with present-day experimental
data (see \cref{sec:pheno.qm.spectrum}).  A linear confining potential
between more fundamental constituents is also needed to explain the
approximate linear relation between the total spin and the squared
mass for certain groups of mesons (\textquote{Regge trajectories}),
which had been noticed even before the advent of the quark model in
the early 1960s.  A simple model of a rotating string first proposed
by Nambu is presented as illustration in
\cref{sec:pheno.trajectories}.  Regge theory, which will be summarized
in \cref{sec:regge}, links the observed trajectories to the forces
between quarks and to the high-energy behavior of the scattering
amplitude. Regge trajectories can help to assign experimentally
observed resonances to states predicted by quark models. States not
fitting the model predictions are generally termed exotic. We discuss
candidates for light exotic states (see \cref{sec:pheno.exotics}) and
show that lattice QCD, providing hints towards the structure of the
meson spectrum from first principles, also predicts the existence of a
group of exotic states called hybrids (see \cref{sec:pheno.lattice}).

\subsection{The Quark Model}
\label{sec:pheno.qm}

In general, the term quark model refers to a class of models, which
describe the properties of hadrons by effective degrees of freedom
carried by valence quarks.  For baryons, the number of valence quarks
is given by the net number of quarks of a given flavor,
$N_{\Pq} - N_{\Paq}$.  As their counterparts, the valence electrons in
atoms, valence quarks determine the quantum numbers, the excitation
spectrum, and the interactions of hadrons.  While the quark structure
and the quantum numbers of hadrons in quark models are determined
based on symmetry considerations and static quark content, the
determination of the excitation spectrum, \ie the masses and widths of
states, requires the inclusion of the dynamics of the interaction,
usually based on effective Hamiltonians.

\subsubsection{Quantum Numbers of Mesons}
\label{sec:pheno.qm.quantum-numbers}

All additive quantum numbers mentioned in \cref{tab:fermions} also
apply to bound systems of quarks.  Mesons carry baryon number
$\mathcal{B} = 0$, \ie in their simplest form they are composed of a
quark and an antiquark of the same (flavorless\footnote{Here,~$\Pqu$
  and~$\Pqd$ are considered the same flavor because of isospin
  symmetry.} or hidden-flavor mesons) or a different flavor (open
flavor mesons).  Containing only additive quantum numbers, both the
Gell-Mann--Nishijima formula \cref{eq:intro.gmn} and the hypercharge
\cref{eq:intro.Y} apply to individual quarks as well as bound systems
of quarks.  In addition to the baryon number and the flavor quantum
numbers, mesons are characterized by the quantum numbers
\JPC,\footnote{They are related to the Poincar\'{e} symmetry of
  relativistic field theories.}  with the total angular momentum
$J = 0, 1, 2\ldots$, and the parity $P = \pm 1$, specifying the
symmetry of the wave function under reflection through a point in
space.  Neutral flavorless mesons are eigenstates of the
particle--antiparticle conjugation operator and thus have a defined
charge conjugation parity $C = \pm 1$.  The concept of $C$-parity can
be extended to include the charged combinations $\Pqu\Paqd$ and
$\Pqd\Paqu$ by introducing the $G$~parity, defined as charge
conjugation followed by a \SI{180}{\degree} rotation in isospin space
about the $y$~axis.  Flavored mesons have one or two of the quantum
numbers strangeness~$\mathsf{S}$, charm~$\mathsf{C}$, or
bottomness~$\mathsf{B}$ different from zero.  All the quantum numbers
mentioned above are conserved in strong interactions.

\begin{figure}[tbp]
  \centering
  \hfill%
  \subfloat[]{%
    \includegraphics[height=0.3\columnwidth]{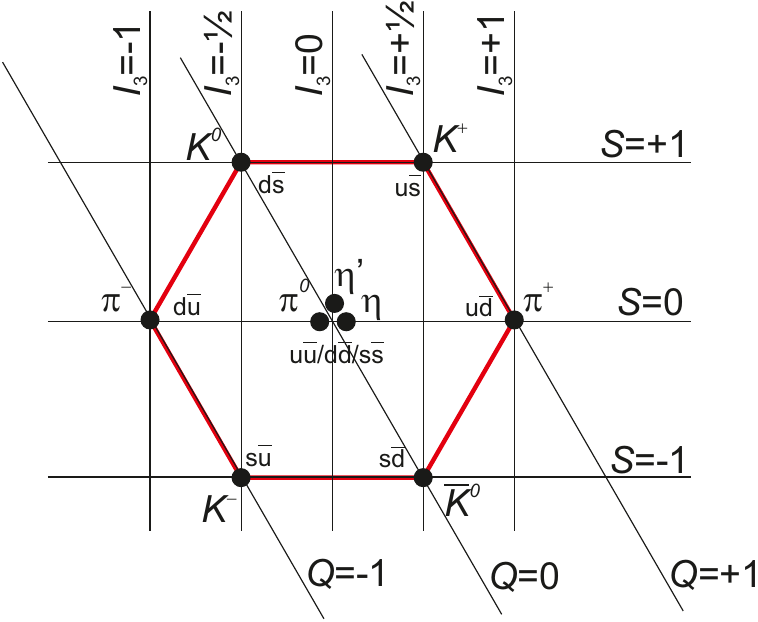}%
    \label{fig:theory.nonet.0-+}%
  }%
  \hfill%
  \subfloat[]{
    \includegraphics[height=0.3\columnwidth]{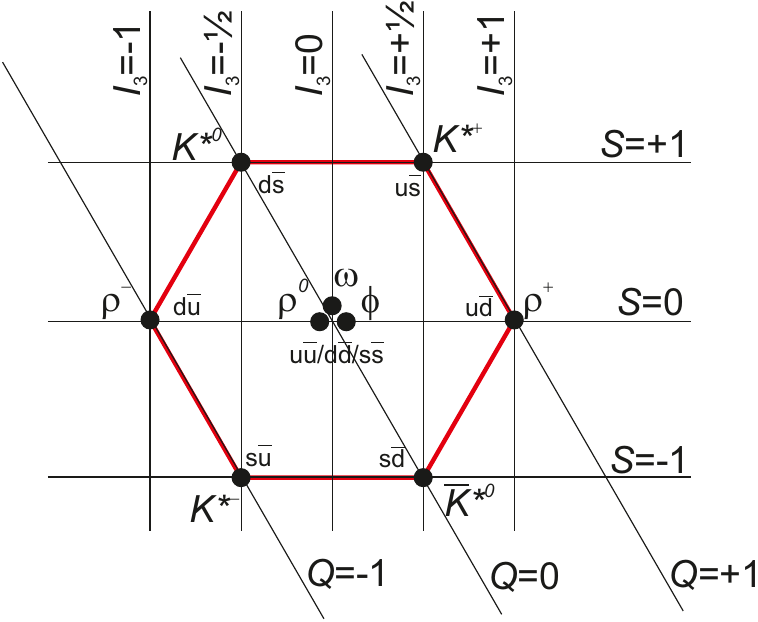}%
    \label{fig:theory.nonet.1--}%
  }%
  \hfill\null%
  \caption{Nonets of ground-state
    \subfloatLabel{fig:theory.nonet.0-+}~pseudoscalar,
    \subfloatLabel{fig:theory.nonet.1--}~vector mesons, classified by
    the $z$~component of the strong isospin~$I_3$ and the
    strangeness~$\mathsf{S}$.}
  \label{fig:theory.nonets}
\end{figure}

Light mesons, as well as light baryons, with a given~$J^P$ can be
classified in multiplets according to their isospin and hypercharge,
or, equivalently, the strangeness~$\mathsf{S}$ (see
\cref{eq:intro.Y}).  \Cref{fig:theory.nonet.0-+,fig:theory.nonet.1--}
show the corresponding nonets of the lightest pseudoscalar
($J^P = 0^-$) and vector ($J^P = 1^-$) mesons, respectively.
This systematics is explained in the quark model of
Gell-Mann~\cite{GellMann:1964nj} and Zweig~\cite{Zweig:1981pd} by
extending the SU(2)$_{\text{iso}}$ symmetry group of the strong isospin
to SU(3)$_{\text{flavor}}$, with the triplet $(\Pqu, \Pqd, \Pqs)$ of
light quarks being the fundamental representation, and combining it
with the symmetry groups SU(2)$_{\text{spin}}$ for spin and
SO(3)$_{\text{orb}}$ for orbital angular momentum between the quarks:
\begin{equation}
  \label{eq:mesons.symmetry}
  \text{SU(3)}_{\text{flavor}}
  \otimes \text{SU(2)}_{\text{spin}}
  \otimes \text{SO(3)}_{\text{orb}}.
\end{equation}
The flavor symmetry is of course only approximate, since
$m_{\Pqs} > m_{\Pqu,\Pqd}$.  In the quark model, mesons are described as bound
states of a quark~$\Pq$ and an antiquark~\Paqpr, while baryons are
composed of three quarks, both forming singlets in color space
SU(3)$_{\text{color}}$.

\begin{figure}[tbp]
  \centering
  \includegraphics[width=0.2\columnwidth]{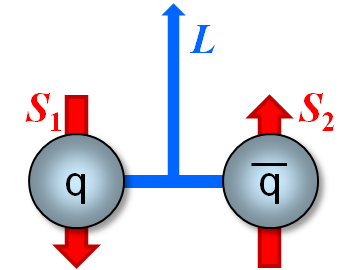}
  \caption{A meson in the quark model.}
  \label{fig:theory.meson}
\end{figure}

In the $L$-$S$ coupling scheme,\footnote{The $L$-$S$ coupling scheme
  is a good basis to characterize systems consisting of only light or
  heavy quarks.  For heavy-light systems $Q\Paq$, heavy-quark symmetry
  implies that in the limit $m_Q \to \infty$ the spin of the heavy
  quark and the total angular momentum~$j$ of the light quark are
  separately conserved~\cite{Isgur:1991wq,Manohar:2000dt},
  corresponding to the $j$-$j$ coupling scheme in atomic physics.}
the spins of the \qqbarpr~pair couple to the total intrinsic spin
$S = 0, 1$.  The total spin~$J$ of the system is the vector sum of the
total intrinsic spin~$S$ and the relative orbital angular momentum~$L$
with quantum number $\abs{L - S} \leq J \leq (L + S)$ (\confer\
\cref{fig:theory.meson}).

The meson nonets for a given \JP combination are obtained by combining
the SU(3)$_{\text{flavor}}$ quark triplet with the antiquark triplet,
resulting in nine possible combinations, transforming like an octet
and a singlet in SU(3)$_{\text{flavor}}$ space:
\begin{equation}
  \label{eq:theory.qqbar_combinations}
  3 \otimes \overline{3} = 8 \oplus 1\eqPunctSpacing,
\end{equation}
where the dimensions of the multiplets are used to label their
irreducible representations.  The flavor wave functions of the
isoscalar (\ie $I = 0$) flavor singlet and octet combinations are
\begin{equation}
  \label{eq:theory.qqbar_isosinglet}
  \psi_1 = \sqrt{\frac{1}{3}} \left( \Pqu\Paqu + \Pqd\Paqd + \Pqs\Paqs \right)
  \quad\text{and}\quad
  \psi_8 = \sqrt{\frac{1}{6}} \left( \Pqu\Paqu + \Pqd\Paqd - 2\Pqs\Paqs \right)\eqPunctSpacing,
\end{equation}
respectively.  The flavor wave functions of the isovector (\ie
$I = 1$) triplet are\footnote{The minus sign that appears in the
  $I_3 = 0$ and~$-1$ components in \cref{eq:theory.qqbar_isovector} is
  a convention, which allows the SU(2)$_{\text{iso}}$ quark doublet and
  the antiquark doublet to transform in exactly the same way under
  isospin rotations or ladder operations.  Such a representation is,
  in general, not possible in SU(3)$_{\text{flavor}}$.}
\begin{equation}
  \label{eq:theory.qqbar_isovector}
  \left( \Pqd\Paqu,
    \sqrt{\frac{1}{2}} \left( \Pqu\Paqu - \Pqd\Paqd \right), -\Pqu\Paqd
    \right)\eqPunctSpacing.
\end{equation}
The states with open strangeness are~$\Pqd\Paqs$, $\Pqu\Paqs$,
$\Pqs\Paqu$, and~$\Pqs\Paqd$.  The SU(3)$_{\text{flavor}}$ singlet and
octet states with $I = 0$ generally mix to produce the physical
mesons~$A$ and~$B$, \ie
\begin{align}
  \label{eq:meson.mixing}
  \begin{pmatrix}
    A \\ B
  \end{pmatrix}
  =
  \begin{pmatrix*}[r]
    \cos\theta & -\sin\theta \\
    \sin\theta & \cos\theta
  \end{pmatrix*}
  \begin{pmatrix}
    \psi_8 \\ \psi_1
  \end{pmatrix}\eqPunctSpacing,
\end{align}
with a mixing angle~$\theta_\text{P}$ between \SI{-10}{\degree} and
\SI{-20}{\degree} for the pseudoscalar ground states $A = \eta$ and
$B = \eta^\prime$, and $\theta_\text{V} = \SI{36.4}{\degree}$ for the
vector ground states $A = \phi$ and
$B = \omega$~\cite{pdg_quark_model:2018}.  The latter are close to ideal
mixing, such that $\phi \approx \Pqs\Paqs$.

Introducing the heavier charm quark into the model extends the
symmetry by one dimension to SU(4)$_{\text{flavor}}$, which, however, is
strongly broken due to the much heavier $\Pqc$~quark.

In the quark model, the quantum numbers~$P$, $C$, and~$G$ of the mesons
are given by
\begin{equation}
  \label{eq:meson.qn}
  P = (-1)^{L + 1},
  \quad
  C = (-1)^{L + S},
  \quad\text{and}\quad
  G = (-1)^{L + S + I}.
\end{equation}
This gives rise to meson states with quantum numbers
\begin{equation}
  \label{eq:meson.qqbar_qn}
  \JPC = 0^{-+},\, 0^{++},\, 1^{--},\, 1^{+-},\,
  1^{++},\, 2^{--},\, 2^{-+},\, 2^{++},\ldots\eqPunctSpacing.
\end{equation}
States with $P = (-1)^J$ are called \emph{natural-parity} states,
those with $P = (-1)^{J+1}$ \emph{unnatural parity} states.  The
\emph{naturality} of a state is defined by
\begin{equation}
  \label{eq:theory.mesons.naturality}
  \natur \equiv P (-1)^J= \pm 1\eqPunctSpacing,
\end{equation}
and is~$+1$ for the natural-parity series states~$0^+$, $1^-$,
$2^+,\ldots$, and~$-1$ for the unnatural-parity series states~$0^-$,
$1^+$, $2^-,\ldots$.  The spectroscopic notation $n^{2S + 1}L_J$ is
used to fully designate a given \qqbarpr state, with~$n = n_r + 1 = 1,\,2,\ldots$
numbering radial excitations ($n_r$ is the number of radial
nodes).\footnote{Note that the convention used in the quark model is
  different from atomic physics, where the principal quantum number
  $n = n_r + L + 1$ with $L$ being the orbital angular momentum
  quantum number.}

The $C$~parity ($\pm 1$) is commonly used not only for neutral
flavorless mesons, which are eigenstates of the particle--antiparticle
conjugation operator, but also for the charged members of an isospin
triplet, or sometimes even for all members of an SU(3)$_{\text{flavor}}$
nonet.  In these cases, it is understood to be defined through the
neutral component of the corresponding multiplet.

\begin{table}[tbp]
  \centering
  \renewcommand{\arraystretch}{1.2}
  \caption{Naming scheme of mesons (inspired
    by~\refCite{Klempt:2007cp}).  The quoted $C$~parity is defined for
    the flavorless states of the corresponding multiplet with
    $I_3 = \mathsf{S} = \mathsf{C} = \mathsf{B} = 0$.}
  \label{tab:theory.mesons.naming}
  \begin{tabular}{l|c|cccc}
    \toprule
    & & $^{2S + 1}L_J = {^1(\text{even})_J}$ & $^1(\text{odd})_J$ &
    $^3(\text{even})_J$ & $^3(\text{odd})_J$ \\
    \qqbarpr content & $I$ &
    $\JPC = (0,2,\ldots)^{-+}$ & $(1,3,\ldots)^{+-}$ & $(1,2,\ldots)^{--}$ &
    $(0,1,\ldots)^{++}$ \\
    \midrule
    $\Pqd\Paqu$, $\Pqu\Paqu-\Pqd\Paqd$, $\Pqu\Paqd$ & 1 &
    $\HepParticle{\pi}{J}{}$ & $\HepParticle{b}{J}{}$ &
    $\HepParticle{\rho}{J}{}$ & $\HepParticle{a}{J}{}$ \\
    $\Pqu\Paqu+\Pqd\Paqd$, $\Pqs\Paqs$ & 0 &
    $\HepParticle{\eta}{J}{},\HepParticle{\eta}{J}{\prime}$ &
    $\HepParticle{h}{J}{},\HepParticle{h}{J}{\prime}$ &
    $\HepParticle{\omega}{J}{},\HepParticle{\phi}{J}{}$ &
    $\HepParticle{f}{J}{},\HepParticle{f}{J}{\prime}$ \\
    $\Pqu\Paqs$, $\Pqd\Paqs$ & $\frac{1}{2}$ & $\HepParticle{K}{J}{}$ &
    $\HepParticle{K}{J}{}$ &
    $\HepParticle{K}{J}{\ast},\HepParticle{K}{J}{}$ &
    $\HepParticle{K}{J}{\ast},\HepParticle{K}{J}{}$ \\
    $\Pqc\Paqc$ & 0 & $\Pcgh$ & \Pch & $\Pgy$ & $\Pcgc$ \\
    $\Pqb\Paqb$ & 0 & $\Pbgh$ & \Pbh & \PgU & $\Pbgc$ \\
    $\Pqc\Paqu$, $\Pqc\Paqd$ & $\frac{1}{2}$ & $\HepParticle{D}{J}{}$
    & $\HepParticle{D}{J}{}$ &
    \HepParticle{D}{J}{\ast},\HepParticle{D}{J}{}
    & \HepParticle{D}{J}{\ast},\HepParticle{D}{J}{} \\
    $\Pqc\Paqs$ & 0 & $\HepParticle{D}{sJ}{}$
    & $\HepParticle{D}{sJ}{}$ &
    \HepParticle{D}{sJ}{\ast},\HepParticle{D}{sJ}{}
    & \HepParticle{D}{sJ}{\ast},\HepParticle{D}{sJ}{} \\
    $\Pqu\Paqb$, $\Pqd\Paqb$ & $\frac{1}{2}$ & $\HepParticle{B}{J}{}$
    & $\HepParticle{B}{J}{}$ &
    \HepParticle{B}{J}{\ast},\HepParticle{B}{J}{}
    & \HepParticle{B}{J}{\ast},\HepParticle{B}{J}{} \\
    $\Pqs\Paqb$ & 0 & $\HepParticle{B}{sJ}{}$
    & $\HepParticle{B}{sJ}{}$ &
    \HepParticle{B}{sJ}{\ast},\HepParticle{B}{sJ}{}
    & \HepParticle{B}{sJ}{\ast},\HepParticle{B}{sJ}{} \\
    $\Pqc\Paqb$ & 0 & $\HepParticle{B}{cJ}{}$
    & $\HepParticle{B}{cJ}{}$ &
    \HepParticle{B}{cJ}{\ast},\HepParticle{B}{cJ}{}
    & \HepParticle{B}{cJ}{\ast},\HepParticle{B}{cJ}{} \\
    \bottomrule
  \end{tabular}
\end{table}

The naming scheme for mesons reflects their quantum numbers, and is
given in \cref{tab:theory.mesons.naming}.
The total spin~$J$ is added as a subscript except for the pseudoscalar
($\JPC = 0^{-+}$) and vector ($1^{--}$) mesons.  Light isoscalar
states, which generally are a mixture of $(\Pqu\Paqu + \Pqd\Paqd)$ and
$\Pqs\Paqs$ (see \cref{eq:meson.mixing}), may be distinguished by
adding a prime ($^\prime$) or by the pair ($\Pgo$, $\Pgf$).  For
mesons with open flavor, the natural-parity series states are tagged
by a star, while the unnatural parity series states do not carry a
star.  Particle states have positive flavor quantum
number~$\mathsf{S}$, $\mathsf{C}$, or~$\mathsf{B}$ (\eg~$\PKz$,
$\PKp$, $\PDz$, $\PDp$), while antiparticle states have negative ones.
Radial excitations of the \qqbarpr~system carry the same name as the
corresponding ground state.

\subsubsection{The Meson Spectrum}
\label{sec:pheno.qm.spectrum}

There is a large variety of different models to calculate the spectrum
of mesons~\cite{Godfrey:1985xj}.  Typically the dynamics of the
quark--antiquark interaction is parameterized by effective Hamiltonians
with the following ingredients:
\begin{itemize}
\item effective masses of constituent, \ie \textquote{dressed},
  valence quarks of $m_{\Pqu} = m_{\Pqd} \approx 0.3\,\GeV$ and
  $m_{\Pqs} \approx 0.5\,\GeV$,
\item an attractive potential $\propto 1/r$ due to one-gluon
  exchange, $r$~being the interquark distance,
\item a long-range confining potential $\propto r$,
\item a chromomagnetic spin--spin interaction
  $\propto \bvec{S}_{1} \cdot \bvec{S}_{2}$ (\confer\
  \cref{fig:theory.meson}).
\end{itemize}

\begin{figure}[p]
  \centering
  \includegraphics[angle=270,width=0.85\columnwidth]{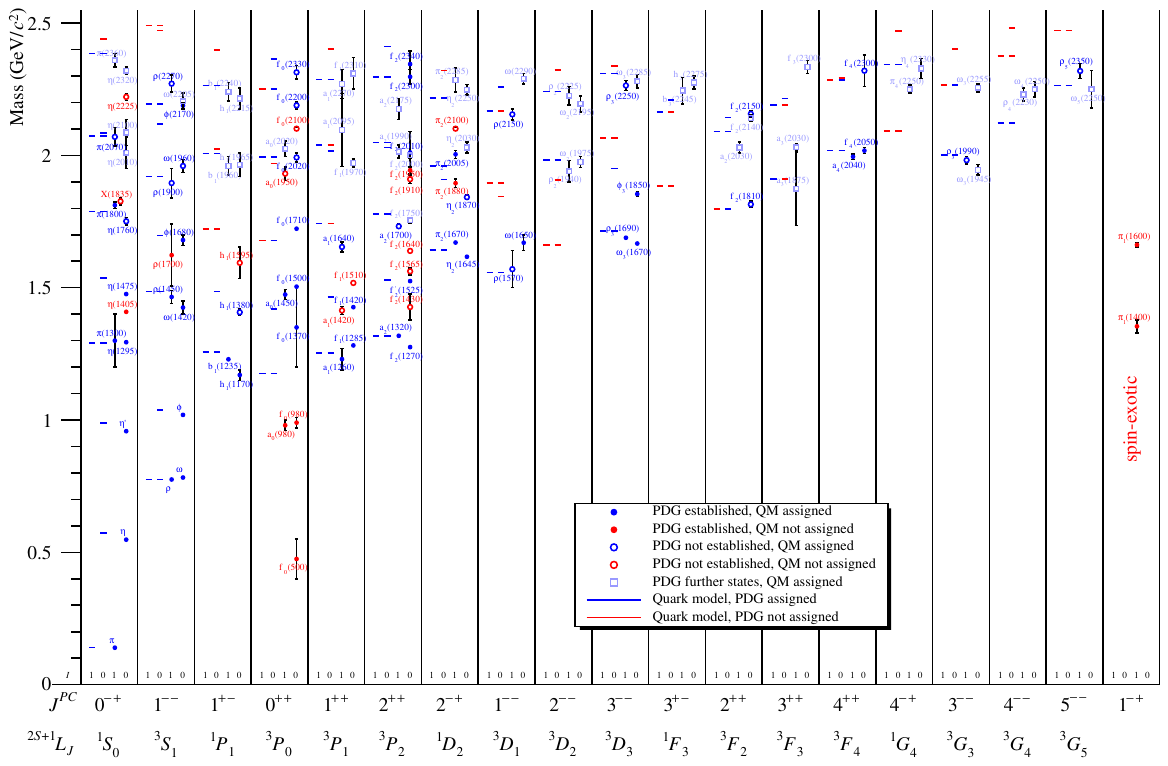}
  \caption{Mass spectrum of light flavorless mesons sorted by their
    \JPC quantum numbers.  Data points indicate the masses of
    experimentally observed mesons with their corresponding
    uncertainties~\cite{Tanabashi:2018zz}. The PDG names are given
    immediately above or below the corresponding data
    point. Horizontal lines show the results for \qqbarpr~states from
    a relativistic quark model~\cite{Ebert:2009ub}, with the
    spectroscopic notation given below the \JPC values.  See text for
    detailed explanation.}
  \label{fig:light_flavorless_spectrum}
\end{figure}

In \cref{fig:light_flavorless_spectrum} the spectrum of light
flavorless mesons calculated in a relativistic quark model
(RQM)~\cite{Ebert:2009ub} is compared with the masses of the
experimentally observed mesons~\cite{Tanabashi:2018zz}. The RQM
spectrum is displayed by horizontal lines.  The energy levels for the
non-strange isotriplet and isosinglet wave functions are displayed by
two lines next to each other, with $I = 1$ shifted to the left and
$I = 0$ to the right, while the pure $\Pqs\Paqs$~combinations with
$I = 0$ are indicated by single lines shifted to the right.  Mixing of
states with $I = 0$ is neglected in this model, except for the
pseudoscalar ground states, where the $\eta$-$\eta^\prime$ mixing
scheme of \refCite{Feldmann:1998vh} is applied.  The corresponding
spectroscopic symbols are given below the \JPC~values.  Experimental
results for all mesons below a mass of $2.5\,\GeV$ appearing in the
full listing of the Particle Data Group (PDG)~\cite{Tanabashi:2018zz}
are displayed as data points, with the uncertainties on the masses as
error bars.  Full dots denote established states, open circles
non-established states, which need confirmation, while open squares
indicate states in the \textquote{Further States} listing of the PDG.
As for the RQM states, the experimental states with $I = 1$ are
shifted to the left, the ones with $I = 0$ to the right.  The
assignment to RQM states follows~\cite{Ebert:2009ub}, except in a few
cases where the PDG entries changed in the meantime. States appearing
in the \textquote{Further States} listing of the PDG are included only
if they were explicitly assigned to RQM states.  Blue data points
indicate states which have been assigned to RQM states, while red data
points designate states which have not been assigned, and are thus
considered potentially supernumerary.  Blue lines indicate assignment,
while red lines designate states, which are expected from the RQM, but
where a corresponding state has not yet been found by experiments.

Despite its conceptual simplicity, the quark model is quite successful
in explaining the overall structure of the observed spectrum of
mesons, especially for the ground states of pseudoscalar and vector
mesons.  However, many radial and orbital excitations of
quark--antiquark systems predicted by the model have not yet been
observed experimentally (red lines).  On the experimental side, there
are noticeable and often well-established states, which lack a
correspondence or a clear assignment to quark model nonets.  Prominent
cases can be found in the scalar sector with $\JPC = 0^{++}$, in the
axial-vector and tensor sectors with $\JPC = 1^{++}$ and $2^{++}$,
respectively, and in the sector with spin-exotic quantum numbers
$\JPC = 1^{-+}$, which will be discussed in more detail below.

\subsection{Regge Trajectories}
\label{sec:pheno.trajectories}

\begin{figure}[tbp]
  \centering
  \includegraphics[width=0.25\textwidth]{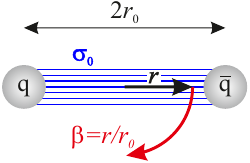}
  \caption{String model of a meson, proposed by Nambu~\cite{Nambu:1978bd} in 
    order to explain the approximate linear
    relation between spin and mass squared for groups of mesons.}
  \label{fig:string_model}
\end{figure}

In parallel to the development of the quark model, it was noted by
Chew and Frautschi~\cite{Chew:1962eu} that hadrons can be arranged in
groups, which exhibit an approximate linear dependence of the spin~$J$
on their mass~$M$ squared,
\begin{equation}
  \label{eq:regge_trajectory}
  J(M) = \alpha(0) + \alpha'\, M^2\eqPunctSpacing,
\end{equation}
with $\alpha(0)$~being the $y$-axis intercept and $\alpha'$~the slope
of the so-called Regge trajectory. This observation provided another
hint that hadrons were composite objects of more fundamental entities.
An attractive interpretation was given by Nambu in terms of a simple
string model~\cite{Nambu:1978bd}. In this model, massless quarks are
bound to the ends of a string (or flux tube) of length~$2 r_0$ with
constant energy density~$\sigma_0$ per unit length (at rest):
\begin{equation}
  \label{eq:eq:theory.string.sigma}
   \sigma_0 = \dif{E_0} / \dif{r}\eqPunctSpacing.
\end{equation}
For massless quarks, the ends of the string may move at the speed of
light with respect to the center of momentum of the string, and the
velocity of the string element at a radius~$r$ is $\beta(r) = r/r_0$
(with $c \equiv 1$), see \cref{fig:string_model} for illustration.
The energy (or mass-equivalent) stored in this string element is
\begin{equation}
  \label{eq:theory.string.de}
  \dif{E} = \gamma(r)\, \dif{E_0}
  = \gamma(r)\, \sigma_0\, \dif{r}\eqPunctSpacing,
\end{equation}
with $\gamma(r) = \left[ 1 - \beta^2(r) \right]^{-1/2}$.  The total
energy in the rotating string is then
\begin{equation}
  \label{eq:theory.string.e}
  E = 2 \sigma_0 \int_0^{r_0}\! \dif{r}\, \gamma(r)
  = \pi \sigma_0 r_0\eqPunctSpacing,
\end{equation}
\ie a potential that increases linearly with the quark
separation~$r_0$, corresponding to a total relativistic mass
\begin{equation}
  \label{eq:theory.string.m}
  M = \pi \sigma_0 r_0 \eqPunctSpacing.
\end{equation}
Similarly, the angular momentum of the rotating string element is
\begin{equation}
  \label{eq:theory.string.dj}
  \dif{J} = r\, \beta(r)\, \gamma(r)\, \dif{E_0}\eqPunctSpacing,
\end{equation}
and the total angular momentum
\begin{equation}
  \label{eq:theory.string.j}
  J = 2 \sigma_0
  \int_0^{r_0}\! \dif{r}\, r\, \beta(r)\, \gamma(r)
  = \frac{M^2}{2 \pi \sigma_0}\eqPunctSpacing.
\end{equation}
The slope of the trajectory in this model is therefore
\begin{equation}
  \label{eq:theory.string.slope}
  \alpha' = \frac{1}{2\pi \sigma_0}\eqPunctSpacing.
\end{equation}

\begin{figure}[tbp]
  \centering
  \includegraphics[width=0.75\textwidth]{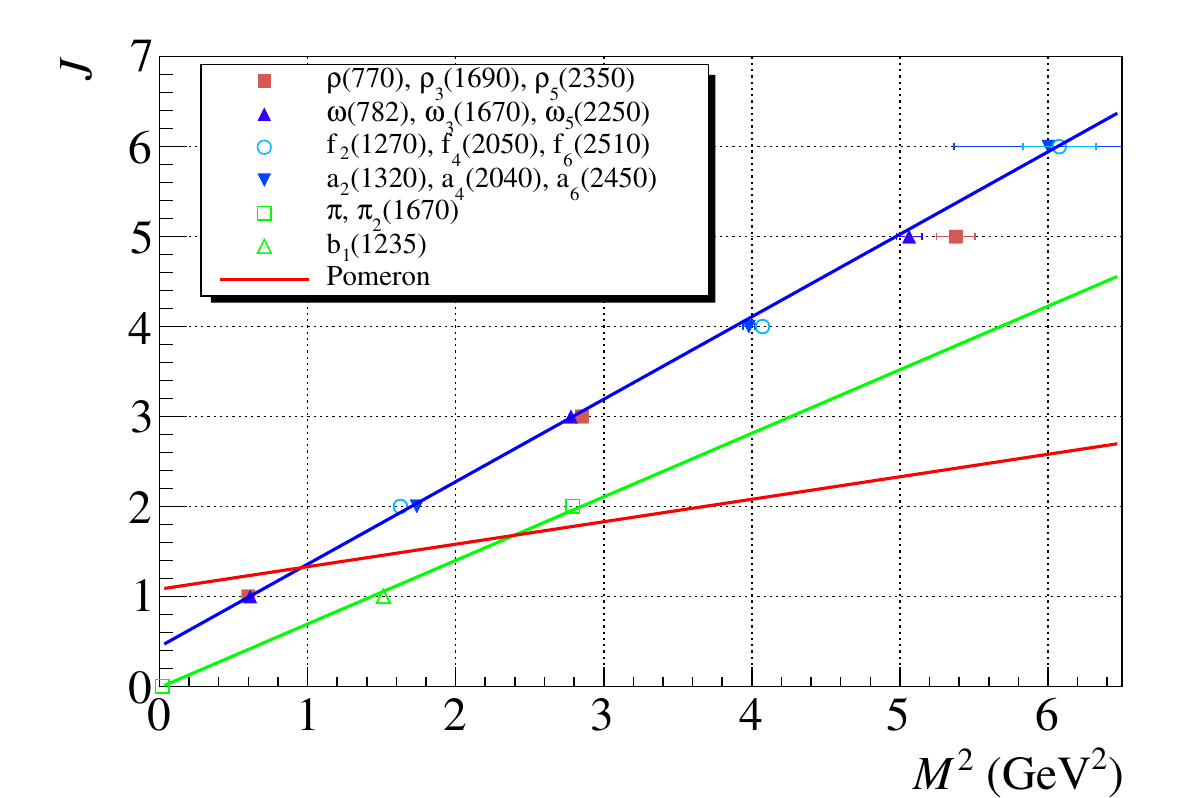}
  \caption{Chew--Frautschi plot exhibiting the linear dependence of
    $\alpha(t) = J$ on the squared mass~$M^2$ of
    radial-ground-state mesons.}
  \label{fig:chew_frautschi}
\end{figure}

\Cref{fig:chew_frautschi} shows a so-called Chew--Frautschi plot,
displaying the experimentally observed mesons
$\left\{ \rho, \rho_3, \rho_5 \right\}$,
$\left\{ \omega, \omega_3, \omega_5 \right\}$,
$\left\{ f_2, f_4, f_6 \right\}$, $\left\{ a_2, a_4, a_6 \right\}$,
$\left\{ \pi, \pi_2 \right\}$, and $\left\{ b_1 \right\}$ in the
($\alpha(t) = J$) versus ($t = M^2$) plane.
All the mesons listed above are characterized in the quark model by
being in the radial ground state ($n = 1$). Within the groups, the
mesons have the same isospin~$I$, intrinsic spin~$S$, and parity~$P$,
while the total spin~$J$ and the orbital momentum~$L$ increase by two
units, respectively.  The first four series apparently fall on a
single trajectory (blue line), characterized by spin $S = 1$, natural
parity and $J = L + S$.  A fit of \cref{eq:regge_trajectory}, gives
$\alpha(0) = \num{0.440(11)}$ and
$\alpha' = \SI{0.917(16)}{(\GeV)^{-2}}$.\footnote{For the fit, we used
  the experimental uncertainties of the meson masses as given
  by~\refCite{Tanabashi:2018zz}, and scaled the resulting parameter
  uncertainties by the reduced~$\chi^2$ of the fit.}  With
\cref{eq:theory.string.slope}, one finds for the string tension
$\sigma_0$, or, equivalently, the slope~$k$ of the linear confining
potential in \cref{eq:intro.potential},
$\sigma_0 = \SI{0.880(16)}{GeV\!\per\femto\meter}$.  This number is in
qualitatively good agreement with the value of $k = 0.72\,\GeV/\fm$
extracted from a fit to the energy levels of heavy quarkonia (see \eg\
\refCite{Barnes:2005pb}).

Similarly, the pion series $\left\{\pi,\pi_2\right\}$ and the~$b_1$
form a separate, almost degenerate trajectory (green line),
characterized by $S = 0$ ($J = L$) and unnatural parity.  Its slope,
however, is slightly different as it is dominated by the ground
state~$\pi$.  This indicates the dual nature of the~$\pi$ as a
ground-state meson and a Goldstone boson of spontaneous chiral
symmetry breaking (see \eg\ \refCite{Scherer:2012xha} and references
therein).  Radially excited meson states are observed to fall on
trajectories with slopes similar to their parent $n = 1$ trajectories,
but shifted to higher masses. They are hence called daughter
trajectories.  Linear trajectories for quark-model meson families can
be observed not only in the $M^2$-$J$ plane, but also in the
$M^2$-$n_r$ plane (see \eg\ \refCite{Ebert:2009ub}), $n_r$~being the
number of radial nodes of the wave function.  In addition,
\cref{fig:chew_frautschi} includes the so-called Pomeron trajectory,
\begin{align}
  \label{eq:pomeron_trajectory}
  \alpha_{\mathbb{P}}(M^2) & = \alpha_{\mathbb{P}}(0) + \alpha_{\mathbb{P}}'
                           M^2\eqPunctSpacing, \\
  \alpha_{\mathbb{P}}(0) & = 1 + \epsilon_{\mathbb{P}}\eqPunctSpacing,
\end{align}
with $\epsilon_{\mathbb{P}} = 0.08$ and
$\alpha_{\mathbb{P}}' = 0.25\,\GeV^{-2}$~\cite{Donnachie:2002xx} (red
line), for which so far no corresponding mesons could be identified
(see \cref{sec:regge} for the interpretation of this
trajectory).

\subsection{Exotic Mesons}
\label{sec:pheno.exotics}

\begin{figure}[tbp]
  \centering
  \subfloat[]{%
    \includegraphics[width=0.1\columnwidth]{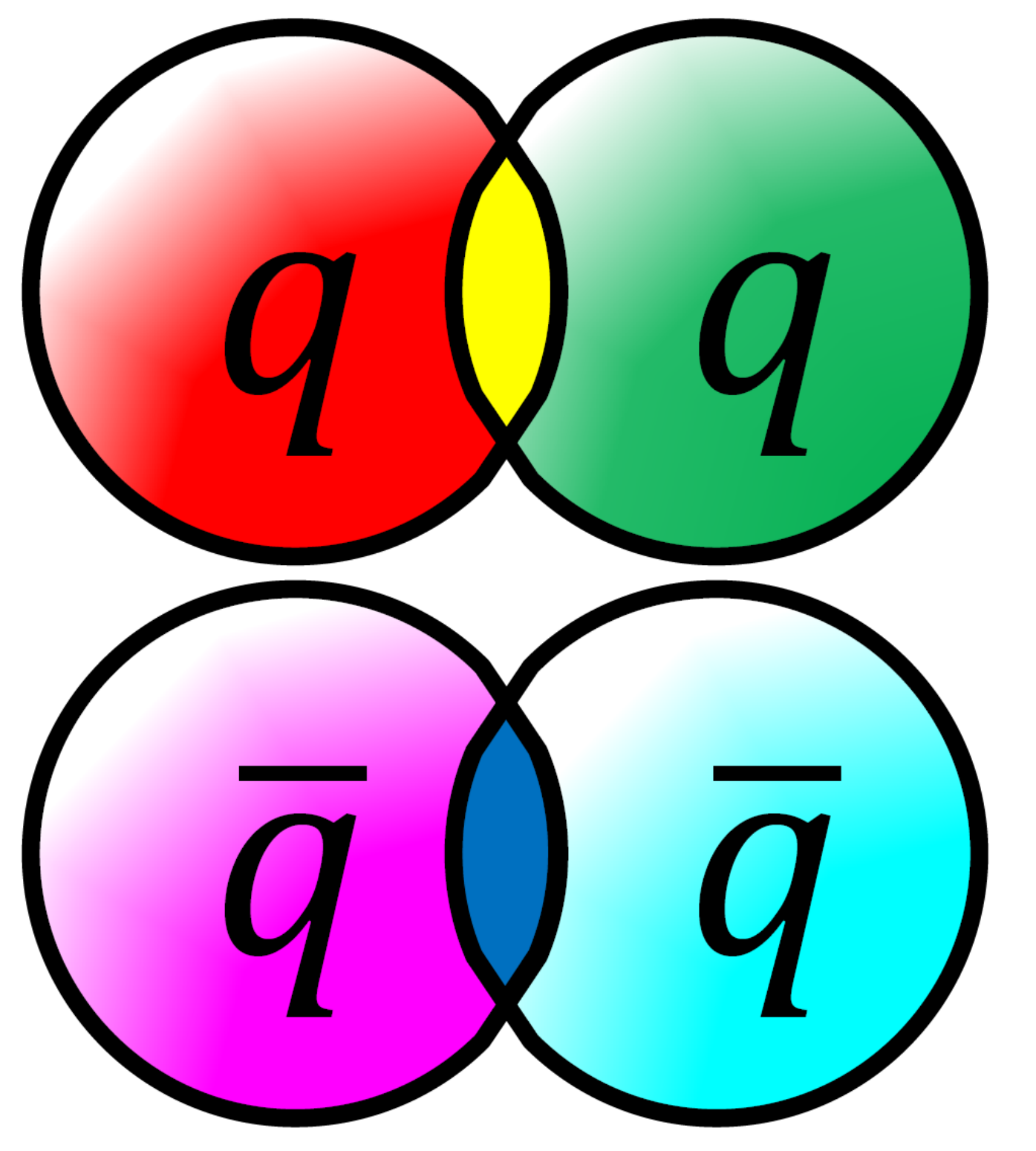}%
    \label{fig:mesons.tetraquark}%
  }%
  \hfill%
  \subfloat[]{%
    \includegraphics[width=0.2\columnwidth]{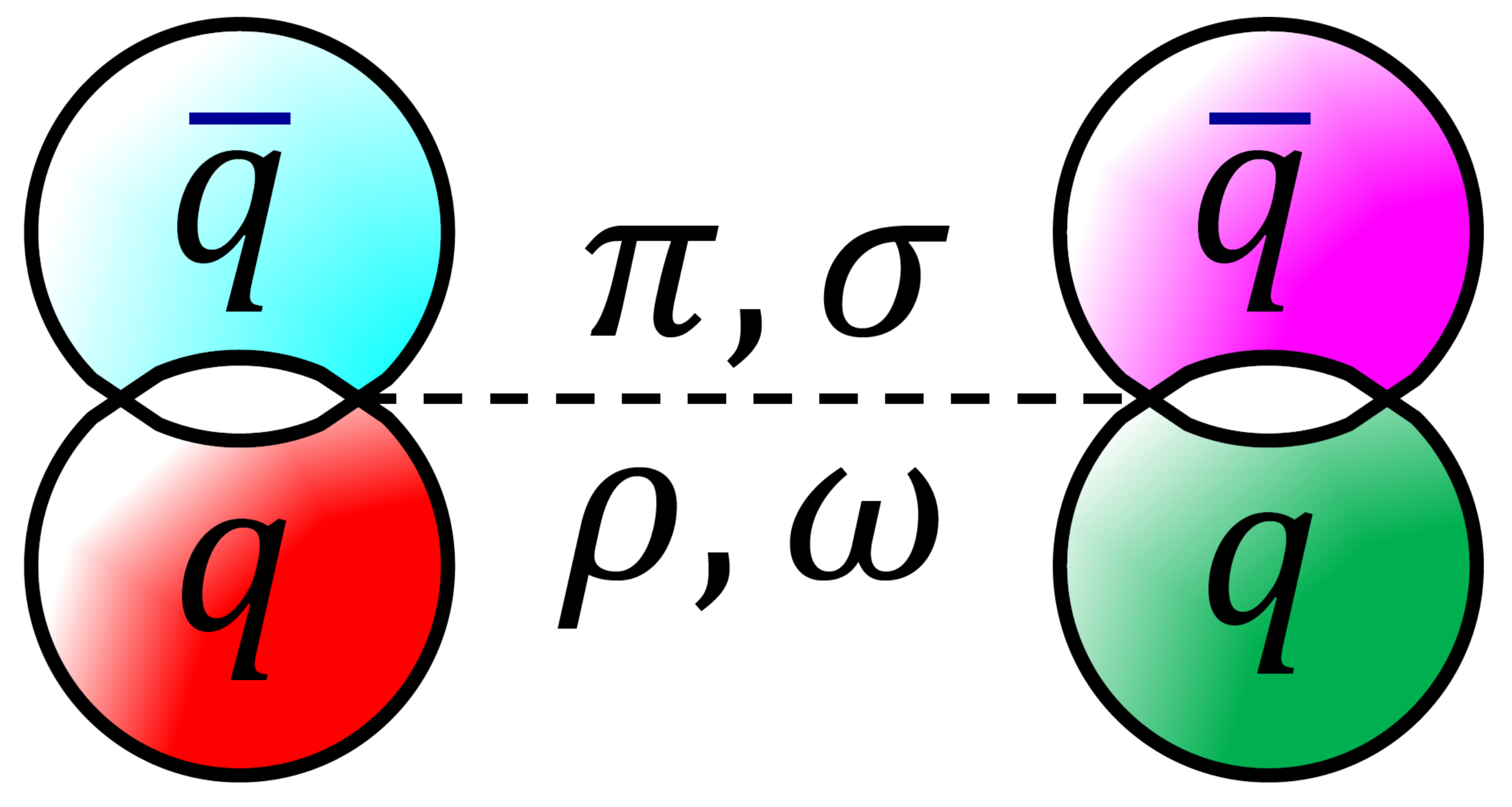}%
    \label{fig:mesons.molecule}%
  }%
  \hfill%
  \subfloat[]{%
    \includegraphics[width=0.2\columnwidth]{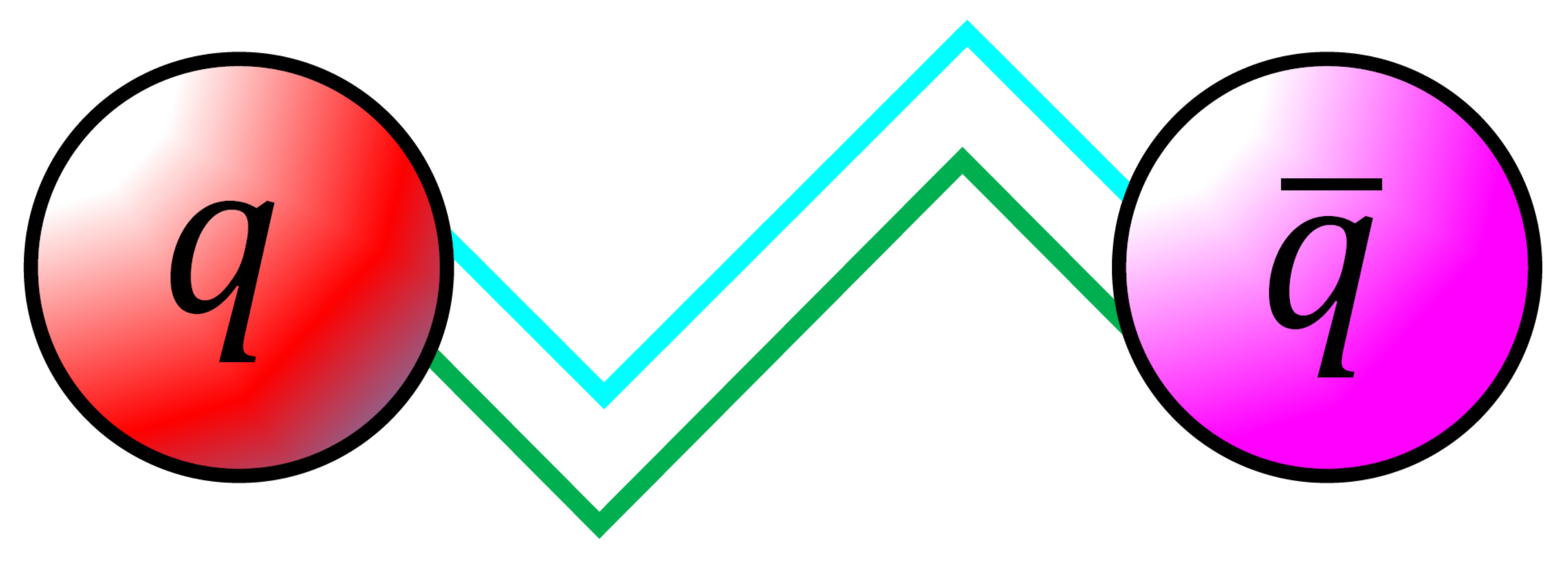}%
    \label{fig:mesons.hybrid}%
  }%
  \hfill%
  \subfloat[]{%
    \includegraphics[width=0.15\columnwidth]{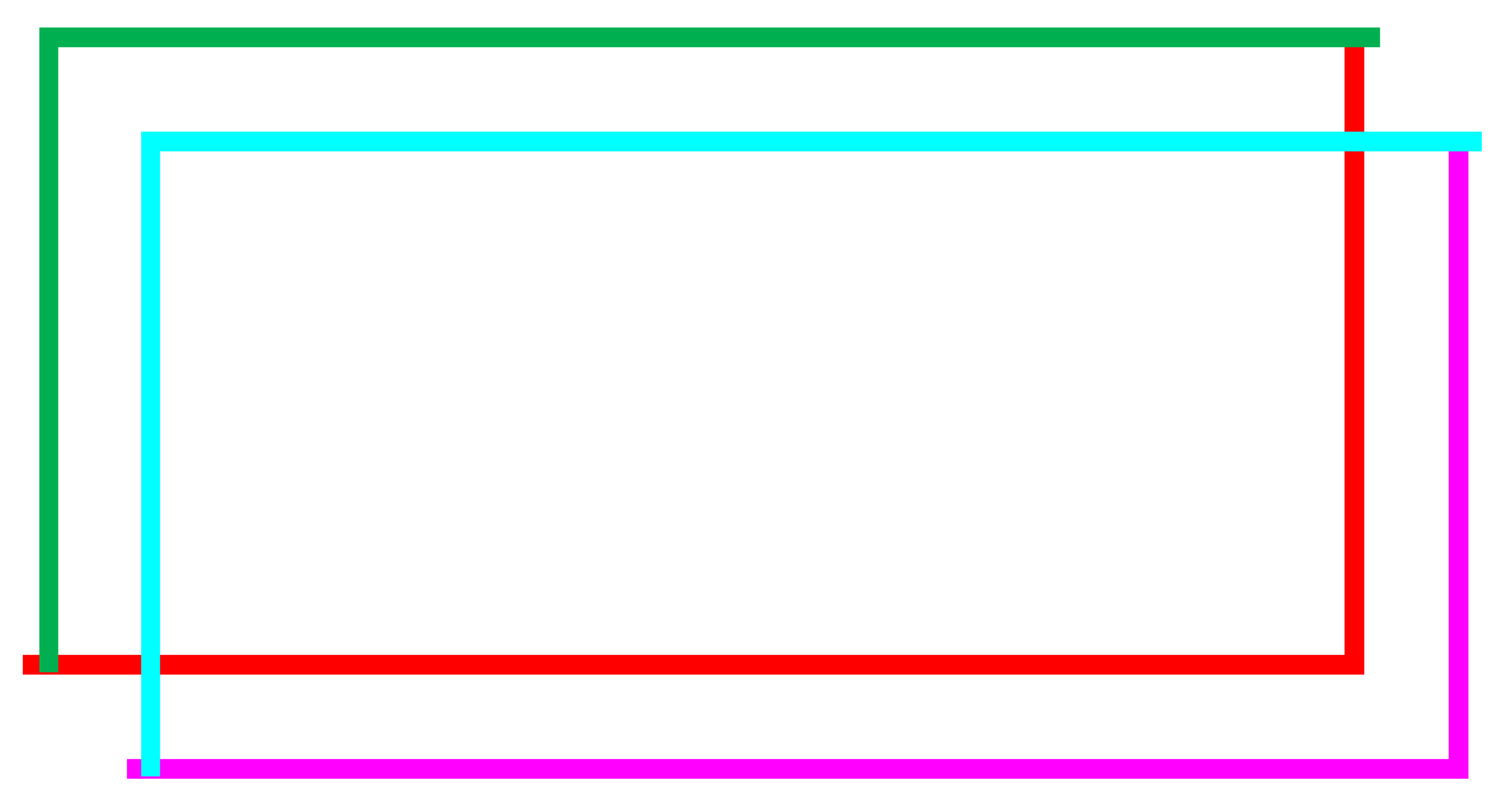}%
    \label{fig:mesons.glueball}%
  }%
  \caption{Examples of non-\qqbarpr configurations of mesons, which
    are allowed by QCD:
    \subfloatLabel{fig:mesons.tetraquark}~tetraquark,
    \subfloatLabel{fig:mesons.molecule}~molecule,
    \subfloatLabel{fig:mesons.hybrid}~hybrid, and
    \subfloatLabel{fig:mesons.glueball}~glueball.}
  \label{fig:mesons.non-qqbar_configurations}
\end{figure}

Mesons which do not fit the simple \qqbarpr~quark model are generally
called \emph{exotic}.  Indeed, QCD allows for a much richer spectrum
of mesons, which includes configurations where more than two valence
quarks contribute to the quantum numbers of the meson. The most
prominent configurations are shown in
\cref{fig:mesons.non-qqbar_configurations}.  \emph{Tetraquarks}
(\cref{fig:mesons.tetraquark}) are color-singlet objects formed by a
color-octet diquark and a color-octet anti-diquark bound by gluon
exchanges.  \emph{Molecular} configurations
(\cref{fig:mesons.molecule}) include two color-singlet \qqbarpr~pairs
bound by long-range meson exchanges.  Owing to the non-Abelian
structure of QCD, additional configurations are possible, in which an
excited gluonic field contributes to the quantum numbers of the meson.
States with a valence color-octet \qqbarpr~pair neutralized in color
by an excited gluon field are termed \emph{hybrids}
(\cref{fig:mesons.hybrid}).  Finally, due to the self-interaction of
gluons, color-singlet states can in principle also be composed
entirely of multiple gluonic excitations without valence quarks; such
objects are generically called glueballs (\cref{fig:mesons.glueball}).

These configurations in principle allow for an extended set of quantum
numbers compared to ordinary \qqbarpr~systems.  States with flavor
quantum numbers~$\abs{I_3}$, $\abs{\mathsf{S}}$, $\abs{\mathsf{C}}$,
or~$\abs{\mathsf{B}}\geq 2$ are called \emph{flavor-exotic}.  No
candidates for such states have been observed to date. Mesons with
\JPC~quantum numbers
\begin{equation}
  \label{eq:mesons.non-qqbar_qn}
  \JPC = 0^{--},\, 0^{+-},\, 1^{-+},\, 2^{+-},\ldots\eqPunctSpacing,
\end{equation}
\ie not fitting to \cref{eq:meson.qn,eq:meson.qqbar_qn}, are called
\emph{spin-exotic}.  If observed, such states are considered strong
candidates for gluonic excitations like hybrids or glueballs (which
are then called \emph{oddballs}), because the gluonic field may carry
quantum numbers different from $0^{++}$.  However, also molecular or
multi-quark states may have spin-exotic quantum numbers.  As can be
seen in \cref{fig:light_flavorless_spectrum}, there are two
light-meson candidate states with spin-exotic quantum numbers listed
by the PDG: the $\pi_1(1400)$ and the
$\pi_1(1600)$~\cite{Tanabashi:2018zz}.  Their nature as a resonance
has been the subject of a long-standing controversy, which has
recently been resolved thanks to the data of the COMPASS experiment.
This is discussed in more detail in \cref{sec:results_1mp}.  The
unambiguous identification of exotic states with the same quantum
numbers as ordinary \qqbarpr~mesons, so-called \emph{crypto-exotic}
mesons, is difficult because they will only appear as supernumerary
states in the observed spectrum for a given \JPC, possibly mixing with
the corresponding conventional \qqbarpr~states.  For such states,
their fitting into Regge trajectories (see
\cref{sec:pheno.trajectories}) can be a useful criterion for their
classification.

Experimental states labeled as \textquote{not assigned} (red data
points) in the columns corresponding to \qqbarpr~configurations in
\cref{fig:light_flavorless_spectrum} are potential candidates for
supernumerary states, however, a full understanding of the
\qqbarpr~spectrum is a prerequisite for claiming such states.  There
is evidence that the light scalar states indeed agree with a
tetraquark interpretation.  The recent observation of the $a_1(1420)$
by COMPASS is another example of a potentially supernumerary state in
the $1^{++}$ sector, which will be discussed in more detail in
\cref{sec:results_1pp}.

The search for exotic states has been a central goal of hadron
spectroscopy in the last
20~years~\cite{Ketzer:2012vn,Meyer:2015eta,Olsen:2017bmm}.  In recent
years, many new and unexpected resonance-like signals have been
observed in the heavy-quark sector. Many of these so-called~$X$, $Y$,
$Z$~states are candidates for exotic configurations of mesons.
Similar studies in the light-quark sector are more challenging due to
the wide and overlapping nature of the known resonances, but have also
revealed possible candidates, in particular for hadrons with explicit
gluonic degrees of freedom.  Still, one expects to find not only
single states, but whole multiplets of states with similar masses,
similar to the \qqbarpr~multiplets.  It remains an open question,
whether these multiplets simply have evaded detection until now or
whether they are not realized in nature as expected.  In order to
settle this fundamental question, large data sets with high
statistical precision are needed.  The unambiguous identification of
exotic states requires experiments with complementary production
mechanisms and the analysis of different final states.  In
\cref{sec:conclusions_outlook}, we will briefly discuss some running
and upcoming high-precision experiments, which are expected to shed
new light on the existence of exotic states in nature.

Last but not least, advanced analysis methods are needed, which
overcome of some of the shortcomings of the traditional partial-wave
techniques.  These include analysis models that reduce or avoid bias
or prejudice, and improved parameterizations of amplitudes that
include reaction models and fulfill fundamental theoretical
constraints such as unitarity, analyticity, and crossing symmetry.
Some important steps in this direction have been pioneered by the
COMPASS experiment in collaboration with JPAC\footnote{The Joint
  Physics Analysis Center (JPAC) is a collaboration of theorists and phenomenologists 
 working on the analysis of hadron physics data~\cite{Mathieu:2016mcy}.} and will be
reported in \cref{sec:results}.

Recently, numerical calculations using the method of lattice QCD (see
\cref{sec:pheno.lattice}) started to make predictions on the multiplet
structure of exotic hadrons, which may be used as a guideline in the
experimental
searches~\cite{Dudek:2010wm,Dudek:2011bn,Dudek:2013yja,Shepherd:2016dni}.

\subsection{Lattice QCD}
\label{sec:pheno.lattice}

Lattice QCD is presently the only available rigorous ab-initio method
that can consistently describe the physics of binding and decay of
hadrons~\cite{pdg_lattice:2018,Gattringer:2010zz,Briceno:2017max}.  It
is a form of lattice gauge theory as proposed by K.~G. Wilson in
\refCite{Wilson:1974sk}, where calculations are performed in a
discretized Euclidean space--time using a hypercubic lattice with
lattice spacing~$a$.  The lattice spacing leads to a momentum cut-off
$\propto 1/a$ (ultraviolet regularization).  In lattice QCD, the
quark fields are placed at the lattice sites, while the gluon gauge
fields are defined on the links that connect neighboring sites.

The calculations are performed numerically by using Monte Carlo
techniques, \ie by sampling possible configurations of the quark and
gluon fields according to the probability distribution given by the
QCD Lagrangian.  This requires large computational resources provided
only by supercomputers.  Because of the employed Monte Carlo approach,
only a finite number of configurations can be considered.  This leads
to statistical uncertainties of the lattice QCD results.  The
calculations are often performed at larger up and down quark masses
than in nature.  This limits the computational cost because it
drastically reduces the number of virtual quark--antiquark loops that
have to be taken into account.  The employed up and down quark masses
are commonly expressed in terms of the resulting (unphysical) pion
mass.

In lattice QCD calculations, the extent of the space--time lattice is
necessarily finite.  In order to obtain physical results, several
limits have to be taken: \one the \emph{continuum limit}, \ie the
extrapolation $a \to 0$, \two the \emph{infinite-volume limit}, \ie
the extrapolation $L \to \infty$, and \three the \emph{physical
  quark-mass limit}, \ie the extrapolation to physical quark masses.
Many present-day lattice calculations are already performed directly
at or very close to the physical values of the quark masses (see \eg\
\refCite{Borsanyi:2014jba}), so that the latter extrapolation becomes
less of an issue and the dominant systematic effects are due to the
finite spatial volume and discrete nature of the lattice.
Unfortunately, lattice QCD calculations of excited hadron resonances
still need to be performed at rather high pion masses as will be
discussed further below.

Lattice QCD calculations have shown that QCD confines
color~\cite{Perantonis:1990dy,Bali:1992ab,Necco:2001xg}.  The method
was also used successfully to calculate various hadron properties, in
particular the masses of ground-state hadrons (see \eg\
\refCite{pdg_quark_model:2018}).  The results match the experimental
values with impressive precision, which shows that QCD indeed seems to
be the correct theory also in the regime where the perturbation
expansion in the strong coupling~$\alpha_{\text{s}}$ does not converge.

Compared to the ground-state hadrons, lattice QCD studies of the
excitation spectrum of hadronic states are still performed further
away from the physical point.  However, tremendous progress has been
made in the development of methods that are now able to extract towers
of highly excited states and to determine the inner structure of these
states (see \eg\ \refCite{Briceno:2017max} and references therein).
The cubic spatial lattice breaks rotational invariance and hence makes
the identification of spin states more difficult.  It was found that
with a sufficiently fine lattice spacing, the effects of the reduced
rotational symmetry of a cubic lattice can be overcome.  By
correlating meson operators with definite spin with the irreducible
representations of cubic rotations, it is possible to make spin
assignments from lattice QCD simulations with a single lattice
spacing.  Another key improvement was the development of variational
methods so that a large base of operators can be used in order to
reliably extract many excited states and to probe their internal
structure.

\begin{figure}[tbp]
  \centering
  \includegraphics[width=\textwidth]{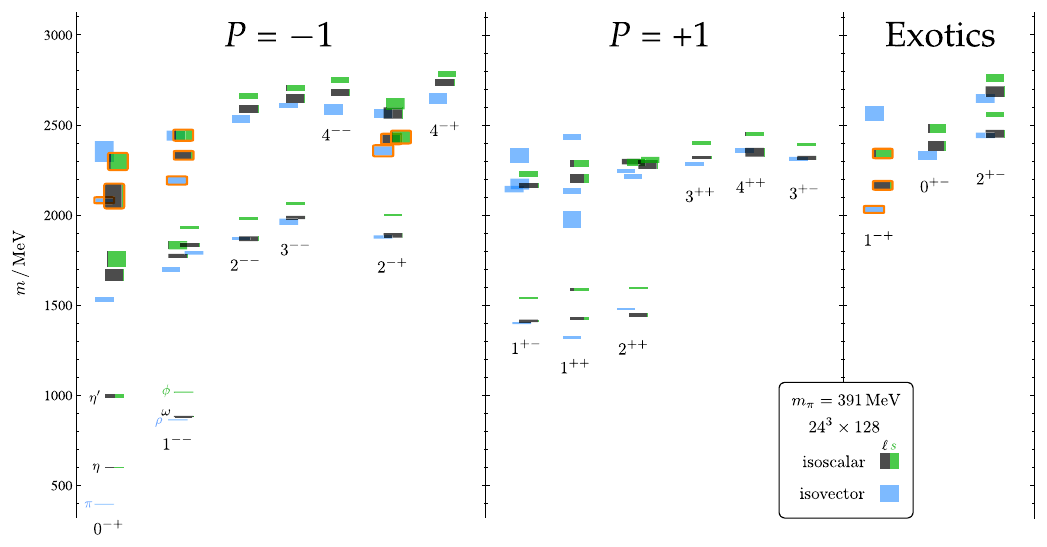}%
  \caption{Spectrum of light non-strange isoscalar (blue boxes) and
    isovector mesons (green/black boxes) from a lattice QCD
    calculation performed on a $24^3 \times 128$ lattice with a
    spatial lattice spacing of \SI{0.12}{fm} and a temporal lattice
    spacing of \SI{0.034}{fm} using unphysical up and down-quark
    masses that correspond to a pion mass of
    \SI{391}{\MeVcc}~\cite{Dudek:2013yja}.  The height of the boxes
    correspond to the statistical uncertainty of the estimated masses.
    For the isoscalar states, the mixing contributions from up and
    down quarks are indicated by the black areas, the contributions
    from strange quarks by the green areas.  Boxes outlined with
    orange represent the lightest states that have a large
    chromomagnetic gluonic component in their wave function.  They can
    be interpreted as the lightest hybrid-meson supermultiplet (see
    \cref{sec:pheno.lattice}).}
  \label{fig:lattice_spectrum_light_ns}
\end{figure}

\Cref{fig:lattice_spectrum_light_ns} shows the spectrum of light
non-strange mesons from the lattice QCD calculation in
\refCite{Dudek:2013yja}.  The spectrum is qualitatively similar to the
one predicted by quark models (see
\cref{fig:light_flavorless_spectrum}), but lattice QCD predicts also
the existence of exotic non-\qqbarpr states that are discussed in
\cref{sec:pheno.exotics}.  The masses of the states depend on the pion
mass used in the calculation.  The lowest pion mass used in
\refCite{Dudek:2013yja} is \SI{391}{\MeVcc}, which is still far away
from the physical point.  Since the simulation did not include
multi-hadron operators, the extracted states are quasi-stable and
their masses are in general not identical to the resonance masses.
For the same reason, this calculation cannot predict widths and decay
modes of the states.  For this, one still has to rely on models for
most of the states.

Lattice QCD calculations are performed by applying periodic spatial
boundary conditions to the quark and gluon fields.  This means that
within the cube of finite volume~$L_s^3$, free particles can only have
discrete three-momenta $\vec{p} = (2\pi/L_s)\, (n_x, n_y, n_z)$ with
integer~$n_i$.  Hence in scattering processes, multi-hadron states are
limited to a discrete spectrum of states, which are the energies of
the eigenstates of the QCD Hamiltonian in a finite box.  The simplest
case is the decay of a state into two (quasi-)stable particles, \eg
two pions.  The energies of the two-particle state depend on the
interactions between these particles.  By inverting this relation, one
can extract information about scattering amplitudes, \eg scattering
phase shifts, from the volume dependence of the discrete spectrum in a
finite volume.  This idea goes back to
M.~L\"{u}scher~\cite{Luscher:1986pf,Luscher:1990ux}.  It was developed
into a general method to calculate scattering amplitudes for all
possible $2 \to 2$ scattering processes of mesons by including
hadron--hadron operators that, for example, represent $\pi\pi$ or
\KKbar systems with defined momenta of the particles (see
\refCite{Briceno:2017max} and references therein).  The resonances,
which may appear in these scattering processes, are extracted by
analytic continuation of the scattering amplitude into the complex
energy-plane, where the resonances appear as poles (see
\cref{sec:scattering}).  Calculations of this kind have been
performed, for example, for the $\pi\pi$ system (see \eg\
\refsCite{Wilson:2015dqa,Briceno:2016mjc}) and the coupled $\pi\eta$
and \KKbar system (see \refCite{Dudek:2016cru}).

The calculations of scattering amplitudes performed so far were mostly
of exploratory nature and were hence performed at rather large pion
masses.  Unfortunately, performing these kind of calculations closer
to the physical point is not only a question of computational cost.
With decreasing pion mass also the kinematical thresholds for three-
and four-hadron channels decrease.  In particular highly excited
states couple strongly to such multi-hadron final states.  However,
the current method is not applicable to these channels.  A complete
finite-volume formalism for three or even more particles would
therefore be a major breakthrough for the calculation of masses and
decay modes of hadron resonances.  Such a formalism is already under
development (see \refCite{Briceno:2017max} and references therein).
\clearpage{}%

\mathversion{normal2}
\clearpage{}%
\section{Strong Interactions}
\label{sec:theory}

Quantum Chromodynamics is well established as the fundamental gauge
field theory of strong interactions in the regime of large momentum
transfers (see \eg\ \refCite{pdg_qcd:2018} and references therein),
where it can be solved using perturbative methods.  The non-Abelian
structure of the SU(3)$_\text{color}$ gauge symmetry group, however,
introduces self-interactions between the gauge bosons, which lead to
an increase of the effective strong coupling~$\alpha_\text{s}$ with
decreasing momentum transfer.  As a consequence, QCD is a strongly
coupled theory in the regime of hadrons, which cannot be solved
perturbatively.  Instead, models and effective theories are employed
to describe the phenomenology and dynamics of hadrons.  As discussed
in \cref{sec:pheno.lattice}, more recently lattice QCD started to
provide data on excited states by numerically investigating the
scattering of light hadrons in a discrete and finite space--time
volume.

Experimentally, hadronic resonances can be studied either in
scattering experiments or in decays (\confer\ \cref{sec:exp}).  The
analysis and interpretation of both experimental and lattice data
requires the implementation of theoretical amplitudes describing the
reaction under study.  In the following subsections, we will review
some of the basic principles of scattering theory based on the
$S$-matrix formulation.  Although $S$-matrix theory, which was
developed long before the advent of QCD, cannot be used to compute the
strong interaction amplitudes from first principles, its fundamental
properties provide important constraints for theoretical amplitudes.

%
\subsection{Scattering Theory}
\label{sec:scattering}

\subsubsection{$S$-Matrix}
\label{sec:scattering.s-matrix}

In a typical scattering experiment, a beam particle~$1$, \eg from an
accelerator, hits a target particle~$2$, which may either be at rest
or also an accelerated particle, and the interaction between the two
particles may result in the production of several new particles
traveling in different directions.  A resonance~$X$ may appear in
direct formation experiments,
\begin{equation}
  \label{eq:formation_reaction}
  \begin{minipage}[c]{0.4\linewidth}
    $1 + 2 \to X \to 3 + 4 +\ldots + (N_f + 2)\eqPunctSpacing,$
  \end{minipage}
  \begin{minipage}[c]{0.4\linewidth}
    \includegraphics[width=\linewidth]{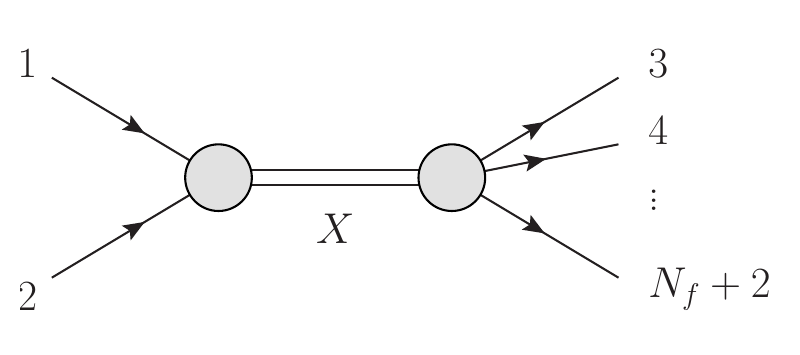}
  \end{minipage}
\eqPunctSpacing.
\end{equation}
or together with a (spectator) recoil particle~$3$,
\begin{equation}
  \label{eq:production_reaction}
  \begin{minipage}[c]{0.5\linewidth}
    $1 + 2 \to 3 + X \to 3 + \left[ 4 + 5 + \ldots + (N_f + 2) \right]\eqPunctSpacing,$
  \end{minipage}
  \begin{minipage}[c]{0.4\linewidth}
    \includegraphics[width=\linewidth]{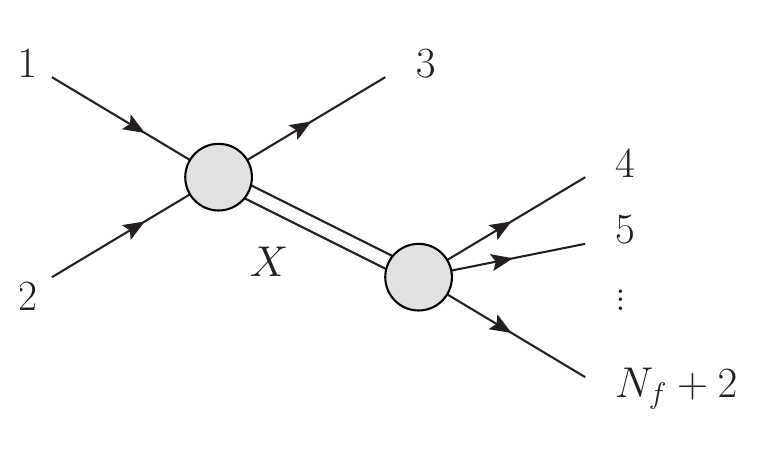}
  \end{minipage}
  \eqPunctSpacing.
\end{equation}

The operator connecting the initial and final states is called the
$S$-matrix.  In order to define the $S$-matrix, we start from an
initial state~$\ket{i}$ at time $t \to -\infty$, usually consisting of
two particles with four-momenta~$p_1$, $p_2$ and total four-momentum
$P_i = p_1 + p_2$, which is scattered into a state~$\ket{i'}$ at time
$t \to +\infty$, containing $N_f$~particles with total four-momentum
$P_f = \sum_{b = 3}^{N_f + 2} p_b$. The asymptotic states consist of
non-interacting particles, implying a short-range interaction such as
the strong force.  The scattering process is described by the
scattering operator~$S$:
\begin{equation}
  \label{eq:scattering_operator}
  \ket{i'} = S \ket{i}\eqPunctSpacing.
\end{equation}
Of all possible states contained in~$\ket{i'}$, the detector projects
out a particular final state~$\ket{f}$.  The probability amplitude for
finding~$\ket{f}$ in~$\ket{i'}$ is given by
\begin{equation}
  \label{eq:s-matrix.element}
  \bracket{f}{i'} = \bra{f} S \ket{i} = S_{fi}\eqPunctSpacing,
\end{equation}
with $S_{fi}$~being the elements of the $S$-matrix.  The probability
of finding a final state~$\ket{f}$, given~$\ket{i}$ as the initial
state, is
\begin{equation}
  \label{eq:s-matrix.probability}
  \left| S_{fi} \right|^2
  = \left| \bra{f} S \ket{i} \right|^2
  = \bra{i} S^\dagger \ket{f} \bra{f} S \ket{i}\eqPunctSpacing.
\end{equation}
The full $S$-matrix therefore connects all possible initial states
with all possible final states, and constitutes a complete description
of all particle interactions.  If~$\ket{f}$ forms a complete set of
orthonormal states, then\footnote{In the case of continuum states, \ie
  states with a continuous momentum variable, the sum also contains an
  integral over all possible momenta for all final states.}
\begin{equation}
  \label{eq:complete_set}
  \sum_f \ket{f} \bra{f} = \mathds{1}\eqPunctSpacing.
\end{equation}
Starting from the initial state~$\ket{i}$, the probability of ending
up in some final state must be unity:
\begin{equation}
  \label{eq:probability_conservation}
  1 = \sum_f \left| \bra{f} S \ket{i} \right|^2
  = \sum_f \bra{i} S^\dagger \ket{f} \bra{f} S \ket{i}
  = \bra{i} S^\dagger S \ket{i}\eqPunctSpacing.
\end{equation}
Since \cref{eq:probability_conservation} must hold for any
state~$\ket{i}$, it follows that $S^\dagger S = \mathds{1}$.  In the
same way, a final state~$\ket{f}$ must have originated from some
initial state out of a complete set~$\ket{i}$, from which one gets
$S\, S^\dagger = \mathds{1}$.  Therefore, as a consequence of the
conservation of probability, $S$~is unitary:
\begin{equation}
  \label{eq:s-matrix.unitarity}
  S^\dagger S = S\, S^\dagger = \mathds{1}\eqPunctSpacing.
\end{equation}

The set~$\ket{i'}$ in \cref{eq:scattering_operator} also contains
states that have not experienced any interaction.  They are usually
separated out by defining the transition matrix element~$T_{fi}$
through\footnote{Some authors use a different definition, \eg with a
  factor of~2 or a minus sign.  We follow the definition
  by~\cite{Eden:1966dnq}.}
\begin{equation}
  \label{eq:t-matrix.element}
  S_{fi} = \delta_{fi} + i T_{fi}\eqPunctSpacing,
\end{equation}
or in matrix form
\begin{equation}
  \label{eq:t-matrix}
  S = \mathds{1} + i T\eqPunctSpacing.
\end{equation}
Inserting \cref{eq:t-matrix} into the unitarity condition
\cref{eq:s-matrix.unitarity}, leads to the general unitarity relation
for~$T$:
\begin{equation}
  \label{eq:t-matrix.unitarity}
  T - T^\dagger = i T^\dagger T = i T\, T^\dagger\eqPunctSpacing,
\end{equation}
or, in terms of the elements of~$T$:
\begin{align}
  \label{eq:t-matrix.element.unitarity}
  \frac{1}{i}\,\left[\bra{f}T\ket{i} - \bra{f}T^\dagger\ket{i}\right]
  &= \frac{1}{i}\,\left[\bra{f}T\ket{i}-\bra{i}T^\ast\ket{f}\right]
  = \frac{1}{i}\, \left(T_{fi} - T_{if}^\ast \right) \notag \\
  &= \bra{f}T^\dagger T\ket{i} = \sum_j \bra{f}T^\dagger\ket{j}\bra{j}T\ket{i} =
    \sum_j \bra{j}T^\ast\ket{f}\bra{j}T\ket{i} \notag \\
  &= \sum_j T_{jf}^\ast\, T_{ji}\eqPunctSpacing,
\end{align}
where we have made use of $(T^\dagger)_{fi} = T_{if}^\ast$.  As in
\cref{eq:complete_set,eq:probability_conservation}, for continuum
states the sum in \cref{eq:t-matrix.element.unitarity} contains an
integral $\int \prod_{c = 1}^{N_j} \diffcb{p_c} / (2 \pi)^3$ over all
possible momenta of the $N_j$~particles in the intermediate
state~$j$.\footnote{We use the indices~$a$, $b$, and~$c$ to enumerate
  the initial-state, final-state, and intermediate-state particles,
  respectively.} It is important to note that the intermediate-state
particles in the unitarity relation are on their mass shells; they are
not virtual particles as they appear \eg\ in Feynman diagrams.  If
\begin{equation}
  \label{eq:t-matrix.sym}
  T_{fi} = T_{if}\eqPunctSpacing,
\end{equation}
which is generally true for $2 \to 2$ reactions if the interaction is
invariant \wrt parity transformation and
time-reversal~\cite{Eden:1966dnq}, we arrive at
\begin{equation}
  \label{eq:t-matrix.element.unitarity.im}
  \frac{1}{i}\, \left( T_{fi} - T_{fi}^\ast \right)
  \equiv 2 \Im T_{fi}
  = \sum_j T_{fj}^\ast\, T_{ji}\eqPunctSpacing.
\end{equation}

Explicitly factoring out the conservation of four-momentum in the
reaction, \ie $P_f = P_i$, the elements of the transition matrix are
usually written as~\cite{Peskin:1995ev}
\begin{equation}
  \label{eq:inv_amp.element}
  i T_{fi} = \left( 2 \pi \right)^4\,
  \delta^{(4)}\!\!\left( P_i - P_f \right)
  \prod_{a \in i} \frac{1}{\sqrt{2E_a}}
  \prod_{b \in f} \frac{1}{\sqrt{2E_b}}\,
  i \mathcal{M}_{fi}
  \eqPunctSpacing,
\end{equation}
with the Lorentz-invariant matrix element or amplitude
$i \mathcal{M}_{fi}$~\cite{Martin:1970xx} that encodes the whole
dynamic content of the transition and that can be calculated \eg from
Feynman rules in perturbation theory for elementary processes
containing quarks and leptons.  For low-energy hadronic interactions,
effective Lagrangians may be employed to calculate the
amplitude~\cite{Meissner:1987ge}.  For light hadrons, chiral
perturbation theory is the low-energy effective theory of QCD (see
\cref{sec:theory.chiPT}).  For high energies, the framework of Regge
theory, discussed in \cref{sec:regge}, provides some guidance on the
asymptotic behavior of scattering amplitudes.  The intermediate
energy-regime is governed by hadronic resonances, for which no
fundamental theory exists, and for which the amplitudes need to be
parameterized using phenomenological models.  Lorentz-invariance
requires the invariant amplitude to be a Lorentz scalar, which may
therefore be written as a function of Lorentz scalars only.  In case
of spinless particles, which we will assume here for simplicity,
$\mathcal{M}_{fi}$~is thus a function of scalar products of
four-momenta only.

The unitarity condition \cref{eq:t-matrix.element.unitarity.im} can be
written in terms of the invariant amplitude~$\mathcal{M}_{fi}$ from
\cref{eq:inv_amp.element} as
\begin{equation}
  \label{eq:inv_amp.unitarity}
  \frac{1}{i}\, \left( \mathcal{M}_{fi} - \mathcal{M}_{fi}^\ast \right)
  \equiv 2 \Im \mathcal{M}_{fi}
  = \sum_j \int\! \dif{\Phi_{N_j}}\, \mathcal{M}_{fj}^\ast\,
  \mathcal{M}_{ji}\eqPunctSpacing,
\end{equation}
where one of the two $\delta$-functions was \textquote{canceled} on
both sides of the equation and the remaining one on the right-hand
side was included in the Lorentz-invariant $N$-body phase-space
element\footnote{We deviate from the convention used
  in~\refCite{pdg_kinematics:2018} by including the factor~$(2 \pi)^4$
  in the phase-space element.} for the intermediate state~$j$:
\begin{equation}
  \label{eq:dPhi_N}
  \dif{\Phi}_{N}(P_i; p_1, \ldots, p_N)
  = (2 \pi)^4\,
  \delta^{(4)}\!\!\left( P_i - \sum_{b = 1}^N p_b \right)
  \prod_{b = 1}^{N} \frac{\diffcb{p_b}}{(2 \pi)^3\, 2E_b}\eqPunctSpacing.
\end{equation}
For $N$~particles with fixed masses in the final state, there are
$3N$~three-momentum variables and 4~constraints from energy and
momentum conservation.  The effective dimension of~$\dif{\Phi}_N$ is
hence~$3N - 4$.  For a two-body final state, $\dif{\Phi}_2$~can be
calculated directly by integrating out the $\delta$-function (see
\cref{eq:dPhi_2.cms}).  For more particles, $\dif{\Phi}_N$ can
be calculated from $\dif{\Phi}_2$ using the phase-space recurrence
relation (see \cref{sec:nbody_phase_space}).  For 2-to-2 scattering,
the unitarity condition \cref{eq:inv_amp.unitarity} can be represented
symbolically as
\begin{equation}
  \vspace{0pt}
  \begin{minipage}[c]{0.65\linewidth}
    \includegraphics[width=0.95\linewidth]{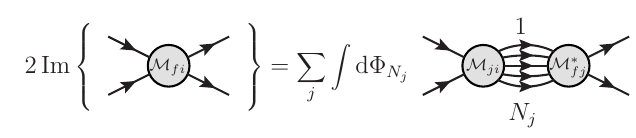}
  \end{minipage}
  \hspace*{-1em}.
  \label{eq:inv_amp.unitarity.diagram}
\end{equation}

\subsubsection{Kinematics of Two-Body Reactions}
\label{sec:scattering.kinematics}

\begin{figure}[tbp]
  \centering
  \includegraphics[width=0.25\columnwidth]{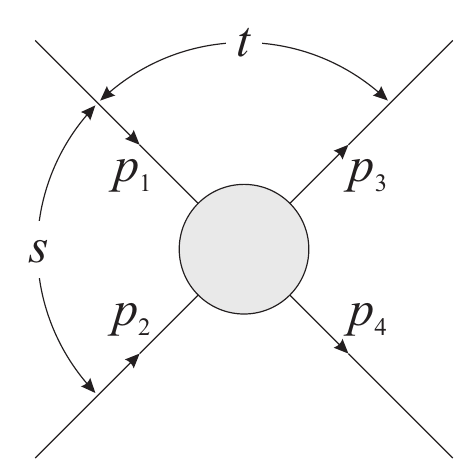}
  \caption{Two-body scattering process $1 + 2 \to 3 + 4$.}
  \label{fig:scattering.2body}
\end{figure}

The decay of a resonance~$X$ into a multi-particle final state is
commonly modeled as a sequence of two-body decays (see
\cref{sec:pwa.analysis_model.decay_amp}).  Therefore, we restrict
ourselves to two-body kinematics in the following.  Consider two
spinless particles with momenta~$\bvec{p}_1$ and~$\bvec{p}_2$ and
masses~$m_1$ and~$m_2$, which scatter to two particles with
momenta~$\bvec{p}_3$ and~$\bvec{p}_4$ and masses~$m_3$ and~$m_4$, as
shown in \cref{fig:scattering.2body}.
The four-momenta are defined as $p_i = (E_i, \bvec{p}_i)$, where
$E_i^2 = m_i^2 + \bvec{p}_i^2$ is the energy of particle~$i$.  The
Lorentz-invariant Mandelstam variables are defined as
\begin{align}
  \label{eq:mandelstam.s}
  s & = \left( p_1 + p_2\right)^2
      = \left( p_3 + p_4\right)^2
      = m_1^2 + m_2^2
      + 2 \left( E_1\, E_2 - \bvec{p}_1 \cdot \bvec{p}_2 \right)\eqPunctSpacing, \\
  \label{eq:mandelstam.t}
  t & = \left( p_1 - p_3 \right)^2
      = \left( p_2 - p_4 \right)^2
      = m_1^2 + m_3^2
      - 2 \left( E_1\, E_3 - \bvec{p}_1 \cdot \bvec{p}_3 \right)\eqPunctSpacing\text{, and} \\
  \label{eq:mandelstam.u}
  u & = \left( p_1 - p_4 \right)^2
      = \left( p_2 - p_3 \right)^2
      = m_1^2 + m_4^2
      - 2 \left( E_1\, E_4 - \bvec{p}_1 \cdot \bvec{p}_4 \right)\eqPunctSpacing.
\end{align}
They satisfy
\begin{equation}
  \label{eq:mandelstam.sum}
  s + t + u = \sum_{i = 1}^4 m_i^2\eqPunctSpacing,
\end{equation}
\ie there are only two independent variables to fully characterize the
scattering process for given masses~$m_i$.  The diagram in
\cref{fig:scattering.2body} not only describes the $s$-channel process
$1 + 2 \to 3 + 4$, but, by reversing the signs of some of the
four-momenta, also the $t$-channel process
$1 + \overline{3} \to \overline{2} + 4$ and the $u$-channel process
$1 + \overline{4} \to \overline{2} + 3$.\footnote{The channels are
  named after their respective energy invariants, as defined by
  \cref{eq:mandelstam.s,eq:mandelstam.t,eq:mandelstam.u}.}

In the $s$-channel center-of-momentum frame, we have
$\bvec{p}_1 = -\bvec{p}_2\eqqcolon \bvec{p} $ and
$\bvec{p}_3 = -\bvec{p}_4 \eqqcolon \bvec{q}$, and thus
\begin{align}
  \label{eq:mandelstam.cms.s}
  s &= \left( E_1 + E_2 \right)^2
      = \left( E_3 + E_4 \right)^2\eqPunctSpacing, \\
  \label{eq:mandelstam.cms.t}
  t &= m_1^2 + m_3^2 - 2 \left( E_1\, E_3
      - \left| \bvec{p}_1 \right|\, \left| \bvec{p}_3 \right|\,
      \cos\theta_s \right)\eqPunctSpacing\text{, and} \\
  \label{eq:mandelstam.cms.u}
  u &= m_1^2 + m_4^2 - 2 \left( E_1\, E_4
      - \left| \bvec{p}_1 \right|\, \left| \bvec{p}_4 \right|\,
      \cos\theta_s \right)\eqPunctSpacing,
\end{align}
with $\theta_s$~the scattering angle between particle~1 and~3 in the
$s$-channel center-of-momentum frame.  The physically allowed region
for the $s$-channel process is
\begin{equation}
  \label{eq:s.physical_region}
  s \geq (m_1 + m_2)^2
  \quad\text{and}\quad
  -1 \leq \cos\theta_s \leq 1\eqPunctSpacing.
\end{equation}
The limiting values for~$t$ are $t_0$~for $\theta_s = 0$ and $t_1$~for
$\theta_s = \pi$:
\begin{equation}
  \label{eq:t.limits}
  t_{0 \atop 1} =
  m_1^2 + m_3^2 - 2 \left( E_1\, E_3
    \mp \left| \bvec{p}_1 \right|\, \left| \bvec{p}_3 \right| \right)\eqPunctSpacing.
\end{equation}
In order to deal with positive values, one usually defines
\begin{equation}
  \label{eq:t.minmax}
  |t|_{\text{min} \atop \text{max}} \coloneqq -t_{0 \atop 1}
\end{equation}
and
\begin{equation}
  \label{eq:tprime}
  t^\prime \coloneqq \left| t \right| - \tMin\eqPunctSpacing.
\end{equation}

\begin{figure}[tbp]
  \centering
  \includegraphics[width=0.4\columnwidth]{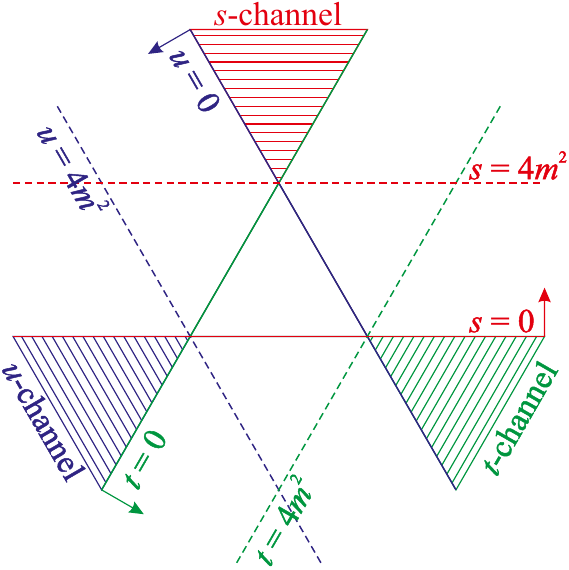}
  \caption{Mandelstam plot of the variables~$s$, $t$, and~$u$.  The
    physically allowed regions for the~$s$-, $t$-, and $u$-channel
    process in the equal-mass case are indicated by the hashed
    triangles.}
  \label{fig:mandelstam_plane}
\end{figure}

For the simple case of equal masses, \ie $m_i = m$ with
$i = 1 \ldots 4$ and $\bvec{p}_1 = \bvec{p}_3 = \bvec{p}$, we have
\begin{align}
  \label{eq:mandelstam.cms.s.equal_masses}
  s &= 4 \left( \bvec{p}^2 + m^2 \right)\eqPunctSpacing, \\
  \label{eq:mandelstam.cms.t.equal_masses}
  t &= -2 \bvec{p}^2 \left( 1 - \cos\theta_s \right)\eqPunctSpacing\text{, and} \\
  \label{eq:mandelstam.cms.u.equal_masses}
  u &= -2 \bvec{p}^2 \left( 1 + \cos\theta_s \right)\eqPunctSpacing.
\end{align}
The boundaries of the physical region for the $s$-channel process are
then given by $s \geq 4m^2$, $t \leq 0$, and $u \leq 0$.  For the
$t$-channel process, they are $t \geq 4m^2$, $u \leq 0$, and
$s \leq 0$, and for the $u$-channel process $u \geq 4m^2$, $s \leq 0$,
and $t \leq 0$.  \Cref{eq:mandelstam.sum} defines a plane in the
Cartesian coordinate system spanned by~$s$, $t$, and~$u$ with normal
vector $(1, 1, 1)$ and distance to the origin of $4 m^2 / \sqrt{3}$.
More commonly, one plots the three axes $s, t, u$ in one plane with
the axes rotated by $60^\circ$,\footnote{In an equilateral triangle,
  the sum of the perpendicular distances from a point to the sides is
  constant, \ie\ \cref{eq:mandelstam.sum} is fulfilled by definition.}
as shown in \cref{fig:mandelstam_plane}.

\subsubsection{Cross Sections}
\label{sec:scattering.xsection}

The differential cross section for the inelastic scattering reaction
$1 + 2 \to 3 + 4 + \ldots + (N_f + 2)$ is given
by~\cite{pdg_kinematics:2018}\footnote{If the colliding particles are
  unpolarized and if the spin states of initial and final-state
  particles are not measured in the experiment, the
  term~$|\mathcal{M}_{fi}|^2$ is assumed to include the incoherent
  averaging and summation over the spin states of the initial and the
  final-state particles, respectively.}
\begin{equation}
  \label{eq:dsigma}
  \dif{\sigma} = \frac{1}{F}\, \abs{\mathcal{M}_{fi}}^2\,
  \dif{\Phi}_{N_f}(p_1 + p_2; p_3, \ldots, p_{N_f + 2})\eqPunctSpacing,
\end{equation}
with the flux factor~$F$, which for a general collinear collision
between two particles is
\begin{equation}
  \label{eq:flux_factor}
  F = 4 \left[ \left( p_1 \cdot p_2\right)^2 - m_1^2\, m_2^2 \right]^{1/2}
  \eqPunctSpacing,
\end{equation}
and the Lorentz-invariant $N_f$-body phase-space
element~$\dif{\Phi}_{N_f}$ (\confer\ \cref{eq:dPhi_N}).

In the center-of-momentum frame for a two-body reaction
$1 + 2 \to 3 + 4$, we have $\sqrt{s} = E_1 + E_2 = E_3 + E_4$,
$\left| \bvec{p}_1 \right| = \left| \bvec{p}_2 \right| \eqqcolon p$
and
$\left| \bvec{p}_3 \right| = \left| \bvec{p}_4 \right| \eqqcolon q$,
and the flux factor in \cref{eq:flux_factor} and the phase-space
element in \cref{eq:dPhi_N} become
\begin{equation}
  \label{eq:flux_factor.cms}
  F = 4 p\, \sqrt{s}
\end{equation}
and
\begin{equation}
  \label{eq:dPhi_2.cms}
  \dif{\Phi}_2 = \frac{1}{4 \pi}\, \frac{q}{\sqrt{s}}\,
  \frac{\dif{\Omega}}{4 \pi}\eqPunctSpacing,
\end{equation}
respectively.  Here, $\dif{\Omega} = \dif{\cos\theta_s}\, \dif{\phi}$
is the solid-angle element around the scattering angle
$\theta_s = \angle{(\bvec{p}_1, \bvec{p}_3)}$.  The momenta~$p$
and~$q$ can be calculated by
\begin{align}
  \label{eq:breakup_mom.cms.i}
  p &= \frac{1}{2\sqrt{s}}\, \lambda^{1/2}(s, m_1^2, m_2^2)\quad\text{and} \\
  \label{eq:breakup_mom.cms.f}
  q &= \frac{1}{2\sqrt{s}}\, \lambda^{1/2}(s, m_3^2, m_4^2)
\end{align}
using the K\"{a}ll\'{e}n function
\begin{align}
  \label{eq:kaellen}
  \lambda(a, b, c) &= a^2 + b^2 + c^2 - 2ab - 2bc - 2ca \nonumber\\
    &= \left[ a - \left( \sqrt{b} + \sqrt{c} \right)^2 \right]
       \left[ a - \left( \sqrt{b} - \sqrt{c} \right)^2 \right]\eqPunctSpacing.
\end{align}
The differential cross section for two-body scattering is then,
according to \cref{eq:dsigma},
\begin{equation}
  \label{eq:dsigma_dOmega.cms}
  \frac{\dif{\sigma}}{\dif{\Omega}}
  = \frac{1}{64 \pi^2\,s}\, \frac{q}{p}\, \abs{\mathcal{M}_{fi}}^2\eqPunctSpacing.
\end{equation}
For spinless particles, the scattering probability is independent of
the azimuthal angle~$\phi$, and we get from
\cref{eq:dsigma_dOmega.cms}, by using
$\dif{\Omega} = \dif{\cos\theta_s}\, \dif{\phi}$ together with
\cref{eq:mandelstam.cms.t} and integrating over the azimuthal
angle~$\phi$,
\begin{equation}
  \label{eq:dsigma_dt.cms}
  \frac{\dif{\sigma}}{\dif{t}}
  = \frac{1}{64 \pi\, s\, p^2}\, \abs{\mathcal{M}_{fi}}^2\eqPunctSpacing.
\end{equation}
For $m_1 = m_2$, this simplifies to
\begin{equation}
  \label{eq:dsigma_dt.cms.equal_masses}
  \frac{\dif{\sigma}}{\dif{t}}
  = \frac{1}{16 \pi\, s\, (s - 4 m^2)}\, \abs{\mathcal{M}_{fi}}^2\eqPunctSpacing.
\end{equation}

The total cross section for the scattering of particles~$1$ and~$2$
can be calculated by integrating \cref{eq:dsigma} over the phase space
and summing over all possible final states,
\begin{equation}
  \label{eq:sigma_tot}
  \sigma_\text{tot}
  = \frac{1}{F}\, \sum_f \int\! \dif{\Phi}_{N_f}\,
  \mathcal{M}_{fi}^\ast\, \mathcal{M}_{fi}\eqPunctSpacing.
\end{equation}
Using the unitarity condition \cref{eq:inv_amp.unitarity}, the
right-hand-side of this equation can be replaced by the imaginary part
of the invariant amplitude for identical initial and final states, \ie
$\ket{f} \equiv \ket{i}$,
\begin{equation}
  \label{eq:optical_theorem}
  \sigma_\text{tot} = \frac{2}{F}\, \Im \mathcal{M}_{ii}\eqPunctSpacing.
\end{equation}
This relation, which connects the total cross section to the imaginary
part of the forward elastic scattering amplitude~$\mathcal{M}_{ii}$ is
known as the \emph{optical theorem}.  The optical theorem can be
represented symbolically as
\begin{equation}
  \vspace{0pt}
  \begin{minipage}[c]{0.9\linewidth}
    \includegraphics[width=\linewidth]{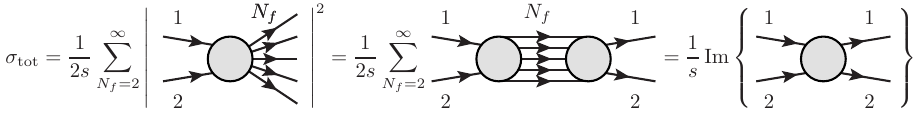}
  \end{minipage}
  .
  \label{eq:optical_theorem.diagram}
\end{equation}
The factor $1 / (2s)$ is the large-$s$ limit of $1 / F$ with $F$~from
\cref{eq:flux_factor.cms}.

\subsubsection{Partial-Wave Expansion and Breit--Wigner Resonances}
\label{sec:scattering.pw_expansion}

For a two-body reaction $1 + 2 \to 3 + 4$, there are 2~independent
variables, which may be chosen to be~$s$ and~$t$:
$\mathcal{M}_{fi} = \mathcal{M}_{fi}(s, t)$.  From
\cref{eq:mandelstam.cms.s.equal_masses,eq:mandelstam.cms.t.equal_masses},
one gets for the equal-mass case
\begin{equation}
  \label{eq:theta.s.cms}
  \cos\theta_s = 1 + \frac{2 t}{s - 4 m^2}\eqPunctSpacing.
\end{equation}
It is clear that $t$~is a linear function of $\cos\theta_s$ at
fixed~$s$.  One can therefore use~$s$ and $z_s \coloneqq \cos\theta_s$
as independent variables.  The invariant amplitude for $s$-channel
scattering of two spinless particles can then be expanded in a
partial-wave series
\begin{equation}
  \label{eq:pw_expansion.s}
  \mathcal{M}_{fi}\big( s, t(s, z_s) \big)
  = \sum_{\ell = 0}^\infty (2 \ell + 1)\,
  t_{\ell,fi}(s)\, P_\ell(z_s)\eqPunctSpacing.
\end{equation}
Here we have assumed that the interaction potential is spherically
symmetric, so that the relative orbital angular momentum~$\ell$
between initial-state and final-state particles, respectively, is a
good quantum number and we can factorize the angular and the energy
dependence into the Legendre polynomials $P_\ell(z_s)$ and the
$s$-channel partial-wave amplitude $t_{\ell,fi}(s)$.  The
$P_\ell(z_s)$ form a complete set of eigenfunctions of~$\ell$ and
encode the angular distribution as a function of the
center-of-momentum scattering angle~$\theta_s$.  They obey the
orthonormality relation
\begin{equation}
  \label{eq:legendre.orthonormality}
  \int_{-1}^1\! \dif{z}\, P_n(z)\, P_m(z)
  = \frac{2 \delta_{n m}}{2 n + 1}\eqPunctSpacing.
\end{equation}
The $t_{\ell,fi}(s)$ are in general complex functions, which describe
the dynamics of the scattering process from initial state~$\ket{i}$ to
final state~$\ket{f}$ in the partial wave~$\ell$.

For $2 \to 2$ scattering, the unitarity condition
\cref{eq:inv_amp.unitarity}, using \cref{eq:dPhi_2.cms}, reads
\begin{equation}
  \label{eq:inv_amp.unitarity.2body}
  2 \Im \mathcal{M}_{fi}
  = \sum_j \int\! \dif{\Phi_2}\, \mathcal{M}_{fj}^\ast\, \mathcal{M}_{ji}
  = \sum_j \frac{q_j}{4 \pi\, \sqrt{s}}
  \int\! \frac{\dif{\Omega}}{4 \pi}\,
  \mathcal{M}_{fj}^\ast\, \mathcal{M}_{ji}\eqPunctSpacing.
\end{equation}
Inserting the partial-wave expansion \cref{eq:pw_expansion.s}
for~$\mathcal{M}_{fi}$, we arrive at the unitarity condition for the
partial-wave amplitudes
\begin{equation}
  \label{eq:pw_amp.unitarity.element}
  \Im t_{\ell,fi}(s)
  = \sum_j \rho_j(s)\, t_{\ell,fj}^\ast(s)\, t_{\ell,ji}(s)\eqPunctSpacing,
\end{equation}
with the two-body phase-space factor
\begin{equation}
  \label{eq:rho.2body.cms}
  \rho_j(s) = \frac{q_j}{8 \pi\, \sqrt{s}} = \frac{1}{2}\, \Phi_2\eqPunctSpacing.
\end{equation}
Here, we have made use of~\cite{Gribov:2009zz}
\begin{equation}
  \label{eq:legendre.addition_theorem}
  \int\! \frac{\dif{\Omega}}{4 \pi}\,
  P_\ell(\cos\theta_1)\, P_n(\cos\theta_2)
  = \frac{\delta_{\ell n}}{2 \ell + 1}\, P_\ell(\cos\theta)\eqPunctSpacing,
\end{equation}
where the integration is performed over all angles of the momentum
vector $\bvec{k}$ of one of the two intermediate-state particles in
the center-of-momentum frame. The angle
$\theta_1 = \angle{(\bvec{p}_1, \bvec{k})}$ denotes the angle between
the momentum vectors of the initial and the intermediate-state
particles in the center-of-momentum frame,
$\theta_2 = \angle{(\bvec{k}, \bvec{p}_3)}$ the angle between
intermediate and final-state particles, and
$\theta = \angle{(\bvec{p}_1, \bvec{p}_3)}$ the angle between initial
and final-state particles.\footnote{The relation can be derived using
  the relation between Legendre polynomials and spherical harmonics
  for spin projection quantum number $m = 0$, \ie
  $Y_\ell^0 = \sqrt{\frac{2 \ell + 1}{4 \pi}}\, P_\ell(\cos\theta)$,
  the addition theorem, and the orthonormality of the spherical
  harmonics~\cite{Arfken:2011zz}.}  For the scattering of equal-mass
particles, the two-body phase-space factor in \cref{eq:rho.2body.cms}
is
\begin{equation}
  \label{eq:rho.equal_masses.cms}
  \rho(s) = \frac{1}{16 \pi}\, \sqrt{\frac{s - 4 m^2}{s}}\eqPunctSpacing.
\end{equation}

Considering a single channel only, \ie for given~$\ket{i}$, $\ket{j}$,
and~$\ket{f}$, the channel indices may be skipped and
\cref{eq:pw_amp.unitarity.element} can be reduced to
\begin{equation}
  \label{eq:pw_amp.unitarity}
  \Im t_\ell(s) = t_\ell^\ast(s)\, \rho(s)\, t_\ell(s)\eqPunctSpacing.
\end{equation}
In the more general case, when $T_{fi} \neq T_{if}$, the
left-hand-side of \cref{eq:t-matrix.element.unitarity} is not equal to
the imaginary part of the amplitude. The unitarity relation for
partial-wave amplitudes then reads
\begin{equation}
  \label{eq:pw_amp.unitarity.general}
  t_\ell(s) - t_\ell^\dagger(s)
  = 2 i\, t_\ell^\dagger(s)\, \rho(s)\, t_\ell(s)\eqPunctSpacing.
\end{equation}

For several coupled two-body channels, \eg the $\eta \pi$-$\eta' \pi$
analysis explained in
\cref{sec:etapi_model:pwa,sec:etapi_model:resonance},
\cref{eq:pw_amp.unitarity,eq:pw_amp.unitarity.general} are to be
understood with $t_\ell(s)$ being a matrix in channel space with the
phase-space matrix $\rho = \diag{(\rho_1, \rho_2, \ldots, \rho_j)}$.
A common parameterization satisfying the unitarity condition
\cref{eq:pw_amp.unitarity} for two-body scattering involving two
channels~$i, j$ is~\cite{pdg_resonances:2018}\footnote{This definition
  is made such that it reduces to the well-known form for
  non-relativistic potential scattering for the case of a single
  channel.}
\begin{equation}
  \label{eq:pw_amp.param.element}
  t_{\ell,ij}(s) =
  \frac{s_{\ell,ij}(s) - \delta_{ij}}{2i\, \sqrt{\rho_{ii}(s)}\,
    \sqrt{\rho_{jj}(s)}}\eqPunctSpacing,
\end{equation}
with $\delta_{ij}$~being the Kronecker delta symbol and
$s_{\ell,ij}$~the elements of a unitary $2 \times 2$ matrix,
\begin{equation}
  \label{eq:pw_amp.param.s}
  s_\ell(s)
  = \begin{pmatrix}
    \eta_\ell\, e^{2i\, \delta_{\ell,1}} &
    i \sqrt{1 - \eta_\ell^2}\, e^{i (\delta_{\ell,1} + \delta_{\ell,2})} \\[0.2cm]
    i \sqrt{1 - \eta_\ell^2}\, e^{i (\delta_{\ell,1} + \delta_{\ell,2})} &
    \eta_\ell\, e^{2i\, \delta_{\ell,2}}
  \end{pmatrix}
  \eqPunctSpacing.
\end{equation}
The phase shifts $\delta_{\ell,1}(s)$, $\delta_{\ell,2}(s)$ and the
inelasticity $\eta_\ell(s) \in [0, 1]$ in the partial wave~$\ell$ are
real numbers.  The $2 \times 2$ matrix of partial-wave amplitudes is
then
\begin{equation}
  \label{eq:pw_amp.param.t}
  t_\ell(s)
  =
  \begin{pmatrix}
    \dfrac{\eta_\ell\, e^{2i\, \delta_{\ell,1}} - 1}{2i\, \rho_{11}} &
    \dfrac{\sqrt{1 - \eta_\ell^2}\, e^{i (\delta_{\ell,1} + \delta_{\ell,2})}}
    {2 \sqrt{\rho_{11}}\, \sqrt{\rho_{22}}} \\[0.3cm]
    \dfrac{\sqrt{1 - \eta_\ell^2}\, e^{i (\delta_{\ell,1} + \delta_{\ell,2})}}
    {2 \sqrt{\rho_{11}}\, \sqrt{\rho_{22}}}
    & \dfrac{\eta_\ell\, e^{2i\, \delta_{\ell,2}} - 1}{2i\, \rho_{22}}
  \end{pmatrix}
  \eqPunctSpacing.
\end{equation}

For inelastic scattering, $\eta_\ell < 1$, while for elastic
scattering $\eta_\ell = 1$.  In the latter case (single-channel
elastic scattering), we can rewrite \cref{eq:pw_amp.param.element} as
\begin{equation}
  \label{eq:pw_amp.param.elastic}
  \rho(s)\, t_\ell(s)
  = \frac{e^{2i\, \delta_\ell(s)} - 1}{2i}
  = e^{i \delta_\ell(s)}\, \sin\delta_\ell(s)
  = \frac{1}{\cot\delta_\ell(s) - i}\eqPunctSpacing.
\end{equation}
The squared amplitude of a particular partial wave goes through a
maximum for $\delta_\ell(s = M_0^2) = (n + 1/2) \pi$ with
$n \in \mathbb{N}$, corresponding to an isolated resonance at
$\sqrt{s} = M_0$. The energy-dependence of the amplitude around the
resonance mass can be investigated by performing a Taylor expansion of
$\cot\delta_\ell(\sqrt{s})$ around this maximum:
\begin{align}
  \label{eq:cot.expansion}
  \cot\delta_\ell(\sqrt{s})
  &= \cot\delta_\ell(M_0) +
    \left( \sqrt{s} - M_0 \right)
    \left. \frac{\dif{\cot\delta_\ell}}{\dif{\sqrt{s}}} \right|_{\sqrt{s} = M_0}
    + \ldots \nonumber\\
  &\simeq 0 + \left( \sqrt{s} - M_0 \right)
    \left( -\frac{2}{\Gamma_0} \right)\eqPunctSpacing,
\end{align}
where the first derivative of $\cot\delta_\ell$ \wrt~$\sqrt{s}$ at
$\sqrt{s} = M_0$ has been defined as $(-2 / \Gamma_0)$.\footnote{This
  relation can formally be derived from the radial solution of the
  non-relativistic Schr\"odinger equation for a spherically symmetric
  potential of finite range~\cite{Perl:1974}.}  Inserting this into
\cref{eq:pw_amp.param.elastic} then gives
\begin{equation}
  \label{eq:bw.nonrel}
  \rho(s)\, t_\ell(s)
  \simeq \frac{\Gamma_0 / 2}{M_0 - \sqrt{s} - i \Gamma_0 / 2}\eqPunctSpacing,
\end{equation}
which is the non-relativistic form of the Breit--Wigner amplitude with
constant width, denoted by the parameter~$\Gamma_0$ (full width at
half maximum).\footnote{A Fourier transformation of a wave function
  decaying exponentially with a time constant~$2 / \Gamma_0$ leads to
  the same result as \cref{eq:bw.nonrel}.}  Using
\begin{equation}
  M_0^2 - s
  = \left( M_0 + \sqrt{s} \right)\, \left( M_0 - \sqrt{s} \right)
  \simeq 2M_0\, \left( M_0 - \sqrt{s} \right)
\end{equation}
for $\sqrt{s} \simeq M_0$, \cref{eq:bw.nonrel} can be written as
\begin{equation}
  \label{eq:bw.rel}
  \rho(s)\, t_\ell(s)
  \simeq \frac{M_0\, \Gamma_0}{M_0^2 - s - i\, M_0\, \Gamma_0}\eqPunctSpacing,
\end{equation}
which is the relativistic, Lorentz-covariant form of the elastic
Breit--Wigner amplitude with constant width.

\begin{figure}[tbp]
  \centering
  \subfloat[]{%
    \includegraphics[width=\threePlotWidth]{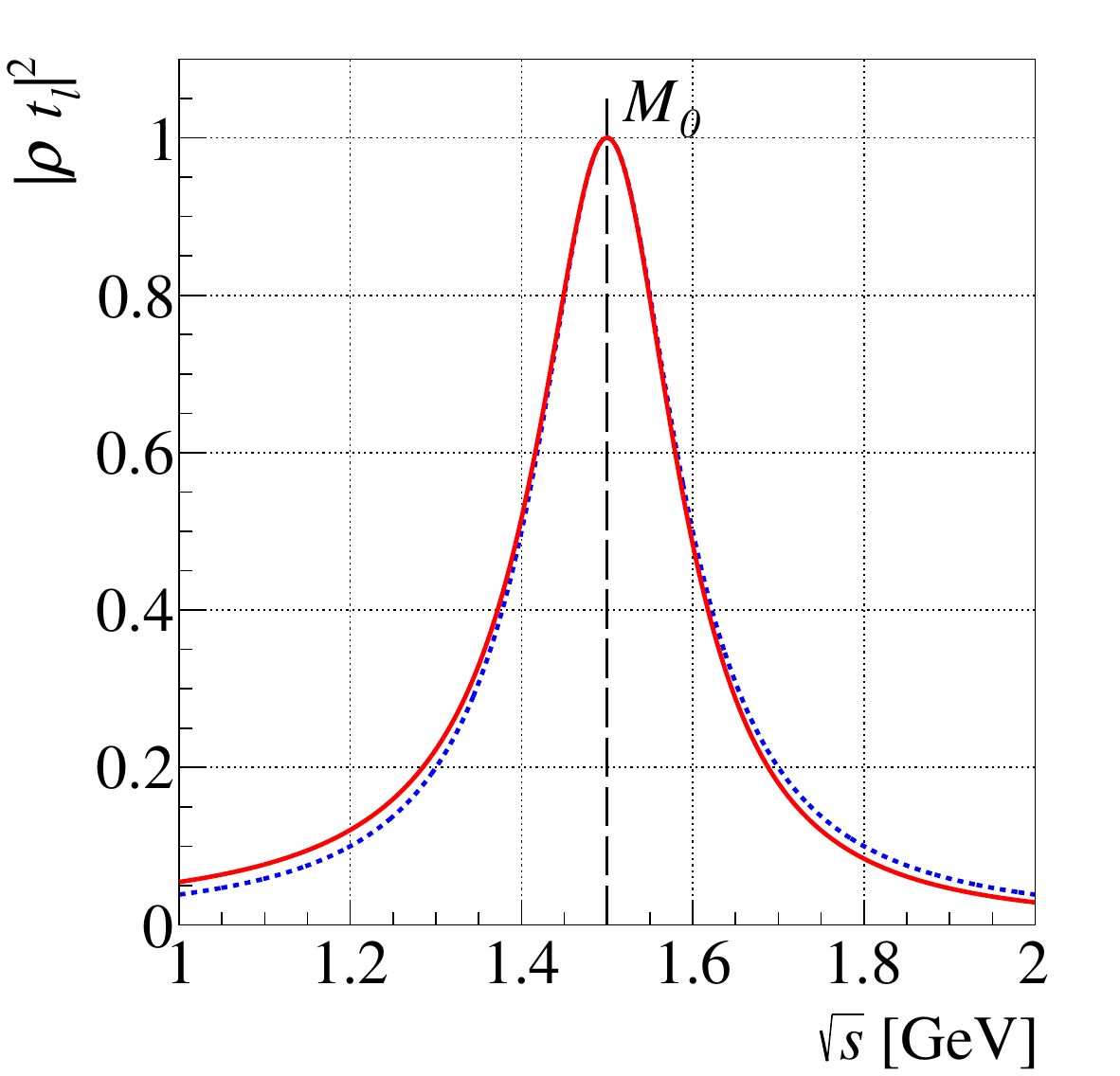}%
    \label{fig:bw.modulus}%
  }%
  \hfill%
  \subfloat[]{%
    \includegraphics[width=\threePlotWidth]{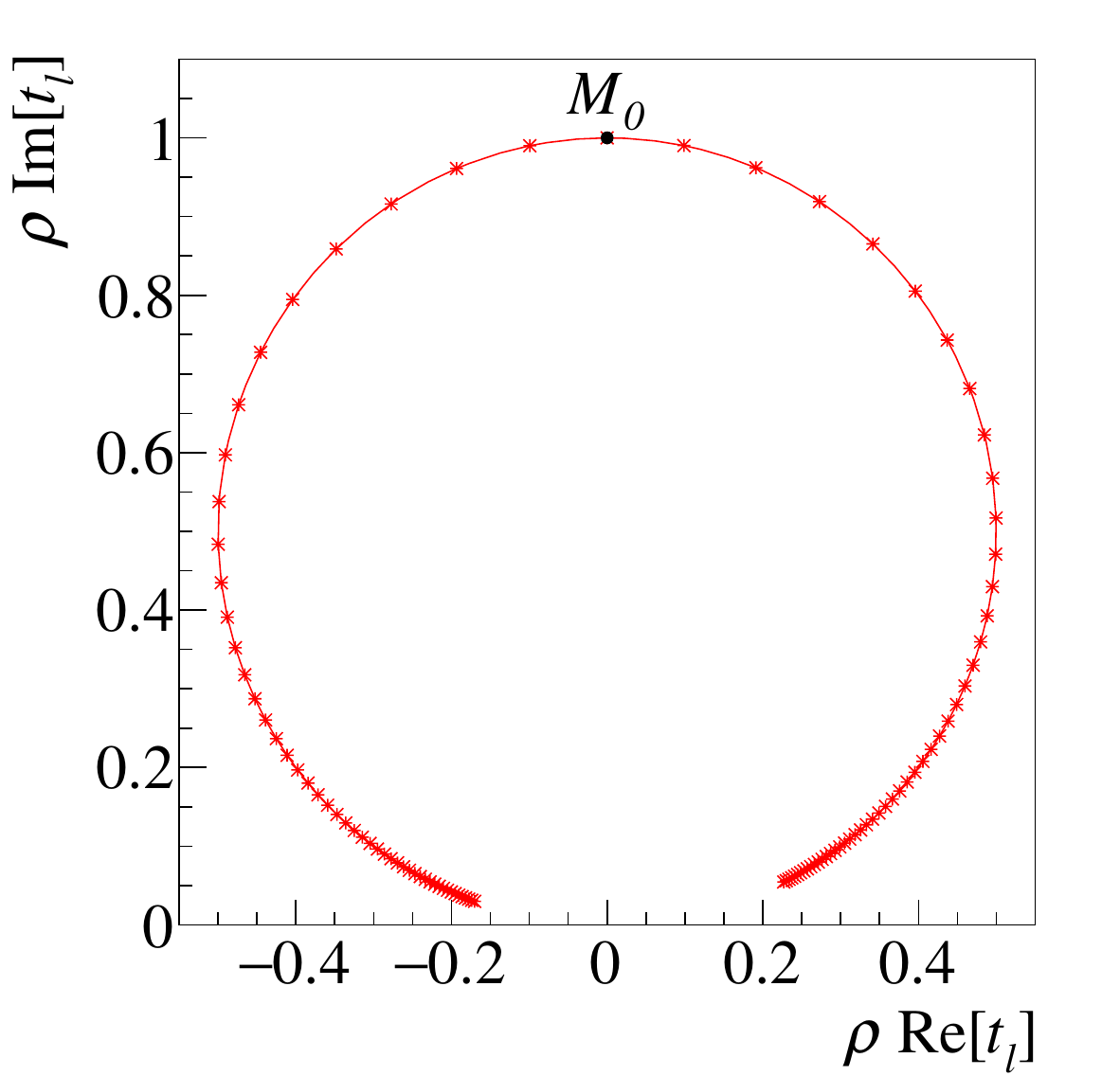}%
    \label{fig:bw.im-re}%
  }%
  \hfill%
  \subfloat[]{%
    \includegraphics[width=\threePlotWidth]{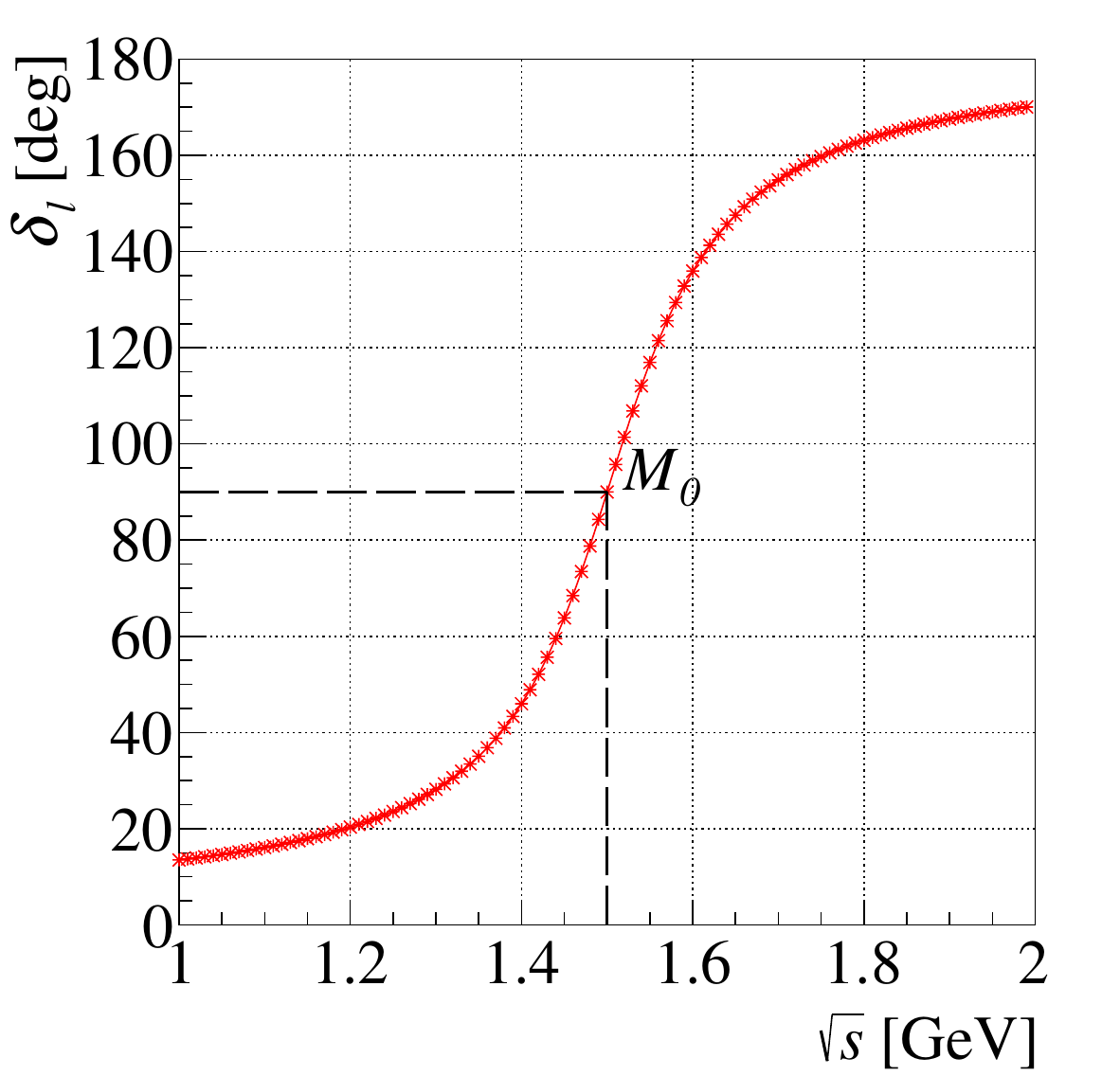}%
    \label{fig:bw.phase}%
  }%
  \caption{Breit--Wigner amplitude for elastic scattering through a
    fictitious resonance in the two-body partial wave with orbital
    angular momentum~$\ell$ with mass $M_0=1500\,\MeV$ and constant
    width $\Gamma_0=200\,\MeV$.
    \subfloatLabel{fig:bw.modulus}~Modulus squared of (red) the
    relativistic, \cref{eq:bw.rel}, and (blue) the non-relativistic
    Breit-Wigner,
    \cref{eq:bw.nonrel}. \subfloatLabel{fig:bw.im-re}~Imaginary versus
    real part (Argand diagram),
    \subfloatLabel{fig:bw.phase}~phase~$\delta_\ell$ as a function
    of~$\sqrt{s}$ for the relativistic Breit-Wigner amplitude.  The
    points in
    \subfloatLabel{fig:bw.im-re}~and~\subfloatLabel{fig:bw.phase} are
    spaced equidistantly in $10\,\MeV$ bins of~$\sqrt{s}$ with
    $s$~increasing in counter-clockwise direction from~1 to
    $2\,\GeV$.}
  \label{fig:bw}
\end{figure}

\Cref{fig:bw} displays the modulus squared, the imaginary versus the
real part (Argand diagram), and the phase of $\rho(s)\, t_\ell(s)$ as
a function of~$\sqrt{s}$.  In \cref{fig:bw.modulus} the relativistic
form is compared to the non-relativistic form.
For elastic scattering, the locus of $\rho(s)\, t_\ell(s)$ follows a
circle of unit diameter around the point $0 + i/2$ in counterclockwise
direction in the Argand diagram shown in \cref{fig:bw.im-re}.  When a
threshold for inelastic processes is passed, we have $\eta_\ell < 1$,
and the locus moves inward from the circle.  The phase
shift~$\delta_\ell$, which is always measured \wrt some
reference\footnote{Compare to the phase shift of a driven oscillator,
  which is measured \wrt the driving force.} undergoes a steep rise
from~0 to~\SI{180}{\degree} when the energy is varied across the
resonance position (see \cref{fig:bw.phase}).  It should be stressed
that the Breit--Wigner approximation for \cref{eq:pw_amp.param.elastic}
in \cref{eq:bw.rel} is only valid for isolated, narrow resonances in
the vicinity of the resonance mass and sufficiently far from
thresholds.  For wider resonances, the opening of the phase space for
the decay products across the resonance width has to be taken into
account by replacing the constant width~$\Gamma_0$ in \cref{eq:bw.rel}
by the dynamic width~$\Gamma(s)$ that will be discussed below:
\begin{equation}
  \label{eq:bw.rel.dyn_width}
  \rho(s)\, t_\ell(s)
  \simeq \frac{M_0\, \Gamma(s)}{M_0^2 - s - i\, M_0\, \Gamma(s)}\eqPunctSpacing.
\end{equation}

\begin{figure}[tbp]
  \centering
  \includegraphics[width=0.25\textwidth]{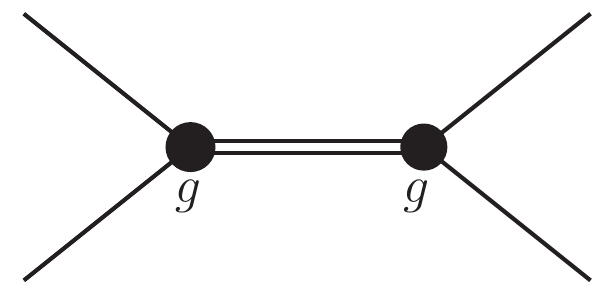}
  \caption{Feynman diagram for the amplitude for scattering of two
    particles via the exchange of a particle, \confer\
    \cref{eq:bw.qft.1}.  The coupling at the vertices is~$g$. }
  \label{fig:resonance_amplitude}
\end{figure}

In quantum field theory, the relativistic Breit--Wigner amplitude with
dynamic width corresponds to the amplitude for scattering of two
particles via the exchange of a particle with mass~$m(s)$ and
four-momentum squared $q^2 = s$, as depicted in
\cref{fig:resonance_amplitude},
\begin{equation}
  \label{eq:bw.qft.1}
  t_\ell(s) = g\, \frac{1}{m^2(s) - q^2}\, g\eqPunctSpacing,
\end{equation}
where $g$~denotes the coupling at the vertices (assumed to be equal
for equal-mass scattering) and $1 / [m^2(s) - q^2]$ is the propagator
term.  For the exchange of unstable particles, we make use of
$m(s) = M_0 - i \Gamma(s) / 2$.  The two-body differential decay width
of an unstable particle with mass~$M_0$ is~\cite{pdg_kinematics:2018}
\begin{equation}
  \label{eq:dGamma.cms}
  \dif{\Gamma}
  = \frac{1}{2M_0}\, \abs{g}^2\, \dif{\Phi_2}\eqPunctSpacing.
\end{equation}
For coupling $g$ independent of~$s$, this yields
$M_0\, \Gamma(s) = \rho(s)\, g^2$ with $\rho(s)$ from
\cref{eq:rho.2body.cms}.  Using
$\left( M_0 - i \Gamma(s) / 2 \right)^2 = M_0^2 - i\, M_0\, \Gamma(s)
- \Gamma(s)^2 / 4 \approx M_0^2 - i\, M_0\, \Gamma(s)$ for
$\Gamma(s) \ll M_0$, \cref{eq:bw.qft.1} then becomes
\begin{equation}
  \label{eq:bw.qft.2}
  t_\ell(s) = \frac{g^2}{M_0^2 - s - i\, \rho(s)\, g^2}\eqPunctSpacing,
\end{equation}
which is equivalent to \cref{eq:bw.rel.dyn_width}, but explicitly
displays the phase-space factor~$\rho(s)$ in the denominator.

Close to the kinematic threshold~$s_\text{thr}$ of the reaction, the
dependence of $\Gamma(s)$ can be estimated from the asymptotic
behavior of the Legendre polynomials in \cref{eq:pw_expansion.s},
$P_\ell(z_s) \proptosim z_s^\ell$ for $z_s\gg 1$.  From
\cref{eq:mandelstam.cms.t.equal_masses}, we find that
\begin{align}
  z_s = \cos{\theta_s} = 1 + \frac{t}{2 \bvec{p}^2}
\end{align}
for equal-mass $2 \to 2$ scattering with $p = q$, implying that
$P_\ell(z_s) \propto 1 / q^{2 \ell}$ for small breakup-momenta~$q$,
which diverges for $q \to 0$, \ie at threshold. In order for the full
scattering amplitude \cref{eq:pw_expansion.s} to remain finite, this
divergence must be cancelled by the appropriate behavior of the
partial-wave amplitudes as $t_{\ell} \propto q^{2 \ell}$.  We hence
obtain for the threshold behavior of \cref{eq:bw.rel.dyn_width},
\begin{equation}
  \label{eq:pw_amp.thr}
  \rho(s)\, t_\ell(s) \propto q^{2 \ell + 1}\eqPunctSpacing,
\end{equation}
where one power of~$q$ obviously arises from the phase-space factor on
the left-hand-side of the equation (see \cref{eq:rho.2body.cms}), and
initial and final state each contribute a barrier factor
of~$q^\ell$. This indicates that partial waves with $\ell > 0$ are
suppressed by the centrifugal barrier close to threshold.\footnote{For
  the more general case of inelastic $2 \to 2$ reactions with
  different masses and different orbital angular momenta~$\ell_i$,
  $\ell_f$ in the initial and finite states, respectively, the
  corresponding partial-wave amplitude behaves as
  $t_\ell \propto p^{\ell_i} q^{\ell_f}$.}  This barrier arises
because the orbital angular momentum is given by~$q$ and the impact
parameter between the particles, which is of the order of the range of
the strong interaction, \ie about $1\,\fm$.  This corresponds via the
Heisenberg uncertainty relation to a range parameter of
$q_R \approx 200\,\MeV$.  At small~$\sqrt{s}$, \ie $q \lesssim q_R$,
higher orbital angular momenta are hence suppressed.

Provided that the denominator in \cref{eq:bw.qft.2} varies slowly
compared to the numerator, \ie the threshold behavior of~$g^2$ follows
the one of~$t_\ell$, we obtain for the threshold behavior of the
dynamic width $\Gamma(s)$
\begin{equation}
  \label{eq:dyn_width_thr}
  \Gamma(s) = \Gamma_0\, \frac{M_0}{\sqrt{s}}\, \left( \frac{q}{q_0} \right)^{2 \ell + 1}\eqPunctSpacing,
\end{equation}
with~$q$ and~$q_0$ being the breakup momenta at~$s$ and~$M_0^2$,
respectively (see \cref{eq:breakup_mom.cms.f}) and $\Gamma_0$~being
the width at $s = M_0^2$.

At high~$\sqrt{s}$, far away from threshold, the behavior in
\cref{eq:pw_amp.thr}, however, is no longer correct, because it would
imply $t_\ell \propto \sqrt{s}^{2 \ell}$, which violates the Froissart
bound~\cite{Froissart:1961ux}. In order to enforce the correct
threshold behavior and damp the indefinite growth at high energies,
model-dependent form factors~$F_\ell$ are commonly introduced so that
\cref{eq:dyn_width_thr} becomes
\begin{equation}
  \label{eq:dyn_width}
  \Gamma(s) = \Gamma_0\, \frac{q}{\sqrt{s}}\, \frac{M_0}{q_0}\, \frac{F_\ell^2(q)}{F_\ell^2(q_0)}\eqPunctSpacing.
\end{equation}
The most popular phenomenological parameterization for these barrier
factors was derived by Blatt and Weisskopf~\cite{Blatt:1952} and later
generalized by von Hippel and Quigg~\cite{VonHippel:1972fg}.  By
solving the non-relativistic Schr\"odinger equation for a spherically
symmetric potential with finite range~$R$ and a sharp boundary, they
showed that the~$F_\ell$ can be expressed in terms of spherical Hankel
functions of the first kind, which are given by
\begin{equation}
  \label{eq:sph_hankel_first}
  h_\ell^{(1)}(x)
  = (-i)^{\ell + 1}\, \frac{e^{i x}}{x}\,
  \sum_{k = 0}^\ell \frac{(\ell + k)!}{(\ell - k)!\, k!}\, \left( \frac{i}{2 x} \right)^k\eqPunctSpacing.
\end{equation}
The~$F_\ell$ are usually written as functions of
$z(s) \coloneqq [q(s) / q_R]^2$, \ie
\begin{equation}
  \label{eq:bw_factor}
  F_\ell^2(z) = \frac{1}{z\, \big| h_\ell^{(1)}(q / q_R) \big|^2}\eqPunctSpacing.
\end{equation}
From \cref{eq:bw_factor,eq:sph_hankel_first} it is apparent that for
$q \to 0$, $F_\ell^2(z) \propto q^{2 \ell}$, consistent with
\cref{eq:pw_amp.thr}.  It is customary to normalize the barrier factor
such that $F_\ell(z = 1) = 1$.  In \cref{sec:barrier_factors}, we list
the normalized barrier factors for~$\ell$ up to~6.

\subsubsection{Analytic Structure of the Amplitude}
\label{sec:scattering.analyticity}

In addition to being a unitary matrix, the $S$-matrix is postulated to
be an analytic function of Lorentz-invariants~\cite{Eden:1966dnq},
that are regarded as complex variables, everywhere in the complex
plane, except for a few well-defined singularities related to the
internal dynamics of the scattering process.  These include poles,
corresponding to stable particles or resonances, and branch points
with cuts required by unitarity. It is these singularities, which
characterize the analytic function. The physical amplitude corresponds
to the boundary values of the analytic function for real-valued
variables, \eg\ the Mandelstam variables for the case of $2\to 2$
scattering of spinless particles.

The postulate of analyticity results from rather general causality
arguments~\cite{Gribov:2009zz}.  It also suggests that the same
complex amplitude, analytically continued from the physical
$s$-channel region to the physical regions of the $t$- and $u$-channel
in the Mandelstam plane in \cref{fig:mandelstam_plane}, correctly
describes also the crossed-channel processes.  This assumption
obviously holds in perturbation theory (Feynman diagrams), and seems
plausible also for the case of hadron scattering.

As mentioned above, it is not possible to calculate the amplitude for
a given process solely based on the properties of the $S$-matrix, \ie
unitarity, analyticity, and crossing symmetry. The theory rather
provides a framework for the development of models and offers
constraints for the corresponding amplitudes, which may or may not be
of concern for a particular application. In any case, the
understanding of the analytic structure of amplitudes is essential for
the interpretation of scattering and lattice QCD data and will be
discussed in the following.

From the unitarity relations \cref{eq:inv_amp.unitarity} or
\cref{eq:pw_amp.unitarity.element}, it is clear that an intermediate
state with $N$~particles of mass $m_1, \ldots, m_N$ can only
contribute to the imaginary part of the amplitude above the
corresponding $N$-particle threshold, \ie if
$s > s_{\text{thr}} \equiv \left( \sum_{i = 1}^N m_i \right)^2$.
Considering for simplicity only particles with equal mass~$m$, the
thresholds are at $4m^2, 9m^2, \ldots$.  Below the three-body
threshold, the imaginary part of the partial-wave amplitude is,
according to \cref{eq:pw_amp.unitarity} for a single $2 \to 2$
channel,
\begin{equation}
  \label{eq:pw_amp.im.1ch}
  \Im t_\ell(s) \equiv \frac{t_\ell(s) - t_\ell^\ast(s)}{2i}
  = \rho(s)\, \abs{t_\ell(s)}^2\eqPunctSpacing.
\end{equation}
In the equal-mass case, the phase-space factor is given by
\cref{eq:rho.equal_masses.cms}.  Hence for real~$s$, the amplitude
$t_\ell(s)$ has a finite imaginary part in the physical region above
the elastic threshold $s_\text{thr} = 4 m^2$, which is the
smallest~$s$ at which a two-particle state can exist.  Below the
elastic threshold the amplitude is purely real.

\begin{figure}[tbp]
  \centering
  \includegraphics[width=0.35\columnwidth]{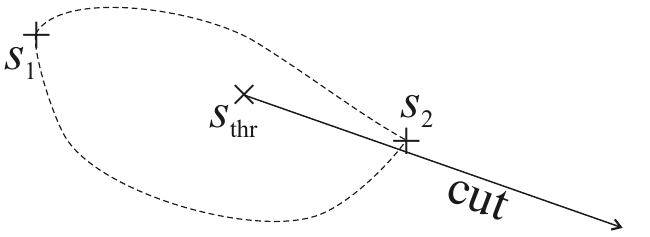}
  \caption{Analytic continuation of the amplitude along different
    paths around a branch point~$s_\text{thr}$.}
  \label{fig:analytic_cont}
\end{figure}

\begin{figure}[tbp]
  \centering
  \includegraphics[width=\columnwidth]{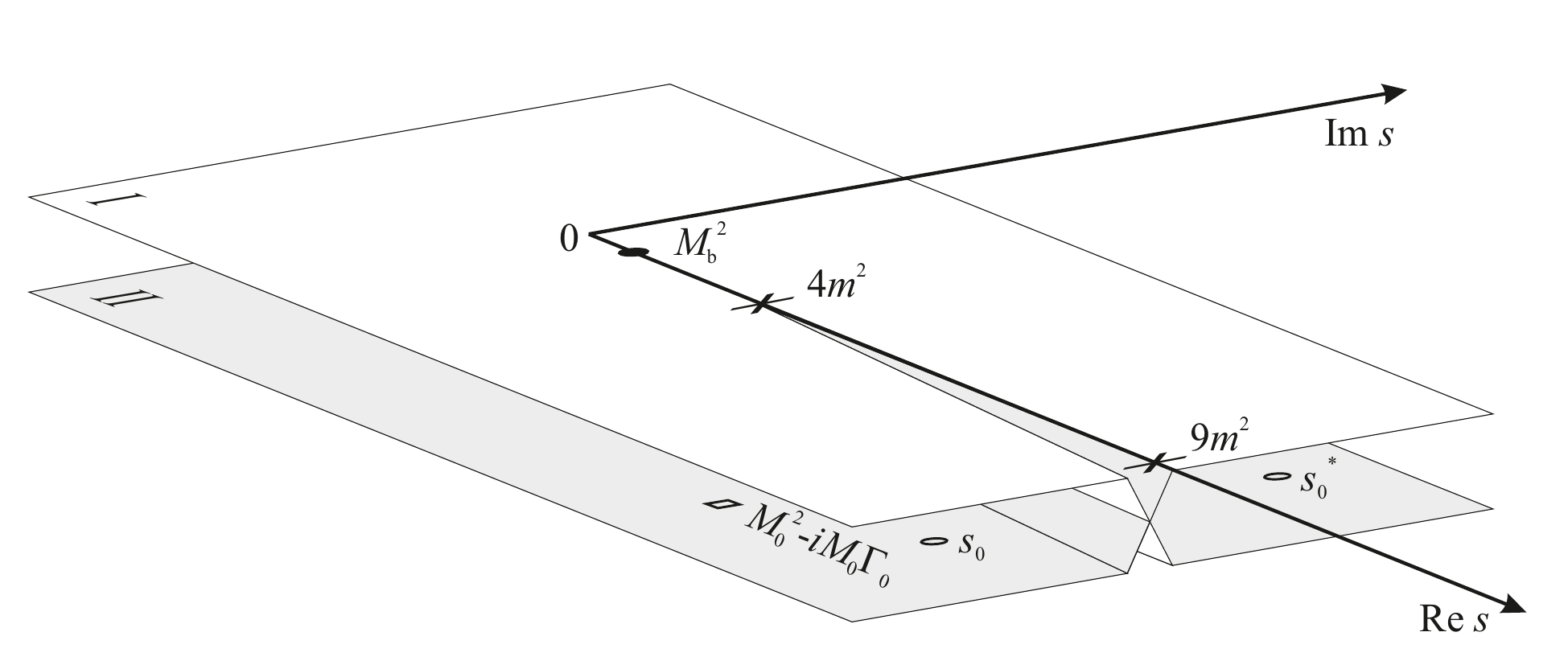}
  \caption{Riemann sheets and singularities of the partial-wave
    amplitude $t_\ell(s)$ in the complex $s$-plane.  The physical
    sheet is labeled~I, the unphysical one~II.  Branch points are
    indicated by crosses.  A conjugate pair of resonance poles on the
    second sheet ($s_0$, $s_0^\ast$) is drawn as open circles, a bound
    state pole on the first sheet ($M_\text{b}^2$) as a filled
    circle.  A single resonance pole corresponding to a Breit--Wigner
    amplitude with mass~$M_0$ and constant width~$\Gamma_0$ (see
    \cref{eq:bw.nonrel,eq:bw.rel}) is indicated by a square.}
  \label{fig:riemann_sheets}
\end{figure}

For complex $s = s_\text{thr} + r\, \exp{i (\theta \pm 2 \pi\, n)}$,
with $n \in \mathbb{N}$, the square root in the numerator of
\cref{eq:rho.equal_masses.cms},
$\sqrt{s - s_\text{thr}} = \sqrt{r}\, \exp{i (\theta / 2 \pm \pi\,
  n)}$, has two different solutions, depending on the choice of~$n$:
\begin{align}
  \label{eq:compl_sqrt_solutions}
  \sqrt{s - s_\text{thr}}
  =
  \begin{cases}
    + \sqrt{r}\, e^{i \theta / 2} & \text{for}~n = 0, 2, 4, \ldots \\
    - \sqrt{r}\, e^{i \theta / 2} & \text{for}~n = 1, 3, 5, \ldots\eqPunctSpacing. \\
  \end{cases}
\end{align}
The threshold~$s_\text{thr}$ is thus a branch point of the amplitude
$t_\ell(s)$; consequently, a cut is attached to the branch point,
indicating that continuing $t_\ell(s)$ from a point~$s_1$ to a
point~$s_2$ along two paths on different sides of the branch point, as
illustrated in \cref{fig:analytic_cont}, will result in different
values of the amplitude.
The different solutions of $t_\ell(s)$ for the same value of~$s$ are
visualized by two planes stacked on top of each other and connected
along the branch cut (Riemann sheets), as shown in
\cref{fig:riemann_sheets}.
Therefore, the amplitude $t_\ell(s)$ has a discontinuity along the
real axis for $s > s_\text{thr}$, \ie the value of $t_\ell(s)$ depends
on whether one approaches the axis from above or from below, which
will be further discussed in \cref{sec:scattering.discontinuities}.
The common convention is that the physical limit for
$s \geq s_\text{thr}$ is reached by approaching the real axis from
above (\textquote{$+i \epsilon$ prescription}),
\begin{equation}
  \label{eq:pw_amp.physical_limit}
  t_\ell(s) = \lim_{\epsilon \to 0+} t_\ell(s + i \epsilon)\eqPunctSpacing.
\end{equation}
The \textquote{physical} sheet (also called sheet~I) is thus defined
by $n = 0, 2, \ldots$ in \cref{eq:compl_sqrt_solutions} and is
indicated in white in \cref{fig:riemann_sheets}.  The second sheet~II
is called \textquote{unphysical}, but, as we will see, may have a
large influence on the amplitude along the physical axis.  The opening
of an additional channel, \eg above the three-particle
threshold~$9m^2$ for equal-mass particles, defines a new branch point
and a corresponding cut.  For $n$~channels, there are $2^n$~Riemann
sheets.  By convention, the cut corresponding to the lowest threshold
is drawn along the positive real $s$~axis, and thus passes through all
other branch points, as indicated in \cref{fig:riemann_sheets}.  This
is called the right-hand (or normal-threshold) cut, which is a
consequence of unitarity when the amplitude is viewed as a function of
complex variables. It has a clear dynamical origin, namely the opening
of the two-particle threshold.  In \cref{eq:rho.equal_masses.cms},
another branchpoint at $s = 0$ appears, which we ignored up to now.
It appears purely because of kinematics and is hence called a
kinematic singularity.  For the case of the phase-space factor, one
can get rid of it by using \eg the Chew--Mandelstam phase-space
function~\cite{Chew:1960iv,Lee:1960zzd,Basdevant:1978tx}, which is an
analytic continuation of \cref{eq:rho.equal_masses.cms}.  In general,
one usually tries to parameterize amplitudes such that they are free
of kinematic singularities and focuses on the dynamic
ones~\cite{Shimada:1978sx,Danilkin:2014cra,Gribov:2009zz}. We will
therefore continue to ignore the branch point at $s = 0$ for our
discussion on the analytic structure of the amplitude.

In addition to the branch points due to unitarity, the
amplitude~$t_\ell$ may contain singularities, corresponding to the
exchange of particles.  Stable bound states appear as poles on the
physical sheet on the real axis below the two-body threshold.
Unstable resonances correspond to poles with a finite imaginary part
in the lower half of the complex $s$-plane. In order for the resonance
to have a noticable effect on the physical amplitude, its pole needs
to be near the physical region. It is therefore located on the second
(or higher) Riemann sheet that is reached from the physical region by
diving down under the threshold cut~\cite{Eden:1971jm}.  The analytic
continuation of the amplitude $t_\ell(s)$ from the physical plane to
the second sheet can be performed using the Schwarz reflection
principle~\cite{Arfken:2011zz}, which states that the analytic
continuation of a complex function $f(z)$, which is analytic in the
upper half of the complex plane ($\Im z > 0$) and which has real
boundary values on part of the real axis, to the lower half of the
complex plane can be done via
\begin{equation}
  \label{eq:schwarz_reflection_principle}
  f(z^\ast) = f^\ast(z)\eqPunctSpacing.
\end{equation}
This implies that, if $t_\ell$~has a pole at complex $s = s_0$, there
must be a corresponding pole at $s = s_0^\ast$, \ie poles outside the
real axis must occur in conjugate pairs for analytic amplitudes.

The Breit--Wigner amplitude with dynamic width in \cref{eq:bw.qft.2}
indeed shows the behavior expected from analyticity: \one~There are no
poles on the first sheet, except for a possible bound-state pole at
$s = M_\text{b}^2$ located on the real axis below the
threshold~$4m^2$, and \two~if there is a complex a pole at $s = s_0$
on the lower half of the second sheet, there is a conjugate pole at
$s = s_0^\ast$ on the upper half, as indicated in
\cref{fig:riemann_sheets}.  It should be noted here that the
Breit--Wigner amplitudes with constant width in
\cref{eq:bw.nonrel,eq:bw.rel} only contain a pole at
$s = M_0^2 - i\, M_0\, \Gamma_0$, \ie on the lower half plane of the
unphysical sheet (see \cref{fig:riemann_sheets}), and not the
conjugate one in the upper half of sheet II.  Hence the Breit--Wigner
amplitudes with constant width violate analyticity.  This may be
acceptable far above threshold, where the pole at~$s_0$ is much closer
to the physical axis than~$s_0^\ast$.\footnote{The pole at~$s_0$ can
  be reached from the physical axis directly by diving under the
  right-hand cut, while in order to reach~$s_0^\ast$, one has to go
  around the branch point at $4 m^2$, as can be seen from
  \cref{fig:riemann_sheets}.}  Therefore, its influence on the
physical amplitude will be much stronger, while the conjugate pole
at~$s_0^\ast$ will have a negligible effect.  This is, however, no
longer true close to threshold, where both poles contribute to the
physical amplitude.\footnote{In fact, it is the interplay of the
  conjugate poles, which makes the amplitude real below threshold.}

\begin{figure}[tbp]
  \centering
  \includegraphics[width=\columnwidth]{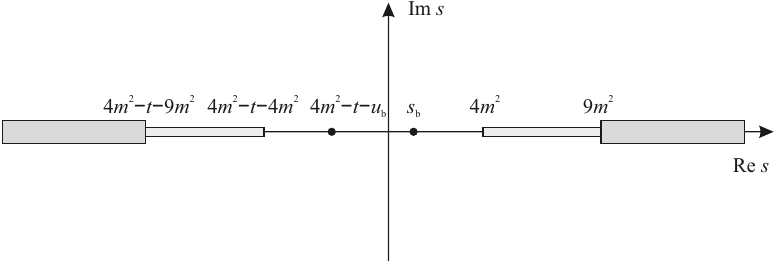}
  \caption{Singularities of the scattering amplitude on the physical
    sheet of the complex $s$-plane for equal-mass $2 \to 2$ scattering
    and fixed~$t$. $s_\text{b}$~and $u_\text{b}$~denote bound-state
    poles in the $s$- and $u$-channel, respectively. The boxes
    represent the branch points and associated unitarity cuts due to
    the opening of two- and three-body channels in the $s$- and $u$-
    reaction channels, respectively.}
  \label{fig:singularities}
\end{figure}

Until now, we have investigated the analytic structure of the
scattering amplitude for the $s$-channel process $1 + 2 \to 3 + 4$.
Analyticity allows us to continue the amplitude from the physically
allowed region for the $s$-channel to the $t$- and $u$-channel
regions.  Analogously to the $s$-channel, where a stable bound state
with mass $M_\text{b} < 4m^2$ gives rise to a pole of the amplitude at
$s = s_\text{b} \equiv M_\text{b}^2$ on the real axis of the physical
sheet, a stable bound state in the $u$-channel leads to a pole at
$u = u_\text{b}$.  The corresponding position of the pole in the
complex $s$~plane is given by \cref{eq:mandelstam.sum}, which for
equal-mass scattering yields $s = 4 m^2 - t - u_\text{b}$ with
$u_\text{b} < 4 m^2$ and $s, t < 0$ (see \cref{fig:mandelstam_plane}).
Similarly, physical thresholds in the $u$-channel lead to branch cuts
in the $s$-plane starting from the two-body branch point $u = 4 m^2$
and extending to infinity along the negative real axis (left-hand
cut), as depicted schematically in \cref{fig:singularities} for
fixed~$t$.
For decreasing $t \leq 0$, the $u$-channel pole and cut move to the
right towards larger values of~$s$ and may even overlap the
$s$-channel singularities.

\subsubsection{Discontinuities}
\label{sec:scattering.discontinuities}

Using the Schwarz reflection principle
\cref{eq:schwarz_reflection_principle}, the partial-wave unitarity
equation \cref{eq:pw_amp.unitarity} can be written as
\begin{equation}
  t_\ell(s) - t_\ell(s^\ast)
  = 2i\, t_\ell(s^\ast)\, \rho(s)\, t_\ell(s)\eqPunctSpacing,
\end{equation}
and, taking the limit onto the physical region,
\begin{equation}
  \label{eq:pw_amp.unitarity.phys_limit}
  t_\ell(s_+) - t_\ell(s_-)
  = 2i\, t_\ell(s_-)\, \rho\, t_\ell(s_+)\eqPunctSpacing,
\end{equation}
where we have defined
$\lim_{\epsilon \to 0+} t_\ell(s + i \epsilon) \eqqcolon t_\ell(s_+)$
and
$\lim_{\epsilon \to 0+} t_\ell(s - i \epsilon) \eqqcolon
t_\ell(s_-)$.\footnote{The phase-space factor is understood to be
  calculated in the limit~$\rho(s_+)$.}  The left-hand-side of
\cref{eq:pw_amp.unitarity.phys_limit} is the difference of the values
of~$t_\ell$ just above and just below the unitarity cut for
$s \geq 4 m^2$, \ie the discontinuity of~$t_\ell$ across the cut:
\begin{equation}
  \label{eq:pw_amp.disc}
  \disc t_\ell(s)
  \coloneqq t_\ell(s_+) - t_\ell(s_-)
  = 2i\, t_\ell(s_-)\, \rho\, t_\ell(s_+)\eqPunctSpacing.
\end{equation}
This derivation of the unitarity relation in terms of the
discontinuity of partial-wave amplitudes started from
\cref{eq:t-matrix.element.unitarity.im}, which relies on
\cref{eq:t-matrix.sym}, and made use of the Schwarz reflection
principle \cref{eq:schwarz_reflection_principle}.  It should be noted
that even if \cref{eq:t-matrix.sym} is not met, the general unitarity
relation \cref{eq:pw_amp.unitarity.general} can be expressed in terms
of the discontinuity of a single analytic amplitude across the
unitarity cut.  This property of scattering amplitudes is called
Hermitian analyticity~\cite{Olive:1962zz,Eden:1966dnq}.

\Cref{eq:pw_amp.disc} can be represented graphically as
\begin{equation}
  \vspace{0pt}
  \begin{minipage}[c]{0.85\linewidth}
    \includegraphics[width=\linewidth]{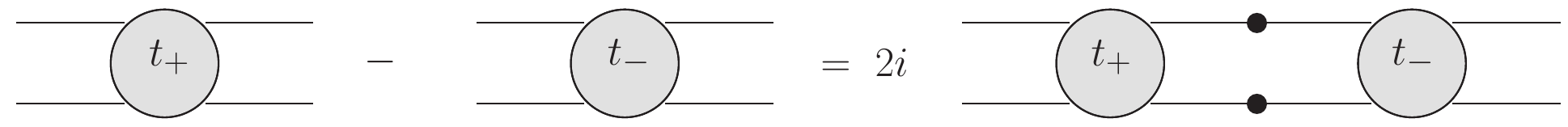}
  \end{minipage}
  \eqPunctSpacing,
  \label{eq:pw_amp.disc.diagram}
\end{equation}
where $t_\pm$ refer to the amplitudes $t_\ell(s_\pm)$, respectively,
and the dots indicate that the intermediate-state particles are on
mass shell.  Expressing the unitarity equations through discontinuity
relations will turn out to be essential to combine unitarity with
analyticity in so-called dispersion relations, as we will show below.

\subsubsection{$K$-Matrix}
\label{sec:scattering.k-matrix}

The unitarity relation \cref{eq:pw_amp.disc} can be cast in a
different form by multiplying both sides on the right by
$t_\ell^{-1}(s_+)$ and on the left by $t_\ell^{-1}(s_-)$:
\begin{equation}
  \label{eq:pw_amp_inv.disc}
  t_\ell^{-1}(s_+) - t_\ell^{-1}(s_-) = -2i\, \rho(s)\eqPunctSpacing.
\end{equation}
Since the discontinuity of~$\rho$ is
$\rho(s_+) - \rho(s_-) = 2\rho(s)$ (\confer\
\cref{eq:compl_sqrt_solutions}), \cref{eq:pw_amp_inv.disc} is
obviously fulfilled by\footnote{It should be noted that the
  phase-space term~$\rho(s)$ in \cref{eq:pw_amp_inv.disc.solution} is
  only one possible choice of functions that have a discontinuity
  of~$2\rho(s)$ across the normal-threshold cut and thus fulfill the
  unitarity condition \cref{eq:pw_amp_inv.disc}. An example are the
  functions introduced by Chew and Mandelstam~\cite{Chew:1960iv}.}
\begin{equation}
  \label{eq:pw_amp_inv.disc.solution}
  t_\ell^{-1}(s) = K_\ell^{-1}(s) - i \rho(s)\eqPunctSpacing,
\end{equation}
where $K_\ell^{-1}$~is a regular function free of the branch point at
$s = 4 m^2$, \ie\ $K_\ell^{-1}(s_+) =
K_\ell^{-1}(s_-)$.  Hence $K_\ell^{-1}$~must be
real.  Solving for~$t_\ell$, we arrive at
\begin{equation}
  \label{eq:pw_amp.k-matrix}
  t_\ell(s)
  = K_\ell(s) \big[ 1 - i\, \rho(s)\, K_\ell(s) \big]^{-1}
  = \big[ 1 - i\, K_\ell(s)\, \rho(s) \big]^{-1}\, K_\ell(s)\eqPunctSpacing.
\end{equation}

Choosing $K_\ell(s) = g^2 / (M_0^2 - s)$, called a
\textquote{bare} pole, where the parameters~$g$ and~$M_0$ do not have
any direct physical meaning, we recover the relativistic Breit--Wigner
amplitude \cref{eq:bw.qft.2}.  The advantage of this so-called
$K$-matrix formalism~\cite{Aitchison:1972ay,Chung:1995dx} is that one
can sum resonance (and possible background) amplitudes in~$K$, with
the resulting~$t_\ell$ still being unitary, in contrast to summing
resonance poles directly in~$t_\ell$.

For more than one channel, \cref{eq:pw_amp.k-matrix} is to be
understood in terms of matrices for $t_\ell$, $K_\ell$, and
$\rho$ in channel space, similar to
\cref{eq:pw_amp.unitarity.element}.  One example of the application of
the $K$-matrix formalism is the Flatt\'e
parameterization~\cite{Flatte:1976xu}, which is commonly used to
describe the \PfZero[980] (\confer\ \cref{sec:3pi_model:pwa}).  It has
two decay channels, $i \coloneqq \pi\pi$ and
$j \coloneqq K \bar{K}$~\cite{Tanabashi:2018zz}, with the threshold
for $K \bar{K}$ being very close to the resonance mass.  Using
\cref{eq:pw_amp.k-matrix} with bare poles
$K_{\ell, ij} = g_i\, g_j / (M_0^2 - s)$, we arrive at
\begin{equation}
  \label{eq:flatte}
  t_{\ell,ij}(s)
  = \frac{g_i\, g_j}{M_0^2 - s - i\, \rho_{ii}(s)\, g_i^2
    - i\, \rho_{jj}(s)\, g_j^2}\eqPunctSpacing.
\end{equation}

\subsubsection{Dispersion Relations}
\label{sec:scattering.disp}

\begin{figure}[tbp]
  \centering
  \includegraphics[width=0.5\textwidth]{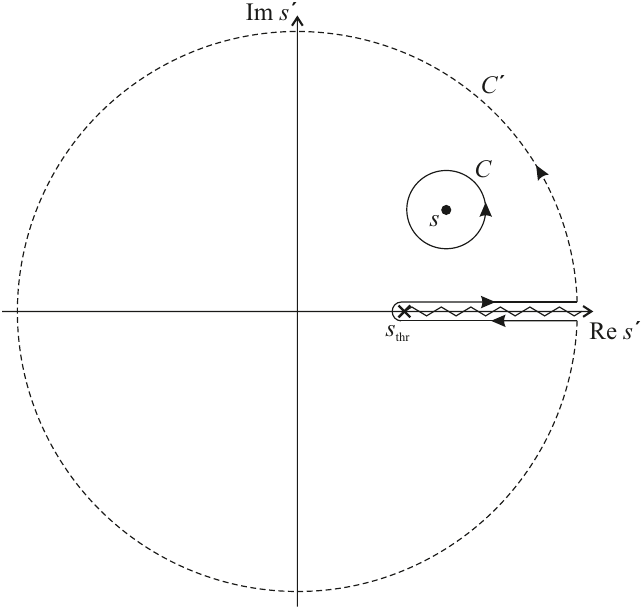}
  \caption{Contours for the Cauchy integral formula
    \cref{eq:cauchy_integral} for the calculation of $f(s)$, which are
    constructed such that $f(s')$ is analytic everywhere inside and on
    the contour.}
  \label{fig:cauchy_contour}
\end{figure}

As we saw in the previous sections, unitarity constrains the imaginary
part of transition amplitudes.  The theory of complex analytic
functions provides a very powerful tool to reconstruct the full
amplitude from its discontinuity by means of dispersion relations.
For an in-depth treatment of dispersion relations in scattering
theory, see \eg~\refsCite{Eden:1966dnq,Weinberg:1995mt}.  Here, we
only review some simple results based on Cauchy's integral
formula~\cite{Arfken:2011zz}, which states that
\begin{equation}
  \label{eq:cauchy_integral}
  f(s) = \frac{1}{2 \pi i}
 \ointctrclockwise\limits_{C} \dif{s'}\, \frac{f(s')}{s' - s}\eqPunctSpacing,
\end{equation}
with $f(s')$ being an analytic function inside and on the closed
contour~$C$, and $s$~any point inside of~$C$.  Since $s' \neq s$ on
the integration path along~$C$, the integral is well-defined.  The
virtue of \cref{eq:cauchy_integral} is that once a function is known
on a closed contour~$C$, it is known everywhere in the region interior
to~$C$.  If $f(s')$ is free of singularities and has only one branch
point at $s' = s_\text{thr}$ and a corresponding cut extending
from~$s_\text{thr}$ to~$+\infty$, we can deform the contour~$C$
to~$C'$ such that its radius goes to infinity and that it excludes the
branch cut, as shown in \cref{fig:cauchy_contour}.
Assuming that $f(s')$ vanishes faster than $1 / s'$ for
$\abs{s} \to \infty$, the circular integral along the dashed line in
\cref{fig:cauchy_contour} does not contribute.  The only remaining
parts of the integral are the integrals along straight lines in
opposite directions above and below the cut, \ie
\begin{equation}
  \label{eq:dispersion_integral}
  f(s)
  = \frac{1}{2 \pi i}
  \int\limits_{s_\text{thr}}^\infty\! \dif{s'}\,
  \frac{f(s_+') - f(s_-')}{s' - s}
  = \frac{1}{2 \pi i}
  \int\limits_{s_\text{thr}}^\infty\! \dif{s'}\,
  \frac{\disc f(s')}{s' - s}\eqPunctSpacing.
\end{equation}
Knowing the discontinuity of $f(s')$ across the normal-threshold cut,
we can reconstruct the full function everywhere in the complex plane
by the dispersion integral \cref{eq:dispersion_integral}.  If $f(s')$
does not fall off quickly enough at ${s'} \to \infty$, the convergence
can be restored by applying \cref{eq:dispersion_integral} to $f(s')$
divided by a polynomial in~$s'$.  Very often, a polynomial of first
order is sufficient, \ie
\begin{equation}
  \label{eq:subtraction}
  \tilde{f}(s') = \frac{f(s')}{s'-  s_1}\eqPunctSpacing,
\end{equation}
with the subtraction point $s_1\neq s$.  This, however, introduces a
singularity in $\tilde{f}(s')$ at $s' = s_1$, which has to be taken
into account when enlarging the contour~$C$ to infinity in
\cref{fig:cauchy_contour}.  The integral of $\tilde{f}(s')$
around~$s_1$ can be calculated using the residue
theorem~\cite{Arfken:2011zz}.  As a result, we get the so-called
once-subtracted dispersion integral,
\begin{equation}
  \label{eq:dispersion_integral_once_sub}
  f(s)
  = f(s_1) + \frac{s - s_1}{2 \pi i}
 \int\limits_{s_\text{thr}}^\infty\! \dif{s'}\,
 \frac{\disc f(s')}{(s' - s_1)(s' - s)}\eqPunctSpacing.
\end{equation}
The subtraction improves convergence of the dispersion integral at the
expense of introducing a new subtraction constant $f(s_1)$, which has
to be determined by external input, \eg from data.\footnote{Even if
  $\lim_{|s'| \to \infty} f(s') = 0$, subtractions may sometimes be
  useful to damp the weight of $f(s')$ at large~$s'$ where it is not
  known precisely, \eg in chiral perturbation theory (see
  \cref{sec:theory.chiPT}).}

\subsubsection{$N$-over-$D$ Method}
\label{sec:scattering.n_over_d}

As discussed in \cref{sec:scattering.k-matrix}, the $K$-matrix
provides a convenient way to take into account the constraints from
unitarity on the partial-wave amplitudes, \ie the right-hand cut.
But, as mentioned in \cref{sec:scattering.analyticity}, partial-wave
amplitudes may in addition have a left-hand cut related to crossed
channels, which are associated with exchange processes.  The
$N$-over-$D$ method allows us to explicitly include the left-hand
singularities of the partial-wave amplitudes (see
\eg~\refCite{Martin:1970xx}).  In order to do so, the partial-wave
amplitude~$t_\ell$, \eg\ the parameterization for single-channel
elastic scattering \cref{eq:pw_amp.param.elastic}, is written as the
quotient of two functions,
\begin{equation}
  \label{eq:n_over_d}
  t_\ell(s)
  = \frac{N_\ell(s)}{D_\ell(s)}
  = N_\ell(s)\, D_\ell^{-1}(s)\eqPunctSpacing,
\end{equation}
such that $D_\ell(s)$ only contains the right-hand cut due to
unitarity, while $N_\ell(s)$ only has the left-hand cut, which depends
on the exchanges in the crossed channels.  Both functions may also
contain possible poles in the respective channels.  Note that poles in
either function can always be represented by zeros of the other.
Often, $D_\ell(s)$ is parameterized to contain universal resonance
poles in the $s$-channel, while $N_\ell(s)$ is a smooth function
representing background processes.  The last expression in
\cref{eq:n_over_d} is also valid in matrix form, \ie\ for multiple
coupled channels~\cite{Bjorken:1960zz}.  Rearranging
\cref{eq:n_over_d} and using unitarity of~$t_\ell$ from
\cref{eq:pw_amp_inv.disc}, the discontinuity of $D_\ell(s)$ across the
right-hand cut can be expressed as
\begin{equation}
  \label{eq:n_over_d.disc_d}
  \disc D_\ell(s)
  \coloneqq D_\ell(s_+) - D_\ell(s_-)
  = \left[ t_\ell^{-1}(s_+) - t_\ell^{-1}(s_-) \right] N_\ell(s)
  = -2i\, \rho(s)\, N_\ell(s)\eqPunctSpacing.
\end{equation}
Making use of \cref{eq:dispersion_integral_once_sub}, the general
solution for \cref{eq:n_over_d.disc_d} can be expressed as a
once-subtracted dispersion integral around the subtraction point
$s_1 = 0$:
\begin{equation}
  \label{eq:n_over_d.dispersion_integral_d}
  D_\ell(s)
  = D_\ell(0) - \frac{s}{\pi}
  \int\limits_{s_\text{thr}}^\infty\! \dif{s'}\,
  \frac{\rho(s')\, N_\ell(s')}{s'\, (s' - s)}\eqPunctSpacing.
\end{equation}
Note that the unitarity condition can still be fulfilled if the
subtraction constant $D_\ell(0)$ is replaced by an analytic function
$D_{\ell,0}(s)$, which does not have a branch point at $s=4m^2$ and no
right-hand cut, \eg $K_\ell^{-1}(s)$ (see
\cref{sec:scattering.k-matrix}).

The connection of the $N$-over-$D$ method to the $K$-matrix discussed
in \cref{sec:scattering.k-matrix} is made by inserting
\cref{eq:n_over_d} into \cref{eq:pw_amp_inv.disc.solution}, which
yields
\begin{equation}
  \label{eq:n_over_d-k_matrix}
  K_\ell^{-1} = (D_\ell + i\, \rho\, N_\ell)\, N_\ell^{-1}\eqPunctSpacing.
\end{equation}

\subsubsection{Final-State Interactions}
\label{sec:scattering.fsi}

\begin{figure}[tbp]
  \centering
  \null\hfill%
  \subfloat[]{%
    \includegraphics[width=0.2\columnwidth]{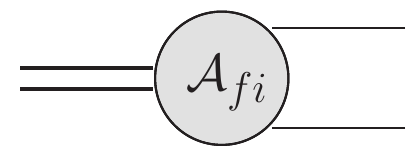}%
    \label{fig:prod_amp}%
  }%
  \hfill%
  \subfloat[]{%
    \includegraphics[width=0.2\columnwidth]{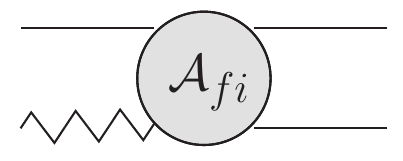}%
    \label{fig:decay_amp}%
  }%
  \hfill\null%
  \caption{Amplitude for the production of a hadronic state, \eg\
    \subfloatLabel{fig:prod_amp}~in the decay of a heavier state or
    \subfloatLabel{fig:decay_amp}~in the interaction of two particles,
    including rescattering.}
  \label{fig:decay_prod_amp}
\end{figure}

Consider the production of a hadronic final state, \eg in the decay of
a heavier state or in the interaction of two particles (\eg
photoproduction or diffractive production), as shown in
\cref{fig:decay_prod_amp}.
The invariant transition amplitude for these processes, which we
denote as~$\mathcal{A}_{fi}$ in order to distinguish it from
$\mathcal{M}_{fi}$, which is used for the hadronic (re)scattering
amplitude in the following, has to satisfy the unitarity relation
\cref{eq:inv_amp.unitarity}.  The right-hand-side of this relation
includes the production of intermediate states~$j$ and the
rescattering of these intermediate states to the final state.
Assuming that the interaction between intermediate and final states
only occurs through the strong interaction (final-state interaction),
we can replace~$\mathcal{A}_{fj}^\ast$ by the scattering
amplitude~$\mathcal{M}_{fj}^\ast$. Performing a partial-wave expansion
of the invariant amplitudes $\mathcal{A}_{fi}$ and $\mathcal{M}_{fj}$,
as in \cref{eq:pw_expansion.s}, we arrive at the unitarity relation
for the partial-wave production amplitude~$a_\ell(s)$ (\confer\
\cref{eq:pw_amp.unitarity}):
\begin{equation}
  \label{eq:pw_amp_prod.unitarity}
  \Im a_\ell(s) = t_\ell^\ast(s)\, \rho(s)\, a_\ell(s)\eqPunctSpacing.
\end{equation}
Here, $t_\ell$~denotes the scattering partial-wave amplitude, which
describes the rescattering processes. In general,
\cref{eq:pw_amp_prod.unitarity} again has to be understood as a matrix
equation containing several channels.  In case of a single open
channel, \ie for elastic rescattering (\eg with $t_\ell$~according to
\cref{eq:pw_amp.param.elastic}), \cref{eq:pw_amp_prod.unitarity}
implies that the phase of the production amplitude is equal to the one
of the elastic rescattering amplitude, since $\Im a_\ell$ has to be
real. This result is known as the Fermi--Watson
theorem~\cite{Watson:1952zz}.

\Cref{eq:pw_amp_prod.unitarity} can be cast into a discontinuity
relation (\confer\ \cref{eq:pw_amp.disc}),
\begin{equation}
  \label{eq:pw_amp_prod.disc}
  \disc a_\ell(s)
  \coloneqq a_\ell(s_+) - a_\ell(s_-)
  = 2i\, t_\ell(s_-)\, \rho\, a_\ell(s_+)\eqPunctSpacing,
\end{equation}
or, graphically,
\begin{equation}
  \vspace{0pt}
  \begin{minipage}[c]{0.85\linewidth}
    \includegraphics[width=\linewidth]{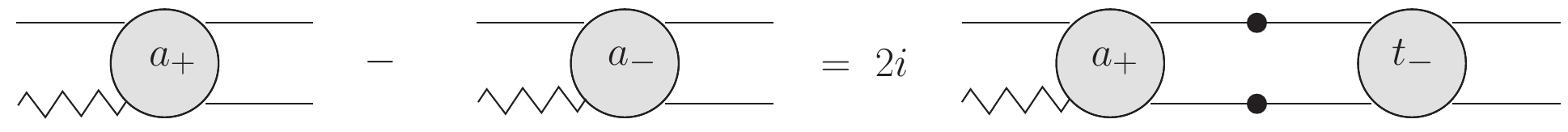}
  \end{minipage}
  \eqPunctSpacing,
  \label{eq:pw_amp_prod.disc.diagram}
\end{equation}
where~$a_\pm$ and~$t_\pm$ refer to $a_\ell(s_\pm)$ and
$t_\ell(s_\pm)$, respectively (\confer\
\cref{eq:pw_amp.disc.diagram}).

A model based on \cref{eq:pw_amp_prod.unitarity} and making use of the
techniques sketched in
\cref{sec:scattering.discontinuities,sec:scattering.k-matrix,sec:scattering.disp,sec:scattering.n_over_d,sec:scattering.fsi}
has been applied to the COMPASS data on $\eta \pi$ and $\eta' \pi$
final states in order to extract pole positions for $\PaTwo$,
$\PaTwo[1700]$, and
$\pi_1(1600)$~\cite{Jackura:2017amb,Rodas:2018owy}, as discussed in
\cref{sec:pwa.unitary_model,sec:etapi_model:resonance,sec:results_1mp,sec:results_2pp}.

%
\subsection{Regge Theory}
\label{sec:regge}

\subsubsection{Motivation}
\label{sec:regge.motivation}

\begin{figure}[tbp]
  \centering
  \null\hfill%
  \subfloat[]{%
    \includegraphics[width=\twoPlotWidth]{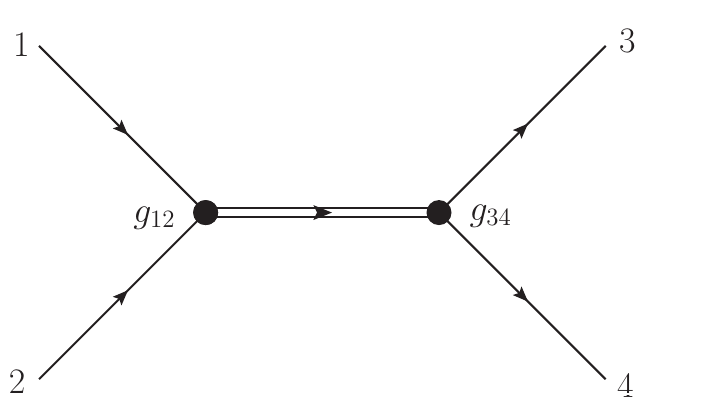}%
    \label{fig:scattering.s-exch}%
  }%
  \hfill%
  \subfloat[]{%
    \includegraphics[width=0.636363\twoPlotWidth]{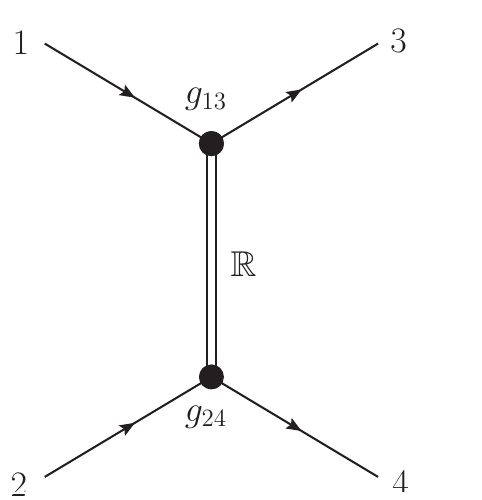}%
    \label{fig:scattering.t-exch}%
  }%
  \hfill\null%
  \caption{\subfloatLabel{fig:scattering.s-exch}~$s$-channel
    scattering $1 + 2 \to 3 + 4$ via the formation of a resonance in
    the intermediate state;
    \subfloatLabel{fig:scattering.t-exch}~$t$-channel scattering via
    the exchange of a Reggeon~\Reg.  The couplings at the
      vertices are denoted by~$g_{ij}$.}
  \label{fig:scattering.exchange}
\end{figure}

The amplitude $\mathcal{M}_{fi}(s, t)$ for the scattering of spinless
particles in the $s$-channel can be expanded in a partial-wave series
according to \cref{eq:pw_expansion.s}.  For small $s > s_\text{thr}$,
the scattering amplitude is dominated by resonances in the
intermediate state (see \cref{fig:scattering.s-exch}), \eg the
$\Delta$~resonance in $\pi^+ p$ scattering for
$\sqrt{s} \approx 1.2\,\GeV$.  Consequently, the partial-wave series
is expected to converge for a limited number of~$\ell$.  For
large~$s$, however, inelastic processes with multi-particle exchanges
dominate (\confer\ \cref{eq:optical_theorem.diagram}), and the
$s$-channel expansion will no longer converge.  There is, however, an
interrelation between the cross section for high-energy $s$-channel
scattering and the exchange of color-singlet clusters of quarks and
gluons in the $t$-channel, as depicted in
\cref{fig:scattering.t-exch}.  It turns out that reactions, for which
the exchange of low-mass particles or resonances is allowed in the
$t$-channel, have considerably higher cross sections than those, for
which the exchanged quantum numbers do not correspond to resonances.
The allowed quantum numbers for the $t$-channel exchanges are
determined by the quantum numbers of the initial and final states of
the corresponding $t$-channel reaction
$1 + \overline{3} \to \overline{2} + 4$ (see
\cref{sec:scattering.kinematics}).  Examples are the reactions
$\PKm \Pp \to \Pgpm \PgSp$ and $\Pap \Pp \to \PagSm \PgSp$, which are
both strongly forward peaked and consequently have a large total cross
section. Regge theory links this experimental observation to the fact
that these reactions may proceed via $t$-channel exchange of~$K^\ast$
and $K$ or $K^\ast$, respectively~\cite{Collins:1984xx}.  This is in
contrast to the reactions $\PKm \Pp \to \Pgpp \PgSm$ and
$\Pap \Pp \to \PagSp \PgSm$, which have a much smaller cross section
with no strong dependence on the scattering
angle~\cite{Loos:1969db,Atherton:1969yy}, because there are no known
resonances with the necessary quantum numbers for $t$-channel
exchange.

According to \cref{eq:compton_wavelength}, the longest-range component
of the force between two hadrons is provided by the exchange of the
lightest color-singlet object carrying the quantum numbers required by
conservation laws.  The exchange of heavier particles will lead to
contributions at shorter ranges.  Therefore, Yukawa's simple one-pion
exchange between interacting hadrons is merely a limiting case of a
much more complicated picture.  Regge theory provides the
generalization of the one-pion exchange model and turned out to be
very successful in describing high-energy scattering processes of
hadrons.

In order to describe $s$-channel scattering processes at large~$s$ and
small~$-t$, we therefore start from a partial-wave expansion in the
$t$-channel in the physically allowed region, \ie $t > 4 m^2$ and
$s < 0$ (equal-mass case),
\begin{equation}
  \label{eq:pw_expansion.t}
  \mathcal{M}_{fi} \big( s(t, z_t), t \big)
  = %
  \sum_{\ell = 0}^\infty (2 \ell + 1)\, t_{\ell,fi}(t)\, P_\ell(z_t)\eqPunctSpacing,
\end{equation}
with
\begin{equation}
  \label{eq:theta.t.cms}
  z_t \coloneqq \cos\theta_t = 1 + \frac{2 s}{t - 4 m^2}
\end{equation}
being the scattering angle in the $t$-channel (\confer\
\cref{eq:theta.s.cms}).

\begin{figure}[tbp]
  \centering
  \includegraphics[width=\textwidth]{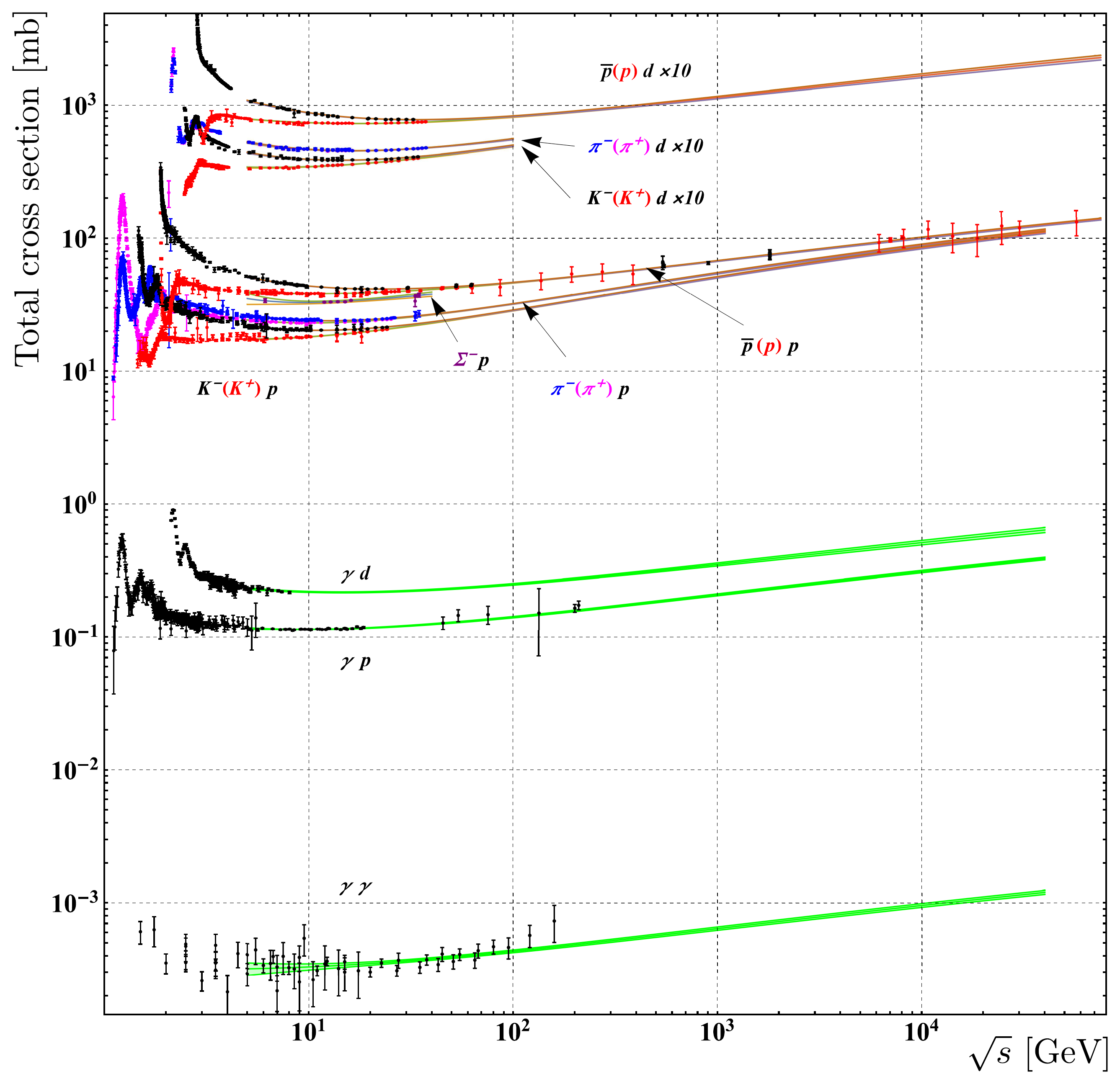}
  \caption{Total cross sections for hadronic, $\gamma \Pp$ and
    $\Pgg \Pgg$~processes as a function of $\sqrt{s}$~from
    \refCite{Patrignani:2016xqp}.}
  \label{fig:sigma_tot}
\end{figure}

Assuming that scattering at large~$s$ proceeds via the exchange of a
single resonance of spin~$J$ in the $t$-channel (as in the
one-pion-exchange approximation), only one partial wave of
\cref{eq:pw_expansion.t} will contribute:
\begin{equation}
  \label{eq:pw_expansion.t.single}
  \mathcal{M}_{fi}(s, t)
  = (2 J + 1)\, t_J(t)\, P_J(z_t)\eqPunctSpacing,
\end{equation}
where we have skipped the channel indices on the right-hand side for
clarity.  At large~$z_t$, $z_t \propto s$ from \cref{eq:theta.t.cms}
and $P_\ell(z_t) \proptosim z_t^\ell$, which gives
$\mathcal{M}_{fi}(s,t) \propto s^J$.  Using the optical theorem
\cref{eq:optical_theorem}, the total cross section is then expected to
scale like $\sigma_\text{tot} \propto s^{J - 1}$ in the high-$s$
limit.  For reactions proceeding via pion exchange, for example, the
cross section would decrease as~$s^{-1}$, while for those involving
a~$\rho(770)$, a constant total cross section would be predicted, and
the exchange of a spin-2 particle would yield a linearly rising cross
section.  This, however, does not match the measured total cross
sections, displayed in \cref{fig:sigma_tot}.
For energies above the resonance region, $\sqrt{s} \gtrsim 2.5\,\GeV$,
the total cross sections first decrease with increasing energy, before
they begin to show a very slow rise at large
$\sqrt{s} \gtrsim 20\,\GeV$.  Instead of a single resonance,
therefore, one has to consider the collective effect of exchanging all
members of a family of resonances in the $t$-channel carrying the
required quantum numbers in order to describe high-$s$ scattering
processes.

The expansion in \cref{eq:pw_expansion.t} in the physical $t$-channel
region, however, cannot be directly applied to high-energy $s$-channel
scattering, because $z_t \propto s$ becomes large for $s \to \infty$.
Since $P_\ell(z_t) \proptosim z_t^\ell$ for large~$z_t$, we would have
$P_\ell(z_t) \proptosim s^\ell$, so that the series would diverge for
large~$s$.  Instead, one has to start in the physical region for the
$t$-channel where the series converges, \ie $t > 4 m^2$ and $s < 0$
(equal-mass case), sum up the series and then analytically continue to
large values of~$s$.

Regge theory provides the mathematical formalism to sum up the
$t$-channel resonances~\cite{Regge:1959mz}.  It requires an extension
of the $t$-channel partial-wave amplitudes $t_\ell(t)$, which enter in
\cref{eq:pw_expansion.t} only for integer values of~$\ell$, to
functions of \emph{complex} angular momentum. We follow the convention
to use the symbol $\alpha$ to denote complex values of angular
momentum, keeping the symbol $\ell$ for integer values. It is thus
assumed that in the complex angular momentum plane, or $\alpha$-plane,
there exists an analytic function $\tpl(\alpha, t)$, which satisfies
\begin{equation}
  \label{eq:tpl_complex}
  \left.\tpl(\alpha, t)\right|_{\alpha=\ell} = t_\ell(t)
  \quad\text{for}~\ell = 0, 1, 2, \ldots\eqPunctSpacing.
\end{equation}
T.~Regge showed that in non-relativistic potential scattering theory
the Schr\"odinger equation for a radially symmetric potential $V(r)$ can
indeed be solved for complex angular momentum~$\alpha$, yielding a
unique analytical extension of the partial-wave amplitudes
$t_\ell(t)$, provided that $\left| \tpl(\alpha, t) \right|$ has a
limited asymptotic behavior for $|\alpha| \to \infty$ (see \eg\
\refsCite{Newton:1964zz,deAlfaro:1965zz}).  For scattering from a
superposition of Yukawa potentials, $\tpl(\alpha, t)$ is found to be
an analytic function except for a number of simple poles in the
complex $\alpha$-plane called Regge poles or Reggeons, which depend
on~$t$ and which we denote as
$\alpha_1(t), \alpha_2(t), \ldots, \alpha_n(t)$.  The term
\textquote{Regge trajectory} is used both for the paths traced out in
the complex $\alpha$-plane by the functions $\alpha_i(t)$ as
$t$~varies and for $\Re[\alpha_i(t)]$ plotted against~$t$.  As long as
a Regge trajectory $\alpha_i(t)$ remains far away from integer
values~$\ell$ of~$\alpha$, there will be little influence of the
corresponding pole on $\tpl(\alpha,t)|_{\alpha = \ell}$. But if one of
the Regge trajectories, $\alpha_n(t)$, passes close to a particular
integer value of $\ell = J$ for a given $t = t_0$, such that
$\Re[\alpha_n(t_0)] = J$ and $\Im[\alpha_n(t_0)] \ll 1$, the amplitude
$\tpl(\alpha,t)|_{\alpha = J}$ will be dominated by this Regge pole
for $t\approx t_0$. The value of~$t_0$, for which
$\alpha_n(t_0)\approx J$, corresponds to the squared mass~$M_J^2$ of a
strong interaction resonance with spin~$J$.\footnote{A bound state of
  spin~$J$ will occur if a Regge trajectory passes \emph{through} the
  point $\alpha = J$ for a value of $t < 4 m^2$~\cite{Perl:1974}.}  A
Regge trajectory passing very close to several integer values of
$\ell = 0,\, 1,\, 2,\ldots$ along the real axis of the complex
$\alpha$-plane as $t$~increases thus predicts a sequence of resonances
with spin $J = 0,\, 1,\, 2,\ldots$ with increasing masses. As was
shown in \cref{sec:pheno.trajectories}, \cref{fig:chew_frautschi}, the
experimentally observed hadrons can indeed be arranged in such
families of resonances, having the same quantum numbers~$I$, $I_3$,
$P$, and~$S$. In addition to non-relativistic scattering theory, this
observation provides another justification for the rather bold
extension to complex angular momentum in Regge theory.

\subsubsection{Regge Poles and Particles}
\label{sec:regge.poles}

In the vicinity of a Regge trajectory $\alpha_n(t)$ passing close to
integer values of $\alpha$, the amplitude corresponding to the
exchange of a whole family of resonances on this trajectory will be
dominated by the Regge pole in the complex $\alpha$-plane, so we can
write it as
\begin{equation}
  \label{eq:tpl.regge_pole}
  \tpl(\alpha, t)
  = \frac{\beta(t)}{\alpha - \alpha_n(t)}\eqPunctSpacing.
\end{equation}
Here, $\beta(t)$ is the residue function at the pole, which determines
the coupling of the pole to the external particles involved in the
scattering process.  By the definition~\cref{eq:tpl_complex},
$\tpl(\alpha, t)$ has to be equal to the $t$-channel partial-wave
amplitude $t_\ell(t)$ for non-negative integer values of~$\alpha$.
The functions $\alpha_i(t)$ and their residues are not predicted by
theory.  They trace out a path in the complex $\alpha$-plane when
$t$~changes (\textquote{Regge trajectories}, \confer\
\cref{sec:pheno.trajectories,sec:regge.motivation}).  For simplicity,
we consider a linear trajectory, as suggested by the Chew--Frautschi
plot in \cref{fig:chew_frautschi},
\begin{equation}
  \label{eq:regge_trajectory.linear}
  \alpha_n(t) = \alpha(0) + \alpha^\prime\, t\eqPunctSpacing,
\end{equation}
with $\alpha_n(t)$ passing close to integer values of $\alpha$ at
$t = M_J^2$, \ie
$\alpha \approx \ell \approx \alpha(0) + \alpha'\, M_J^2$.  Then,
according to \cref{eq:tpl_complex}, the $t$-channel partial-wave
amplitude is
\begin{equation}
  \label{eq:pw_amp.t.regge}
  t_\ell(t) = \left.\tpl(\alpha, t)\right|_{\alpha = \ell}
  \simeq \frac{\beta(t)}{\alpha'\, \left( M_J^2 - t \right)}\eqPunctSpacing.
\end{equation}

Substituting \cref{eq:tpl.regge_pole} into \cref{eq:pw_expansion.t}
yields the following expression for the partial-wave expansion of the
invariant amplitude in the $t$-channel:
\begin{equation}
  \label{eq:pw_expansion.t.regge}
  \mathcal{M}_{fi}\big( s(t, z_t), t \big)
  = \sum_{\ell = 0}^\infty (2 \ell + 1)\,
  \frac{\beta(t)}{\ell - \alpha(t)}\, P_\ell(z_t)\eqPunctSpacing.
\end{equation}
This expression can now be analytically continued to the $s$-channel
physical region, making use of $z_t \propto s$ (see
\cref{eq:theta.t.cms}) and hence $P_\ell(z_t) \proptosim z_t^\ell$ for
large~$s$ and $t < 0$ fixed.  We then arrive at the characteristic
power-law behavior of the invariant amplitude as a function of~$s$ for
large~$s$ and fixed~$t$, originating from the exchange of a single
Regge trajectory of particles or resonances~\cite{Collins:1984xx}:
\begin{equation}
  \label{eq:pw_expansion.t.regge.asym}
  \mathcal{M}_{fi}\big( s(t, z_t), t \big)
  \proptosim \sum_{\ell = 0}^\infty
  (2\ell+1) \frac{\beta(t)}{\ell-\alpha(t)}\, z_t^\ell
  \proptosim \beta(t)\, s^{\alpha(t)}\eqPunctSpacing.
\end{equation}

General properties of the $S$-matrix suggest that the Regge pole
residue $\beta(t)$ factorizes into
\begin{equation}
  \label{eq:regge.residue.factorization}
  \beta(t) = g_{13}(t)\, g_{24}(t)\eqPunctSpacing,
\end{equation}
with $g_{13}(t)$ and $g_{24}(t)$ being the couplings at the upper and
lower Reggeon-exchange vertex in \cref{fig:scattering.t-exch},
respectively.  It has been shown that Regge factorization holds as
long as only a single Regge trajectory contributes to the process and
no other singularities like cuts are
present~\cite{Collins:1977jy,Barone:2002cv}.

\subsubsection{Differential and Total Cross Section}
\label{sec:regge.cross_section}

Regge theory, according to \cref{eq:pw_expansion.t.regge.asym},
predicts that for large~$s$ and $t < 0$ the amplitude of the
$s$-channel reaction produced by a Regge pole in the $t$-channel has
the form
\begin{equation}
  \label{eq:inv_amp.regge.asym}
  \mathcal{M}_{fi} = h(t) \, \left( \frac{s}{s_0} \right)^{\alpha(t)}
  = h(t)\, e^{\alpha(t)\,\ln(s/s_0)}\eqPunctSpacing,
\end{equation}
where $h(t)$ collects all factors depending on~$t$, and a mass scale
$s_0 \coloneqq 1\,\GeV^2$ of the order of hadron masses has been
introduced to make the radix dimensionless.  The differential cross
section for a two-body process, \cref{eq:dsigma_dt.cms.equal_masses},
then asymptotically becomes
\begin{equation}
  \label{eq:dsigma_dt.regge.asym}
  \frac{\dif{\sigma}}{\dif{t}} =
  \frac{1}{16\pi s^2}\,\abs{\mathcal{M}_{fi}}^2
  = \frac{\abs{h(t)}^2 s_0^2}{16\pi}\,
  \left( \frac{s}{s_0} \right)^{2 \alpha(t) - 2} \eqPunctSpacing.
\end{equation}
with $\alpha(t)$ being the leading Regge trajectory,\footnote{In case
  of several Regge poles, the corresponding amplitudes need to be
  summed up. The trajectory that passes closest to real integer values
  of $\alpha$ is the dominant one and called \textquote{leading
    trajectory}.}  which can be exchanged in the process.
Experimentally, it is known that the differential cross section falls
roughly exponentially with~$\abs{t}$ for small~$\abs{t}$.  The
function $h(t)$ can thus be approximated by an exponential form
factor, $h(t) \simeq g\, e^{bt/2}$, with $g$~being a coupling constant
and $b$~an exponential slope~\cite{Perl:1974}.  Substituting
\cref{eq:regge_trajectory.linear} into \cref{eq:dsigma_dt.regge.asym}
yields
\begin{align}
  \label{eq:dsigma_dt.regge.2}
  \frac{\dif{\sigma}}{\dif{t}}
  &= \frac{g^2 s_0^2}{16\pi}\, e^{bt}
    \left( \frac{s}{s_0} \right)^{2 \alpha(0) - 2}
    \left( \frac{s}{s_0} \right)^{2 \alpha'\, t} \nonumber \\
  &= \frac{g^2 s_0^2}{16\pi}\,
    \left( \frac{s}{s_0} \right)^{2 \alpha(0) - 2}
    \exp{\left\{ \left[ b + 2 \alpha'\,
    \ln{\left( \frac{s}{s_0} \right)} \right]\, t \right\}}\eqPunctSpacing.
\end{align}
Because of the slope~$\alpha' > 0$ of the Regge trajectory, the
forward peak toward $\abs{t} = \tmin$ of the differential cross
section is expected to become sharper (\textquote{shrinks}) as $\ln s$
increases.

\begin{figure}[tbp]
  \centering
  \includegraphics[width=0.5\textwidth]{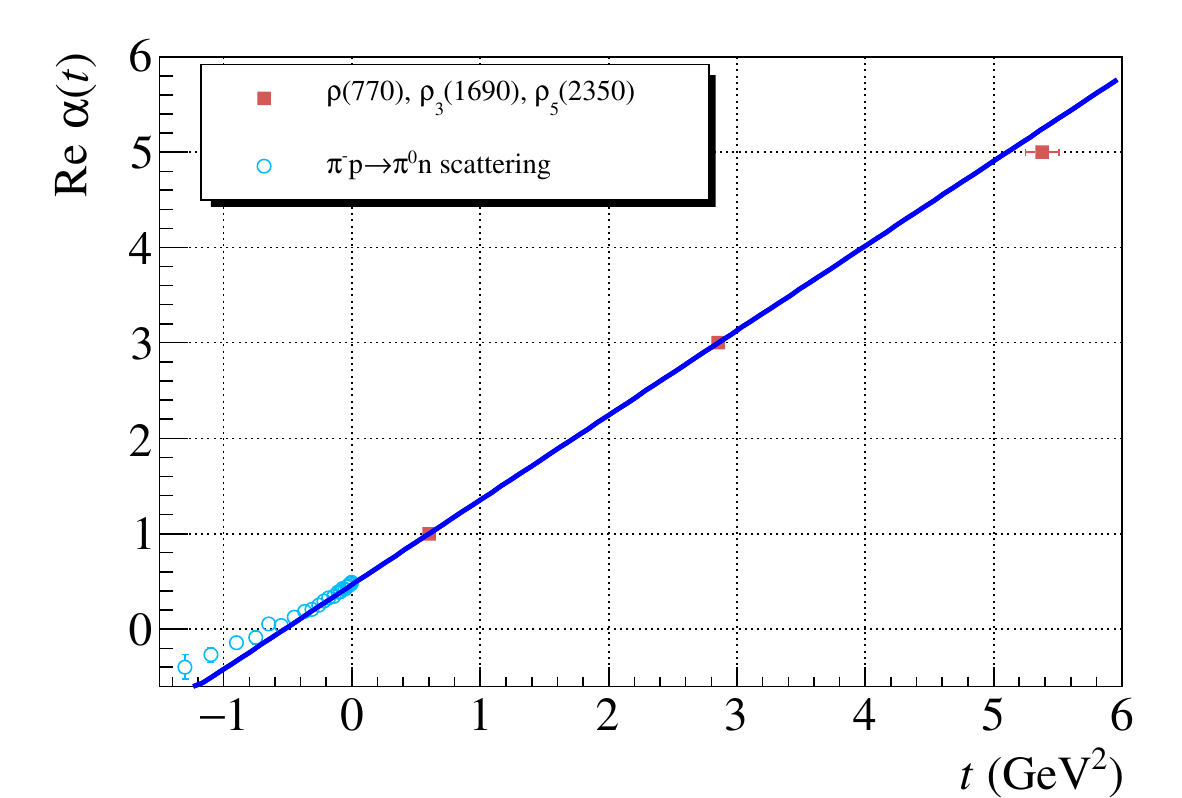}
  \caption{The $\Pgr$~trajectory determined from physical particle
    masses for $t > 0$, and extrapolated to $t < 0$. The data points
    at $t < 0$ are determined by measuring the cross section
    $\dif{\sigma} / \dif{t}$ of the reaction
    $\Pgpm \Pp \to \Pgpz \Pn$~\cite{Barnes:1976ek}.}
  \label{fig:regge.rho_trajectory}
\end{figure}

The trajectory $\alpha(t)$ at high~$s$ and $t < 0$ can be determined
by measuring the differential cross section $\dif{\sigma} / \dif{t}$
as a function of~$s$, and plotting $\ln(\dif{\sigma} / \dif{t})$
versus $\ln{s}$ for fixed values of~$t$.  An example is the
charge-exchange reaction $\Pgpm \Pp \to \Pgpz \Pn$, which in the
$t$-channel reaction $\Pgpm \Pgpz \to \Pap \Pn$ has quantum numbers
$I = 1$, $P = -1$, and $G = +1$, and thus proceeds via exchange of the
$\Pgr$~trajectory.  The measurement of the differential cross section
therefore gives the continuation of the $\Pgr$~trajectory to negative
values of~$t$~\cite{Barnes:1976ek}.  \Cref{fig:regge.rho_trajectory}
shows the scattering data from \refCite{Barnes:1976ek} together with
an extrapolation of the linear $\Pgr$~trajectory fitted to the masses
of $\left\{ \rho, \rho_3, \rho_5 \right\}$ (\confer\
\cref{fig:chew_frautschi}).
The data can be reasonably well described with a single Regge pole,
although some deviations at large values of~$-t$ can be seen, which
may be due to additional contributions to the amplitude beyond the
leading $\rho$~trajectory.

The total cross section is related to the imaginary part of the
forward elastic scattering amplitude via the optical theorem
\cref{eq:optical_theorem}.  Taking only the leading pole of the Regge
amplitude at $t = 0$, we get for large~$s$
\begin{equation}
  \label{eq:sigma_tot.regge}
  \sigma_\text{tot}
  \sim \frac{1}{s}\, \Im \mathcal{M}_{ii}(s, 0)
  \sim s^{\alpha(0) - 1}\eqPunctSpacing.
\end{equation}
\Cref{fig:sigma_tot} suggests that the total cross sections remain
remarkably constant over a wide range of energies for
$\sqrt{s} \gtrsim 10\,\GeV$.  As has been shown in
\cref{fig:chew_frautschi}, all of the known meson trajectories have
$\alpha(0) \lesssim 0.5$, leading to a decreasing cross section with
increasing~$\sqrt{s}$.  In order to obtain an asymptotically constant
total cross section at high energies, a new trajectory called the
Pomeranchuk trajectory, which has $\alpha_\Pom(0) \simeq 1$, was
introduced~\cite{Chew:1961ev}:
\begin{equation}
  \label{eq:trajectory.pomeron}
  \alpha_\Pom(t) = 1 + \epsilon_\Pom + \alpha'_\Pom t\eqPunctSpacing.
\end{equation}
Fitting the $\sqrt{s}$~dependence of hadronic total cross sections,
one obtains values of~$\epsilon_\Pom$ ranging from $0.081$ to
$0.112$~\cite{Amos:1989at,Donnachie:1992ny,Abe:1993xy}.  The slope is
determined by fitting the small-$t$ dependence of the elastic $pp$
scattering cross section at fixed~$\sqrt{s}$, which yields
$\alpha'_\Pom = (0.25 \pm
0.02)\,\GeV^{-2}$~\cite{Jaroszkiewicz:1974ep}.  The trajectory
$\alpha_\Pom(t)$ has been included as a red line in
\cref{fig:chew_frautschi}.  In reaction graphs it is commonly
represented by the exchange of a virtual particle called the Pomeron
($\Pom$).  Processes proceeding via the exchange of the Pomeron are
generally called \emph{diffractive} processes.  The remaining slow
logarithmic rise of the cross section can be attributed to multiple
Pomeron poles~\cite{Cudell:2001pn}.  For elastic scattering
$1 + 2 \to 1 + 2$, the corresponding $t$-channel process has a
particle and its antiparticle in the initial and final states,
respectively, \ie $1 + \overline{1} \to 2 + \overline{2}$.  In order
to be valid for general diffractive reactions, the exchanged
trajectory must have vacuum quantum numbers, \ie
\begin{equation}
  \label{eq:pomeron.qn}
  \mathcal{B} = Q = I = \mathsf{S} = \mathsf{C} = \mathsf{B} = 0
  \quad\text{and}\quad
  P = G = C = \mathcal{S} = +1\eqPunctSpacing,
\end{equation}
with~$\mathcal{S}$ the so-called signature, which appears in the full
expression for the amplitude from Regge theory (see \eg\
\refCite{Collins:1984xx}) and implies that the amplitude has poles
only for even ($\mathcal{S} = +1$) or odd ($\mathcal{S} = -1$) integer
values of $\alpha(t)$, and thus for even or odd partial waves.  The
$f$~mesons carry the required quantum numbers, but the $y$-axis
intercept~$\alpha(0)$ of the corresponding Regge trajectory is too low
(see \cref{fig:chew_frautschi}).

The Pomeron trajectory has $\alpha_{\Pom}(0) \simeq 1$, but there is
no pole and thus no corresponding physical particle near $t = 0$,
because $J = 1$ corresponds to negative signature, whereas the Pomeron
is postulated to have positive signature~\cite{Collins:1977jy}.  The
lowest-mass physical particle of the Pomeron trajectory could have
$\alpha_{\Pom} = J = 2$, but has not been identified uniquely until
now.  It seems plausible to assume that the flavorless exchange in
elastic scattering is dominated by gluon ladders, in contrast to
ordinary Regge trajectories, which include the exchange of valence
quarks.  This might also explain why the Pomeron trajectory has a
different slope compared to all other meson Regge trajectories (see
\cref{fig:chew_frautschi}).  According to
\cref{eq:theory.string.slope}, a different slope also corresponds to a
different string tension for the Pomeron.  Glueballs are discussed as
good candidates for physical particle states on the Pomeron
trajectory.

\subsection{Duality in Hadron Interactions}
\label{sec:duality}

So far, we have treated the two-body $s$-channel scattering reaction
$1+2 \to 3+4$ either via scattering through intermediate $s$-channel
resonances (see \cref{sec:scattering}) or via the exchange of
resonances in the $t$-channel (see \cref{sec:regge}), suggesting that
the $s$-channel resonance mechanism dominates at low energies, while
the $t$-channel exchanges are responsible for the high-energy behavior
of the amplitude (\confer\ also \cref{fig:scattering.exchange}).
These two pictures, however, are not independent of each other, as the
discussion on experimental cross sections for certain reactions in
\cref{sec:regge.motivation} has shown.  The two approaches are rather
two alternative or \emph{dual} ways of describing the same dynamics.
One cannot easily combine the two pictures by simply adding the
corresponding amplitudes without violating unitarity.  The amplitude
for a given $s$-channel reaction is in principle fully determined by
summing up the infinite partial-wave series \cref{eq:pw_expansion.s}
in the $s$-channel.  This, however, is technically difficult, because,
at least at high center-of-momentum energies like at COMPASS, many
high-spin waves will still contribute significantly to the sum.
Therefore, in practice, one has to \emph{truncate} the $s$-channel
partial-wave expansion at some value of~$\ell$.  The contributions
from higher-spin partial waves are then often taken into account by
effective background terms, which are represented by $t$-channel
exchanges.  One example is the Deck-background (see
\cref{fig:exp.production.deck}), which is further discussed in
\cref{sec:exp.prod_reactions}.  It should be noted, however, that this
is an \emph{effective} description of the amplitude, which results
from the truncation of the partial-wave series.  The splitting is of
course not unique and sometimes reaction-dependent.  This may result
in reaction-dependent resonance parameters~\cite{pdg_resonances:2018}.
The pole positions in the complex $s$-plane and their residues (\ie\
couplings), however, should be universal and reaction-independent.
Care has to be taken in order not to violate the unitarity condition
for the $S$-matrix \cref{eq:s-matrix.unitarity} when including
background terms in a model.  This has been studied extensively for
two-body or quasi-two-body scattering (see \eg\
\refsCite{Aitchison:1972ay,Basdevant:1977ya,Basdevant:1978tx}), and is
currently being worked on for three-body scattering (see \eg\
\refsCite{Fleming:1964zz,Niecknig:2015ija,Mai:2017wdv,Mikhasenko:2019vhk}
and references therein).

\subsection{Chiral Perturbation Theory}
\label{sec:theory.chiPT}

At present, QCD is compatible with all strong-interaction phenomena
that are observed at high energies, \ie in the region of asymptotic
freedom.  However, there still exists no analytical method to solve
the QCD Lagrangian at low energies, \ie in the confinement regime.

The masses of the three light quarks~$u$, $d$, and~$s$ are small
compared to the masses of light hadrons.  In the limit of zero masses
of the light quarks, the left-handed and right-handed quark fields
decouple from each other in the QCD Lagrangian.  This leads to a
$\text{SU(3)}_{\text{flavor}, L} \times \text{SU(3)}_{\text{flavor},
  R}$ symmetry, which is called \emph{chiral symmetry}.  This symmetry
can be exploited to construct the low-energy effective field theory of
QCD, where, instead of quarks and gluons, hadrons are the relevant
degrees of freedom.  This approach is called Chiral Perturbation
Theory (\chiPT), which allows the calculation of amplitudes of
strong-interaction processes by performing a systematic expansion in
powers of the particle momenta (for details see \eg\
\refsCite{Meissner:1987ge,Scherer:2012xha} and references therein).
In this regard, it differs from the usual perturbative approach, which
is a power-series expansion in the coupling constant of the
interaction.  The energy scale, defining the applicability of \chiPT,
is defined by the mass of the pion.  The \chiPT expansion contains
universal and process-independent low-energy constants.  The values of
these constants have to be determined by experiments or by other
theoretical calculations, such as lattice QCD.  The leading-order
mesonic Lagrangian contains two low-energy constants: the pion-decay
constant and the scalar quark condensate.  The most general
next-to-leading-order Lagrangian contains 10~low-energy
constants~\cite{Gasser:1983yg}.

The spontaneous breaking of the chiral symmetry in the QCD vacuum
leads to the occurrence of (almost) massless pseudoscalar
Nambu--Goldstone bosons.  They are identified with the observed~$\pi$,
$K$, and $\eta$~mesons, which have smaller masses than the lightest
vector meson, the \Pprho.  In \chiPT, the small masses of the $\pi$,
$K$, and $\eta$~mesons originate from the non-zero quark masses, which
break the chiral symmetry explicitly.

Chiral perturbation theory has become a well-established method for
describing low-energy interactions of the pseudoscalar octet mesons
(see \eg\ \refCite{Scherer:2012xha}).  However, its applicability is
limited to low-energy processes, typically below the threshold for
production of resonances.  For the calculation of resonance
production, one must revert to lattice QCD (see
\cref{sec:pheno.lattice}) or models.
\clearpage{}%

\mathversion{normal}
\clearpage{}%
\section{Experimental Methods}
\label{sec:exp}

This section covers the experimental methods at our disposal to study
the excitation spectrum of light mesons at COMPASS.  After a
discussion of the relevant physical processes in \cref{sec:reactions},
we will describe the setup of the COMPASS experiment in
\cref{sec:compass}, which is optimized to measure scattering reactions
of high-energy hadron beams off a stationary target.

\subsection{Production of Excited Light Mesons at COMPASS}
\label{sec:reactions}
\label{sec:exp.prod_reactions}

A variety of experimental approaches has been developed to produce
excited light-meson states and to study their
properties~\cite{Klempt:2007cp}.  Such resonances can be produced
either in scattering experiments through the interaction of two
colliding particles or as intermediate states in the decay of heavier
states.  Scattering experiments can be further categorized in
production and formation experiments.  In formation experiments,
resonances are formed in the $s$-channel without a recoil particle.
The invariant mass and the quantum numbers of the intermediate and the
final state are fixed by the initial state.  Hence formation
experiments only yield final states with non-exotic quantum numbers.
In production experiments, the total energy is shared between a
multi-meson final state and a recoil particle.  Hence measurements at
a single center-of-momentum energy are sufficient to cover a wide mass
range of the multi-meson final state.  Since the quantum numbers of
the produced multi-meson system are restricted only by the
conservation laws of the interaction, both non-exotic and exotic
states (see \cref{sec:pheno.exotics}) can be produced.  At
high~$\sqrt{s}$, production reactions via $t$-channel exchanges
dominate the cross section.  The dominant contributions in the COMPASS
kinematic domain will be described in this section.

COMPASS is a production experiment where light mesons are produced in
inelastic scattering of a $\pi^-$ beam off proton or nuclear targets.
At the COMPASS beam momentum of \SI{190}{\GeVc}, which corresponds to
a center-of-momentum energy~$\sqrt{s}$ of approximately \SI{19}{GeV},
the total $\pi^- p$ cross section is approximately
\SI{24}{\milli\barn}~\cite{Carroll:1974yx,Carroll:1978vq} (see also
\cref{fig:sigma_tot}), whereas the elastic $\pi^- p$ cross section is
only about \SI{3}{\milli\barn}~\cite{Ljung:1976iv,Schiz:1979rh}.  This
means that most of the scattering processes are inelastic.  In these
inelastic processes, particles may be produced in various ways.
COMPASS has performed exclusive measurements of various produced final
states.  In this paper, we report on the $\etaPi\, p$, $\etaPrPi\, p$,
and $\threePi\, p$ final states.  For the process
$\pi^- + p \to \threePi\, p$, for example, a total cross section of
\SI{635(61)}{\micro\barn} was measured at a similar beam energy of
\SI{205}{GeV}~\cite{Bingham:1974pm}, which corresponds to
\SI{18(2)}{\percent} of the total 4-prong cross section\footnote{This
  is the cross section for exclusive events with 4~charged outgoing
  tracks.} of about \SI{3.5}{\milli\barn}~\cite{Ljung:1976iv} and to
about \SI{2.6}{\percent} of the total cross section.

\begin{figure}[tbp]
  \centering
  \includegraphics[scale=1]{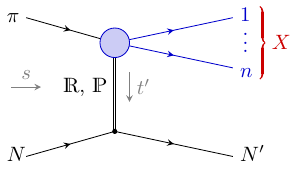}
  \caption{Production of an $n$-body system~$X$ via diffraction of a
    pion beam on a target nucleon~$N$.}
  \label{fig:pion_diffraction}
\end{figure}

In order to study excited mesons, we select processes, where the
target proton scatters elastically (see \cref{sec:exp}).  In these
reactions, the produced final-state particles hence originate from the
beam particle (see \cref{fig:pion_diffraction}).  These processes of
the form $\pi^- + p \to X + p$ are called \emph{single-diffraction
  dissociation processes}\footnote{The name diffraction originates
  from the resemblance of hadronic scattering processes with the
  diffraction of light waves by a black disk.  For both phenomena, the
  measured intensity is characterized by a dominant forward peak, the
  diffraction peak, that is accompanied by a series of minima and
  maxima at characteristic values of~$t$ and the scattering angle,
  respectively, that depend on the size and shape of the object that
  the beam scatters off.}  and are characterized by the relation
$s \gg m_X^2 \gg t$, where $s$~and~$t$ are the Mandelstam variables of
the scattering process as defined in
\cref{eq:mandelstam.s,eq:mandelstam.t} and $m_X$~is the invariant mass
of the $X$~system.  For a \SI{205}{GeV} $\pi^-$ beam, the pion
single-diffraction cross section for the inclusive reaction
$\pi^- + p \to X + p$ was measured to about
\SI{1.9(2)}{\milli\barn}~\cite{Winkelmann:1973bg}.  For the channel
$X = \threePi$, for example, it was found that about 2/3 of the total
cross section of \SI{635(61)}{\micro\barn} quoted above is due to
dissociation of the beam pion, whereas about 1/3 is due to
dissociation of the target proton~\cite{Bingham:1974pm}.

At the center-of-momentum energy considered here, the strong
interaction between the beam and the target particle is described by
$t$-channel exchange of Reggeons (see \cref{sec:regge}).  Since the
target proton remains intact in the reaction, only exchanges with
$Q = 0$, $I = 0, 1$, and $Y = 0$ are possible. Ordinary meson Regge
trajectories like the~$\rho$, $\omega$, or~$f_2$ are suppressed due to
the high center-of-momentum energy (see
\cref{sec:regge.cross_section}).  Hence, the reaction is dominated by
Pomeron exchange.

Because of the approximately exponential behavior of the differential
cross section $\dif{\sigma} / \dif{t}$ in \cref{eq:dsigma_dt.regge.2},
diffractive reactions are characterized by low values of the reduced
four-momentum transfer squared~$t'$ (see \cref{eq:tprime}).  We
therefore limit our analyses to the range $t' < \SI{1}{\GeVcsq}$.  Due
to the small~$t'$, the $X$~system is produced at small scattering
angles and carries most of the beam energy.  This means that the
$X$~system is strongly forward boosted.  In contrast, the recoiling
target proton has only low momentum and is emitted under large angles
\wrt the beam axis.  Hence particles emitted from the target vertex
are kinematically well separated from the particles of the $X$~decay.
This fact is also referred to as \emph{rapidity gap}.  We require a
rapidity gap in the trigger condition of the experiment (see
\cref{sec:exp}).

\begin{figure}[tbp]
  \centering
  \hfill%
  \subfloat[]{%
    \includegraphics[scale=1]{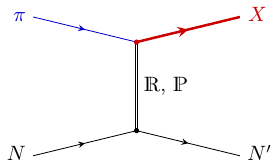}%
    \label{fig:exp.production.diff}%
  }%
  \hfill%
  \subfloat[]{%
    \includegraphics[scale=1]{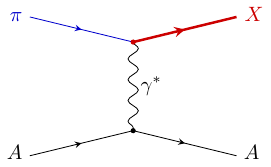}%
    \label{fig:exp.production.coulomb}%
  }%
  \hfill\null%
  \\
  \hfill
  \subfloat[]{%
    \includegraphics[scale=1]{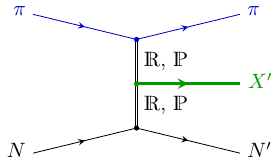}%
    \label{fig:exp.production.central}%
  }%
  \hfill
  \subfloat[]{%
    \includegraphics[scale=1]{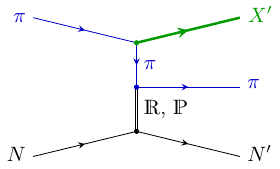}%
    \label{fig:exp.production.deck}%
  }%
  \hfill\null%
  \caption{Production mechanisms of mesons for the COMPASS pion beam.
    \subfloatLabel{fig:exp.production.diff}~Diffractive production of
    an excited meson~$X$ by exchange of a Reggeon or Pomeron with the
    target nucleon~$N$.
    \subfloatLabel{fig:exp.production.coulomb}~Production of an
    excited meson~$X$ by exchange of a quasi-real photon from the
    Coulomb field of the target nucleus~$A$.
    \subfloatLabel{fig:exp.production.central}~and~\subfloatLabel{fig:exp.production.deck}:
    Examples for double-Regge exchange processes that constitute
    protential background for the processes in
    \subfloatLabel{fig:exp.production.diff}~and~\subfloatLabel{fig:exp.production.coulomb},
    if $X'$~represents a subsystem of the final state that $X$~decays
    into.  \Eg, for $X \to 3\pi$, $X'$~would represent the $2\pi$
    subsystem.  For \subfloatLabel{fig:exp.production.diff},
    \subfloatLabel{fig:exp.production.central}, and
    \subfloatLabel{fig:exp.production.deck}, the target could also be
    a nucleus.}
  \label{fig:exp.production}
\end{figure}

At low~$m_X$, the produced system~$X$ is dominated by meson
resonances.  \Cref{fig:exp.production.diff} shows the diagram of such
a reaction.  The produced intermediate states~$X$ are very short-lived
and dissociate via the strong interaction into the measured
forward-going $n$-body final state that consists mostly of~$\pi$, $K$,
$\eta$, and~$\eta'$.

The final states \etaPi, \etaPrPi, and \threePi considered here have
$\IG = 1^-$ quantum numbers identical to those of the beam pion.
Hence only states of the~$a_J$ and $\pi_J$~families, which carry \JPC
quantum numbers of $J^{++}$ and $J^{-+}$, respectively, are accessible
(see \cref{tab:PDG_mesons_2018}).  This means in particular, that only
spin-exotic resonances with $\JPC = 1^{-+},\, 3^{-+},\, \ldots$
quantum numbers can be studied.  Interestingly, model
calculations~\cite{Meyer:2015eta} and lattice QCD calculations (see
\cref{fig:lattice_spectrum_light_ns}) predict the lightest spin-exotic
state to have $\JPC = 1^{-+}$.

In addition to the dominant Pomeron exchange also so-called
\emph{double-Regge exchange processes} contribute to the scattering
cross section.  In these processes, two Reggeons are exchanged and the
produced forward-going particles do not originate from the decay of a
common resonant intermediate state~$X$.  Hence for a given measured
final state, double-Regge exchange processes constitute irreducible
non-resonant background contributions to the resonance production in
\cref{fig:exp.production.diff}.  However, resonances still may appear
in subsystems~$X'$ of the final state.  Various Reggeons may be
exchanged and
\cref{fig:exp.production.central,fig:exp.production.deck} show only
two examples of whole set of possible diagrams.

The exchanged Reggeons affect the kinematic distribution of the
corresponding events.  There are two special cases, which play a role
in the analysis of COMPASS data.  In \emph{central-production
  reactions}, both beam and target particles scatter elastically and a
resonance is produced by the fusion of two Reggeons as shown in
\cref{fig:exp.production.central}.  Such processes are characterized
kinematically by a fast scattered beam particle, which usually carries
a larger momentum than the other final-state particles.  Hence
backgrounds from central production to diffractive reactions are
usually small because they can be separated kinematically.  Another
class of double-Regge exchange processes is the so-called \emph{Deck
  process}~\cite{Deck:1964hm}, where the resonant intermediate state
is produced at the beam vertex and a pion is exchanged.  The pion
interacts with the target particle via Pomeron exchange and emerges as
a final-state particle (see \cref{fig:exp.production.deck}).  The Deck
process is kinematically less well separable from single-diffraction
and therefore plays an important role in the analysis.

\begin{figure}[tbp]
  \centering
  \subfloat[]{%
    \includegraphics[scale=1]{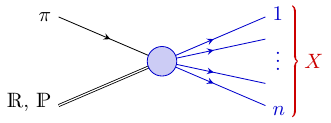}%
    \label{fig:exp.production_factorized}%
  }%
  \\
  \hfill%
  \subfloat[]{%
    \includegraphics[scale=1]{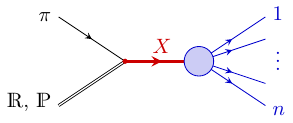}%
    \label{fig:exp.production_factorized_s}%
  }%
  \hfill%
  \subfloat[]{%
    \includegraphics[scale=1]{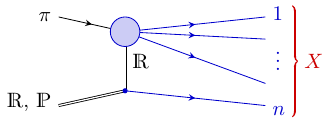}%
    \label{fig:exp.production_factorized_t}%
  }%
  \hfill\null%
  \caption{\subfloatLabel{fig:exp.production_factorized}~Subprocess of
    the pion single-diffraction process shown in
    \cref{fig:pion_diffraction}.  The blue blob in
    \subfloatLabel{fig:exp.production_factorized} contains $s$-channel
    processes that proceed via intermediate states~$X$ as shown
    in~\subfloatLabel{fig:exp.production_factorized_s}.  In addition,
    also $t$-channel exchange processes contribute.
    \subfloatLabel{fig:exp.production_factorized_t}~shows an example
    for such a process.  Depending on the final state, the blue blob
    in~\subfloatLabel{fig:exp.production_factorized_t} may contain
    further $s$-channel subprocesses, \ie intermediate states~$X'$, as
    well as $t$-channel subprocesses.}
  \label{fig:exp.production_factorized_s_vs_t}
\end{figure}

Regge theory allows us to factorize the beam vertex in
\cref{fig:pion_diffraction} from the target vertex.  In the
partial-wave analysis model that will be discussed in
\cref{sec:pwa.analysis_model}, we consider only the subprocess
$\pi^- + \Reg, \Pom \to 1 + \ldots + n$ as shown in
\cref{fig:exp.production_factorized}.  We decompose the $n$-body
system into partial waves by inserting a complete set of intermediate
states~$X$ with well defined quantum numbers as shown in
\cref{fig:exp.production_factorized_s}.  This is equivalent to
resonance production as shown in \cref{fig:exp.production.diff}.  In
the considered subprocess, it corresponds to $s$-channel scattering.
In addition, also $t$-channel processes may contribute.
\Cref{fig:exp.production_factorized_t} shows, as an example, one of
the possible $t$-channel processes, which correspond to the
non-resonant processes shown in
\cref{fig:exp.production.central,fig:exp.production.deck}.  As was
discussed in \cref{sec:duality}, $s$~and $t$-channel processes are
related by duality.  The applied truncated partial-wave expansion in
the $s$-channel provides only an effective description of the data and
the partial-wave amplitudes that are estimated from the data contain
contributions from $s$~as well as $t$-channel processes that need to
be taken into account when modeling the partial-wave amplitudes in
terms of resonances, which will be discussed in
\cref{sec:pwa.res_fit}.

\begin{figure}[tbp]
  \centering
  \includegraphics[scale=1]{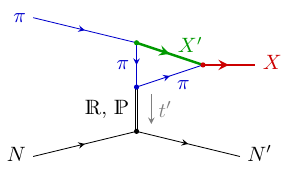}
  \caption{Example diagram for resonance production via rescattering
    in the Deck process.}
  \label{fig:non_res_rescattering}
\end{figure}

Via rescattering, the $t$-channel processes discussed above may also
contribute to resonance production.  \Cref{fig:non_res_rescattering}
shows, as an example, the corresponding lowest-order diagram for a
Deck process in \cref{fig:exp.production.deck}.  Typically, these
diagrams exhibit a different dependence on~$t'$ than the diagrams for
direct resonance production in \cref{fig:exp.production.diff}.

In addition to the processes described above, which are manifestations
of the strong interaction, also electromagnetic processes may
contribute.  In these so-called \emph{Primakoff reactions}, the beam
particle scatters off quasi-real photons in the Coulomb field of the
target particle.  Since the cross section is proportional to~$Z^2$,
where $Z$~is the charge number of the target material, and the photons
are mostly quasi-real, these processes contribute significantly only
for nuclear targets with large~$Z$ and at extremely low
$t' < \SI{e-3}{\GeVcsq}$ at COMPASS energies.  Primakoff reactions
allow us to study $\pi^-$-$\gamma$ reactions, which are difficult to
measure otherwise.  At low values of the $\pi^-$-$\gamma$
center-of-momentum energy, we test predictions from chiral
perturbation theory (see \cref{sec:theory.chiPT}); at higher values,
we study photoproduction of resonances.  Both topics will be covered
in \cref{sec:results_3pic_primakoff}.  The latter process is shown in
\cref{fig:exp.production.coulomb} and we use it to measure the partial
widths of radiative decays $X \to \pi^- + \gamma$, which due to their
smallness are difficult to measure directly.

\subsection{The COMPASS Experiment}
\label{sec:compass}

COMPASS~\cite{Abbon:2007pq} is a large-acceptance, high-resolution
double-magnetic spectrometer, located at the M2 beam line of the 
CERN\footnote{European Organization for Nuclear Research, Geneva, Switzerland.}
Super Proton Synchrotron (SPS).  The beam line delivers high-energy
and high-intensity secondary hadron and tertiary muon beams as well as
a low-intensity electron beam.  This makes COMPASS a unique laboratory
to investigate non-perturbative aspects of QCD related to the
structure and the spectroscopy of hadrons.

The partonic structure of nucleons is studied using hard
electromagnetic reactions of naturally polarized $160\,\GeV/c$ beam
muons with polarized target nucleons.  As polarized targets, solid
$^6$LiD and NH$_3$ have been used in the past, inserted in a
superconducting solenoid magnet providing an extremely homogeneous
magnetic field of $2.5\,\T$ over a length of $130\,\Cm$ along the beam
axis.

Strong or electromagnetic interactions of the beam hadrons with
different targets give access to the excitation spectrum of hadrons.
For this part of the COMPASS physics program, large event samples of
diffractive, central, and Primakoff production reactions of hadronic
beam particles into final states containing charged and neutral
particles have been gathered in several data taking periods in the
years 2004, 2008, 2009, and 2012.  Mesonic resonances and in
particular spin-exotic states are being investigated in a variety of
final states.  In this review, we will focus on \etaPi, \etaPrPi, and
\threePi.  A complete account of the configuration of COMPASS for the
measurements with hadron beams in 2008 and 2009 can be found in
\refCite{Abbon:2014aex}. In the following subsections, a short
overview of the apparatus is given and some of the key equipments for
the spectroscopy program are explained in more detail.\footnote{For
  the data taken with the lead target, which are described in
  \cref{sec:results,sec:results_3pic_primakoff}, the setup was
  slightly different in a few details, which can be found in
  \refCite{Abbon:2007pq}.}

\subsubsection{Setup for Hadron Beams}
\label{sec:compass.overview}

\begin{figure}[tbp]
  \centering
  \includegraphics[width=\textwidth]{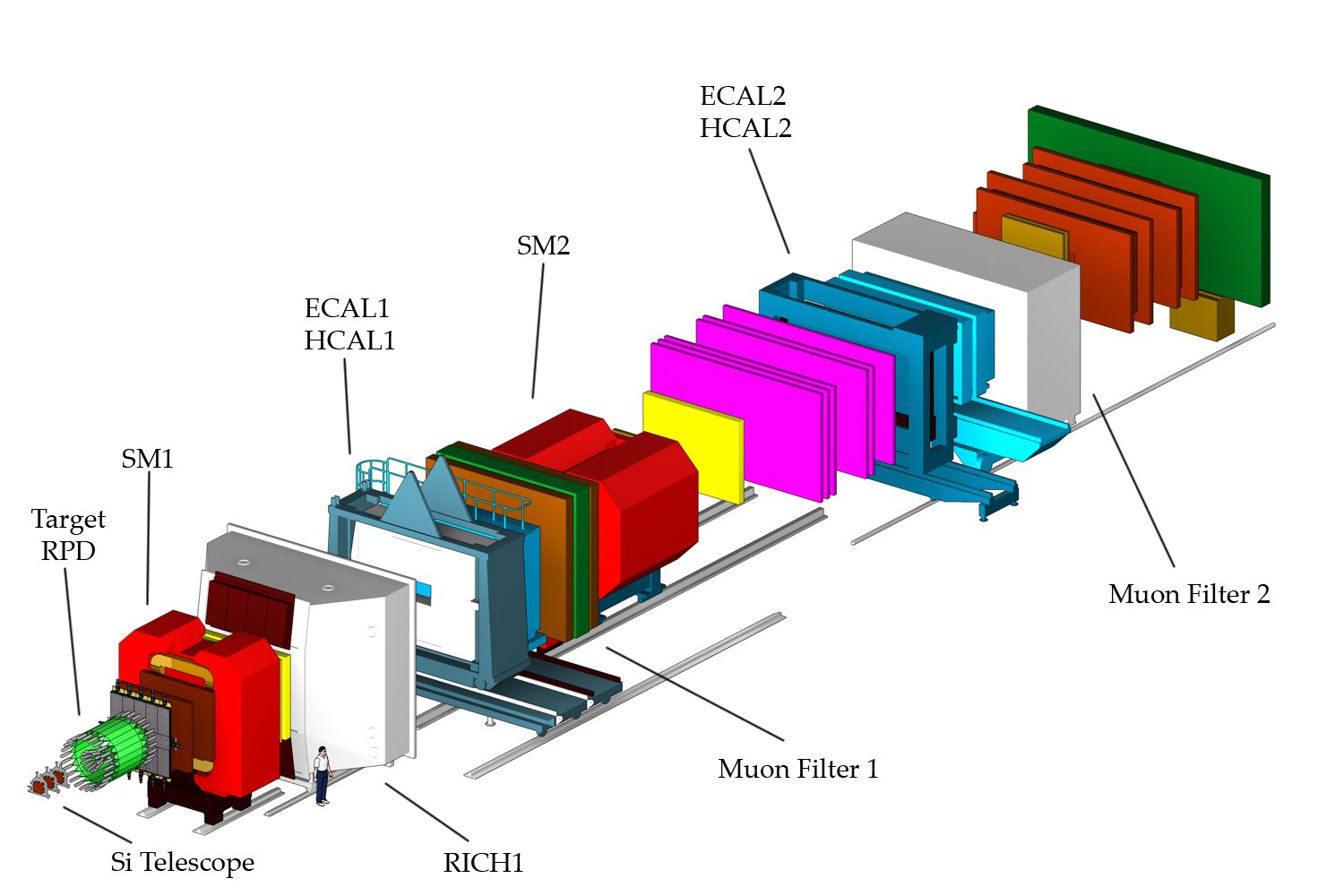}
  \caption{Three-dimensional view of the COMPASS setup for
    measurements with hadron beams~\cite{Abbon:2014aex}.  The
    different colors indicate different detector types.  The CEDAR
    detectors located about $30\,\m$ upstream of the target are not
    shown.}
  \label{fig:layout.3d}
\end{figure}

\begin{figure}[tbp]
  \centering
  \includegraphics[width=0.6\textwidth]{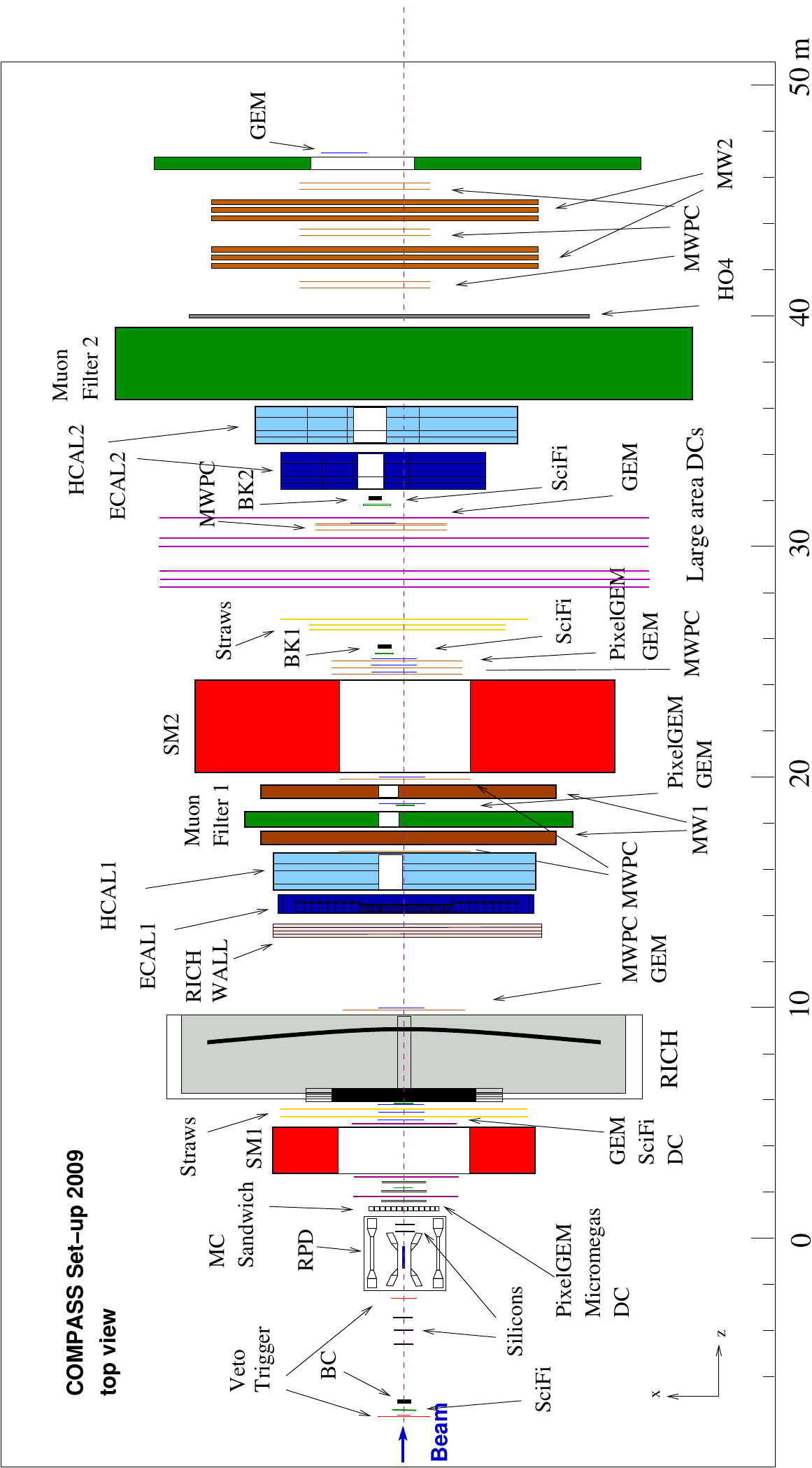}
  \caption{Top view of the COMPASS setup for data taking with hadron
    beams~\cite{Abbon:2014aex}.  The labels indicate the various
    detectors, as referenced throughout this paper.}
  \label{fig:layout_setup}
\end{figure}

The COMPASS fixed-target setup is logically divided into four main
sections along the beam direction.  The first section comprises the
\emph{beam line} located upstream of the COMPASS target and the
detectors which identify the incoming beam particles (CEDAR) and
measure their timing and their coarse positions (Scintillating Fibers,
Beam Counter, Veto Detectors).  The \emph{target region} includes the
various targets used for the measurements with hadron beams (liquid
hydrogen and nuclear targets Pb, W, and Ni) and the detectors
immediately surrounding them (Silicon Microstrip Vertex Detector,
Recoil Proton Detector, Sandwich Veto Detector).  The third section,
called \emph{Large-Angle Spectrometer} (LAS) includes the first dipole
magnet (SM1), the tracking detectors up- and downstream of it, and the
RICH1 detector for particle identification.  The fourth part, called
\emph{Small-Angle Spectrometer} (SAS), is located downstream of the
LAS.  It is built around the second dipole magnet (SM2) and includes
several types of tracking detectors.  Both LAS and SAS comprise a pair
of electromagnetic and hadronic calorimeters, and a muon filter.
\Cref{fig:layout.3d,fig:layout_setup} show a three-dimensional and a
top view of the COMPASS setup, respectively.

\subsubsection{Beam Line}
\label{sec:compass.beamline}

The COMPASS setup is located at the end of the $1.13\,\km$ long M2
beam line of the CERN SPS accelerator.  High-intensity secondary
hadron beams are produced by slow extraction of 9~to~$12\EE{12}$
protons per accelerator cycle with $400\,\GeV/c$ momentum from the SPS
onto a $500\,\mm$ thick production target made of Beryllium.
Particles emitted from the production target are captured using a set
of 6~high-gradient, large-acceptance quadrupoles.  Momentum and charge
selection is performed in several dipole magnets and collimators along
the beam line, with the goal to achieve a momentum resolution below
$1\%$.  The maximum beam momentum for the used beam-line optics is
$225\,\GeV/c$; momenta up to $280\,\GeV/c$ can be reached with a
slightly larger momentum spread.  The results reported in this review
have been obtained with a hadron beam of $(190\pm 1.4)\,\GeV/c$
momentum and an intensity of $5\EE{6}\,\s^{-1}$ during the extraction
from the SPS, which is typically $9.6\,\s$ long with a cycle time of
$30$~to~$48\,\s$.  At $190\,\GeV/c$ momentum, the positive hadron beam
contains $p$ ($74.6\%$), $\pi^+$ ($24\%$), and $K^+$ ($1.4\%$), the
negative beam consists mainly of $\pi^-$ ($96.8\%$), but also contains
small fractions of $K^-$ ($2.4\%$) and $\pbar$ ($0.8\%$).

Switching between different beams is done fully automatic by loading
the corresponding beam elements configuration. This operation
typically takes about thirty minutes.  During data taking with hadron
beams only the trajectory of the incident beam particle is measured,
but not the magnitude of its momentum. The Beam Momentum Station
(BMS), an equipment used for the determination of the incident
momentum during measurements with a muon beam, is moved out of the
hadron beam in order to minimize the material budget along the beam
path.  The momentum spread of the hadron beam arriving at the target
is $1\%$.  About $30\,\m$ upstream of the COMPASS target, two
ChErenkov Differential counters with Achromatic Ring focus (CEDAR)
identify and tag each incoming beam hadron.

\subsubsection{Target Region}
\label{sec:compass.target}

\begin{figure}[tbp]
  \centering
  \includegraphics[width=\textwidth]{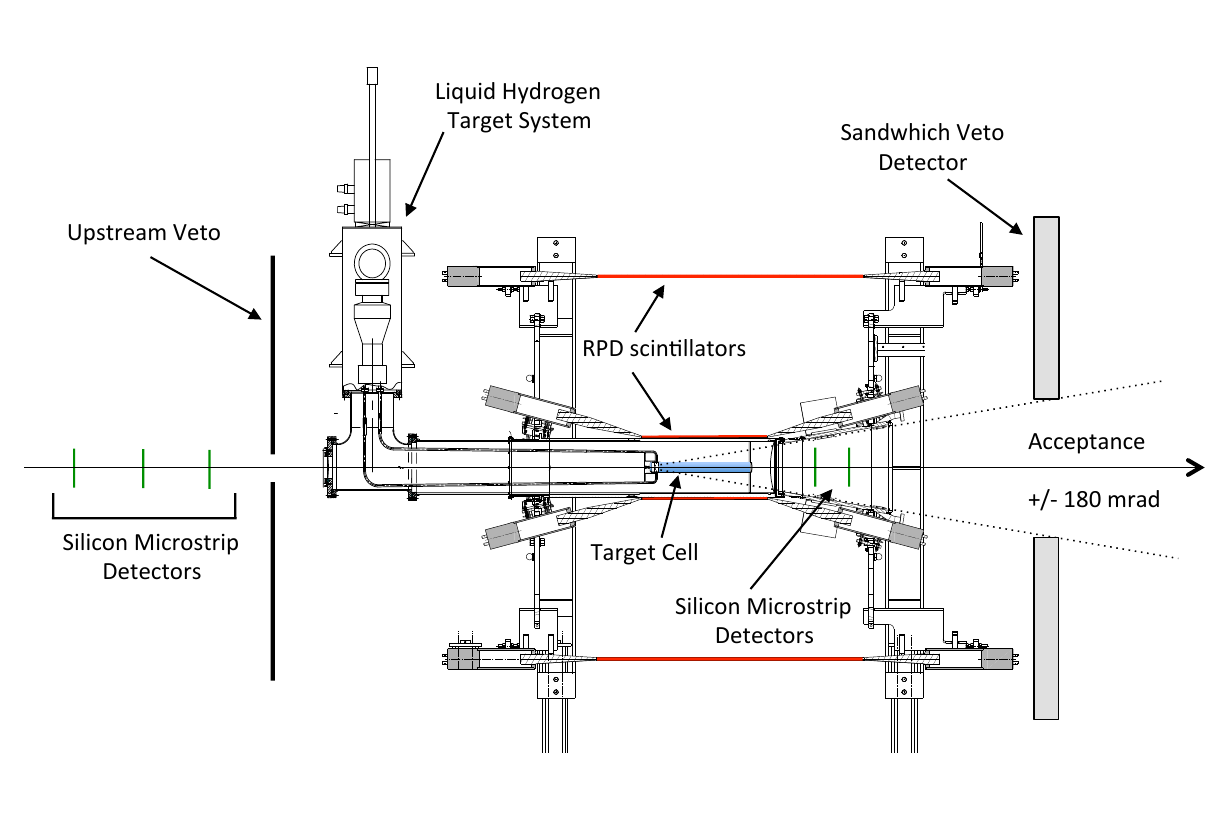}
  \caption{Side view of the target region with the liquid-hydrogen
    target~\cite{Abbon:2014aex}.}
  \label{fig:side-target}
\end{figure}

Most of the data with hadron beams were collected using a cylindrical
liquid-hydrogen target (see \cref{fig:side-target}) with a diameter of
$35\,\mm$ and a length of $400\,\mm$ along the beam direction,
corresponding to $5.5\%$ of a nuclear interaction length and $4.5\%$
of a radiation length.  The target cell is made of $125\,\mum$ thick
Mylar and surrounded by a cryostat tube made of $1.8\,\mm$ thick
aluminum, both faces closed by $250\,\mum$ thick Mylar windows.
Inside the cell, the hydrogen is kept just at the phase transition
point at a pressure of $1200\,\mbar$, resulting in a small fraction of
evaporated gas at the top.  In order to allow for convection and heat
exchange, the cylinder axis is tilted \wrt the beam axis by lowering
the downstream end by $1\,\mm$.  During part of the data taking, the
liquid-hydrogen target was removed and replaced with a specially
designed solid-target holder, onto which several solid targets with
different atomic numbers (Ni, Pb, W) and different thicknesses were
mounted.

The recoil-proton trigger and the momentum-conservation criterion
applied in the event selection suppress such events on average by
about an order of magnitude. The remaining contributions consist
predominantly of low-mass $N^\ast$ produced at large $t'$. In
diffractive reactions, target and beam vertex factorize, so that these
events are expected to have only little effect on the production of
the three-pion final state.

A dedicated time-of-flight detector, called Recoil Proton Detector
(RPD)~\cite{Bernhard:2690563}, was installed around the
target.  This detector serves a two-fold purpose.  First, it was used
in the trigger to select single-diffractive events with a recoiling
proton.  Second, the information from this detector is used in the
offline analysis~(see \cref{sec:compass.select}) to fully reconstruct
the produced final states and to ensure the exclusivity of the
processes under investigation by measuring the velocity and the energy
loss of particles recoiling from the target at angles between
$50^{\circ}$ and $90^{\circ}$ \wrt the beam axis ($z$~axis).  With
this information, double-diffraction events with simultaneous
excitation of the target particle (see \cref{sec:exp.prod_reactions})
were suppressed by about one order of magnitude.  The PRD consists of
two concentric barrels (\textquote{rings}) made of plastic
scintillator slabs (12~for the inner ring with radius $12\,\Cm$,
24~for the outer ring with radius $75\,\Cm$), with a length of
$50\,\Cm$ and $115\,\Cm$, respectively.  Each slab is read out from
both sides by photomultiplier tubes, providing position resolutions
along~$z$ of $\sigma_z = 2.7\,\Cm$ and $5.0\,\Cm$ for the inner and
outer ring slabs, respectively.  RPD straight radial track candidates
are reconstructed by combining hits from the inner ring elements with
the corresponding outer ring elements in the azimuthal range covered
by the inner element. The momenta of the recoil particles are
calculated from the measured time-of-flight using the proton-mass
hypothesis, applying a momentum correction to take into account the
energy loss in the material crossed.  Protons recoiling from the
target at large angles can be identified by their energy loss in the
outer ring up to a velocity of $\beta = 0.4$.  A lower momentum
cut-off of $270\,\MeV/c$ for detectable recoil protons is given by the
energy loss in the target walls and the inner ring.  This translates
into a minimum momentum transfer of $|t| = 0.07\,(\GeV/c)^2$. For
measurements with nuclear targets, the recoil particle is mostly
stopped inside the target and hence does not reach the recoil
detector.

For the precise determination of the interaction vertex position, the
target is surrounded by 5~stations of Silicon Microstrip Detectors.
Three stations are mounted upstream of the target.  Together with a
scintillating-fiber counter (SciFi1) installed about $7\,\m$ upstream
of the target, these detectors measure the beam particle trajectory
and its time before entering the target.  Two other silicon stations
are positioned immediately downstream of the target, still inside the
RPD as indicated in \cref{fig:side-target}, to measure the tracks of
outgoing charged particles.  One station comprises two silicon
detectors, each read out by microstrips with a pitch of about
$50\,\mum$ on both sides, oriented in perpendicular directions.  In
order to resolve hit ambiguities, the strips of the two detectors in a
station form a stereo angle of $\pm 2.5^\circ$ \wrt the horizontal and
vertical axes.  The detectors are operated at a temperature of
$200\,\K$ in order to minimize effects of radiation damage and to
improve the signal-to-noise ratio.  The vertex resolution achieved
with this system is typically of the order of $1\,\mm$ along the beam
axis and $15\,\mum$ in the plane transverse to the beam, but of course
depends strongly on the number of outgoing charged particles and their
angles.

Immediately downstream of the RPD, a Sandwich Veto
Detector~\cite{Schluter:2011rv} was installed in order to veto events
with charged or neutral particles emitted at large polar angles with
respect to the incident beam, thus enhancing single-diffractive
reactions by requiring a rapidity gap between the forward-going
system~$X$ and the recoil proton.  Including the detector in the
trigger logic increased the purity of the diffractive trigger by a
factor of about~\num{3.5}.  The detector consists of a sandwich of
five layers of $5\,\mm$ thick lead plates, reinforced by $1\,\mm$
steel plates on each side, interspersed by five layers of plastic
scintillators, the first three with a thickness of $10\,\mm$, the last
two with a thickness of $5\,\mm$.  The light from the scintillators is
extracted by wavelength-shifting fibers and read out by
photomultiplier tubes.  The efficiency of the detector was determined
to be $98\%$ for minimum ionizing particles.  It has a transverse size
of $2 \times 2\,\m^2$ with an inner hole matching the opening angle of
the RPD and the angular acceptance of the spectrometer.

\subsubsection{Large-Angle Spectrometer}
\label{sec:compass.las}

The Large-Angle Spectrometer (LAS) connects immediately downstream of
the target region and includes the detectors up- and downstream of the
first dipole spectrometer magnet (SM1), located at a distance of about
$4\,\m$ downstream of the target center and operated at a field
integral of $1\,\T\m$.  Its large aperture matches the opening of the
RPD and the Sandwich Veto.  Tracking is performed using different
types of planar gaseous detectors, depending on the surface to be
covered, the expected rates and the required resolutions.  The
large-area tracking is performed by drift chambers, drift tubes, and
multi-wire proportional chambers (MWPCs) with active areas up to
$5\,\m^2$.  All of them have central inactive areas with diameters of
$20$~to $30\,\Cm$, covered by Micropattern Gaseous Detectors with
higher rate capabilities and better resolutions.  Upstream of SM1,
this region is covered by Micromegas (Micromesh Gaseous Structure)
detectors, optimized for operation in hadron
beams~\cite{Abbon:2014aex}.  Downstream of SM1, GEM (Gas Electron
Multiplier) detectors with a two-coordinate
readout~\cite{Altunbas:02a} are used for this purpose.  Since both
GEMs and Micromegas feature strip-patterned anodes, they have a
central inactive region of $5\,\Cm$ diameter in order to avoid
excessive pile-up of signals on the central strips due to beam
particles.  For tracking directly in and close to the beam region, GEM
detectors with pixelized anodes (PixelGEMs)~\cite{Ketzer:07a} are
installed in replacement of all but one scintillating-fiber counters
previously used with the muon beam.  The momentum resolution for
charged particles with momenta from~$1$ up to about $15\,\GeV/c$,
detected only in the LAS, was determined to be between $1.1$~and
$1.5\%$.

Charged hadrons are identified in a Ring-Imaging Cherenkov detector
(RICH) filled with $\text{C}_4\text{F}_{10}$ gas~\cite{Abbon:2006bb}.
The Cherenkov light emitted by charged particles passing the radiator
gas is focused by a large mirror onto two different types of photon
detectors located outside of the geometrical acceptance of the
spectrometer.  In the peripheral region, MWPCs with CsI photocathodes
are used, while the central region is instrumented with multi-anode
photomultipliers.  The RICH can be used to distinguish pions and kaons
with momenta between~$5$ and $43\,\GeV/c$ at a level of $2.5\sigma$ or
better.

\begin{figure}[tbp]
  \centering
  \subfloat[]{%
    \includegraphics[width=0.557\textwidth,valign=m]{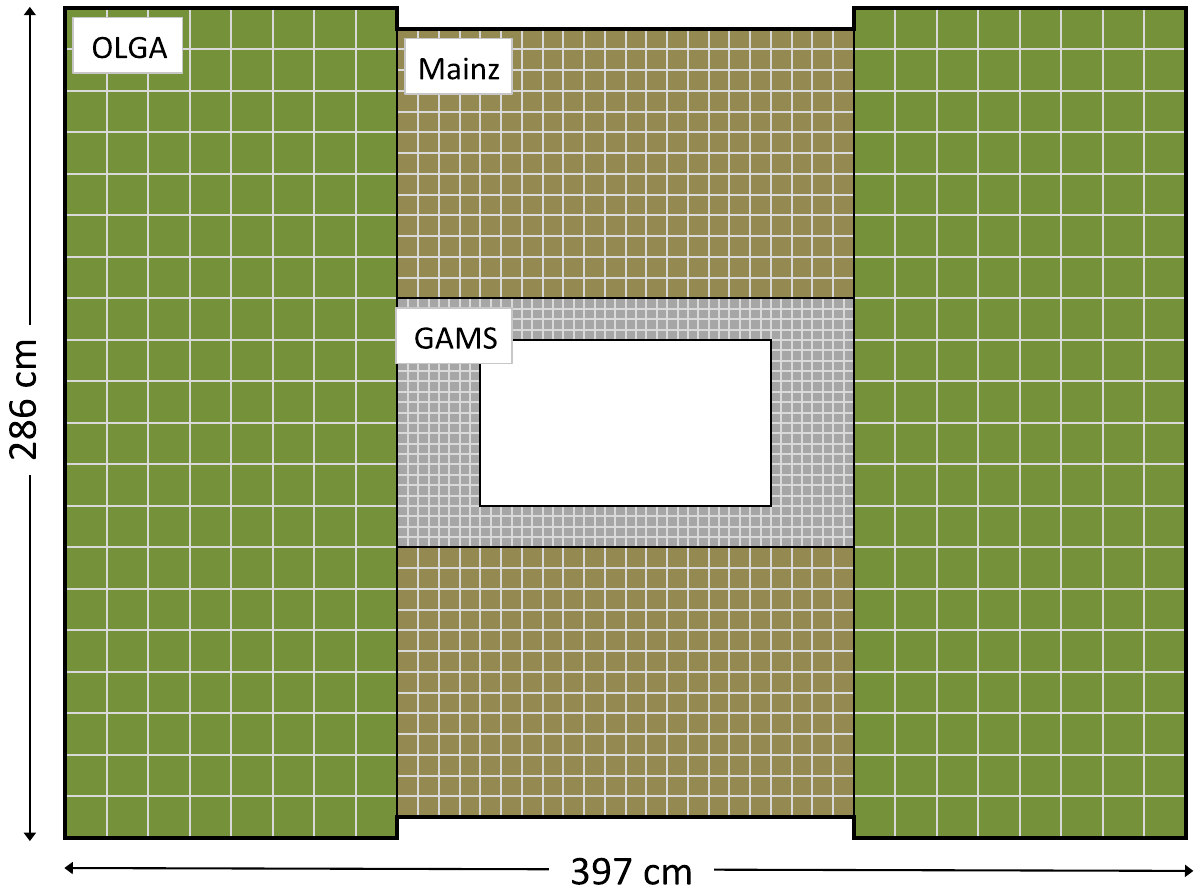}%
    \label{fig:ecal1-structure}%
  }%
  \hfill%
  \subfloat[]{%
    \includegraphics[width=0.343\textwidth,valign=m]{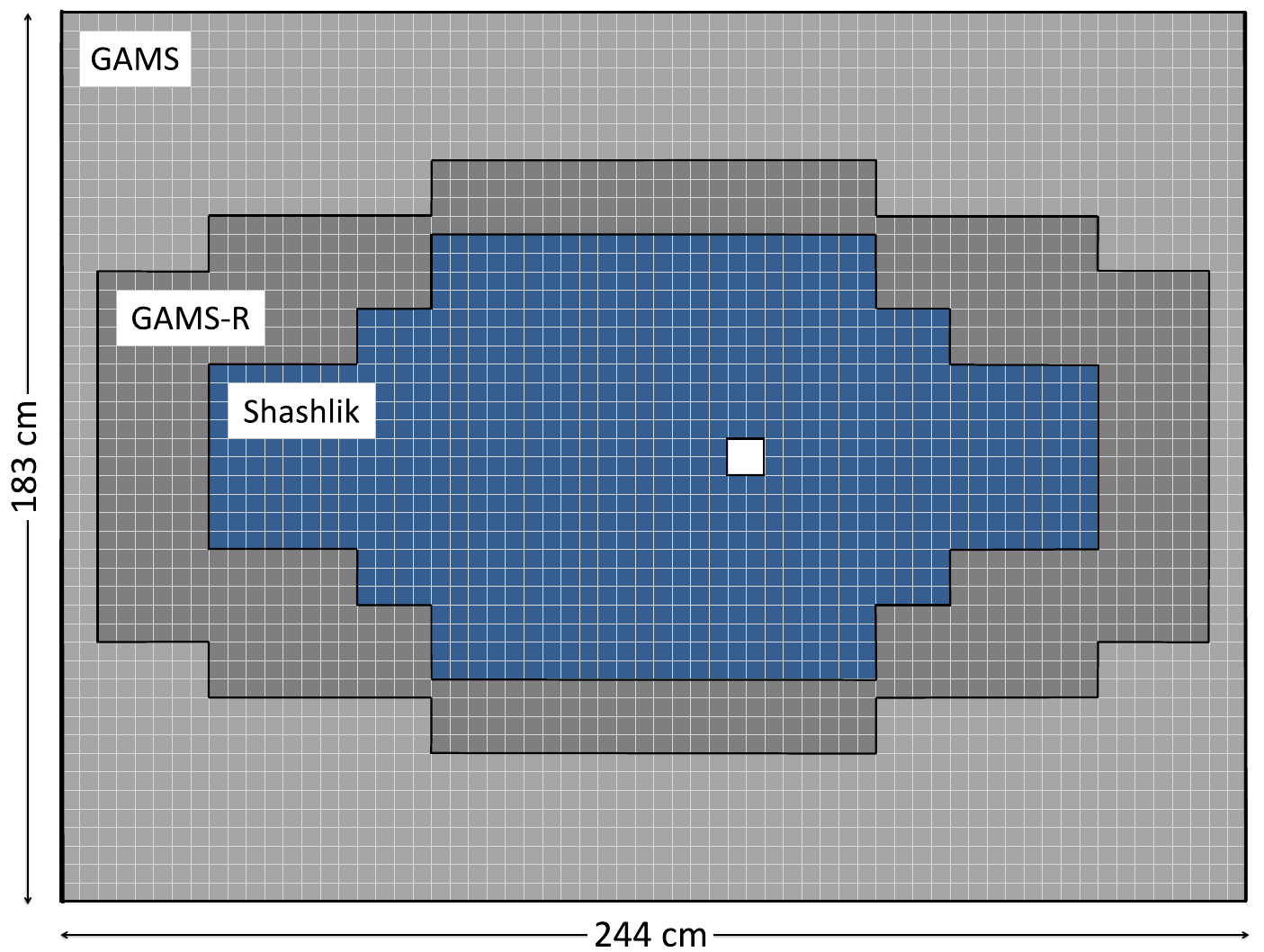}%
    \vphantom{\includegraphics[width=0.557\textwidth,valign=m]{fig27a}}%
    \label{fig:ecal2-structure}%
  }%
  \caption{\subfloatLabel{fig:ecal1-structure}~Configuration of
    ECAL1~\cite{Abbon:2014aex}.  The central area is equipped with
    GAMS modules of size $3.83 \times 3.83 \times 45\,\Cm^3$.  The
    MAINZ modules (size $7.5 \times 7.5 \times 36\,\Cm^3$) are
    installed above and below the GAMS area.  The OLGA modules (size
    $14.1 \times 14.1 \times 47\,\Cm^3$) cover the outer left and
    right regions.  The names refer to the experiments where these
    modules were originally used.
    \subfloatLabel{fig:ecal2-structure}~Configuration of ECAL2
    (corrected from~\cite{Abbon:2014aex}).  The outer and intermediate
    regions are equipped with GAMS and radiation-hardened GAMS
    modules, respectively, of the same size as the corresponding
    modules in ECAL1.  The inner region is equipped with Shashlik
    sampling modules of $39\,\Cm$ length.  The transverse sizes of all
    three types of modules are identical.  The central beam hole of
    $2 \times 2$ modules can be seen as a white spot.}
  \label{fig:ecal-structure}
\end{figure}

Photons and electrons emitted at large angles are detected and
identified in the first electromagnetic calorimeter (ECAL1), which
consists of 1500~lead-glass modules of different types (see
\cref{fig:ecal1-structure}). The Cherenkov light created by the
charged particles in a shower is detected by photomultiplier tubes
mounted to the downstream end face of each module. For neutral pions
reconstructed from photon pairs detected in ECAL1, a mass resolution
of $8.8\,\MeV/c^2$ has been achieved after careful calibration of all
modules.

Directly downstream of ECAL1 a hadronic calorimeter is installed to
measure the energy of charged and neutral hadrons.  Its 480 modules
consist of 40~Fe/scintillator layers, each $20\,\Cm$ and $5\,\Cm$
thick, adding up to $4.8$ nuclear interaction lengths.  The light
produced in the active medium is measured by photomultiplier tubes
(PMT).

The LAS is completed by a Muon Detection System consisting of a
$60\,\Cm$ thick iron absorber (Muon Filter~1) sandwiched between
large-area drift-tube detectors (MW1).

\subsubsection{Small-Angle Spectrometer}
\label{sec:compass.sas}

Particles emitted from the target at angles smaller than
$\pm 30\,\mrad$ and charged particles with momenta larger than about
$15\,\GeV/c$ pass through central rectangular holes in the
calorimeters and the muon absorber of the LAS and enter the
Small-Angle Spectrometer (SAS).  It includes the second spectrometer
magnet (SM2), which is positioned about $22\,\m$ downstream of the
target and has a higher bending power than SM1 with a field integral
of $4.4\,\T\m$.  Similarly to the LAS, tracking of charged particles
is performed using planar gaseous detectors of different active areas
and resolutions: MWPCs and drift chambers for the large-area tracking,
GEMs for the small-area tracking, scintillating fibers and PixelGEMs
for the tracking in the beam.  For charged particles detected both in
the LAS and the SAS the momentum resolution is determined to be
between~$0.3$ and $0.4\%$.

Photons are detected in the second electromagnetic calorimeter ECAL2,
which is made of 3068~modules of three different types (see
\cref{fig:ecal2-structure}): homogeneous lead-glass modules in the
outer part and so-called Shashlik Pb/scintillator sampling modules in
the inner part, used for their improved radiation hardness.  The mass
resolution for~$\pi^0$ reconstructed from photon pairs detected in
ECAL2 is $3.9\,\MeV/c^2$.

ECAL2 is followed immediately by the second hadronic calorimeter
HCAL2, comprising 220~modules made of 36~Fe/scintillator layers read
out by PMTs.  For both calorimeters in the SAS, the acceptance is
maximized by reducing the size of the central hole so that only the
beam passes through.

A Muon Detection System comprising a $2.40\,\m$ thick concrete
absorber (Muon Filter~2) and large-area drift-tube detectors (MW2)
completes the spectrometer setup.

\subsubsection{Trigger and Data Acquisition}
\label{sec:compass.trigger}

The trigger system of COMPASS serves several purposes. \one~It selects
candidates for physics (or calibration) events in a high-intensity
beam with a time delay of $\lesssim 1\,\mus$, \two~it initiates the
readout of all detectors, \three~it distributes a precise reference
clock to all front-end electronics modules with a frequency of
$38.88\,\MHz$, and \four~it distributes a unique event identification
to allow for the merging of the data streams from the individual
detector front-ends.

\begin{figure}[tbp]
  \centering
  \includegraphics[width=0.8\textwidth]{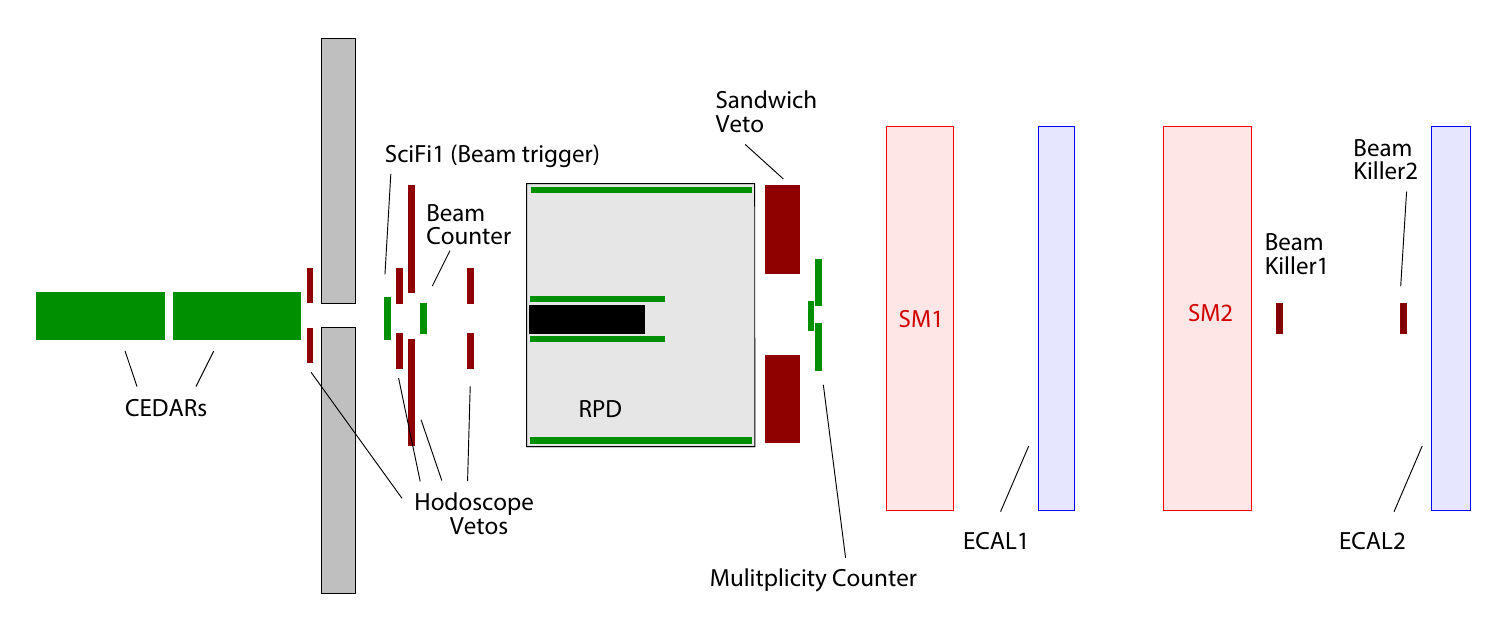}
  \caption{Arrangement of trigger elements in the spectrometer
    (schematic side view, not to scale)~\cite{Abbon:2014aex}.}
  \label{fig:trigger.setup}
\end{figure}

Candidate physics events for hadron spectroscopy are identified by
dedicated trigger elements~\cite{Bernhard:2690563} detecting
the incoming beam particle, a veto system discarding interactions
outside of the target or outside of the spectrometer acceptance, and
detectors which enhance the different final states under
investigation. \Cref{fig:trigger.setup} shows a schematic view of the
hardware elements contributing to physics triggers for hadron
spectroscopy.

Beam particles are detected by a coincidence of the analog sum of
signals from SciFi1 (vertical fibers) and a scintillator disk with
$3.2\,\Cm$ diameter (Beam Counter) installed $50\,\Cm$ downstream of
SciFi1.  These elements also serve as a time reference for the trigger
with a resolution of $(450\pm 50)\,\ns$.  The veto system comprises
two scintillator disks of $3.5\,\Cm$ diameter (Beam Killers) installed
$25\,\m$ and $33\,\m$ downstream of the target, their coincidence
inhibiting non-interacting beam particles, the Sandwich Veto (see
\cref{sec:compass.target}), and three scintillator hodoscopes located
upstream of the target.

Diffractive reactions in the liquid-hydrogen target with a recoiling
proton are selected by requiring signals in both rings of the RPD (see
\cref{sec:compass.target}).  Target pointing in azimuthal direction is
assured by requiring coincidences in matching segments of inner and
outer rings.  Pions and $\delta$-electrons are suppressed by
correlating the energy loss in the inner and outer ring using
combinations of signals with two different thresholds.  The
requirement of a signal in both rings selects events with a squared
momentum transfer larger than $\abs{t} \approx 0.07\,(\GeV/c)^2$.  In
order to also record reactions with smaller values of $\abs{t}$, a
Multiplicity Counter positioned downstream of the RPD in front of the
Sandwich Veto was used for part of the data taking with nuclear
targets. This counter selects events with more than a given number
(one or two) of minimum ionizing charged particles in the final state,
irrespective of $\abs{t}$.  Events with a high-energy photon in the
final state, \eg\ via the Primakoff reaction
$\pi^- + A \to \pi^- \gamma + A$, are selected by triggering on
photons with energies larger than a given threshold (40~or $60\,\GeV$)
detected in the central part of the ECAL2 calorimeter.
Last but not least, a trigger on incoming beam kaons using the signals
from both CEDARs is used.

Adding up all different triggers, the rate of recorded events at
nominal beam intensity of $5\EE{6}\,\s^{-1}$ amounts to 3~to
$4\EE{4}\,\s^{-1}$ with an average event size of $40\,\text{kB}$. This
large instantaneous data rate is averaged out to about
$300\,\text{MB}/\s$ over the typical SPS spill structure by using
large buffers at every stage of the data acquisition chain.  The
detector signals are collected by dedicated front-end electronics and
are digitized in custom-made TDC\footnote{Time-to-digital converter;
  measures signal time.} or sampling ADC\footnote{Analog-to-digital
  converter; measures signal amplitude.}  modules.  The digitized data
from several detector front-ends are then transmitted to about
150~data concentrator modules located in VME\footnote{The Versa Module
  Eurocard standard defines a bus protocol as well as mechanical and
  electrical interfaces.}  crates for easy control and power supply.
From there the merged data from several detectors are transmitted via
optical fibers to Readout Buffer PCs.  These distribute the data from
a given event through a Gbit Ethernet switch to Event Builder PCs,
which temporarily store the raw data before they are transferred to
permanent storage in the CERN data center.  The Event Builder PCs also
run online monitoring and calibration tasks.

\subsubsection{Event Reconstruction}
\label{sec:compass.reco}

The entities for data recording and processing are so-called
\emph{runs} which typically contain data from 100~SPS spills. During
data taking, the raw data are split up in files of $1\,\text{GB}$ size
each with about 1200~such files per run.  The reconstruction of the
raw data is performed using the software framework CORAL (COMPASS
Reconstruction Algorithm)~\cite{Abbon:2007pq}.  The first steps of the
reconstruction are \one~decoding and mapping, \ie the translation of
raw data to hits in terms of physical detector channels, including
amplitude and time information if applicable, and \two~clustering, \ie
the combination of hits adjacent in space to reconstruction clusters.

The next steps of the reconstruction depend on the detector type.  The
information from the tracking detectors is used to reconstruct
charged-particle tracks and to determine their momenta from the
bending by the two spectrometer magnets.  This step requires a precise
knowledge of the relative positions in space of sensitive detector
elements, \eg wires, strips, or pixels.  These are determined by an
iterative software alignment procedure using tracks recorded at low
beam intensity both with spectrometer magnets off and on.  The
association of detector clusters to track candidates is performed by
first searching for straight track segments in regions without or
negligible magnetic field and without large amounts of material.
These segments are then bridged across the magnets and absorber walls
by extrapolation and matching within a variable road width and by
comparing the track segments to a dictionary of reconstructable
tracks.  The precise determination of the track momentum and of the
charge sign is finally done using a Kalman filter for track fitting.

The vertexing algorithm attempts to combine beam-particle tracks
reconstructed in the beam telescope and any number of
final-state-particle tracks reconstructed in the spectrometer to form
primary vertices.  The optimal vertex positions are determined using a
Kalman filter which progressively tests the contribution of each track
to the overall~$\chi^2$ of the vertex.  Secondary vertices from decays
of long-lived unstable particles are constructed from pairs of two
oppositely charged spectrometer-tracks, which do not contain a beam
track.

Clusters in the electromagnetic calorimeters may contain showers from
several incident particles overlapping in space and time.  In order to
disentangle the individual contributions, several shower profiles,
which have been pre-determined from single electron or photon events,
are fitted to the reconstructed clusters in the plane transverse to
the beam direction, until a good description is reached.  In order to
optimize the energy resolution, the cell responses are calibrated
using~$\pi^0$ reconstructed from events containing at least two
photons.

\subsubsection{Offline Event Selection}
\label{sec:compass.select}

The events selected by the hardware triggers (see
\cref{sec:compass.trigger}) contain reactions of various beam-particle
species into a variety of final states.  The analysis of the data set
proceeds by selecting exclusive events corresponding to a given
reaction channel.  The final states may comprise both charged and
neutral particles.

\begin{figure}[tbp]
  \centering
  \subfloat[]{%
    \includegraphics[height=0.3\textwidth]{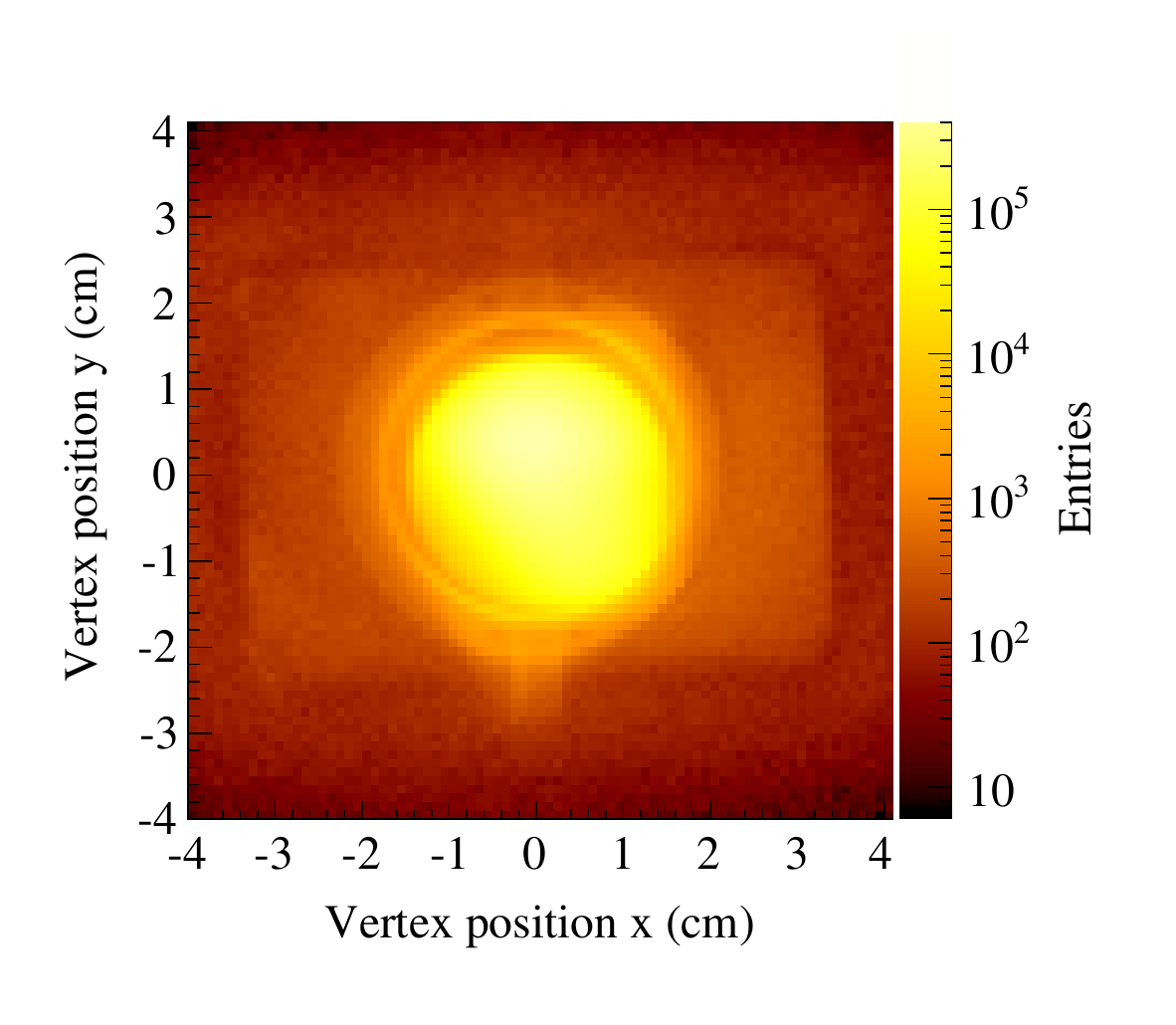}%
    \label{fig:vertex.xy}%
  }%
  \hfill%
  \subfloat[]{%
    \raisebox{1.25ex}{%
      \includegraphics[height=0.3\textwidth]{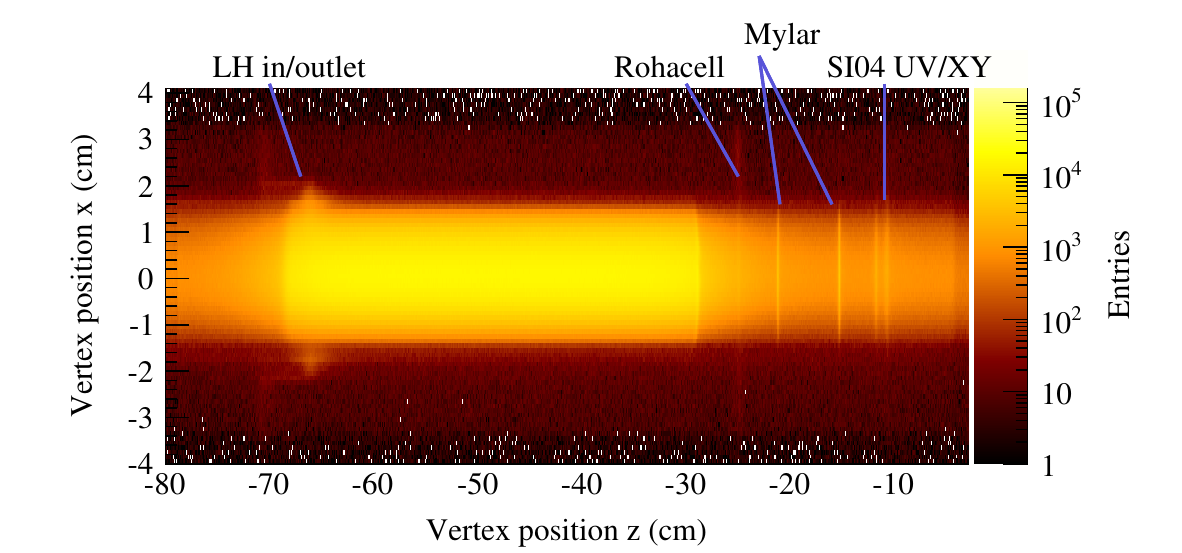}%
    }%
    \label{fig:vertex.xz}%
  }%
  \caption{Spatial distribution of the primary vertices for the
    liquid-hydrogen target for events with three forward-going charged
    particles~\cite{Abbon:2014aex}, \subfloatLabel{fig:vertex.xy}~in
    the plane perpendicular to the beam direction and
    \subfloatLabel{fig:vertex.xz}~in the horizontal plane with
    $z$~along the beam.}
  \label{fig:vertex}
\end{figure}

\begin{figure}[tbp]
  \centering
  \includegraphics[width=0.45\textwidth]{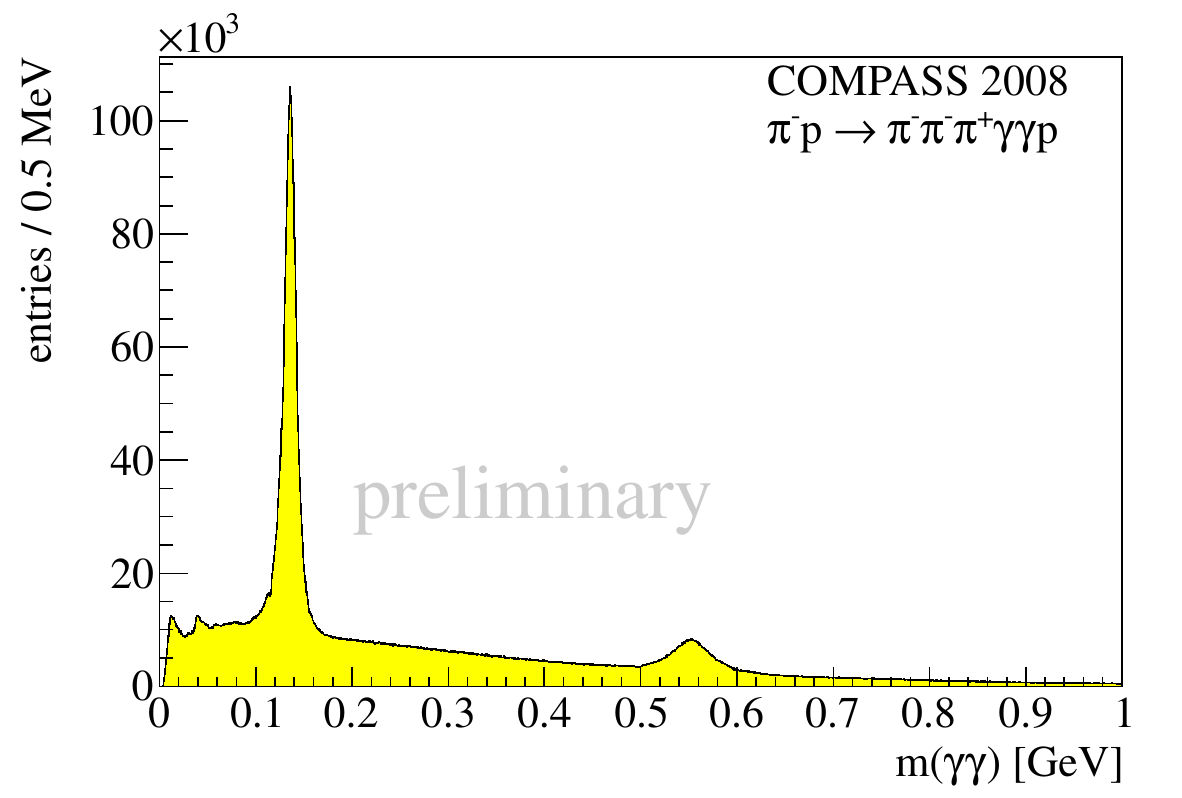}
  \caption{Invariant mass of photon pairs from the reaction
    $\pi^- + p \to \threePi \gamma \gamma +
    p$~\cite{Schluter:2011dt}.\protect\footnotemark}
  \label{fig:mgg}
\end{figure}

\begin{figure}[tbp]
  \centering
  \null\hfill%
  \subfloat[]{%
    \includegraphics[width=\twoPlotWidth]{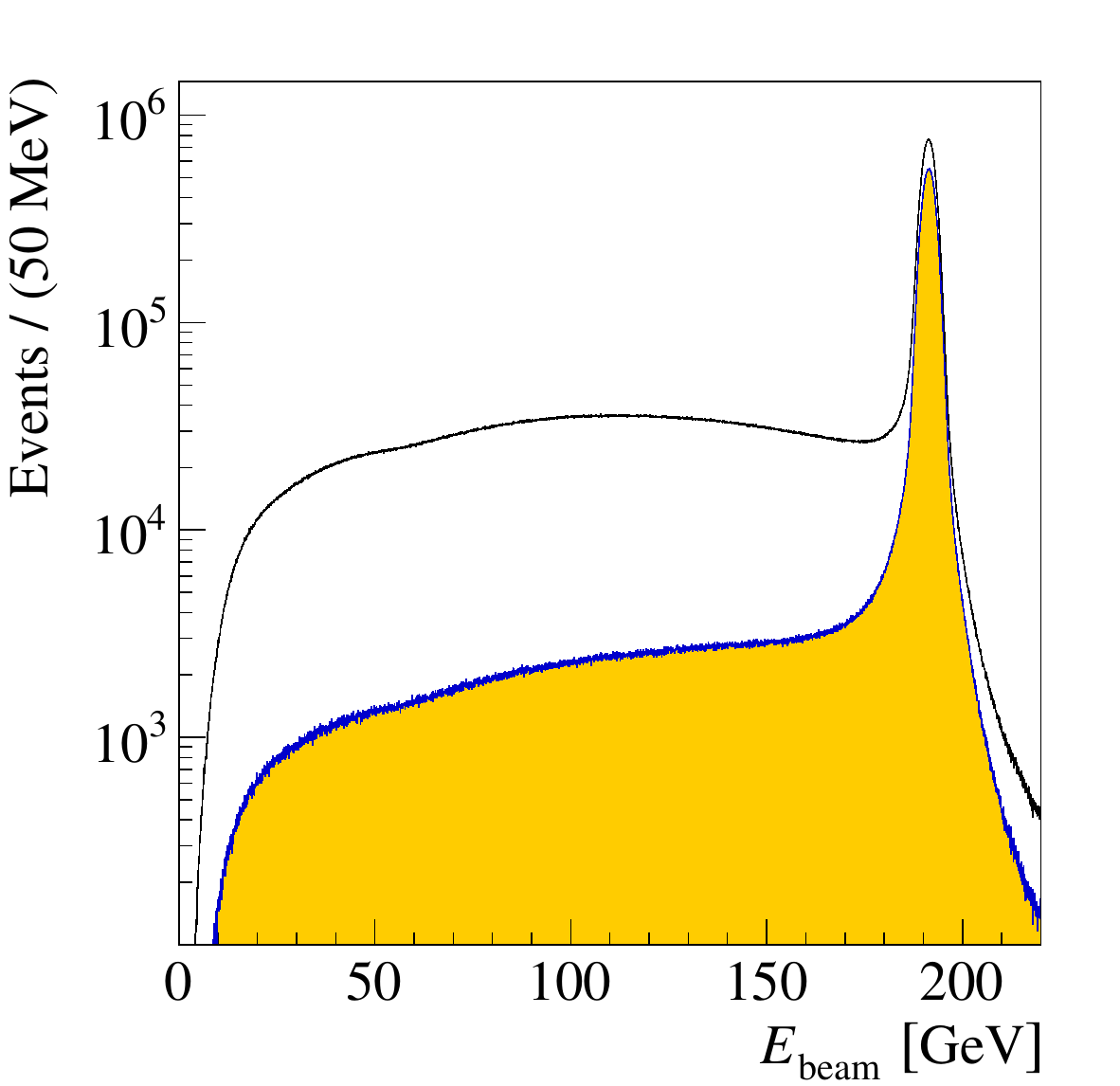}%
    \label{fig:exclusivity.full}%
  }%
  \hfill%
  \subfloat[]{%
    \includegraphics[width=\twoPlotWidth]{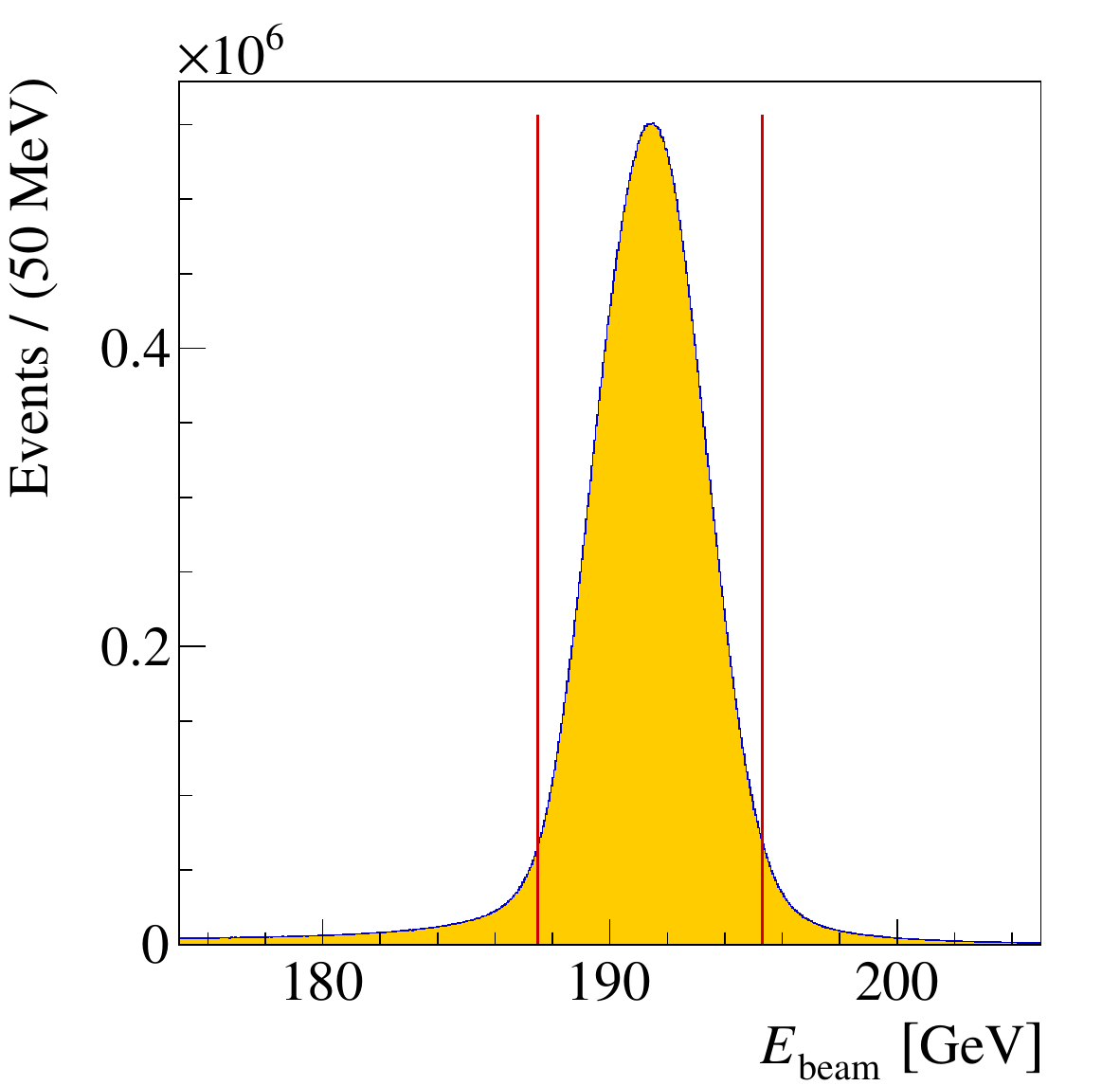}%
    \label{fig:exclusivity.zoom}%
  }%
  \hfill\null%
  \caption{Total energy of all reconstructed final-state particles for
    the reaction $\pi^- + p \to \threePi + p$~\cite{Adolph:2015tqa},
    \subfloatLabel{fig:exclusivity.full}~full energy range, without
    (empty histogram) and with coplanarity cut (full histogram),
    \subfloatLabel{fig:exclusivity.zoom}~zoomed range with the
    vertical lines indicating the accepted range around the nominal
    beam energy.}
  \label{fig:exclusivity}
\end{figure}

\begin{figure}[tbp]
  \centering
  \includegraphics[width=0.65\textwidth]{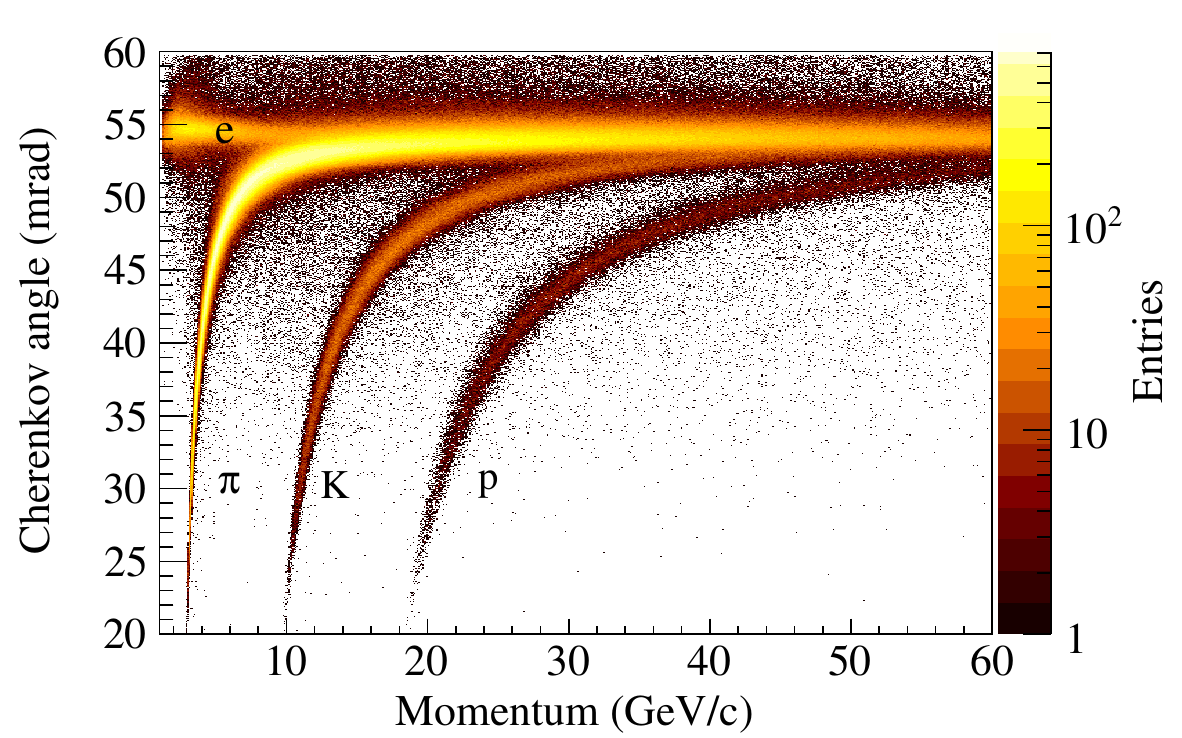}
  \caption{Cherenkov angle for reconstructed rings versus particle
    momentum for C$_4$F$_{10}$ radiator gas~\cite{Abbon:2014aex}.}
  \label{fig:rich.angle-vs-mom}
\end{figure}

In the following, we discuss the criteria for offline event selection,
which are common to all analyses.  As example, the steps are
illustrated for $\etaOrPrPi\,p$ and $\threePi\,p$ final states
produced by incoming~$\pi^-$ scattering on a proton target.
\begin{enumerate}
\item Events from a given physics trigger are selected by requiring
  that the corresponding trigger bit is set for the event.
\item A primary vertex (see \cref{sec:compass.reco}) between the beam
  particle and the forward-going charged particles is required to be
  formed in a fiducial volume within the target.  Charge conservation
  between incoming and outgoing particles needs to be fulfilled.  In
  case more than one vertex candidate has been found by the vertexing
  algorithm, usually the one with the better $\chi^2$~value is chosen.
  \Cref{fig:vertex} shows the spatial distribution of primary vertices
  for events with three forward-going charged particles.
\item In case of final states containing neutral short-lived particles
  decaying to photons, \eg $\pi^0$ or $\eta$, photon showers
  reconstructed by the calorimeters are selected by requiring that
  \begin{enumerate}
  \item no charged-particle track is associated with the shower,
  \item the time difference between the beam track and the calorimeter
    shower does not exceed a certain maximum value,
  \item the shower energies are above certain thresholds (typically
    $1\,\GeV$ for ECAL1 and $4\,\GeV$ for ECAL2),
  \item the shower is reconstructed in a geometrical position that is
    not obstructed from the target by a large amount of material.
  \end{enumerate}
  Based on the energy of the reconstructed photons, their
  three-momenta are determined by assuming that they originate from
  the primary vertex.  Final-state $\pi^0$ or $\eta$ are selected by
  combining photon pairs and applying a cut on the two-photon
  invariant mass.  \Cref{fig:mgg} shows the two-photon invariant mass
  distribution from the reaction
  $\pi^-+ p \to \threePi \gamma \gamma + p$.
\item Transverse momentum conservation at the interaction vertex is
  ensured by requiring exactly one recoil track detected in the RPD
  that is back-to-back with the forward-going system reconstructed in
  the spectrometer in the plane perpendicular to the beam direction
  (\textquote{coplanarity cut}).
\item Energy conservation is enforced by requiring that the energy sum
  of all final-state particles corresponds to the nominal beam energy
  within a window given by the momentum spread of the beam
  (\textquote{exclusivity cut}).  \Cref{fig:exclusivity} shows the
  total energy calculated from all reconstructed final-state particles
  including the recoil proton for the reaction
  $\pi^- + p \to \threePi + p$.
\item Depending on the reaction under study, the beam particle may be
  required to be tagged or vetoed by the CEDAR detectors.
\item In the same way, tagging and vetoing of forward-going
  final-state particles by the RICH may be required.  In
  \cref{fig:rich.angle-vs-mom}, the Cherenkov angles for reconstructed
  rings are shown as a function of the particle momenta.  The bands
  correspond to different particle types and can be used for
  identification.
\item Further kinematical cuts, \eg on the rapidity of the fastest
  particle, defined by\footnote{Here, the $z$~direction is defined by
    the measured direction of the beam particle.}
  \begin{equation}
    \label{eq:rapidity}
    y = \frac{1}{2}\, \ln{\frac{E + p_z}{E - p_z}}\eqPunctSpacing,
  \end{equation}
  or on the Feynman-$x$ variable, defined by
  \begin{equation}
    \label{eq:xF}
    x_\text{F} \coloneqq \frac{p_z}{p_{z\,\text{max}}}
    \approx \frac{2 p_{z\,\text{CM}}}{\sqrt{s}}\eqPunctSpacing,
  \end{equation}
  may be applied in order to select or remove events from production
  reactions with distinct kinematics, \eg central production (see
  \cref{sec:exp.prod_reactions}).
\end{enumerate}

\footnotetext{\Cref{fig:mgg} is an unpublished auxiliary plot from the
  analysis presented in \refCite{Adolph:2014rpp}.}

\clearpage{}%
\clearpage{}%
\section{Partial-Wave Analysis Formalism}
\label{sec:pwa}

\subsection{Analysis Model}
\label{sec:pwa.analysis_model}

\subsubsection{Ansatz}
\label{sec:pwa.analysis_model.ansatz}

As discussed in \cref{sec:exp.prod_reactions}, we measure at COMPASS
single-diffraction dissociation reactions, which are inelastic
scattering processes of the form
\begin{equation}
  \label{eq:diffractive_dissociation}
  a + b \to (1 + 2 + \ldots + n) + c\eqPunctSpacing.
\end{equation}
In these reactions, a high-energetic beam hadron~$a$ interacts
strongly with a target hadron~$b$ thereby producing an $n$-body
hadronic final state $(1, 2, \ldots, n)$ and a target recoil~$c$.

\begin{figure}[b]
  \centering
  \includegraphics[width=\threePlotWidth]{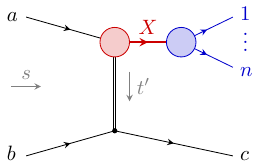}
  \caption{Production of an intermediate state~$X$ by excitation of
    beam particle~$a$ via $t$-channel exchange with target
    particle~$b$.  The intermediate state decays into an $n$-body
    hadronic final state $(1, 2, \ldots, n)$.}
  \label{fig:diffractive_nbody}
\end{figure}

Based on \cref{eq:dsigma} in \cref{sec:scattering.xsection}, we
construct a model for the cross section of
reaction~\eqref{eq:diffractive_dissociation} by assuming that the
$n$-body final state is produced via $t$-channel exchange processes
and by performing an $s$-channel partial-wave expansion of the
$n$-body system by inserting a complete set of intermediate states~$X$
with well defined quantum numbers.  We hence subdivide
process~\eqref{eq:diffractive_dissociation} into two subprocesses:
\one~an inelastic two-body scattering reaction $a + b \to X + c$ and
\two~the subsequent decay of state~$X$ into the $n$-body final state,
\ie $X \to 1 + 2 + \ldots + n$.  This is shown in
\cref{fig:diffractive_nbody}.
Subprocess~\one is described in terms of the invariant mass~$m_X$ of
the $n$-body system and the two Mandelstam variables $s$~and~$t$,
where $s$~is the squared center-of-momentum energy of the $(a, b)$
system (see \cref{eq:mandelstam.s}) and $t$~the transferred squared
four-momentum (see \cref{eq:mandelstam.t}).  For convenience, we use
instead of~$t$ the positive definite reduced squared four-momentum
transfer~$t'$ as defined in \cref{eq:tprime}.  For a fixed-target
experiment such as COMPASS, $p_a$~and~$p_b$ are constant and hence the
center-of-momentum energy $\sqrt{s}$ of the scattering reaction is
fixed.  In this case, the kinematic distribution of the final-state
particles depends on~$m_X$, $t'$, and a set of $(3n - 4)$ additional
phase-space variables (see \cref{sec:scattering.s-matrix}) represented
by~$\tau_n$.  These phase-space variables fully describe the $n$-body
decay and will be defined in
\cref{sec:pwa.analysis_model.decay_amp,sec:pwa.analysis_model.coordsys}.

Analogous to subdividing the scattering process, we split the phase
space of the $(n + 1)$ outgoing particles into a two-body phase space
for $X$~and~$c$ (see \cref{eq:dPhi_2.cms}) and an $n$-body phase space
for the decay $X \to 1 + 2 + \ldots + n$ (see \cref{eq:dPhi_N}).  For
any given choice of the phase-space variables~$\tau_n$, the Dirac
delta function in \cref{eq:dPhi_N} that represents four-momentum
conservation can be integrated out so that the differential
phase-space element can be written as
\begin{equation}
  \label{eq:dLIPS_dens}
  \dif{\Phi_n}(m_X, \tau_n)
  = \rho_n(m_X, \tau_n)\, \dif{\tau_n}\eqPunctSpacing,
\end{equation}
where the phase-space factor $\rho_n(m_X, \tau_n)$ represents the
density of states in the variables~$m_X$ and~$\tau_n$.\footnote{For
  example, the two-body phase-space element can be expressed in terms
  of the polar and azimuthal angles of one of the two particles in
  their center-of-momentum frame, \ie
  $\tau_2 = \Omega = (\vartheta, \phi)$, then using
  \cref{eq:dPhi_2.cms} and $\sqrt{s} = m_X$ we get
  \begin{equation}
    \dif{\Phi_2}(m_X, \tau_2) = \rho_2(m_X, \tau_2)\, \dif{\tau_2}
    \quad\text{with}\quad
    \rho_2(m_X, \tau_2) = \frac{1}{(4\pi)^2}\, \frac{q}{m_X}\eqPunctSpacing.
  \end{equation}}  A convenient
choice for~$\tau_n$ is presented in \cref{sec:nbody_phase_space}.

Splitting the phase space and using
\cref{eq:tprime,eq:dsigma,eq:flux_factor}, we can write the cross
section for reaction~\eqref{eq:diffractive_dissociation} as
\begin{multline}
  \label{eq:cross_section}
  \dif{\sigma_{a + b \to (1 + \ldots + n) + c}} \\
  = \frac{1}{16\pi}\, \frac{1}{\left[ s - (m_a + m_b)^2 \right]\! \left[ s - (m_a - m_b)^2 \right]}\,
  \left| \mathcal{M}_{fi} \right|^2\, \dif{t'}\, \frac{2 m_X}{2\pi}\, \dif{m_X}\,
  \dif{\Phi_n}(p_X; p_1, p_2, \ldots, p_n)\eqPunctSpacing.
\end{multline}
Here, we have expressed the flux factor in terms of~$s$ and we have
written the two-body phase space of the $(X, c)$ system (see
\cref{eq:dPhi_2.cms}) in terms of~$t'$.  We also have used the fact
that for an unpolarized reaction, $\mathcal{M}_{fi}$ is independent of
the azimuthal angle of~$X$ about the beam axis in the
center-of-momentum frame of the reaction $a + b \to X + c$.  The
factor $2 m_X\, \dif{m_X}$ arises due to the splitting of the phase
space.

The intensity distribution, \ie the distribution of the events that
are produced in reaction~\eqref{eq:diffractive_dissociation}, is
\begin{equation}
  \label{eq:intensity_def}
  \mathcal{I}(m_X, t', \tau_n)
  \coloneqq \frac{\dif{N}}{\dif{m_X}\, \dif{t'}\, \dif{\Phi_n}(m_X, \tau_n)}
  \propto \frac{\dif{\sigma_{a + b \to (1 + \ldots + n) + c}}}
               {\dif{m_X}\, \dif{t'}\, \dif{\Phi_n}(m_X, \tau_n)}
  \propto m_X\, \left| \mathcal{M}_{fi}(m_X, t', \tau_n) \right|^2\eqPunctSpacing,
\end{equation}
where $N$~is the number of produced events and
$\dif{\Phi_n}(m_X, \tau_n)$ is given by \cref{eq:dLIPS_dens}.  On the
right-hand side of \cref{eq:intensity_def}, all constant factors have
been dropped.  It is worth noting that $\mathcal{I}$~is differential
in the Lorentz-invariant $n$-body phase-space element and is thus
independent of the particular choice of the phase-space
variables~$\tau_n$ because the phase-space density
$\rho_n(m_X, \tau_n)$ in \cref{eq:dLIPS_dens} contains the
corresponding Jacobian.  The intensity $\mathcal{I}$ essentially
represents the deviation of the kinematic distribution of the produced
final-state particles from a pure phase-space distribution.  It is
therefore a direct measure for
$\left| \mathcal{M}_{fi}(m_X, t', \tau_n) \right|^2$.

We first construct a model for the amplitude
$\mathcal{M}_{a + b \to X + c}$ for the hypothetical process, where
only a single intermediate state~$X$ is produced and decays into the
measured $n$-body final state.  In
\cref{sec:pwa.analysis_model.extension}, we will extend this model to
the case of several intermediate resonances.  Since production and
decay of the intermediate state~$X$ are independent, we can factorize
the amplitude into three parts: \one an amplitude
$\mathcal{P}_{a + b \to X + c}(m_X, t')$ that describes the production
of $X$, \two a dynamical amplitude $\mathcal{D}_X(m_X)$ that describes
the propagation of $X$, and \three the decay amplitude
$\widetilde{\Psi}_{X \to \text{FS}}(m_X, \tau_n)$ that describes the
decay of~$X$ into the $n$-body final state via a particular decay
chain.  Therefore,
\begin{equation}
  \label{eq:ampl_ansatz_single_res}
  \mathcal{M}_{a + b \to (X \to \text{FS}) + c}(m_X, t', \tau_n)
  = \mathcal{P}_{a + b \to X + c}(m_X, t')\,
  \mathcal{D}_X(m_X)\, \widetilde{\Psi}_{X \to \text{FS}}(m_X, \tau_n)\eqPunctSpacing.
\end{equation}

\subsubsection{Decay Amplitude}
\label{sec:pwa.analysis_model.decay_amp}

The decay amplitude $\widetilde{\Psi}_{X \to \text{FS}}(m_X, \tau_n)$
in \cref{eq:ampl_ansatz_single_res} is calculated using the isobar
model~\cite{Hansen:1973gb,Herndon:1973yn}.  In this model, the
$X$~decay is described as a chain of successive two-body decays via
additional intermediate resonances, the so-called \emph{isobars},
which appear in the subsystems of the $n$-body final
state.\footnote{For example, the decay $X^- \to \threePi$ may proceed
  via $\rho(770)^0$ as an intermediate $\pi^-\pi^+$ resonance, \ie
  $X^- \to \Pprho^0 \pi^- \to \threePi$.  The usage of the term isobar
  for the intermediate states has historical reasons.  The isobar
  model was first introduced by Lindenbaum and Sternheimer
  in~\refCite{Lindenbaum:1957ec} to describe excited intermediate
  nucleon resonances, which they called isobars in analogy to nuclear
  physics.}  It is also assumed that the outgoing particles of the
two-body decays do not interact with each other, \ie final-state
interactions are neglected.

The fundamental building block for the construction of
$\widetilde{\Psi}_{X \to \text{FS}}$ is the two-body decay amplitude
$\mathcal{A}_r^{J_r\, M_r}$.  It describes the propagation of a
resonance~$r$ with spin~$J_r$ and spin projection~$M_r$ \wrt a chosen
quantization axis and the decay of~$r$ into particles 1~and~2.  The
two-body decay amplitude can be calculated in the $r$~rest frame using
the helicity
formalism~\cite{Jacob:1959at,Martin:1970xx,Chung:1971ri,Richman:1984gh,Kutschke:1996,Salgado:2013dja}.
The two daughter particles have spins~$J_1$ and~$J_2$ and are
described in the helicity basis, where the quantization axes are the
directions of the momenta of particles 1~and~2.  In the $r$~rest
frame, the momenta of particles 1~and~2 are by definition back to back
and have a fixed magnitude~$q$, which is given by
\cref{eq:breakup_mom.cms.f,eq:kaellen}.  Hence the kinematics of the
decay $r \to 1 + 2$ is completely defined by the polar
angle~$\vartheta_r$ and the azimuthal angle~$\phi_r$ of the momentum
of one of the daughter particles.

The daughter particles~1 and~2 are described by the two-particle
plane-wave center-of-momentum helicity state
$\ket{\vec{p}_1, \vec{p}_2; \lambda_1, \lambda_2}$.  Here, $\lambda_1$
and $\lambda_2$ are the helicities of the two daughter particles and
$\vec{p}_1 = -\vec{p}_2 \eqqcolon \vec{q}$.  Since $\abs{\vec{q}}$ as
given by \cref{eq:breakup_mom.cms.f,eq:kaellen} is constant, the
quantum state can be written as
$\ket{\vartheta_r, \phi_r; \lambda_1, \lambda_2}$.  The amplitude for
the propagation and the decay of resonance~$r$ with mass~$m_r$ and
spin state $\ket{J_r, M_r}$ is
\begin{equation}
  \label{eq:2_body_decay_amp_ansatz}
  \mathcal{A}_r^{J_r\, M_r}(m_r, \vartheta_r, \phi_r)
  = \mathcal{D}_r(m_r)\, \sum_{\lambda_1, \lambda_2}
  \braket{\vartheta_r, \phi_r; \lambda_1, \lambda_2 | \hat{T}(m_r) | J_r, M_r}\eqPunctSpacing,
\end{equation}
where $\mathcal{D}_r(m_r)$ describes the propagation of~$r$ and
$\hat{T}(m_r)$ is the transition operator of the decay $r \to 1 + 2$.
The coherent summation over all allowed daughter helicities in
\cref{eq:2_body_decay_amp_ansatz} is performed only in the case, where
particles~1 and~2 appear as intermediate states in the decay chain of
the~$X$.  If one or both daughters are (quasi-stable) final-state
particles and their helicity is not measured, the summation over the
respective helicities has to be performed incoherently at the
intensity level.

We expand the helicity amplitude into partial waves by inserting a
complete set of angular-momentum helicity states
$\ket{J, M; \lambda_1, \lambda_2}$, which describe two-particle states
with definite total angular momentum~$J$.\footnote{Completeness means
  that
  \begin{equation}
    \sum_{J, M} \sum_{\lambda_1, \lambda_2}
    \ket{J, M; \lambda_1, \lambda_2}\, \bra{J, M; \lambda_1, \lambda_2}
    = \mathds{1}\eqPunctSpacing.
  \end{equation}}  Applying
angular-momentum conservation, the helicity amplitude reads
\begin{multline}
  \label{eq:2_body_decay_amp_ansatz_2}
  \mathcal{A}_r^{J_r\, M_r}(m_r, \vartheta_r, \phi_r) \\
  = \mathcal{D}_r(m_r)\,
  \sum_{\lambda_1, \lambda_2}
  \braket{\vartheta_r, \phi_r; \lambda_1, \lambda_2 | J_r, M_r; \lambda_1, \lambda_2}\,
  \braket{J_r, M_r; \lambda_1, \lambda_2 | \hat{T}(m_r) | J_r, M_r}\eqPunctSpacing.
\end{multline}

We expand the helicity amplitude further into states
$\ket{J_r, M_r; L_r, S_r}$, which describe two-particle states that
have definite relative orbital angular momentum~$L_r$ between the two
particles and where the spins of the two particles couple to the total
intrinsic spin~$S_r$.  This yields
\begin{multline}
  \label{eq:2_body_decay_amp_ansatz_3}
  \mathcal{A}_r^{J_r\, M_r\, L_r\, S_r}(m_r, \vartheta_r, \phi_r)
  = \sum_{\lambda_1, \lambda_2}
  \Overbrace{\braket{\vartheta_r, \phi_r; \lambda_1, \lambda_2 | J_r, M_r; \lambda_1, \lambda_2}\,
    \braket{J_r, M_r; \lambda_1, \lambda_2 | J_r, M_r; L_r, S_r}}%
  {\displaystyle{\text{angular part}}} \\
  \times
  \Underbrace{\mathcal{D}_r(m_r)\, \braket{J_r, M_r; L_r, S_r | \hat{T}(m_r) | J_r, M_r}}%
  {\displaystyle{\text{dynamical part}}}\eqPunctSpacing.
\end{multline}
The two-body decay amplitude factorizes into an angular part, which is
given by first principles and completely defined by the
angular-momentum quantum numbers of the involved particles, and a
dynamical part, which describes the dependence of the amplitude on the
invariant mass~$m_r$ of the $(1, 2)$ system and needs to be
modeled.\footnote{This is analogous to the case of a two-particle
  system in a central potential, where the two-body wave function can
  be factorized into an angular part that is completely determined by
  the angular-momentum quantum numbers and a radial part that depends
  on the shape of the potential.}

The first scalar product in the angular part of
\cref{eq:2_body_decay_amp_ansatz_3} represents the angular
distribution of the daughter particles in the $r$~rest frame.  It is
given by~\cite{Martin:1970xx,Chung:1971ri}
\begin{equation}
  \label{eq:ang_dist}
  \braket{J_r, M_r; \lambda_1, \lambda_2 | \vartheta_r, \phi_r; \lambda_1, \lambda_2}
  = \sqrt{\frac{2 J_r + 1}{4 \pi}}\,
  D_{M_r\; \lambda}^{J_r}(\phi_r, \vartheta_r, 0)\eqPunctSpacing,
\end{equation}
which is derived in \cref{sec:ang_dist}.  The appearing Wigner
$D$-function
$D_{M'\; M}^{J}(\alpha, \beta, \gamma)$~\cite{Wigner:1931,Wigner:1959}
represents the transformation property of a spin state $\ket{J, M}$
under an arbitrary active rotation $\hat{\mathcal{R}}$, which is
defined by the three Euler angles~$\alpha$, $\beta$, and~$\gamma$.  We
use the $y$-$z$-$y$ convention for the Euler angles that is defined,
for example, in \refCite{Rose:1957}.  Details about the Wigner
$D$-function can be found in \cref{sec:wigner_D_function}.  The
quantum number~$\lambda$ is discussed below.

The second scalar product in the angular part of
\cref{eq:2_body_decay_amp_ansatz_3} is the so-called recoupling
coefficient.  It connects the two-particle angular-momentum states in
the helicity and the $L$-$S$-coupling representations and is given
by~\cite{Martin:1970xx,Chung:1971ri}
\begin{equation}
  \label{eq:recoupling_coeff}
  \braket{J_r, M_r; L_r, S_r | J_r, M_r; \lambda_1, \lambda_2}
  = \sqrt{\frac{2 L_r + 1}{2 J_r + 1}}\,
  \Underbrace{\clebsch{J_1}{\lambda_1}{J_2}{-\lambda_2}{S_r}{\lambda}}%
  {\displaystyle{\text{spin--spin coupling}}}~
  \Underbrace{\clebsch{L_r}{0}{S_r}{\lambda}{J_r}{\lambda_r}}%
  {\displaystyle{\text{spin--orbit coupling}}}\eqPunctSpacing.
\end{equation}
Here, two Clebsch--Gordan coefficients appear: one for the coupling of
the spins of the two daughter particles to the total intrinsic
spin~$S_r$ and one for the coupling of~$L_r$ and~$S_r$ to~$J_r$.  The
two Clebsch--Gordan coefficients define the spin projection~$\lambda$
of the $(1, 2)$~system and the spin projection~$\lambda_r$ of~$r$
using, without loss of generality, the direction of particle~1 as the
quantization axis:
\begin{equation}
  \label{eq:lambda}
  \lambda_r = \lambda = \lambda_1 - \lambda_2\eqPunctSpacing.
\end{equation}
Since the orbital angular momentum~$L_r$ in the decay is by
construction perpendicular to the momenta of particles~1 and~2, it has
no projection onto the helicity quantization axis.

Inserting \cref{eq:ang_dist,eq:recoupling_coeff} into
\cref{eq:2_body_decay_amp_ansatz_3}, we get
\begin{multline}
  \newcommand*{\setH}{\vphantom{\sqrt{\frac{2L + 1}{4\pi}}}}
  \label{eq:2_body_decay_amp}
  \mathcal{A}_r^{J_r\, M_r\, L_r\, S_r}(m_r, \vartheta_r, \phi_r)
  = \Underbrace{\sqrt{\frac{2L_r + 1}{4\pi}}}{\mathclap{\substack{\text{normalization\vphantom{bp}}}}}~\;
  \Overbrace{%
    \Underbrace{\mathcal{D}_r(m_r) \setH}{\mathclap{\substack{\text{propagator\vphantom{bp}} \\ \text{term\vphantom{ft}}}}}~
    \Underbrace{\alpha_{r \to 1 + 2} \setH}{\mathclap{\substack{\substack{\text{coupling\vphantom{bp}}}}}}~
    \Underbrace{F_{L_r}(m_r) \setH}{\mathclap{\substack{\text{barrier\vphantom{bp}} \\ \text{factor\vphantom{ft}}}}}}%
    {\text{dynamical part}}
  \\
  \newcommand*{\setH}{\vphantom{\sqrt{\frac{2L + 1}{4\pi}}}}
  \times \sum_{\lambda_1, \lambda_2}
  \Overbrace{%
    \clebsch{J_1}{\lambda_1}{J_2}{-\lambda_2}{S_r}{\lambda}\,
    \clebsch{L_r}{0}{S_r}{\lambda}{J_r}{\lambda_r}\,
    D_{M_r\; \lambda_r}^{J_r \text{*}}(\phi_r, \vartheta_r, 0)}%
  {\text{angular part}}
  \\
  \newcommand*{\setH}{\vphantom{\sqrt{\frac{2L + 1}{4\pi}}}}
  \times
  \Underbrace{\mathcal{A}_1^{J_1\, \lambda_1\, L_1\, S_1}(m_1, \vartheta_1, \phi_1) \setH}%
  {\mathclap{\substack{\text{two-body decay} \\ \text{amplitude of particle 1}}}}~
  \Underbrace{\mathcal{A}_2^{J_2\, \lambda_2\, L_2\, S_2}(m_2, \vartheta_2, \phi_2) \setH}%
  {\mathclap{\substack{\text{two-body decay} \\ \text{amplitude of particle 2}}}}\eqPunctSpacing.
\end{multline}
Here, we have modeled the dynamical part in terms of the propagator
term $\mathcal{D}_r(m_r)$, the complex-valued coupling
$\alpha_{r \to 1 + 2}$, which describes strength and relative phase of
the decay mode, and the barrier factor $F_{L_r}(m_r)$ (see
\cref{eq:bw_factor}), which describes the suppression of higher
orbital angular momentuma~$L_r$ between the two daughter particles at
low~$m_r$.\footnote{Note that in order to ease notation, we have
  omitted in \cref{eq:2_body_decay_amp} the implicit dependence
  of~$\mathcal{D}_r$ and~$F_{L_r}$ on the daughter masses~$m_1$
  and~$m_2$.  These dependences become important when particles~1
  and/or~2 are not final-state particles but isobar resonances.}  The
propagator term will be discussed in
\cref{sec:pwa.analysis_model.dyn_amp}.  It is worth stressing that we
assume the coupling $\alpha_{r \to 1 + 2}$ that appears at the decay
vertex to be independent of~$m_r$.  Note that due to the chosen
conventions, the complex-conjugate $D$-function appears in
\cref{eq:2_body_decay_amp}.  Compared to
\cref{eq:2_body_decay_amp_ansatz_3}, we added the two-body decay
amplitudes $\mathcal{A}_1^{J_1\, \lambda_1\, L_1\, S_1}$ and
$\mathcal{A}_2^{J_2\, \lambda_2\, L_2\, S_2}$ of particles~1 and~2,
which are different from unity only if the respective daughter
particle is also unstable and decays into further particles, \ie when
the daughter appears as an isobar in the decay chain.  In this case,
the decay amplitude of the respective daughter particle has the same
form as \cref{eq:2_body_decay_amp}.  To simplify notation, we assume
spinless final-state particles in the remaining text, which is true
for all final states discussed in this paper.

\subsubsection{Parameterization of Propagator Terms}
\label{sec:pwa.analysis_model.dyn_amp}

As was shown in \cref{sec:scattering.pw_expansion}, the propagator
terms $\mathcal{D}_r(m_r)$ of the isobar resonances in
\cref{eq:2_body_decay_amp} can be approximated by relativistic
Breit--Wigner amplitudes with constant width as in \cref{eq:bw.rel},
\ie
\begin{equation}
  \label{eq:BW_const_width}
  \mathcal{D}_r^\text{BW}(m_r; m_0, \Gamma_0)
   = \frac{m_0\, \Gamma_0}{m_0^2 - m_r^2 - i\, m_0\, \Gamma_0}\eqPunctSpacing,
\end{equation}
where $m_0$~and~$\Gamma_0$ are nominal mass and total width of the
resonance~$r$.\footnote{The numerator in \cref{eq:BW_const_width}
  depends on the chosen normalization. The decay amplitudes are
  independent of this choice because they are normalized separately
  (see \cref{sec:pwa_cells.normalization}).}

However, in \cref{sec:scattering.pw_expansion} it was also shown that
\cref{eq:BW_const_width} is a good approximation only for narrow
resonances.  For wider resonances, a better approximation is the
relativistic Breit--Wigner amplitude with dynamic width as in
\cref{eq:bw.rel.dyn_width}, \ie
\begin{equation}
  \label{eq:BW_mass-dep_width}
  \mathcal{D}_r^\text{BW}(m_r; m_0, \Gamma_0)
   = \frac{m_0\, \Gamma_0}{m_0^2 - m_r^2 - i\, m_0\, \Gamma(m_r)}
  \quad\text{with}\quad
  \Gamma(m_r)
  = \sum_j^{\mathclap{\substack{\text{decay} \\ \text{modes}}}}
  \Gamma_j\, \frac{q_j}{m_r}\, \frac{m_0}{q_{j, 0}}\, \frac{F_{L_j}^2(q_j)}{F_{L_j}^2(q_{j, 0})}\eqPunctSpacing,
\end{equation}
where we have generalized \cref{eq:dyn_width} to the multi-channel
case.  This parameterization takes into account the opening of the
decay phase space for the decay modes~$j$ of resonance~$r$ across the
resonance width and the centrifugal barrier due to the orbital angular
momenta~$L_j$ in the respective decay modes via the barrier
factors~$F_{L_j}$ (see \cref{eq:bw_factor}).  Note that compared to
\cref{eq:bw.rel.dyn_width} the numerator in $\mathcal{D}_r^\text{BW}$
is constant.  The phase space factor from \cref{eq:bw.rel.dyn_width}
does not appear because we model the partial-wave amplitude and the
barrier factor~$F_{L_j}$ is already taken into account as a separate
factor in \cref{eq:2_body_decay_amp}.  It is important to note that
the decay amplitude in \cref{eq:2_body_decay_amp} is proportional to
$F_{L_j}$ and not to $F_{L_j}^2$ as is the case for the partial-wave
amplitude in \cref{eq:bw.rel.dyn_width}.  This is because the decay
amplitude describes the decay of a resonance, where the barrier factor
enters only via the decay vertex, and not a $2 \to 2$ scattering
process, where the barrier factor enters in addition via the
production vertex.

The dynamic total width is given by the sum of the phase-space volumes
of all decay modes, weighted by their partial widths~$\Gamma_j$.  In
\cref{eq:BW_mass-dep_width}, we assume that all decay modes of~$r$ are
two-body decays that are characterized by a relative orbital angular
momentum~$L_j$ between the two daughter particles, which are assumed
to be stable.  The two-body phase-space volume is proportional to
$q_j / m_r$ (see \cref{eq:dPhi_2.cms}), where
$q_j(m_r; m_{j, 1}, m_{j, 2})$ is the magnitude of the two-body
breakup momentum in the $r$~rest frame as given by
\cref{eq:breakup_mom.cms.f,eq:kaellen} with $m_{j, 1}$ and $m_{j, 2}$
being the masses of the daughter particles of decay mode~$j$.  Since
$q_{j, 0} \coloneqq q_j(m_0)$, the dynamic width is normalized such
that
\begin{equation}
  \label{eq:mass-dep_width_norm}
  \Gamma(m_0)
  = \sum_j^{\mathclap{\substack{\text{decay} \\ \text{modes}}}} \Gamma_j\eqPunctSpacing.
\end{equation}
Although we assume in \cref{eq:BW_mass-dep_width} that the daughter
particles of~$r$ are stable, this parameterization is often also
applied to cases where at least one of the daughter particles is
unstable.  Such two-body approximations neglect the effect of the
finite width(s) of the unstable daughter particle(s).\footnote{Note
  that \cref{eq:BW_mass-dep_width,eq:breakup_mom.cms.f,eq:kaellen} are
  not a good approximation anymore if subthreshold contributions are
  important, \ie if the decay of~$r$ proceeds via the low-mass tail(s)
  of the unstable daughter particle(s).}

The Breit--Wigner parameterizations in
\cref{eq:BW_const_width,eq:BW_mass-dep_width} do not take into account
coupled-channel effects due to multiple decay modes of a resonance.
In cases, where the threshold of a decay channel is close to the
resonance, the next best approximation beyond the Breit--Wigner
amplitude is usually the Flatt\'e parameterization in \cref{eq:flatte}.
More general approaches for the parameterization of propagator terms
that respect at least analyticity and two-body unitarity, such as
$K$-matrix approaches (see \cref{sec:scattering.k-matrix}) are usually
more difficult to employ.  One of the difficulties is that these
amplitudes typically have unknown parameters that need to be estimated
from data.  As will be discussed in
\cref{sec:pwa_cells.likelihood_fit}, we cannot allow the decay
amplitudes $\widetilde{\Psi}_{X \to \text{FS}}$ to have any free
parameters because otherwise the problem becomes prohibitively
expensive in terms of computational resources.

\subsubsection{Coordinate Systems}
\label{sec:pwa.analysis_model.coordsys}

The angles that describe the two-body decays in the $X$~decay chain
and that enter the Wigner $D$-function in \cref{eq:2_body_decay_amp},
are defined in the respective rest frame of the parent particles using
right-handed coordinate systems.  For high-energy scattering reactions
with beam particle~$a$, target particle~$b$, which is at rest in the
laboratory frame, and target recoil~$c$, the decay of~$X$ is usually
described in the Gottfried--Jackson~(GJ) frame.  In this reference
frame, the direction of the beam particle defines the \zGJ~axis and
the \yGJ~axis is given by the normal of the production plane:
$\yGJv \propto \hat{p}_a^\text{\,lab} \times \hat{p}_X^\text{\,lab}
\propto \hat{p}_c^\text{\;GJ} \times \hat{p}_a^\text{\;GJ}$, where
unit vectors are indicated by circumflexes.  Since in the GJ~frame
$X$~is at rest, the momenta of its two daughter particles are back to
back.  Thus the angular distribution is described by the polar
angle~\thetaGJ and the azimuthal angle~\phiGJ of one of the daughter
particles.  The choice of this analyzer is a question of convention,
but care has to be taken that the analyzer is chosen consistently for
all considered decay chains.

The decays of the isobars are described in the respective helicity
reference frames~(HF), which are constructed recursively by boosting
from the rest frame of the parent particle of the respective isobar
into the isobar rest frame.  The coordinate system is defined by
taking the \zHF~axis along the original direction of motion of the
isobar, \ie opposite to the direction of motion of its parent particle
in the isobar rest frame, and
$\yHFv \propto \hat{z}_\text{parent} \times \zHFv$, where
$\hat{z}_\text{parent}$ is the direction of the $z$~axis in the parent
rest frame.  In the helicity frame, the two daughter particles of the
isobar are emitted back to back, so that the angular distribution is
described by the polar angle~\thetaHF and the azimuthal angle~\phiHF
of one of the daughters.  Again, the choice of this analyzer is a
question of convention but has to be consistent across different decay
chains.

\begin{figure}[tb]
  \centering
  \includegraphics[width=0.65\textwidth]{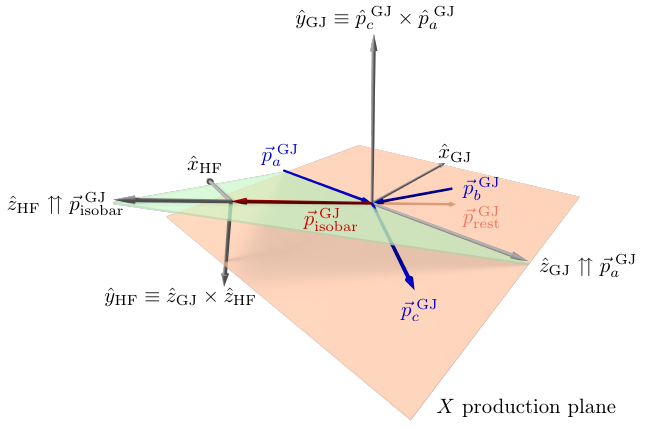}
  \caption{Definition of the Gottfried--Jackson~(GJ) and helicity~(HF)
    reference frames for the reaction $a + b \to X + c$ with
    $X$~decaying into an isobar and a rest, which may be either a
    final-state particle or again an isobar.  Here, $a$~is the beam,
    $b$~the target, and $c$~the recoil particle.  Unit vectors are
    indicated by circumflexes. (Adapted from
    \refCite{Adolph:2015tqa})}
  \label{fig:pwa.coordsys}
\end{figure}

\Cref{fig:pwa.coordsys} illustrates the definition of the
Gottfried--Jackson and helicity reference frames for the decay
$X \to \text{isobar} + \text{bachelor}$ at the beginning of an
$n$-body decay chain with $n \geq 3$.  Here, the direction of the
isobar is chosen as the analyzer for the $X$~decay.

\subsubsection{Examples for Decay Amplitudes}
\label{sec:pwa.analysis_model.examples}

\paragraph{Decay into Two Spinless Final-State Particles}
\label{sec:pwa.analysis_model.examples.2body}

The simplest example is the decay of~$X$ into a two-body final state
of spinless particles, \ie $X \to 1 + 2$.\footnote{This could be, for
  example, $X \to \eta \pi$ or $\eta' \pi$, where $\eta$, $\eta'$,
  and~$\pi$ are considered as quasi-stable particles.}  In this case,
the amplitude $\widetilde{\Psi}_{X \to 1 + 2}(m_X, \tau_2)$ for the
$X$~decay is given by \cref{eq:2_body_decay_amp} with $r = X$ and
$\mathcal{A}_1 = \mathcal{A}_2 = 1$.  In addition, we have to take
into account that the propagator term $\mathcal{D}_X(m_X)$ for~$X$ is
already accounted for as a factor in \cref{eq:ampl_ansatz_single_res}.
With $J_1 = J_2 = 0$ and $\lambda_1 = \lambda_2 = 0$, the decay
amplitude hence reads
\begin{multline}
  \label{eq:decay_amp_2body}
  \widetilde{\Psi}_{X \to 1 + 2}(m_X,
  \Overbrace{\thetaGJ, \phiGJ \vphantom{\widetilde{\Psi}}}%
  {\displaystyle{\eqqcolon \tau_2}})
  = \sqrt{\frac{2 L_X + 1}{4\pi}}\,
  \alpha_{X \to 1 + 2}\,
  F_{L_X}(m_X) \\
  \times \clebsch{0}{0}{0}{0}{S_X}{\lambda}\,
  \clebsch{L_X}{0}{S_X}{\lambda}{J}{\lambda_X}\,
  D_{M\; \lambda_X}^{J \text{*}}(\phiGJ, \thetaGJ, 0)\eqPunctSpacing.
\end{multline}
The only non-vanishing Clebsch--Gordan coefficients are the ones, where
the total spin~$S_X$ of the $(1, 2)$~system and its spin
projection~$\lambda$ are both zero and as a consequence
$\lambda_X = \lambda = 0$ and $L_X = J$.  Since $\lambda_X = 0$, the
Wigner $D$-function reduces to the spherical harmonics $Y_J^M$
according to
\begin{equation}
  \label{eq:d_func_sph_harm}
  D_{M\; 0}^J(\phiGJ, \thetaGJ, 0)
  = \sqrt{\frac{4\pi}{2 J + 1}}\, Y_J^{M \text{*}}(\thetaGJ, \phiGJ)\eqPunctSpacing.
\end{equation}
Therefore,
\begin{equation}
  \label{eq:decay_amp_2body_2}
  \widetilde{\Psi}_{X \to 1 + 2}(m_X, \thetaGJ, \phiGJ)
  = \alpha_{X \to 1 + 2}\,  F_J(m_X)\, Y_J^M(\thetaGJ, \phiGJ)\eqPunctSpacing.
\end{equation}

\paragraph{Decay into Three Spinless Final-State Particles}
\label{sec:pwa.analysis_model.examples.3body}

For $n$-body decays of~$X$ with $n \geq 3$, the decay amplitude
$\widetilde{\Psi}_{X \to \text{FS}}(m_X, \tau_n)$ is calculated by
recursive application of \cref{eq:2_body_decay_amp}.  The simplest
example for a three-body decay is the decay $X \to r + 3$ with
$r \to 1 + 2$, where 1, 2, and 3 are spinless particles.  Here, an
intermediate isobar resonance~$r$ appears in the
$(1, 2)$~subsystem.\footnote{A concrete example would be
  $X^- \to \pi^-\pi^-\pi^+$, where the two indistinguishable $\pi^-$
  need to be symmetrized as described in
  \cref{sec:pwa.analysis_model.symmetrization}.}  Applying
\cref{eq:2_body_decay_amp} recursively, the amplitude for such a decay
reads
\begin{multline}
  \label{eq:decay_amp_3body}
  \widetilde{\Psi}_{X \to 1 + 2 + 3}(m_X,
  \Overbrace{\thetaGJ, \phiGJ, m_r, \thetaHF, \phiHF \vphantom{\widetilde{\Psi}}}%
  {\displaystyle{\eqqcolon \tau_3}})
  = \sqrt{\frac{2 L_X + 1}{4\pi}}\,
  \alpha_{X \to r + 3}\,
  F_{L_X}(m_X) \\
  \times \sum_{\lambda_r}
  \Overbrace{\clebsch{J_r}{\lambda_r}{0}{0}{S_X}{\lambda}}%
  {\displaystyle{= \delta_{J_r S_X}\, \delta_{\lambda_r \lambda}}}\,
  \clebsch{L_X}{0}{S_X}{\lambda}{J}{\lambda_X}\,
  D_{M\; \lambda_X}^{J \text{*}}(\phiGJ, \thetaGJ, 0) \\[-1.5ex]
  \times
  \Underbrace{\mathcal{D}_r(m_r)\,
    \alpha_{r \to 1 + 2}\,
    F_{J_r}(m_r)\,
    Y_{J_r}^{\lambda_r}(\thetaHF, \phiHF)}%
  {\mathclap{\displaystyle{= \mathcal{A}_r^{J_r\, \lambda_r\, L_r\, S_r}(m_r, \thetaHF, \phiHF)~\text{as given by \cref{eq:2_body_decay_amp}}}}}\eqPunctSpacing.
\end{multline}
The first part of the decay amplitude describes the decay
$X \to r + 3$ and corresponds to the amplitude
$\mathcal{A}_X^{J\, M\, L_X\, S_X}(m_X, \thetaGJ, \phiGJ)$ as given by
\cref{eq:2_body_decay_amp}, with the only difference that the
propagator term $\mathcal{D}_X(m_X)$ for~$X$ does not appear here
since it is already explicitly contained as a factor in
\cref{eq:ampl_ansatz_single_res}.  In \cref{eq:decay_amp_3body},
$\lambda_r$~is the helicity of~$r$ in the Gottfried--Jackson rest frame
of~$X$.  Since $J_3 = 0$ and hence $\lambda_3 = 0$, the total
spin~$S_X$ of the $(r, 3)$~system is equal to~$J_r$ and
$\lambda = \lambda_X = \lambda_r$.  The last equality arises because
$L_X$~has no component along the quantization axis in the helicity
frame.  The second part of \cref{eq:decay_amp_3body} describes the
decay $r \to 1 + 2$.  Since both final-state particles are spinless,
the same arguments apply as for the decay $X \to 1 + 2$ discussed in
the paragraph above.  \Cref{eq:decay_amp_3body} can thus be simplified
to
\begin{multline}
  \label{eq:decay_amp_3body_2}
  \widetilde{\Psi}_{X \to 1 + 2 + 3}(m_X, \thetaGJ, \phiGJ, m_r, \thetaHF, \phiHF)
  = \sqrt{\frac{2 L_X + 1}{4\pi}}\,
  \alpha_{X \to r + 3}\,
  F_{L_X}(m_X)\,
  \mathcal{D}_r(m_r)\,
  \alpha_{r \to 1 + 2}\,
  F_{J_r}(m_r) \\
  \times \sum_{\lambda_r}
  \clebsch{L_X}{0}{J_r}{\lambda_r}{J}{\lambda_r}\,
  D_{M\; \lambda_r}^{J \text{*}}(\phiGJ, \thetaGJ, 0)\,
  Y_{J_r}^{\lambda_r}(\thetaHF, \phiHF)\eqPunctSpacing.
\end{multline}

\subsubsection{Symmetrization of the Decay Amplitudes}
\label{sec:pwa.analysis_model.symmetrization}

If the $n$-body final state contains indistinguishable particles, the
decay amplitude $\widetilde{\Psi}_{X \to \text{FS}}$ has to be
symmetrized accordingly.  For mesonic final states,
$\widetilde{\Psi}_{X \to \text{FS}}$ has to be totally symmetric under
exchange of any of the indistinguishable final-state mesons (Bose
symmetry).  The symmetrized decay amplitude is constructed by summing
the amplitudes for all $N_\text{perm}$ permutations of the
indistinguishable final-state mesons.  For each permutation~$k$ of the
final-state four-momenta, the phase-space variables~$\tau_{n, k}$ are
calculated and then used to calculate the decay amplitude
$\widetilde{\Psi}_{X \to \text{FS}}^{k}(m_X, \tau_{n, k})$.  Using
these amplitudes, the Bose symmetrized decay amplitude is given by
\begin{equation}
  \label{eq:decay_amp_bose}
  \widetilde{\Psi}_{X \to \text{FS}}^\text{sym}
  = \frac{1}{\sqrt{N_\text{perm}}} \sum_{k = 1}^{N_\text{perm}}
  \widetilde{\Psi}_{X \to \text{FS}}^{k}(m_X, \tau_{n, k})\eqPunctSpacing.
\end{equation}

For some final states, the same isobar resonance may appear in
different subsystems of the final-state particles.  In this case, the
relative magnitudes and phases of the corresponding decay amplitudes
are fixed by the respective isospin Clebsch--Gordan
coefficients.\footnote{For example, in the $\pi^-\pi^0\pi^+$ final
  state, which can be produced in charge-exchange $\pi^-$-$p$
  scattering, the $\rho(770)$ isobar appears in the $\pi^-\pi^0$,
  $\pi^0\pi^+$, and $\pi^-\pi^+$ subsystems.  The decay amplitudes for
  $X^0 \to \rho(770)^-\, \pi^+$, $\rho(770)^+\, \pi^-$, and
  $\rho(770)^0\, \pi^0$ are related by isospin Clebsch--Gordan
  coefficients (see \eg\ \refCite{Cashmore:1973na}).}  Also if one
compares or combines data from decay channels that are related by
isospin symmetry, isospin Clebsch--Gordan coefficients need to be taken
into account.\footnote{For example, for the $\pi^-\pi^-\pi^+$ and
  $\pi^-\pi^0\pi^0$ final states different isospin Clebsch--Gordan
  coefficients appear depending on whether the two-pion isobar
  resonance is isoscalar or isovector.}  In addition, the analyzers
for the angular distributions have to be chosen consistently for the
final states.

To simplify notation, we redefine
$\widetilde{\Psi}_{X \to \text{FS}}(m_X, \tau_n)$ to represent the
Bose- and isospin-symmetrized decay amplitude in the remaining text.

\subsubsection{Extension of the Model to Several Intermediate States}
\label{sec:pwa.analysis_model.extension}

The decay amplitude in \cref{eq:decay_amp_bose} is completely defined
by the quantum numbers of~$X$, the quantum numbers and propagator
terms of the isobars, the total intrinsic spins, and the orbital
angular momenta in the decay chain.  If final-state particles have
spin, their helicities appear as additional parameters, which for
unpolarized reactions are summed over incoherently, \ie at the
intensity level.  In diffractive-dissociation reactions, the possible
quantum numbers for~$X$ are limited only by the conservation laws of
the strong interaction.  As discussed in
\cref{sec:pheno.qm.quantum-numbers}, the quantum numbers of a light
meson are defined by isospin~$I$, spin~$J$, parity~$P$, and spin
projection~$M$.  If $X$~is a non-strange meson, also the $G$~parity is
defined; if it is in addition neutral, also the
$C$~parity.\footnote{By convention, for charged non-strange~$X$ the
  $C$~parity of the neutral member of the isospin triplet is quoted
  instead of the $G$~parity.}  In case $X$~is a strange meson,
neither~$G$ nor $C$~parity are defined and the state is characterized
by its strangeness quantum number~$\mathsf{S}$.

For convenience, we introduce an index~$i$ that summarizes all
information required to calculate the decay amplitude (except possible
helicities of the final-state particles):
\begin{equation}
  \label{eq:wave_index}
  i \coloneqq \big\{ I^{(G)}\, \JP\, M;
  \text{isobars, total intrinsic spins, orbital angular momenta} \big\}\eqPunctSpacing.
\end{equation}
A particular index~$i$ represents a \emph{partial wave} and
$|\widetilde{\Psi}_i(m_X, \tau_n)|^2$ represents the
$(3n - 4)$-dimensional kinematic distribution of the final-state
particles for this partial wave for a given value of~$m_X$.

With the above definitions, the amplitude in
\cref{eq:ampl_ansatz_single_res}, that describes the production of a
specific~$X$ and its decay chain which are both defined by~$i$, reads
\begin{equation}
  \label{eq:ampl_single_res}
  \mathcal{M}_{a + b \to (X \to \text{FS})_i + c}(m_X, t', \tau_n)
  = \mathcal{P}_{a + b \to X + c}(m_X, t')\,
  \mathcal{D}_X(m_X)\, \widetilde{\Psi}_i(m_X, \tau_n)\eqPunctSpacing.
\end{equation}
The decay amplitude $\widetilde{\Psi}_i$ as given by
\cref{eq:decay_amp_bose,eq:2_body_decay_amp} contains for each of the
$N_\text{vertex}$ two-body decay vertices a complex-valued coupling
$\alpha_{\text{parent} \to \text{daughter}_1 + \text{daughter}_2}$,
which is in general unknown.  Assuming that these couplings are
independent of~$m_X$ or any of the phase-space variables, they can be
pulled out of each two-body amplitude
$\mathcal{A}_r^{J_r\, M_r\, L_r\, S_r}$ so that
\begin{equation}
  \label{eq:decay_amp_redef}
  \widetilde{\Psi}_i(m_X, \tau_n)
  \coloneqq \Underbrace{\left[\prod_{k = 1}^{N_\text{vertex}} \alpha_k \right]}%
                       {\displaystyle{\eqqcolon \alpha_i}}\,
  \overline{\Psi}_i(m_X, \tau_n)\eqPunctSpacing.
\end{equation}
Here, $\alpha_k$~is the coupling at the $k$th two-body decay vertex in
the decay chain.  Assuming that we know the propagator terms for all
isobars, the decay amplitude $\overline{\Psi}_i(m_X, \tau_n)$ defined
in \cref{eq:decay_amp_redef} is calculable and does not contain any
unknown parameters.

Usually, also the amplitude $\mathcal{P}_{a + b \to X + c}(m_X, t')$
that describes the production of~$X$ in \cref{eq:ampl_single_res} is
only partly known.  As was discussed in \cref{sec:regge}, the strong
interaction in $t$-channel scattering processes can be described as
the exchange of Reggeons.  At high beam energies, Pomeron exchange
becomes dominant (see \cref{sec:exp.prod_reactions}).  Using Regge
theory, one can usually construct an approximate amplitude
$\mathcal{P}(m_X, t')$ that at least models the average production
probability of an intermediate state with mass~$m_X$ as a function
of~$t'$.  The unknown details of the beam-Reggeon vertex are
factorized into a so-called \emph{coupling amplitude}
$\mathcal{C}_{X, i}(t')$, which also absorbs the unknown~$\alpha_k$
from \cref{eq:decay_amp_redef}.  The coupling amplitude therefore
depends not only on the $X$~quantum numbers but also on the decay
chain and hence on~$i$.  With these definitions, the amplitude for the
scattering process in \cref{eq:ampl_single_res} reads
\begin{equation}
  \label{eq:ampl_single_res_redef}
  \mathcal{M}_{a + b \to (X \to \text{FS})_i + c}(m_X, t', \tau_n)
  = \mathcal{P}(m_X, t')\, \mathcal{C}_{X, i}(t')\,
  \mathcal{D}_X(m_X)\, \overline{\Psi}_i(m_X, \tau_n)\eqPunctSpacing.
\end{equation}

Up to now we considered only the hypothetical case of a single
intermediate resonance~$X$ with a single decay chain.  However,
usually several resonances~$X$ with the same quantum numbers may
contribute to the same decay chain, \ie to the same partial wave~$i$.
Since these resonances appear as intermediate states, they may
interfere and the corresponding amplitudes have to be summed
coherently.  In addition, different partial waves, \ie resonances with
different quantum numbers and different decay chains, may contribute
to the intermediate states~$X$.  Since these resonances appear as
intermediate states and eventually decay into the same final-state
particles, the corresponding partial-wave amplitudes may also
interfere and hence have to be added coherently.\footnote{In
  \cref{sec:pwa_cells.rank}, we will discuss possible sources of
  incoherence and how to include them in the analysis model.
  Incoherences also arise in the analysis of data on quasi-real
  photoproduction via the Primakoff reaction.  They will be discussed
  in \cref{sec:results_3pic_primakoff}.}  Therefore, the total
amplitude for the production of the $n$-body final state taking into
account all appearing intermediate resonances~$X$ and all their
various decay chains is
\begin{equation}
  \label{eq:ampl_multi_res}
  \mathcal{M}_{fi}(m_X, t', \tau_n)
  = \mathcal{P}(m_X, t')\, \sum_i^{N_\text{waves}} \bigg[ \sum_{k\, \in\, \mathbb{S}_i}
  \mathcal{C}_{k i}(t')\, \mathcal{D}_k(m_X) \bigg]\, \overline{\Psi}_i(m_X, \tau_n)\eqPunctSpacing,
\end{equation}
where the index~$k$ enumerates the various contributing $n$-body
resonances~$X$.  The inner sum runs over the subsets $\mathbb{S}_i$ of
the indices of those resonances that contribute to partial wave~$i$.
The same resonance may appear in several waves and several resonances
may appear in the same wave.  Depending on the context, the
partial-wave index~$i$ has two meanings.  It either represents
directly the set of quantum numbers that define a partial-wave
amplitude as in \cref{eq:wave_index} or it enumerates all waves
consecutively, \ie $i$~is an integer that uniquely identifies a
specific wave as in \cref{eq:ampl_multi_res}.

\Cref{eq:ampl_multi_res} is a model for the $(2 + 3n - 4)$-dimensional
intensity distribution $\mathcal{I}(m_X, t', \tau_n)$ as defined in
\cref{eq:intensity_def}.  In principle, the best possible approach
would be to fit \cref{eq:ampl_multi_res} directly to the data.  Such
an approach is often called \emph{global fit}.  However, in many cases
it is impractical or even impossible to perform such a global fit.
One of the reasons for this is that the set~$\{k\}$ of resonances that
contribute to the intermediate states~$X$ is usually unknown, as are
the corresponding resonance parameters.  It is the goal of the
analysis to determine the resonance content and the resonance
parameters from the data.  Often, the analysis is guided by
information from the kinematic distributions, in particular from the
invariant mass distribution of the $n$-body final state, and by
results from previous experiments.  As was discussed in
\cref{sec:exp.prod_reactions}, the identification of resonances in the
data is complicated by contributions from non-resonant components that
have to be taken into account.  This will be discussed further in
\cref{sec:pwa.res_fit}.  Also the values of the coupling amplitudes
$\{\mathcal{C}_{k i}(t')\}$ are in general unknown and have to be
inferred from data.  Up to now there are no models that are able to
describe the details of the $t'$~dependence of the coupling
amplitudes.  Finally, even if a realistic model of the form of
\cref{eq:ampl_multi_res} could be formulated, it is in most cases
computationally too expensive to fit it to the high-dimensional data
(see \cref{sec:pwa_cells.likelihood_fit}).

Given the large data samples of recent experiments, it has proven
advantageous to pursue an opposite analysis approach.  Instead of
formulating the ultimate model of the form of \cref{eq:ampl_multi_res}
that has to describe all details of the whole data sample, one tries
to reduce the required modeling and hence the model dependence to a
minimum.  This approach is especially attractive in cases where the
analyzed data samples are so large that systematic uncertainties
dominate the total uncertainties.  Commonly, a two-stage approach is
used to separate the decomposition of the data into partial-wave
amplitudes with well-defined $X$~quantum numbers and decay chains, \ie
the modeling of the $\tau_n$~dependence of~$\mathcal{I}$, from the
extraction of the intermediate $n$-body resonances, \ie the modeling
of the~$m_X$ and $t'$~dependence of $\mathcal{I}$.  As will be
discussed further in \cref{sec:pwa_cells.discussion} below, this
two-stage approach circumvents or at least separates many of the
issues and limitations that would be encountered in global fits.

In the first analysis stage, which is often called \emph{partial-wave
  decomposition} or \emph{mass-independent fit}, we make as little
assumptions as possible about the intermediate states~$X$.  To this
end, we start from \cref{eq:intensity_def,eq:ampl_multi_res} and
collect all the \apriori unknown quantities into the so-called
\emph{transition amplitudes} $\overline{\mathcal{T}_i}(m_X, t')$:
\begin{equation}
  \label{eq:intensity_model}
  \mathcal{I}(m_X, t', \tau_n)
  = \bigg| \sum_i^{N_\text{waves}} \Overbrace{\sqrt{m_X}\, \mathcal{P}(m_X, t')\,
    \bigg[ \sum_{k\, \in\, \mathbb{S}_i} \mathcal{C}_{k i}(t')\, \mathcal{D}_k(m_X) \bigg]}%
  {\displaystyle{\eqqcolon \overline{\mathcal{T}_i}(m_X, t')}}\,
  \overline{\Psi}_i(m_X, \tau_n) \bigg|^2\eqPunctSpacing.
\end{equation}
Here, we have implicitly rescaled the coupling amplitudes
$\mathcal{C}_{k i}(t')$ such that they include the proportionality
factor between~$\mathcal{I}$ and the
$m_X\, \left| \mathcal{M}_{fi}(m_X, t', \tau_n) \right|^2$ term in
\cref{eq:intensity_def}.  The goal of the first analysis stage is to
determine the set $\{\overline{\mathcal{T}_i}(m_X, t')\}$ of
transition amplitudes from the data.  This is possible because we can
calculate the decay amplitudes $\overline{\Psi}_i(m_X, \tau_n)$ for a
given value of~$m_X$ assuming that we know the propagator terms of all
intermediate isobar resonances that appear in
\cref{eq:intensity_model} via \cref{eq:2_body_decay_amp} precisely and
without any free parameters.  In order to avoid modeling of the
$m_X$~and $t'$~dependence of the transition amplitudes, the
partial-wave decomposition is performed in small $(m_X, t')$ cells.
This yields binned approximations to the
$\overline{\mathcal{T}_i}(m_X, t')$ and will be explained in
\cref{sec:pwa_cells}.  Because of this binning approach, the first
analysis stage makes no assumptions about the resonance content of the
partial waves.  Based on the isobar model for the $n$-body decay
of~$X$, the data are decomposed into partial-wave amplitudes with
well-defined quantum numbers that are given by the partial-wave
index~$i$ as defined in \cref{eq:wave_index}.  One may think of the
partial-wave decomposition as merely a transformation of the data from
four-momentum space of the final-state particles into the space of
transition amplitudes.  This yields a representation of the data in
terms of intensities and relative phases of transition amplitudes that
allows for a direct interpretation in terms of resonances.  As was
discussed in \cref{sec:exp.prod_reactions}, the extracted partial-wave
amplitudes also contain contributions from non-resonant processes that
have to be taken into account by the resonance model (see discussion
below).  Since the partial-wave decomposition takes into account the
detection and reconstruction efficiency of the experimental setup (see
\cref{sec:pwa_cells.likelihood_fit}) and thanks to the chosen
normalization (see \cref{sec:pwa_cells.normalization}), the results
can be directly compared across experiments.  It is important to note
that the model in \cref{eq:intensity_model} describes the density of
events differential in all phase-space variables.  As a consequence,
the partial-wave decomposition is able to extract the interference
even between waves~$i$ and~$j$ that have different quantum numbers and
hence correspond to orthogonal decay amplitudes, for which
$\int \dif{\Phi_n}(m_X, \tau_n)\, \overline{\Psi}_i(m_X, \tau_n)\,
\overline{\Psi}_j^*(m_X, \tau_n) = 0$.\footnote{This is in contrast to
  Dalitz-plot analyses of three-body decays of spinless mesons, where
  such interferences are inaccessible, because one integrates over the
  three Euler angles that define the spatial orientation of the
  three-body system leaving only two of the five phase-space
  variables.}

In the second analysis stage, which is often called
\emph{resonance-model fit} or \emph{mass-dependent fit}, we use the
definition of the transition amplitudes in \cref{eq:intensity_model}
to construct a model that describes the $m_X$~and $t'$~dependence of a
selected subset of transition amplitudes in terms of resonant and
non-resonant components.  This will be explained in
\cref{sec:pwa.res_fit}.  It is important to note that the
resonance-model fit does not have to describe \emph{all} transition
amplitudes that are extracted from the data, but may focus on selected
partial waves.  This is in particular important for the analysis of
the high-energy scattering reactions considered here, where often
sizable non-resonant components contribute to~$X$ (see
\cref{sec:exp.prod_reactions}).  These non-resonant contributions
project into all transition amplitudes and often dominate the
intensity of high-spin waves.  Such waves are only important to
describe the data but are uninteresting when it comes to extracting
resonances.  They are therefore not included in resonance-model fits.

\subsection{Stage~I: Partial-Wave Decomposition in Kinematic Cells}
\label{sec:pwa_cells}

The decomposition of the data into partial-wave amplitudes with
well-defined quantum numbers and isobar decay chains constitutes the
first stage of the analysis.  In this stage, the~$m_X$ and
$t'$~dependence of the unknown transition
amplitudes~$\overline{\mathcal{T}_i}$, which are defined in
\cref{eq:intensity_model}, is extracted from the data by subdividing
the data sample into narrow bins in these two kinematic variables.  We
assume that the~$m_X$ and $t'$~bins are narrow enough, such that we
can in good approximation neglect the~$m_X$ and $t'$~dependence within
the $(m_X, t')$ cells.  Hence within a given $(m_X, t')$ cell, the
intensity distribution in \cref{eq:intensity_model} is a function only
of the set of phase-space variables represented by $\tau_n$.  The
other two kinematic variables, $m_X$~and~$t'$, appear as constant
parameters:
\begin{equation}
  \label{eq:intensity_model_bin}
  \mathcal{I}(\tau_n; m_X, t')
  = \bigg| \sum_i^{N_\text{waves}} \overline{\mathcal{T}_i}(m_X, t')\,
  \overline{\Psi}_i(\tau_n; m_X) \bigg|^2\eqPunctSpacing.
\end{equation}
Compared to a functional description of the $m_X$~and $t'$~dependence
of the transition amplitudes, the binned approach does not require a
model for the $n$-body resonances and the non-resonant contributions.
It therefore does not make any assumptions about the resonance content
of the transition amplitudes.  The only model dependence that enters
at this point is the truncation of the partial-wave expansion, \ie the
set of waves used in \cref{eq:intensity_model_bin}.  This issue will
be discussed in \cref{sec:pwa_cells.discussion}.  In addition, the
kinematic cells are statistically independent so that the partial-wave
decomposition can be performed in parallel.  A caveat of the binning
approach is that the method introduces a large number of free
parameters---the values of the transition amplitudes in each
$(m_X, t')$ cell---that need to be determined from data.  For that
reason, this approach is only applicable to sufficiently large data
sets.

In order to make the PWA model more realistic,
\cref{eq:intensity_model_bin} still needs to be extended in order to
take into account \one possible incoherent background contributions,
\two incoherences caused by the spin states of the target and recoil
particles, and \three parity conservation in the strong-interaction
scattering process.

\subsubsection{Incoherent Background Contributions}
\label{sec:pwa_cells.flat}

The analyzed data are usually contaminated by misreconstructed or
partially reconstructed events that are similar to the signal process
and hence fulfill the event selection criteria.  We model this
background by a distribution that is isotropic in phase space.
Therefore, these events have a constant probability density over all
phase-space elements.  In order to account for such events, we
incoherently add a component to \cref{eq:intensity_model_bin}, the
so-called \emph{flat wave}.  The corresponding decay amplitude
$\overline{\Psi}_\text{flat}$ is constant and, without loss of
generality, we set $\overline{\Psi}_\text{flat} \equiv 1$ so that
\begin{equation}
  \label{eq:intensity_model_bin_flat}
  \mathcal{I}(\tau_n; m_X, t')
  = \bigg| \sum_i^{N_\text{waves}} \overline{\mathcal{T}_i}(m_X, t')\,
  \overline{\Psi}_i(\tau_n; m_X) \bigg|^2 + |\overline{\mathcal{T}}_{\!\!\text{flat}}(m_X, t')|^2\eqPunctSpacing.
\end{equation}
Since only the intensity of the flat wave enters in
\cref{eq:intensity_model_bin_flat}, the phase of this wave is
immeasurable.  Thus $\overline{\mathcal{T}}_{\!\!\text{flat}}$ is
chosen to be real-valued.

Background contributions from other processes, \eg from non-resonant
components, for which the final-state particles are correlated and are
hence distributed anisotropically in phases space, do not contribute
strongly to the flat wave and usually contaminate the transition
amplitudes of the other waves in the PWA model (see
\cref{sec:exp.prod_reactions}).  These contributions have to be taken
into account in the resonance-model fit, which constitutes the second
stage of the analysis (see \cref{sec:pwa.res_fit}).

\subsubsection{Spin-Density Matrix and Rank}
\label{sec:pwa_cells.rank}

In order to avoid potential complications from nuclear effects, often
protons are used as targets.  Since protons have spin~1/2, the
absolute value squared $|\mathcal{M}_{fi}|^2$ of the scattering
amplitude in \cref{eq:cross_section} has to be averaged over the two
spin states of the target proton and summed over the spin states of
the recoiling proton, assuming an unpolarized target and spinless beam
particles.  Due to parity conservation and rotational invariance, the
scattering amplitude depends only on the relative orientation of the
spin states of the target and the recoil proton.  Hence the cross
section consists of two incoherent parts: one for spin-flip at the
target vertex and one for spin-non-flip.  Additional incoherent terms
may arise if the target proton is excited in the scattering process.
Also performing the partial-wave decomposition over wide $t'$~ranges
may lead to effective incoherence, if the transition amplitudes have
different dependences on~$t'$.  Depending on the center-of-momentum
energy, multiple exchange processes might contribute to the scattering
process, which may lead to additional incoherence.

A way to include these incoherences into the analysis model in
\cref{eq:intensity_model_bin_flat} is the introduction of an
additional index~$r$ for the transition amplitudes that is summed over
incoherently:
\begin{equation}
  \label{eq:intensity_model_bin_rank_rho}
  \begin{aligned}
    \mathcal{I}(\tau_n; m_X, t')
    &= \sum_{r = 1}^{N_r} \bigg| \sum_i^{N_\text{waves}} \overline{\mathcal{T}_i}^r(m_X, t')\,
    \overline{\Psi}_i(\tau_n; m_X) \bigg|^2 + |\overline{\mathcal{T}}_{\!\!\text{flat}}(m_X, t')|^2 \\
    &= \sum_{i, j}^{N_\text{waves}} \overline{\Psi}_i(\tau_n; m_X)\, \bigg[
    \Underbrace{\sum_{r = 1}^{N_r} \overline{\mathcal{T}_i}^r(m_X, t')\, \overline{\mathcal{T}_j}^{r \text{*}}(m_X, t')}%
    {\displaystyle{\eqqcolon \overline{\varrho}_{ij}(m_X, t')}} \bigg]\,
    \overline{\Psi}_j^\text{*}(\tau_n; m_X) + |\overline{\mathcal{T}}_{\!\!\text{flat}}(m_X, t')|^2\eqPunctSpacing.
  \end{aligned}
\end{equation}
In \cref{eq:intensity_model_bin_rank_rho}, we introduce the Hermitian
and positive-semidefinite spin-density matrix
$\bm{\overline{\varrho}}(m_X, t')$ that completely describes the
intermediate state~$X$ in terms of combinations of pure quantum
states.  The elements $\overline{\varrho}_{ij}(m_X, t')$ of the
spin-density matrix represent the actually measurable observables.
The number~$N_r$ of transition amplitudes per partial wave corresponds
to the rank of the spin-density matrix.  For most reactions, $N_r$~is
significantly smaller than the mathematically allowed maximum, which
is given by the dimension of the spin-density matrix, \ie
by~$N_\text{waves}$.\footnote{Since the spin-density matrix is
  Hermitian and positive-semidefinite, it is diagonalizable and has
  real-valued eigenvalues.  Of these eigenvalues, $N_r$~are positive
  and $N_\text{wave} - N_r$ are zero.}  We employ the parameterization
of Chung and Trueman~\cite{Chung:1974fq} for the spin-density matrix,
which is explained in \cref{sec:chung_trueman} and reduces the number
of free real-valued parameters that need to be determined from the
data to the minimum, which is $N_r\, (2 N_\text{waves} - N_r)$.

Assuming a single production mechanism and neglecting other sources of
incoherence, the maximum rank $N_r$ of the spin-density matrix is
determined by the spins $J_a$, $J_b$, and $J_c$ of beam, target, and
recoil particle, respectively.  Together, these particles have
\begin{equation}
  \label{eq:multiplicity}
  N_\text{mult} = (2 J_a + 1)\, (2 J_b + 1)\, (2 J_c + 1)
\end{equation}
different unobserved spin states.\footnote{We assume here that all
  particles are massive.}  Taking into account that some of the spin
states are related by parity and rotational invariance, the maximum
rank is
\begin{equation}
  \label{eq:max_rank}
  N_r = \begin{cases}
    \dfrac{N_\text{mult} + 1}{2} & \text{if $N_\text{mult}$ is odd}\eqPunctSpacing, \\
    \dfrac{N_\text{mult}}{2} & \text{if $N_\text{mult}$ is even}\eqPunctSpacing.
  \end{cases}
\end{equation}
For example, the scattering process $\pi + p \to X + p$ is described
by a spin-density matrix with maximum rank of $N_r = 2$, corresponding
to spin-flip and spin-non-flip processes at the target vertex, as
discussed above.  If, hypothetically, the target proton would be
excited into an $N(1520)$ resonance with spin 3/2, the maximum rank
would increase to $N_r = 4$.  It is worth noting that
\cref{eq:multiplicity,eq:max_rank} define only the \emph{maximum} rank
of the spin-density matrix.  Depending on the scattering process and
the data sample, a lower rank might be sufficient to describe the
data.  For example, diffractive scattering of spinless beam particles
off nucleons is dominated by spin-non-flip
processes~\cite{Donnachie:2002xx,Odorico:1970ev}.  Hence for these
processes, a rank-1 spin-density matrix yields in most cases a
satisfactory description of the data, in particular if other sources
for incoherence such as target excitations or integration over large
$t'$~intervals are avoided.  A different approach that is studied at
the VES~experiment is to use a spin-density matrix with maximum rank
and then to analyze the transition amplitudes that correspond to the
largest eigenvalue of the spin-density matrix~\cite{Kachaev:2016yph}.
The rank considerations for photo- and leptoproduction are discussed
in \refCite{Salgado:2013dja}.  If multiple production mechanisms
contribute to the analyzed process, \cref{eq:multiplicity,eq:max_rank}
apply separately to each contribution.  For incoherent contributions,
the total maximum rank is given by the sum of the maximum rank values
for each contribution.

\subsubsection{Parity Conservation and Reflectivity}
\label{sec:pwa_cells.reflectivity}

Although parity violation is not forbidden in strong interactions by
any known fundamental principle~\cite{Kim:2008hd,Hook:2018dlk}, all
experimental results are in agreement with the assumption that the
strong interaction conserves parity.  A convenient way of
incorporating parity conservation into the PWA model in
\cref{eq:intensity_model_bin_rank_rho} is to consider the scattering
subprocess $a + b \to X + c$ in the so-called \emph{reflectivity
  basis}~\cite{Chung:1974fq}.  We choose the coordinate system that we
use to describe the quantum spin states of the particles such that the
quantization axis ($z$~axis) lies in the production plane.  A possible
choice for such a quantization axis is the beam axis, like it is used
in the definition of the Gottfried--Jackson frame in
\cref{sec:pwa.analysis_model.coordsys}.  If the process
$a + b \to X + c$ is parity conserving, the scattering amplitude is
invariant under space inversion.  Such a parity transformation flips
the direction of all particle momenta.  Since the four particles lie
in the production plane, this can be undone by performing, in addition
to the space inversion, a rotation by~\SI{180}{\degree} about the
production plane normal, which is taken as the $y$~axis.  The combined
transformation is represented by the reflectivity operator~\reflOp and
corresponds to a reflection through the production plane, which leaves
the momenta of those particles that lie in that plane unchanged.  A
single-particle spin state $\ket{J^P, M}$ with momentum in the
production plane, spin~$J$, intrinsic parity~$P$, and spin
projection~$M$ transforms under~\reflOp as follows
\begin{equation}
  \label{eq:refl_canonstate}
  \reflOp \ket{J^P, M} = P\, (-1)^{J - M}\, \ket{J^P, {-M}}\eqPunctSpacing.
\end{equation}
We can therefore construct eigenstates to \reflOp from linear
combinations of canonical states with spin projections of opposite
sign:
\begin{equation}
  \label{eq:refl_eigenstate}
  \ket{J^P, M^\refl}
  \coloneqq \mathcal{N}_M\, \Big[ \ket{J^P, M} - \refl\,
  \Underbrace{P\, (-1)^{J - M}\, \ket{J^P, {-M}}}{\displaystyle{= \reflOp \ket{J^P, M}}} \Big]\eqPunctSpacing,
\end{equation}
where we choose the normalization factor to be
\begin{equation}
  \label{eq:refl_eigenstate_norm}
  \mathcal{N}_M =
  \begin{cases}
    1 / \sqrt{2} & \text{for $M > 0$}\eqPunctSpacing, \\
    1 / 2        & \text{for $M = 0$}\eqPunctSpacing, \\
    0            & \text{for $M < 0$}\eqPunctSpacing.
  \end{cases}
\end{equation}
This choice ensures that in the reflectivity basis the multiplicity of
the spin state of $2J + 1$ remains unchanged.  The reflectivity
eigenstate defined in \cref{eq:refl_eigenstate} is characterized by
the spin-projection quantum number $M$ and the eigenvalue $\refl^*$ of
the reflectivity operator, \ie
\begin{equation}
  \reflOp \ket{J^P, M^\refl} = \refl^*\, \ket{J^P, M^\refl}\eqPunctSpacing.
\end{equation}
The reflectivity \refl is $\pm 1$ for bosons.\footnote{For fermions,
  the reflectivity is $\pm i$.}  According to
\cref{eq:refl_eigenstate_norm}, there are no states with $M < 0$ in
the reflectivity basis.  States with spin projection $M = 0$ vanish,
if $\refl = P\, (-1)^J$.  Hence for given~\JP, there exists only one
state with $\refl = P\, (-1)^{J + 1}$ and $M = 0$.  For each $M > 0$,
two states with $\refl = \pm 1$ exist, so that in total the
multiplicity of the spin state is $2J + 1$, as in the canonical basis.

Using \cref{eq:refl_eigenstate}, we can define rotation function for
the reflectivity eigenstates:
\begin{align}
  \prescript{\refl}{}{D}_{M'\; M}^{J}(\alpha, \beta, \gamma)
  &\coloneqq \bra{J^P, M'^\refl}\, \hat{\mathcal{R}}(\alpha, \beta, \gamma)\, \ket{J^P, M} \nonumber \\
  &= \mathcal{N}_M\, \Big[ \bra{J^P, M}\, \hat{\mathcal{R}}\, \ket{J^P, M}
    - \refl^*\, P\, (-1)^{J - M}\, \bra{J^P, {-M}}\, \hat{\mathcal{R}}\, \ket{J^P, M} \Big] \nonumber \\
  \label{eq:D_func_refl}
  &= \mathcal{N}_M\, \Big[ D_{M'\; M}^{J}(\alpha, \beta, \gamma)
    - \refl^*\, P\, (-1)^{J - M}\, D_{{-M'}\; M}^{J}(\alpha, \beta, \gamma) \Big]\eqPunctSpacing.
\end{align}
Although these functions are not a representation of the rotation
group, they still have properties similar to those of the fundamental
$D$-functions (see \cref{sec:wigner_D_function}).  In particular, the
$\prescript{\refl}{}{D}_{M'\; M}^{J}$ form an orthogonal function
system.  In order to calculate the $X$~decay amplitude in the
reflectivity basis, we replace the Wigner $D$-function
$D_{M\; \lambda_X}^J(\phiGJ, \thetaGJ, 0)$ in the two-body decay
amplitude for~$X$ by \cref{eq:D_func_refl}.

Describing the spin state of~$X$ in the reflectivity basis also
changes the structure of the spin-density matrix.  In order to fully
define a wave in the reflectivity basis, we have to specify in
addition to the wave index~$i$, which contains the spin-projection
quantum number $0 \leq M \leq J$ (see \cref{eq:wave_index}), also the
reflectivity quantum number~$\refl = \pm 1$.  Analogous to the
definition of the spin-density matrix in
\cref{eq:intensity_model_bin_rank_rho}, where $-J \leq M \leq +J$, we
can write the spin-density matrix in the reflectivity basis:
\begin{equation}
  \label{eq:spin-dens_refl_def}
  \overline{\varrho}_{ij}^{\refl \refl'}
  = \sum_{r = 1}^{N_r} \overline{\mathcal{T}}_i^{r \refl}\,
  \overline{\mathcal{T}}_j^{r \refl' \text{*}}\eqPunctSpacing.
\end{equation}
It is important to note that $\overline{\varrho}_{ij}^{\refl \refl'}$
has the same number of elements as $\overline{\varrho}_{ij}$ in
\cref{eq:intensity_model_bin_rank_rho}.

It is shown in \refCite{Chung:1974fq} that due to parity conservation
and rotational invariance, the spin-density matrix in the reflectivity
basis assumes a block-diagonal form \wrt\ \refl, \ie
\begin{equation}
  \label{eq:spin-dens_refl}
  \overline{\varrho}_{ij}^{\refl \refl'}
  = \begin{pmatrix}
    \overline{\varrho}_{ij}^{+ +} & 0 \\
    0 & \overline{\varrho}_{ij}^{- -}
  \end{pmatrix}\eqPunctSpacing.
\end{equation}
This means that all interference terms of transition amplitudes with
opposite reflectivity quantum numbers are zero.

In the reflectivity basis, the PWA model in
\cref{eq:intensity_model_bin_rank_rho} can therefore be written as
\begin{equation}
  \label{eq:intensity_model_bin_rank_rho_refl}
  \begin{aligned}
    \mathcal{I}(\tau_n; m_X, t')
    &= \sum_{\refl = \pm 1} \sum_{r = 1}^{N_r} \bigg| \sum_i^{N_\text{waves}} \overline{\mathcal{T}_i}^{r \refl}(m_X, t')\,
    \overline{\Psi}_i^\refl(\tau_n; m_X) \bigg|^2 + |\overline{\mathcal{T}}_{\!\!\text{flat}}(m_X, t')|^2 \\
    &= \sum_{\refl = \pm 1} \sum_{i, j}^{N_\text{waves}} \overline{\Psi}_i^\refl(\tau_n; m_X)\,
    \overline{\varrho}_{ij}^\refl(m_X, t')\,
    \overline{\Psi}_j^{\refl \text{*}}(\tau_n; m_X)
    + |\overline{\mathcal{T}}_{\!\!\text{flat}}(m_X, t')|^2\eqPunctSpacing,
  \end{aligned}
\end{equation}
where we use the Chung--Trueman parameterization (see
\cref{sec:chung_trueman}) for the two submatrices
\begin{equation}
  \label{eq:spin-dens_refl_2}
  \overline{\varrho}_{ij}^\refl
  \coloneqq \overline{\varrho}_{ij}^{\refl \refl}
  = \sum_{r = 1}^{N_r} \overline{\mathcal{T}}_i^{r \refl}\,
  \overline{\mathcal{T}}_j^{r \refl \text{*}}\eqPunctSpacing.
\end{equation}
In \cref{eq:intensity_model_bin_rank_rho_refl}, the
$\overline{\Psi}_i^\refl$ are the decay amplitudes in the reflectivity
basis, where the Wigner $D$-function for the $X$~decay is replaced by
the function in \cref{eq:D_func_refl}.

Up to this point, we just transformed from one complete set of states
to another.  The formulation in the reflectivity basis in
\cref{eq:spin-dens_refl,eq:spin-dens_refl_2} is completely equivalent
to the formulation in the canonical basis in
\cref{eq:intensity_model_bin_rank_rho}.  An important advantage of the
formulation in the reflectivity basis is that, at high $\sqrt{s}$ and
neglecting corrections of order $1 / s$, the reflectivity quantum
number~\refl of~$X$ corresponds to the naturality \naturEx (see
\cref{eq:theory.mesons.naturality}) of the exchange particle in the
scattering
process~\cite{Gottfried:1964nx,Cohen-Tannoudji:1968eoa,Mathieu:2019fts}.
Note that this relation is only valid for spinless beam particles and
unpolarized target and recoil particles.  For beam particles with
spin, \cref{eq:refl_eigenstate,eq:refl_eigenstate_norm} have to be
extended to also take into account the spin state of the beam
particle.  For the case of photoproduction, this has been worked out
in \refCite{Mathieu:2019fts}.

Depending on the scattering process, amplitudes with certain \naturEx
values may be suppressed.  Since $\refl \equiv \naturEx$, the
corresponding partial-wave amplitudes are also suppressed.  This is,
for example, the case for scattering processes of hadrons at high
energies, which are dominated by Pomeron exchange (see
\cref{sec:exp.prod_reactions}).  Since the Pomeron has positive
naturality,\footnote{This is also true for the $f_2$~exchange, which
  may contribute as well.}  partial-wave amplitudes with $\refl = -1$
that correspond to unnatural-parity exchange are suppressed.  As a
consequence, PWA models for these reactions require much less waves
with negative than with positive reflectivity in order to describe the
data.  Therefore, the dimension of
$\overline{\varrho}_{ij}^{\refl = -1}$ is much smaller than that of
$\overline{\varrho}_{ij}^{\refl = +1}$.  This corresponds to a reduced
number of free parameters of the PWA model.  The different number of
waves in the two reflectivity sectors is taken into account by the
replacement
\begin{equation}
  \label{eq:N_waves_refl}
  N_\text{waves} \to N_\text{waves}^\refl
\end{equation}
in \cref{eq:intensity_model_bin_rank_rho_refl}.

Since different reflectivity values correspond to different exchange
particles and hence different production mechanisms, also the
effective rank~$N_r$ of the spin-density matrix may be different for
the two values of~\refl.  We incorporate this into the PWA model in
\cref{eq:intensity_model_bin_rank_rho_refl} by the replacement
\begin{equation}
  \label{eq:N_rank_refl}
  N_r \to N_r^\refl\eqPunctSpacing.
\end{equation}

\subsubsection{Normalization}
\label{sec:pwa_cells.normalization}

An important technical issue is the normalization of the transition
and decay amplitudes.  A consistent normalization allows us to extract
yields of resonances and to compare the transition amplitudes of
different waves in a PWA model as well as across different analyses
and experiments.

In order to derive a normalization scheme, we go back to the
definition of the intensity~$\mathcal{I}$ in \cref{eq:intensity_def}
as the number of produced events per unit in~$m_X$, $t'$, and $n$-body
phase-space volume.  By integrating~$\mathcal{I}$ over the volume of
the $n$-body phase space of the final-state particles, we get the
density of produced events differential in~$m_X$ and~$t'$:
\begin{equation}
  \label{eq:event_dens_m_t}
  \frac{\dif{N}}{\dif{m_X}\, \dif{t'}}
  = \int\! \dif{\Phi_n}(\tau_n; m_X)\, \mathcal{I}(\tau_n; m_X, t')\eqPunctSpacing.
\end{equation}
Integrating \cref{eq:event_dens_m_t} over the $(m_X, t')$ cell, in
which the partial-wave decomposition is performed, yields the number
of events in that cell as predicted by the model:
\begin{equation}
  \label{eq:events_pred_m_t_def}
  N_\text{pred}(m_X, t')
  = \int_{m_{X, 1}}^{m_{X, 2}}\! \dif{\widetilde{m}_X} \int_{t'_1}^{t'_2}\! \dif{\widetilde{t}'}
  \frac{\dif{N}}{\dif{\widetilde{m}_X}\, \dif{\widetilde{t}'}}\eqPunctSpacing.
\end{equation}
Here, $(m_{X, 1}, t'_1)$ and $(m_{X, 2}, t'_2)$ define the borders of
the $(m_X, t')$ cell.  In our binned analysis approach, we neglect the
$m_X$~and~$t'$ dependence of $\dif{N} / (\dif{m_X}\, \dif{t'})$ within
the $(m_X, t')$ cell, so that the integration in
\cref{eq:events_pred_m_t_def} is trivial and equivalent to multiplying
the integrand with the respective bin widths.

We define the normalization of the intensity by demanding\footnote{In
  special cases, where one has to use wide $m_X$~bins, the integration
  of the decay amplitudes in \cref{eq:events_pred} has to be performed
  also over~$m_X$ (see \eg \cref{sec:results_3pic_primakoff}).}
\begin{align}
  \nonumber
  N_\text{pred}(m_X, t')
  &= \int\! \dif{\Phi_n}(\tau_n; m_X)\, \mathcal{I}(\tau_n; m_X, t') \\
  \label{eq:events_pred}
  &\!\begin{multlined}
    = \sum_{\refl = \pm 1} \Bigg\{ \sum_i^{N_\text{waves}^\refl}
    \sum_{r = 1}^{N_r^\refl} \big| \overline{\mathcal{T}}_i^{r \refl}(m_X, t') \big|^2
    \int\! \dif{\Phi_n}(\tau_n; m_X)\, \big| \overline{\Psi}_i^\refl(\tau_n; m_X) \big|^2 \\
    {} + 2 \sum_{i, j; i < j}^{N_\text{waves}^\refl}
    \Re\!\bigg[ \sum_{r = 1}^{N_r^\refl} \overline{\mathcal{T}}_i^{r \refl}(m_X, t')\, \overline{\mathcal{T}}_j^{r \refl \text{*}}(m_X, t')
    \int\! \dif{\Phi_n}(\tau_n; m_X)\, \overline{\Psi}_i^\refl(\tau_n; m_X)\, \overline{\Psi}_j^{\refl \text{*}}(\tau_n; m_X) \bigg] \Bigg\} \\
    {} + |\overline{\mathcal{T}}_{\!\!\text{flat}}(m_X, t')|^2 \int\! \dif{\Phi_n}(\tau_n; m_X)\eqPunctSpacing.
  \end{multlined}
\end{align}
Here we have absorbed the~$m_X$ and $t'$~bin widths into the
normalization of~$\mathcal{I}$.  We also have used
\cref{eq:intensity_model_bin_rank_rho_refl,eq:N_waves_refl,eq:N_rank_refl,eq:event_dens_m_t}.

\Cref{eq:events_pred} fixes the unit of~$\mathcal{I}$ to number of
produced events predicted by the model.  However, it still leaves room
for an arbitrary factor that can be shifted between the transition and
decay amplitudes.  In order to also fix the unit of the transition
amplitudes, we normalize the decay amplitudes to the diagonal elements
of the so-called \emph{integral matrix}\footnote{See also
  \cref{sec:integral_matrices}.}
\begin{equation}
  \label{eq:int_matrix_def}
  I_{ij}^\refl(m_X)
  \coloneqq \int\! \dif{\Phi_n}(\tau_n; m_X)\,
  \overline{\Psi}_i^\refl(\tau_n; m_X)\, \overline{\Psi}_j^{\refl \text{*}}(\tau_n; m_X)\eqPunctSpacing.
\end{equation}
This is a convenient choice because it makes the normalized decay
amplitudes independent of the normalization of the propagator terms
that enter the decay amplitudes (see
\cref{sec:pwa.analysis_model.decay_amp,sec:pwa.analysis_model.dyn_amp}).
The normalized decay amplitudes are defined by\footnote{Since we set
  the decay amplitude $\overline{\Psi}_\text{flat}$ of the flat wave
  to unity, the corresponding normalized decay amplitude is
  \begin{equation}
    \label{eq:decay_amp_flat_norm}
    \Psi_\text{flat}(\tau_n; m_X)
    \coloneqq \frac{1}{\sqrt{I_\text{flat}(m_X)}}
    \quad\text{with}\quad
    I_\text{flat}(m_X) = V_n(m_X)\eqPunctSpacing,
  \end{equation}
  where~$V_n$ is the $n$-body phase-space volume:
  \begin{equation}
    \label{eq:phase-space_vol}
    V_n(m_X)
    \coloneqq \int\! \dif{\Phi_n}(\tau_n; m_X)\eqPunctSpacing.
  \end{equation}}
\begin{equation}
  \label{eq:decay_amp_norm}
  \Psi_i^\refl(\tau_n; m_X)
  \coloneqq \frac{\overline{\Psi}_i^\refl(\tau_n; m_X)}{\sqrt{I_{ii}^\refl(m_X)}}
\end{equation}
so that
\begin{equation}
  \label{eq:int_matrix_norm}
  \int\! \dif{\Phi_n}(\tau_n; m_X)\, \Psi_i^\refl(\tau_n; m_X)\, \Psi_j^{\refl \text{*}}(\tau_n; m_X)
  = \frac{I_{ij}^\refl(m_X)}{\sqrt{I_{ii}^\refl(m_X)\, I_{jj}^\refl(m_X)}}
\end{equation}
and in particular
\begin{equation}
  \label{eq:decay_amp_norm_int}
  \int\! \dif{\Phi_n}(\tau_n; m_X)\, |\Psi_i^\refl(\tau_n; m_X)|^2 = 1\eqPunctSpacing.
\end{equation}
In order to leave $N_\text{pred}$ in \cref{eq:events_pred} unchanged,
the transition amplitudes are normalized according to\footnote{In a
  similar way, the transition amplitude of the flat wave is normalized
  based on \cref{eq:decay_amp_flat_norm,eq:phase-space_vol} according
  to
  \begin{equation}
    \label{eq:trans_amp_flat_norm}
    \mathcal{T}_\text{flat}(m_X, t')
    \coloneqq \overline{\mathcal{T}}_\text{flat}(m_X, t')\, \sqrt{V_n(m_X)}\eqPunctSpacing.
  \end{equation}}
\begin{equation}
  \label{eq:trans_amp_norm}
  \mathcal{T}_i^{r \refl}(m_X, t')
  \coloneqq \overline{\mathcal{T}}_i^{r \refl}(m_X, t')\, \sqrt{I_{ii}^\refl(m_X)}\eqPunctSpacing.
\end{equation}

Expressing \cref{eq:events_pred} in terms of the normalized transition
and decay amplitudes yields
\begin{align}
  N_\text{pred}(m_X, t')
  &= \sum_{\refl = \pm 1} \Bigg\{ \sum_i^{N_\text{waves}^\refl}
    \sum_{r = 1}^{N_r^\refl} \big| \mathcal{T}_i^{r \refl}(m_X, t') \big|^2 \nonumber \\
  &\qquad{} + 2 \sum_{i, j; i < j}^{N_\text{waves}^\refl}
    \Re\!\bigg[ \sum_{r = 1}^{N_r^\refl} \mathcal{T}_i^{r \refl}(m_X, t')\, \mathcal{T}_j^{r \refl \text{*}}(m_X, t')\,
    \frac{I_{ij}^\refl(m_X)}{\sqrt{I_{ii}^\refl(m_X)\, I_{jj}^\refl(m_X)}} \bigg] \Bigg\} \nonumber \\
  &\qquad{} + |\mathcal{T}_\text{flat}(m_X, t')|^2 \nonumber \\
  &= \sum_{\refl = \pm 1} \Bigg\{ \sum_i^{N_\text{waves}^\refl}
    \Underbrace{\varrho_{ii}^\refl(m_X, t')\vphantom{\frac{I_j^\refl}{\sqrt{I_j^\refl}}}}%
    {\displaystyle{\text{intensities}}}
    + \sum_{i, j; i < j}^{N_\text{waves}^\refl}
    \Underbrace{2 \Re\!\bigg[ \varrho_{ij}^\refl(m_X, t')\,
    \frac{I_{ij}^\refl(m_X)}{\sqrt{I_{ii}^\refl(m_X)\, I_{jj}^\refl(m_X)}} \bigg]}%
    {\displaystyle{\text{overlaps}}} \Bigg\} \nonumber \\
  \label{eq:events_pred_norm_amp}
  &\qquad{} + |\mathcal{T}_\text{flat}(m_X, t')|^2\eqPunctSpacing,
\end{align}
where we have used \cref{eq:int_matrix_norm,eq:decay_amp_norm_int} and
we have introduced the normalized spin-density matrix analogous to
\cref{eq:spin-dens_refl_2}
\begin{equation}
  \label{eq:spin-dens_norm}
  \varrho_{ij}^\refl(m_X, t')
  = \sum_{r = 1}^{N_r^\refl} \mathcal{T}_i^{r \refl}(m_X, t')\,
  \mathcal{T}_j^{r \refl \text{*}}(m_X, t')\eqPunctSpacing.
\end{equation}
\Cref{eq:events_pred_norm_amp} provides an interpretation for the
normalized spin-density matrix elements.  The diagonal
elements~$\varrho_{ii}^\refl$ are the \emph{partial-wave intensities},
\ie the expected number of produced events in wave~$i$ with
reflectivity \refl.\footnote{This number does not include interference
  effects between the waves.  For an experiment with acceptance
  different from unity, it corresponds to the acceptance-corrected
  number of events (see \cref{sec:pwa_cells.likelihood_fit}).}  The
off-diagonal elements~$\varrho_{ij}^\refl$, which contain information
about the relative phase between waves~$i$ and~$j$, contribute to the
so-called \emph{overlaps}, which are the number of events originating
from the interference between waves~$i$ and~$j$ with reflectivity
\refl (see also \cref{sec:pwa_cells.observables}).

Note that the normalizations in
\cref{eq:decay_amp_norm,eq:decay_amp_flat_norm,eq:trans_amp_norm,eq:trans_amp_flat_norm}
only define the \emph{relative} normalization of the transition
amplitudes of the various partial waves.  In order to fix the
\emph{absolute} normalization, the additional normalization condition
in \cref{eq:events_pred_norm_amp} is required.  This condition is
applied implicitly at the stage of the maximum likelihood fit that
will be discussed in \cref{sec:pwa_cells.likelihood_fit}.

Using the normalized transition and decay amplitudes in
\cref{eq:intensity_model_bin_rank_rho_refl}, the final formula for the
intensity reads
\begin{equation}
  \label{eq:intensity_model_final}
  \begin{aligned}
    \mathcal{I}(\tau_n; m_X, t')
    &= \sum_{\refl = \pm 1} \sum_{r = 1}^{N_r^\refl} \bigg| \sum_i^{N_\text{waves}^\refl}
    \mathcal{T}_i^{r \refl}(m_X, t')\,
    \Psi_i^\refl(\tau_n; m_X) \bigg|^2 + |\mathcal{T}_\text{flat}(m_X, t')|^2 \\
    &= \sum_{\refl = \pm 1} \sum_{i, j}^{N_\text{waves}^\refl} \Psi_i^\refl(\tau_n; m_X)\,
    \varrho_{ij}^\refl(m_X, t')\,
    \Psi_j^{\refl \text{*}}(\tau_n; m_X) + |\mathcal{T}_\text{flat}(m_X, t')|^2\eqPunctSpacing.
  \end{aligned}
\end{equation}

\subsubsection{Unbinned Extended Maximum Likelihood Fit}
\label{sec:pwa_cells.likelihood_fit}

The maximum likelihood method is used to estimate unknown parameter
values of a statistical model by maximizing the likelihood
function~$\mathcal{L}$, which is the joint probability density of the
data set given the parameter values~\cite{Fisher:1922saa}.  For a
given data set $\vec{x} \coloneqq (x_1, \ldots, x_N)^T$ of
$N$~independent random variables\footnote{Depending on the
  measurement, these variables may in turn be vectors in a
  multi-dimensional data space.  In the case of partial-wave analysis,
  each~$x_k$ corresponds to the measured phase-space
  variables~$\tau_{n, k}$ of event~$k$.} each following the same
probability density function $f(x; \vec{\theta})$ with $m$~parameters
$\vec{\theta} \coloneqq (\theta_1, \ldots, \theta_m)^T$ with unknown
values, the likelihood function is
\begin{equation}
  \label{eq:likelihood_def}
  \mathcal{L}(\vec{\theta}; \vec{x})
  = \prod_{k = 1}^N f(x_k; \vec{\theta})\eqPunctSpacing.
\end{equation}
The maximum likelihood estimate~$\hat{\vec{\theta}}$ for the
parameters is given by those parameter values that maximize the
likelihood function, \ie
\begin{equation}
  \label{eq:mle}
  \hat{\vec{\theta}} = \argmax_{\vec{\theta}} \mathcal{L}(\vec{\theta}; \vec{x})\eqPunctSpacing.
\end{equation}
In the above equation, $\mathcal{L}$~is a function of~$\vec{\theta}$
for a given, \ie fixed,~$\vec{x}$.  It is important to note that
$\mathcal{L}$~is the probability density function of~$\vec{x}$ for
given~$\vec{\theta}$.  However, $\mathcal{L}$~is in general \emph{not}
the probability density function of~$\vec{\theta}$ for
given~$\vec{x}$.  In other words, maximizing the probability to
observe the data \wrt~$\vec{\theta}$, as expressed in \cref{eq:mle},
does not necessarily yield the most probable parameter values.  Also,
$\mathcal{L}$~does not need to be normalized \wrt~$\vec{\theta}$.  It
can be shown that the maximum likelihood estimate $\hat{\vec{\theta}}$
in \cref{eq:mle} is consistent, unbiased, and efficient under rather
general conditions in the asymptotic limit, where the data-set size
$N \to \infty$~\cite{James:2006,Kendall:2010}.  Another advantage of
the maximum likelihood method is that \cref{eq:mle} does not require
any binning of the data.  Therefore, the method is applicable also to
high-dimensional data, where binned approaches quickly becomes
prohibitively expensive in terms of computational
resources.\footnote{This is already the case for the PWA of three-body
  final states, for which the phase-space in an $(m_X, t')$ cell is
  five-dimensional.}

If the number~$N$ of data points is not predetermined but is a result
of the measurement and therefore a random variable, the maximum
likelihood principle can be extended.  In the case of counting
experiments, events are produced randomly in time with constant
average rates and hence the number of produced events follows the
Poisson distribution with the expected number of events~$\lambda$.
For this case, the extended likelihood function is
\begin{equation}
 \label{eq:likelihood_ext_def}
 \mathcal{L}_\text{ext}(\vec{\theta}, \lambda; \vec{x}, N)
 = \Underbrace{\frac{\lambda^N\, e^{-\lambda}}{N!} \vphantom{\prod_{k = 1}^N}}{\substack{\displaystyle{\text{Poisson}} \\ \displaystyle{\text{distribution}}}}
 \Underbrace{\prod_{k = 1}^N f(x_k; \vec{\theta})}%
 {\quad\displaystyle{= \mathcal{L}(\vec{\theta}; \vec{x})}}\eqPunctSpacing.
\end{equation}
This approach was first proposed by Fermi and is discussed in more
detail in \refsCite{Orear:1958zz,Orear:1982zz,Barlow:1990vc}.  In
\refCite{Barlow:1990vc}, it is shown that the extended maximum
likelihood estimator inherits the desired asymptotic statistical
properties of consistency, unbiasedness, and efficiency from the
maximum likelihood estimator.

We apply the extended maximum likelihood method to partial-wave
analysis in order to estimate the values of the set
$\{\mathcal{T}_i^{r \refl}\}$ of transition amplitudes in our PWA
model in \cref{eq:intensity_model_final}.  This approach has been
pioneered by Ascoli~\etal~\cite{Ascoli:1970xi,Ascoli:1973nj}.  The
partial-wave decomposition is performed independently in $(m_X, t')$
cells and within each cell, we neglect the dependence of the intensity
on~$m_X$ and~$t'$.  In order to simplify notation, we from here on
consider a specific $(m_X, t')$ cell and leave off the $m_X$~and
$t'$~dependence in the formulas below.

Within a given $(m_X, t')$ cell, our model in
\cref{eq:intensity_model_final} describes the $(3n - 4)$-dimensional
$\tau_n$~distribution of the produced events, \ie the events a
hypothetical perfect detector with unit acceptance would measure.  In
practice, the \emph{acceptance}~\acc of the detector
setup\footnote{The term acceptance is used here in a broad sense and
  includes all effects that affect the detection efficiency such as
  the geometry of the detector setup as well as the efficiencies of
  the detectors in the setup, of the reconstruction, and of the event
  selection.} is smaller than unity and \acc~depends on the kinematic
variables~$m_X$, $t'$, and~$\tau_n$.\footnote{Depending on the
  detector setup, the acceptance might depend on additional kinematic
  variables.  Hence \acc~represents the acceptance integrated over all
  these variables.}  To obtain a model for the actual intensity
distribution measured by the detector, we have to weight
\cref{eq:intensity_model_final} by the detector acceptance.  By
normalizing this model and using \cref{eq:dLIPS_dens}, we get the
probability density function of the measured events in the
$\tau_n$~space, \ie
\begin{equation}
  \label{eq:data_pdf}
  f(\tau_n; \{\mathcal{T}_i^{r \refl}\})
  = \frac{\rho_n(\tau_n)\, \acc(\tau_n)\, \mathcal{I}(\tau_n; \{\mathcal{T}_i^{r \refl}\})}{\int\! \dif{\tau_n'}\, \rho_n(\tau_n')\, \acc(\tau_n')\,
    \mathcal{I}(\tau_n'; \{\mathcal{T}_i^{r \refl}\})}\eqPunctSpacing,
  ~\text{where}\quad
  \int\! \dif{\tau_n}\, f(\tau_n; \{\mathcal{T}_i^{r \refl}\}) = 1\eqPunctSpacing.
\end{equation}
The above equation assumes that detector resolution effects that lead
to a smearing of the variables~$\tau_n$ are negligible.  If this is
not the case, the likelihood function has to be constructed from the
observed, \ie, smeared probability density function
\begin{equation}
  \label{eq:data_pdf_smeared}
  f'(\tau_n') = \int\! \dif{\tau_n}\, S(\tau_n', \tau_n)\, f(\tau_n)\eqPunctSpacing,
\end{equation}
where $S(\tau_n', \tau_n)$ is the smearing function of the detector
system, for which an analytical form is usually not known.  Correcting
for smearing requires more elaborate and computationally more
expensive methods (see \eg\ \refCite{Schmidt:1992si}).

Using \cref{eq:data_pdf}, we can construct the extended likelihood
function for our PWA model analogous to \cref{eq:likelihood_ext_def}:
\begin{equation}
  \label{eq:likelihood_ext}
  \mathcal{L}_\text{ext}(\{\mathcal{T}_i^{r \refl}\}; \{\tau_{n, k}\}, N)
  = \frac{(N_\text{pred}^\text{meas})^N\, e^{-N_\text{pred}^\text{meas}}}{N!} \prod_{k = 1}^N
  \frac{\rho_n(\tau_{n, k})\, \acc(\tau_{n, k})\, \mathcal{I}(\tau_{n, k}; \{\mathcal{T}_i^{r \refl}\})}%
  {\int\! \dif{\tau_n}\, \rho_n(\tau_n)\, \acc(\tau_n)\,
    \mathcal{I}(\tau_n; \{\mathcal{T}_i^{r \refl}\})}\eqPunctSpacing.
\end{equation}
In order to find the maximum likelihood estimate for the set
$\{\mathcal{T}_i^{r \refl}\}$ of the transition amplitudes, we have to
maximize \cref{eq:likelihood_ext} \wrt these parameters.

Note that the conventional likelihood function,
\begin{equation}
  \label{eq:likelihood}
  \mathcal{L}(\{\mathcal{T}_i^{r \refl}\}; \{\tau_{n, k}\})
  = \prod_{k = 1}^N
  \frac{\rho_n(\tau_{n, k})\, \acc(\tau_{n, k})\, \mathcal{I}(\tau_{n, k}; \{\mathcal{T}_i^{r \refl}\})}%
  {\int\! \dif{\tau_n}\, \rho_n(\tau_n)\, \acc(\tau_n)\,
    \mathcal{I}(\tau_n; \{\mathcal{T}_i^{r \refl}\})}\eqPunctSpacing,
\end{equation}
without the Poisson term is invariant under scaling of all transition
amplitudes by a common factor $C \in \mathbb{R}$, \ie under the
substitution $\mathcal{I} \to C^2\, \mathcal{I}$.  The maximization of
the additional Poisson term in \cref{eq:likelihood_ext} enforces
\begin{equation}
  \label{eq:abs_norm_cond}
  N_\text{pred}^\text{meas}(\{\mathcal{T}_i^{r \refl}\}) = N\eqPunctSpacing.
\end{equation}
In order to normalize the intensity distribution~$\mathcal{I}$ such
that it is given in number of produced events, we set
\begin{equation}
  \label{eq:events_exp}
  N_\text{pred}^\text{meas}(\{\mathcal{T}_i^{r \refl}\})
  = \int\! \dif{\tau_n}\, \rho_n(\tau_n)\, \acc(\tau_n)\,
  \mathcal{I}(\tau_n; \{\mathcal{T}_i^{r \refl}\})\eqPunctSpacing.
\end{equation}
Via \cref{eq:abs_norm_cond}, this fixes the arbitrary scaling
factor~$C$, \ie the absolute normalization of the transition
amplitudes (see \cref{sec:pwa_cells.normalization}), and ensures also
that \cref{eq:events_pred,eq:events_pred_norm_amp} are
fulfilled.\footnote{An alternative approach, which is often used in
  Dalitz-plot analyses, is to maximize the conventional likelihood in
  \cref{eq:likelihood} while fixing one of the transition amplitudes
  to $\abs{\mathcal{T}_i^{r \refl}} = 1$ in order to fix the
  normalization.  However, in this approach one looses the information
  about absolute yields of resonances.}  Using \cref{eq:events_exp}
the extended likelihood function in \cref{eq:likelihood_ext}
simplifies to
\begin{equation}
  \label{eq:likelihood_ext_2}
  \mathcal{L}_\text{ext}(\{\mathcal{T}_i^{r \refl}\}; \{\tau_{n, k}\}, N)
  = \frac{e^{-N_\text{pred}^\text{meas}}}{N!}
  \prod_{k = 1}^N \rho_n(\tau_{n, k})\, \acc(\tau_{n, k})\,
  \mathcal{I}(\tau_{n, k}; \{\mathcal{T}_i^{r \refl}\})\eqPunctSpacing.
\end{equation}

Since the actual value of the likelihood function at the maximum is
irrelevant for the parameter estimation, we can simplify
\cref{eq:likelihood_ext_2} further by dropping all constant factors
that are independent of the transition amplitudes.\footnote{It is
  important to note that in particular the factor
  \begin{equation*}
    \prod_{k = 1}^N \rho_n(\tau_{n, k})\, \acc(\tau_{n, k})
  \end{equation*}
  in \cref{eq:likelihood_ext_2} that contains the phase-space and
  acceptance weights for the measured events is independent of the
  transition amplitudes and is hence a constant in the maximization
  procedure.}  It is convenient to consider the
logarithm of $\mathcal{L}_\text{ext}$.  Since the logarithm is a
strictly monotonous function, it leaves the position of the maximum of
$\mathcal{L}_\text{ext}$ in the space of the transition amplitudes
unchanged.  However, the logarithm converts the product over the
measured events into a sum and thus makes the expression numerically
easier to treat.  Using \cref{eq:events_exp,eq:intensity_model_final},
we arrive at the final expression of the likelihood function that is
maximized in order to estimate the transition amplitudes:
\begin{multline}
    \label{eq:likelihood_ext_final}
  \ln \mathcal{L}_\text{ext}(\{\mathcal{T}_i^{r \refl}\}; \{\tau_{n, k}\}, N) \\
  \begin{aligned}
    &= \sum_{k = 1}^N \ln \mathcal{I}(\tau_{n, k}; \{\mathcal{T}_i^{r \refl}\}) - N_\text{pred}^\text{meas}(\{\mathcal{T}_i^{r \refl}\}) \\
    &= \sum_{k = 1}^N
    \ln \Bigg[ \sum_{\refl = \pm 1} \sum_{r = 1}^{N_r^\refl} \bigg| \sum_i^{N_\text{waves}^\refl}
    \mathcal{T}_i^{r \refl}\, \Psi_i^\refl(\tau_{n, k}) \bigg|^2 + |\mathcal{T}_\text{flat}|^2 \Bigg] \\
    &\qquad{} - \sum_{\refl = \pm 1} \sum_{r = 1}^{N_r^\refl} \sum_{i, j}^{N_\text{waves}^\refl}
    \mathcal{T}_i^{r \refl}\, \mathcal{T}_j^{r \refl \text{*}}
    \Underbrace{\int\! \dif{\tau_n}\, \rho_n(\tau_n)\, \acc(\tau_n)\, \Psi_i^\refl(\tau_n)\, \Psi_i^{\refl \text{*}}(\tau_n)}%
    {\displaystyle{\eqqcolon \prescript{\text{acc}\!}{}{I}_{ij}^\refl}}
    - |\mathcal{T}_\text{flat}|^2 \Underbrace{\int\! \dif{\tau_n}\, \rho_n(\tau_n)\, \acc(\tau_n)}%
    {\displaystyle{\eqqcolon \prescript{\text{acc}\!}{}{I}_\text{flat}}}\eqPunctSpacing.
  \end{aligned}
\end{multline}
The maximum likelihood estimate for the transition amplitudes is given
by
\begin{equation}
  \label{eq:mle_trans_amp}
  \{\hat{\mathcal{T}}_i^{r \refl}\}
  = \argmax_{\{\mathcal{T}_i^{r \refl}\}} \Big[ \ln \mathcal{L}_\text{ext}(\{\mathcal{T}_i^{r \refl}\};
  \{\tau_{n, k}\}, N) \Big]\eqPunctSpacing.
\end{equation}
The numerical procedure used to determine the
$\{\hat{\mathcal{T}}_i^{r \refl}\}$ and their uncertainties is
discussed in \cref{sec:max_likelihood_procedure}.

Since the transition amplitudes are independent of~$\tau_n$, they can
be pulled out of the normalization integral in \cref{eq:events_exp}.
In \cref{eq:likelihood_ext_final}, this leads to an integral matrix
$\prescript{\text{acc}\!}{}{I}_{ij}^\refl$ that is similar to the
integral matrix $I_{ij}^\refl$ in \cref{eq:int_matrix_def}.  The only
differences between the two matrices are that in
$\prescript{\text{acc}\!}{}{I}_{ij}^\refl$ the normalized decay
amplitudes appear in the integrand (see \cref{eq:decay_amp_norm}) and
that the integration is performed over the accepted phase space.  As
was discussed in \cref{sec:pwa.analysis_model}, the decay
amplitudes do not contain any free parameters.  We can therefore
calculate the decay amplitudes as well as the integral matrices
$I_{ij}^\refl$ and $\prescript{\text{acc}\!}{}{I}_{ij}^\refl$ using
Monte Carlo integration techniques before maximizing the likelihood
function.  This is described in \cref{sec:integral_matrices}.  Since
the integral matrices are computationally very expensive, this reduces
the time to compute the likelihood function by several orders of
magnitude.  It is only due to this fact that the maximization
procedure becomes actually feasible in terms of computational
resources.  However, allowing no free parameters in the decay
amplitudes is a severe limitation.  This means in particular that the
dynamical amplitudes of all isobar resonances in the PWA model have to
be known.  In many analyses, this becomes an important source of
systematic uncertainties.  This issue will be discussed
further in \cref{sec:pwa_cells.discussion}.  In
\cref{sec:pwa_cells:freed_isobar}, we will present a novel
method that circumvents most of these issues.

\subsubsection{Observables}
\label{sec:pwa_cells.observables}

The spin-density matrices~$\varrho_{ij}^\refl$ (see
\cref{eq:spin-dens_norm}), which are extracted from the data by
maximizing the likelihood function in \cref{eq:likelihood_ext_final}
independently in the $(m_X, t')$ cells, contain all information
obtainable about the intermediate states~$X$.  Based on
$\varrho_{ij}^\refl(m_X, t')$, we can define a number of observables
that are useful to characterize the result of a partial-wave
decomposition and to search for resonance signals.

In \cref{eq:events_pred_norm_amp}, we defined the number of produced
events $N_\text{pred}(m_X, t')$ that the model predicts for the given
$(m_X, t')$ cell.  This number can be expressed in terms of the
\emph{partial-wave intensities},
\begin{equation}
  \label{eq:intens_def}
  \intens_i^\refl(m_X, t')
  \coloneqq \varrho_{ii}^\refl(m_X, t')
  = \sum_{r = 1}^{N_r^\refl} \big| \mathcal{T}_i^{r \refl}(m_X, t') \big|^2\eqPunctSpacing,
\end{equation}
and the \emph{overlaps},
\begin{equation}
  \label{eq:overlap_def}
  \ovl_{ij}^\refl(m_X, t')
  \coloneqq 2 \Re\!\bigg[ \varrho_{ij}^\refl(m_X, t')\,
  \frac{I_{ij}^\refl(m_X)}{\sqrt{I_{ii}^\refl(m_X)\, I_{jj}^\refl(m_X)}} \bigg]\eqPunctSpacing,
\end{equation}
of all pairs of waves, \ie
\begin{equation}
  \label{eq:events_pred_int_ovl}
  N_\text{pred}(m_X, t')
  = \sum_{\refl = \pm 1} \Bigg\{ \sum_i^{N_\text{waves}^\refl} \intens_i^\refl(m_X, t')
  + \sum_{i, j; i < j}^{N_\text{waves}^\refl} \ovl_{ij}^\refl(m_X, t') \Bigg\}
  + \intens_\text{flat}(m_X, t')\eqPunctSpacing.
\end{equation}
The partial-wave intensities in \cref{eq:intens_def} correspond to the
diagonal elements of the spin-density matrix.  Due to the chosen
normalization (see
\cref{sec:pwa_cells.normalization,sec:pwa_cells.likelihood_fit}), the
intensities are given in terms of number of produced events in
wave~$i$ with reflectivity~\refl.  If a resonance is present in a
partial wave, the $m_X$~dependence of the intensity of this waves
often exhibits a Breit--Wigner-shaped peak similar to the one shown in
\cref{fig:bw.modulus}.  The overlaps as defined in
\cref{eq:overlap_def} are real-valued and represent the interference
terms between waves~$i$ and~$j$ with reflectivity~\refl.  They are
also expressed in terms of number of produced events.\footnote{Note
  that although the overlap is given in units of number of events, it
  is a signed number.  For constructive interference, the overlap is
  positive; for destructive interference it is negative.}

It is often useful to limit the two sums in
\cref{eq:events_pred_int_ovl} to a selected subset~$\mathbb{S}$ of
partial waves.  In this case, \cref{eq:events_pred_int_ovl} yields the
number of produced events $N_\text{pred}^\mathbb{S}$ in these waves in
the given $(m_X, t')$ cell.  This number takes into account all mutual
interference terms of the waves in~$\mathbb{S}$.  For example, it is
often interesting to calculate the number of produced events in
subsets of waves that have the same \JPCMrefl quantum numbers but
different decay chains.  Studying these so-called \emph{spin totals}
as a function of~$m_X$ and/or~$t'$ often better reveals possible
resonance signals because statistical fluctuations are reduced.

The contribution of an individual partial wave to the data sample is
quantified by the \emph{relative intensity}, which is defined as the
ratio of the partial-wave intensity $\intens_i^\refl(m_X, t')$ in
\cref{eq:intens_def} and the total number of produced events
$N_\text{pred}(m_X, t')$ in
\cref{eq:events_pred_int_ovl}.\footnote{The relative intensities are
  equivalent to the so-called \emph{fit fractions} often quoted for
  Dalitz-plot analyses.}  Often, the relative intensity is calculated
by summing $\intens_i^\refl(m_X, t')$ and $N_\text{pred}(m_X, t')$
over the same~$m_X$ and/or $t'$~range:
\begin{equation}
  \label{eq:rel_intens_def}
  \relintens_i^\refl
  \coloneqq \dsum^{m_X\,\text{bins}}~\dsum^{t'\,\text{bins}} \intens_i^\refl(m_X, t')
  \Big/
  \Big[ \dsum^{m_X\,\text{bins}}~\dsum^{t'\,\text{bins}} N_\text{pred}(m_X, t') \Big]\eqPunctSpacing.
\end{equation}
Although the relative intensity of a given wave includes---where
applicable---the effect of self-interference due to Bose
symmetrization (see \cref{sec:pwa.analysis_model.symmetrization}), it
does \emph{not} include the overlaps, \ie the interference effects of
this wave with any of the other waves in the PWA model.  Consequently,
the relative intensities of all waves in the PWA model will in general
not sum to unity.  The difference of this sum from unity is a measure
for the overall strength of the interference in the model.

Another important observable is the relative phase
$\phase_{ij}^\refl(m_X, t')$ between two waves $i$~and~$j$ with
reflectivity~\refl.  It is given by the off-diagonal element of the
spin-density matrix
\begin{equation}
  \label{eq:phase_def}
  \phase_{ij}^\refl(m_X, t')
  \coloneqq \arg\!\big[ \varrho_{ij}^\refl(m_X, t') \big]\eqPunctSpacing,
  ~\text{\ie}\quad
  \varrho_{ij}^\refl(m_X, t')
  = \big| \varrho_{ij}^\refl(m_X, t') \big|\, e^{i\, \phase_{ij}^\refl(m_X, t')}\eqPunctSpacing.
\end{equation}
Note that for $N_r^\refl = 1$, \ie full coherence of all partial
waves,
\begin{equation}
  \label{eq:phase_rank_1}
  \phase_{ij}^\refl(m_X, t')
  = \arg\!\big[ \mathcal{T}_i^\refl(m_X, t') \big]
  - \arg\!\big[ \mathcal{T}_j^\refl(m_X, t') \big]\eqPunctSpacing.
\end{equation}
If a resonance is present in a partial wave, the phase relative to a
second wave that has no resonance typically grows with rising $m_X$ by
about $180^\circ$ across the resonance, \ie the partial wave exhibits
a \emph{phase motion}.\footnote{If the second wave also contains a
  resonance, the phase motion might be reduced or even completely
  compensated if the two resonances have similar masses and widths.}
\Cref{fig:bw.phase} shows, as an example, the phase motion of a
Breit--Wigner resonance.  The amplitude can also be visualized in the
complex plane as shown in \cref{fig:bw.im-re}.  In these so-called
Argand diagrams, a single Breit--Wigner resonance appears as
counter-clockwise circular structure.

It is important to note that although the overlap $\ovl_{ij}^\refl$
between two waves may be zero because of the orthogonality of the
decay amplitudes (see \cref{sec:wigner_D_function}), the corresponding
off-diagonal element $\varrho_{ij}^\refl$ of the spin-density matrix
in general does not vanish and hence the two waves still have a
well-defined phase, which characterizes their interference.  Since the
spin-density matrix has a block-diagonal structure \wrt the
reflectivity, partial-wave amplitudes with different~\refl do not
interfere and therefore relative phases between such waves are
undefined.

In \cref{sec:additional_observables}, we will introduce additional
observables that are useful in certain cases.

\subsubsection{Comparison of Partial-Wave Analysis Model and Data}
\label{sec:pwa.verification_of_model}

The goodness of the PWA fit that was described in
\cref{sec:pwa_cells.likelihood_fit} can be assessed by generating
Monte Carlo pseudo data that are distributed according to the
acceptance-weighted PWA model using the maximum-likelihood estimates
for the transition amplitudes from \cref{eq:mle_trans_amp}.  Since the
intensity distribution $\mathcal{I}(m_X, t', \tau_n)$ as defined in
\cref{eq:intensity_def} describes the distribution of the events
relative to a distribution that is uniform in phase-space, we can
generate such a pseudo-data sample in a three-step procedure: \one~We
generate Monte Carlo events that are uniformly distributed in the
$n$-body phase space using, for example, \cref{eq:n-body_dLIPS} (see
\refsCite{James:1968gu,Block:1991} for more details).  \two~These
events are processed through the detector simulation, reconstruction,
and event selection chain.\footnote{To save computation time, one can
  reuse the Monte Carlo data sample that was generated to calculate
  the integral matrix $\prescript{\text{acc}\!}{}{I}_{ij}^\refl$ in
  \cref{eq:int_matrix_acc_mc}.}  \three~The reconstructed and selected
phase-space events are weighted by the intensity model in
\cref{eq:intensity_model_final} using the maximum likelihood estimate
for the transition amplitudes.  The weights from the model intensity
are applied using accept--reject sampling (see \eg\
\refCite{Casella:2004}) in order to obtain a pseudo-data sample that
consists of events with unit weight.

For a good PWA fit, distributions obtained from these weighted Monte
Carlo events are expected to reproduce the real-data distributions.
It is in particular mandatory to verify the agreement of the
$\tau_n$~distributions for all $(m_X, t')$ cells.  For an $n$-body
final state, these distributions are $(3n - 4)$-dimensional so that
for $n \geq 3$ usually only one- or two-dimensional projections can be
compared in a meaningful way (see \eg Section~IV.E in
\refCite{Adolph:2015tqa}).  It is important to note that the
phase-space events are weighted with the product of the detector
acceptance $\acc(m_X, t', \tau_n)$ and the PWA model
$\mathcal{I}(m_X, t', \tau_n)$.  Therefore, observed disagreements
between weighted Monte Carlo and real-data distributions can in
general be caused by an unrealistic model or by an inaccurate
description of the apparatus in the Monte Carlo simulation.  One
cannot distinguish the two effects based on the weighted Monte Carlo
data alone.  The detector model used in the Monte Carlo simulation
hence needs to be verified using other physics processes.  Assuming
that the detector description is sufficiently accurate, disagreements
\eg in the invariant mass distributions of the isobars can be a sign
that isobars might be missing and/or that some of the employed
parameterizations for the isobar amplitudes (see
\cref{sec:pwa.analysis_model.dyn_amp}) do not agree with the data.
Similarly, disagreements in the angular distributions can hint at
missing partial waves with certain spin, spin-projection, or orbital
angular momentum quantum numbers.

To study properties of the PWA model, one often wants to generate
Monte Carlo pseudo-data samples that are distributed according to a
PWA model $\mathcal{I}(m_X, t', \tau_n)$ with known parameters.  This
can be achieved by slightly modifying the Monte Carlo procedure
described above.  One starts again with Monte Carlo events that are
uniformly distributed in the $n$-body phase-space but one skips
step~\two, \ie one does not weight the events with the acceptance.
Based on these pseudo data one can perform PWA fits with the same or
altered PWA models\footnote{In these PWA fits, the acceptance is of
  course unity.}  in order to study, for example, the
distinguishability of certain partial-wave amplitudes.  If the
acceptance is very non-uniform in any of the kinematic variables that
enter the PWA model or if detector resolution effects in any of these
variables are large, the distinguishability of certain partial-wave
amplitudes might be degraded.\footnote{For example, at very low
  $t' \lesssim 10^{-3}\,(\GeV/c)^2$ the resolution of the azimuthal
  angle~\phiGJ in the Gottfried--Jackson frame (see
  \cref{sec:pwa.analysis_model.coordsys}) becomes so low in the
  COMPASS data, that waves with opposite reflectivity quantum number
  cannot be distinguished anymore (see
  \cref{sec:results_3pic_primakoff}).}  Such degradations can be
studied using a similar procedure as above, where the phase-space
events are weighted with the acceptance and the acceptance correction
is applied in the likelihood function as for real data.

\subsubsection{Discussion of the Partial-Wave Analysis Method}
\label{sec:pwa_cells.discussion}

The PWA model presented in \cref{sec:pwa.analysis_model,sec:pwa_cells}
makes a number of assumptions that have practical consequences.  Here,
we focus on those aspects that are most relevant for the
interpretation of the results that will be presented in
\cref{sec:results}.  Further details are discussed \eg in
\refCite{Hansen:1973gb}.

An important practical issue is the truncation of the partial-wave
expansion in \cref{eq:intensity_model_final}, \ie the decision which
waves to include in the PWA model and which ones to leave out.
Although the quantum numbers of the most dominantly produced
intermediate states~$X$ may be known from previous experiments, in
situations where the analyzed data set is about an order of magnitude
larger than any of the existing ones---like it is, for example, the
case for the COMPASS \threePi proton-target data (see
\cref{sec:3pi_model:pwa})---this knowledge is usually insufficient to
construct a realistic PWA model.

\begin{figure}[tbp]
  \centering
  \hfill%
  \subfloat[][]{%
    \includegraphics[width=\threePlotWidth]{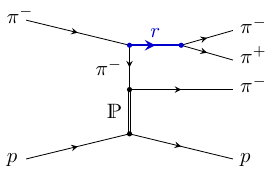}%
    \label{fig:deck}%
  }%
  \hfill%
  \subfloat[][]{%
    \includegraphics[width=\threePlotWidth]{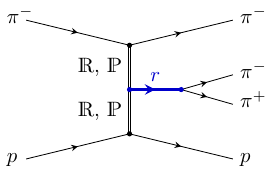}%
    \label{fig:central_prod}%
  }%
  \hfill\null%
  \caption{Examples for irreducible non-resonant contributions to
    diffractive production of $3\pi$ on a proton target.
    \subfloatLabel{fig:deck}~Process proposed by
    R.~T.~Deck~\cite{Deck:1964hm}, where the beam pion dissociates
    into a \twoPi isobar~$r$ and a bachelor $\pi^-$, followed by
    diffractive scattering of the $\pi^-$ off the target proton.
    \subfloatLabel{fig:central_prod}~Process, where both the beam and
    target particles scatter elastically via exchange of
    Reggeons, which fuse to produce a \twoPi system.}
  \label{fig:double_regge}
\end{figure}

In contrast to formation experiments, where the maximum spin of the
intermediate state is limited by the break-up momentum and therefore
by the center-of-momentum energy, in formation experiments, the
high-energy $t$-channel exchange processes may in principle generate
arbitrarily high spins.  Even though the production of intermediate
states is expected to be suppressed with increasing spin, there is no
clear cut-off.  Also irreducible and interfering contributions from
non-resonant processes are often contained in the data.  For
$t$-channel exchange processes, the largest non-resonant contributions
come from double-Regge exchange processes (see
\cref{sec:exp.prod_reactions}).  \Cref{fig:double_regge} shows two
examples for such processes in \threePi production.  Since in these
processes, no $3\pi$ intermediate states with well-defined quantum
numbers appear, the non-resonant contributions typically project into
all partial waves.  Some high-spin partial waves are even dominated by
non-resonant contributions.  Such waves are important in order to
describe the data but are uninteresting when it comes to extracting
resonances.  So, in order to estimate the transition amplitudes from
the data as was described in \cref{sec:pwa_cells.likelihood_fit}, one
also needs to determine the set~$\{i\}$ of partial waves that enters
in \cref{eq:intensity_model_final} from the data.  The analysis
problem hence turns into a more difficult model-selection problem,
which is complicated by the high dimensionality of the data and the
correlations of the partial-wave amplitudes due to their mutual
interference.

For $n$-body final states with $n \geq 3$, another important
ingredient of the PWA model that is part of the model-selection
problem is the set of isobar resonances.  Although signals in the
invariant mass distributions of the respective subsystems of the
final-state particles help to identify the contributing isobars (see
\eg\ \cref{fig:kin_distr} in \cref{sec:3pi_model:pwa}), this
information is in some cases not enough to unambiguously identify all
isobars.\footnote{For example, some isobar resonances have different
  quantum numbers but similar resonance parameters and therefore may
  appear as indistinguishable peaks in the invariant mass
  distribution.  In addition, broad isobars and/or isobars that
  contribute only weakly to the data are difficult to find in the
  invariant mass distributions.}  As was explained in
\cref{sec:pwa_cells.likelihood_fit}, the decay amplitudes must not
have any free parameters because otherwise the computation of the
likelihood function becomes very expensive.  Thus we do not only have
to decide which isobar resonances to include in the PWA model but also
which parameterizations and resonance parameters describe them best.
The model building hence requires input from other measurements.  In
most cases, the dynamical amplitudes of the isobar resonances are
approximated by Breit--Wigner amplitudes as in
\cref{eq:BW_const_width,eq:BW_mass-dep_width} and the PDG world
averages are used for the resonance parameters.  However, for some
isobar resonances, the parameters are not precisely known and/or it is
unclear which parameterization to use for their dynamical amplitude.
For example, the isoscalar resonances with $\JPC = 0^{++}$ quantum
numbers, \ie the \PfZero* states, that appear in $\pi \pi$ and
$K \bar{K}$ subsystems are notoriously difficult to describe (see \eg\
\refCite{pdg_scalars:2018}).  For high-precision data, the issue is
exacerbated by possible effects from final-state interactions, which
may distort the effective dynamical amplitudes of the isobar
resonances (see \eg\ \refsCite{Niecknig:2012sj,Danilkin:2014cra}).
The restriction of the conventional PWA approach to parameter-free
decay amplitudes makes it difficult to implement and test models that
take into account such effects.  A crude way of checking, whether the
used parameters of the isobar resonances deviate from the data, are
so-called likelihood scans.  Here, one performs several PWA fits with
different values for selected isobar parameters and compares the
maximum values of the likelihood function for each parameter set.
Using the asymptotic normality of the maximum of the likelihood
function in the resonance parameters, one can estimate isobar
parameters that better fit the data (see \eg\ \refCite{Teige:1996fi}).
However, with this method one can optimize only a few free parameters
simultaneously.  A more model-independent approach that is able to
extract the amplitudes of subsystems with well-defined quantum numbers
from the data will be presented in \cref{sec:pwa_cells:freed_isobar}.

Another issue that complicates the wave-set selection is the
non-orthogonality of some partial-wave amplitudes.  In principle, the
orthogonality of the Wigner $D$-functions that appear in the decay
amplitudes (see \cref{sec:pwa.analysis_model.decay_amp}) also makes
the decay amplitudes for waves $i$~and~$j$ orthogonal, unless the
quantum numbers of the two waves that determine the $D$-functions are
the same (see \cref{sec:wigner_D_function}).  For orthogonal decay
amplitudes, the corresponding off-diagonal elements of the phase-space
integral matrix $I_{ij}^\refl$ in \cref{eq:int_matrix_def}
vanish. Depending on the analyzed final state, the orthogonality of
the decay amplitudes may be broken by the Bose symmetrization in
\cref{eq:decay_amp_bose}.  A measure for the (non-)orthogonality are
the normalized off-diagonal elements
\begin{equation}
  \label{eq:int_matrix_norm_def}
  \prescript{\text{norm}\!}{}{I}_{ij}^\refl(m_X)
  \coloneqq \frac{I_{ij}^\refl(m_X)}{\sqrt{I_{ii\vphantom{j}}^\refl(m_X)\, I_{jj}^\refl(m_X)}}
\end{equation}
of the integral matrix in \cref{eq:int_matrix_def}.  This matrix is
actually a Gram matrix for the scalar product of two partial-wave
amplitudes that is represented by the phase space integral in
\cref{eq:int_matrix_def}.  A modulus of~1 of an off-diagonal element
of $\prescript{\text{norm}\!}{}{I}_{ij}^\refl$ would mean that the
corresponding partial-wave amplitudes are mathematically
indistinguishable, \ie linearly dependent.  A PWA model containing
such a pair of waves would be ill-defined and would lead to unphysical
fit results for these waves.  In practice, due to the finiteness of
the measured data set, such distinguishability issues may already
arise for waves, for which
$|\prescript{\text{norm}\!}{}{I}_{ij}^\refl|$ is close to unity.  This
is usually the case for wave sets that include in addition to the
ground-state isobar resonances also the corresponding radially excited
states.  At low~$m_X$, \ie well below thresholds that correspond to
the nominal masses of an isobar and its radial excitation, the
dynamical amplitudes of the two isobar resonances are dominated by
phase space and are hence very similar.  In this case, the elements of
$\prescript{\text{norm}\!}{}{I}_{ij}^\refl$ that correspond to waves
with the same quantum numbers have large moduli.  This is a general
problem that always appears when an isobar and its radial excitation
are included in the PWA model as separate partial-wave amplitudes.  It
is a consequence of the binning in~$m_X$.  In order to resolve the
arising ambiguities, one usually makes the PWA model discontinuous
in~$m_X$, \ie one introduces $m_X$~thresholds, below which certain
waves are excluded from the PWA model.  In practice, one would hence
exclude waves with radially excited isobars at low masses.  This means
that we have to determine the wave set in principle for each $m_X$~bin
individually.\footnote{In principle, the wave set also depends on~$t'$
  because the intensity of a wave is approximately proportional
  to~$(t')^{|M|}$, so that waves with spin projection $M > 0$ are
  suppressed at low~$t'$ (see \cref{sec:pwa.res_fit_obs}).}  This also
takes into account the different size of the data samples in the
various $m_X$~bins.

Up to now, the wave-set selection was performed \textquote{by hand},
\eg by performing the PWA with a large wave set and iteratively
removing waves with small intensities.  Although this method often
yields satisfactory results (as is shown in \cref{sec:results}), this
approach has a number of caveats and limitations.  For large data sets
that require large wave sets, it is not a well-defined procedure and
is hence prone to observer bias and difficult to document and
reproduce.  It is also often very time consuming.  In addition,
systematic uncertainties due to the choice of the wave set are
difficult to estimate.

A more systematic approach that is currently under study is the
\emph{regularization} of the log-likelihood function in
\cref{eq:likelihood_ext_final}.  This means that one adds a
regularization term to the log-likelihood function that introduces
additional constraints on the fit parameters, \ie
\begin{equation}
  \label{eq:likelihood_reg}
  \ln \widetilde{\mathcal{L}}(\{\mathcal{T}_i^{r \refl}\})
  = \ln \mathcal{L}_\text{ext}(\{\mathcal{T}_i^{r \refl}\})
  + \ln \mathcal{L}_\text{reg}(\{\mathcal{T}_i^{r \refl}\})\eqPunctSpacing.
\end{equation}
The idea is to perform the maximization of the regularized
log-likelihood function
$\ln \widetilde{\mathcal{L}}(\{\mathcal{T}_i^{r \refl}\})$ using a
systematically constructed set of all possible partial waves up to a
cut-off criterion and to choose the regularization term
$\ln \mathcal{L}_\text{reg}(\{\mathcal{T}_i^{r \refl}\})$ such that
partial-wave amplitudes that are statistically consistent with zero
are suppressed, while partial-wave amplitudes far away from zero are
influenced only negligibly by $\ln \mathcal{L}_\text{reg}$.  This way,
one could determine from the data an optimal wave set that describes
the data well.

Up to now, several such approaches were studied using Monte Carlo
pseudo data with known partial-wave content.  The authors of
\refCite{Guegan:2015mea} applied the so-called LASSO
method,\footnote{LASSO stands for \textquote{least absolute shrinkage
    and selection operator} and is a regularization method that was
  first proposed by R.~Tibshirani in \refCite{Tibshirani:1996}.} where
$\mathcal{L}_\text{reg}$ has Laplacian form in
$\big| \mathcal{T}_i^{r \refl} \big|$, so that
\begin{equation}
  \label{eq:lasso}
  \ln \mathcal{L}_\text{reg}(\{\mathcal{T}_i^{r \refl}\}; \lambda)
  = -\lambda \sum_{\refl = \pm 1} \sum_{r = 1}^{N_r^\refl} \sum_i^{N_\text{waves}^\refl}
  \big| \mathcal{T}_i^{r \refl} \big|\eqPunctSpacing.
\end{equation}
This effectively suppresses partial waves with small intensities but
also potentially biases waves with large intensities.  We studied
independently the regularization with $\mathcal{L}_\text{reg}$ having
a Cauchy form in
$\big| \mathcal{T}_i^{r \refl}
\big|$~\cite{Bicker:2016xlx,msc_thesis_drotleff,msc_thesis_kaspar},
\ie
\begin{equation}
  \label{eq:cauchy}
  \ln \mathcal{L}_\text{reg}(\{\mathcal{T}_i^{r \refl}\}; \Gamma)
  = -\sum_{\refl = \pm 1} \sum_{r = 1}^{N_r^\refl} \sum_i^{N_\text{waves}^\refl}
  \ln \left[ 1 + \frac{\big| \mathcal{T}_i^{r \refl} \big|^2}{\Gamma^2} \right]\eqPunctSpacing.
\end{equation}
This term pulls the amplitudes of partial waves with small intensities
toward zero, but leaves the waves with large intensities nearly
unaffected.

Both regularization approaches yield promising results when applied to
Monte Carlo pseudo data.  The application to real data is currently
under study.  A general problem is that the regularization terms have
parameters, \ie $\lambda$~in \cref{eq:lasso} and $\Gamma$~in
\cref{eq:cauchy}, that need to be determined.  The authors of
\refCite{Guegan:2015mea} propose to apply information
criteria~\cite{Akaike:1974,Schwarz:1978tpv} to determine the optimal
parameter values, but more studies are needed to verify the
applicability of these criteria.  Currently, the regularization of the
likelihood function seems to be an interesting approach that is worth
to be studied in more detail.  The method would make the wave-set
selection reproducible and the bias that is introduced in form of the
regularization term explicit.  By applying different regularization
terms and parameter values, one could study the wave-set dependence of
the PWA result.  The method makes it also possible to study, for
example, the dependence of the PWA result on the set of isobars, on
the parameterizations of the dynamical amplitudes of the isobar
resonances, and on the inclusion of partial waves with higher spin.
In most cases, such studies are impossible to perform if the waves set
is selected by hand.

An additional issue is the choice of the formalism that is employed to
calculate the decay amplitudes.  As was discussed in
\cref{sec:pwa.analysis_model.decay_amp}, we employ the helicity
formalism together with simple parameterizations for the propagator
terms of the isobar resonances and the orbital angular-momentum
barrier factors (see
\cref{sec:pwa.analysis_model.dyn_amp,sec:scattering.pw_expansion}).
The disadvantage of this approach is that the extracted partial-wave
amplitudes have a complicated transformation behavior under Lorentz
transformations and---more importantly---do no satisfy all constraints
required by analyticity.  An alternative approach is to formulate the
decay amplitudes in a covariant
formalism~\cite{Zemach:1968zz,Zemach:1965zz,Scadron:1969rw,Chung:1993da,Filippini:1995yc,Anisovich:2006bc,Chung:2007nn},
often referred to as \emph{covariant tensor formalism}.  In a recent
comparison of both approaches, the authors of
\refCite{Mikhasenko:2017rkh}, however, find that some of the covariant
approaches violate crossing symmetry.  They also propose improved
parameterizations for the dynamical parts of the helicity amplitudes
that are more consistent with the relativistic $S$-matrix principles
(see also \cref{sec:pwa.ref_fit_discussion}).  Applying these
amplitudes to COMPASS data is work in progress.

\subsubsection{Freed-Isobar Partial-Wave Analysis}
\label{sec:pwa_cells:freed_isobar}

As was discussed in
\cref{sec:pwa_cells.likelihood_fit,sec:pwa_cells.discussion}, in the
conventional PWA method the decay amplitudes must not have any free
parameters because otherwise the computation of the likelihood
function would become too expensive in the minimization procedure.
Consequently, the choice of the parameterizations and parameters of
the dynamical amplitudes of the isobar resonances induces significant
systematic uncertainties that are difficult to estimate.  In addition,
in the conventional PWA approach it is difficult to take into account
decays via ground \emph{and} excited states of isobar resonances.
Partial-wave amplitudes that have the same angular quantum numbers but
describe decays via isobars with the same \IGJPC quantum numbers are
non-orthogonal because of the non-vanishing overlaps of the dynamic
amplitudes of the isobar states.  As was discussed in
\cref{sec:pwa_cells.discussion}, the non-orthogonality may lead to
distinguishability issues between these partial-wave amplitudes.
Hence yields of resonances can only be measured reliably from the
coherent sum of these amplitudes.  A notoriously difficult case, where
both aspects---not well-known parameterizations and several excited
states---come together, is, for example, the sector of isoscalar
scalar isobars.  Decays via such isobars may involve the \PfZero[500],
\PfZero[980], \PfZero[1370], \PfZero[1500], or even higher-lying
states.  In order to circumvent these limitations and in addition
render the analysis less model-dependent, members of the COMPASS
analysis group developed the novel so-called \emph{freed-isobar PWA
  method}.  This method is inspired by the so-called model-independent
PWA method that was developed by the authors of
\refCite{Aitala:2005yh} in order to study three-body decays of heavy
mesons.  Here, we briefly sketch the freed-isobar method for the case
of three spinless final-state particles as was discussed in
\cref{sec:pwa.analysis_model.examples.3body}.  More details can be
found in \refsCite{Adolph:2015tqa,Krinner:2017dba,Krinner:2018}.  The
extension of the method to final states with more particles is
straight forward.  Without loss of generality, we assume that the
three-body decay proceeds via the two-step process $X \to r + 3$ with
$r \to 1 + 2$.  We assume in addition a rank-1 spin-density matrix.

The conventional PWA model uses fixed parameterizations for the
propagator terms $\mathcal{D}_r(m_r)$ of the isobar resonances in
order to calculate the decay amplitude in \cref{eq:decay_amp_3body_2}.
To simplify notation, we assume that $m_X$~is constant and omit all
constant and $m_X$-dependent factors in the decay amplitude that are
fixed by the normalization in \cref{eq:decay_amp_norm}.  Taking into
account \cref{eq:wave_index,eq:decay_amp_redef}, we rewrite
\cref{eq:decay_amp_3body_2} as
\begin{equation}
  \label{eq:decay_amp_3body_3}
  \Psi_i^\refl(\Underbrace{\thetaGJ, \phiGJ, m_r, \thetaHF, \phiHF}{\displaystyle{\eqqcolon \tau_3}})
  = F_{L_X}(m_r)\, \mathcal{D}_r(m_r)\, F_{J_r}(m_r)\, \mathcal{K}_i^\refl(\anglesGJ, \anglesHF)\eqPunctSpacing.
\end{equation}
Here, the index~$i$ represents the quantum numbers of the three-body
partial wave and those of the isobar resonance~$r$,
$\mathcal{D}_r(m_r)$ is the propagator term of the isobar
resonance~$r$ (see \cref{sec:pwa.analysis_model.dyn_amp}), and
$F_{L_X}(m_r)$ and $F_{J_r}(m_r)$ are the centrifugal-barrier factors
for the $X \to r + 3$ and the $r \to 1 + 2$ decay, respectively (see
\cref{eq:bw_factor}).\footnote{Note that although $m_X$~is constant,
  the barrier factor~$F_{L_X}$ still depends on~$m_r$ via the breakup
  momentum of the $X \to r + 3$ decay that appears in
  \cref{eq:bw_factor}.}  The amplitude
$\mathcal{K}_i^\refl(\anglesGJ, \anglesHF)$ collects all terms that
depend on the angles $\anglesGJ \coloneqq (\cosThetaGJ, \phiGJ)$ in
the Gottfried--Jackson frame and the angles
$\anglesHF \coloneqq (\cosThetaHF, \phiHF)$ in the helicity frame (see
\cref{sec:pwa.analysis_model.coordsys}).  Taking into account the Bose
symmetrization of the final-state particles according to
\cref{eq:decay_amp_bose}, the decay amplitude in a particular
$m_X$~bin is\footnote{To simplify notation, we disregard the constant
  $1 / \sqrt{N_\text{perm}}$ factor, which is irrelevant because of
  the normalization in \cref{eq:decay_amp_norm}.}
\begin{equation}
  \label{eq:freed-isobar_ansatz}
  \Psi_i^{\refl, \text{sym}}(\{\tau_{3, k}\})
  = \sum_{k = 1}^{N_\text{perm}}
  F_{L_X}(m_{r, k})\, \mathcal{D}_r(m_{r, k})\, F_{J_r}(m_{r, k})\,
  \mathcal{K}_i^\refl(\Omega^\text{GJ}_k, \Omega^\text{HF}_k)\eqPunctSpacing,
\end{equation}
where the index~$k$ enumerates the indistinguishable permutations of
the final-state particles.\footnote{In cases where decay amplitudes
  are related by isospin symmetry, the symmetrization expression in
  \cref{eq:freed-isobar_ansatz} becomes more complicated.  However,
  the symmetrized decay amplitude still remains a linear combination
  of terms of the form of \cref{eq:decay_amp_3body_3}, which is the
  property relevant for the freed-isobar method.}

In our novel freed-isobar method, we replace the fixed
parameterizations for $\mathcal{D}_r(m_r)$ by a set of piece-wise
constant amplitudes that fully cover the kinematically allowed mass
range for~ $m_r$, \ie
\begin{equation}
  \label{eq:decay_amp_step_func}
  \widecheck{\mathcal{D}}_i^\refl(m_r)
  = \sum_l^{\mathclap{\text{$m_r$ bins}}} \mathscr{T}_{i, l}^\refl\, \Pi_{l, r}(m_r)\eqPunctSpacing.
\end{equation}
Here, the index~$l$ runs over the $m_r$~bins.  These bins are defined
by sets of window functions $\big\{\Pi_{l, r}(m_r)\big\}$ that are
non-zero only in a narrow $m_r$~interval given by the set
$\{m_{r, l}\}$ of bin borders, \ie
\begin{equation}
  \label{eq:dynamic_step_func}
  \Pi_{l, r}(m_r) = \begin{cases}
    1 & \text{if}~m_{r, l} \leq m_r < m_{r, l + 1}\eqPunctSpacing, \\
    0 & \text{otherwise}\eqPunctSpacing.
  \end{cases}
\end{equation}
In general, the $m_r$~bins may be non-equidistant so that the bin
width $\delta m_{r, l} = m_{r, l + 1} - m_{r, l}$ may depend on the
mass region of the $(1, 2)$ subsystem.  The $\mathscr{T}_{i, l}^\refl$
are a set of unknown complex numbers that together determine the
freed-isobar amplitude $\widecheck{\mathcal{D}}_i^\refl(m_r)$.  This
approach is conceptually similar to the binning in~$m_X$ and~$t'$ in
the conventional PWA.  Note that the freed-isobar amplitude depends on
the partial-wave index~$i$ and the reflectivity~\refl, \ie the model
permits different $\widecheck{\mathcal{D}}_i^\refl(m_r)$ for different
intermediate states~$X$.  This is in contrast to the conventional PWA
approach, where the \emph{same} isobar parameterization is used for
\emph{all} partial-wave amplitudes, where the corresponding isobar
resonance appears.  Since the analysis is performed independently in
$(m_X, t')$ cells, $\widecheck{\mathcal{D}}_i^\refl(m_r)$ is also
allowed to change as a function of~$m_X$ and~$t'$.

For a given $(m_X, t')$ cell, the model for the intensity distribution
in \cref{eq:intensity_model_final} contains terms of the
form\footnote{Since we are considering a rank-1 spin-density matrix,
  we leave off the rank index~$r$.}
\begin{equation}
  \label{eq:freed-isobar_intensity_term}
  \mathcal{T}_i^\refl\, \Psi_i^{\refl, \text{sym}}(\{\tau_{3, k}\})
  = \mathcal{T}_i^\refl \sum_{k = 1}^{N_\text{perm}}
  \Bigg[~\;\sum_l^{\mathclap{\text{$m_r$ bins}}} \mathscr{T}_{i, l}\, \Pi_{l, r}(m_{r, k}) \Bigg]\,
  F_{L_X}(m_{r, k})\, F_{J_r}(m_{r, k})\, \mathcal{K}_i^\refl(\Omega^\text{GJ}_k, \Omega^\text{HF}_k)\eqPunctSpacing,
\end{equation}
where we have used
\cref{eq:freed-isobar_ansatz,eq:decay_amp_step_func}.  We absorb the
unknown amplitudes $\mathscr{T}_{i, l}^\refl$ into the transition
amplitude $\mathcal{T}_i^\refl$ via the redefinition
\begin{equation}
  \label{eq:freed-isobar_prod_amp}
  \widecheck{\mathcal{T}}_{i, l}^\refl \coloneqq \mathcal{T}_i^\refl\, \mathscr{T}_{i, l}^\refl\eqPunctSpacing.
\end{equation}
Note that the transition amplitude now also depends on the $m_r$~bin
index~$l$.  Using \cref{eq:freed-isobar_prod_amp}, we can write the
intensity distribution in \cref{eq:intensity_model_final} as
\begin{equation}
  \label{eq:freed-isobar_intensity_bin}
  \mathcal{I}
  = \sum_{\refl = \pm 1} \bigg| \sum_i^{N_\text{waves}^\refl} ~\; \sum_l^{\mathclap{\text{$m_r$ bins}}}
  \widecheck{\mathcal{T}}_{i, l}^\refl\, \widecheck{\Psi}_{i, l}^{\refl, \text{sym}}(\{\tau_{3, k}\}) \bigg|^2
  + |\mathcal{T}_\text{flat}|^2\eqPunctSpacing,
\end{equation}
where we have redefined the Bose-symmetrized decay amplitude to
include the $m_r$~window function, \ie
\begin{equation}
  \label{eq:freed-isobar_decay_amp}
  \widecheck{\Psi}_{i, l}^{\refl, \text{sym}}(\{\tau_{3, k}\})
  \coloneqq
  \sum_{k = 1}^{N_\text{perm}}
  \Pi_{l, r}(m_{r, k})\, F_{L_X}(m_{r, k})\, F_{J_r}(m_{r, k})\,
  \mathcal{K}_i^\refl(\Omega^\text{GJ}_k, \Omega^\text{HF}_k)\eqPunctSpacing.
\end{equation}
Note that the decay amplitude now also depends on the $m_r$~bin
index~$l$.

In \cref{eq:freed-isobar_intensity_bin}, the additional coherent sum
over the $m_r$~bin index~$l$ appears in the same way as the sum over
the partial-wave index~$i$.  Therefore, each $m_r$~mass bin can be
treated like an independent partial wave.  By defining a new
freed-isobar partial-wave index $\widecheck{i} \coloneqq \{i, l\}$,
\cref{eq:freed-isobar_intensity_bin} becomes mathematically equivalent
to \cref{eq:intensity_model_final}.  This means that exactly the same
maximum likelihood fit procedure as was described in
\cref{sec:pwa_cells.likelihood_fit} can be employed to determine the
$\big\{\widecheck{\mathcal{T}}_{i, l}^\refl\big\}$.  This also
includes the pre-calculation of the integral matrices of the decay
amplitudes.  It is actually this property of the freed-isobar approach
that makes it practically applicable, in contrast to allowing free
parameters in the dynamical isobar amplitudes, which is usually
prohibitively expensive in terms of computational resources.  Another
important property of the freed-isobar approach is that not all waves
in the PWA model have to be parameterized using the freed-isobar
amplitudes in \cref{eq:decay_amp_step_func}.  Depending on the size of
the analyzed data sample, one usually employs freed-isobar amplitudes
only for a subset of waves while for the remaining waves in the PWA
model the conventional fixed parameterizations of the isobar
amplitudes are used.  Depending on how many decay amplitudes are
parameterized using freed-isobar amplitudes and how many $m_r$~bins
are used for the freed-isobar amplitudes, the computation cost for a
freed-isobar PWA grows only moderately, typically by about an order of
magnitude.

Performing such a freed-isobar PWA in $(m_X, t')$ cells, yields
transition amplitudes $\widecheck{\mathcal{T}}_i^\refl(m_X, t', m_r)$
that now depend not only on~$m_X$ and~$t'$ but also on~$m_r$ (via the
$m_r$~bin index~$l$).  As is clear from
\cref{eq:freed-isobar_prod_amp}, the freed-isobar transition
amplitudes contain information about the intermediate states~$X$ as
well as about the respective isobar subsystems.  For each freed-isobar
wave and each $(m_X, t')$ cell, the method yields an \Argand ranging
in $m_r$ from $m_1 + m_2$ to $m_X - m_3$.  It is important to note
that by parameterizing the $m_r$~dependence of the decay amplitude for
a wave~$i$ using the freed-isobar amplitude in
\cref{eq:decay_amp_step_func}, we do not make any assumptions about
the resonance content of the $(1, 2)$~subsystem.  The freed-isobar PWA
determines from the data the total amplitude of all intermediate
isobar states with given \IGJPC quantum numbers in the three-body
partial wave defined by~$i$ and~\refl.  This
amplitude~$\widecheck{\mathcal{T}}_i^\refl$ hence includes all isobar
resonances, potential non-resonant contributions, as well as possible
distortions due to final-state interactions.  As a consequence, the
method avoids the issue of non-orthogonal partial-wave amplitudes for
decays via isobar states with the same \IGJPC quantum numbers that
appear in the conventional PWA as was discussed above.  The reduced
model dependence of the freed-isobar method and the obtained
additional information about the correlation of the dynamic isobar
amplitude with the dynamic amplitude of the~$X$ come at the price of a
considerably larger number of fit parameters compared to the
conventional PWA.  Thus even for large data sets, the freed-isobar
approach can only be applied to a subset of the partial waves in the
PWA model.  For the other partial waves, the conventional fixed isobar
parameterizations are used.

An additional complication arises if a freed-isobar PWA model contains
more than one freed-isobar partial wave with the same three-body
$\JPCMrefl$ quantum numbers.  In this case, mathematical ambiguities,
so-called zero modes, may arise at the amplitude level.  How to
identify and resolve these ambiguities is described in
\refsCite{Krinner:2017dba,Krinner:2018}.

Results from a first freed-isobar PWA of COMPASS \threePi data, that
focuses in particular on the \PfZero* isobars mentioned above, will be
discussed in
\cref{sec:3pi_model:pwa,sec:results_0mp,sec:results_1pp,sec:results_2mp}.

\subsection{Stage~II: Resonance-Model Fit of Spin-Density Matrix}
\label{sec:pwa.res_fit}

The first analysis stage, \ie the partial-wave decomposition of the
data as described in \cref{sec:pwa_cells}, yields as a result the
spin-density matrices $\varrho_{ij}^\refl$ in the $(m_X, t')$ cells.
This set of spin-density matrices serves as the input for the second
analysis stage, where we want to identify $n$-body resonances that
contribute to certain partial-wave amplitudes and determine their
parameters.  To this end, we formulate a model
$\widehat{\varrho}_{ij}^\refl(m_X, t')$\footnote{In the following
  text, modeled quantities are distinguished from their measured
  counterparts by a hat (\enquote{$\widehat{\phantom{m}}$}).} that
describes the dependence of the spin-density matrix elements on~$m_X$
and~$t'$ in terms of resonant and non-resonant components.  The
parameters and yields of the resonances included in this model are
determined by a fit to the measured spin-density matrices.  In this
\emph{resonance-model fit}, which is also referred to as
\emph{mass-dependent fit}, we exploit that resonances have a
characteristic signature.  In the simplest case, they appear as
Breit--Wigner-shaped peaks in the partial-wave intensities that are
accompanied by $180^\circ$ phase motions relative to non-resonating
waves (see \eg\ \cref{fig:bw}).  In reality, this simple picture often
does not hold.  The behavior of the spin-density matrix elements is
usually more complicated because multiple resonances, \eg a ground
state and corresponding radially excited states, may appear in a given
partial wave.  In general, the resonances overlap and interfere so
that resonance peaks might shift, disappear, or---in the case of
destructive interference---might even turn into dips in the intensity
distribution.\footnote{This fact is expressed in a concise way by the
  common saying: \enquote{Not every peak is a resonance and not every
    resonance is a peak} (see also \refCite{Dalitz:1970ga}).}  In
diffractive-dissociation reactions, this interference pattern is
further complicated by additional coherent contributions from
non-resonant double-Regge exchange processes (see
\cref{sec:exp.prod_reactions,sec:pwa_cells.discussion,fig:exp.production.central,fig:exp.production.deck}).
The different $t'$~dependences of these contributions lead to a
$t'$-dependent interference.  Hence in many cases, the information
from partial-wave intensities alone is insufficient to extract
resonances and their parameters reliably from the data.  It is
therefore a great advantage that the partial-wave decomposition is
performed at the amplitude level.  The parameters of the resonances
are hence constrained not only by the measured intensities of the
partial waves but also by the mutual interference terms of the waves.
This greatly improves the sensitivity for potential resonance signals.

As was discussed in \cref{sec:pwa.analysis_model.extension}, an
advantage of the two-stage analysis approach is that the resonance
model does not need to describe all partial-wave amplitudes that are
included in the PWA model.  The representation of the data in terms of
the spin-density matrix allows us to model only selected matrix
elements.  Usually, one selects a subset of waves and models all
elements of the corresponding spin-density submatrix.  The criteria
that are used to select the waves that are included in the
resonance-model fit depend strongly on the analyzed channel.  In
general, one tries to select waves with clear signals of known
resonances plus a few \enquote{interesting} waves, the idea being that
the known resonances act as reference amplitudes, against which the
resonant amplitudes in the interesting waves can interfere.  When
searching for resonance signals in the partial-wave amplitudes, one
has to take into account that similar to the resonance peaks in the
intensity distributions also the phase motions are distorted by the
presence of other resonances.  If, for example, waves~$i$ and~$j$ with
reflectivity~\refl each contain a resonance, the rising motion of
their relative phase $\phase_{ij}^\refl$ (see \cref{eq:phase_def}) due
to the resonance in wave~$i$ may be compensated by the falling motion
due to the other resonance in wave~$j$.  In the extreme case when the
two resonances have similar parameters, this cancellation may become
nearly complete.  It is therefore often difficult to observe
undisturbed phase motions of resonances.  This effect is particularly
pronounced in the high-$m_X$ regions, where broad excited states
appear in most or even all waves in the resonance-model fit.

\subsubsection{Resonance Model}
\label{sec:pwa.res_model}

In order to formulate the resonance model, we go back to
\cref{eq:intensity_model} in \cref{sec:pwa.analysis_model.extension},
where we defined the transition amplitudes.  Using this definition,
and taking into account the reflectivity basis (see
\cref{sec:pwa_cells.reflectivity}) and possible incoherences in terms
of the rank of the spin-density matrix (see
\cref{sec:pwa_cells.rank}), we can write the model for the transition
amplitudes as
\begin{align}
  \label{eq:res_model_amp_wave}
  \widehat{\mathcal{T}}_i^{r \refl}(m_X, t')
  = \sqrt{I_{ii}^\refl(m_X)}\, \sqrt{m_X}\,
  \mathcal{P}^\refl(m_X, t')\,
  \sum_{k\, \in\, \mathbb{S}_i^\refl} \mathcal{C}_{k i}^{r \refl}(t')\,
  \mathcal{D}_k^{r \refl}(m_X, t'; \zeta_k^{r \refl})\eqPunctSpacing.
\end{align}
Here, the factor $\sqrt{I_{ii}^\refl(m_X)}$ appears due to the chosen
normalization of the transition amplitudes in
\cref{eq:trans_amp_norm}.  The amplitude $\mathcal{P}^\refl(m_X, t')$
models the average production probability of an intermediate state
with mass~$m_X$ as a function of~$t'$.  It hence is independent of the
wave index~$i$, but it depends in the most general case on the
reflectivity~\refl.  This is because different values of~\refl
correspond to different exchange particles and hence different
production processes. Likewise, the coupling amplitudes
$\mathcal{C}_{k i}^{r \refl}(t')$ depend on~\refl.  They depend in
addition on the rank index~$r$ in order to model potential
incoherences.  Together with $\mathcal{P}^\refl(m_X, t')$, the
coupling amplitudes describe the production of the resonant and
non-resonant components that are represented by the dynamical
amplitudes $\mathcal{D}_k^{r \refl}(m_X, t'; \zeta_k^{r \refl})$.
Concerning the latter amplitudes we extended the model \wrt\
\cref{eq:intensity_model} such that in the most general case also the
dynamical amplitudes may depend on~\refl and on~$r$ and that they may
have an explicit $t'$~dependence.  The reason for this will become
clear in the discussion of the parameterization of the non-resonant
components below.

In \cref{eq:res_model_amp_wave}, we parameterize the transition
amplitude of a given partial wave~$i$ as a coherent sum of the
amplitudes of so-called \emph{wave components} that can be either
resonances or non-resonant components.  The wave components are
represented by the dynamical amplitudes, which are enumerated by the
index~$k$.  The same wave component, \eg a resonance, may appear in
several waves.  Hence the sum in \cref{eq:res_model_amp_wave} runs
over the set~$\mathbb{S}_i^\refl$ of those wave components that we
assume to appear in wave~$i$ with reflectivity~\refl.  The dynamical
amplitudes depend on a set of \emph{shape parameters} that we denote
by~$\zeta_k^{r \refl}$.  For example, in the case of a Breit--Wigner
resonance, these parameters are mass and width of that resonance.  It
is important to note that for resonances the dynamical amplitudes and
their shape parameters are independent of the partial-wave index~$i$,
the rank index~$r$, and the reflectivity~\refl.  Thus if a resonance
component~$\mathcal{D}_k^\text{R}$ appears in several partial waves,
which \eg represent different decay chains or \Mrefl~states of this
resonance, the same shape parameters~$\zeta_k^\text{R}$ are used in
these waves.

The coupling amplitudes $\mathcal{C}_{k i}^{r \refl}(t')$ in
\cref{eq:res_model_amp_wave} collect two kinds of unknowns: \one the
product $\alpha_i$ of the complex-valued couplings that appear at the
two-body decay vertices in the $n$-body isobar decay chain of wave~$i$
(see \cref{eq:decay_amp_redef})\footnote{The couplings and hence their
  product are independent of~$t'$.} and \two the details of the
beam-Reggeon vertex, \ie the $t'$~dependence of the production
strength and phase of wave component~$k$ in wave~$i$ with
reflectivity~\refl.  Currently, theoretical models are not detailed
enough in order to parameterize these $t'$~dependences.  Therefore,
binned approximations of the $\mathcal{C}_{k i}^{r \refl}(t')$ are
extracted from the data by fitting the resonance model to all
$t'$~bins simultaneously and leaving the values of each coupling
amplitude in each $t'$~bin as free parameters to be determined by the
fit.  The resonance model hence parameterizes mostly the
$m_X$~dependence of the spin-density matrix, but is model-independent
\wrt the $t'$~dependence in the sense that the amplitude of each
component in each wave is allowed to have a different $t'$~dependence.
An additional advantage of this binned approach is that the resonant
and non-resonant wave components can be better disentangled because
their amplitudes have usually different dependences on~$t'$.  The
caveat of such a $t'$-resolved resonance-model fit is the large number
of free parameters.

Assuming factorization of production, propagation, and decay of the
intermediate $n$-body resonances, the dynamical amplitudes
$\mathcal{D}_k^\text{R}(m_X; \zeta_k^\text{R})$ of the resonances
should be independent of~$r$, \refl, and~$t'$.  The latter constraint
is built into the model by using the same resonance shape
parameters~$\zeta_k^\text{R}$, \ie masses and widths of the
resonances, across all $t'$~bins.  Only strength and phase of each
resonance component, represented by the coupling amplitudes
$\mathcal{C}_{k i}^{r \refl}(t')$, can be chosen freely by the fit in
each $t'$~bin.  Since the $t'$~dependence of an amplitude of a
resonant component is determined by the production mechanism,
factorization of production and decay means that the $t'$~dependence
of the amplitude should be the same in different decay chains of that
resonance for a given spin projection \Mrefl.  This can be exploited
to reduce the number of parameters by fixing the $t'$~dependence of
the amplitude of resonance~$k$ in wave~$j$, which is represented by
$\mathcal{C}_{k j}^{r \refl}(t')$, to the $t'$~dependence of that
resonance in wave~$i$:\footnote{The waves~$i$ and~$j$ must have the
  same \JPCMrefl quantum numbers.  If more than two waves in the
  resonance-model fit fulfill this criterion, always the same
  reference wave~$i$ is used in \cref{eq:branching_amp}.  Usually, the
  wave that corresponds to the most dominant decay chain of
  resonance~$k$ is chosen as the reference wave~$i$.}
\begin{equation}
  \label{eq:branching_amp}
  \mathcal{C}_{k j}^{r \refl}(t')\
  = \prescript{}{j}{\mathcal{B}}_{k i}^\refl\, \mathcal{C}_{k i}^{r \refl}(t')\eqPunctSpacing.
\end{equation}
The $t'$-independent complex-valued proportionality factors
$\prescript{}{j}{\mathcal{B}}_{k i}^\refl$, which are determined by
the fit, are called \emph{branching amplitudes} and encode the
relative strength and phase of the two decay chains~$i$ and~$j$ of
resonance~$k$.

The dynamical amplitudes
$\mathcal{D}_k^\text{R}(m_X; \zeta_k^\text{R})$ of the resonance
components with the shape parameters
$\zeta_k^\text{R} \coloneqq (m_k, \Gamma_k)$ are often parameterized
using Breit--Wigner amplitudes.  The parameterization used for the
total width depends on how well the decay modes of the resonance are
known.  If the branching fractions of the resonance are not well-known
or even unknown, the constant-width approximation in
\cref{eq:BW_const_width} is used.  If the branching fractions are
known sufficiently well, \cref{eq:BW_mass-dep_width} is employed.
This parameterization takes into account the opening of the phase
space over the resonance by modeling the decay as an incoherent sum of
two-body decays of quasi-stable particles, \ie by neglecting the width
of any isobars that appear in the decays.  \Cref{eq:BW_mass-dep_width}
is only a good approximation in the $m_X$~range well above the nominal
thresholds of the respective decay channels.  Close to or below these
thresholds, the finite width of the isobar(s) cannot be neglected and
more advanced parameterizations have to be used (see \eg Section~IV.A
in \refCite{Akhunzyanov:2018lqa}).

The dynamical amplitudes of the non-resonant components are in general
not well known so that empirical parameterizations are employed, which
are often developed in a data-driven approach.  In general, we cannot
assume factorization of production and decay for the non-resonant
components and also \cref{eq:branching_amp} does not hold.  Thus one
usually includes for each wave a separate non-resonant component with
independent shape parameters. Consequently, the dynamical amplitude of
a non-resonant component may depend on the reflectivity~\refl.  In the
most general case, the non-resonant terms may also have incoherent
components so that the dynamical amplitude may depend on the rank
index~$r$.  A common assumption that is built into the employed
parameterizations for these dynamical amplitudes is that their phase
is independent of~$m_X$.  Inspired by \refCite{tornqvist:1995kr}, in
many analyses a Gaussian in the two-body break-up momentum~$q_i$ of
the $X$~decay in wave~$i$ is used,
\ie\footnote{\label{footnote:dyn_amp_non_res}To simplify notation, we
  omit the indices~$k$, $r$, and~\refl from the shape parameters.}
\begin{equation}
  \label{eq:dyn_amp_non_res_simple}
  \prescript{\text{NR}}{}{\mathcal{D}_k^{r \refl}}(m_X; \Underbrace{b, c_0}%
  {\displaystyle{\mathclap{\hspace*{2em}\eqqcolon \prescript{\text{NR}}{}{\zeta_k^{r \refl}}}}})
   = \Big[ \frac{m_X - m_\text{thr}}{m_\text{norm}} \Big]^b\, e^{-c_0\, q_i^2(m_X)}\eqPunctSpacing.
\end{equation}
Here, we have added the term in square brackets, which approximates
the phase-space opening.  The constant $m_\text{thr}$ is set to a
value around the kinematical threshold and $m_\text{norm}$ is usually
set to $1\,\GeV/c^2$.  Since for~$m_X$ values close to or below the
two-body threshold, \cref{eq:breakup_mom.cms.f,eq:kaellen} are not
applicable to calculate $q_i(m_X)$, more advanced parameterizations
are used, which effectively take into account the width of the isobar
resonance that appears in wave~$i$ (see Section~IV.A in
\refCite{Akhunzyanov:2018lqa} for details).  In general, the shape of
the $m_X$~distribution of a non-resonant component may change
with~$t'$ so that the dynamical amplitude may have an explicit
$t'$~dependence.  In such cases, it was found that an extension of
\cref{eq:dyn_amp_non_res_simple} of the
form\cref{footnote:dyn_amp_non_res}
\begin{equation}
  \label{eq:dyn_amp_non_res}
  \prescript{\text{NR}}{}{\mathcal{D}_k^{r \refl}}(m_X, t'; \Underbrace{b, c_0, c_1, c_2}%
  {\displaystyle{\mathclap{\hspace*{2em}\eqqcolon \prescript{\text{NR}}{}{\zeta_k^{r \refl}}}}})
   = \Big[ \frac{m_X - m_\text{thr}}{m_\text{norm}} \Big]^b\, e^{-\big( c_0 + c_1\, t' + c_2\, t'^2 \big)\, q_i^2(m_X)}
\end{equation}
yields a better description of the data.

For diffractive-dissociation reactions, the average production
amplitude $\mathcal{P}^\refl(m_X, t')$ can be modeled using Regge
theory (see \cref{sec:regge}).  At high energies, these reactions are
dominated by $t$-channel Pomeron (\Pom) exchange (see
\cref{sec:exp.prod_reactions}).  We follow the phenomenological Regge
approach developed in~\refCite{Ataian:1991gn} to describe
central-production reactions.  The cross section for this reaction is
proportional to the so-called \emph{Pomeron flux}
factor~\cite{Ingelman:1999np}
\begin{equation}
  \label{eq:pom_flux}
  F_{\Pom p}(x_\Pom, t')
  \propto \frac{e^{-b_\Pom\, t'}}{x_\Pom^{2 \alpha_\Pom(t') - 1}}\eqPunctSpacing,
  ~\text{where}\quad
  x_\Pom
  \approx \frac{m_X^2}{s}
\end{equation}
is the fraction of the longitudinal proton momentum carried by the
Pomeron in the center-of-momentum frame of the reaction.  Mandelstam
$s$~is the squared center-of-momentum energy of the beam-target
system, $b_\Pom$~is the slope parameter of the $t'$~dependence of the
cross section for Pomeron-exchange reactions, and $\alpha_\Pom(t')$ is
the Regge trajectory of the Pomeron (see
\cref{eq:regge_trajectory.linear}).\footnote{Here, we neglect \tMin so
  that $t' \approx -t$.  A discussion about the choice of the values
  for $\alpha_0$ and $\alpha'$ can be found in
  \refCite{Adolph:2015tqa}.}  The Pomeron flux in \cref{eq:pom_flux}
can be interpreted as the probability for emission of a Pomeron by the
proton.  In the limit of $\alpha_0 = 1$ and
$\alpha' = 0\,(\GeV/c)^{-2}$, this probability is proportional to
$1 / x_\Pom$ and hence proportional to $1 / m_X^2$.  This dependence
can be explained, for example, by a Bremsstrahlungs-like
mechanism~\cite{Dederichs:1989gk,Faessler:1993gn}.  We assume that the
Pomeron flux factor in \cref{eq:pom_flux} is universal and apply it to
the single-Pomeron exchange process of diffractive dissociation as it
was done \eg in \refCite{Cox:2000jt}.  We absorb the normalization and
the factor $e^{-b_\Pom\, t'}$ into the coupling amplitudes
$\mathcal{C}_{k i}^{r \refl}(t')$ so that the production probability
is given by
\begin{equation}
  \label{eq:prod_prob}
  |\mathcal{P}^\refl(m_X, t')|^2
  = \frac{1}{x_\Pom^{2 \alpha_\Pom(t') - 1}}
  \approx \bigg[ \frac{s}{m_X^2} \bigg]^{2 \alpha_\Pom(t') - 1}\eqPunctSpacing.
\end{equation}

\subsubsection{$t'$-Dependent Observables}
\label{sec:pwa.res_fit_obs}

Since the resonance-model fit is performed simultaneously in several
$t'$~bins, we can extract the $t'$~dependence of production strength
and phase for each wave component in the model.  This information is
contained in the coupling amplitudes
$\mathcal{C}_{k i}^{r \refl}(t')$.

From the intensity~$\intens_i^\refl$ of partial wave~$i$ with
reflectivity~\refl as defined in \cref{eq:intens_def}, we can derive
an expression for the intensity of a wave component in this wave for a
given $(m_X, t')$ cell by using the resonance model in
\cref{eq:res_model_amp_wave}:
\begin{multline}
  \label{eq:res_model_int_wave}
  \intens_i^\refl(m_X, t')
  = \sum_{k\, \in\, \mathbb{S}_i^\refl}
  \Underbrace{I_{ii}^\refl(m_X)\, m_X\, \big| \mathcal{P}^\refl(m_X, t') \big|^2\,
    \sum_{r = 1}^{N_r^\refl}
    \big| \mathcal{C}_{k i}^{r \refl}(t') \big| ^2 \big|\mathcal{D}_k^{r \refl}(m_X, t'; \zeta_k^{r \refl}) \big|^2}%
  {\displaystyle{\text{intensity of wave component~$k$}}} \\
  + \sum_{\substack{k, l\, \in\, \mathbb{S}_i^\refl \\ k < l}}
  \Underbrace{I_{ii}^\refl(m_X)\, m_X\, \big| \mathcal{P}^\refl(m_X, t') \big|^2\,
    2\Re\!\bigg[ \sum_{r = 1}^{N_r^\refl}
    \mathcal{C}_{k i}^{r \refl}(t')\, \mathcal{C}_{l i}^{r \refl \text{*}}(t')\,
    \mathcal{D}_k^{r \refl}(m_X, t'; \zeta_k^{r \refl})\, \mathcal{D}_l^{r \refl \text{*}}(m_X, t'; \zeta_l^{r \refl}) \bigg]}%
  {\displaystyle{\text{overlap of wave components~$k$ and~$l$}}}\eqPunctSpacing.
\end{multline}
This expression has a structure analogous to
\cref{eq:events_pred_norm_amp} and we can define the
intensity~$\intens_{k i}^\refl$ of wave component~$k$ as
\begin{equation}
  \label{eq:intens_comp_def}
  \intens_{k i}^\refl(m_X, t')
  \coloneqq I_{ii}^\refl(m_X)\, m_X\, \big| \mathcal{P}^\refl(m_X, t') \big|^2\,
  \sum_{r = 1}^{N_r^\refl}
  \big| \mathcal{C}_{k i}^{r \refl}(t') \big|^2
  \big|\mathcal{D}_k^{r \refl}(m_X, t'; \zeta_k^{r \refl}) \big|^2\eqPunctSpacing.
\end{equation}
Due to the chosen normalization, $\intens_{k i}^\refl(m_X, t')$
corresponds to the expected number of produced events in wave
component~$k$ in wave~$i$ with reflectivity~\refl in a given
$(m_X, t')$ cell (see \cref{sec:pwa_cells.normalization}).

By integrating \cref{eq:intens_comp_def} over the full $m_X$~range
under consideration, we get the \emph{$t'$~spectrum} of wave
component~$k$:
\begin{align}
  \mathcal{I}_{ki}^\refl(t')
  &= \frac{\dif{\intens_{k i}^\refl}}{\dif{t'}} \nonumber \\
  \label{eq:t_spectrum}
  &= \frac{1}{\Delta t'} \sum_{r = 1}^{N_r^\refl}
  \big| \mathcal{C}_{k i}^{r \refl}(t') \big| ^2
  \int_{m_\text{min}}^{m_\text{max}}\! \dif{m_X}\,
  I_{ii}^\refl(m_X)\, m_X\, \big| \mathcal{P}^\refl(m_X, t') \big|^2\,
  \big|\mathcal{D}_k^{r \refl}(m_X, t'; \zeta_k^{r \refl}) \big|^2\eqPunctSpacing.
\end{align}
Here, we divide by the $t'$~bin width $\Delta t'$ in order to take
into account the non-equidistant $t'$~binning.  The $m_X$~integration
range $[m_\text{min}, m_\text{max}]$ is usually given by the fit range
that is used for the given partial wave.

In diffractive reactions, the $t'$~spectra of most wave components
exhibit an approximately exponential decrease with increasing~$t'$ in
the range $t' \lesssim 1\,(\GeV/c)^2$.  This behavior can be explained
in the framework of Regge theory (see \cref{eq:dsigma_dt.regge.2}).
The $t'$~spectra of partial-wave amplitudes with a spin projection of
$M \neq 0$ along the beam direction are modified by an additional
factor~$(t')^{|M|}$.  This factor is given by the forward limit of the
Wigner $D$-functions~\cite{Perl:1974} and suppresses the intensity
toward small~$t'$.  Following \cref{eq:dsigma_dt.regge.2}, we
therefore parameterize the $t'$~spectra by the model
\begin{equation}
  \label{eq:t_spectrum_model}
  \widehat{\mathcal{I}}_{ki}^\refl(\tpr; A_{ki}^\refl, b_{ki}^\refl)
  = A_{ki}^\refl \cdot \big( \tpr \big)^{\abs{M}} \cdot e^{-b_{ki}^\refl\, \tpr}\eqPunctSpacing.
\end{equation}
This parameterization has two free real-valued parameters: the
magnitude parameter~$A_{ki}^\refl$ and the \emph{slope
  parameter}~$b_{ki}^\refl$.  The parameters are estimated by
performing a $\chi^2$~fit, where the model function
$\widehat{\mathcal{I}}_{ki}^\refl(\tpr)$ is integrated over each
$t'$~bin and compared to the measured value
$\mathcal{I}_{ki}^\refl(\tpr)$ in \cref{eq:t_spectrum}.

In addition to the $t'$~spectrum, we can also extract the relative
phase between the coupling amplitudes of wave component~$k$ in
wave~$i$ and wave component~$l$ in wave~$j$,
\begin{equation}
  \label{eq:phase_comp_def}
  \phase_{ki, lj}^\refl(t')
  \coloneqq \arg\!\bigg[ \sum_{r = 1}^{N_r^\refl} \mathcal{C}_{k i}^{r \refl}(t')\,
  \mathcal{C}_{l j}^{r \refl \text{*}}(t') \bigg]\eqPunctSpacing.
\end{equation}
This quantity is called \emph{coupling phase} and is conceptually
similar to the relative phase in \cref{eq:phase_def}.  Note that for a
rank-1 spin-density matrix, \ie full coherence of all partial waves,
\begin{equation}
  \label{eq:phase_comp_rank1}
  \phase_{ki, lj}^\refl(t')
  = \arg\!\big[ \mathcal{C}_{k i}^\refl(t') \big] - \arg\!\big[ \mathcal{C}_{l j}^\refl(t') \big]\eqPunctSpacing.
\end{equation}

If the coupling amplitudes of a resonance in different decay chains
are constrained via \cref{eq:branching_amp}, the corresponding
coupling phases are by definition independent of~$t'$ and are given by
$\arg\!\big[\prescript{}{j}{\mathcal{B}}_{k i}^\refl\big]$.  The
interpretation of the values of a coupling phase is complicated by the
fact that the coupling amplitude is the product of the actual
production amplitude of wave component~$k$ and of the
coupling~$\alpha_{k i}$, which in turn is the product of the
complex-valued couplings that appear at each two-body decay vertex in
the $n$-body isobar decay chain of wave~$i$ (see
\cref{eq:decay_amp_redef}).  The coupling phases of resonances are
expected to be approximately independent of~$t'$, if a single
production mechanism dominates~\cite{Daum:1979sx,Daum:1980ay}.  This
is in general not true for non-resonant components, which are produced
by different mechanisms.  Relative to resonances, the coupling phases
of the non-resonant components usually change with~$t'$.  This leads
in general to a $t'$-dependent interference pattern, which often
causes a $t'$-dependent shift of resonance peaks, especially for waves
with large non-resonant components.  So if the shape of the measured
$m_X$~distribution of the intensity of a partial-wave amplitude
changes with~$t'$, it is a sign of contributions from non-resonant
components.

\subsubsection{Fit Method}
\label{sec:pwa.res_fit_method}

We construct the model for the spin-density submatrix of those waves
that are selected for the resonance-model fit by using
\cref{eq:spin-dens_norm,eq:res_model_amp_wave}:
\begin{align}
  \widehat{\varrho}_{ij}^\refl(m_X, t')
  &= \sum_{r = 1}^{N_r^\refl} \widehat{\mathcal{T}}_i^{r \refl}(m_X, t')\, \widehat{\mathcal{T}}_j^{r \refl \text{*}}(m_X, t') \nonumber \\
  \label{eq:res_model_spin-dens}
  &\!\begin{multlined}
    = \sqrt{I_{ii}^\refl(m_X)\, I_{jj}^\refl(m_X)}\, m_X\, \big| \mathcal{P}^\refl(m_X, t') \big|^2 \\
    \qquad\times \sum_{r = 1}^{N_r^\refl}
    \bigg[ \sum_{k\, \in\, \mathbb{S}_i^\refl} \mathcal{C}_{k i}^{r \refl}(t')\,
    \mathcal{D}_k^{r \refl}(m_X, t'; \zeta_k^{r \refl}) \bigg]\,
    \bigg[ \sum_{l\, \in\, \mathbb{S}_j^\refl} \mathcal{C}_{l j}^{r \refl}(t')\,
    \mathcal{D}_l^{r \refl}(m_X, t'; \zeta_l^{r \refl}) \bigg]^\text{*}\eqPunctSpacing.
  \end{multlined}
\end{align}
The parameters of the resonance model, \ie the set
$\{\mathcal{C}_{k i}^{r \refl}(t')\}$ of coupling amplitudes and the
set $\{\zeta_k^{r \refl}\}$ of shape parameters, are estimated by
fitting the model $\widehat{\varrho}_{ij}^\refl(m_X, t')$ to the
spin-density matrices that are measured in the $(m_X, t')$ cells (see
\cref{sec:pwa_cells}).  In order to perform this fit, we represent the
Hermitian spin-density matrix by a real-valued matrix
$\Lambda_{ij}^\refl(m_X, t')$ of the same dimension.  The elements
of~$\Lambda_{ij}^\refl$ are defined by the upper triangular part of
the spin-density matrix as follows
\begin{equation}
  \label{eq:res_spin-dens_redef}
  \Lambda_{ij}^\refl(m_X, t') =
  \begin{cases}
    \Re\!\big[ \varrho_{ij}^\refl(m_X, t') \big] & \text{for}~i < j\eqPunctSpacing, \\[1.1ex]
    \Im\!\big[ \varrho_{ji}^\refl(m_X, t') \big] & \text{for}~i > j\eqPunctSpacing, \\[1.1ex]
    \varrho_{ii}^\refl(m_X, t')                  & \text{for}~i = j\eqPunctSpacing.
  \end{cases}
\end{equation}
This means that the upper off-diagonal elements
of~$\Lambda_{ij}^\refl$ are the real parts of the interference terms,
the lower off-diagonal elements are the imaginary parts of the
interference terms, and the diagonal elements are the partial-wave
intensities.  In an analogous way, we define
$\widehat{\Lambda}_{ij}^\refl(m_X, t')$ for our resonance model in
\cref{eq:res_model_spin-dens}.

In order to quantify the deviation of the resonance model
$\widehat{\Lambda}_{ij}^\refl(m_X, t')$ from the measured data
$\Lambda_{ij}^\refl(m_X, t')$, we sum the squared Pearson's
residuals~\cite{Pearson:1900} of all matrix elements for all
$(m_X, t')$ cells, \ie
\begin{equation}
  \label{eq:fit_chi2}
  \chi^2
  = \sum_{\refl = \pm 1}\; \sum_{i, j}^{N_\text{waves}^\refl}\; \sum^{\text{$t'$ bins}}\; \sum^{(\text{$m_X$ bins})_{ij}}
  \bigg[ \frac{\Lambda_{ij}^\refl(m_X, t')
    - \widehat{\Lambda}_{ij}^\refl(m_X, t')}{\sigma_{ij}^\refl(m_X, t')} \bigg]^2\eqPunctSpacing.
\end{equation}
Here, $N_\text{waves}^\refl$~is the number of partial waves with
reflectivity~\refl that are included in the fit model and
$\sigma_{ij}^\refl(m_X, t')$ is the statistical uncertainty of the
matrix element $\Lambda_{ij}^\refl(m_X, t')$ as determined in the
partial-wave decomposition stage (see
\cref{sec:max_likelihood_procedure}).  The sum in \cref{eq:fit_chi2}
runs over all $t'$~bins and those $m_X$~bins that lie within the
chosen fit ranges.  For the off-diagonal interference terms
$\Lambda_{ij}^\refl$, the $m_X$~range is given by the intersection of
the fit ranges for the intensities of waves~$i$ and~$j$.  The best
estimate for the model parameters is determined by minimizing the
$\chi^2$~function using the MIGRAD algorithm of the MINUIT
package~\cite{James:1975dr,minuit}.\footnote{For high-precision data,
  the resonance model usually is not able to reproduce all details of
  the data.  The residual deviations of the model from the data often
  induce a multi-modal behavior of the minimization.
  \RefCite{Akhunzyanov:2018lqa} discusses strategies on how to resolve
  this multi-modality in order to obtain the best physical solution.}

It is important to note that although we use the notation~$\chi^2$ for
the minimized quantity in \cref{eq:fit_chi2}, the minimum~$\chi^2$
does in general not follow a $\chi^2$~distribution.  This means that
the expectation value of the minimum is not the number of degrees of
freedom (n.d.f.).  Also the deviation of the minimum from the n.d.f.\
is not directly interpretable as an absolute measure for the goodness
of the fit.  This is because we assume in \cref{eq:fit_chi2} that the
elements of $\Lambda_{ij}^\refl(m_X, t')$ are all statistically
independent.  Although this assumption is true for matrix elements
from different $m_X$ or $t'$~bins, it is in general not true for
matrix elements within a given $(m_X, t')$ cell.  They could be
correlated due to statistical correlations of the transition
amplitudes in the PWA fit.  In principle, these correlations are known
because in the partial-wave decomposition that is performed in the
first analysis stage also the covariance matrix of the transition
amplitudes $\mathcal{T}_i^{r \refl}(m_X, t')$ is estimated (see
\cref{eq:mle_cov_est}).  However, the propagation of this information
to the covariance matrix of the elements of $\Lambda_{ij}^\refl$ is
not well defined.  This is because the spin-density matrix has more
free real-valued parameters than the set of transition amplitudes.  In
an $(m_X, t')$ cell, the resonance-model fit minimizes the distance of
the model to
$\big( [N_\text{waves}^{\refl = +1}]^2 + [N_\text{waves}^{\refl =
  -1}]^2 \big)$ data points, which are the elements of
$\Lambda_{ij}^\refl$.  In contrast, the set of transition amplitudes
from the partial-wave decomposition represents only
$\big( N_r^{\refl = +1}\, [2 N_\text{waves}^{\refl = +1} - N_r^{\refl
  = +1}] + N_r^{\refl = -1}\, [2 N_\text{waves}^{\refl = -1}
-N_r^{\refl = -1} ] \big)$ data points (see
\cref{sec:pwa_cells.rank}).\footnote{Only in the case of maximum rank,
  \ie for $N_r^\refl = N_\text{waves}^\refl$, the set of transition
  amplitudes has the same number of real-valued parameters as
  $\Lambda_{ij}^\refl$.}  For a rank-1 spin density matrix, this
actually leads to analytical relations among the spin-density matrix
elements of waves~$i$, $j$, $k$, and~$l$ of the form
\begin{equation}
  \label{eq:spin_dens_r1_corr}
  \varrho_{i j}\, \varrho_{k l}
  = \varrho_{i l}\, \varrho_{k j}\eqPunctSpacing,
\end{equation}
which are also not taken into account by
\cref{eq:fit_chi2}.\footnote{For the case of a rank-1 spin-density
  matrix, Appendix~C in \refCite{Akhunzyanov:2018lqa} provides
  formulations of the $\chi^2$~function that correctly take into
  account all correlations.  However, in practice these
  $\chi^2$~functions may yield unphysical results due to tensions of
  the model with the data.  The $\chi^2$~function in
  \cref{eq:fit_chi2} turns out to be more robust against such
  effects.}

The covariance matrix of the fit parameters of the resonance model is
estimated from the inverse of the Hessian matrix at the
minimum~$\chi^2$.  Due to the not-accounted-for correlations of the
elements of the spin-density matrices, this may not be a good
approximation.  The uncertainty estimates for the fit parameters of
the resonance model can be verified, for example, by employing
resampling techniques, such as the
jackknife~\cite{Quenouille:1949,Tukey:1958} or the
bootstrap~\cite{Efron:1979,Wolter:1985,Shao:1995}.  Using these
techniques, one generates a set of~$\Lambda_{ij}^\refl$ that is
distributed according to the covariance matrix of the transition
amplitudes.  For each $\Lambda_{ij}^\refl$ in this set, a
resonance-model fit is performed using \cref{eq:fit_chi2}.  This
yields a set of fit results, from which the probability distribution
of the fit parameters and hence their covariance matrix can then be
estimated.

\subsubsection{Discussion of the Resonance Model}
\label{sec:pwa.ref_fit_discussion}

The resonance model that was described in \cref{sec:pwa.res_model} has
a number of potential caveats and
limitations~\cite{pdg_resonances:2018}. In general, the employed
Breit--Wigner amplitudes are good approximations only for isolated
resonances far above the kinematical thresholds of their decay modes.
In addition, using a sum of Breit--Wigner amplitudes to describe the
total amplitude of a set of resonances with the same quantum numbers
is a good approximation only if the resonances are well-separated so
that they have little overlap.  Otherwise unitarity, \ie the
conservation of probability, may be violated.  For resonances with
multiple decay modes, Breit--Wigner parameterizations do also neglect
coupled-channel effects caused by unitarity constraints.  These
effects may become large when the kinematical thresholds of one or
more of these decays are in the mass range of the resonance.  In
practice, the limitations discussed above render Breit--Wigner
parameters model- and process-dependent.

Another potential weakness of our model is the decomposition of the
partial-wave amplitudes into a sum of resonant and non-resonant
components, which is not unique~\cite{pdg_resonances:2018} and hence
introduces additional process and model dependence of the resonance
parameters.  In principle, this decomposition would be constrained by
unitarity, which is, however, not taken into account here.  Due to the
absence of realistic models, the parameterizations employed for the
non-resonant wave components (see
\cref{eq:dyn_amp_non_res_simple,eq:dyn_amp_non_res}) are only
phenomenological.  Also, the model assumption that the phase of the
non-resonant amplitudes is independent of~$m_X$ may not be well
justified especially for cases where these amplitudes exhibit
pronounced peaks in their intensity distribution.  Although the
parameters of the non-resonant components are in principle nuisance
parameters, the choice of their parameterizations turns out to be one
of the dominant sources of systematic uncertainty.

A striking advantage of the resonance model employed here is its
simplicity, both from a conceptual and computational point of view.
The results can also be compared directly to previous analyses of
diffractive meson production that employed similar models.  Some of
the potential issues mentioned above are expected to be mitigated by
our approach of simultaneously fitting multiple $t'$~bins while
forcing the resonance parameters to be the same across all $t'$~bins.
This puts strong constraints on the resonant and non-resonant wave
components.  Additional constraints that help to reduce systematic
effects come from including as many decay chains of a resonance as
possible into the resonance-model fit and requiring the same resonance
parameters in these decay chains.  Fitting a large spin-density
submatrix with many waves with well-defined resonance signals has a
similar effect because the resonance-model fit takes into account the
whole spin-density submatrix with all its interference terms.  The
number of these interference terms grows with the number of selected
waves squared.  Thus the statistic in \cref{eq:fit_chi2} that is
employed to estimate the fit parameters effectively gives more weight
to the interference terms than to the intensities.  In practice, this
often helps to stabilize the fit as the model reproduces the phase
motions usually more accurately than the intensities.

In order to overcome the caveats and limitations discussed above,
improved models that respect the fundamental physical principles and
symmetries need to be applied in order to extract process-independent
resonance properties, \ie the resonance poles.  This will be described
in more detail in \cref{sec:pwa.unitary_model} below.  However, there
is an additional aspect that complicates the identification of
resonances. An observed intensity peak with an associated phase motion
of about $180^\circ$ does not uniquely define a resonance pole.  Under
special circumstances, scattering amplitudes may have singularities
that exhibit the same experimental signature without being related to
resonances.  A possible example is the \PaOne[1420] signal, which
could be the logarithmic singularity of a triangle diagram in a
rescattering process of the daughter
particles~\cite{Ketzer:2015tqa,Aceti:2016yeb}.  This will be discussed
further in \cref{sec:results_1pp}.

\subsubsection{Analytical Unitary Model}
\label{sec:pwa.unitary_model}

In principle, strong-interaction scattering and decay processes are
described by the relativistic $S$-matrix, which we discussed in
\cref{sec:scattering.s-matrix}.  As described there, the $S$-matrix is
constrained by unitarity, crossing symmetry, and analyticity.  As
shown in \cref{sec:scattering.analyticity}, the partial-wave
amplitudes must be analytic functions in the complex plane of the
Mandelstam variable~$s$ up to \one~branch cuts that are caused by the
opening of scattering channels with increasing~$s$ and \two~poles that
are caused either by bound states or by resonances.  The location of a
resonance pole in the complex $s$-plane and its residue represent the
actual universal resonance properties, which one wants to extract from
data.  However, the construction of amplitudes that describe
multi-body decays of hadrons and that fulfill all $S$-matrix
constraints is a formidable task and has so far only been achieved in
a few cases (see \eg\ \refsCite{Niecknig:2012sj,Niecknig:2015ija}).
In the past, two-body $K$-matrix approaches that respect at least
analyticity and two-body unitarity (\confer\
\cref{sec:scattering.k-matrix}) were applied in some analyses (see
\eg\ \refsCite{Daum:1980ay,Amelin:1995gu}).

A general framework for a systematic analysis of diffractive meson
production at COMPASS using reaction models based on the principles of
the relativistic $S$-matrix is currently under development in
collaboration with the Joint Physics Analysis Center
(JPAC)~\cite{Mikhasenko:2016mox,Jackura:2016llm,Mikhasenko:2017jtg,Mikhasenko:2018bzm}.
The goal is to extend the Breit--Wigner based fit models discussed in
\cref{sec:pwa.res_fit} to analytic models constrained by unitarity,
which allows the extraction of resonance pole positions.  In this
section, we will briefly sketch the model which was successfully
applied to the \etaPi and \etaPrPi COMPASS data.  The fit results and
the extracted pole positions for the \PpiOne, \PaTwo, and
\PaTwo[1700]~\cite{Jackura:2017amb,Rodas:2018owy} will be summarized
in \cref{sec:results_1mp,sec:results_2pp}.

\begin{figure}[t]
  \centering
  \subfloat[]{%
    \includegraphics[width=.4\columnwidth]{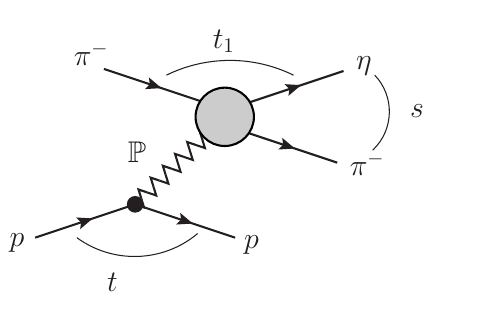}%
    \label{fig:etapi_reaction}%
  }%
  \\
  \subfloat[]{%
    \includegraphics[width=.7\columnwidth]{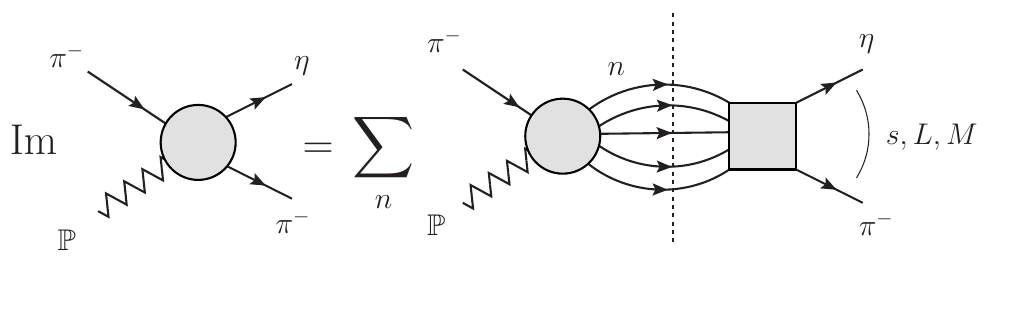}%
    \label{fig:etapi_unitarity}%
  }%
  \caption{\subfloatLabel{fig:etapi_reaction}~Diffractive-dissociation
    reaction $\pi^- + p \to \etaOrPrPi + p$ with $t$-channel Pomeron
    (\Pom) exchange.  \subfloatLabel{fig:etapi_unitarity}~The
    $\pi^- + \Pom \to \etaOrPrPi$ amplitudes are expanded in
    partial-wave amplitudes $a_{J, i}(s)$ in the $s$-channel of the
    \etaOrPrPi system with $J = L$ and $i = \{\etaPi, \etaPrPi\}$.
    Unitarity relates the imaginary part of the production
    partial-wave amplitude $a_{J, i}(s)$ (gray disc) to final-state
    interactions that include all kinematically allowed intermediate
    states~$n$, described by the scattering partial-wave amplitudes
    $t_{J, ij}(s)$ (gray box).  From \refCite{Jackura:2017amb},
    \confer also \cref{eq:pw_amp_prod.disc.diagram}.}
  \label{fig:etapi_diagram}
\end{figure}

As discussed in \cref{sec:exp.prod_reactions}, the diffractive
production of the \etaOrPrPi final state proceeds dominantly via
Pomeron (\Pom) exchange (see \cref{fig:etapi_reaction}).
In \cref{sec:regge}, we have shown that the concept of Pomeron
exchange emerges from Regge theory, which allows us to factorize the
$p$-\Pom vertex from the $\pi$-\Pom vertex. In the model, we hence
only consider the process $\pi + \Pom \to \etaOrPrPi$, which we
describe (\confer\ \cref{sec:scattering.fsi}) by $2 \to 2$
partial-wave production amplitudes $a_{J, i}(s)$, with
$i = \{\etaPi, \etaPrPi\}$, $J$~the angular momentum of the \etaOrPrPi
system, and $s$~the \etaOrPrPi mass squared.  Note that this notation
of invariants differs from the one introduced in
\cref{sec:pwa.analysis_model}.  In principle, the
amplitudes~$a_{J, i}$ also depend on the Pomeron helicity and
virtuality~$t$.  Due to the steeply falling $t'$~spectrum in the
analyzed range from \SIrange{0.1}{1.0}{\GeVcsq}, we use an effective
value $t'_\text{eff} = \SI{0.1}{\GeVcsq}$.  The Pomeron is modeled as
an effective exchange particle with spin~1, in accordance with the
dominance of $M = 1$ partial waves in the data and the approximate
constancy of hadron cross sections at high energies (\confer\
\cref{fig:chew_frautschi} and \cref{sec:regge}).

According to \cref{eq:pw_amp_prod.unitarity}, unitarity relates the
production partial-wave amplitudes $a_{J, i}(s)$ to the scattering
partial-wave amplitudes $t_{J, ij}(s)$ that describe the rescattering
of the final-state particles, as displayed graphically in
\cref{fig:etapi_unitarity}.  Unitarity also constrains the scattering
partial-wave amplitudes, as given by \cref{eq:pw_amp.unitarity}, so we
have
\begin{align}
  \label{eq:etapi.pw_amp.unitarity}
  \Im a_{J, i}(s)
  = \sum_k \rho_k\, t_{J, ik}^\ast(s)\, a_{J, k}(s)
  \quad\text{and}\quad
  \Im t_{J, ij}(s)
  = \sum_k \rho_k\, t_{J, ik}^\ast(s)\, t_{J, kj}(s)\eqPunctSpacing.
\end{align}

It is known that the barrier factors (\confer\ \cref{eq:pw_amp.thr})
produce kinematic singularities of the
amplitudes~\cite{Martin:1970xx,Mikhasenko:2017rkh,Pilloni:2018kwm}. We
therefore define the reduced partial-wave amplitudes~$\hat{a}_J$
and~$\hat{t}_J$ by explicitly factoring out the barrier factors:
\begin{align}
  a_{J, i}
  \eqqcolon p^{J-1}\, q_i^J\, \hat{a}_{J, i}
  \quad\text{and}\quad
  t_{J, ij}
  \eqqcolon q_i^{J}\, q_j^{J}\, \hat{t}_{J, ij}\eqPunctSpacing,
\end{align}
with $p$~being the momentum of the beam~$\pi$ in the $\pi$-$\Pom$
center-of-momentum frame and $q_i$~the breakup momentum of the \etaPi
or \etaPrPi system.  In~$a_{J, i}$, the barrier factor is one unit
less for the incoming momentum~$p$ because of the effective
vector-nature of the Pomeron.  The reduced amplitudes are
parameterized using the $N$-over-$D$ method described in
\cref{sec:scattering.n_over_d}, \cref{eq:n_over_d}:
\begin{align}
  \label{eq:etapi.n_over_d}
  \hat{t}_{J, ij}(s)
  = \sum_k N_{J, ik}(s)\, D_{J, kj}^{-1}(s)
  \quad\text{and}\quad
  \hat{a}_{J, i}(s)
  = \sum_k n_{J, k}(s)\, D_{J, ki}^{-1}(s)\eqPunctSpacing.
\end{align}
The matrix $D_J(s)$ has only the right-hand cuts due to unitarity and
is assumed to be universal, while $N_J(s)$ and $n_J(s)$ have only
left-hand cuts, which depend on the exchange processes.  They are
assumed to be smooth functions of~$s$.  As was shown in
\cref{sec:scattering.n_over_d}, unitarity leads to a relation
between~$D_J$ and~$N_J$, \ie $\Im D_J(s) = -\rho(s)\, N_J(s)$, which
can be solved by the once-subtracted dispersion integral
\cref{eq:n_over_d.dispersion_integral_d}:
\begin{equation}
  \label{eq:etapi_dispersion_integral.d}
  D_{J, ij}(s)
  = D_{J, 0, ij}(s) - \frac{s}{\pi} \int\limits_{s_i}^\infty\! \dif{s'}\,
  \frac{\rho_{ij}(s')\, N_{J, ij}(s')}{s'\, (s' - s)}\eqPunctSpacing,
\end{equation}
with $s_i$~being the threshold for channel~$i$.  The subtraction
constant is parameterized as an inverse $K$-matrix
$(K_{J, ij}^{-1}(s))$.  The numerator functions $\rho(s)\, N_J(s)$ and
$n_J(s)$ are parameterized by smooth polynomials with poles to
represent the left-hand cuts (see
\refsCite{Jackura:2017amb,Rodas:2018owy} for details).

\Cref{fig:int_eta_p_wave_fit,fig:int_etaprime_p_wave_fit,fig:phase_eta_p_d_wave_fit,fig:phase_etaprime_p_d_wave_fit}
in \cref{sec:results_1mp} show the result of a coupled-channel fit
employing the unitary model described above to the COMPASS \etaPi and
\etaPrPi data.  For the $D$~wave, two poles in the $K$-matrix
for~$D_{J, 0, ij}(s)$ are needed to describe the data.  These poles
correspond to the \PaTwo and the \PaTwo[1700].  In the $P$~wave, in
contrast, it turns out that both the structure around \SI{1.4}{\GeVcc}
in the \etaPi final state and the one around \SI{1.6}{\GeVcc} in the
\etaPrPi final state can be described by a single pole slightly below
\SI{1.6}{\GeVcc}.  The analysis thus confirms the \PpiOne[1600], while
it puts strong doubts on the existence of the \PpiOne[1400].  For a
comparison of the result from the pole search with the ones from the
Breit--Wigner analyses, see the ideograms in
\cref{fig:ideogram_pi1_1600,fig:ideogram_a2_1320,fig:ideogram_a2_1700}
in \cref{sec:results_by_qn}.
\clearpage{}%
\clearpage{}%
\section{Results from Pion-Diffraction Data}
\label{sec:results}

Since the strong interaction conserves isospin and $G$~parity,
intermediate states~$X$ that are produced in diffractive dissociation
of pions are restricted to $\IG = 1^-$ quantum numbers, \ie only
resonances of the~$a_J$ and $\pi_J$~families with spin~$J$ can be
produced.  \Cref{tab:PDG_mesons_2018} in \cref{sec:res_par_aJ_piJ}
lists the known~$a_J$ and $\pi_J$~states and their parameters in the
mass region below \SI{2.2}{\GeVcc} according to the
PDG~\cite{Tanabashi:2018zz}.

In this section, we will present results of those COMPASS analyses
that have been published so far.  We will briefly characterize the
analyzed data samples in \cref{sec:result_data_sets} and describe the
models employed for the partial-wave decomposition and the
resonance-model fits in \cref{sec:analysis_models}.  In
\cref{sec:results_by_qn}, we will discuss the results from the
analyses organized by the \JPC quantum numbers of the intermediate
state~$X$.

\subsection{COMPASS Data Samples}
\label{sec:result_data_sets}

The COMPASS collaboration has so far published results from the
analysis of four data samples taken with a \SI{190}{\GeVc}
$\pi^-$~beam in the kinematic range
\SIvalRange{0.1}{t'}{1.0}{\GeVcsq}: \etaPim and \etaPrPim
diffractively produced on a proton target~\cite{Adolph:2014rpp} and
\threePi diffractively produced on a solid-lead~\cite{Alekseev:2009aa}
and a proton
target~\cite{Adolph:2015pws,Adolph:2015tqa,Akhunzyanov:2018lqa}.

\subsubsection{The \etaPim and \etaPrPim Data Samples}
\label{sec:data_set_etapi}

\begin{figure}[!b]
  \centering
  \hfill%
  \subfloat[]{%
    \includegraphics[width=\twoPlotWidth]{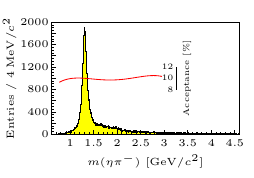}%
    \label{fig:mass_eta_pim}%
  }%
  \hfill%
  \subfloat[]{%
    \includegraphics[width=\twoPlotWidth]{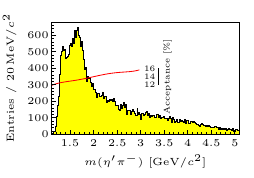}%
    \label{fig:mass_etaprime_pim}%
  }%
  \hfill\null%
  \caption{Invariant mass spectra (not acceptance corrected) of
    diffractively produced \subfloatLabel{fig:mass_eta_pim}~\etaPim
    and
    \subfloatLabel{fig:mass_etaprime_pim}~\etaPrPim~\cite{Adolph:2014rpp}.
    The continuous red curves show the acceptances in the analyzed
    kinematic ranges.}
  \label{fig:mass_eta_etaprime_pim}
\end{figure}

For the \etaPim data sample, the~$\eta$ is reconstructed via its decay
$\eta \to \pi^-\pi^+\pi^0$ with $\pi^0 \to \gamma \gamma$.
The~$\eta'$ in the \etaPrPim data sample is reconstructed via its
decay $\eta' \to \pi^- \pi^+ \eta$ with $\eta \to \gamma \gamma$.
Therefore, exclusive \etaPim and \etaPrPim events are both selected
from the same final state $\threePi \gamma \gamma$ (see
\refCite{Adolph:2014rpp} and \cref{sec:compass.select}).  A kinematic
fit is applied to the $\gamma \gamma$ system constraining its mass to
the nominal mass of $\pi^0$ and $\eta$, respectively.  This yields
sharp $\eta$~and $\eta'$~peaks with widths of \SIrange{3}{4}{\MeVcc}
in the invariant mass distributions of the $\twoPi \gamma \gamma$
system.  Applying cuts of $\pm \SI{10}{\MeVcc}$ around the nominal
$\eta$~and $\eta'$~mass values, we obtain \num{116000}~\etaPim and
\num{39000} \etaPrPim events in the analyzed range
\SIvalRange{0.1}{t'}{1.0}{\GeVcsq}.  The \etaPim and \etaPrPim
invariant mass spectra for these events are shown in
\cref{fig:mass_eta_etaprime_pim}.
The $m_\etaPi$~spectrum is dominated by a narrow peak at
\SI{1.3}{\GeVcc}, which is the \PaTwo (see \cref{sec:results_2pp}).
This peak appears much weaker in the $m_\etaPrPi$~spectrum, which
instead is dominated by a broad peak at \SI{1.7}{\GeVcc} that is
related to the \PpiOne[1600] (see \cref{sec:results_1mp}).

The overall $t'$~distributions are well described by a simple
exponential model of the form (\confer
\cref{eq:dsigma_dt.regge.2,eq:t_spectrum_model})
\begin{equation}
  \label{eq:tspectrum_etapi}
  \frac{\dif{N}}{\dif{t'}}
  \propto t'\, e^{-b\, t'}\eqPunctSpacing,
\end{equation}
with slope parameters of $b = \SI{8.45}{\perGeVcsq}$ for the \etaPim
final state and of $b = \SI{8.0}{\perGeVcsq}$ for the \etaPrPim final
state.  In both cases, we observe only a weak mass dependence of~$b$.

\subsubsection{The \threePi Data Samples}
\label{sec:data_set_3pi}

\begin{figure}[tbp]
  \centering
  \hfill%
  \subfloat[]{%
    \includegraphics[width=0.38\textwidth]{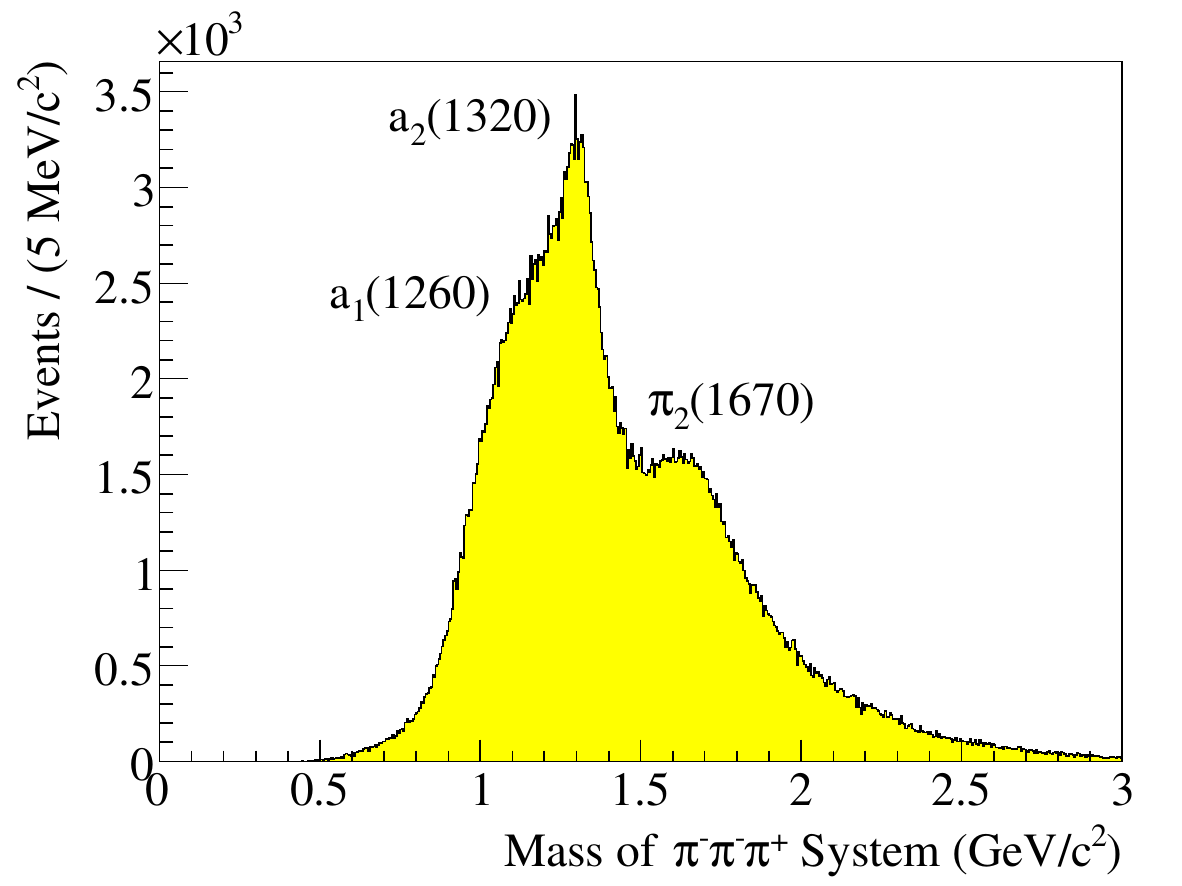}%
    \label{fig:mass_3pi_lead}%
  }%
  \hfill%
  \subfloat[]{%
    \includegraphics[width=\threePlotWidth]{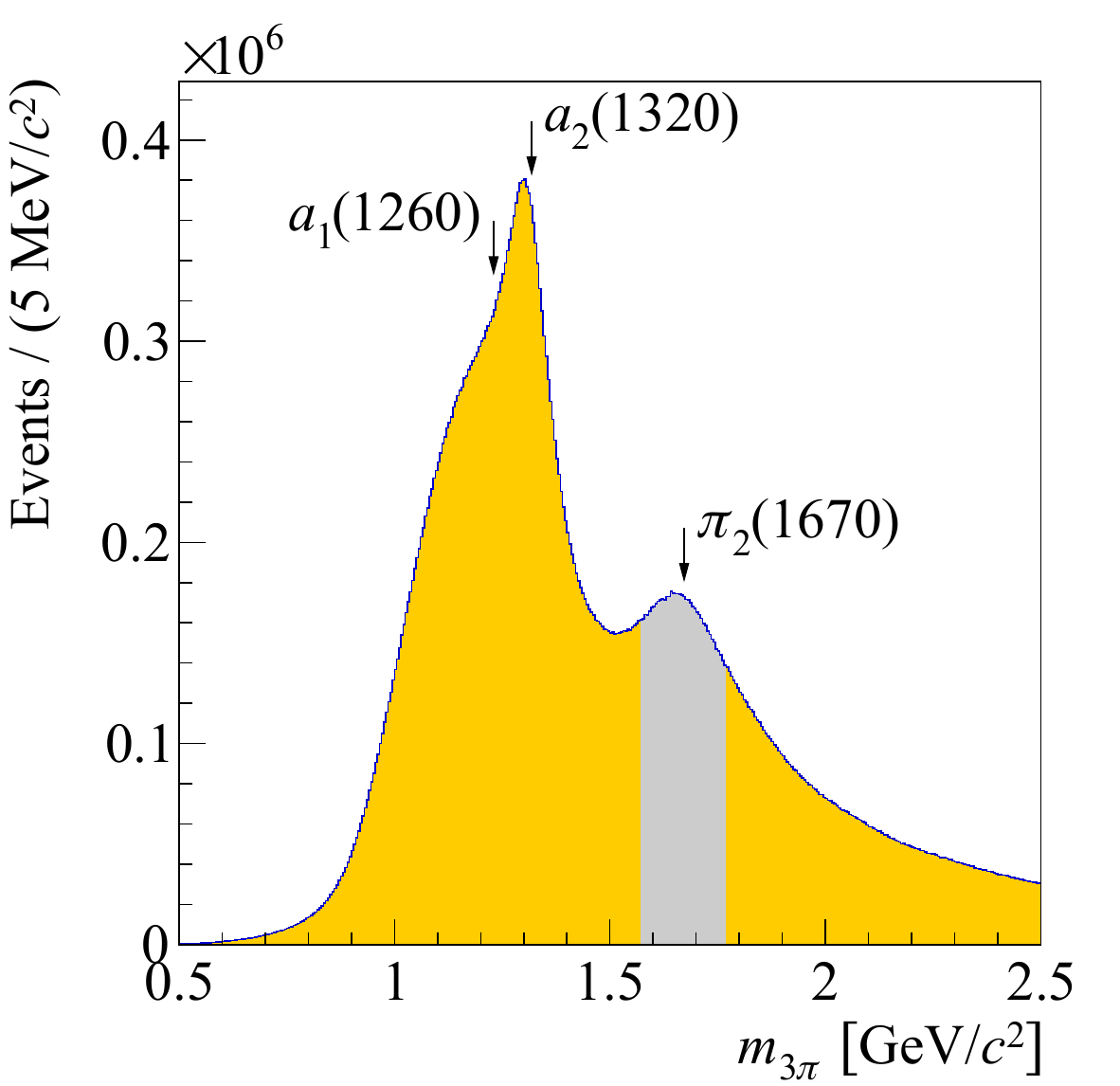}%
    \label{fig:mass_3pi_proton}%
  }%
  \hfill\null%
  \caption{Invariant mass spectra (not acceptance corrected) of
    diffractively produced \threePi using
    \subfloatLabel{fig:mass_3pi_lead}~a solid-lead
    target~\cite{Alekseev:2009aa} and
    \subfloatLabel{fig:mass_3pi_proton}~a liquid-hydrogen
    target~\cite{Adolph:2015tqa}.  The shaded region
    in~\subfloatLabel{fig:mass_3pi_proton} indicates the
    \mThreePi~range used for the Dalitz plot in
    \cref{fig:dalitz_plot_pi2}.}
  \label{fig:mass_3pi}
\end{figure}

\begin{figure}[tbp]
  \centering
  \includegraphics[width=\threePlotWidth]{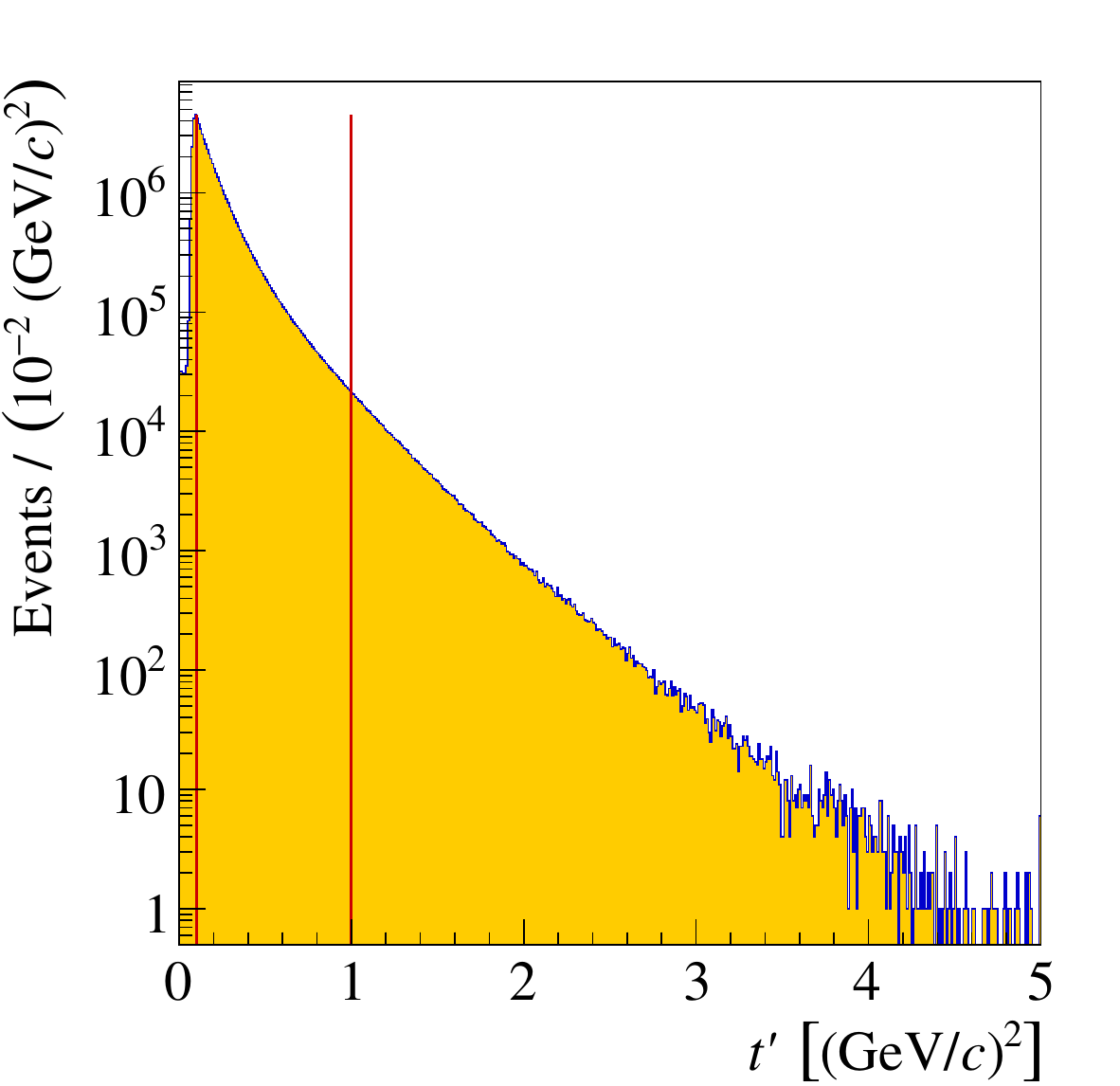}
  \caption{Distribution of the squared four-momentum transfer~$t'$
    (not acceptance corrected) of diffractively produced \threePi
    using a liquid-hydrogen target~\cite{Adolph:2015tqa}.  The
    vertical lines indicate the $t'$~range used in the PWA.}
  \label{fig:tprime}
\end{figure}

The \threePi data samples are substantially larger than the
\etaOrPrPim data samples.  After event selection (see
\refsCite{Alekseev:2009aa,Adolph:2015tqa} and
\cref{sec:compass.select}), we obtain \num{420000} exclusive events
for the solid-lead target and \num{46e6} exclusive events for the
liquid-hydrogen target in the analyzed three-pion mass range
\SIvalRange{0.5}{\mThreePi}{2.5}{\GeVcc}.  For the lead-target data,
the target recoil was not measured so that the contribution of
non-exclusive backgrounds is larger in this sample.  In the analyzed
range \SIvalRange{0.1}{t'}{1.0}{\GeVcsq}, the scattering off the lead
nucleus is predominantly of incoherent nature, \ie the beam pions
scatter off individual nucleons within the lead nucleus.  Thus the
lead-target data are expected to be similar to the proton-target data.
\Cref{fig:mass_3pi} shows the \mThreePi~distributions for the two data
samples.  They exhibit pronounced structures that correspond to the
well known resonances \PaOne, \PaTwo, and \PpiTwo (see
\cref{sec:results_1pp,sec:results_2pp,sec:results_2mp}).  As shown in
\cref{fig:tprime}, the $t'$~distribution follows an approximately
exponential behavior in the analyzed $t'$~range.

\subsection{Analysis Models}
\label{sec:analysis_models}

All three final states considered here (see
\cref{sec:result_data_sets}) were already studied by previous
experiments using similar scattering reactions with lower pion-beam
energies.  Since some of the COMPASS data samples, in particular the
\threePi proton-target sample, are significantly larger than those
obtained by previous experiments, we extend and refine the analysis
models used in previous analyses.

\subsubsection{\etaOrPrPim Partial-Wave Decomposition}
\label{sec:etapi_model:pwa}

Exclusive diffractive production of \etaPi and \etaPrPi final states
was studied by the VES experiment using a \SI{37}{\GeVc} $\pi^-$~beam
scattering off a solid-beryllium
target~\cite{Beladidze:1993km,Dorofeev:2001xu} and by the BNL E852
experiment using an \SI{18}{\GeVc} $\pi^-$~beam on a liquid-hydrogen
target~\cite{Chung:1999we,Ivanov:2001rv}.  At COMPASS, the higher beam
energy enables us to access higher invariant masses of the produced
final states and higher values of the orbital angular momentum between
the two final-state particles.

For final states consisting of two pseudoscalars, the total spin~$J$
of the intermediate state~$X$ is identical to the relative orbital
angular momentum~$L$ between the two daughter particles.  The
parity~$P$ of the state is given by $(-1)^L$.  Therefore, only states
with positive naturality (see \cref{sec:pheno.qm.quantum-numbers}) are
accessible.  In diffractive reactions, the isospin~$I$ and the
$C$~parity of~$X$ are given by the isospin and $C$~parity of the beam
particle, \ie $I = 1$ and $C = +1$ for a pion beam.  This means that
partial waves with odd~$L$ correspond to spin-exotic quantum numbers.
In the reflectivity basis (see \cref{sec:pwa_cells.reflectivity}), the
combination $\Mrefl = 0^+$ of the spin-projection quantum number~$M$
and the reflectivity~\refl is forbidden for natural-parity states.
Thus $\JPC = 0^{++}$ waves, which may contain~\PaZero* resonances, may
only appear with $\Mrefl = 0^-$ quantum numbers that correspond to
unnatural-parity exchange.  At high beam energies, Pomeron exchange
dominates and unnatural-parity exchange processes are suppressed.
Therefore, $\Mrefl = 1^+$ waves are expected to dominate and
$\refl = -1$ waves should be suppressed.  Often, the spectroscopic
notation $LM_\refl$ is used to designate two-pseudoscalar partial
waves.  \Eg, partial waves with $\JPCMrefl = 0^{++}0^-$, $1^{-+}0^-$,
$1^{-+}1^-$, $1^{-+}1^+$, and $2^{++}2^+$ quantum numbers correspond
to $S0_{-}$, $P0_{-}$, $P1_{-}$, $P1_{+}$, and $D2_{+}$, respectively.

The rough partial-wave structure of the data is already visible in the
angular distributions in the Gottfried--Jackson frame, where we use the
direction of the \etaOrPr as the analyzer (see
\cref{sec:pwa.analysis_model.coordsys}).  As already observed by
previous
experiments~\cite{Beladidze:1993km,Chung:1999we,Ivanov:2001rv,Dorofeev:2001xu},
the \phiGJ distributions of both final states follow an approximate
$\sin^2\phiGJ$ pattern~\cite{Schluter:2012mep,Schluter:2014tja} over
the full analyzed mass range.  This corresponds to the dominance of
$\Mrefl = 1^+$ waves (see also discussion below), which is consistent
with the fact that the overall $t'$~spectrum is described by
\cref{eq:tspectrum_etapi}, which includes an explicit $t'$~factor
(\confer also \cref{eq:dsigma_dt.regge.2,eq:t_spectrum_model}).

\begin{figure}[tbp]
  \centering
  \hfill%
  \subfloat[]{%
    \includegraphics[width=\twoPlotWidth]{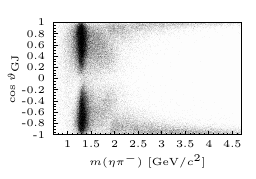}%
    \label{fig:massVsCosTh_eta_pim}%
  }%
  \hfill%
  \subfloat[]{%
    \includegraphics[width=\twoPlotWidth]{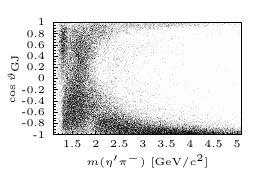}%
    \label{fig:massVsCosTh_etaprime_pim}%
  }%
  \hfill\null%
  \caption{Distribution of the cosine of the polar angle of the
    \etaOrPr in the Gottfried--Jackson frame (not acceptance corrected;
    see \cref{sec:pwa.analysis_model.coordsys}) as a function of the
    invariant mass of \subfloatLabel{fig:massVsCosTh_eta_pim}~the
    \etaPim and \subfloatLabel{fig:massVsCosTh_etaprime_pim}~the
    \etaPrPim system~\cite{Adolph:2014rpp}.  Here, $\cosThetaGJ = 1$
    corresponds to \etaOrPr emission along the beam direction.}
  \label{fig:massVsCosTh}
\end{figure}

\begin{figure}[tbp]
  \centering
  \hfill%
  \subfloat[]{%
    \includegraphics[width=\threePlotWidth]{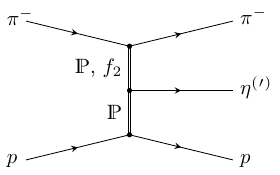}%
    \label{fig:eta_etaprime_multi_regge_backward}%
  }%
  \hfill%
  \subfloat[]{%
    \includegraphics[width=\threePlotWidth]{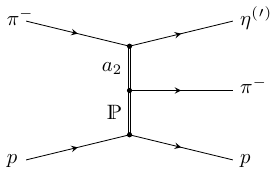}%
    \label{fig:eta_etaprime_multi_regge_forward}%
  }%
  \hfill\null%
  \caption{Possible diagrams for non-resonant production of
    \etaOrPrPim.
    In~\subfloatLabel{fig:eta_etaprime_multi_regge_backward}, a
    backward \etaOrPr (opposite to the beam direction in the
    \etaOrPrPi center-of-momentum frame) is produced by Pomeron or
    $f_2$~exchange;
    in~\subfloatLabel{fig:eta_etaprime_multi_regge_forward}, a forward
    \etaOrPr (along the beam direction) is produced by
    $a_2$~exchange.}
  \label{fig:eta_etaprime_multi_regge}
\end{figure}

The dominant spin content of the data is visible in
\cref{fig:massVsCosTh}, which shows the correlation of \cosThetaGJ of
the \etaOrPr with the \etaOrPrPim invariant mass.  In the
\SI{1.3}{\GeVcc} mass region of the \etaPim data, the
\cosThetaGJ~distribution exhibits a two-bump structure that is
characteristic for a $D$~wave.  This is consistent with the dominant
\PaTwo signal (see also \cref{fig:mass_eta_pim}).  In contrast, the
modulation of the angular distribution is weaker in the \etaPrPim data
as is the \PaTwo signal.  In the \SI{2}{\GeVcc} mass region of the
\etaPim data, the \cosThetaGJ distribution exhibits four bumps, which
indicates the presence of a $G$~wave.  This structure is disentangled
by the PWA and explained as the \PaFour (see \cref{sec:results_4pp}).
Again, this structure is much weaker in the \etaPrPim data.  For
masses above \SI{2}{\GeVcc} the character of the \cosThetaGJ
distributions changes drastically.  For both final states, the
distributions show narrow peaks at $\cosThetaGJ = \pm 1$, which
correspond to \etaOrPr going along the beam direction in the
Gottfried--Jackson frame, \ie forward, or against it, \ie backward.
The peaks become sharper with increasing mass.  This behavior
corresponds to a rapidity gap that widens with increasing mass and
indicates the presence of large contributions from non-resonant
processes.  Clearly, high-spin partial waves are needed to describe
such \cosThetaGJ distributions.  The non-resonant processes are
dominated by double-Regge exchanges (see
\cref{sec:exp.prod_reactions,sec:pwa_cells.discussion,fig:exp.production.central,fig:exp.production.deck}).
\Cref{fig:eta_etaprime_multi_regge} shows possible double-Regge
exchange diagrams for the \etaOrPrPim final state.  These processes
are similar to the well-known non-resonant Deck process in diffractive
$3\pi$~production~\cite{Deck:1964hm,Ascoli:1974hi} and to central
production of \etaOrPr in $pp$~collisions as, for example, observed by
the WA102 experiment~\cite{Barberis:1998ax}.  These irreducible
background processes typically have one leading final-state particle.
For a two-body final state, this means that the other final-state
particle is relatively slow, leading to forward--backward peaks in the
\cosThetaGJ distribution that become sharper with increasing mass of
the two-body system.

In the high-mass region, the \cosThetaGJ distributions of both systems
show a significant forward--backward asymmetry, \ie we observe an
excess of \etaOrPr in the backward direction.  This asymmetry is not
caused by detector effects.  In the \etaPrPim data, the backward
excess is more pronounced over the whole mass range.  Since even-$L$
waves have only symmetric \cosThetaGJ distributions, the
forward--backward asymmetry indicates the presence of spin-exotic
odd-$L$ waves, predominantly those with lowest possible~$L$, \ie
$P$~waves.  In case of \etaPrPim, contributions of odd-$L$ partial
waves that are comparable in strength to those of the even-$L$ partial
waves are needed in order to describe the backward enhancement.  In
the \etaPim data, the \cosThetaGJ asymmetry flips from a backward
$\eta$~excess at high masses to a forward excess in the \PaTwo mass
region.  In the \etaPrPim data, this effect is partly masked by the
broad enhancement around \SI{1.6}{\GeVcc} that is caused by the
$P$~wave, which is much stronger in the \etaPrPim channel as compared
to \etaPim.

The PWA model used for the partial-wave decomposition is based on
\cref{eq:intensity_model_final}.  The two-body decay amplitudes
$\Psi_{LM}^\refl(\thetaGJ, \phiGJ)$ in the reflectivity basis are
given by
\cref{eq:decay_amp_2body_2,eq:decay_amp_redef,eq:D_func_refl,eq:decay_amp_norm}.
In order to account for non-\etaOrPr incoherent backgrounds, we extend
the two-body PWA model to a four-body model for the
$\threePi (\gamma \gamma)$ system, where the $(\gamma \gamma)$
subsystem is either a~$\pi^0$ in case of~\etaPim or an~$\eta$ in case
of~\etaPrPim.  Instead of selecting a narrow $\twoPi \gamma \gamma$
mass window around the nominal~$\etaOrPr$~mass, the analysis is
performed requiring at least one $\twoPi \gamma \gamma$ combination
inside a wide $m_{\twoPi \gamma \gamma}$ interval of
$\pm \SI{25}{\MeVcc}$ around the nominal \etaOrPr mass.  In order to
take into account the three-body decay of \etaOrPr, we extend the
two-body decay amplitudes that are given by the spherical harmonics by
a factor~$f_{\etaOrPr}$ that describes the measured~\etaOrPr peak
shape and the Dalitz-plot distribution of the three-body
\etaOrPr~decay.\footnote{The angular-momentum barrier factor
  $F_J(m_{\etaOrPrPi})$ in \cref{eq:decay_amp_2body_2} depends in
  principle also on the reconstructed \etaOrPr mass via the breakup
  momentum that appears in \cref{eq:bw_factor}.  However, since the
  width of the \etaOrPr mass peak is given by the detector resolution,
  the barrier factor is set to a constant for a given
  $m_{\etaOrPrPi}$~bin and is hence absorbed by the normalization in
  \cref{eq:decay_amp_norm}.}  The latter was taken from
\refsCite{Ambrosino:2008ht,Dorofeev:2006fb}.  The \etaOrPr peak shape
is determined by the detector resolution and is hence assumed to be a
Gaussian.  The decay amplitudes depend on the four-body phase-space
variables~$\tau_4$ and are given by
\begin{equation}
  \label{eq:decay_amp_eta_etaprime_pi}
  \hspace*{-8em}
  \Psi^\refl_{LM}(\tau_4; m_{\etaOrPrPi})
  = f_{\etaOrPr}(p_{\pi^-}, p_{\pi^+}, p_{\gamma\gamma})\,
  \Underbrace{\Big[ Y_L^M(\thetaGJ, \phiGJ) - \refl\, Y_L^{M \text{*}}(\thetaGJ, \phiGJ) \Big]}%
  {\hspace*{-5em}\mathrlap{\displaystyle{%
        \propto P_{LM}(\cosThetaGJ) \times
        \begin{cases}
          \sin(M\, \phiGJ) & \text{if $\refl = +1$}\eqPunctSpacing, \\
          \cos(M\, \phiGJ) & \text{if $\refl = -1$}\eqPunctSpacing.
        \end{cases}}}}\eqPunctSpacing.
\end{equation}
Here, $p_{\pi^\pm}$ and $p_{\gamma\gamma}$ are the four-momenta of the
three \etaOrPr decay daughters and the $P_{LM}(\cosThetaGJ)$ are the
associated Legendre polynomials.  From
\cref{eq:decay_amp_eta_etaprime_pi} it now clear, that the
above-mentioned observed $\sin^2\phiGJ$ distribution is in agreement
with the dominance of $\Mrefl = 1^+$ waves.

The decay amplitude $\Psi_\text{flat}(\tau_4)$ of the flat wave (see
\cref{sec:pwa_cells.flat}) is isotropic in the
$\threePi (\gamma \gamma)$ four-body phase space.  It hence absorbs
the incoherent non-\etaOrPr backgrounds.  We account for the
combinatorial background due to the two possible
$\twoPi \gamma \gamma$ combinations by summing the total intensities
that correspond to the two combinations in the PWA
model.\footnote{This approach differs from the Bose symmetrization
  discussed in \cref{sec:pwa.analysis_model.symmetrization} in that we
  do not allow for self-interference.  We add the intensities because
  $\eta$ and $\eta'$ have very small widths so that the two
  $\twoPi \gamma \gamma$ combinations are in principle
  distinguishable.}

Most previous partial-wave analyses used PWA models that contained
waves only up to $L = 2$, except the one in \refCite{Ivanov:2001rv}
where in addition a $L = 4$ wave was introduced.  The PWA model for
the COMPASS data contains all positive-reflectivity waves with $L = 1$
to~6 and $M = 1$.  For the \etaPi data, we include in addition a wave
with $L = 2$ and $\Mrefl = 2^+$.  We also include the $L = 0$ wave
with $\Mrefl = 0^-$ for both channels.  Since this is the only
negative-reflectivity wave in the model, its amplitude is added
incoherently to the intensity.  Our four-body PWA approach is able to
separate the incoherent non-\etaOrPr backgrounds, which are absorbed
by the flat wave, from the incoherent negative-reflectivity
\etaOrPrPim contributions.  The PWA fits require only a weak $S$-wave
amplitude with $\refl = -1$.  This wave contributes \SI{0.5}{\percent}
to the total \etaPim intensity and \SI{1.1}{\percent} to the total
\etaPrPim intensity and contains mainly incoherent \etaOrPrPim
background.

One of the crucial assumptions of our PWA model is that it uses a
rank-1 spin-density matrix, \ie we assume full coherence of the
partial-wave amplitudes.  As was discussed in
\cref{sec:pwa_cells.rank}, incoherence of partial waves may arise from
contributions with and without proton helicity flip, or from different
$t'$~dependences of the amplitudes over the broad analyzed $t'$~range.
It is known that the spin-density matrix elements may have
significantly different $t'$~dependences as is, for example, the case
for the \threePi proton-target data. The \etaOrPrPim data are
dominated by $M = 1$ amplitudes with similar
$t'$~dependences~\cite{Schluter:2012mep}.  Since waves with $M = 0$
and~2 are strongly suppressed, integration over~$t'$ does not
introduce large incoherences.

The PWA of two-pseudoscalar final states may suffer from discrete
mathematical
ambiguities~\cite{Martin:1978jn,Sadovsky:1991hm,Chung:1997qd}, which
can be expressed in terms of so-called
Barrelet-zeros~\cite{Barrelet:1971pw}.  For our \etaOrPrPim data, the
ambiguities are drastically reduced, because we do not observe any
significant $\refl = -1$ contributions except for the $S$~wave.  In
the case of \etaPim, no ambiguities appear because the PWA model
contains the $D$~wave with $\Mrefl = 2^+$.  For \etaPrPim, ambiguities
occur when we extend the PWA model beyond the dominant $L = 1$, 2, and
4 waves.  We resolve them by requiring continuous behavior of the
dominant partial-wave amplitudes and of the Barrelet zeros as a
function of $m_{\etaPrPi}$.  From the acceptable solutions, which all
agree within statistical uncertainties, we select the one with the
smallest $L = 3$ contribution.

We performed the partial-wave decomposition using the model described
above in \SI{40}{\MeVcc} wide bins of the \etaOrPrPim mass from
threshold up to \SI{3}{\GeVcc}.  The spin-exotic waves with higher
orbital angular momenta of $L = 3$ and~5 have small but statistically
significant intensities in both final states.  This is also true for
the $L = 6$ wave, which has conventional \JPC quantum numbers.  These
three waves were never included in previous analyses.  Each of them
contributes less than \SI{1}{\percent} to the total \etaPim and
\etaPrPim intensities.

The results from our PWA of the \etaOrPrPim data show that the
relative phases of all partial waves change only little with mass for
$m_{\etaOrPrPi} \gtrsim \SI{2.2}{\GeVcc}$.  In addition, the relative
phases between the waves with spin projection $M = 1$ are compatible
with \SI{0}{\degree} in both final states.  This is consistent with
the results of the partial-wave decomposition of model amplitudes for
double-Regge exchange processes, where the relative phases between the
partial waves are found to depend only weakly on the mass of the
produced hadronic final state and are often close to
\SI{0}{\degree}.\footnote{For multi-body final states such as
  \threePi, these relative phases may be \SI{0}{\degree} or
  \SI{180}{\degree} depending on the choice of the analyzer in the
  calculation of the decay amplitudes (see
  \cref{sec:pwa.analysis_model.coordsys}).}

In order to compare the strengths of the partial waves in the \etaPim
and \etaPrPim final states, we scale the intensities of the \etaPim
waves in each $m_{\etaPi}$~bin by the relative kinematic factor
\begin{equation}
  \label{eq:etaprime_eta_r}
  c(m_{\etaPi}; L)
  = B \cdot
  \bigg[ \frac{q(m_{\etaPi}; m_{\eta'}, m_\pi)\hfill}{q(m_{\etaPi}; m_\eta, m_\pi)} \bigg]^{2L + 1}\eqPunctSpacing.
\end{equation}
Here, $q$~is the two-body breakup momentum as given by
\cref{eq:breakup_mom.cms.f,eq:kaellen} and $m_{\etaOrPr}$~and
$m_\pi$~are the nominal masses of~\etaOrPr and $\pi^-$, respectively.
The kinematic factor takes into account the different phase-space and
angular-momentum barrier factors of the two final states and the
branching-fraction ratio of $B = 0.746$ for~$\eta$ and
$\eta'$~decaying into $\twoPi \gamma \gamma$~\cite{pdg:2013}.  In
\cref{eq:etaprime_eta_r}, we have used the fact that the transition
rate for the two-body decay of a point-like particle is expected to be
proportional to
$q^{2L +
  1}$~\cite{Peters:1995jv,Abele:1997dz,Bramon:1997va}.\footnote{This
  corresponds to the simplest possible expression for the
  centrifugal-barrier factor (see \cref{eq:pw_amp.thr}) and is
  consistent with the decay amplitude in \cref{eq:decay_amp_2body_2}
  being proportional to the barrier factor~$F_J$. Hence for $q \to 0$,
  the decay amplitude is proportional to $q^L$.}

A feature of our data, which will be discussed further in
\cref{sec:results_1mp,sec:results_2pp,sec:results_4pp}, is that the
intensity distributions of the even-$L$ \etaPim waves with $L = 2$
and~4 that are scaled by \cref{eq:etaprime_eta_r} are in astonishing
agreement with the corresponding intensity distributions of the
\etaPrPim waves (see \cref{sec:results_2pp,sec:results_4pp}).  This
means that the apparent differences of the unscaled intensity
distributions are of purely kinematical origin.  In contrast, a strong
enhancement of the \etaPrPim over the scaled \etaPim partial-wave
intensities is observed for the spin-exotic odd-$L$ partial waves, in
particular for the $L = 1$ wave (see \cref{sec:results_1mp}).  Since
the amplitudes of the high-spin waves with $L = 3$, 5, and~6 have
large statistical uncertainties, they are not shown here.  However,
they follow the same trend as the lower-spin even-$L$ and odd-$L$
waves.

\subsubsection{\etaOrPrPim Resonance-Model Fits}
\label{sec:etapi_model:resonance}

The spin-density matrices (see \cref{eq:spin-dens_norm}), which have
been extracted from the \etaPim and \etaPrPim data in the partial-wave
decomposition stage described in \cref{sec:etapi_model:pwa}, serve as
input for the resonance-model fit.  As was discussed in
\cref{sec:pwa.analysis_model.extension}, for this fit we select a
subset of waves.  This set includes the $D$- and $G$-wave amplitudes
with $\Mrefl = 1^+$.  For the \etaPim final state, we include in
addition the $D$-wave amplitude with $\Mrefl = 2^+$.  These amplitudes
exhibit clear resonance signals, \ie intensity peaks with associated
phase motions, of the well-known \PaTwo and \PaFour.  We include in
addition the spin-exotic $P$-wave amplitude with $\Mrefl = 1^+$.  The
resonance content of this wave is discussed controversially (see
\cref{sec:results_1mp}).

The $D$-wave amplitudes are described by a coherent sum of
Breit--Wigner amplitudes for the \PaTwo and the \PaTwo[1700], the
$G$-wave amplitudes by an \PaFour Breit--Wigner amplitude, and the
$P$-wave amplitudes by a \PpiOne* Breit--Wigner amplitude.  Except for
the \PaTwo, we use constant-width relativistic Breit--Wigner
amplitudes, \ie\ \cref{eq:BW_const_width}.  The parameterization of
the \PaTwo dynamic total width assumes that the total width is
saturated by the two dominant decay modes $\Pprho \pi$ and
$\eta \pi$~\cite{Beladidze:1993km,Alekseev:2009aa} (see also Eq.~(25)
in \refCite{Akhunzyanov:2018lqa}).  In addition, we add for each wave
a coherent non-resonant amplitude of the form of
\cref{eq:dyn_amp_non_res_simple} with parameter $b = 0$.

We fit the resonance model to the real and imaginary parts of the
elements of the spin-density submatrices of the selected waves by
minimizing the distance measure defined in \cref{eq:fit_chi2}.
Results from separate fits to the \etaPim and the \etaPrPim data are
consistent with each other and with those from a combined fit of both
data sets.  Statistical uncertainties of the measured \PaTwo and
\PaFour resonance parameters (see
\cref{sec:results_2pp,sec:results_4pp}) are much smaller than the
systematic uncertainties and are hence neglected.  The systematic
uncertainties are estimated based on several studies.  These studies
include changing the fit ranges, changing the parameterization of the
background components, and excluding the background component in
certain waves.  In these studies, we find that the resonance
parameters of the \PaTwo[1700] and of the \PpiOne* strongly depend on
the resonance model and on the fit range.  Hence these parameters
cannot be reliably extracted from the data.  However, the \PpiOne*
parameters are compatible with previous measurements.  In addition, we
find that a combined fit of the \etaPim and \etaPrPim data with the
same \PpiOne* resonance parameters in both channels also yields a
satisfactory description of the data.
This point will be discussed again in \cref{sec:results_1mp}.

Our inability to reliably extract the parameters of the \PaTwo[1700]
and the \PpiOne* shows the limitations of our simple
sum-of-Breit--Wigner approach.  Applying improved analytic and unitary
models, which were developed by the JPAC collaboration and were
discussed in \cref{sec:pwa.unitary_model}, to the COMPASS \etaOrPrPi
data yields well-defined pole positions for \PaTwo, \PaTwo[1700], and
\PpiOne[1600] that are less model-dependent.  These results
demonstrate the superiority of this approach.  They will be discussed
in more detail in \cref{sec:results_1mp,sec:results_2pp}.

\subsubsection{\threePi Partial-Wave Decomposition}
\label{sec:3pi_model:pwa}

Compared to the PWA of two-body final states (see \eg\
\cref{sec:etapi_model:pwa}), the partial-wave decomposition of
three-body final states requires considerably more modeling.  As was
discussed in \cref{sec:pwa_cells.discussion}, the development of an
optimal PWA model for multi-body decays is a non-trivial and
challenging task.  This is in particular true for the COMPASS \threePi
data sample taken with a liquid-hydrogen target, which is about an
order of magnitude larger than any data sample used in previously
published analyses (see \eg\
\refsCite{Daum:1980ay,Amelin:1995gu,Chung:2002pu,Dzierba:2005jg,Alekseev:2009aa}).
Hence the employed PWA model has to be significantly more detailed
than the models used in previous analyses in order to achieve a good
description of the data.

\begin{figure}[tbp]
  \centering
  \hfill%
  \subfloat[][]{%
    \includegraphics[width=\threePlotWidth]{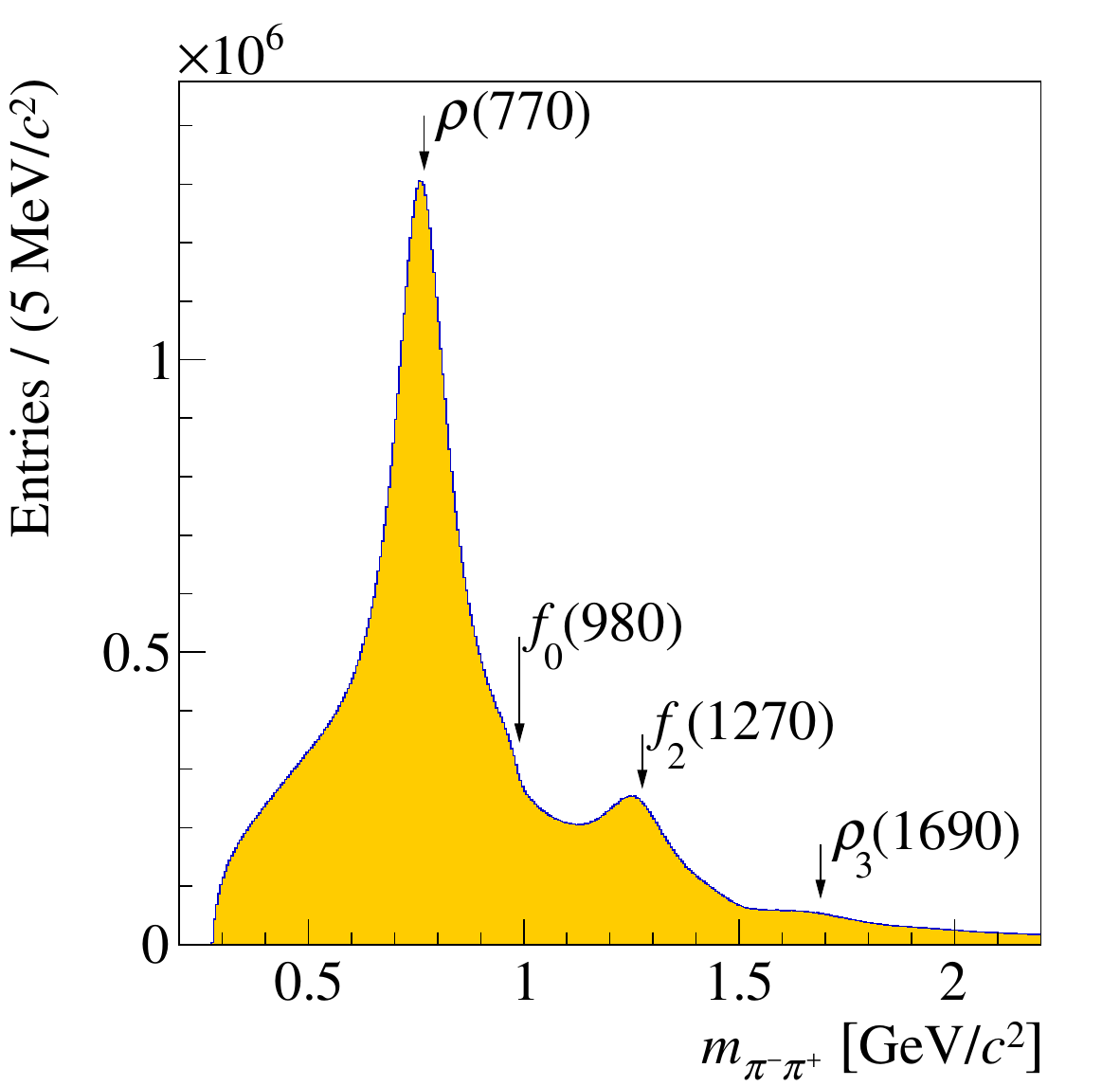}%
    \label{fig:mass_spectrum_2pi}%
  }%
  \hfill%
  \subfloat[][]{%
    \includegraphics[width=\threePlotWidthTwoD]{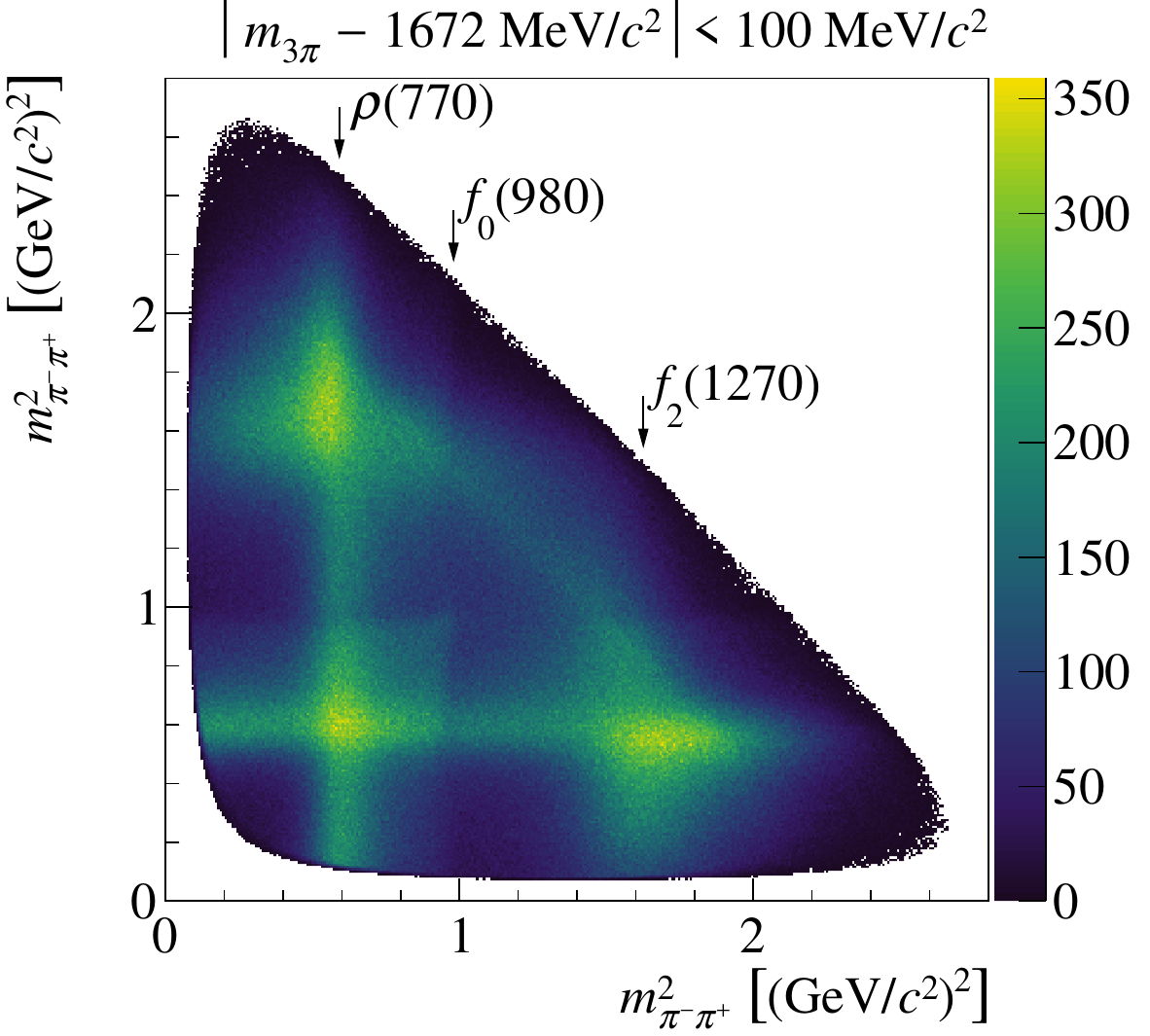}%
    \label{fig:dalitz_plot_pi2}
  }%
  \hfill\null%
  \caption{Kinematic distributions for the \threePi proton-target
    data~\cite{Adolph:2015tqa}.
    \subfloatLabel{fig:mass_spectrum_2pi}~Invariant mass distribution
    of the \twoPi subsystem (two entries per event).  The labels
    indicate the position of $2\pi$~resonances.
    \subfloatLabel{fig:dalitz_plot_pi2}~Dalitz plot for the
    \SI{100}{\MeVcc} wide \mThreePi~region around the \PpiTwo as
    indicated by the shaded region in \cref{fig:mass_3pi_proton}.
    This region exhibits signals for $\Pprho \pi$, $\PfZero[980] \pi$,
    and $\PfTwo \pi$ decays.}
  \label{fig:kin_distr}
\end{figure}

To construct the wave set, we first have to determine the set of
two-pion isobar resonances that appear in the data and we have to
choose the parameterizations for their propagator terms (see
\cref{sec:pwa.analysis_model.dyn_amp}).  Since there are no known
resonances in the $\pi^-\pi^-$ channel, which has $I = 2$ and is thus
flavor-exotic, we consider only \twoPi isobar resonances.  From the
\twoPi mass spectrum in \cref{fig:mass_spectrum_2pi} and from Dalitz
plots like the one in \cref{fig:dalitz_plot_pi2}, we can already infer
that we have to include \Pprho, \PfZero[980], and \PfTwo as isobars.
The slight enhancement at about $\mTwoPi = \SI{1.7}{\GeVcc}$ in
\cref{fig:mass_spectrum_2pi} could be due to \PrhoThree, \Pprho[1700],
or \PfZero[1710].  In our PWA model, we include only the \PrhoThree,
which is known from previous experiments (see \eg\
\refCite{Chung:2002pu}) to appear in the \twoPi subsystem of
diffractively produced $3\pi$~final states.  Due to ambiguities of the
partial-wave amplitudes that may arise when the PWA model contains
radially excited isobar resonances (see
\cref{sec:pwa_cells.discussion}), we include neither the \Pprho[1700]
nor the \PfZero[1710].  We confirm that the \PfZero[1710] does not
contribute significantly to the data by performing a co-called
freed-isobar PWA (see \cref{sec:pwa_cells:freed_isobar}), the results
of which are described in more detail below.  For the proton-target
data, we include one wave with the \PfZero[1500] isobar in the range
$\mThreePi > \SI{1.7}{\GeVcc}$.  The PWA model for both \threePi data
samples contains in addition the \PfZero[500], which is also known
as~$\sigma$.  Ignoring possible contributions from excited \Pprho* and
\PfTwo* states and from isobar states with $J > 3$, we arrive at sets
of 6~and 5~isobars that we include in the PWA models for the proton-
and lead-target data, respectively.  The \Pprho, \PfTwo,
\PfZero[1500], and \PrhoThree are described using relativistic
Breit--Wigner amplitudes (see \cref{sec:pwa.analysis_model.dyn_amp} and
Section~IV.A in \refCite{Adolph:2015tqa} for details).  The
description of the \PfZero[500] and \PfZero[980] isobars is more
difficult.  The distortion of the \PfZero[980] line shape due to its
closeness to the \KKbar threshold is taken into account by using a
Flatt\'{e}
parameterization~\cite{Flatte:1976xu,Flatte:1976xv,Ablikim:2004wn}.
The \PfZero[500] is extremely broad and is thus not well described by
a Breit--Wigner amplitude.  We describe this state effectively by a
parameterization of the amplitude of the $S$~wave in
$\pi \pi \to \pi \pi$ and $KK$ scattering~\cite{Au:1986vs} modified to
separate out the \PfZero[980] (see Section~IV.A in
\refCite{Adolph:2015tqa} for details).  In the text below, we denote
this isobar amplitude by \pipiS.

\begin{figure}[tbp]
  \centering
  \hfill%
  \subfloat[][]{%
    \includegraphics[width=\threePlotWidth]{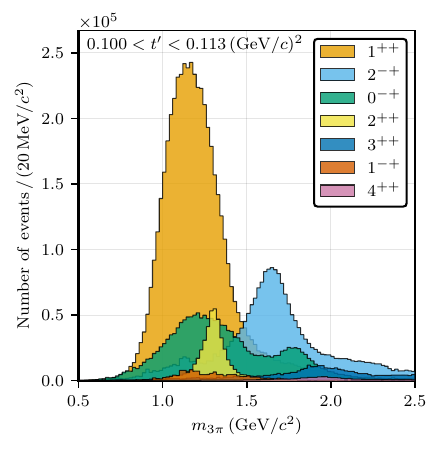}%
    \label{fig:3pi_spin_totals_tbin1}%
  }%
  \hfill%
  \subfloat[][]{%
    \includegraphics[width=\threePlotWidth]{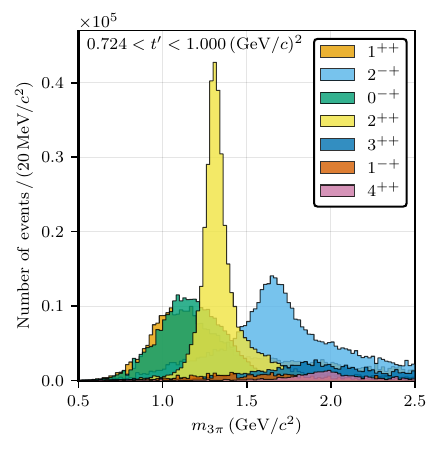}%
    \label{fig:3pi_spin_totals_tbin11}%
  }%
  \hfill\null%
  \caption{Representation of the fit result
    from~\refCite{Adolph:2015tqa}: intensities of the coherent sums of
    partial-wave amplitudes with the same \JPC quantum numbers
    (encoded by different colors) and positive reflectivity as
    obtained by the 88-wave PWA fit of the \threePi proton-target data
    \subfloatLabel{fig:3pi_spin_totals_tbin1}~for the lowest and
    \subfloatLabel{fig:3pi_spin_totals_tbin11}~for the highest
    $t'$~bin.}
  \label{fig:3pi_spin_totals}
\end{figure}

In our model for the intensity distribution in
\cref{eq:intensity_model_final}, the decay amplitudes are given by
\cref{eq:decay_amp_3body_2,eq:decay_amp_bose}.  Since the final-state
particles are spinless, the $3\pi$ partial waves are completely
defined by the short-hand notation \wave{J}{PC}{M}{\refl}{r}{L}, where
\JPCMrefl are the quantum numbers of the $X^-$~intermediate state,
$r$~is the isobar resonance with well-defined quantum numbers and
propagator term, and $L$~is the relative orbital angular momentum in
the decay $X^- \to r + \pi^-$ (\confer\ \cref{eq:wave_index}).  Based
on the set of selected isobar resonances described above, the number
of possible partial waves is largely determined by the maximum allowed
spin~$J$ of~$X^-$ and the maximum allowed~$L$.  For the proton-target
data, we have constructed a set of 128~waves, which---in accordance
with Pomeron dominance---includes mainly positive-reflectivity waves
with spin $J \leq 6$, orbital angular momentum $L \leq 6$, and spin
projection $M = \numlist{0;1;2}$.  From the result of a PWA performed
using this wave set, we have derived a smaller wave set by eliminating
structureless waves with relative intensities below approximately
\num{e-3} in an iterative process~\cite{Haas:2014bzm}.  This yields a
final set of 88~partial waves, which consists of 80~waves with
reflectivity $\refl = +1$, seven waves with $\refl = -1$ and one
non-interfering flat wave representing three uncorrelated pions.  This
is the largest wave set that was used so far in an analysis of
diffractively produced~$3\pi$.  In order to avoid ambiguities between
partial-wave amplitudes (see \cref{sec:pwa_cells.discussion}), we have
to exclude 27~waves in the low-\mThreePi~region.  The corresponding
\mThreePi~thresholds were carefully tuned for each wave and are given
in Table~IX in Appendix~A of \refCite{Adolph:2015tqa}.  In order to
illustrate the fit result, we show in \cref{fig:3pi_spin_totals} the
intensities of sums of positive-reflectivity amplitudes that have the
same \JPC quantum numbers.  Note that the composition of the data in
terms of \JPC changes substantially with~$t'$, in particular for the
$1^{++}$ and $2^{++}$ waves.  For the less precise lead-target data, a
smaller PWA model with only 42~waves is used~\cite{Alekseev:2009aa},
which is a subset of the 88-wave set.

Another important difference between the PWA models for the proton-
and the lead-target data is how the $t'$~dependence is taken into
account.  Using the 42-wave set, the PWA of the lead-target data was
performed independently in 50~\mThreePi~bins with \SI{40}{\MeVcc}
width integrating over the analyzed $t'$~range from
\SIrange{0.1}{1.0}{\GeVcsq}. In this $t'$~range, it is assumed that
the beam pions scatter off individual nucleons in the lead nucleus,
\ie that the scattering off the nucleus is incoherent.  The results
are therefore expected to be similar to the ones obtained from the
proton-target data.  The $t'$~dependences of the partial-wave
intensities were incorporated into the PWA model in
\cref{eq:intensity_model_final} via the replacement
\begin{equation}
  \label{eq:intensity_model_lead}
  \mathcal{T}_i^{r \refl}(m_X, t') \to \mathcal{T}_i^{r \refl}(m_X)\, f_i^\refl(t')\eqPunctSpacing,
\end{equation}
with empirical real-valued functions
\begin{equation}
  \label{eq:tdep_model_lead}
  f_i^\refl(t'; b_i^\refl, A_{i, 1}^\refl, A_{i, 2}^\refl, A_{i, 3}^\refl)
  = \bigg[ (t')^{\abs{M}}\, \Big( A_{i, 1}^\refl\, e^{-b_i^\refl\, t'}
  + A_{i, 2}^\refl\, e^{-\SI{15}{\perGeVcsq} t'} + A_{i, 3}^\refl\, e^{-\SI{3}{\perGeVcsq} t'}
  \Big) \bigg]^{\frac{1}{2}}
\end{equation}
that are inspired by \cref{eq:dsigma_dt.regge.2,eq:t_spectrum_model}.
For each wave, the values of the four free parameters, $b_i^\refl$,
$A_{i, 1}^\refl$, $A_{i, 2}^\refl$, and~$A_{i, 3}^\refl$, were
estimated by first performing the PWA in narrow bins of~$t'$ and in
three wide \mThreePi~ranges and then fitting \cref{eq:tdep_model_lead}
to the extracted partial-wave intensities.  This approach assumes that
the shapes of the $t'$~spectra of the partial waves are largely
independent of~\mThreePi and also does not take into account possible
$t'$~dependences of the relative phases between the partial waves.
The much larger proton-target data sample allows us to perform the PWA
using a two-dimensional binning with 100~\mThreePi bins with
\SI{20}{\MeVcc} width and 11~non-equidistant $t'$~bins, where the
$t'$~bins are chosen to contain approximately equal number of events.
For each of the resulting \num{1100} $(\mThreePi, t')$ cells, an
independent PWA fit is performed using the 88-wave PWA model (see
\cref{sec:pwa_cells}).  Because of the binning in~$t'$, we do not have
to assume any model for the $t'$~dependences of the partial-wave
amplitudes but instead extract this information from the data.  This
$t'$-binned analysis reveals a complicated \mThreePi~dependence of the
$t'$~spectra.  For many waves, the $t'$~spectra are not well described
by a single exponential and the slope parameters differ significantly
between waves and also between different \mThreePi~regions in the same
wave.  Hence the model in
\cref{eq:intensity_model_lead,eq:tdep_model_lead} that is used in the
analysis of the lead-target data may not be sufficient to describe all
features of the data.

The different ways how the $t'$~dependence is taken into account by
the PWA models also affect the rank of the spin-density matrix that is
required to describe the data.  For the $t'$-binned analysis of the
proton-target data, a rank-1 spin-density submatrix for the 80~waves
with $\refl = +1$ is sufficient to describe the data, \ie
$N_r^{(\refl = +1)} = 1$ in
\cref{eq:intensity_model_final}.\footnote{This corresponds to full
  coherence of the partial waves with $\refl = +1$.  For the seven
  waves with $\refl = -1$, a rank-2 spin-density submatrix is found to
  describe the data best, \ie $N_r^{(\refl = -1)} = 2$ in
  \cref{eq:intensity_model_final}.}  In contrast, a rank-2
spin-density matrix is used for the analysis of the lead-target data,
\ie $N_r^{(\refl = +1)} = N_r^{(\refl = -1)} = 2$.  Most of the
additional incoherence originates probably from the fact that the PWA
is performed integrating over the analyzed $t'$~range and that the
model in \cref{eq:intensity_model_lead,eq:tdep_model_lead} does not
take into account the $t'$~dependences of the partial-wave phases as
well as contributions from reactions where the beam pion scatters off
larger fractions of the nucleus.

As already mentioned above, we also performed a freed-isobar PWA (see
\cref{sec:pwa_cells:freed_isobar}) for the proton-target data.  In the
first analysis of this kind, we replace a subset of seven
$3\pi$~partial waves with fixed parameterizations for the \pipiS,
\PfZero[980], and \PfZero[1500] isobars and $\JPCMrefl = 0^{-+}\,0^+$,
$1^{++}\,0^+$, and $2^{-+}\,0^+$ quantum numbers of the three-body
system by the three partial waves \wave{0}{-+}{0}{+}{\pipiSF}{S},
\wave{1}{++}{0}{+}{\pipiSF}{P}, and
\wave{2}{-+}{0}{+}{\pipiSF}{D}~\cite{Adolph:2015tqa}.  These waves use
a free parameterization of the dynamical isobar amplitude for the
$\JPC = 0^{++}$ isobars according to \cref{eq:decay_amp_step_func},
which we denote by \pipiSF.  The freed-isobar amplitudes contain the
total $0^{++}$ isobar amplitudes including all resonant and
non-resonant components for the given $3\pi$ quantum numbers and the
analyzed $(\mThreePi, t')$ cell and are model-independent \wrt the
parameterizations of the $0^{++}$ isobars.  Due to the much larger
number of fit parameters, the freed-isobar PWA is performed using only
50~bins in~\mThreePi and 4~bins in~$t'$.  The results of the
conventional and the freed-isobar PWA are qualitatively in agreement
(see Section~VI in \refCite{Adolph:2015tqa}).  This validates the
fixed parameterizations of the $0^{++}$ isobars that are employed in
the conventional PWA and shows that in the conventional PWA waves
decaying into $\pipiS \pi$ and $\PfZero[980] \pi$ can be well
separated.

\subsubsection{\threePi Resonance-Model Fits}
\label{sec:3pi_model:resonance}

The goal of the resonance-model fit of the \threePi data is to study
isovector resonances of the~$a_J$ and $\pi_J$~families with masses up
to about \SI{2}{\GeVcc}.  In addition to establishing the existence of
the resonances, we want to determine their masses and total widths,
the relative strengths and phases of their decay modes, and their
production parameters, \ie their $t'$~spectra and the relative phases
between their coupling amplitudes as a function of~$t'$ (see
\cref{sec:pwa.res_fit_obs}).  The latter information is not obtainable
in the $t'$-integrated PWA approach that was used for the lead-target
data.

As was discussed in \cref{sec:pwa.analysis_model.extension}, the
resonance-model fit is performed on a selected subset of waves that
exhibit clear signals of well-known resonances, \ie resonance peaks
that are associated with phase motions.  If possible, we include waves
that represent different decay modes and different $M$~states of these
resonances.  These waves are intended to act as reference amplitudes,
against which the resonant amplitudes in more interesting waves can
interfere.  These latter waves exhibit signals of less well-known
excited states or have controversial resonance content such as the
spin-exotic \wave{1}{-+}{1}{+}{\Pprho}{P} wave.

The input for the resonance-model fit are the spin-density
submatrices~$\varrho_{ij}^\refl$ of the selected waves (see
\cref{eq:spin-dens_norm}) that have been extracted from the data in
the first analysis stage by performing the partial-wave decomposition
independently in bins of~\mThreePi and for the proton-target data also
in bins of~$t'$ (see \cref{sec:3pi_model:pwa}).  For the lead-target
data, we select a subset of 6~waves out of the 42~waves in the PWA
model~\cite{Alekseev:2009aa}; for the proton-target data we select
14~out of the 88~waves in the PWA
model~\cite{Akhunzyanov:2018lqa}.\footnote{The 6~waves selected for
  the lead-target data are a subset of the 14~waves selected for the
  proton-target data.}  The latter one is the so far largest wave set
that is consistently described in a single resonance-model fit of
diffractively produced~$3\pi$.

The selected sets of 6~and 14~waves contain signals of the well-known
$3\pi$~resonances \PaOne, \PaTwo, \PpiTwo, \Pppi[1800], and \PaFour,
which appear as peaks in the respective partial-wave intensities and
as phase motions in the relative phases of these waves.  The 14-wave
set includes in addition waves with signals for the well-known
\PpiTwo[1880].  This set also includes a clear resonance-like signal
of the novel \PaOne[1420], which was discovered in an earlier analysis
of the same COMPASS proton-target data in \refCite{Adolph:2015pws}.
This earlier analysis used the same 88-wave PWA fit result but the
resonance-model fit included only three partial waves.  The 14-wave
set also contains signals of the less well-known states \PaOne[1640]
and \PaTwo[1700].  It turns out that the proton-target data require a
third \PpiTwo* resonance, the \PpiTwo[2005], which according to the
PDG requires confirmation~\cite{Tanabashi:2018zz}.  For both data
samples, the waves selected for the resonance-model fit include the
spin-exotic \wave{1}{-+}{1}{+}{\Pprho}{P} wave.  The appearance of a
potential \PpiOne[1600] resonance in this wave is
disputed~\cite{Adams:1998ff,Chung:2002pu,Dzierba:2005jg}.  In total,
the resonance model in \cref{eq:res_model_spin-dens} contains
6~resonances for the 6-wave fit of the lead-target data and
11~resonances for the 14-wave fit of the proton-target data.  All
resonances are described using Breit--Wigner amplitudes as in
\cref{eq:BW_const_width,eq:BW_mass-dep_width}.  For the \PaOne and the
\PaTwo, we use dynamic widths (see Section~IV.A.1 in
\refCite{Akhunzyanov:2018lqa} for details).  For the other resonances,
we use constant widths as in \cref{eq:BW_const_width}.  In addition to
the resonant components, we include for each wave a separate coherent
non-resonant component (see \cref{sec:pwa.res_model}).  The resonance
models for the lead- and proton-target data are summarized in
\cref{tab:fitmodels}.  For the lead-target data, the resonance model
in \cref{eq:res_model_spin-dens} has rank~2 to match the rank of the
spin-density matrix from the $t'$-integrated PWA.  The 14-wave
resonance-model fit of the proton-target data is performed
simultaneously in 11~$t'$ bins using a rank-1 resonance
model.\footnote{This resonance model has \num{722}~real-valued free
  parameters, which are constrained by \num{76505} data points that
  enter the $\chi^2$~function in \cref{eq:fit_chi2}.  It is worth
  noting that only 51~of the 722~fit parameters correspond to shape
  parameters~$\{\zeta_k^\text{R}\}$ of resonant components, \ie masses
  and widths.  The vast majority of the fit parameters in
  \cref{eq:res_model_spin-dens} are the real and imaginary parts of
  the coupling amplitudes $\{\mathcal{C}_{k i}^{r \refl}(t')\}$ in the
  11~$t'$ bins.}

%
\begin{table}[tbp]
  \renewcommand{\arraystretch}{1.2}
  \centering
  \caption{Resonance models used to describe the elements of the
    spin-density matrix of selected partial waves, which are extracted
    from the \threePi data, using \cref{eq:res_model_spin-dens}.  For
    the lead-target data, 6~waves (highlighted in blue) are selected
    and described using a resonance model containing 6~resonances
    (highlighted in blue) and a coherent non-resonant component for
    each wave~\cite{Alekseev:2009aa}.  For the proton-target data, the
    listed 14~waves are selected.  They are described using a
    resonance model containing 11~resonances and a coherent
    non-resonant component for each wave~\cite{Akhunzyanov:2018lqa}.}
  \label{tab:fitmodels}
  \begin{footnotesize}
    \begin{tabular}{$l^p{0.26\textwidth}}
      \toprule
      \rowstyle{\bfseries}
      Partial wave                   & Resonance(s) \\
      \midrule
      \color{kBlue+1}\wave{0}{-+}{0}{+}{\PfZero}{S} & \mbox{\color{kBlue+1}\Pppi[1800]} \\[1.2ex]

      \color{kBlue+1}\wave{1}{++}{0}{+}{\Pprho}{S}  & \mbox{{\color{kBlue+1}\PaOne},~\PaOne[1640]} \\
      \wave{1}{++}{0}{+}{\PfZero}{P} & \PaOne[1420] \\
      \wave{1}{++}{0}{+}{\PfTwo}{P}  & \mbox{\PaOne,~\PaOne[1640]} \\[1.2ex]

      \color{kBlue+1}\wave{1}{-+}{1}{+}{\Pprho}{P}  & \mbox{\color{kBlue+1}\PpiOne[1600]} \\[1.2ex]

      \color{kBlue+1}\wave{2}{++}{1}{+}{\Pprho}{D}  & \rdelim\}{3}{\linewidth}[~\mbox{{\color{kBlue+1}\PaTwo},~\PaTwo[1700]}] \\
      \wave{2}{++}{2}{+}{\Pprho}{D}  & \\
      \wave{2}{++}{1}{+}{\PfTwo}{P}  & \\[1.2ex]

      \wave{2}{-+}{0}{+}{\Pprho}{F}  & \rdelim\}{4}{\linewidth}[~\mbox{{\color{kBlue+1}\PpiTwo},~\PpiTwo[1880],~\PpiTwo[2005]}] \\
      \color{kBlue+1}\wave{2}{-+}{0}{+}{\PfTwo}{S}  & \\
      \wave{2}{-+}{1}{+}{\PfTwo}{S}  & \\
      \wave{2}{-+}{0}{+}{\PfTwo}{D}  & \\[1.2ex]

      \color{kBlue+1}\wave{4}{++}{1}{+}{\Pprho}{G}  & \rdelim\}{2}{\linewidth}[~\mbox{\color{kBlue+1}\PaFour}] \\
      \wave{4}{++}{1}{+}{\PfTwo}{F}  & \\[1.2ex]

      \bottomrule
    \end{tabular}
  \end{footnotesize}
\end{table}
 
\begin{figure}[tbp]
  \centering
  \hfill%
  \subfloat[][]{%
    \includegraphics[width=\threePlotWidth]{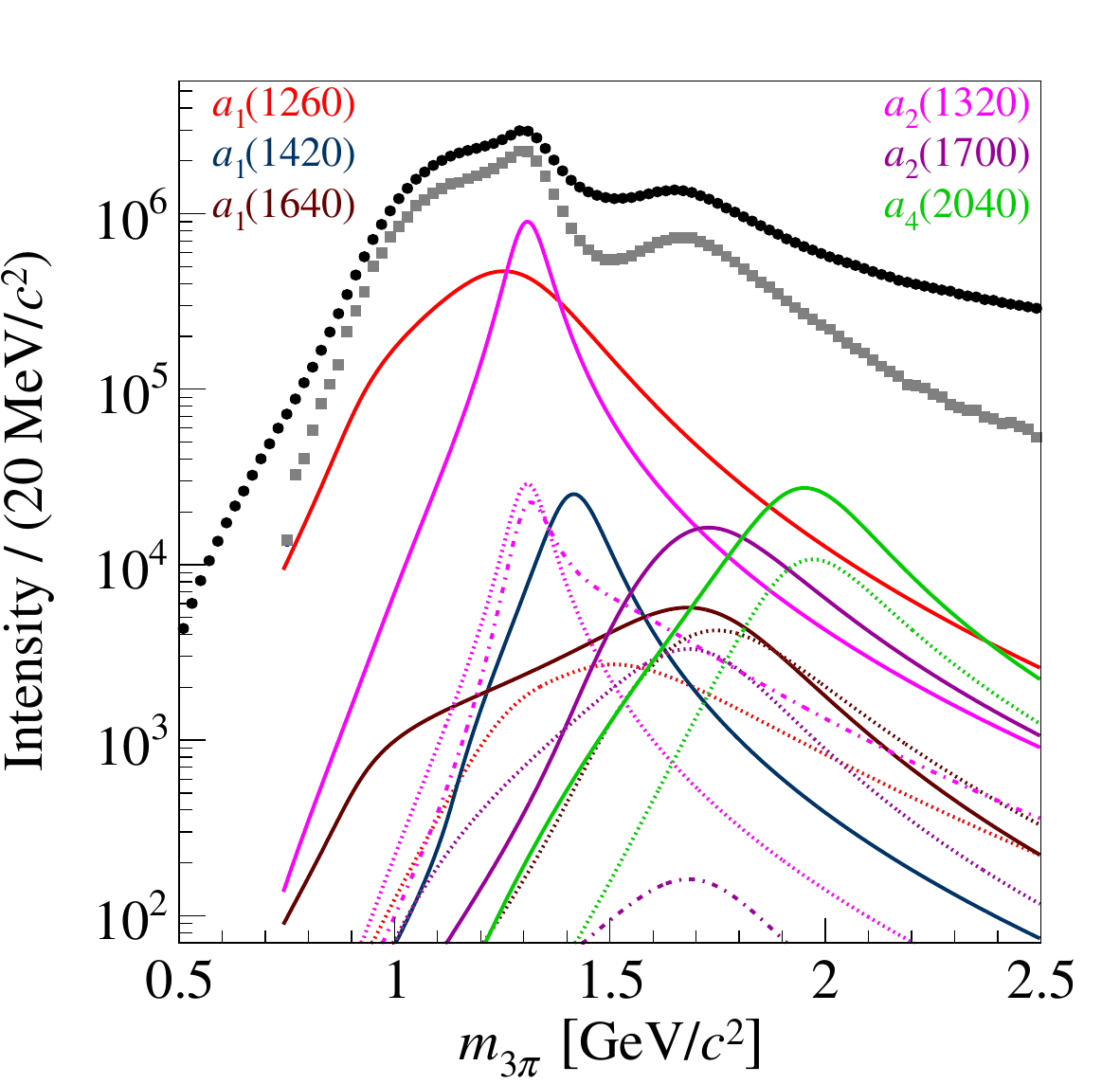}%
    \label{fig:int:a}%
  }%
  \hfill%
  \subfloat[][]{%
    \includegraphics[width=\threePlotWidth]{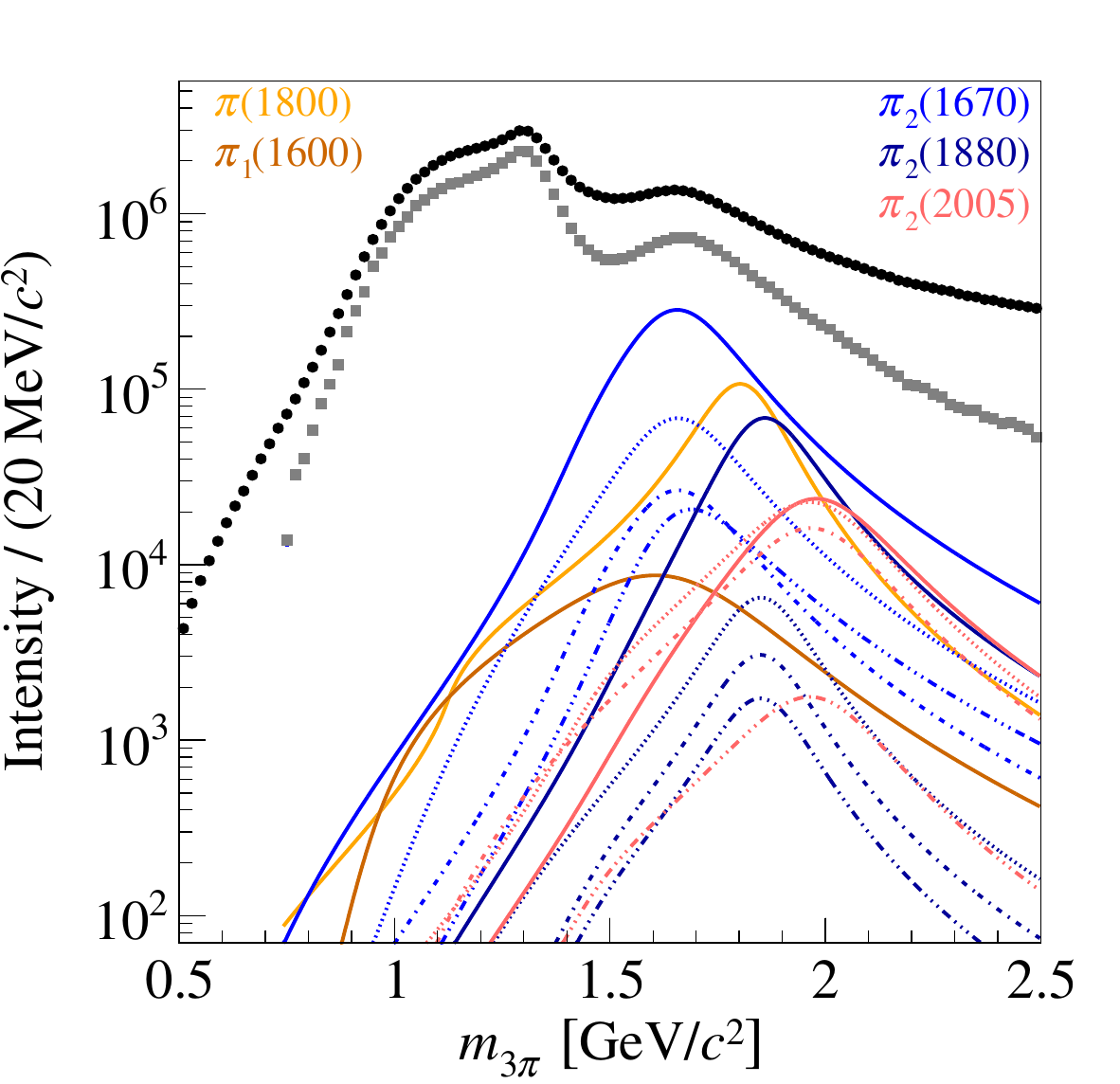}%
    \label{fig:int:p}%
  }%
  \hfill\null%
  \\
  \null\hfill%
  \subfloat[][]{%
    \includegraphics[width=\threePlotWidth]{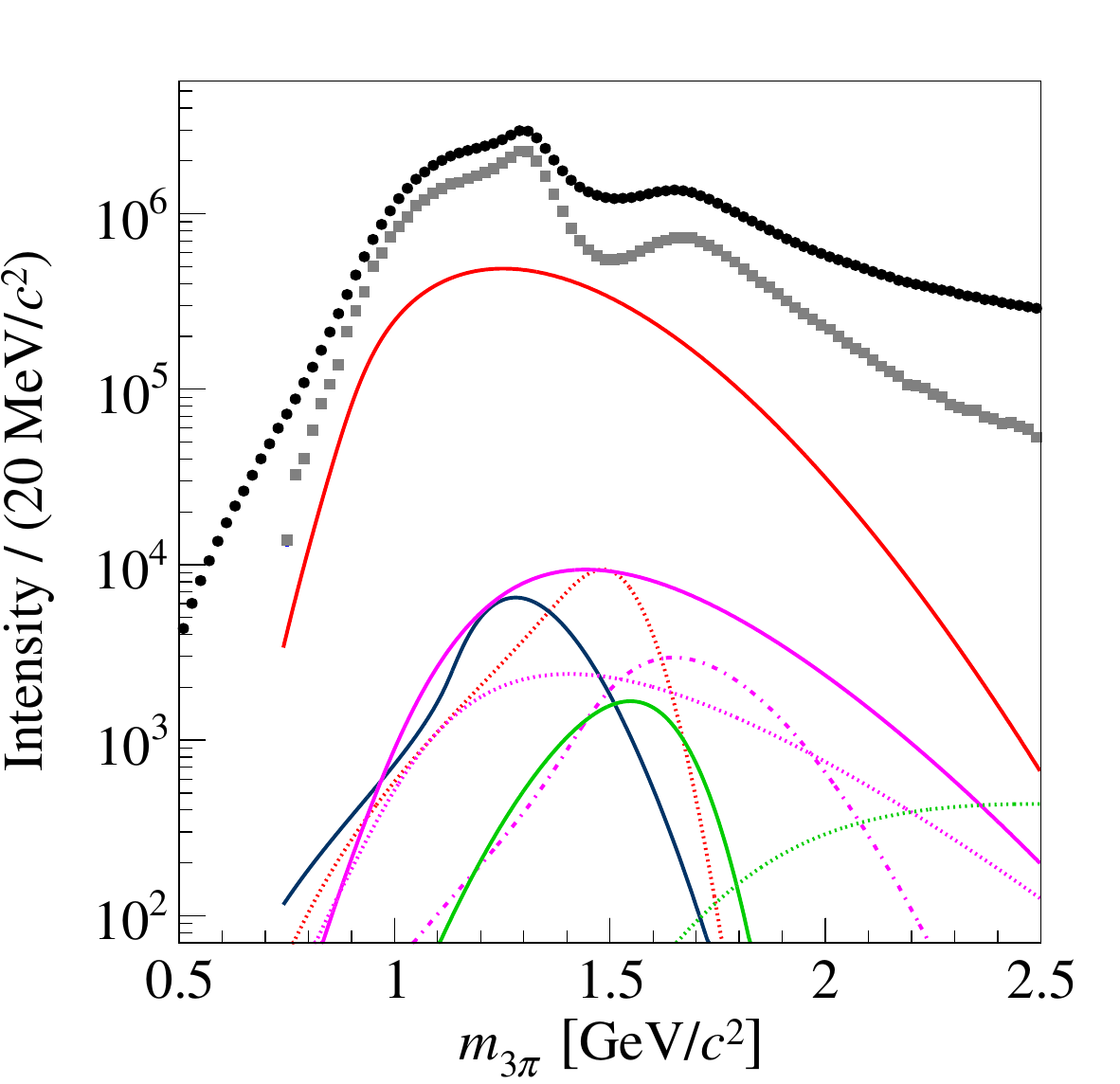}%
    \label{fig:int:a_nonres}%
  }%
  \hfill%
  \subfloat[][]{%
    \includegraphics[width=\threePlotWidth]{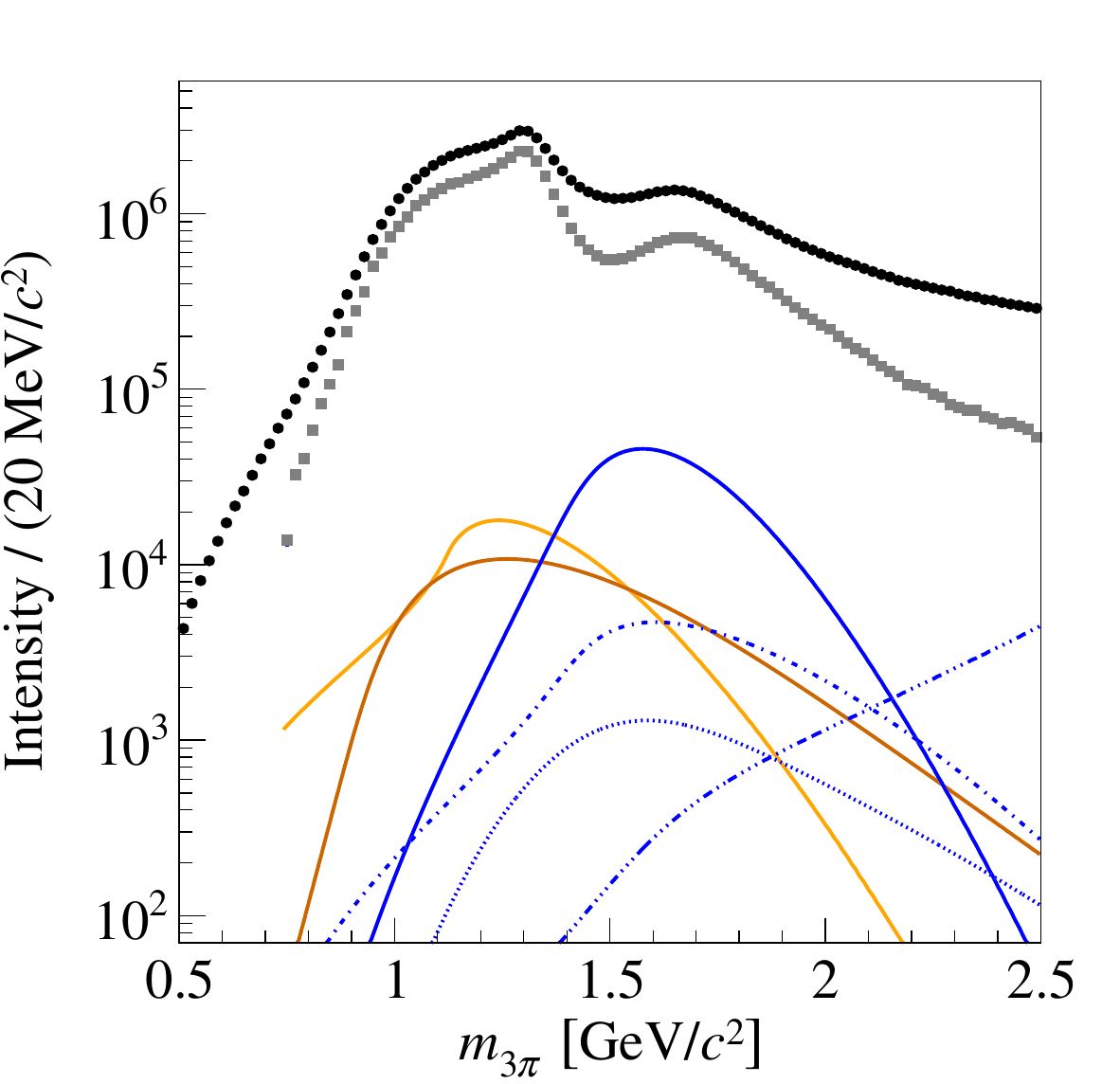}%
    \label{fig:int:p_nonres}%
  }%
  \hfill\null%
  \caption{Summary of the result of the 14-wave resonance-model fit of
    the \threePi proton-target data from
    \refCite{Akhunzyanov:2018lqa}: The intensities of the wave
    components in the resonance model as a function of~\mThreePi
    (colored curves) are compared to the total intensity of the
    coherent sum of all 88~partial-wave amplitudes in the PWA model
    (black points) and to the intensity of the coherent sum of the
    14~partial-wave amplitudes selected for the resonance-model fit
    (gray squares).  (Top row) Intensity distributions of the
    11~resonances that are included in the resonance model (left:
    $a_J$~resonances, right: $\pi_J$~resonances). Different colors
    encode different resonances. The same resonance may appear in up
    to four partial waves that correspond to different decay modes.
    These different decay modes are encoded by different line styles,
    which are assigned according to the height of the respective
    resonance peak.  The line shapes of the resonances differ in the
    various decay modes because of different phase space.  (Bottom
    row) Intensity distributions of the 14~non-resonant components
    included in the resonance model, one in each wave.  Color and line
    style are defined by the dominant resonance in the respective wave
    as shown in the top row.}
  \label{fig:summary:int}
\end{figure}

\Cref{fig:summary:int} shows the intensities of the resonant and
non-resonant wave components from the 14-wave resonance-model fit of
the proton-target data as a function of~\mThreePi (colored curves)
together with the intensity of the coherent sum of the 14~selected
partial-wave amplitudes (gray squares) and the total intensity of the
PWA model (black points).  Integrated over the analyzed
\mThreePi~range, the coherent sum of the 14~selected partial-wave
amplitudes contributes \SI{57.9}{\percent} to the total intensity.
This value is similar to the intensity sum of the 14~waves, which
corresponds to \SI{56.8}{\percent} of the total intensity.  Thus the
net effect of the interferences between the 14~partial-wave amplitudes
is slightly constructive.  It is worth noting that the intensities of
the wave components cover a large dynamic range of more than three
orders of magnitude.  The largest contributions to the intensity come
from the \PaOne and the non-resonant term in the
\wave{1}{++}{0}{+}{\Pprho}{S} wave (continuous red curves in
\cref{fig:int:a,fig:int:a_nonres}).  Since the resonance model
contains only waves with positive reflectivity, all wave components,
in particular the resonant and non-resonant components, do interfere.
For waves with larger intensities, this interference is mostly
constructive.

Due to the highly precise proton-target data, the uncertainties of the
resonance parameters from the 14-wave fit are completely dominated by
systematic effects.  The corresponding uncertainties are estimated
based on extensive systematic studies.  The parameters of the various
resonances have vastly different systematic uncertainties.  This
mainly reflects two aspects of our data: \one the large dynamic range
of the intensities of the resonances in the selected waves as is
illustrated in \cref{fig:int:a,fig:int:p} and \two the vastly
different strength of the non-resonant components relative to the
resonances.  As a consequence, we can determine the parameters of
\PaTwo, \PaOne[1420], \PpiTwo, \Pppi[1800], and \PaFour with high
accuracy.  In contrast, the parameters of \PaOne, \PpiOne[1600],
\PaOne[1640], and \PpiTwo[2005] have relatively large systematic
uncertainties.

A particularly important systematic effect is caused by the choice of
the parameterization of the non-resonant contributions (see
\cref{sec:pwa.res_model}).  The Deck process~\cite{Deck:1964hm} is a
dominant source of non-resonant contributions (see
\cref{sec:exp.prod_reactions,sec:pwa_cells.discussion,fig:deck}).  In
order to study the dependence of our fit result on the
parameterization of the non-resonant component, we performed a study,
where the shapes of the non-resonant components are determined from a
model for the Deck process (see Appendix~B in
\refCite{Akhunzyanov:2018lqa} for details).  Based on this model, we
generate Monte Carlo pseudo-data and perform a PWA fit like for the
real data using the 88-wave PWA model.  The empirical
parameterizations of the non-resonant components are then replaced by
the square root of the intensity distributions as obtained from the
partial-wave decomposition of the Deck Monte Carlo data.  We will
discuss the results from this study in
\cref{sec:results_1pp,sec:results_1mp} below.  More details of the
performed systematic studies are discussed in Section~V and Appendix~D
in \refCite{Akhunzyanov:2018lqa}.

\subsection{Results}
\label{sec:results_by_qn}

In the following sections, we present selected results from the
partial-wave decompositions and the resonance-model fits of the
COMPASS \etaPim, \etaPrPim, and the two \threePi data samples.  These
results are ordered by the \JPC quantum numbers of the states.  Since
the \threePi proton-target data sample is by far the largest one, we
discuss results from these data in more detail.

\subsubsection{The $\JPC = 0^{-+}$ Sector }
\label{sec:results_0mp}

The PDG lists currently five isovector states with
$\JPC = 0^{-+}$~\cite{Tanabashi:2018zz}: \Pppi*, \Pppi[1300],
\Pppi[1800], \Pppi[2070], and \Pppi[2360] (see also
\cref{fig:light_flavorless_spectrum}).  Although the first three
states are considered well established, the parameters of the
\Pppi[1300] are not well known.  The \Pppi[2070] and the \Pppi[2360]
are \textquote{further states} and require confirmation.

The \Pppi[1300] is considered the first radial excitation of the pion,
\ie the \termSym{2}{1}{S}{0} quark-model state, by the
PDG~\cite{pdg_quark_model:2018} and also the quark model from
\refCite{Ebert:2009ub} that is discussed in
\cref{sec:pheno.qm.spectrum}.  The \Pppi[1300] was reported in
$\Pprho \pi$ and $\pipiS \pi$ decay modes, although the latter one is
contradicted by a measurement by the Crystal Barrel
Collaboration~\cite{Abele:2001js}.  The \Pppi[1800] has a mass that is
consistent with the second radial excitation of the pion, \ie the
\termSym{3}{1}{S}{0} quark-model state, but it exhibits a peculiar
decay pattern.  The \Pppi[1800] decays mostly into~$3\pi$ and
experiments have reported $\PfZero[500] \pi$, $\PfZero[980] \pi$, and
$\PfZero[1370] \pi$ decays, \ie decays into $\JPC = 0^{++}$
isobars~\cite{Tanabashi:2018zz}.  Surprisingly, the decay into
$\Pprho \pi$ is not seen~\cite{Bellini:1982ec}.  For the $3\pi$~final
state, also the decay into $\PfZero[1500] \pi$ is not
observed~\cite{Chung:2002pu}, although it is seen in the
$\eta \eta \pi$ final
state~\cite{Amelin:1995fg,Anisovich:2001hj,Eugenio:2008zza}.  In the
latter final state, however, the $\PfZero[1370] \pi$ decay is not
observed~\cite{Eugenio:2008zza}.

The 88-wave PWA model used to analyze the COMPASS \threePi
proton-target data contains five waves with $\JPC = 0^{-+}$ that
correspond to the five decay modes $\pipiS \pi$, $\Pprho \pi$,
$\PfZero[980] \pi$, $\PfTwo \pi$, and $\PfZero[1500] \pi$.  The waves
with \pipiS, \Pprho, and \PfZero[980] isobars have comparatively large
relative intensities of \SI{8.0}{\percent}, \SI{3.5}{\percent}, and
\SI{2.4}{\percent}, respectively.  In contrast, the waves with \PfTwo
and \PfZero[1500] isobars have relative intensities of only
\SI{0.2}{\percent}, and \SI{0.1}{\percent}, respectively.

\begin{figure}[tbp]
  \centering
  \subfloat[][]{%
    \includegraphics[width=\threePlotWidth]{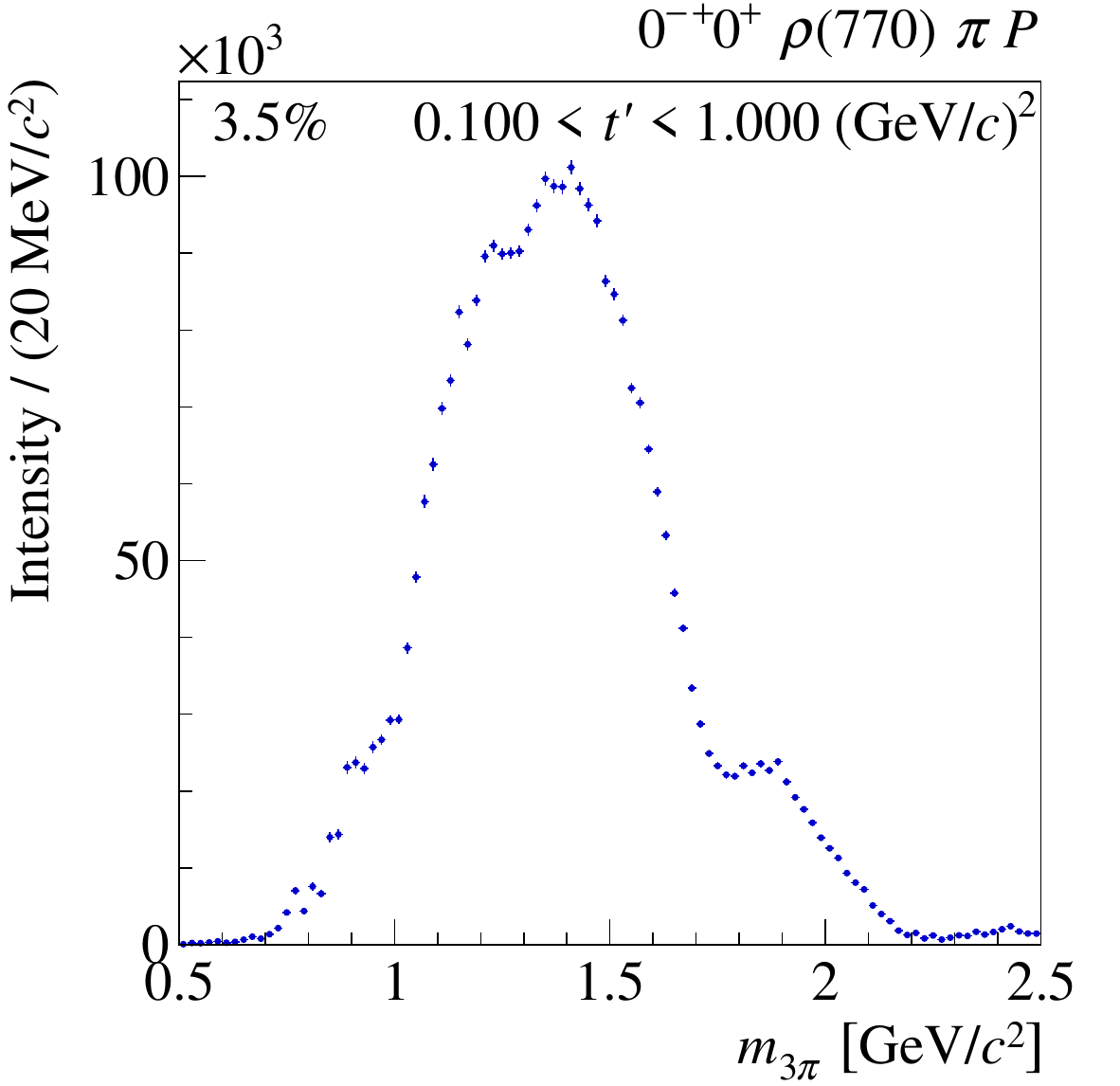}%
    \label{fig:int_0mp_rho}%
  }%
  \hfill%
  \subfloat[][]{%
    \includegraphics[width=\threePlotWidth]{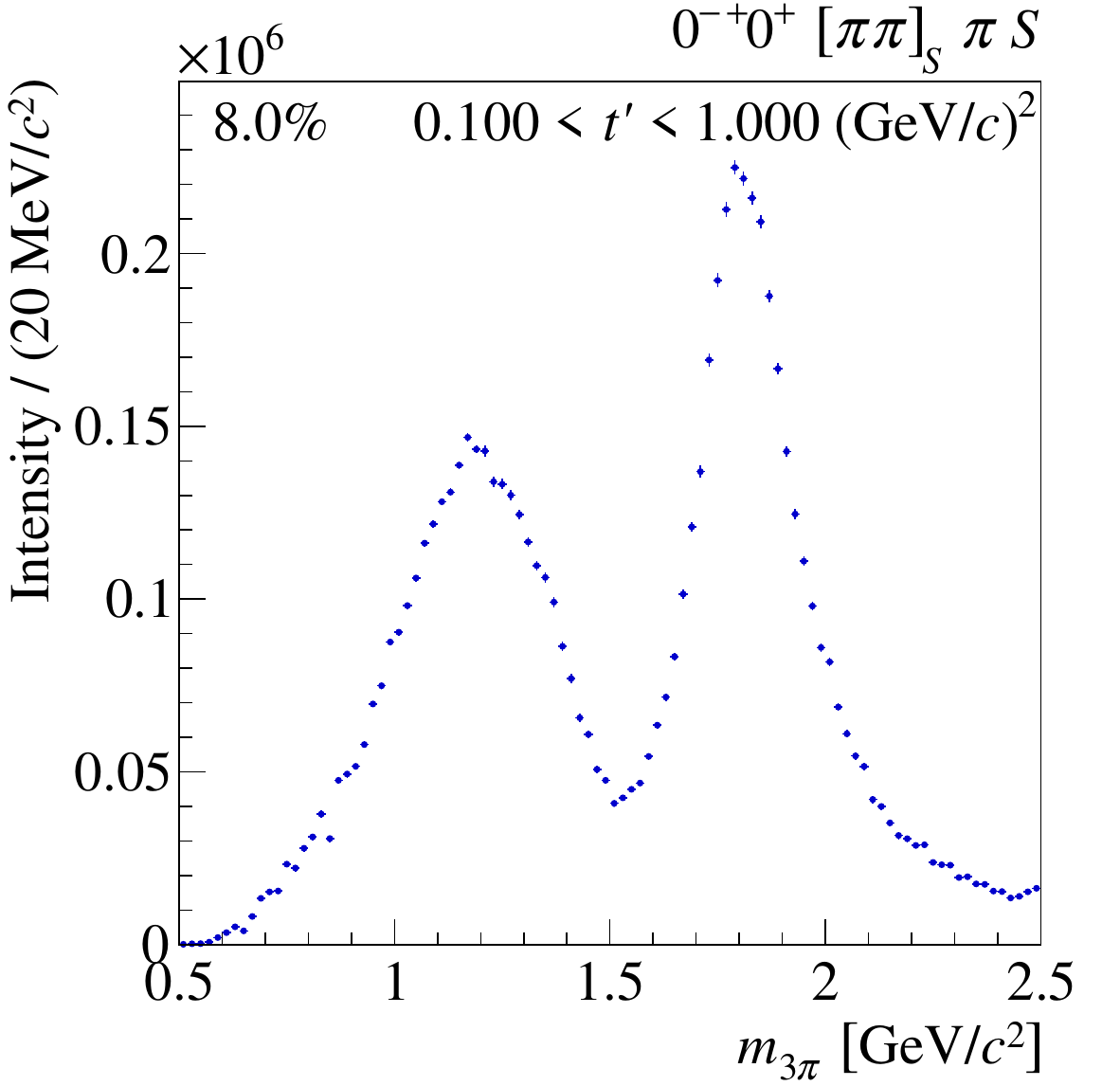}%
    \label{fig:int_0mp_pipiS}%
  }%
  \hfill%
  \subfloat[][]{%
    \includegraphics[width=\threePlotWidth]{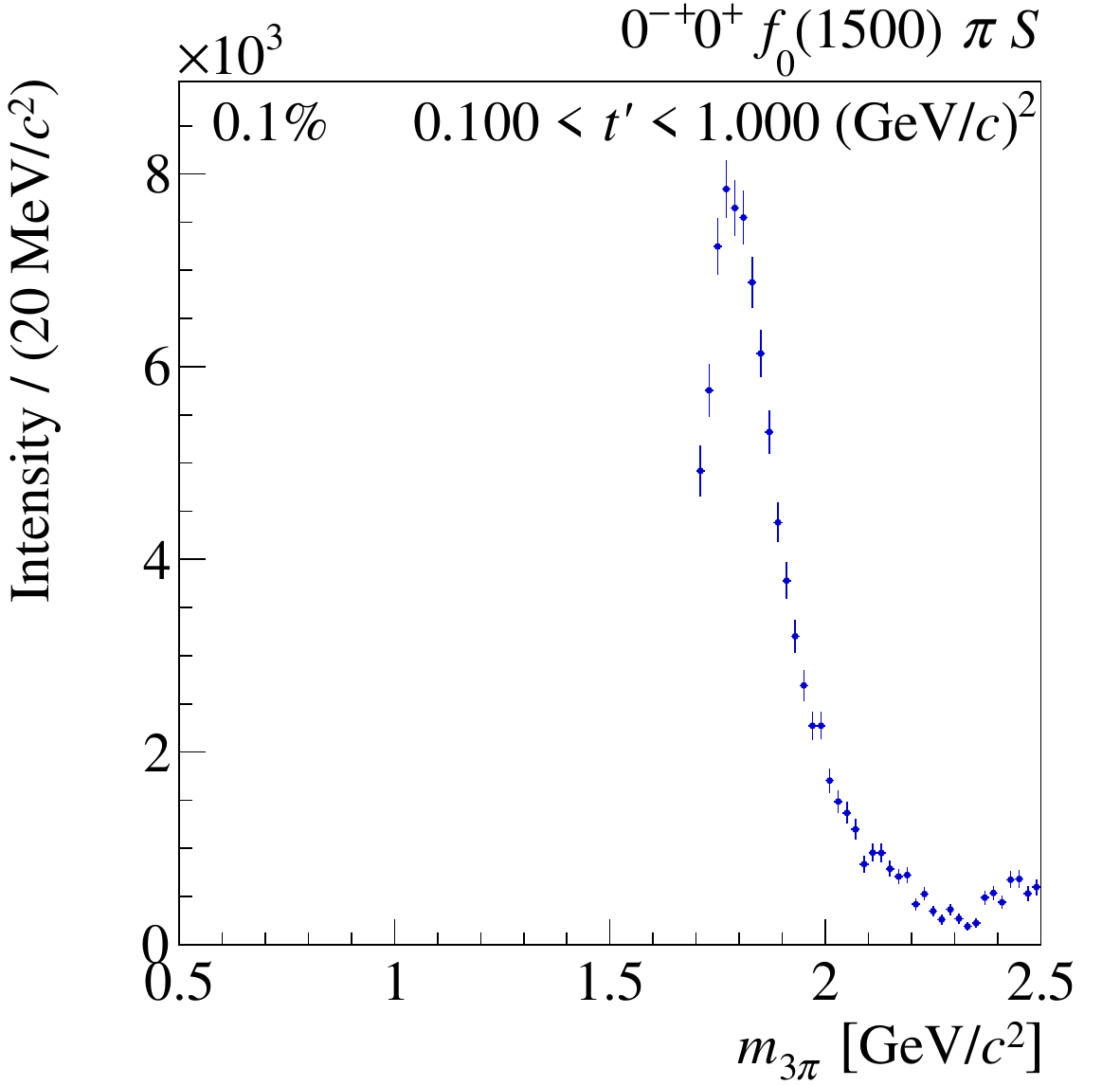}%
    \label{fig:int_0mp_f01500}%
  }%
  \caption{Intensity distributions, summed over the 11~$t'$~bins, for
    \subfloatLabel{fig:int_0mp_rho}~the \wave{0}{-+}{0}{+}{\Pprho}{P},
    \subfloatLabel{fig:int_0mp_pipiS}~the
    \wave{0}{-+}{0}{+}{\pipiS}{S}, and
    \subfloatLabel{fig:int_0mp_f01500}~the
    \wave{0}{-+}{0}{+}{\PfZero[1500]}{S} wave in the \threePi
    proton-target data.  From
    \refsCite{Adolph:2015tqa,Adolph:2015tqa_suppl}.}
  \label{fig:intensities_0mp}
\end{figure}

We observe broad enhancements in the \SI{1.3}{\GeVcc} \mThreePi~region
in the intensity distributions of the $\Pprho \pi$ and $\pipiS \pi$
waves, which may contain the \Pppi[1300] (see
\cref{fig:int_0mp_rho,fig:int_0mp_pipiS}).  However, for both waves
the shape of the intensity distribution around \SI{1.3}{\GeVcc}
changes strongly with~$t'$.  This indicates large non-resonant
contributions, which makes the determination of the resonance
parameters difficult.  This is probably one of the reasons, for the
large spread of the \Pppi[1300] parameters as measured by previous
experiments using similar reactions~\cite{Tanabashi:2018zz} and the
resulting large uncertainty of the PDG estimate (see
\cref{tab:PDG_mesons_2018}).  First attempts to describe the
$\Pprho \pi$ wave in the COMPASS \threePi proton-target data (see
\cref{fig:int_0mp_rho}) by a resonance model failed, because our model
is not able to reproduce the observed intensity distribution.
Improving the resonance model is the topic of future research.

\begin{figure}[tbp]
  \centering
  \subfloat[][]{%
    \includegraphics[width=\threePlotWidth]{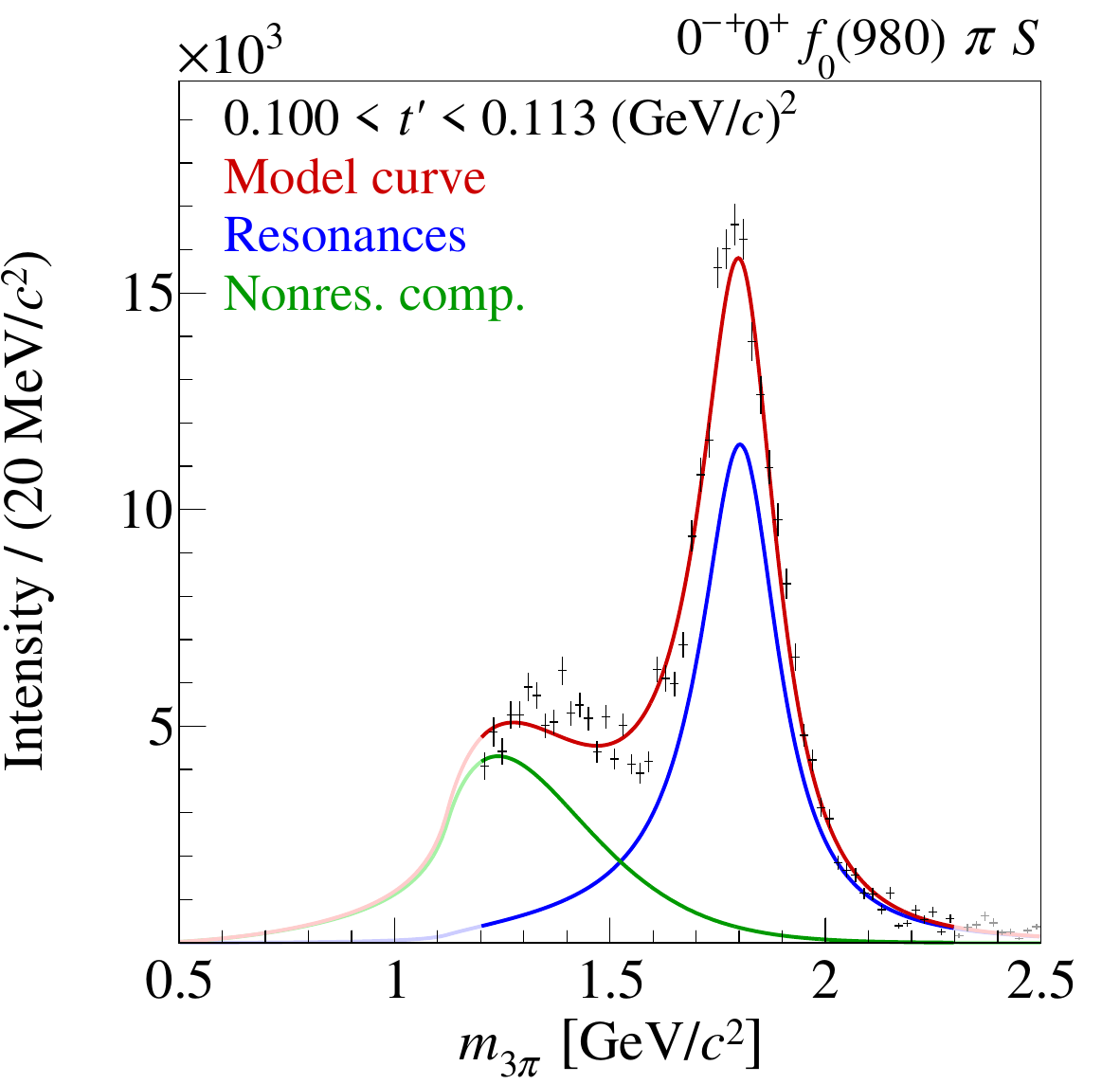}%
    \label{fig:intensity_0mp_f0_tbin1}%
  }%
  \hfill%
  \subfloat[][]{%
    \includegraphics[width=\threePlotWidth]{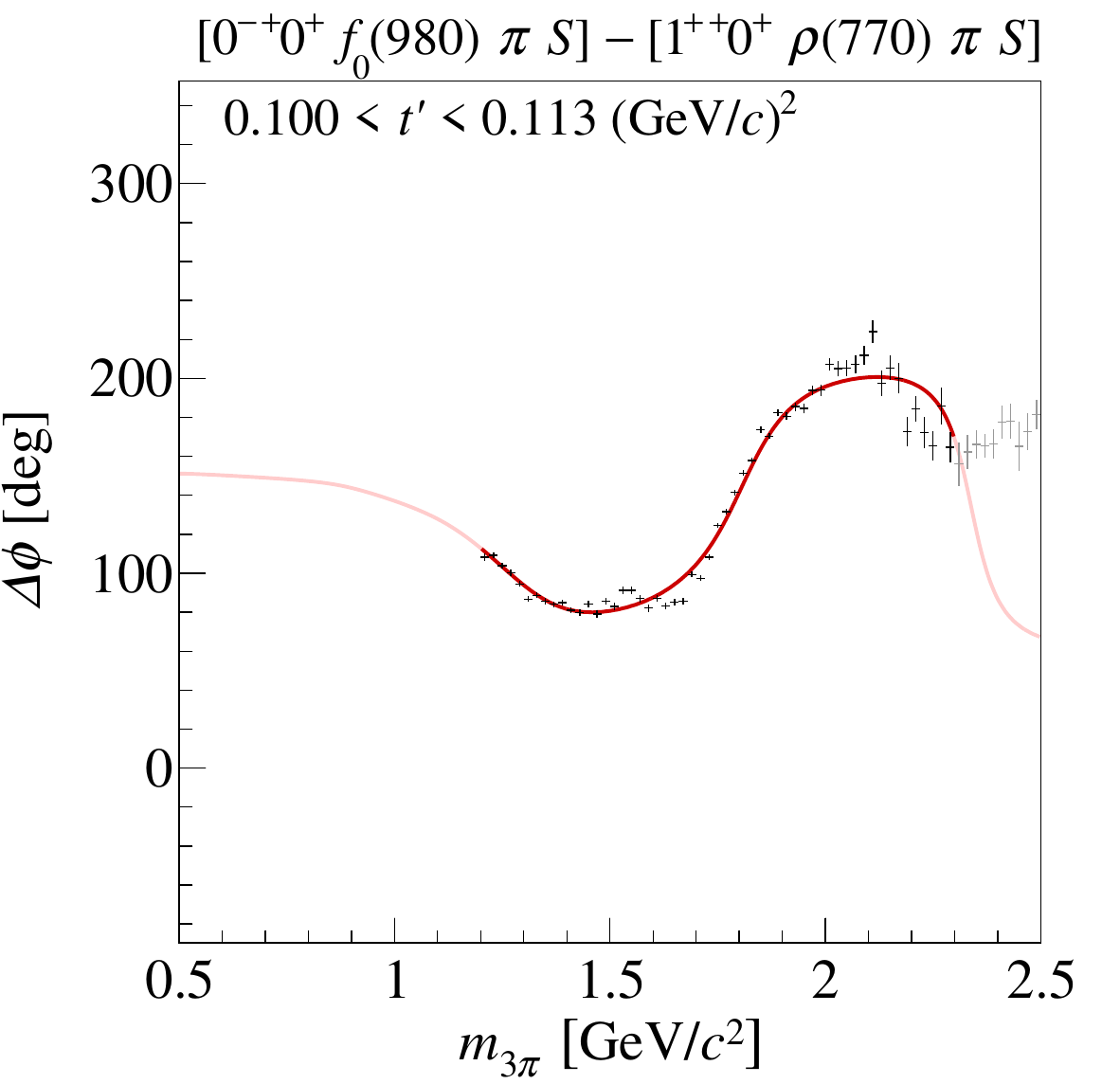}%
    \label{fig:phase_0mp_f0_1pp_rho_tbin1}%
  }%
  \hfill%
  \subfloat[][]{%
    \includegraphics[width=\threePlotWidth]{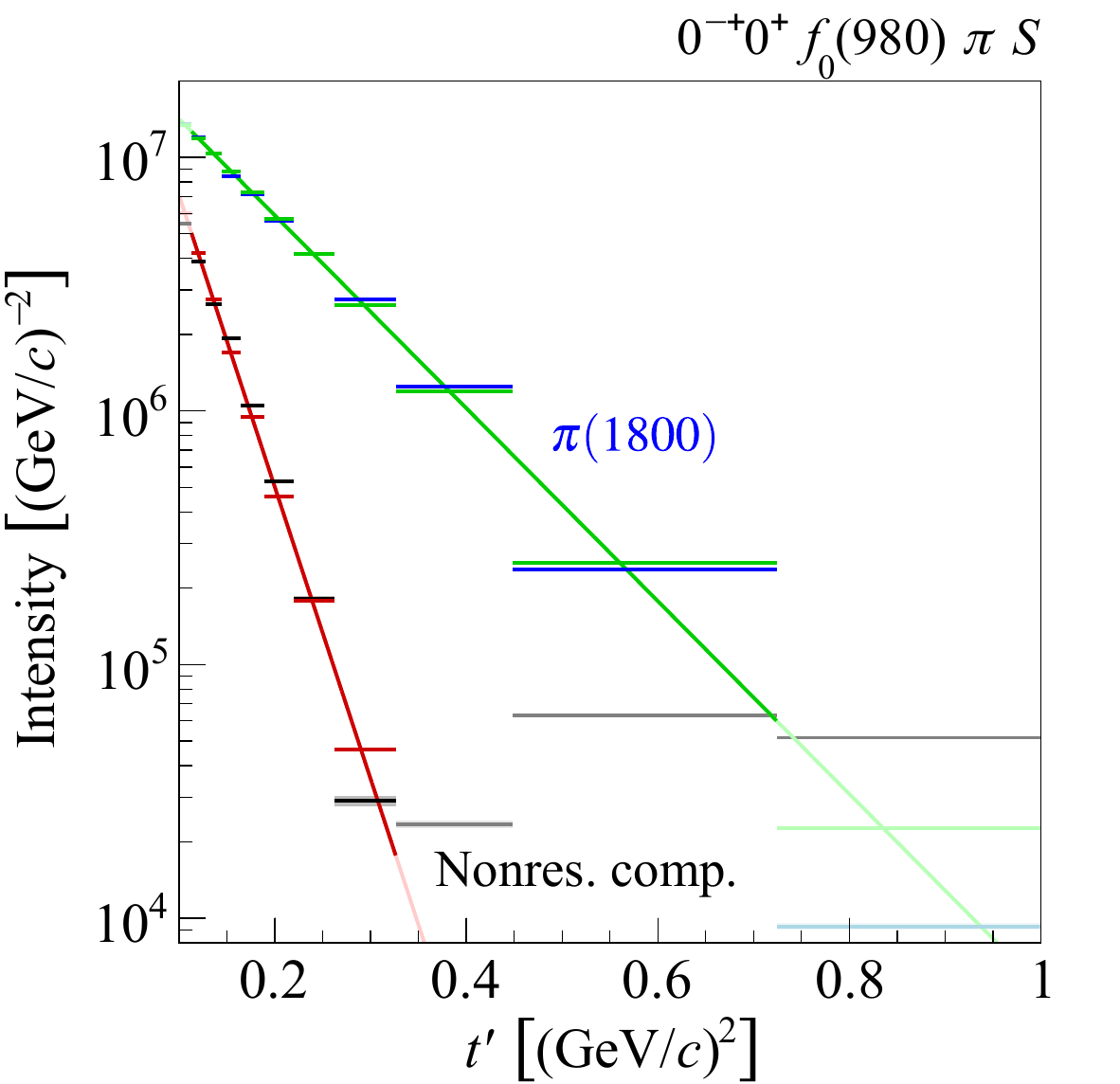}%
    \label{fig:tspectrum_0mp_f0}%
  }%
  \caption{Amplitude of the \wave{0}{-+}{0}{+}{\PfZero[980]}{S} wave
    in the lowest $t'$~bin of the \threePi proton-target
    data~\cite{Akhunzyanov:2018lqa}.
    \subfloatLabel{fig:intensity_0mp_f0_tbin1}~Intensity distribution.
    \subfloatLabel{fig:phase_0mp_f0_1pp_rho_tbin1}~Phase of this wave
    \wrt the \wave{1}{++}{0}{+}{\Pprho}{S} wave.  The data points
    represent the result of the partial-wave decomposition in the
    first analysis stage (see \cref{sec:pwa_cells,sec:3pi_model:pwa}).
    The red curve represents the result of the resonance-model fit
    (see \cref{sec:pwa.res_fit,sec:3pi_model:resonance}).  The red
    curve is the coherent sum of the wave components, which are
    represented by the other curves: (blue curve) \Pppi[1800]
    resonance; (green curve) non-resonant contribution.  The
    extrapolations of the model and the wave components beyond the
    fitted \mThreePi~range are shown in lighter colors.
    \subfloatLabel{fig:tspectrum_0mp_f0}~$t'$~spectra of the two
    components in the \wave{0}{-+}{0}{+}{\PfZero[980]}{S} wave as
    given by \cref{eq:t_spectrum}.  In each $t'$~bin, horizontal lines
    indicate the central value of the yield of the respective model
    component.  The horizontal extent of the lines indicate the width
    of the $t'$~bins.  The statistical uncertainty of the yield is
    represented by the height of a shaded box around the central value
    (for most bins invisibly small).  The \Pppi[1800] component is
    shown as blue lines and light blue boxes, and the non-resonant
    component as black lines and gray boxes.  The red and green
    horizontal lines represent the integrals of the model function
    over the $t'$~bins and can be directly compared to the data.  The
    red and green curves represent fits using
    \cref{eq:t_spectrum_model}.  Extrapolations of the model beyond
    the fitted $t'$~range are shown in lighter colors.}
  \label{fig:intensity_phase_0mp}
\end{figure}

The results of our analysis of the \threePi proton-target data confirm
the small branching of the \Pppi[1800] into $\Pprho \pi$.  The
intensity distribution of the \wave{0}{-+}{0}{+}{\Pprho}{P} wave in
\cref{fig:int_0mp_rho} exhibits only a small dip-like structure in the
\SI{1.8}{\GeVcc} region, which could be due to the \Pppi[1800], but
the spectrum is dominated by a broad peak around \SI{1.4}{\GeVcc}.
Consistent with previous experiments, we observe clear \Pppi[1800]
peaks in the $\pipiS \pi$ and $\PfZero[980] \pi$ waves (see
\cref{fig:int_0mp_pipiS,fig:intensity_0mp_f0_tbin1}).  However,
contrary to the result of \refCite{Chung:2002pu}, we also observe a
\Pppi[1800] peak in the $\PfZero[1500] \pi$ wave (see
\cref{fig:int_0mp_f01500}).  The signal in this wave is about a
factor~25 smaller than the one in the $\PfZero[980] \pi$ wave and
hence might have been undetectable in previous experiments.

\begin{figure}[tbp]
  \centering
  \subfloat[][]{%
    \includegraphics[width=\threePlotWidthTwoD]{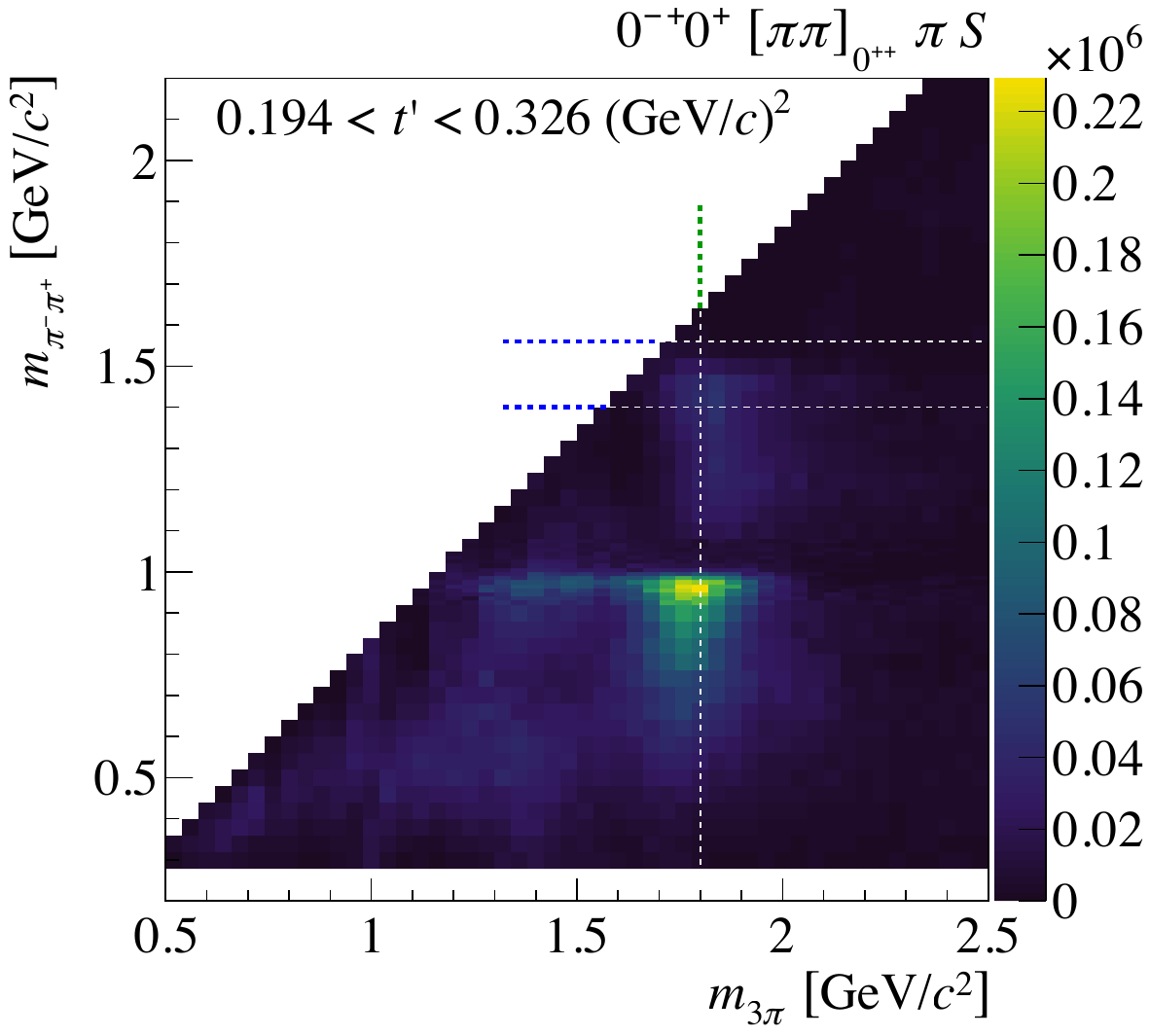}%
    \label{fig:0mp_pipiS_highT_2d}
  }%
  \hfill%
  \subfloat[][]{%
    \includegraphics[width=\threePlotWidth]{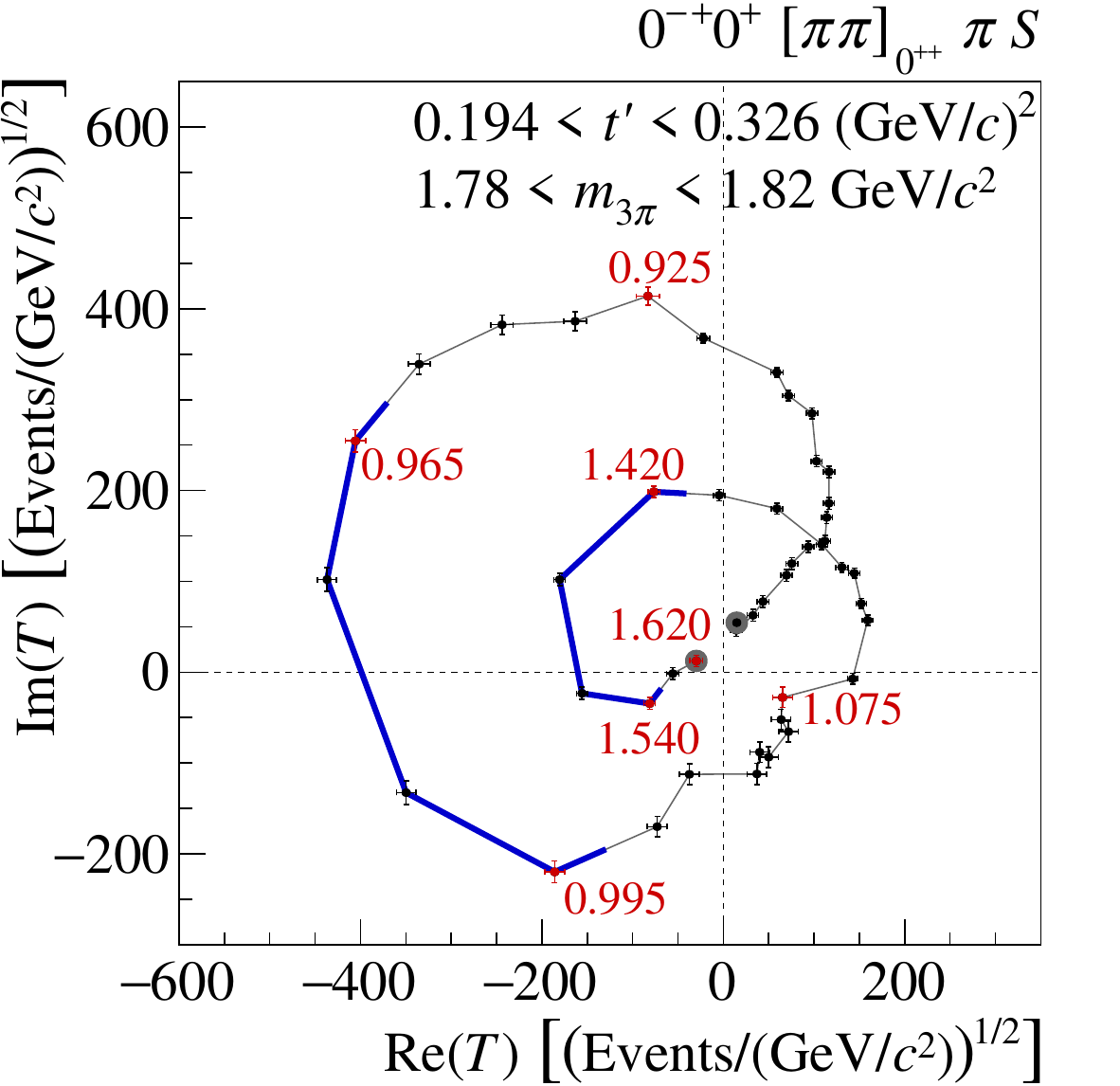}%
    \label{fig:0mp_pipiS_1.8_highT_argand}
  }%
  \hfill%
  \subfloat[][]{%
    \includegraphics[width=\threePlotWidth]{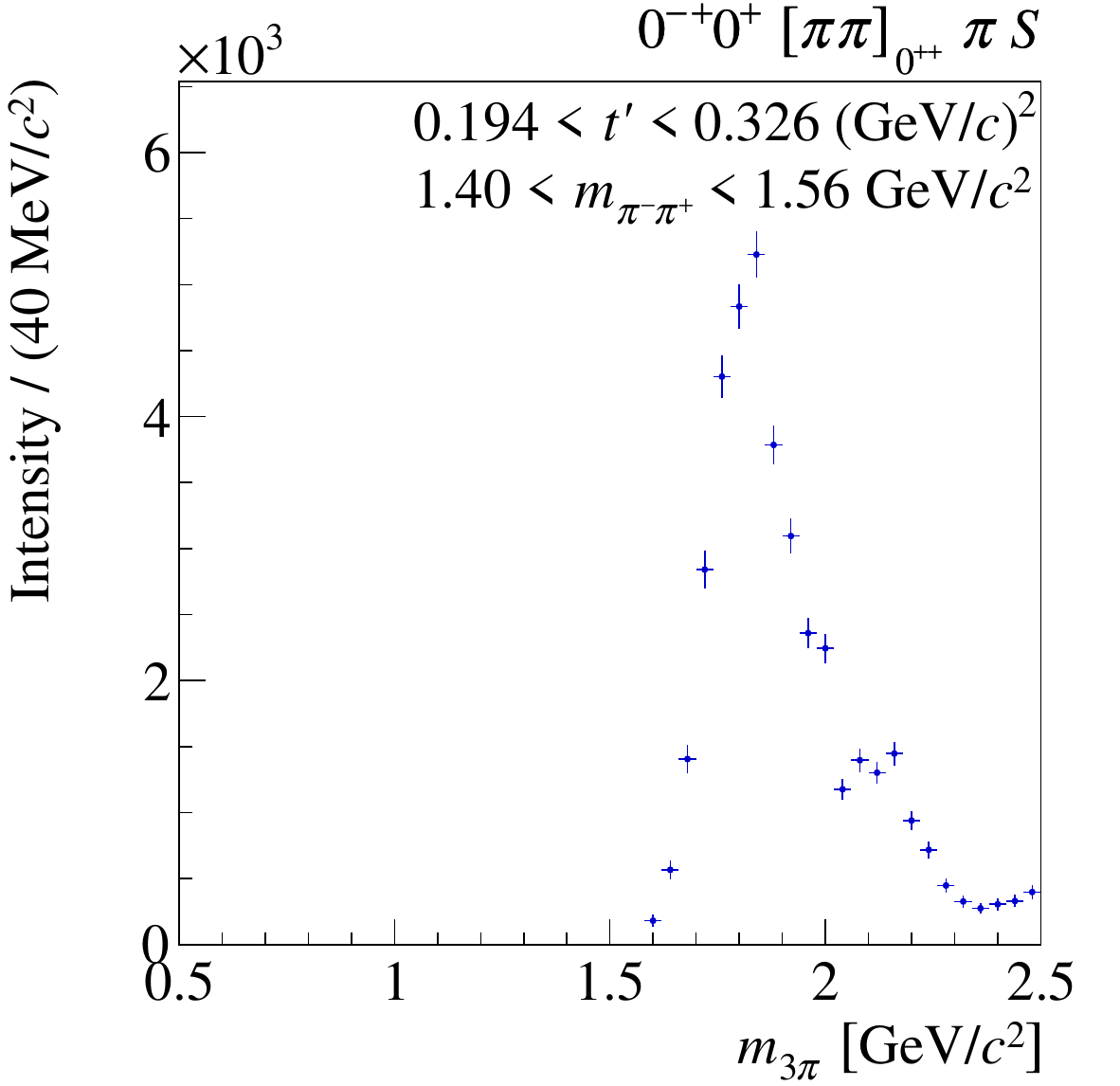}%
    \label{fig:0mp_pipiS_highT_f01500}
  }%
  \caption{Amplitude of the \wave{0}{-+}{0}{+}{\pipiSF}{S} wave with
    the freed-isobar amplitude \pipiSF in an intermediate $t'$~bin of
    the \threePi proton-target data~\cite{Adolph:2015tqa} (see
    \cref{sec:pwa_cells:freed_isobar,sec:3pi_model:pwa}).
    \subfloatLabel{fig:0mp_pipiS_highT_2d}~Two-dimensional
    representation of the intensity of the
    \wave{0}{-+}{0}{+}{\pipiSF}{S} wave as a function of~\mTwoPi
    and~\mThreePi.
    \subfloatLabel{fig:0mp_pipiS_1.8_highT_argand}~\Argand of the
    \pipiSF freed-isobar amplitude as a function of~\mTwoPi for the
    \mThreePi~bin at the \Pppi[1800] mass as indicated by the vertical
    dashed line in~\subfloatLabel{fig:0mp_pipiS_highT_2d}.  The
    crosses with error bars are the result of the PWA fit.  The
    numbers in the \Argand show the corresponding \mTwoPi~values.  The
    data points are connected by lines indicating the order of
    increasing~\mTwoPi.  The line segments highlighted in blue
    correspond to the \mTwoPi~ranges around the \PfZero[980] from
    \SIrange{960}{1000}{\MeVcc} and around the \PfZero[1500] from
    \SIrange{1400}{1560}{\MeVcc}.  The $2\pi$ mass is binned in
    \SI{10}{\MeVcc} wide intervals around the \PfZero[980] and in
    \SI{40}{\MeVcc} wide intervals elsewhere.  The overall phase of
    the \Argand is fixed by the \wave{1}{++}{0}{+}{\Pprho}{S} wave.
    \subfloatLabel{fig:0mp_pipiS_highT_f01500}~Intensity as a function
    of~\mThreePi summed over the selected \mTwoPi~interval around the
    \PfZero[1500] as indicated by the pair of horizontal dashed lines
    in~\subfloatLabel{fig:0mp_pipiS_highT_2d}.}
  \label{fig:pipi_s_wave_0mp_highT}
\end{figure}

The resonant nature of the \PfZero[980] and the \PfZero[1500] in the
\Pppi[1800] decay is confirmed by the results of the freed-isobar PWA
(see \cref{sec:pwa_cells:freed_isobar,sec:3pi_model:pwa}).
\Cref{fig:0mp_pipiS_highT_2d} shows the correlation of the
\mThreePi~intensity distribution of the \wave{0}{-+}{0}{+}{\pipiSF}{S}
wave with the \mTwoPi~intensity distribution of the freed-isobar
amplitude with $\JPC = 0^{++}$.  The distribution exhibits a clear
peak at $\mThreePi \approx \SI{1.8}{\GeVcc}$ and
$\mTwoPi \approx \SI{1.0}{\GeVcc}$.  In the \Pppi[1800]
\mThreePi~region, a circular resonance structure appears in the
\PfZero[980] region in the \Argand (see
\cref{fig:0mp_pipiS_1.8_highT_argand}).  A second circular resonance
structure appears in the \PfZero[1500] region.  The
\mThreePi~intensity distribution in the \PfZero[1500] region, as shown
in \cref{fig:0mp_pipiS_highT_f01500}, exhibits a clear \Pppi[1800]
peak and is similar to the corresponding distribution from the
conventional PWA with fixed parameterizations for the isobar
resonances shown in \cref{fig:int_0mp_f01500}.  These results show
that also in the $3\pi$~final state the \Pppi[1800] decays into
$\PfZero[1500] \pi$, as it is expected based on the observation of
this decay mode in the $\eta \eta \pi$ final state.

It is worth noting, that the \Argand in
\cref{fig:0mp_pipiS_1.8_highT_argand} shows no significant signal of a
narrow \PfZero[1370],\footnote{The PDG estimates only a range for the
  Breit--Wigner width of the \PfZero[1370] of
  \SIrange{200}{500}{\MeVcc}~\cite{Tanabashi:2018zz}.} although
$\PfZero[1370] \pi$ was claimed as a $3\pi$~decay mode of the
\Pppi[1800] by previous
experiments~\cite{Bellini:1982ec,Amelin:1995gu}.  In addition, the VES
experiment measured the branching-fraction ratio
$\text{BF}[\Pppi[1800] \to \PfZero[980] \pi] / \text{BF}[\Pppi[1800]
\to \PfZero[1370] \pi] = \num{1.7(13)}$~\cite{Amelin:1995gu}.  A
reason for this discrepancy could be the strong dependence of
$\PfZero[500] \pi$ and $\PfZero[1370] \pi$ branching fractions on the
PWA model, in particular on the parameterizations employed for the
dynamical amplitudes of the overlapping $0^{++}$ isobars.  Also in
other \mThreePi~slices and for other \JPC quantum numbers of the
$3\pi$ system (see \eg
\cref{fig:PIPIS_1pp_above_res_argand,fig:PIPIS_2mp_1.9_argand} and
Section~VI in \refCite{Adolph:2015tqa}), we do not observe significant
signals of a narrow \PfZero[1370] in the extracted isobar amplitudes.
Our results from the freed-isobar PWA hence call into question the
existence of the \PfZero[1370] as a narrow $\pi\pi$ resonance.

\begin{figure}[tbp]
  \centering
  \subfloat[]{%
    \includegraphics[width=0.5\textwidth]{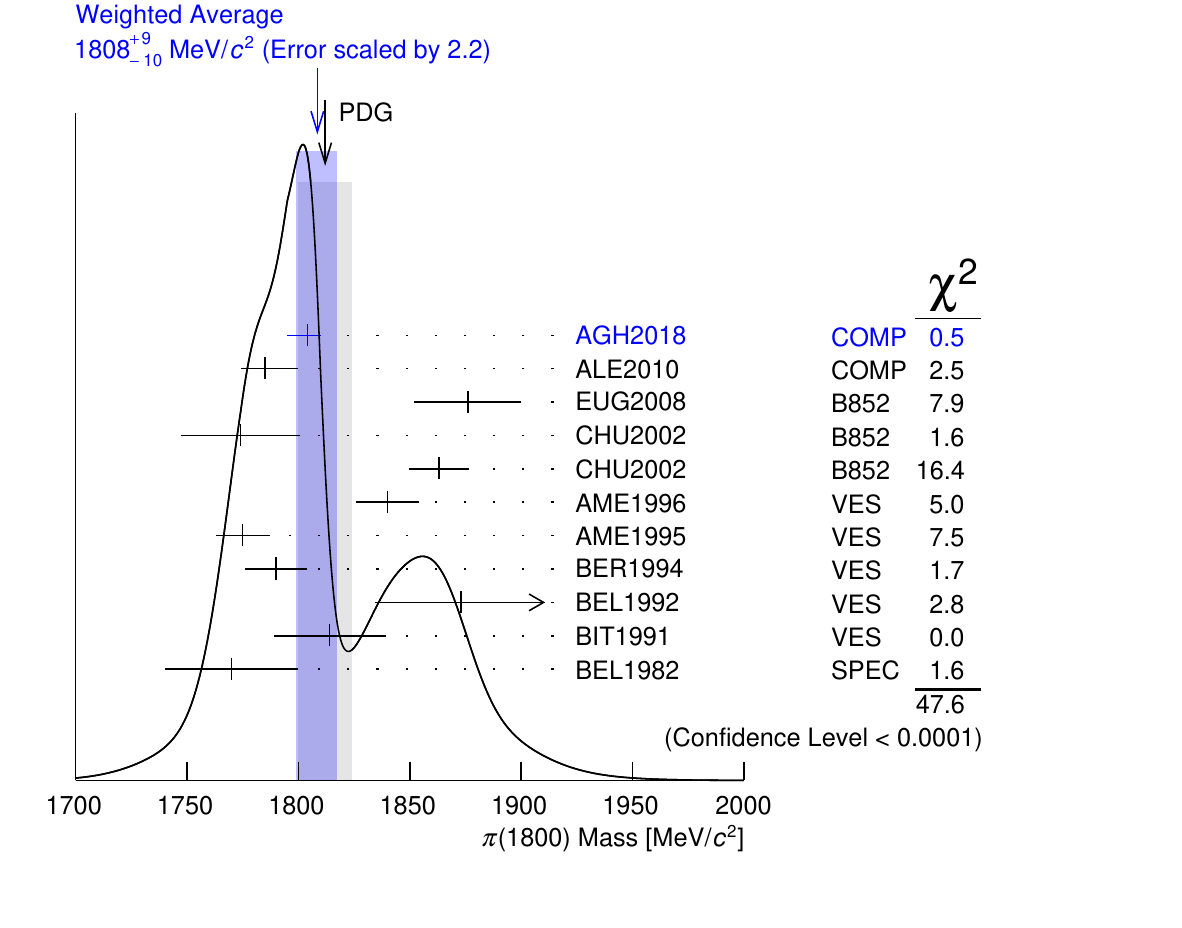}%
    \label{fig:ideogram_pi_1800_mass}%
  }%
  \subfloat[]{%
    \includegraphics[width=0.5\textwidth]{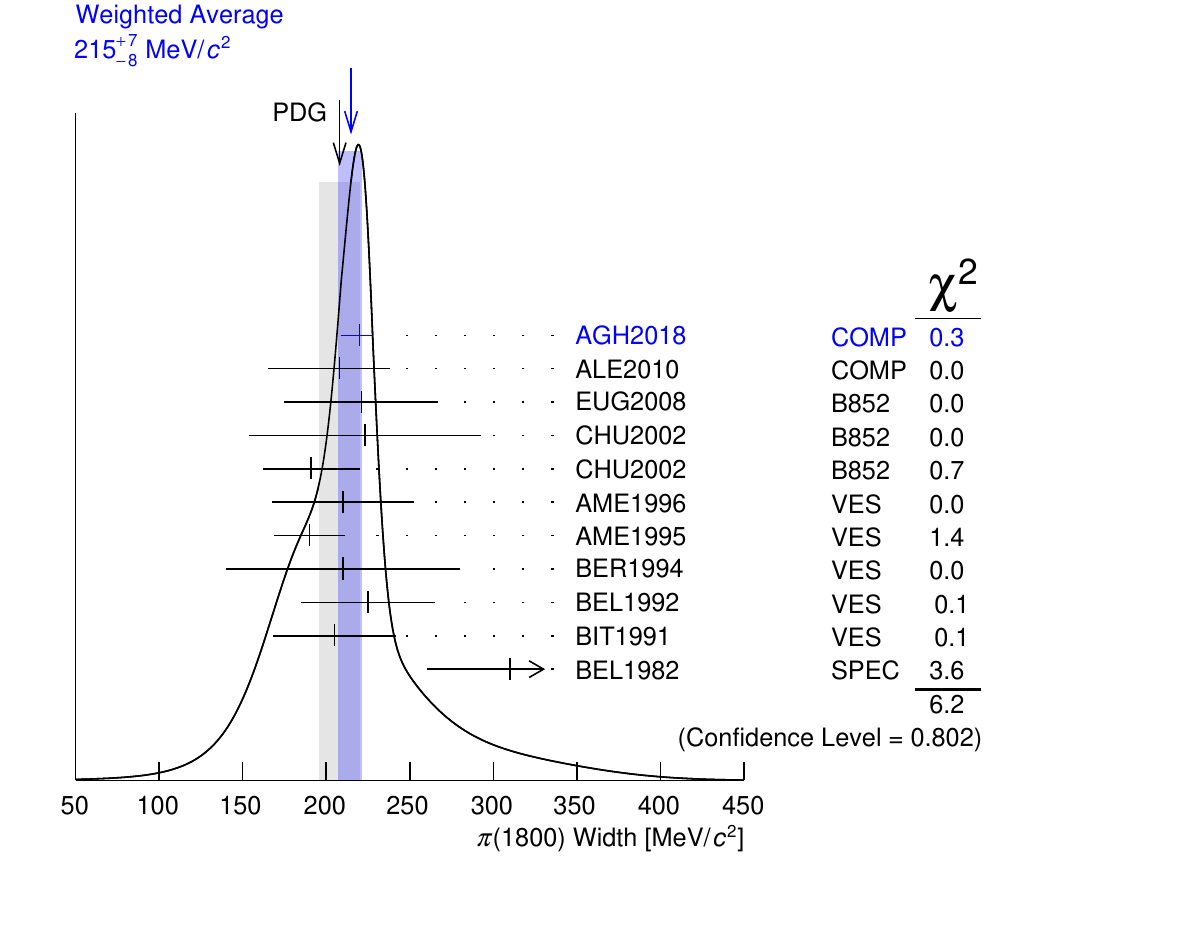}%
    \label{fig:ideogram_pi_1800_width}%
  }%
  \caption{Ideograms for \subfloatLabel{fig:ideogram_pi_1800_mass}~the
    mass and \subfloatLabel{fig:ideogram_pi_1800_width}~the width of
    the \Pppi[1800] as defined by the Particle Data Group (see
    Section~5.2.2 in \refCite{Tanabashi:2018zz}).  The Breit--Wigner
    parameters obtained from fits of the COMPASS \threePi
    proton-target (AGH2018,~\cite{Akhunzyanov:2018lqa}) and
    lead-target data (ALE2010,~\cite{Alekseev:2009aa}) are compared to
    previous measurements.  Values that are based on COMPASS data and
    that are not yet included in the present PDG
    averages~\cite{Tanabashi:2018zz} are shown in blue.  The weighted
    averages of all data points (blue arrows and blue-shaded bands)
    are compared to the present PDG averages (black arrows and
    gray-shaded bands).}
  \label{fig:ideogram_pi_1800}
\end{figure}

We find the cleanest \Pppi[1800] signal in the $\PfZero[980] \pi$
wave, which is the only $0^{-+}$ wave included in the resonance-model
fit.  \Cref{fig:intensity_0mp_f0_tbin1} shows the intensity
distribution of the \wave{0}{-+}{0}{+}{\PfZero[980]}{S} wave in the
lowest $t'$~bin.  It exhibits a pronounced peak at \SI{1.8}{\GeVcc}
that is accompanied by a rapid phase motion (see
\cref{fig:phase_0mp_f0_1pp_rho_tbin1}).  Although all final-state
particles in the $\PfZero[980] \pi$ wave are in relative $S$~waves,
the amplitude of this wave is reliably extracted from the data and is
found to be robust against changes of the PWA model.  The resonance
model (red curve) contains one resonance term for the \Pppi[1800]
(blue curve) and one non-resonant term (green curve).  From the fit,
we obtain the Breit--Wigner parameters
$m_{\Pppi[1800]} = \SIaerr{1804}{6}{9}{\MeVcc}$ and
$\Gamma_{\Pppi[1800]} =
\SIaerr{220}{8}{11}{\MeVcc}$~\cite{Akhunzyanov:2018lqa}.  Our
measurement of the \Pppi[1800] parameters is the so far most accurate
and in good agreement with the PDG world
average~\cite{Tanabashi:2018zz} as well as with our result of
$m_{\Pppi[1800]} = \SIsaerrs{1785}{9}{12}{6}{\MeVcc}$ and
$\Gamma_{\Pppi[1800]} = \SIsaerrs{208}{22}{21}{37}{\MeVcc}$ from the
analysis of the \threePi lead-target data~\cite{Alekseev:2009aa}.  The
ideograms in \cref{fig:ideogram_pi_1800} show that the measured
\Pppi[1800] width values are in good agreement, whereas the mass
values have a large spread and fall into two clusters centered at
about \SI{1780}{\MeVcc} and \SI{1860}{\MeVcc}.  Our new value from the
analysis of the \threePi proton-target data (top point in
\cref{fig:ideogram_pi_1800_mass}) falls inbetween these two clusters.
It is interesting to note that the measurement that has the largest
discrepancy \wrt the weighted average of the \Pppi[1800] mass in
\cref{fig:ideogram_pi_1800_mass} was obtained from a resonance-model
fit of the \SI{1.8}{\GeVcc} region of the
\wave{0}{-+}{0}{+}{\sigma}{S} amplitude in an analysis of BNL~E852
data on diffractively produced \threePi~\cite{Chung:2002pu}.  Although
this wave exhibits a clear \Pppi[1800] signal also in our data (see
\cref{fig:int_0mp_pipiS}), it contains an additional broad enhancement
at about \SI{1.3}{\GeVcc}, which may contain the \Pppi[1300].
However, the strong $t'$~dependence of the shape of the intensity
distribution in this mass range suggests large non-resonant
components.  In order to reliably measure the \Pppi[1800] resonance
parameters in this wave, the interference of the \Pppi[1800] with the
other wave components that dominate the \SI{1.3}{\GeVcc} region would
need to be taken into account, like in our fit of the \Pppi[1800] in
the \wave{0}{-+}{0}{+}{\PfZero[980]}{S} wave.  This, however, was not
done in the analysis in \refCite{Chung:2002pu}.

The low-mass shoulder in \cref{fig:intensity_0mp_f0_tbin1} is well
described by the non-resonant component.  We do not find evidence for
a possible \Pppi[1300] component in the
\wave{0}{-+}{0}{+}{\PfZero[980]}{S} wave (see Section~VI.A.2 in
\refCite{Akhunzyanov:2018lqa} for details).  Interestingly, the
non-resonant component (black lines in \cref{fig:tspectrum_0mp_f0})
has an extremely large slope parameter value of
\SIaerr{26}{6}{5}{\perGeVcsq} and its $t'$~spectrum exhibits a dip at
$t' \approx \SI{0.3}{\GeVcsq}$.  This behavior is strikingly different
from that of all other resonant components in our model and is in
agreement with the absence of a significant \Pppi[1300] component in
the $\PfZero[980] \pi$ wave.  Similar dip-like $t'$~spectra were
observed in other inelastic diffraction-dissociation reactions, for
example in the reaction
$n + p \to p \pi^- + p$~\cite{Biel:1975eu,Biel:1976zu}, and were
explained by a cancellation of double-Regge exchange processes and
direct resonance production (see \eg\
\refsCite{CohenTannoudji:1976tj,Hayot:1977uk,Kaidalov:1979jz}).  In
contrast, the $t'$~spectrum of the \Pppi[1800], which is shown as blue
lines in \cref{fig:tspectrum_0mp_f0}, has an approximately exponential
behavior with a slope parameter value of
\SIaerr{8.8}{0.7}{0.3}{\perGeVcsq}, which is consistent with a
resonance.

\subsubsection{The $\JPC = 1^{++}$ Sector }
\label{sec:results_1pp}

Currently, six \PaOne* states are known~\cite{Tanabashi:2018zz} (see
also \cref{fig:light_flavorless_spectrum}).  However, only the \PaOne
ground state is considered an established state by the PDG.  The
\PaOne[1420] and \PaOne[1640] are omitted from the summary table.  The
\PaOne[1930], the \PaOne[2095], and the \PaOne[2270] are listed as
\textquote{further states}.

Although the existence of the \PaOne as the isovector $\JPC = 1^{++}$
ground state, \ie the \termSym{1}{3}{P}{1} quark-model state, is well
established, the parameters of the \PaOne are not well known.
Depending on the analyzed process and the employed parameterizations,
the measured values of the \PaOne parameters differ
substantially~\cite{pdg_a1_1260:2006}.  The reported values for the
\PaOne mass cover a wide range from
\SI{1041(13)}{\MeVcc}~\cite{Gavillet:1977kx} to
\SIerrs{1331}{10}{3}{\MeVcc}~\cite{Asner:1999kj}; the values for the
\PaOne width range from \SI{230(50)}{\MeVcc}~\cite{Gavillet:1977kx} to
\SIerrs{814}{36}{13}{\MeVcc}~\cite{Asner:1999kj}.  Due to the large
spread of the measured parameter values, the PDG does not perform an
average but only gives estimates of
$m_{\PaOne} = \SI{1230(40)}{\MeVcc}$ and
$\Gamma_{\PaOne} =
\SIrange{250}{600}{\MeVcc}$~\cite{Tanabashi:2018zz}.

\begin{figure}[tbp]
  \centering
  \hfill%
  \subfloat[][]{%
    \includegraphics[width=\threePlotWidth]{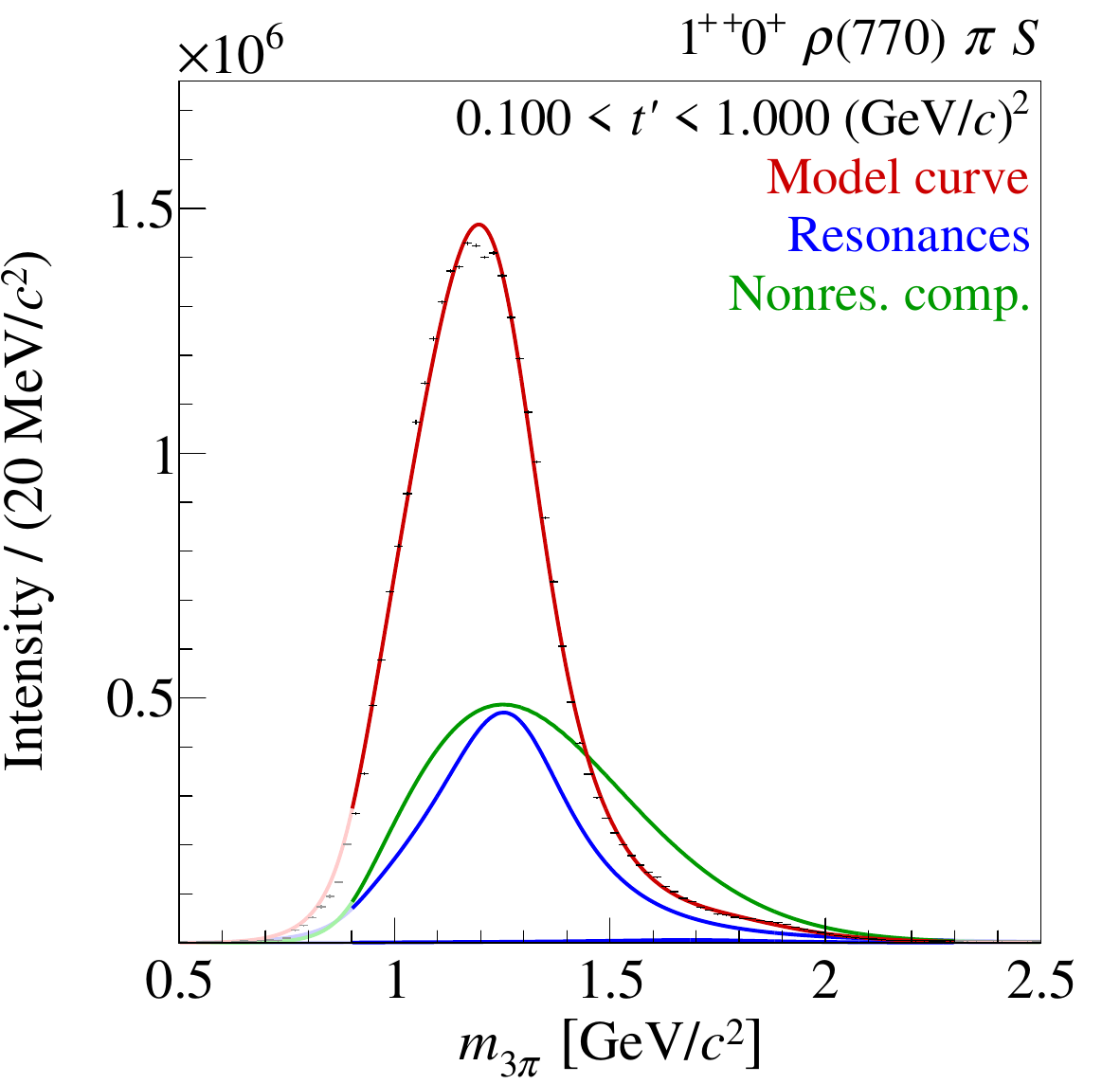}%
    \label{fig:intensity_1pp_rho_tsum}%
  }%
  \hfill%
  \subfloat[][]{%
    \includegraphics[width=\threePlotWidth]{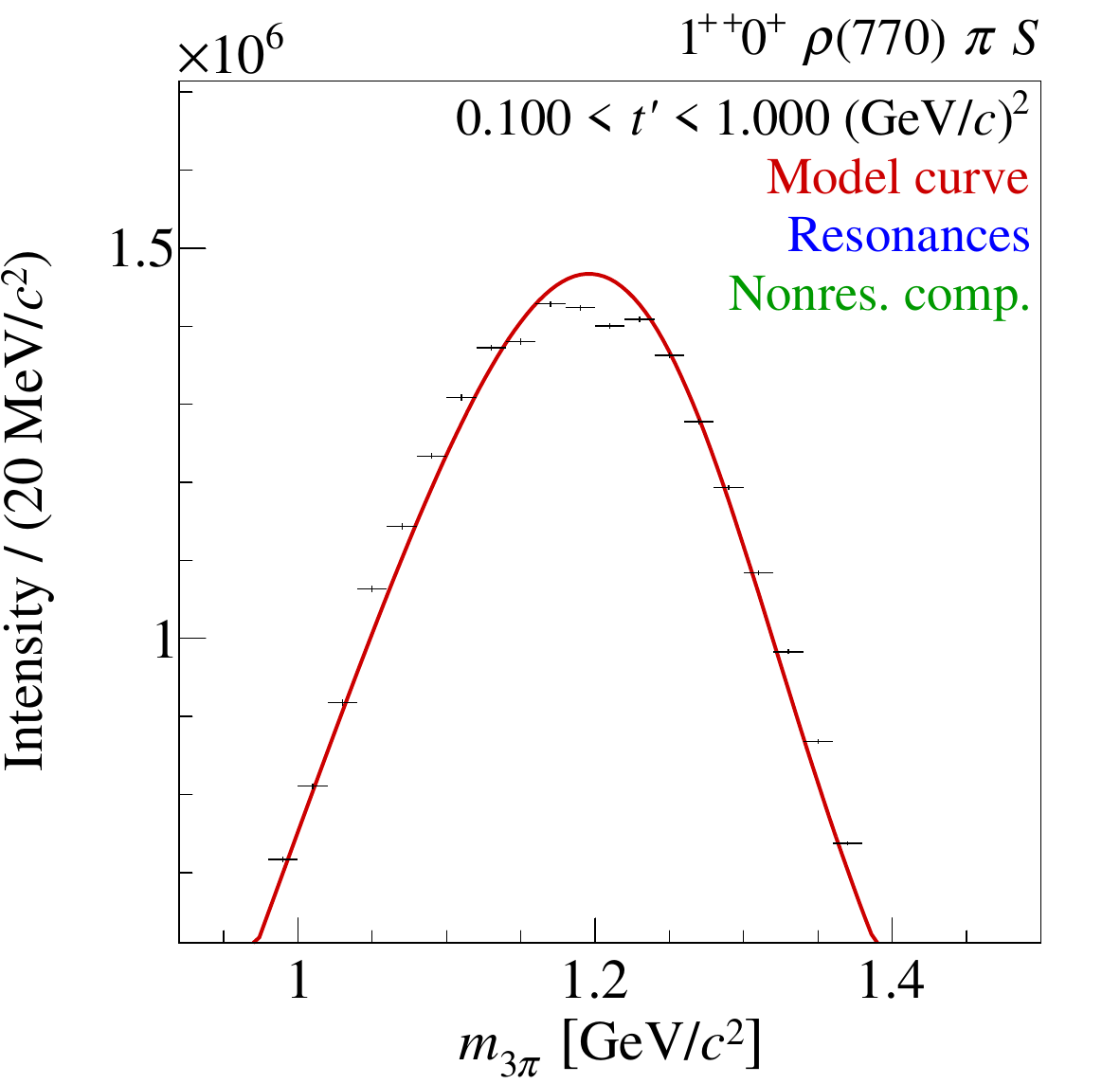}%
    \label{fig:intensity_1pp_rho_tsum_zoom}%
  }%
  \hfill\null%
  \caption{Intensity distribution of the \wave{1}{++}{0}{+}{\Pprho}{S}
    wave in the \threePi proton-target data summed over the
    11~$t'$~bins~\cite{Akhunzyanov:2018lqa}.  The curves represent the
    result of the resonance-model fit.  The model and the wave
    components are represented as in \cref{fig:intensity_phase_0mp}.
    The dominant resonant component is the \PaOne; the \PaOne[1640] is
    barely visible.
    \subfloatLabel{fig:intensity_1pp_rho_tsum_zoom}~shows a zoomed
    view of~\subfloatLabel{fig:intensity_1pp_rho_tsum}.}
  \label{fig:intensity_1pp_rho}
\end{figure}

The \PaOne decays mainly to~$3\pi$. The $\Pprho \pi$ $S$-wave decay
mode is the most dominant one with a branching fraction of
\SI{60.19}{\percent}~\cite{Asner:1999kj}.  The branching fractions
into $\sigma \pi$ and $\PfZero[1370] \pi$ are also large, whereas
those into the $\Pprho \pi$ $D$~wave and $\PfTwo \pi$ are small.  This
is consistent with our \threePi lead- and proton-target data, where
the \PaOne in the \wave{1}{++}{0}{+}{\Pprho}{S} wave is the by far
largest resonance signal (see
\cref{fig:3pi_spin_totals,fig:intensity_1pp_rho_tsum}).  Hence the
\wave{1}{++}{0}{+}{\Pprho}{S} wave plays a special role in the
resonance-model fits.  In the highly precise proton-target data, the
statistical uncertainties of the \wave{1}{++}{0}{+}{\Pprho}{S}
amplitude are extremely small, so that the resonance model has
difficulties to describe all details of the data.  This applies in
particular to the peak region in the intensity distribution of this
wave as shown in \cref{fig:intensity_1pp_rho_tsum_zoom}.  Hence the
\wave{1}{++}{0}{+}{\Pprho}{S} amplitude has a large contribution of
about \SI{25}{\percent} to the minimum value of the $\chi^2$~function
of the fit model (see \cref{eq:fit_chi2}).  The deviations of the
model from the data induce a multi-modal behavior of the minimization
procedure and comparatively large systematic uncertainties of the
resonance parameters of the \PaOne and of other resonances (see
Sections~IV.B and~V in \refCite{Akhunzyanov:2018lqa} for details).

\begin{figure}[tbp]
  \centering
  \subfloat[][]{%
    \includegraphics[width=\threePlotWidth]{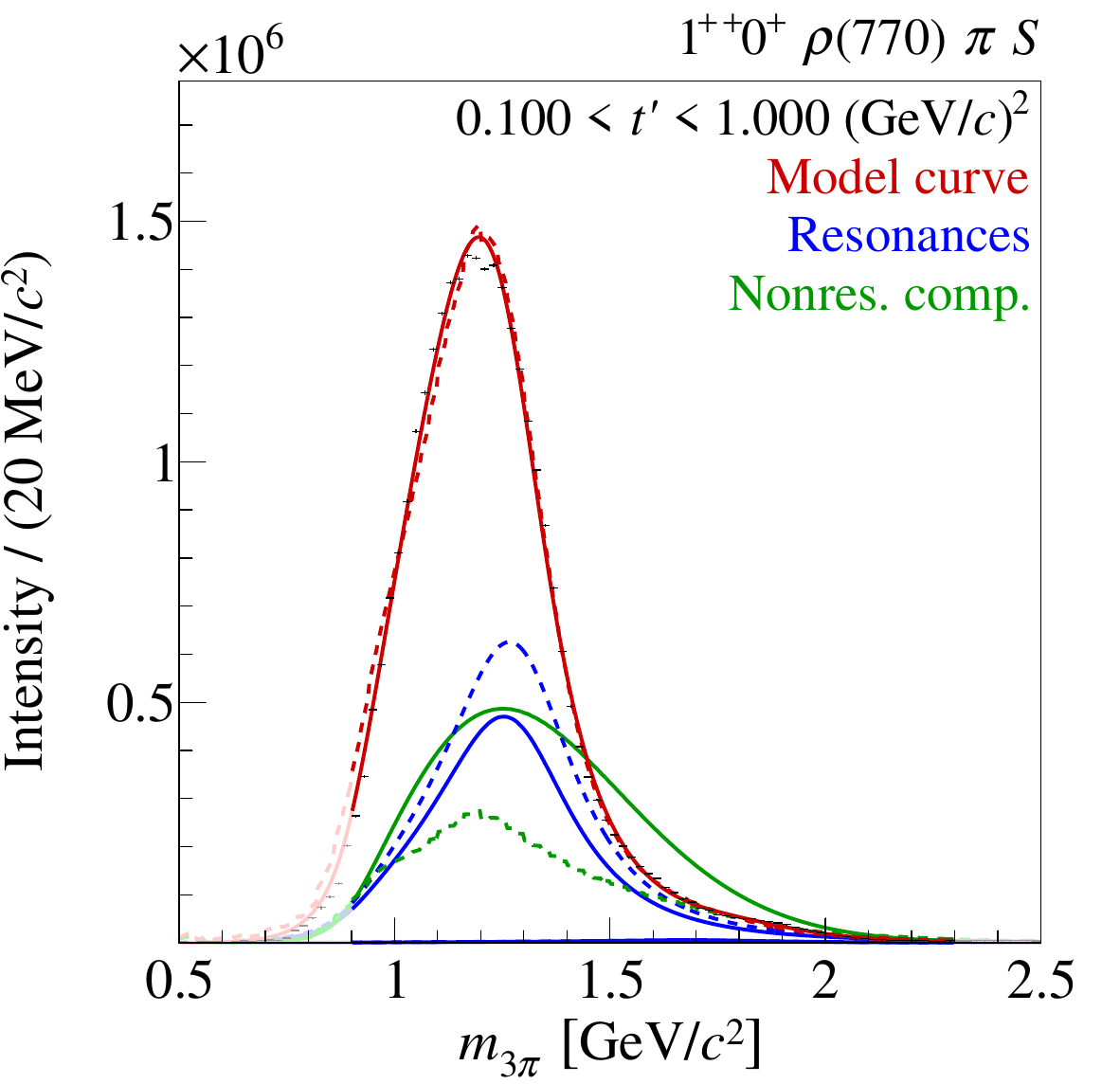}%
    \label{fig:intensity_1pp_rho_tbin1}%
  }%
  \hfill%
  \subfloat[][]{%
    \includegraphics[width=\threePlotWidth]{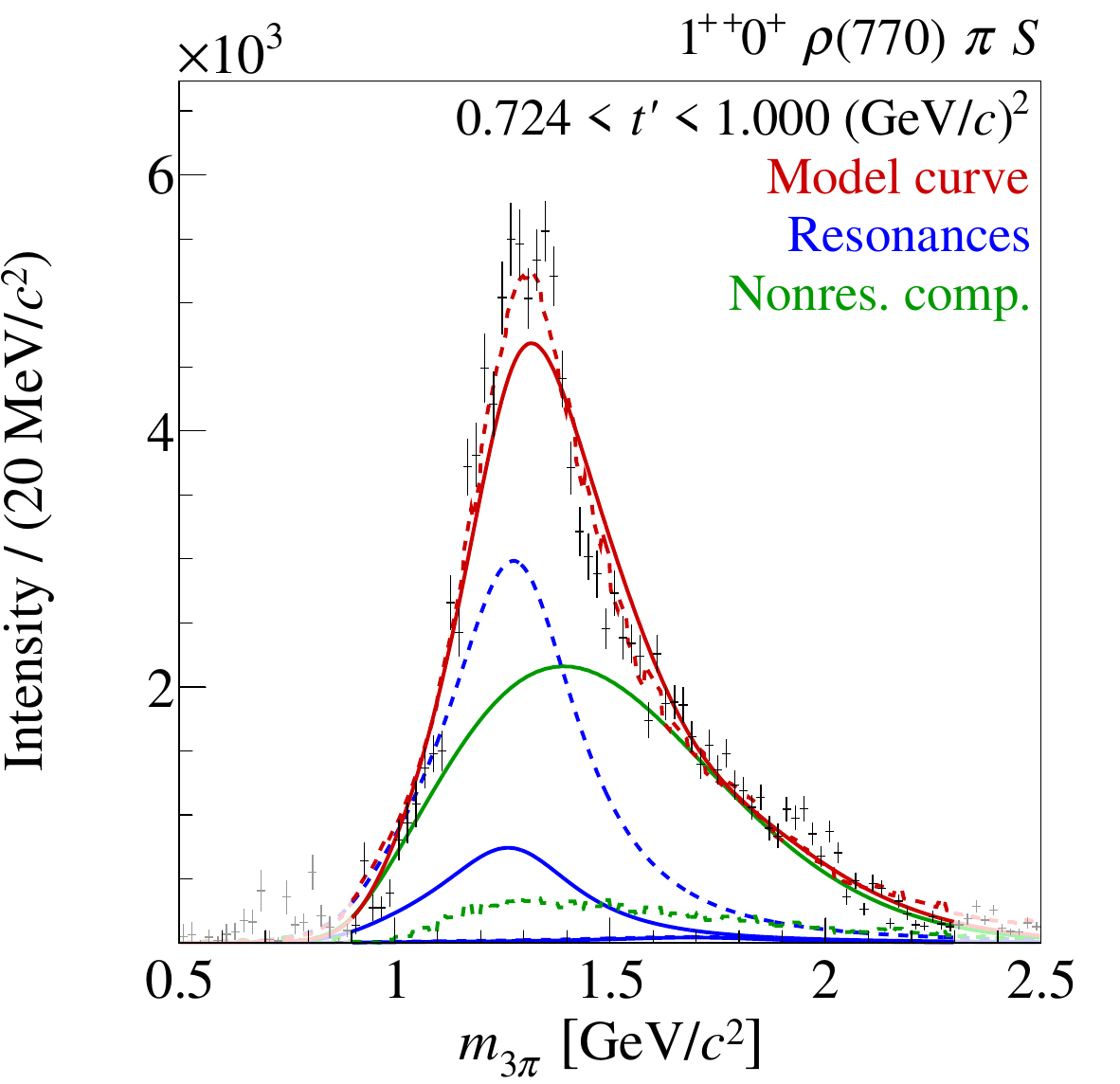}%
    \label{fig:intensity_1pp_rho_tbin11}%
  }%
  \hfill%
  \subfloat[][]{%
    \includegraphics[width=\threePlotWidth]{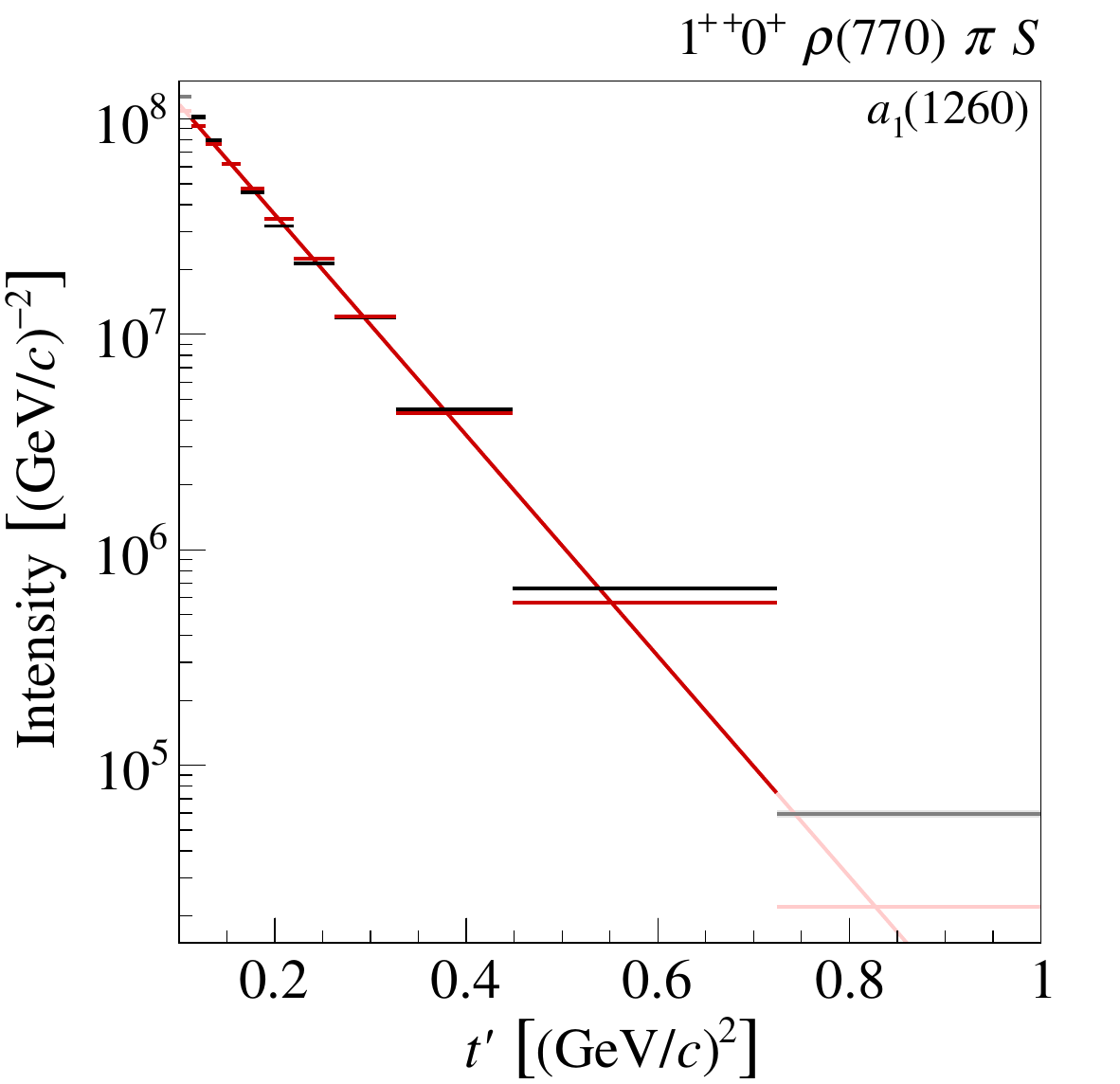}%
    \label{fig:tspectrum_1pp_rho}%
  }%
  \caption{Intensity distribution of the \wave{1}{++}{0}{+}{\Pprho}{S}
    wave \subfloatLabel{fig:intensity_1pp_rho_tbin1}~in the lowest and
    \subfloatLabel{fig:intensity_1pp_rho_tbin11}~in the highest
    $t'$~bin of the \threePi proton-target
    data~\cite{Akhunzyanov:2018lqa}.  The curves represent the result
    of two resonance-model fits.  The model and the wave components
    are represented as in \cref{fig:intensity_phase_0mp}.  The
    dominant resonant component is the \PaOne; the \PaOne[1640] is
    barely visible.  The result of the main resonance-model fit is
    represented by the continuous curves.  The dashed curves represent
    the result of a fit, where the empirical parameterizations of the
    non-resonant components are replaced by the square root of the
    intensity distributions of the partial-wave decomposition of Deck
    Monte Carlo data.  \subfloatLabel{fig:tspectrum_1pp_rho}~Similar
    to \cref{fig:tspectrum_0mp_f0}, but showing the $t'$~spectrum of
    the \PaOne in the \wave{1}{++}{0}{+}{\Pprho}{S} wave.}
  \label{fig:tdep_1pp_rho}
\end{figure}

One of the challenges in describing the \wave{1}{++}{0}{+}{\Pprho}{S}
intensity distribution is that the peak in the \PaOne region changes
its position and shape as a function of~$t'$.  This behavior is
illustrated in
\cref{fig:intensity_1pp_rho_tbin1,fig:intensity_1pp_rho_tbin11}.  At
low~$t'$, the peak is located at about \SI{1.15}{\GeVcc} and shifts
toward higher masses with increasing~$t'$, up to about
\SI{1.30}{\GeVcc} in the highest $t'$~bin.  In addition, the peak
narrows significantly.  This demonstrates the necessity to perform a
$t'$-resolved analysis.  It is also a sign that contributions from
non-resonant processes to this partial-wave amplitude are large.
Indeed, the resonance-model fit finds a non-resonant component that is
comparable in strength to the \PaOne in the low and intermediate
$t'$~range and that is even dominant at high~$t'$.  The $t'$~spectrum
of the \PaOne in the \wave{1}{++}{0}{+}{\Pprho}{S} wave exhibits an
approximately exponential dependence on~$t'$ (see
\cref{fig:tspectrum_1pp_rho}).  However, the slope parameter has a
value of \SIaerr{11.8}{0.9}{4.2}{\perGeVcsq}, which is larger than
what one would expect for a resonance.  The slope parameter value is
similar to the one of the non-resonant component in this wave.  This
might be a hint that our resonance model is not able to completely
separate the \PaOne from the non-resonant component.

Since the \wave{1}{++}{0}{+}{\Pprho}{S} wave has a large contribution
from non-resonant processes, the fit result depends strongly on the
parameterization employed for the non-resonant component.  We study
this model dependence by performing a resonance-model fit, where the
empirical parameterizations of the non-resonant components (see
\cref{eq:dyn_amp_non_res_simple,eq:dyn_amp_non_res}) are replaced by
the square root of the intensity distributions obtained from the
partial-wave decomposition of Deck Monte Carlo data (see
\cref{sec:3pi_model:resonance}).  The result of this fit is shown as
dashed curves in
\cref{fig:intensity_1pp_rho_tbin1,fig:intensity_1pp_rho_tbin11}.  The
model describes the \wave{1}{++}{0}{+}{\Pprho}{S} intensity
distributions well.  At high~$t'$, the peak is described even better
than in the main fit (see \cref{fig:intensity_1pp_rho_tbin11}), where
we used the empirical parameterization in \cref{eq:dyn_amp_non_res}
for the non-resonant component.  The shapes of the non-resonant
component from the Deck Monte Carlo are for all $t'$~bins similar to
those found in the main fit, but the yields are considerably smaller.
In turn, the \PaOne yields are significantly larger, in particular at
high~$t'$.

From the 14-wave resonance-model fit to the \threePi proton-target
data, we extract the Breit--Wigner parameters
$m_{\PaOne} = \SIaerr{1299}{12}{28}{\MeVcc}$ and
$\Gamma_{\PaOne} = \SI{380(80)}{\MeVcc}$~\cite{Akhunzyanov:2018lqa}.
The large systematic uncertainties are mostly due to the issues
discussed above.  From the 6-wave fit to the \threePi lead-target
data, we obtain $m_{\PaOne} = \SIsaerrs{1255}{6}{7}{17}{\MeVcc}$ and
$\Gamma_{\PaOne} =
\SIsaerrs{367}{9}{28}{25}{\MeVcc}$~\cite{Alekseev:2009aa}.  In both
analyses, the \PaOne is described by a Breit--Wigner amplitude as in
\cref{eq:BW_mass-dep_width} with a parameterization of the dynamic
width that takes into account the variation of the $\Pprho \pi$ decay
phase space across the resonance width~\cite{Bowler:1987bj} (see also
Eq.~(24) in \refCite{Akhunzyanov:2018lqa}).  Our mass values are
compatible with the PDG estimate and our width values lie close to the
center of the range estimated by the PDG~\cite{Tanabashi:2018zz}.

\begin{figure}[tbp]
  \centering
  \subfloat[][]{%
    \includegraphics[width=\threePlotWidth]{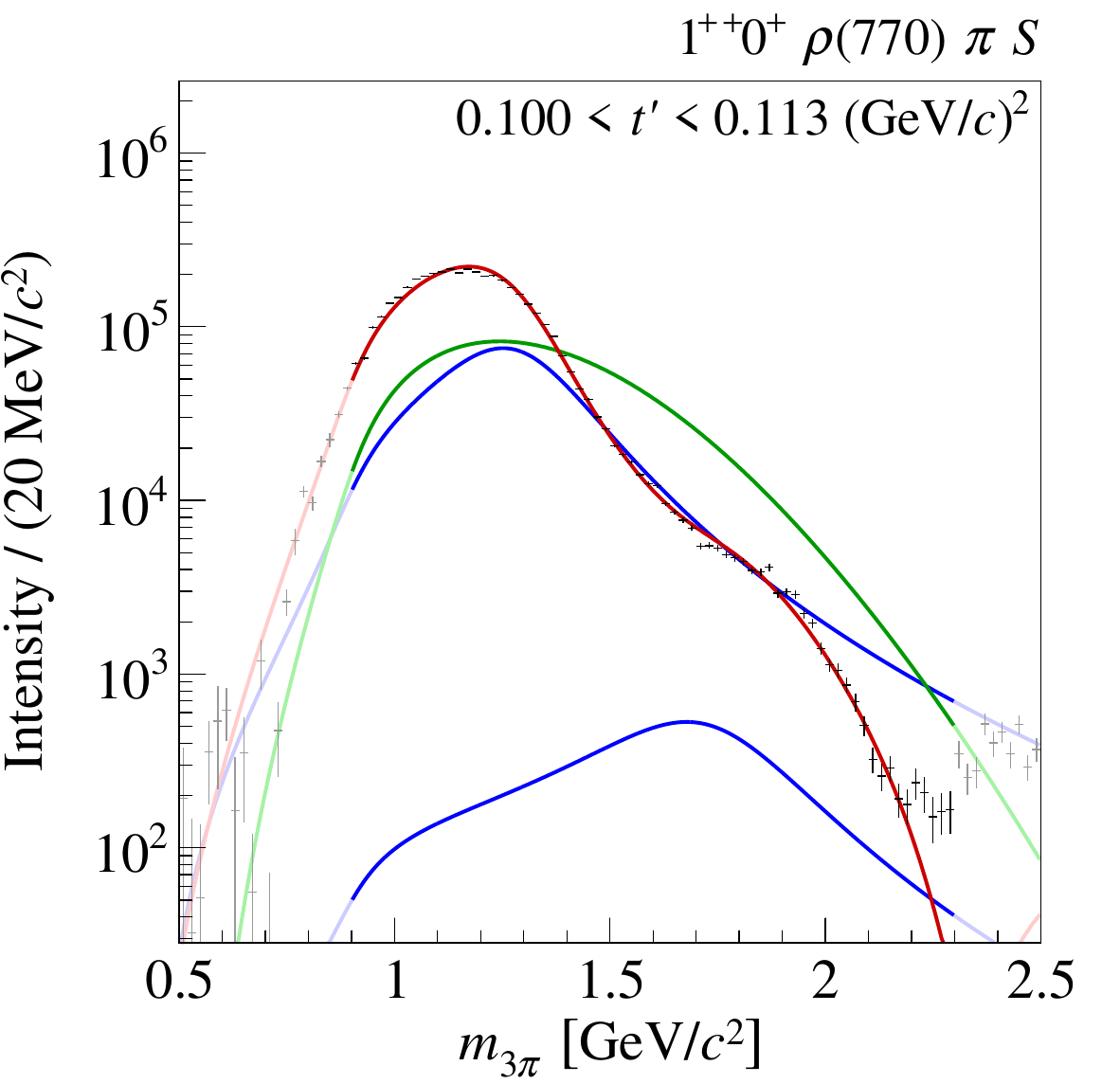}%
    \label{fig:intensity_1pp_rho_tbin1_log}%
  }%
  \hfill%
  \subfloat[][]{%
    \includegraphics[width=\threePlotWidth]{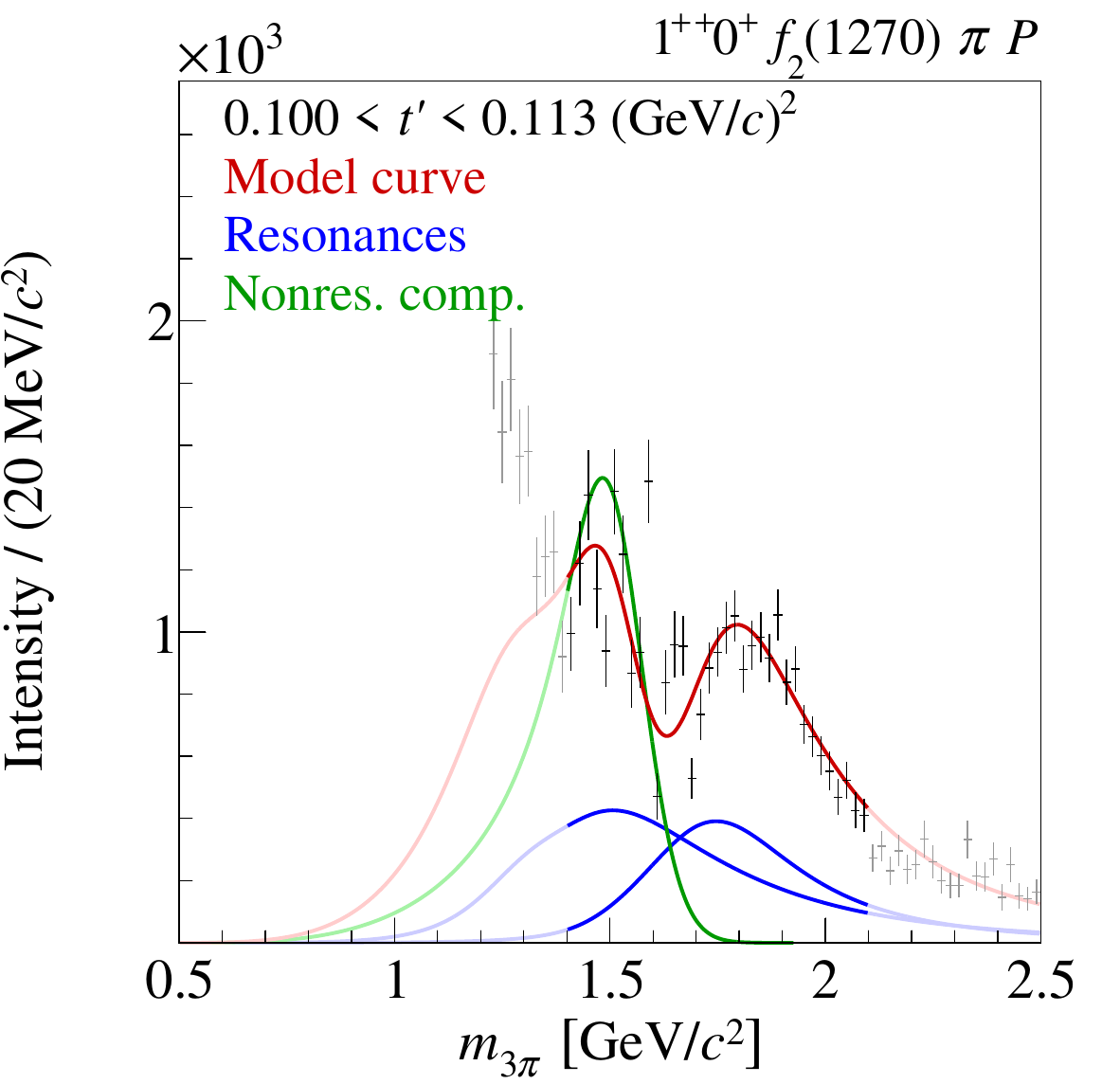}%
    \label{fig:intensity_1pp_f2_tbin1}%
  }%
  \hfill%
  \subfloat[][]{%
    \includegraphics[width=\threePlotWidth]{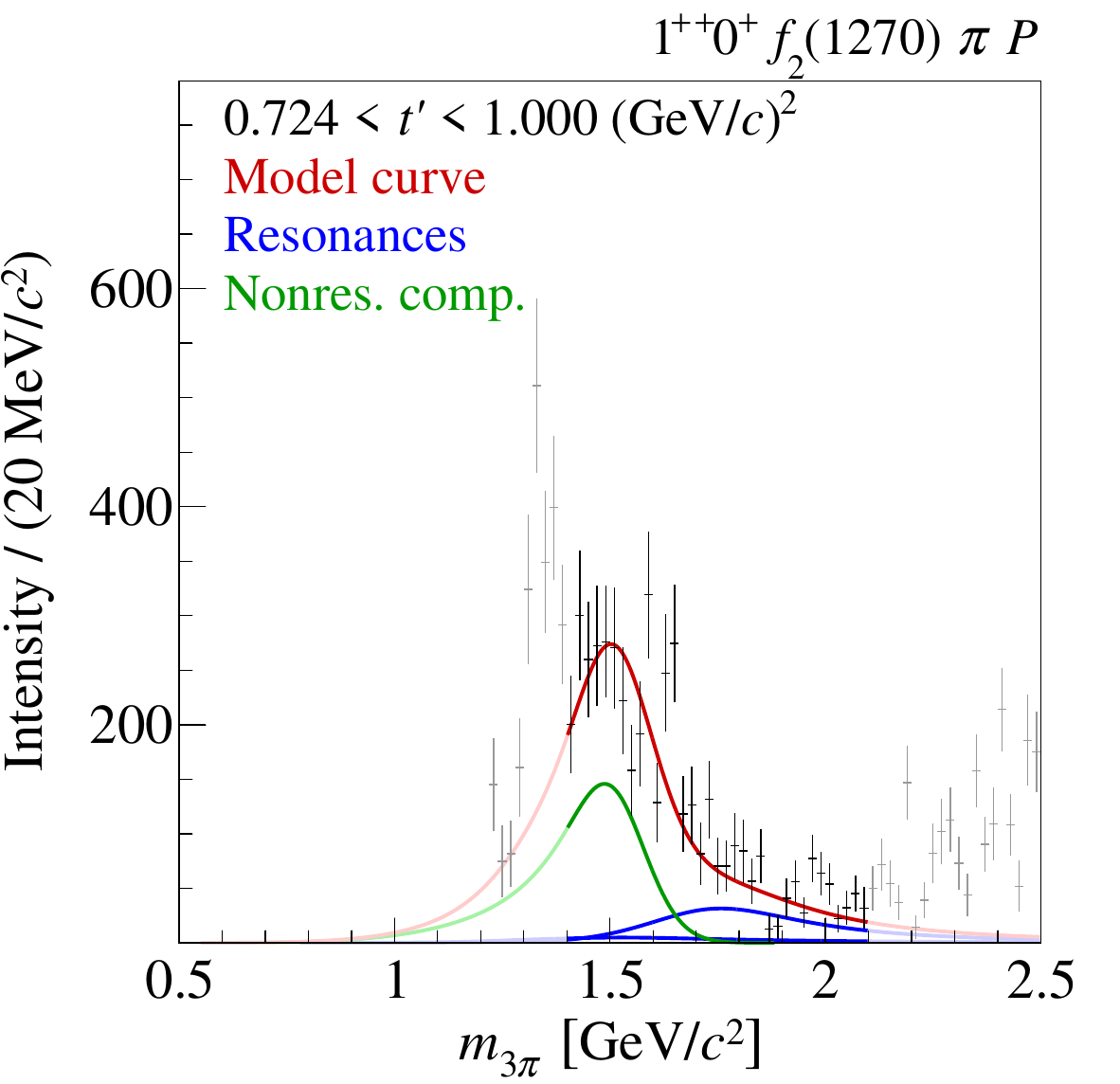}%
    \label{fig:intensity_1pp_f2_tbin11}%
  }%
  \caption{Signals of the \PaOne[1640] in the \threePi proton-target
    data~\cite{Akhunzyanov:2018lqa}.
    \subfloatLabel{fig:intensity_1pp_rho_tbin1_log}~Intensity
    distribution of the \wave{1}{++}{0}{+}{\Pprho}{S} wave in the
    lowest $t'$~bin (same as \cref{fig:intensity_1pp_rho_tbin1} but in
    logarithmic scale).
    \subfloatLabel{fig:intensity_1pp_f2_tbin1}~and~\subfloatLabel{fig:intensity_1pp_f2_tbin11}:
    Intensity distributions of the \wave{1}{++}{0}{+}{\PfTwo}{P} wave
    in the lowest and highest $t'$~bin, respectively.  The curves
    represent the result of the resonance-model fit.  The model and
    the wave components are represented as in
    \cref{fig:intensity_phase_0mp}.}
  \label{fig:intensity_1pp_rho_f2}
\end{figure}

\begin{figure}[tbp]
  \centering
  \subfloat[]{%
    \includegraphics[width=0.5\textwidth]{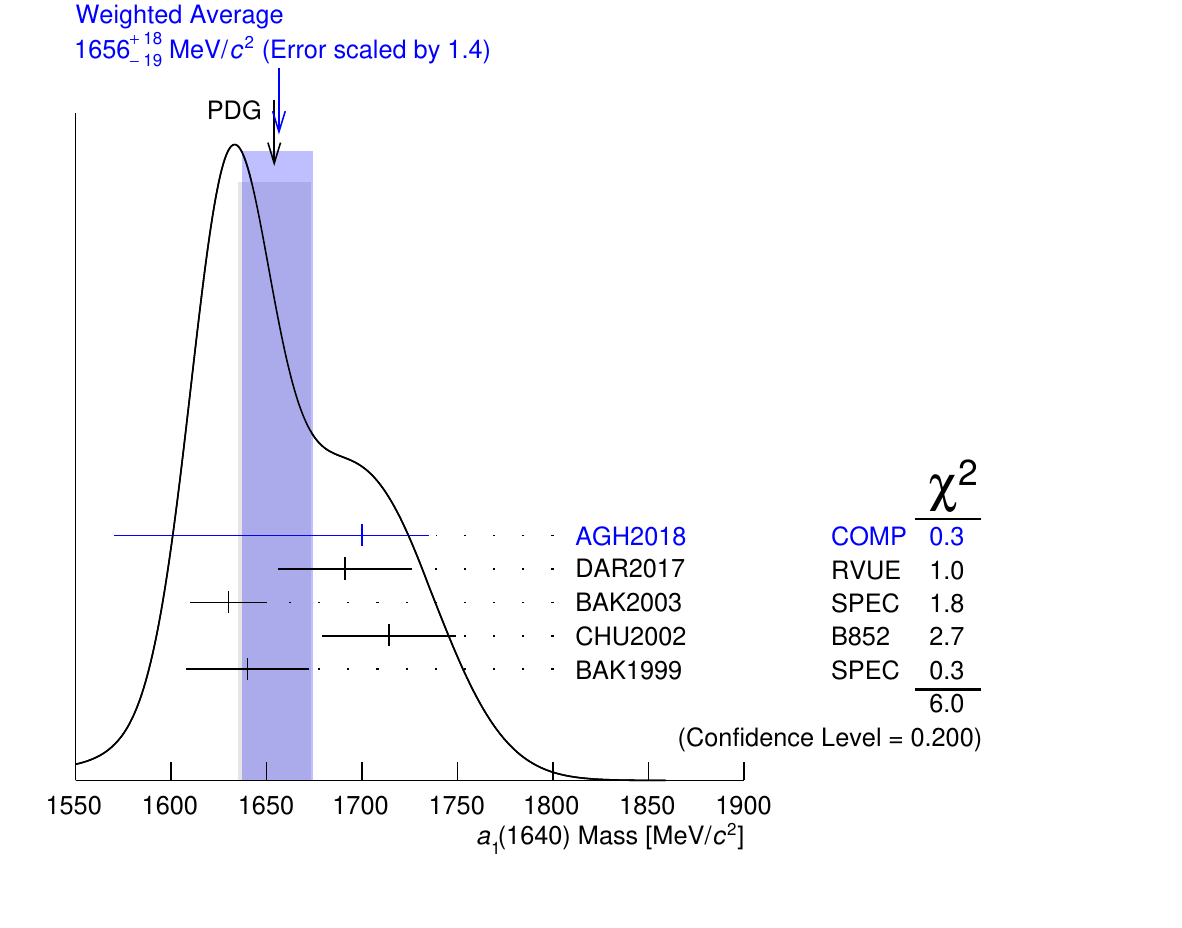}%
    \label{fig:ideogram_a1_1640_mass}%
  }%
  \subfloat[]{%
    \includegraphics[width=0.5\textwidth]{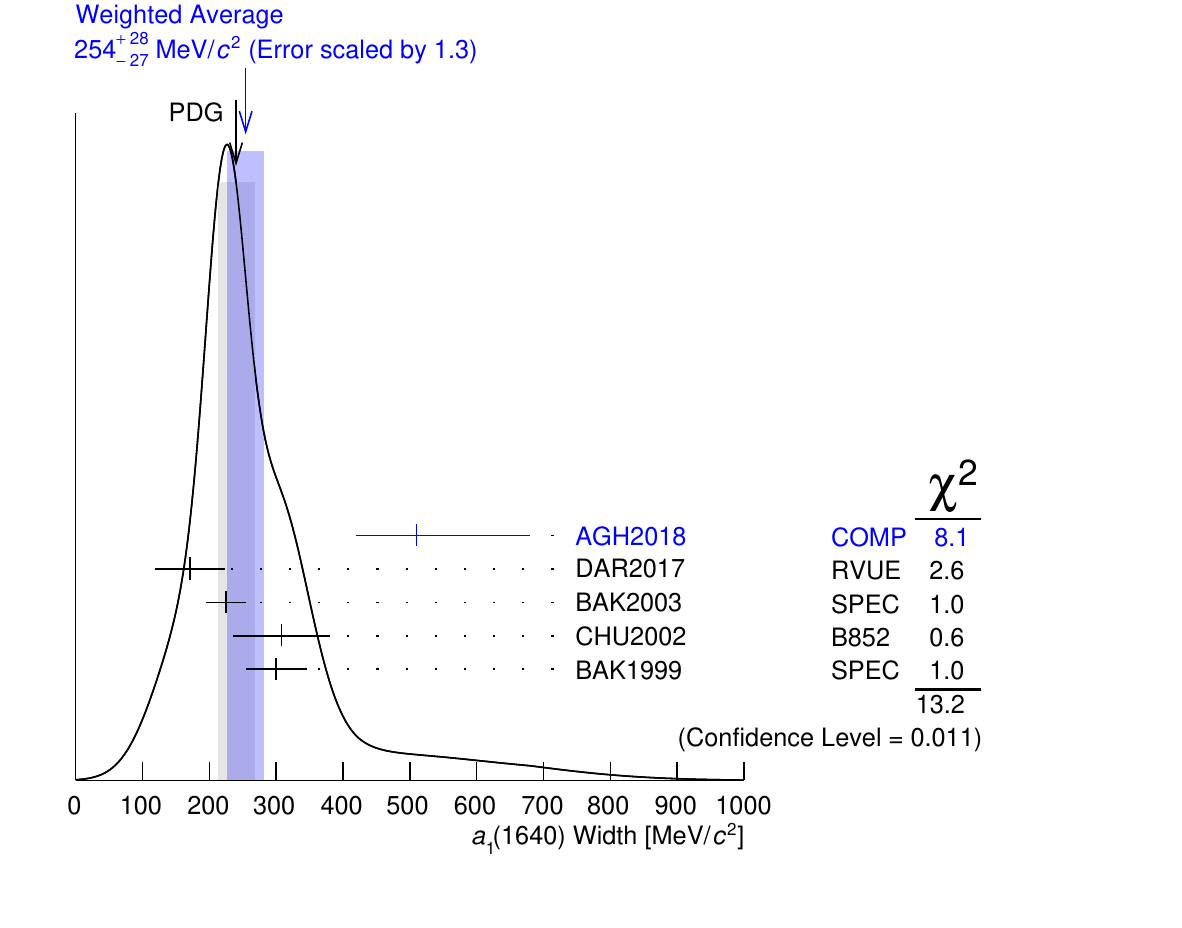}%
    \label{fig:ideogram_a1_1640_width}%
  }%
  \caption{Ideograms similar to the ones in
    \cref{fig:ideogram_pi_1800} but for
    \subfloatLabel{fig:ideogram_a1_1640_mass}~the mass and
    \subfloatLabel{fig:ideogram_a1_1640_width}~the width of the
    \PaOne[1640].  The Breit--Wigner parameters obtained from a fit of
    the COMPASS \threePi proton-target data
    (AGH2018,~\cite{Akhunzyanov:2018lqa}) are compared to previous
    measurements~\cite{Tanabashi:2018zz}.}
  \label{fig:ideogram_a1_1640}
\end{figure}

A good candidate for the first radial excitation of the \PaOne, \ie
for the \termSym{2}{3}{P}{1} quark-model state, is the \PaOne[1640].
In the \threePi data, most of the $1^{++}$ waves are dominated by the
ground-state \PaOne signal.  This is true in particular for the
\wave{1}{++}{0}{+}{\Pprho}{S} wave.  In this wave, the \PaOne[1640]
signal is about two orders of magnitude smaller than the \PaOne signal
(see \cref{fig:intensity_1pp_rho_tbin1_log} and \confer\
\cref{fig:intensity_1pp_rho_tbin1}).  Since the $\Pprho \pi$ $S$~wave
is the only $1^{++}$ wave in the 6-wave resonance-model fit of the
\threePi lead-target data, no \PaOne[1640] component was included.
The \PaOne signal is suppressed in waves with heavier isobars, \eg in
$\PfTwo \pi$ and $\PrhoThree \pi$ waves.  In the \threePi
proton-target data, we observe an \PaOne[1640] signal in the
\wave{1}{++}{0}{+}{\PfTwo}{P} wave, which was included in the 14-wave
resonance-model fit (see
\cref{fig:intensity_1pp_f2_tbin1,fig:intensity_1pp_f2_tbin11}).  We
obtain \PaOne[1640] Breit--Wigner parameters with large uncertainties:
$m_{\PaOne[1640]} = \SIaerr{1700}{35}{130}{\MeVcc}$ and
$\Gamma_{\PaOne[1640]} =
\SIaerr{510}{170}{90}{\MeVcc}$~\cite{Akhunzyanov:2018lqa}.  In
\cref{fig:ideogram_a1_1640}, we compare these values to the
measurements that are included in the PDG
average~\cite{Tanabashi:2018zz}.  Our mass value is in agreement with
previous measurements.  However, our width value is higher.  This
discrepancy might be at least in part due to the limitations of our
analysis model.

\begin{figure}[tbp]
  \centering
  \subfloat[][]{%
    \includegraphics[width=\threePlotWidth]{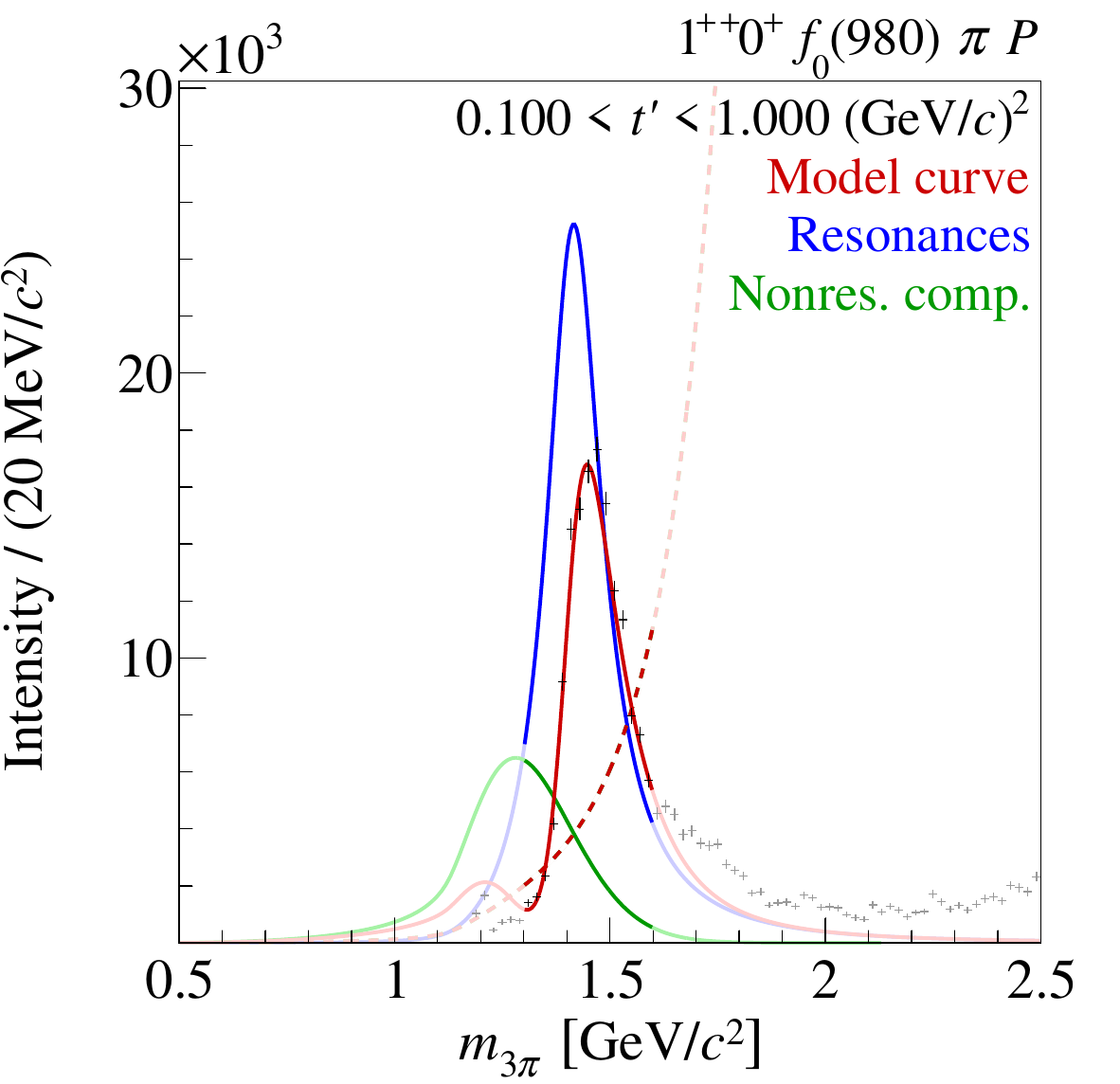}%
    \label{fig:intensity_1pp_f0_no_a1_1420}%
  }%
  \hfill%
  \subfloat[][]{%
    \includegraphics[width=\threePlotWidth]{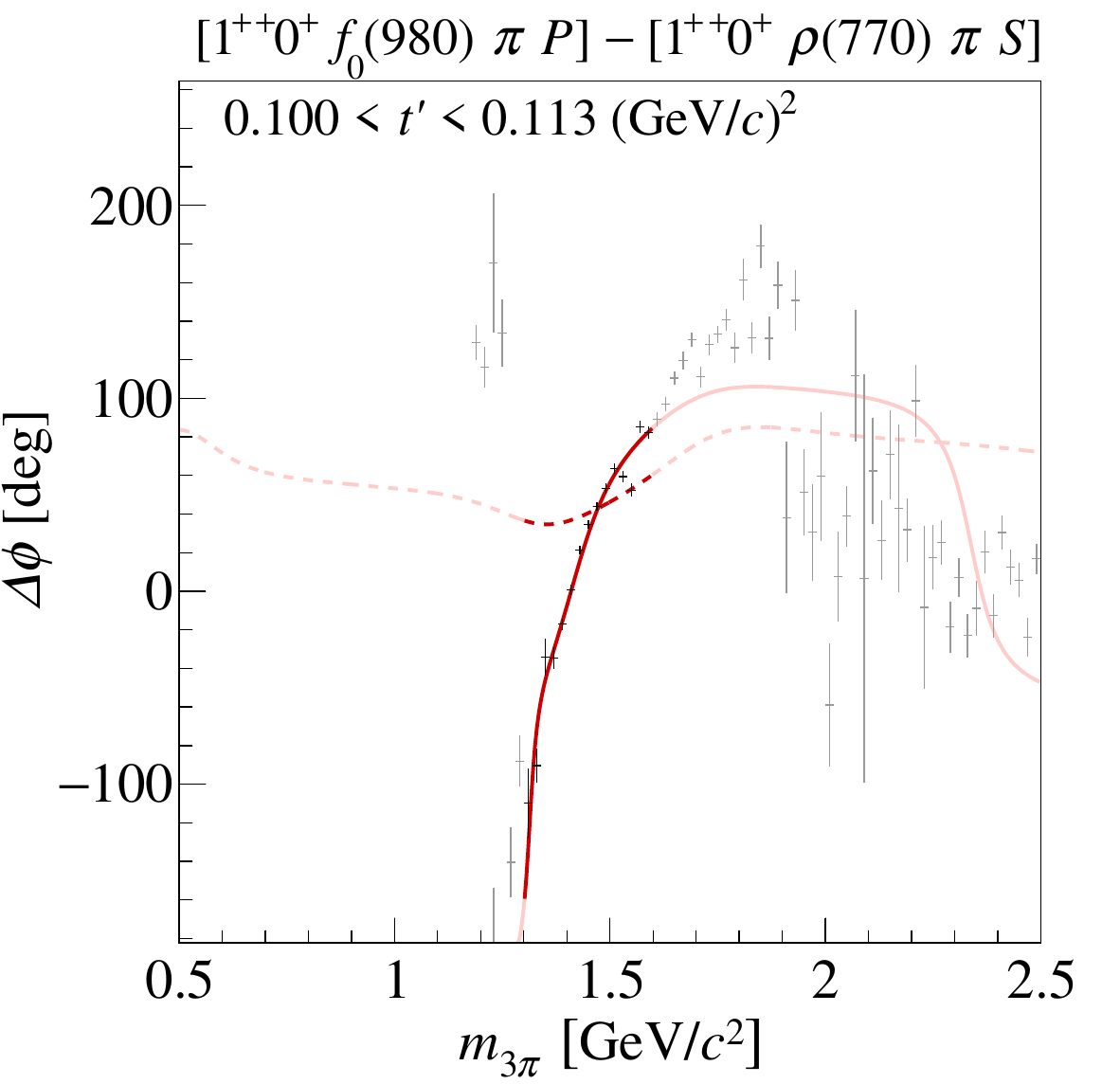}%
    \label{fig:phase_1pp_f0_1pp_rho_tbin1_no_a1_1420}%
  }%
  \hfill%
  \subfloat[][]{%
    \includegraphics[width=\threePlotWidth]{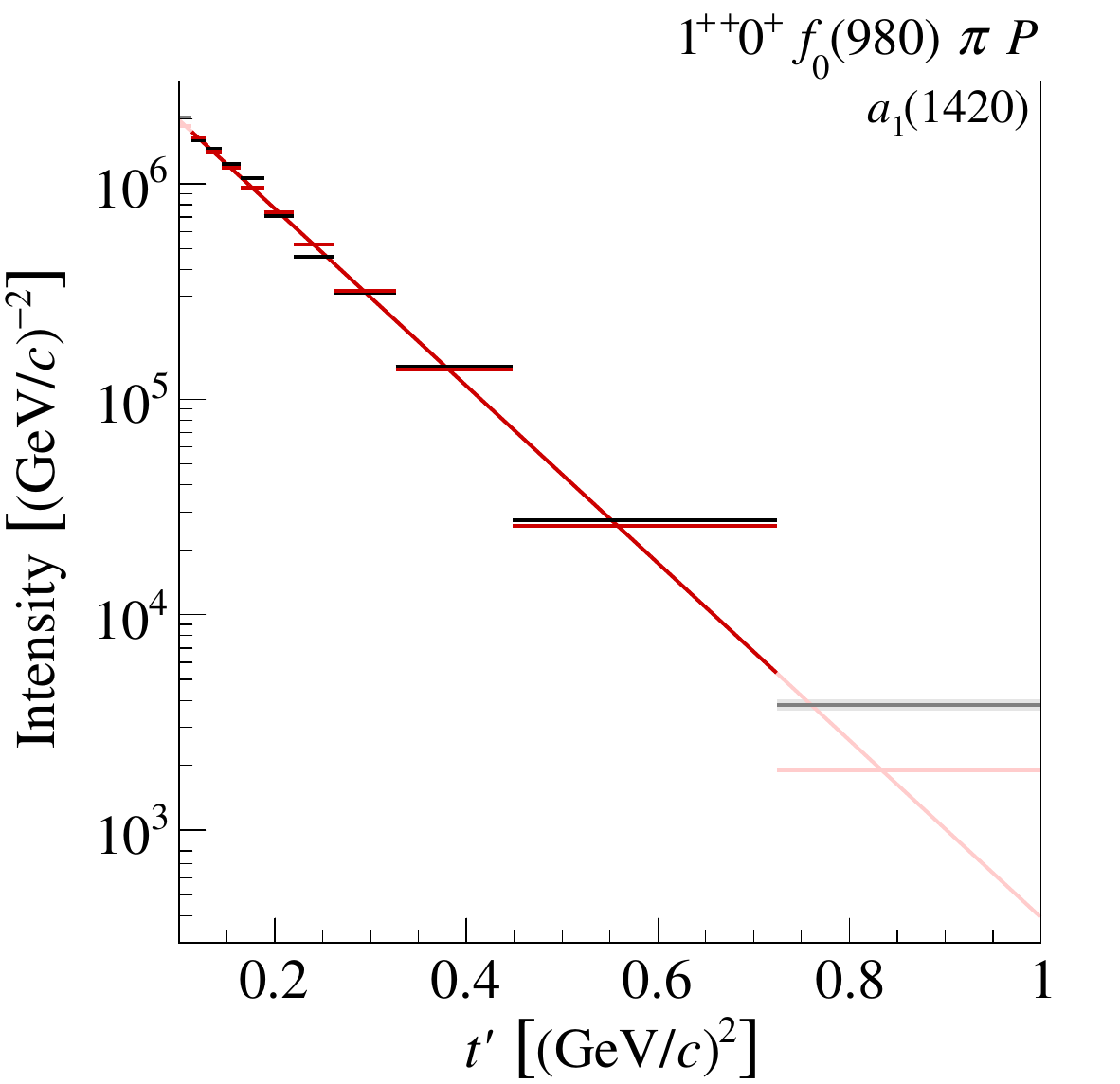}
    \label{fig:a1_1420_t_spectrum}
  }%
  \caption{\subfloatLabel{fig:intensity_1pp_f0_no_a1_1420}~Intensity
    distribution of the \wave{1}{++}{0}{+}{\PfZero[980]}{P} wave in
    the \threePi proton-target data summed over the
    11~$t'$~bins~\cite{Akhunzyanov:2018lqa}.
    \subfloatLabel{fig:phase_1pp_f0_1pp_rho_tbin1_no_a1_1420}~Phase of
    this wave \wrt the \wave{1}{++}{0}{+}{\Pprho}{S} wave in the
    lowest $t'$~bin.  The curves represent the result of two
    resonance-model fits.  The model and the wave components are
    represented as in \cref{fig:intensity_phase_0mp} except that the
    blue curve represents the \PaOne[1420].  The result of the main
    resonance-model fit is represented by the continuous curves.  The
    dashed curves represent the result of a fit, where the
    \PaOne[1420] component is removed from the resonance model, \ie
    where we try to model the data using only a non-resonant
    component.  \subfloatLabel{fig:a1_1420_t_spectrum}~Similar to
    \cref{fig:tspectrum_0mp_f0}, but showing the $t'$~spectrum of the
    \PaOne[1420] in the \wave{1}{++}{0}{+}{\PfZero[980]}{P} wave.}
  \label{fig:intensity_phase_1pp_f0_no_a1_1420}
\end{figure}

In addition to the known states discussed above, we have discovered a
surprising resonance-like signal, the \PaOne[1420], in the
\wave{1}{++}{0}{+}{\PfZero[980]}{P} wave in the COMPASS \threePi
proton-target data~\cite{Adolph:2015pws}.  The observed
$\PfZero[980] \pi$ decay mode is peculiar.  Only few light mesons are
known to decay into \PfZero[980], among them only two isovector
mesons, the
\Pppi[1800]~\cite{Amelin:1995gu,Chung:2002pu,Adolph:2015tqa,Akhunzyanov:2018lqa}
and the \PpiTwo[1880]~\cite{Adolph:2015tqa} (see
\cref{sec:results_1pp,sec:results_2mp}).  Hence the 88-wave PWA model
for the proton-target data includes only four waves with an
\PfZero[980] isobar (see Table~IX in Appendix~A of
\refCite{Adolph:2015tqa}), two of which, the $0^{-+}\,0^+$ and the
$1^{++}\,0^+$ wave, are included in the 14-wave resonance-model fit.
The coherent sum of the four \PfZero[980] waves, which have all
positive reflectivity, contributes only \SI{3.3}{\percent} to the
total intensity.  Most of this intensity is due to the
\wave{0}{-+}{0}{+}{\PfZero[980]}{S} wave (see \cref{sec:results_0mp}).
The \wave{1}{++}{0}{+}{\PfZero[980]}{P} wave contributes only
\SI{0.3}{\percent} to the total intensity.  This small relative
intensity is the main reason why previous experiments were not able to
observe the \PaOne[1420] signal.  As shown in
\cref{fig:intensity_1pp_f0_no_a1_1420}, the intensity distribution of
the \wave{1}{++}{0}{+}{\PfZero[980]}{P} wave exhibits a narrow peak at
about \SI{1.4}{\GeVcc} that is associated by a rapid phase motion of
about \SI{180}{\degree} \wrt other waves, in particular \wrt the
dominant \wave{1}{++}{0}{+}{\Pprho}{S} wave as shown in
\cref{fig:phase_1pp_f0_1pp_rho_tbin1_no_a1_1420}.  Both features are
robust against changes of the PWA model and other systematic effects
(see Section~IV.F and Appendix~B in \refCite{Adolph:2015tqa}).

\begin{figure}[tbp]
  \centering
  \hfill%
  \subfloat[][]{%
    \includegraphics[width=\threePlotWidthTwoD]{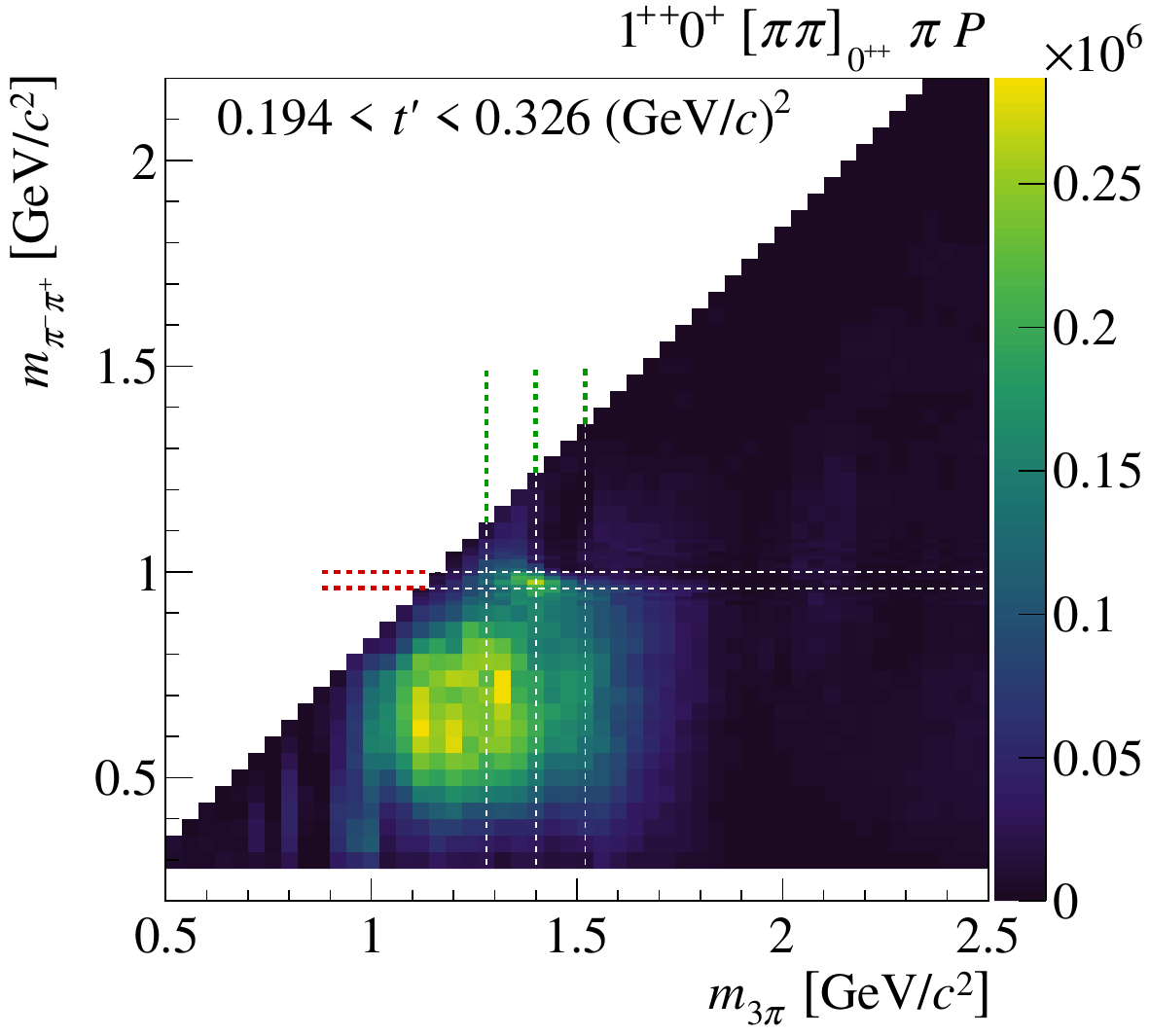}%
    \label{fig:PIPIS_1pp_2D}%
  }%
  \hfill%
  \subfloat[][]{%
    \includegraphics[width=\threePlotWidth]{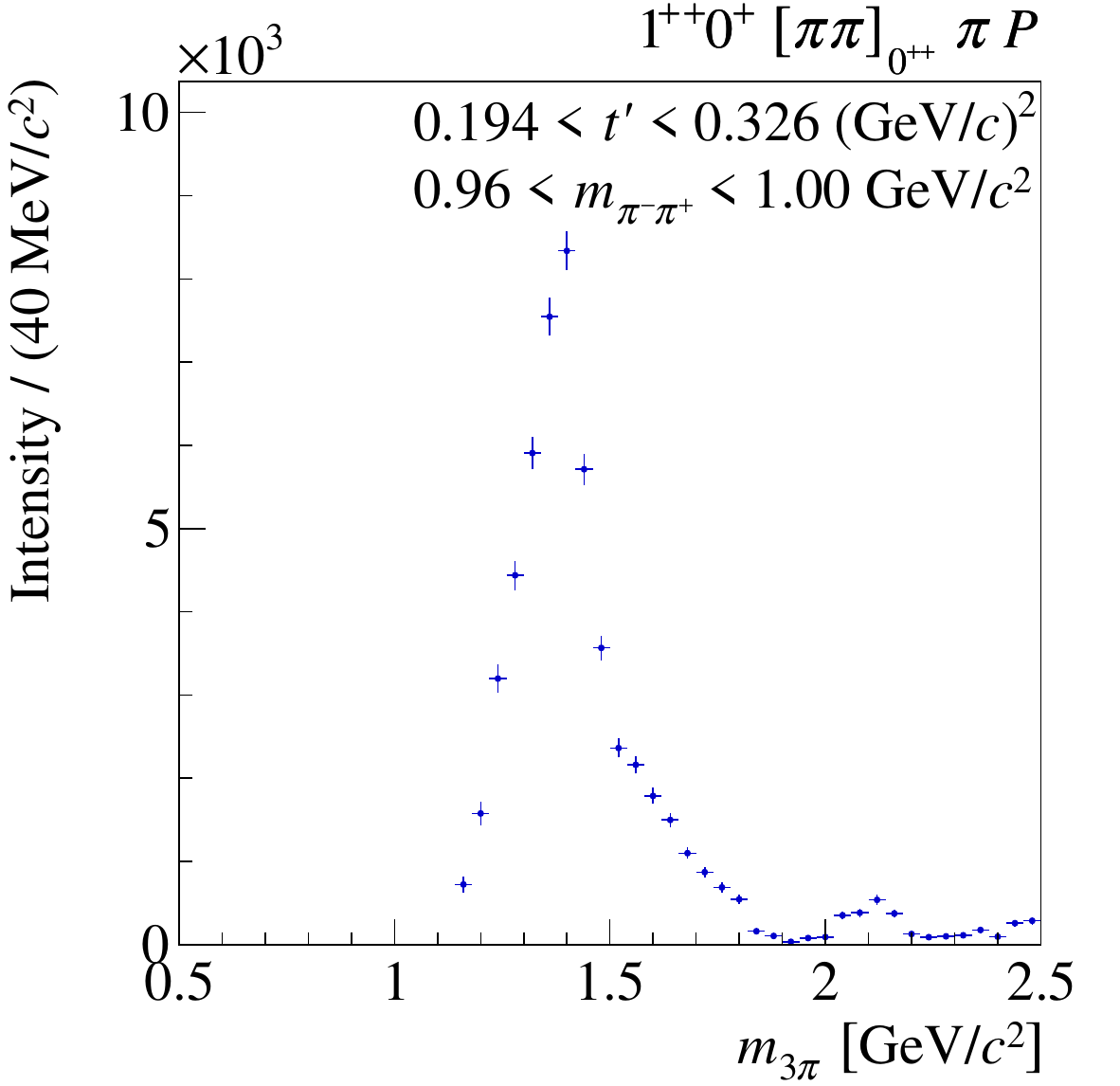}%
    \label{fig:PIPIS_1pp_980}%
  }%
  \hfill\null%
  \\
  \subfloat[][]{%
    \includegraphics[width=\threePlotWidth]{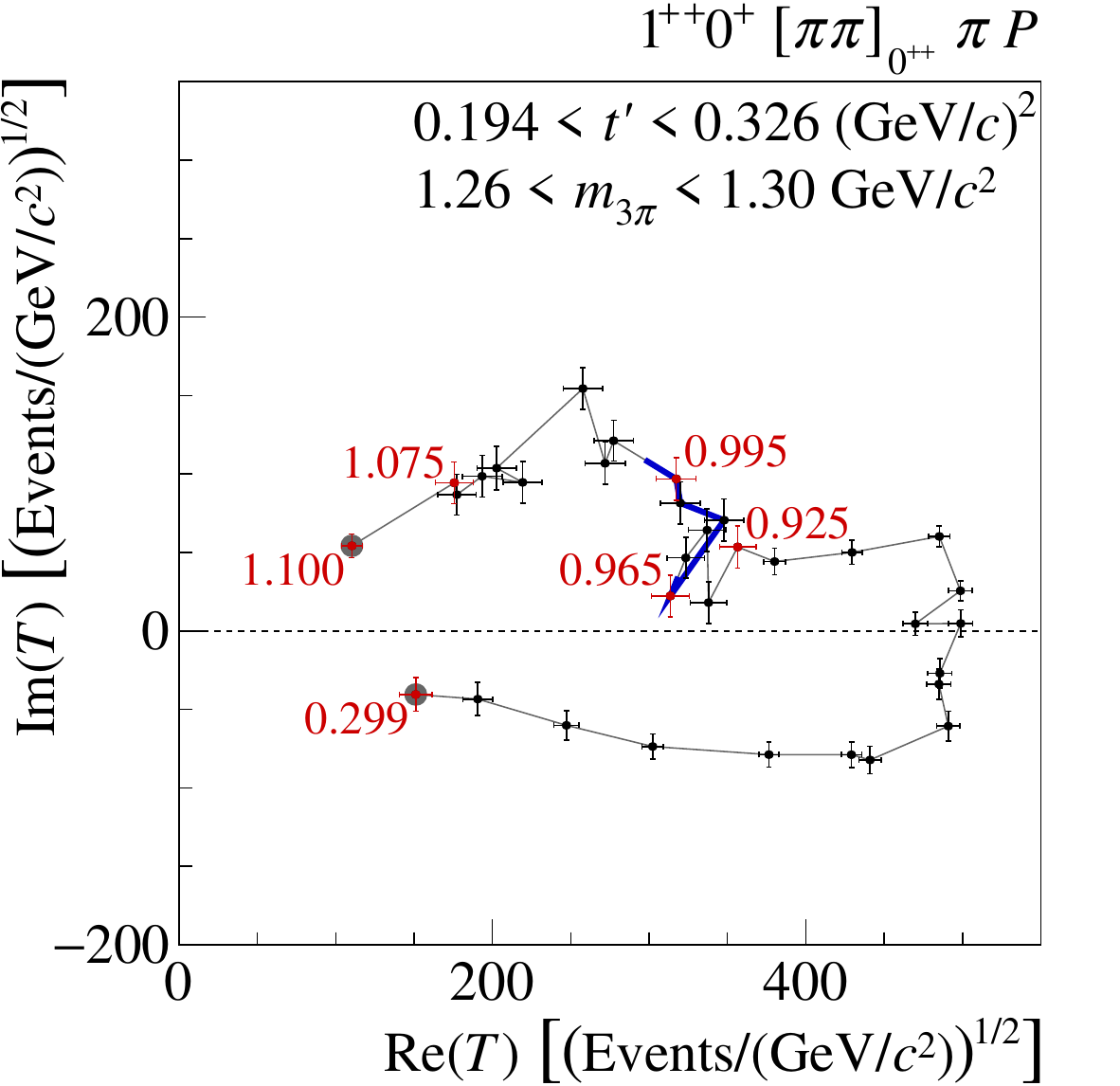}%
    \label{fig:PIPIS_1pp_below_res_argand}%
  }%
  \hfill%
  \subfloat[][]{%
    \includegraphics[width=\threePlotWidth]{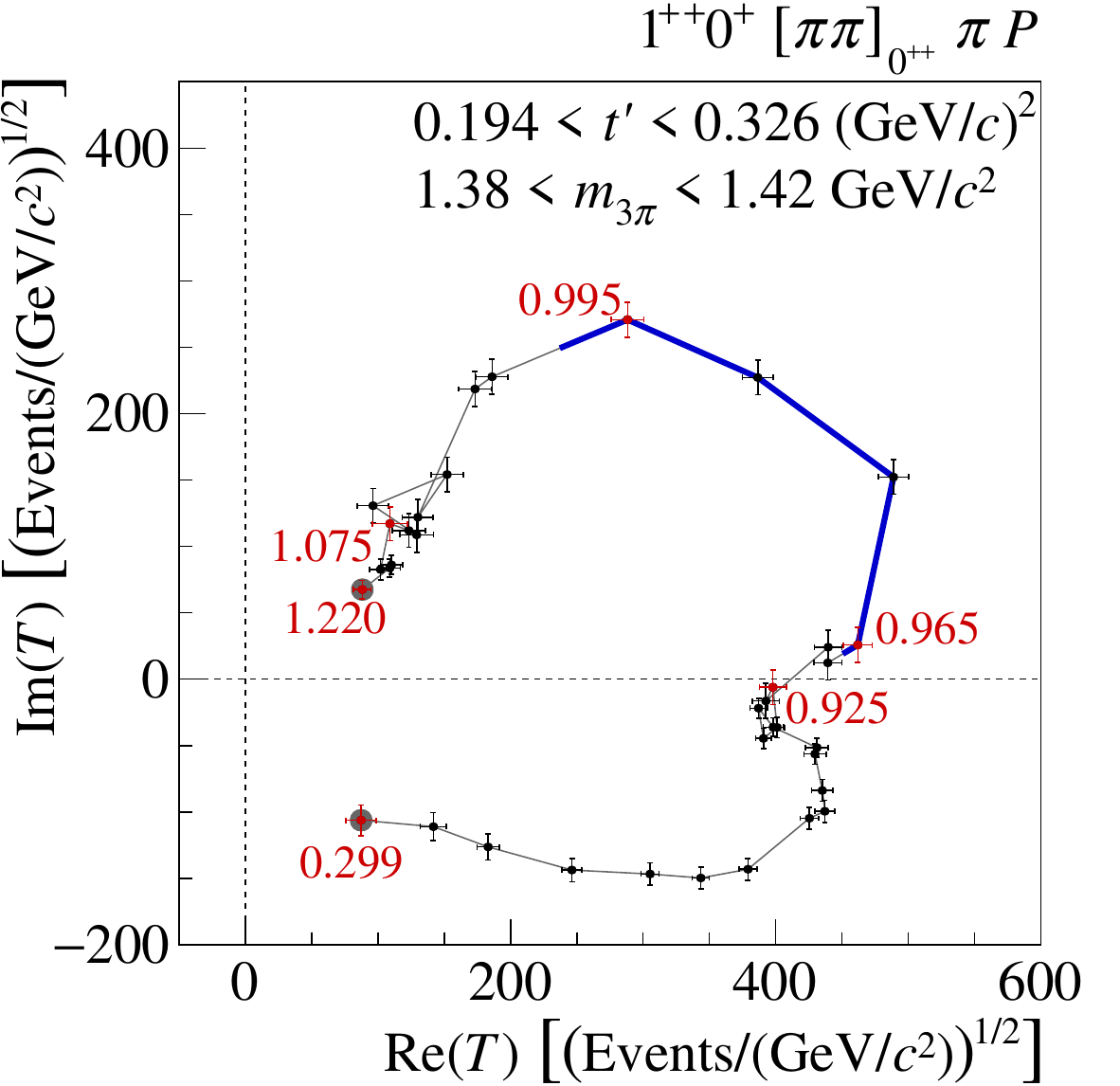}%
    \label{fig:PIPIS_1pp_at_res_argand}%
  }%
  \hfill%
  \subfloat[][]{%
    \includegraphics[width=\threePlotWidth]{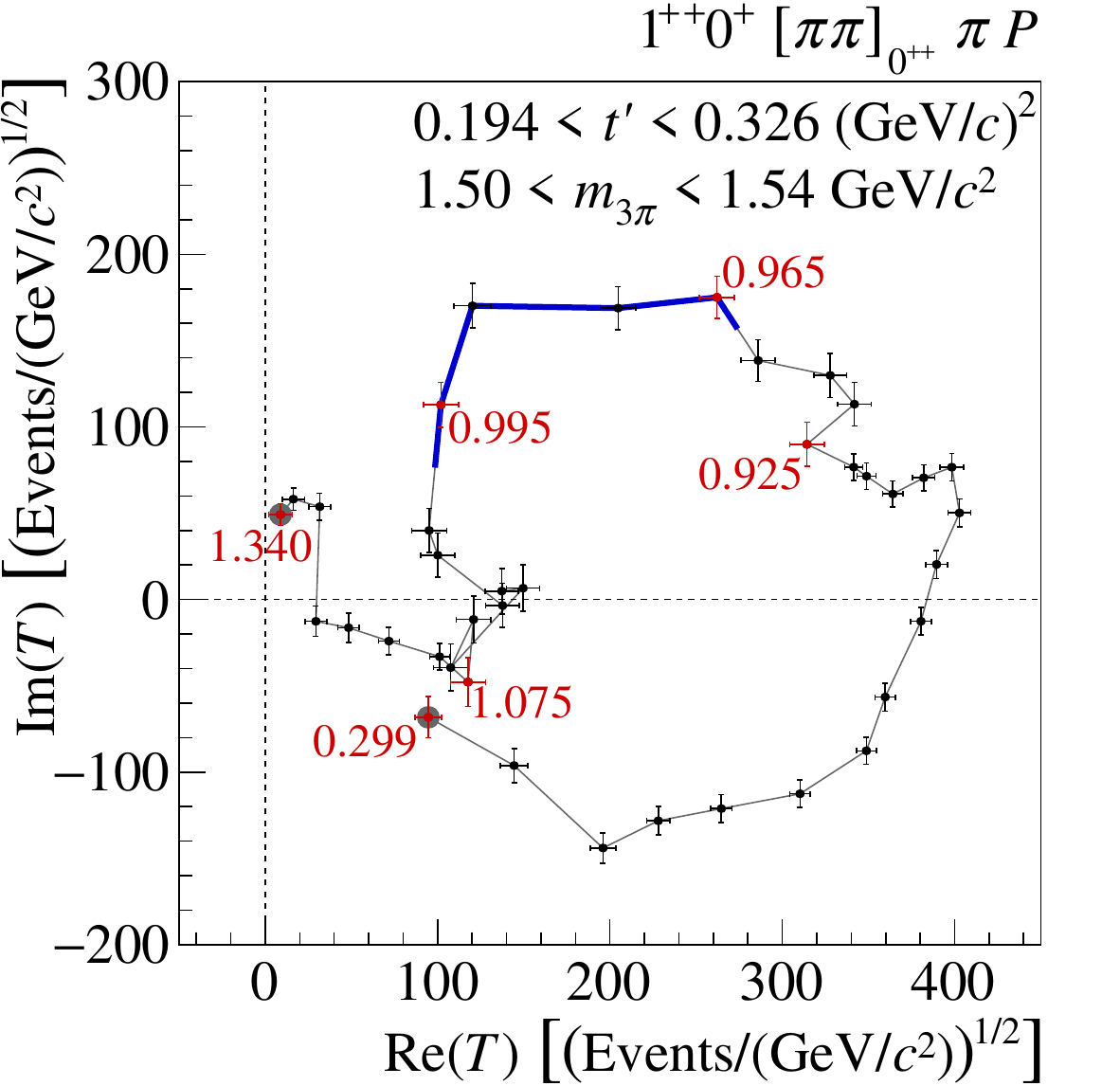}%
    \label{fig:PIPIS_1pp_above_res_argand}%
  }%
  \caption{Similar to \cref{fig:pipi_s_wave_0mp_highT} but for the
    \wave{1}{++}{0}{+}{\pipiSF}{P} wave with the freed-isobar
    amplitude \pipiSF in an intermediate $t'$~bin of the \threePi
    proton-target data~\cite{Adolph:2015tqa}.
    \subfloatLabel{fig:PIPIS_1pp_2D}~Intensity of the
    \wave{1}{++}{0}{+}{\pipiSF}{P} wave as a function of~\mTwoPi
    and~\mThreePi.  \subfloatLabel{fig:PIPIS_1pp_980}~Intensity as a
    function of~\mThreePi summed over the selected \mTwoPi~interval
    around the \PfZero[980] as indicated by the pair of horizontal
    dashed lines
    in~\subfloatLabel{fig:PIPIS_1pp_2D}. \subfloatLabel{fig:PIPIS_1pp_below_res_argand},~\subfloatLabel{fig:PIPIS_1pp_at_res_argand},
    and~\subfloatLabel{fig:PIPIS_1pp_above_res_argand}:~\Argands of
    the \pipiSF freed-isobar amplitude for \mThreePi~bins below, at,
    and above the \PaOne[1420] mass, respectively, as indicated by the
    vertical dashed lines in~\subfloatLabel{fig:PIPIS_1pp_2D}.}
  \label{fig:pipis_2D_1pp}
\end{figure}

The \PaOne[1420] signal also appears in the freed-isobar PWA (see
\cref{sec:pwa_cells:freed_isobar,sec:3pi_model:pwa}) in the
\wave{1}{++}{0}{+}{\pipiSF}{P} wave.  \Cref{fig:PIPIS_1pp_2D} shows
the correlation of the \mThreePi~intensity distribution of this wave
with the \mTwoPi~intensity distribution of the freed-isobar amplitude
with $\JPC = 0^{++}$.  A clear peak is found at
$\mThreePi \approx \SI{1.4}{\GeVcc}$ and
$\mTwoPi \approx \SI{1.0}{\GeVcc}$.  The \mThreePi~intensity
distribution in the \PfZero[980] region, as shown in
\cref{fig:PIPIS_1pp_980}, exhibits a clear \PaOne[1420] peak that is
similar to the one in the conventional PWA (\confer\
\cref{fig:intensity_1pp_f0_no_a1_1420}).  The resonant nature of the
\twoPi subsystem at the \PaOne[1420] mass is proven by the \Argand in
\cref{fig:PIPIS_1pp_at_res_argand}, which exhibits a clear circular
resonance structure in the highlighted \PfZero[980] region.  We
observe a continuous evolution of the \Argands with~\mThreePi, where
the circular \PfZero[980] structure rotates counterclockwise \wrt its
center due to the phase motion caused by the \PaOne[1420] signal
thereby confirming its resonance-like nature.  As an example, we show
in
\cref{fig:PIPIS_1pp_below_res_argand,fig:PIPIS_1pp_at_res_argand,fig:PIPIS_1pp_above_res_argand}
the \Argands measured \wrt the \wave{1}{++}{0}{+}{\Pprho}{S} wave for
three \mThreePi~bins.  The results from the freed-isobar PWA therefore
confirm the \PaOne[1420] signal and prove in particular that it is not
an artificial structure caused by the parameterizations that are used
for the $\JPC = 0^{++}$ isobars in the conventional 88-wave PWA fit.

The resonance features of the \PaOne[1420] signal were first
established in a much simpler resonance-model fit that was based on
the same 88-wave PWA result but included only three
waves~\cite{Adolph:2015pws}.  The estimated Breit--Wigner parameters
$m_{\PaOne[1420]} = \SIaerr{1414}{15}{13}{\MeVcc}$ and
$\Gamma_{\PaOne[1420]} = \SIaerr{153}{8}{23}{\MeVcc}$ from this 3-wave
fit are consistent with the parameters
$m_{\PaOne[1420]} = \SIaerr{1411}{4}{5}{\MeVcc}$ and
$\Gamma_{\PaOne[1420]} = \SIaerr{161}{11}{14}{\MeVcc}$ that are
obtained in the 14-wave fit~\cite{Akhunzyanov:2018lqa}.  In both
analyses, a Breit--Wigner amplitude describes the \PaOne[1420] peak
well.  This is also true for the observed phase motions of the
\wave{1}{++}{0}{+}{\PfZero[980]}{P} wave (see \eg\
\cref{fig:phase_1pp_f0_1pp_rho_tbin1_no_a1_1420}).  Another feature of
the data that supports the resonance interpretation of the
\PaOne[1420] signal is the approximately exponential behavior of its
$t'$~spectrum (see \cref{fig:a1_1420_t_spectrum}) with a slope
parameter of \SIaerr{9.5}{0.6}{1.0}{\perGeVcsq}, which is a value in
the range that is expected for resonances.

\begin{figure}[tbp]
  \centering
  \centering
  \includegraphics[width=\twoPlotWidth]{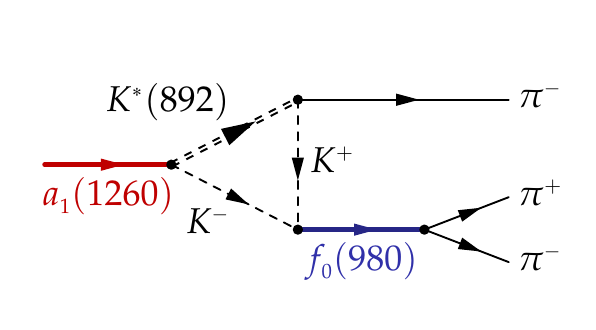}
  \caption{Rescattering diagram proposed in
    \refsCite{Ketzer:2015tqa,Aceti:2016yeb} to explain the
    \PaOne[1420] signal.}
  \label{fig:a1_1420_triangle}
\end{figure}

The interpretation of the \PaOne[1420] signal is still unclear.  It is
too close in mass and too narrow in order to be the radial excitation
of the ground-state \PaOne, \ie the \termSym{2}{3}{P}{1} quark-model
state.  Also the low intensity of the \PaOne[1420] signal, which is
about 20~times smaller than the one of the \PaOne in the
\wave{1}{++}{0}{+}{\Pprho}{S} wave speaks against a $\ket{\qqbar}$
interpretation.  It is peculiar that we find the \PaOne[1420] signal
only in the $\PfZero[980] \pi$ decay mode.  The \PfZero[980] is known
to have a large \ssbar component and is interpreted by some models as
a tetraquark, a molecular state, or a mixture of
both~\cite{Amsler:2004ps,pdg_scalars:2018}.  Another interesting
aspect of the \PaOne[1420] is that its mass is suspiciously close to
the $K \PKstarbar$ threshold.

Several interpretations were proposed for the \PaOne[1420].  Based on
its mass, it could be the isospin partner of the \PfOne[1420].  The
much smaller width of the \PfOne[1420] of only \SI{54.9
  (26)}{\MeVcc}~\cite{Tanabashi:2018zz} could be explained by its
strong coupling to $K \PKstarbar$, which has a much smaller phase
space than the decay $\PaOne[1420] \to \PfZero[980] \pi$.  The
molecular model for the \PfOne[1420] proposed in
\refCite{Longacre:1990uc} could possibly be extended to the isovector
case.  The \PaOne[1420] signal was also described as a mixed state of
a \qqbar and a tetraquark component~\cite{Wang:2014bua} and as a
tetraquark with mixed flavor symmetry~\cite{Chen:2015fwa}.  In
addition, calculations based on the AdS/QCD correspondence find
isovector tetraquarks with masses similar to that of the
\PaOne[1420]~\cite{Gutsche:2017oro,Nielsen:2018uyn}.  The authors of
\refCite{Gutsche:2017twh} studied the two-body decay rates for the
modes $\PaOne[1420] \to \PfZero[980] \pi$ and
$\PaOne[1420] \to K \PKstarbar$ for four-quark configurations using
the covariant confined quark model.  They found that a molecular
configuration is preferred over a compact diquark--antidiquark state.

\begin{figure}[tbp]
  \centering
  \includegraphics[width=0.75\textwidth]{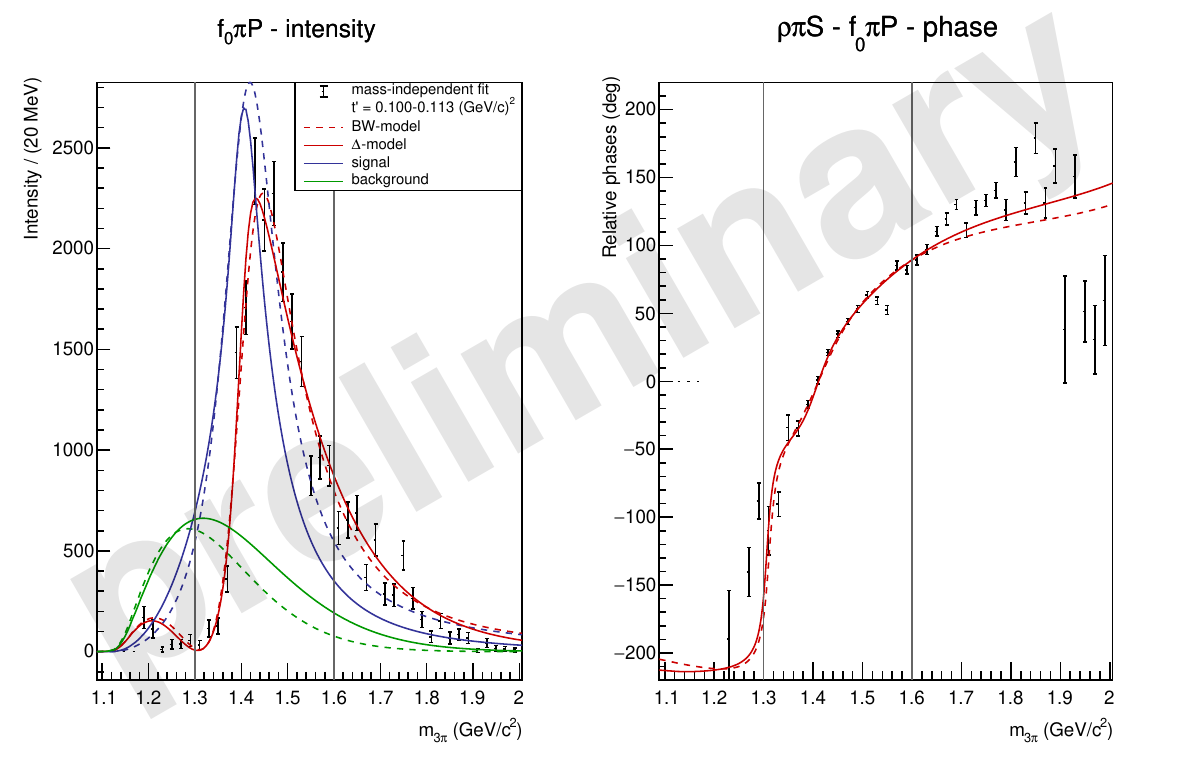}
  \caption{Intensity distribution of the
    \wave{1}{++}{0}{+}{\PfZero[980]}{P} wave (left) and phase of this
    wave \wrt the \wave{1}{++}{0}{+}{\Pprho}{S} wave (right), both in
    the lowest $t'$~bin of the \threePi proton-target data (black
    points with error bars; same as in
    \cref{fig:phase_1pp_f0_1pp_rho_tbin1_no_a1_1420}).  The curves
    represent the result of two 3-wave resonance-model fits (see
    text).\protect\footnotemark\ The vertical lines indicate the fit
    range.  The model and the wave components are represented as in
    \cref{fig:intensity_phase_0mp}.  The dashed curves represent the
    result of a fit, where the \PaOne[1420] is described by a
    Breit--Wigner amplitude (blue dashed curve) as in the 14-wave
    resonance-model fit shown in
    \cref{fig:intensity_phase_1pp_f0_no_a1_1420}.  The continuous
    curves represent a fit, where we use the amplitude of the triangle
    rescattering diagram shown in \cref{fig:a1_1420_triangle} (blue
    continuous curve) instead of the Breit--Wigner amplitude.}
  \label{fig:intensity_phase_1pp_f0_triangle}
\end{figure}

In addition to the resonance interpretations discussed above, other
explanations do not require an additional \PaOne* resonance.
Basdevant and Berger proposed resonant rescattering corrections in the
Deck process as an
explanation~\cite{Basdevant:2015ysa,Basdevant:2015wma}, whereas the
authors of \refCite{Ketzer:2015tqa} suggested an anomalous triangle
singularity in the rescattering diagram for
$\PaOne \to K \PKstarbar \to K \PKbar \pi \to \PfZero[980] \pi$, which
is shown in \cref{fig:a1_1420_triangle}.  The results of the latter
calculation were confirmed in \refCite{Aceti:2016yeb}.  In order to
study how well the amplitude of the triangle diagram describes the
data in comparison to the Breit--Wigner amplitude, we have performed
two resonance-model fits with a reduced wave set of only three waves:
\wave{1}{++}{0}{+}{\Pprho}{S}, \wave{1}{++}{0}{+}{\PfZero[980]}{P},
and \wave{2}{++}{1}{+}{\Pprho}{D}.  In the first fit (dashed curves in
\cref{fig:intensity_phase_1pp_f0_triangle}), we use a Breit-Wigner
amplitude for the \PaOne[1420] (blue dashed curve), like in the
14-wave fit discussed above.  In the second fit (continuous curves in
\cref{fig:intensity_phase_1pp_f0_triangle}), the Breit-Wigner
amplitude is replaced by the amplitude of the triangle diagram
(continuous blue curve).  From these fits, we find that the amplitude
of the triangle diagram describes the data equally well as the
Breit--Wigner model.  In the case of a triangle singularity, the
production rate of the \PaOne[1420] would be completely determined by
that of the \PaOne.  Therefore, the slope parameters of the two wave
components would be equal.  Unfortunately, in our analysis the
systematic uncertainties of the slope parameters are too large in
order to draw a conclusion.  Hence more detailed studies are still
needed in order to distinguish between the different models for the
\PaOne[1420].

\footnotetext{Publication in preparation.}

\subsubsection{The $\JPC = 1^{-+}$ Sector }
\label{sec:results_1mp}

The PDG lists three light-meson states with spin-exotic quantum
numbers~\cite{Tanabashi:2018zz}: \PpiOne[1400], \PpiOne[1600], and
\PpiOne[2015], which are all isovector states with $\JPC = 1^{-+}$
quantum numbers (see also \cref{fig:light_flavorless_spectrum}).  The
\PpiOne[1400] has so far been observed only in the \etaPi final
state~\cite{Alde:1988bv,Aoyagi:1993kn,Thompson:1997bs,Abele:1998gn,Abele:1999tf,Chung:1999we,Adams:2006sa}
with the exception of the Obelix experiment, which claims the
observation of the \PpiOne[1400] in the $\Pprho \pi$ decay
mode~\cite{Salvini:2004gz}.  The \PpiOne[2015] is listed by the PDG as
a \textquote{Further State} and has been observed only by the BNL E852
experiment in the $\PbOne \pi$ and $\PfOne \pi$ final
states~\cite{Kuhn:2004en,Lu:2004yn}.  The \PpiOne[1600] is the best
established of the three known spin-exotic light-meson candidates.  It
has been observed by several experiments in various decay modes.  The
COMPASS, E852, and VES experiments have studied high-energy inelastic
scattering reactions of pion beams on nuclear targets and have
reported \PpiOne[1600] signals in the
$\Pprho
\pi$~\cite{Adams:1998ff,Chung:2002pu,Alekseev:2009aa,Akhunzyanov:2018lqa},
$\PetaPr \pi$~\cite{Ivanov:2001rv,Amelin:2005ry},
$\PbOne \pi$~\cite{Lu:2004yn,Amelin:2005ry}, and
$\PfOne \pi$~\cite{Kuhn:2004en,Amelin:2005ry} decay modes.  In an
analysis of Crystal Barrel data on the reaction
$\ppbar \to \omega \pi^+ \pi^- \pi^0$, the authors of
\refCite{Baker:2003jh} reported evidence for the \PpiOne[1600] in the
$\PbOne \pi$ decay mode.  The CLEO-c experiment has studied the decays
of the charmonium state~$\chi_{c1}$ to $\eta \pi^+ \pi^-$ and
$\eta' \pi^+ \pi^-$~\cite{Adams:2011sq}.  They found evidence for an
exotic signal in the $\eta' \pi$ subsystem consistent with the
\PpiOne[1600] signal seen in other production mechanisms.  A recent
summary of all measurements can be found in \refCite{Meyer:2015eta}.

Although significant intensity was observed by previous experiments in
the \etaPi and \etaPrPi $P$~waves with $\Mrefl = 1^+$, surprisingly
the resonance content of the two $P$~waves seems to be different: the
\PpiOne[1400] is seen only in \etaPi, whereas the \PpiOne[1600] is
seen only in \etaPrPi.  In addition, the VES group observed a
significant suppression of the squared matrix element, \ie the
intensity divided by the two-body phase-space volume
$q_{\etaOrPrPi} / m_{\etaOrPrPi}$ (\confer\ \cref{eq:etaprime_eta_r}),
of the $P$~wave with $\Mrefl = 1^+$ in \etaPi compared to
\etaPrPi~\cite{Beladidze:1993km}.  This favors a possible hybrid
($q\bar{q}g$) or four-quark interpretation of the \PpiOne[1600] and
disfavors a hybrid interpretation of the
\PpiOne[1400]~\cite{Close:1987aw,Iddir:1988jd,Iddir:1988jc,Chung:2002fz}.
The \PpiOne[1600] is consistent with the lightest hybrid state
predicted by lattice QCD (see \cref{sec:pheno.lattice}) and models
(see \eg\ \refCite{Meyer:2015eta}).  In contrast, the \PpiOne[1400] is
too light for a hybrid state.  Four-quark interpretations of the
\PpiOne[1400] (\eg the one in \refCite{Chung:2002fz}) suffer from the
fact that they predict additional SU(3)$_\text{flavor}$ partner states
that have not been observed so far.  It should also be noted that
resonance-like signals were observed mostly in \etaOrPrPim systems
that have the same charge as the beam particle.  Such final states are
produced predominantly via the exchange of Pomerons, which have vacuum
quantum numbers and are commonly expected to have largely gluonic
content.  In contrast, in charge-exchange reactions no spin-exotic
resonance signal was found in the $\etaPrPi^0$ final state by the VES
experiment~\cite{Amelin:2004ns}, while analyses of $\etaPi^0$ data
from the BNL E852 experiment yielded contradictory
results~\cite{Dzierba:2003fw,Adams:2006sa}.  A coupled-channel
analysis of the \etaPim and \etaPrPim data from the BNL E852
experiment found no evidence for the
\PpiOne[1400]~\cite{Szczepaniak:2003vg}.

\begin{figure}[p]
  \hfill%
  \subfloat[][]{%
    \includegraphics[width=0.4\textwidth,valign=m]{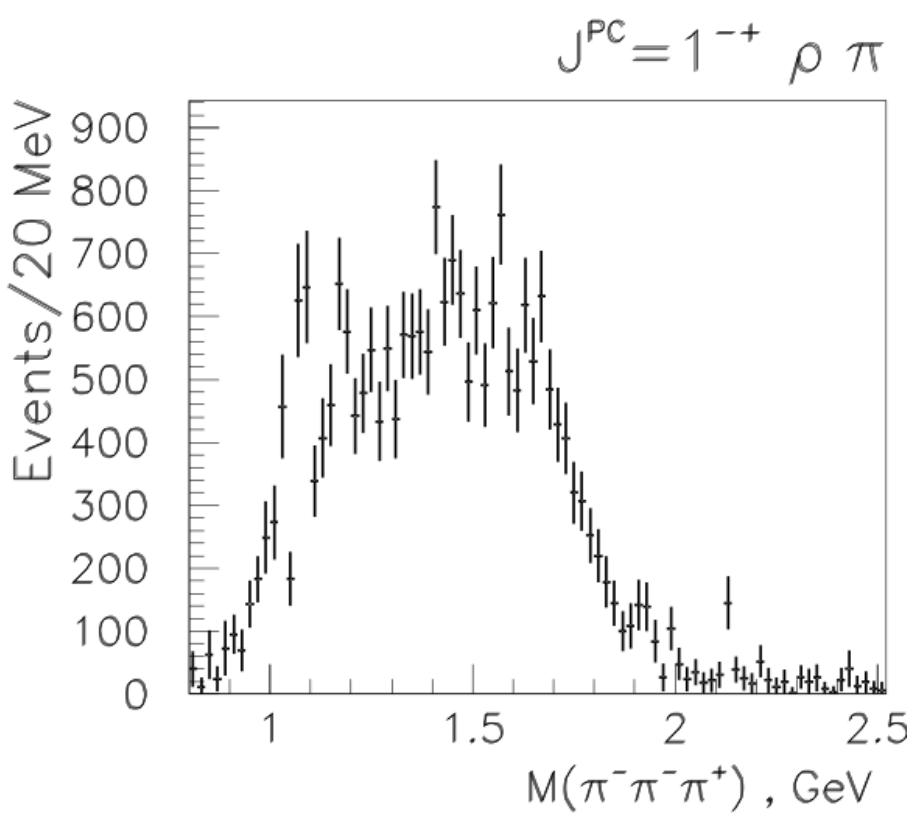}%
    \label{fig:1mp_ves}%
  }%
  \hfill%
  \subfloat[][]{%
    \includegraphics[width=1.1\threePlotWidth,valign=m]{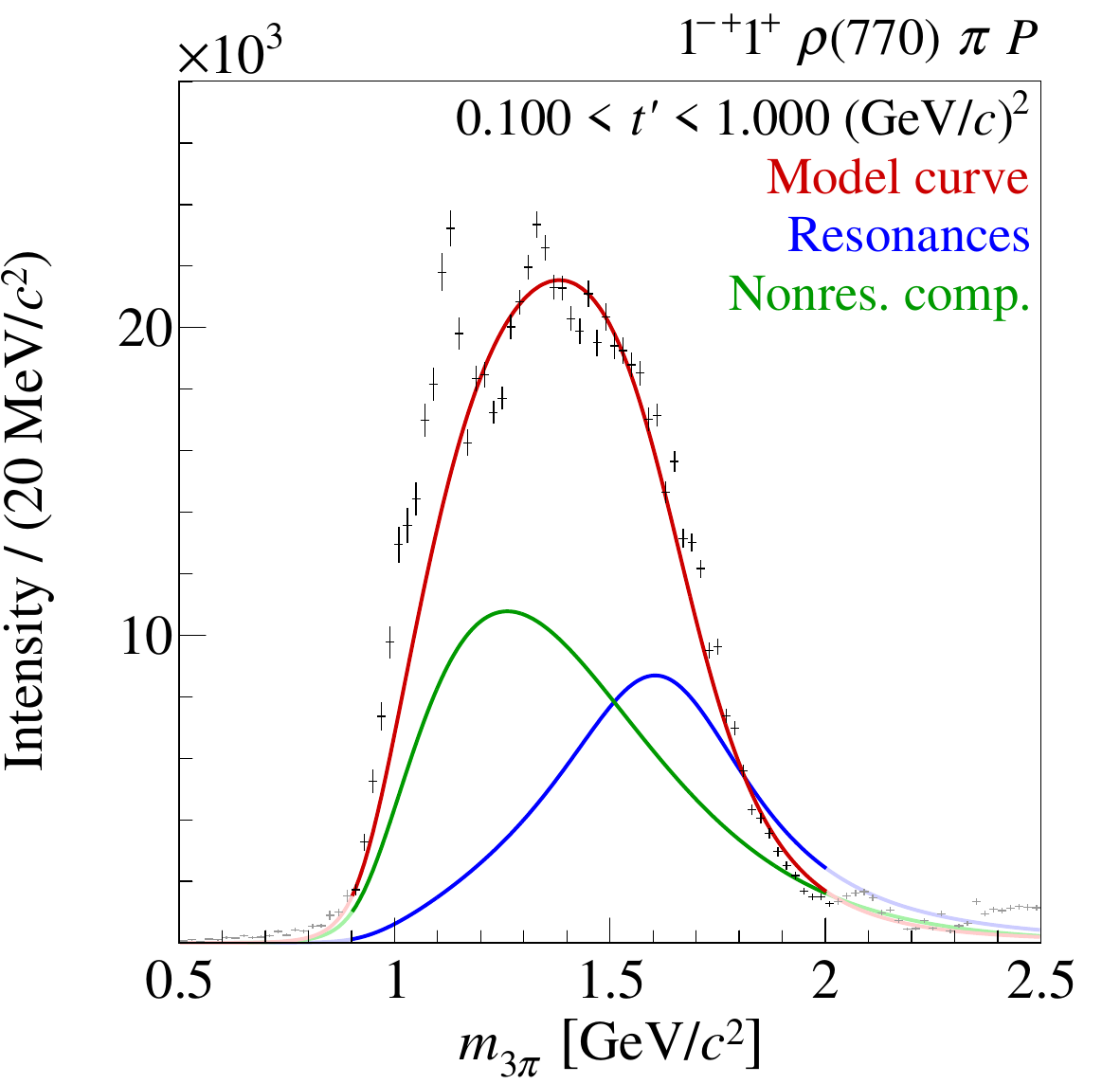}%
    \vphantom{\includegraphics[width=0.4\textwidth,valign=m]{fig57a}}%
    \label{fig:1mp_compass_tsummed}%
  }%
  \hfill\null%
  \\
  \null\hfill%
  \subfloat[][]{%
    \includegraphics[width=0.34\textwidth,valign=m]{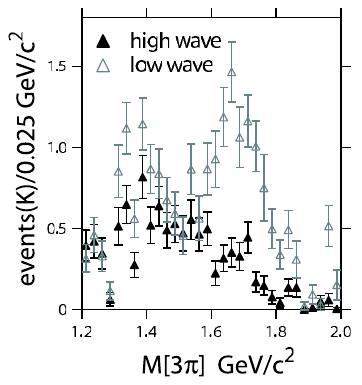}%
    \label{fig:1mp_e852}%
  }%
  \hfill%
  \subfloat[][]{%
    \includegraphics[width=1.05\threePlotWidth,valign=m]{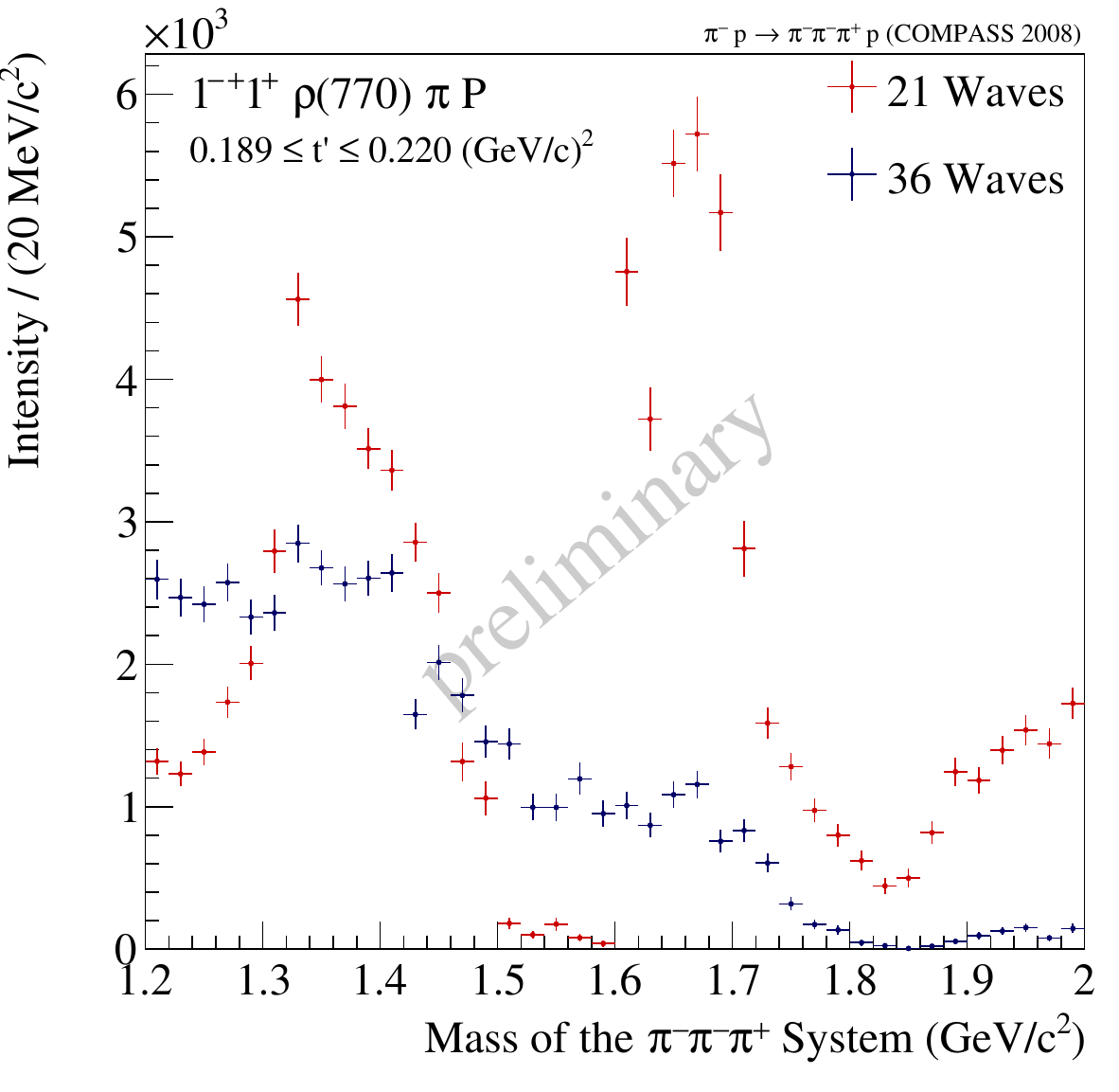}%
    \vphantom{\includegraphics[width=0.34\textwidth,valign=m]{fig57c}}%
    \label{fig:1mp_e852_compass}%
  }%
  \hfill\null%
  \\
  \null\hfill%
  \subfloat[][]{%
    \includegraphics[width=0.4\textwidth,valign=m]{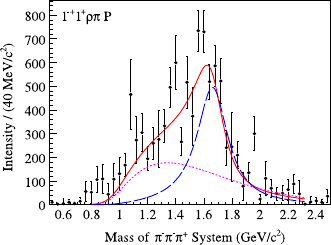}%
    \vphantom{\includegraphics[width=1.1\threePlotWidth,valign=m]{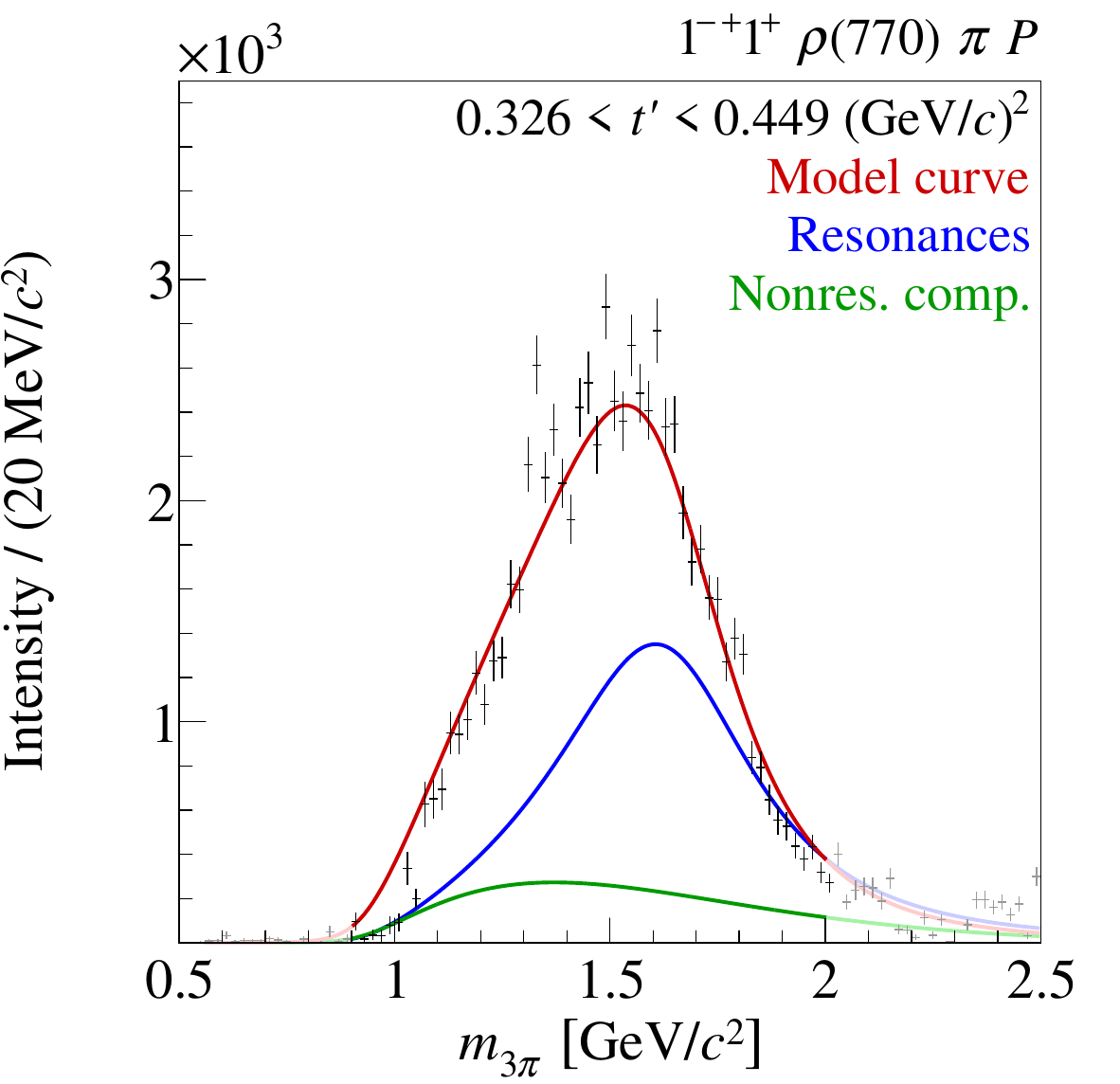}}%
    \label{fig:1mp_compass_pb}%
  }%
  \hfill%
  \subfloat[][]{%
    \includegraphics[width=1.1\threePlotWidth,valign=m]{fig57f}%
    \label{fig:1mp_compass_tbin10}%
  }%
  \hfill\null%
  \caption{Comparison of intensity distributions of the
    \wave{1}{-+}{1}{+}{\Pprho}{P} wave obtained by different
    experiments measuring the diffractive dissociation of pion beams
    into \threePi.  Note the different \mThreePi~ranges.
    \subfloatLabel{fig:1mp_ves}~Result of an analysis of VES
    data~\cite{Zaitsev:2000rc}.
    \subfloatLabel{fig:1mp_compass_tsummed}~Result of the analysis of
    COMPASS proton-target data~\cite{Akhunzyanov:2018lqa}. The
    intensity is summed over the analyzed 11~$t'$~bins.
    \subfloatLabel{fig:1mp_e852}~Results of an analysis of BNL E852
    data using two different PWA models~\cite{Dzierba:2005jg} (see
    text).  \subfloatLabel{fig:1mp_e852_compass}~Result of an analysis
    of the COMPASS proton-target data using the same two PWA models as
    in~\subfloatLabel{fig:1mp_e852}.\protect\footnotemark\ The 21-wave
    set corresponds to \textquote{low wave}
    in~\subfloatLabel{fig:1mp_e852}, the 36-wave set to
    \textquote{high wave}.
    \subfloatLabel{fig:1mp_compass_pb}~and~\subfloatLabel{fig:1mp_compass_tbin10}:
    Results of analyses of COMPASS data:
    \subfloatLabel{fig:1mp_compass_pb}~using a solid-lead target and
    integrating over
    \SIvalRange{0.1}{t'}{1.0}{\GeVcsq}~\cite{Alekseev:2009aa} and
    \subfloatLabel{fig:1mp_compass_tbin10}~using a proton target and
    selecting the kinematic range
    \SIvalRange{0.326}{t'}{0.449}{\GeVcsq}~\cite{Akhunzyanov:2018lqa}.}
  \label{fig:1mp_comparison}
\end{figure}

Although \PpiOne[1600] signals were claimed in various final states
measured by several experiments, in most of these analyses the
resonance interpretation of the \PpiOne[1600] relies on model
assumptions and alternative explanations could not be ruled out
completely.  Hence the experimental situation is actually rather
unclear.  Particularly controversial is the \PpiOne[1600] signal in
the $\Pprho \pi$ channel.  The BNL E852 experiment used an
\SI{18}{\GeVc} pion beam incident on a proton target and was the first
to claim a signal for $\PpiOne[1600] \to \Pprho \pi$ based on a PWA
performed in the kinematic range
\SIvalRange{0.1}{t'}{1.0}{\GeVcsq}~\cite{Adams:1998ff,Chung:2002pu}.
The VES experiment used a \SI{37}{\GeVc} pion beam on a
solid-beryllium target and performed the PWA in the kinematic range
\SIvalRange{0.03}{t'}{1.0}{\GeVcsq}.  They also observed significant
intensity in the \wave{1}{-+}{1}{+}{\Pprho}{P} wave (see
\cref{fig:1mp_ves}).  However, they found that the intensity
distribution in this wave depends significantly on the PWA model and
hence concluded that the wave is contaminated by intensity that leaks
from the dominant $1^{++}$ waves into the $1^{-+}$
wave~\cite{Zaitsev:2000rc}.  They neither excluded nor claimed the
existence of the \PpiOne[1600].  A later analysis of a BNL E852 data
sample that was about an order or magnitude larger than the one used
in \refsCite{Adams:1998ff,Chung:2002pu} came to the conclusion that
there is no \PpiOne[1600] signal in the $\Pprho \pi$
channel~\cite{Dzierba:2005jg}.  The authors of
\refCite{Dzierba:2005jg} performed systematic studies to find the
optimal wave set and found that in the original BNL E852 analysis in
\refsCite{Adams:1998ff,Chung:2002pu} a number of important waves were
missing in the PWA model.  When they included these waves in the PWA
model, the peak at about \SI{1.6}{\GeVcc} in the
\wave{1}{-+}{1}{+}{\Pprho}{P} wave disappeared.  However, the slow
phase motions \wrt other waves remained.  This is shown in
\cref{fig:1mp_e852} for the kinematic range
\SIvalRange{0.18}{t'}{0.23}{\GeVcsq}, where the \textquote{low wave}
points are the result of a PWA fit performed with the smaller wave set
from \refsCite{Adams:1998ff,Chung:2002pu} and the \textquote{high
  wave} points are the result of a PWA fit performed with the larger
wave set from \refCite{Dzierba:2005jg}.  Based on this observation,
the authors of \refCite{Dzierba:2005jg} concluded that there is no
evidence for a \PpiOne[1600] in this wave.  For the discussion below
it is important to note that this conclusion was based on a PWA
performed in the range $t' < \SI{0.53}{\GeVcsq}$.  In contrast to the
analysis in \refCite{Dzierba:2005jg}, the analysis of the \threePi
data from the COMPASS experiment using a solid-lead target showed
again evidence for a \PpiOne[1600] signal~\cite{Alekseev:2009aa} (see
\cref{fig:1mp_compass_pb}) although the PWA model contained even more
partial waves than the one used in \refCite{Dzierba:2005jg}.

\footnotetext{Publication in preparation.}

\begin{figure}[tbp]
  \centering
  \subfloat[][]{%
    \includegraphics[width=\threePlotWidth]{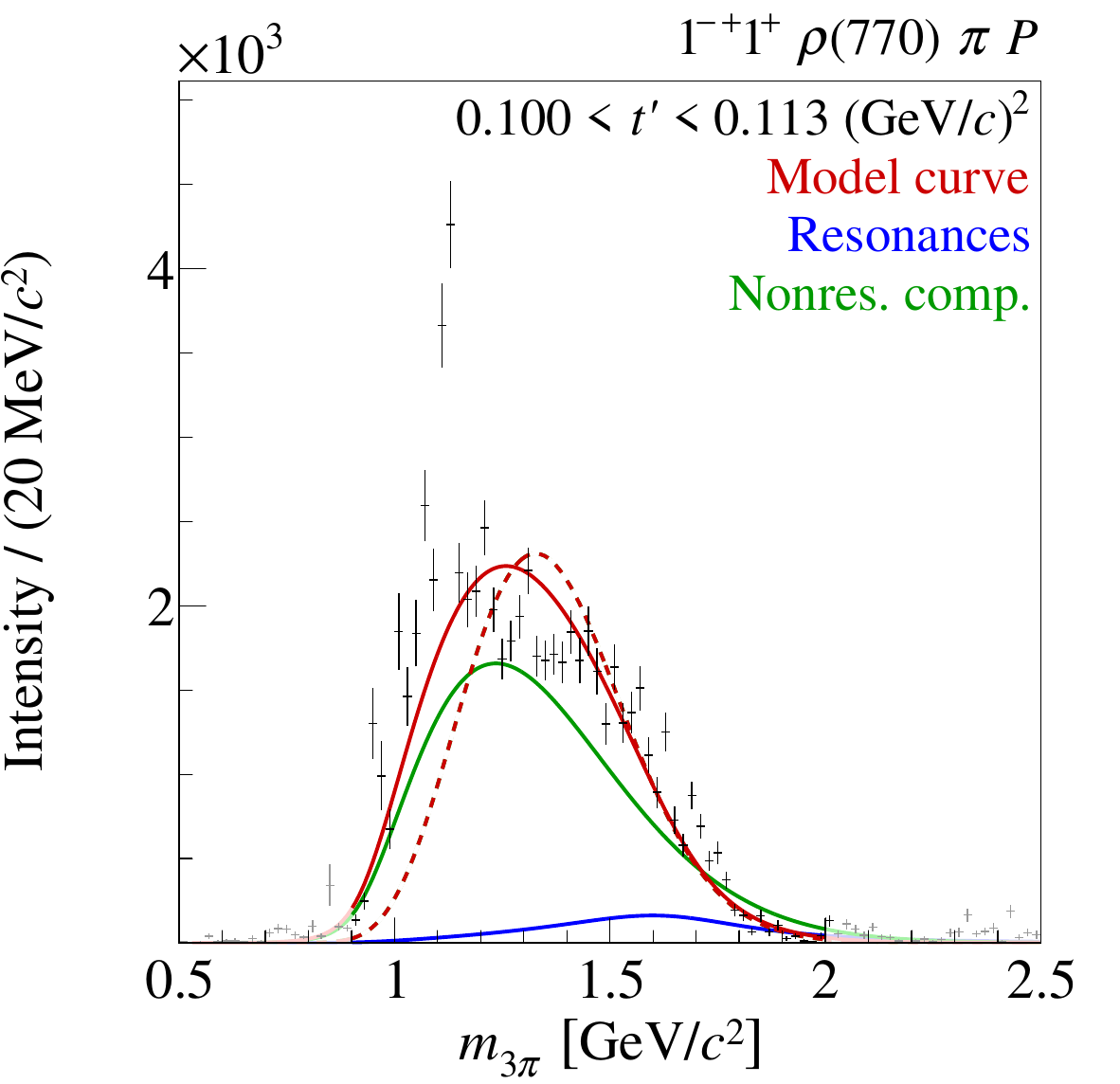}
    \label{fig:intensity_1mp_tbin1}%
  }%
  \hfill%
  \subfloat[][]{%
    \includegraphics[width=\threePlotWidth]{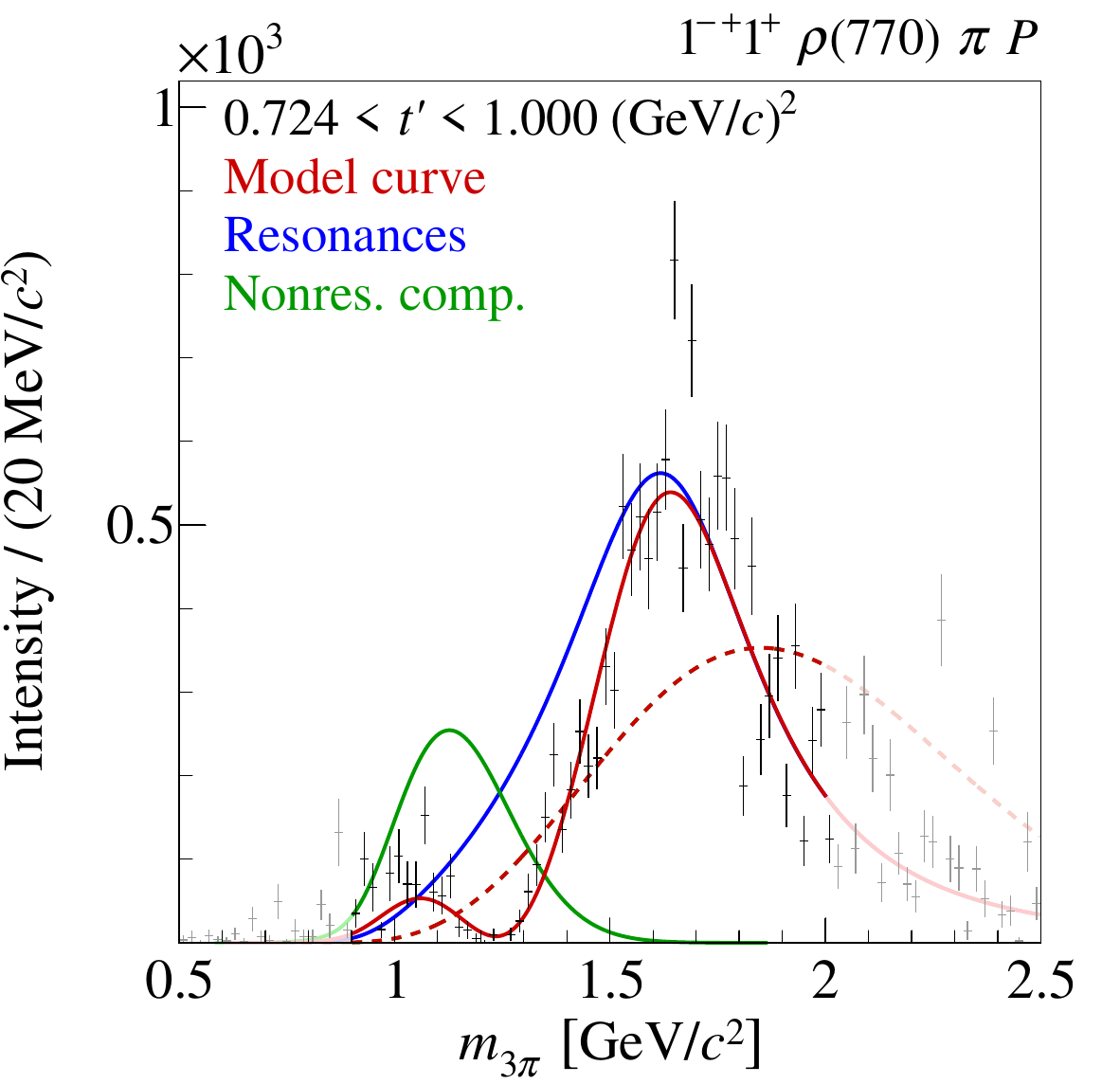}%
    \label{fig:intensity_1mp_tbin11}%
  }%
  \hfill%
  \subfloat[][]{%
    \includegraphics[width=\threePlotWidth]{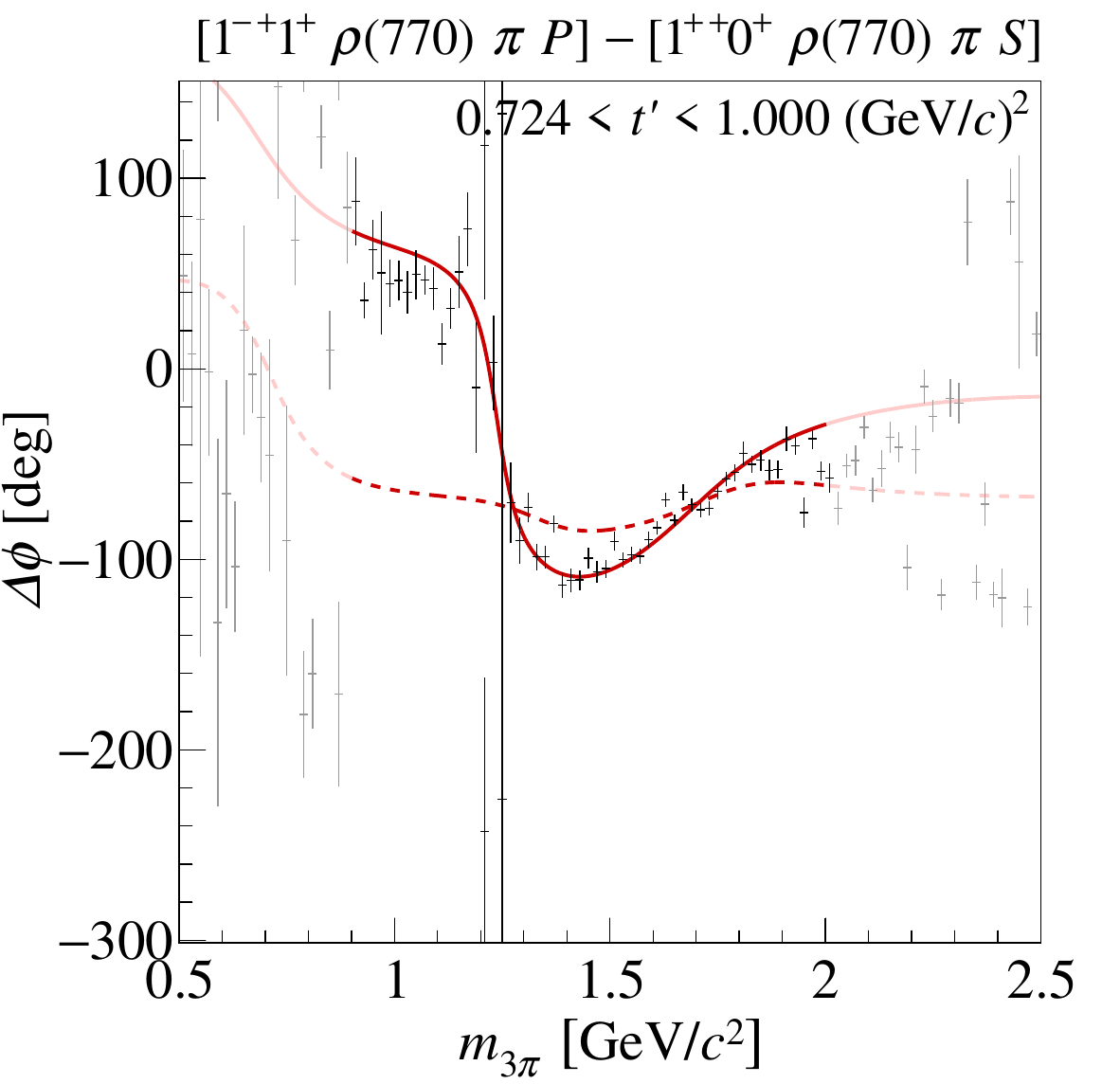}%
    \label{fig:phase_1mp_tbin11}%
  }%
  \caption{\subfloatLabel{fig:intensity_1mp_tbin1}~and~\subfloatLabel{fig:intensity_1mp_tbin11}:
    Intensity distributions of the \wave{1}{-+}{1}{+}{\Pprho}{P} wave
    in the COMPASS \threePi proton-target data in the lowest and
    highest $t'$~bin, respectively~\cite{Akhunzyanov:2018lqa}.
    \subfloatLabel{fig:phase_1mp_tbin11}~Phase of this wave relative
    to the \wave{1}{++}{0}{+}{\Pprho}{S} wave in the highest $t'$~bin.
    The curves represent the result of two resonance-model fits.  The
    model and the wave components are represented as in
    \cref{fig:intensity_phase_0mp} except that the blue curves
    represent the \PpiOne[1600].  The result of the main
    resonance-model fit is represented by the continuous curves.  The
    dashed curves represent the result of a fit, where the
    \PpiOne[1600] component is removed from the resonance model, \ie
    where we try to model the data using only a non-resonant
    component.}
  \label{fig:1mp}
\end{figure}

The 88-wave set used to analyze the COMPASS \threePi proton-target
data also includes the \wave{1}{-+}{1}{+}{\Pprho}{P} wave, which has a
relative intensity of \SI{0.8}{\percent}.  As shown in
\cref{fig:intensity_1mp_tbin1,fig:intensity_1mp_tbin11}, the shape of
the intensity distribution of this wave has a surprisingly strong
dependence on~$t'$.  At low~$t'$, the intensity distribution exhibits
a broad structure with a maximum at about \SI{1.2}{\GeVcc} (see
\cref{fig:intensity_1mp_tbin1}).  This structure becomes narrower with
increasing~$t'$ and the maximum moves to higher masses, such that a
peak emerges at about \SI{1.6}{\GeVcc} in the two highest $t'$~bins
(see \eg \cref{fig:intensity_1mp_tbin11}).  This $t'$~dependence of
the intensity distribution illustrates the necessity for performing
the analysis in narrow $t'$~bins.  It also indicates that, in addition
to potential resonant components, this wave contains large
contributions from non-resonant processes.  This is consistent with
the fact that we do not observe large phase motions \wrt other waves
in the \SI{1.6}{\GeVcc} region (see \eg\ \cref{fig:phase_1mp_tbin11}).

The strong modulation of the shape of the $1^{-+}$ intensity
distribution with~$t'$ is successfully reproduced by the resonance
model as a $t'$-dependent interference of a \PpiOne[1600] Breit--Wigner
amplitude as in \cref{eq:BW_const_width} and a non-resonant component
parameterized using \cref{eq:dyn_amp_non_res}.  The dashed curves in
\cref{fig:1mp} represent the result of a fit, where the \PpiOne[1600]
resonance component is omitted from the model so that only the
non-resonant component remains in this wave.  Although at low~$t'$,
the intensity distribution is still reproduced roughly by the model
(see \cref{fig:intensity_1mp_tbin1}), it fails to reproduce the phases
(see \eg\ \cref{fig:phase_1mp_tbin11}) and the intensity distributions
at higher~$t'$ (see \cref{fig:intensity_1mp_tbin11}).  This
demonstrates that the COMPASS \threePi proton-target data require a
resonant component in the $1^{-+}$ wave.  The resonance interpretation
is supported by the fact that the coupling phases of the \PpiOne[1600]
component relative to the other resonances in the model (see
\cref{eq:phase_comp_def}) depend only weakly on~$t'$.  In contrast,
the coupling phase of the $1^{-+} $non-resonant component \wrt the
\PpiOne[1600] changes strongly with~$t'$, as does the shape of the
non-resonant component (see Section~VII in
\refCite{Akhunzyanov:2018lqa} for more details).

\begin{figure}[tbp]
  \centering
  \subfloat[][]{%
    \includegraphics[width=\threePlotWidth]{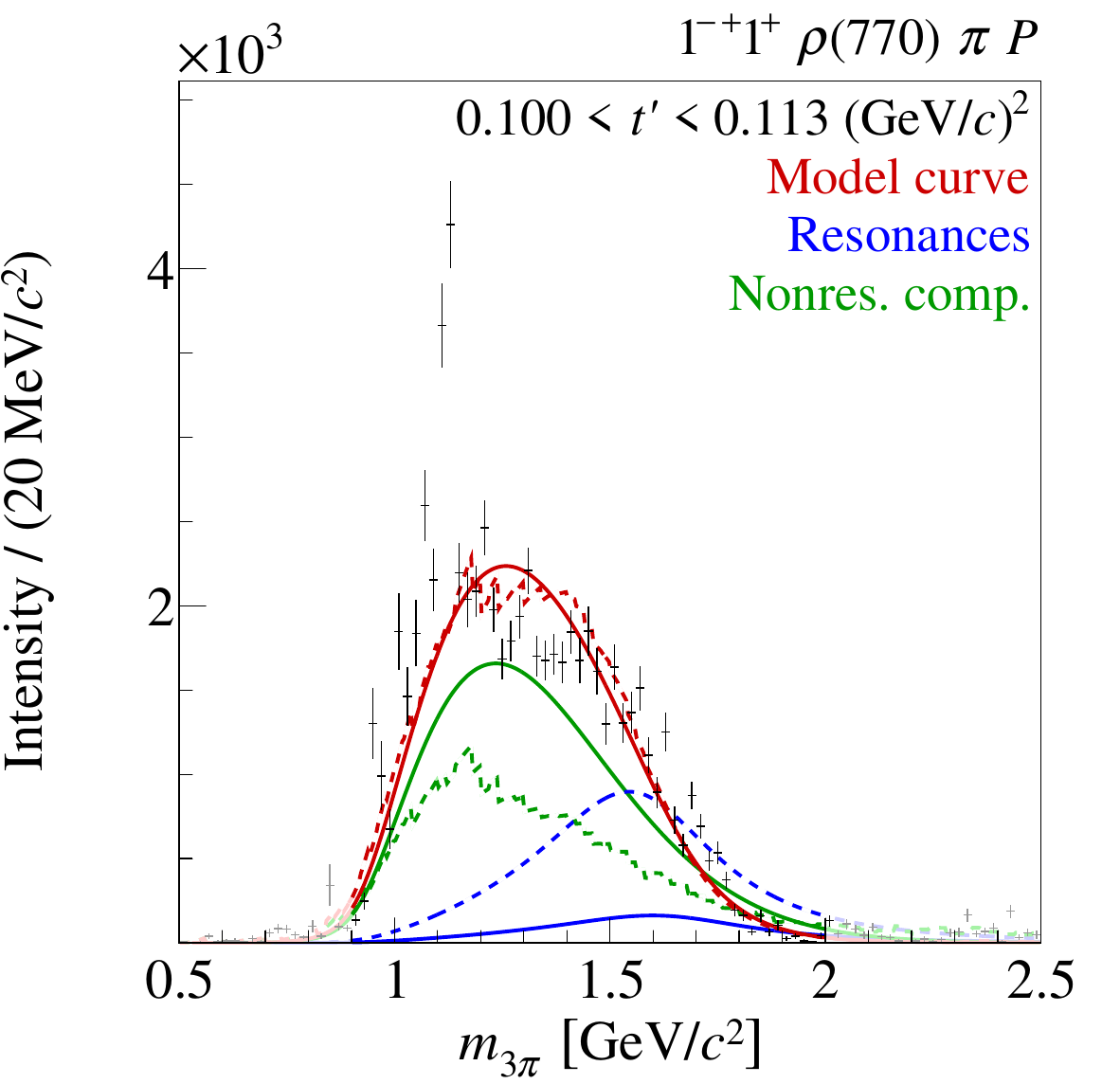}
    \label{fig:intensity_1mp_deck_tbin1}%
  }%
  \hfill%
  \subfloat[][]{%
    \includegraphics[width=\threePlotWidth]{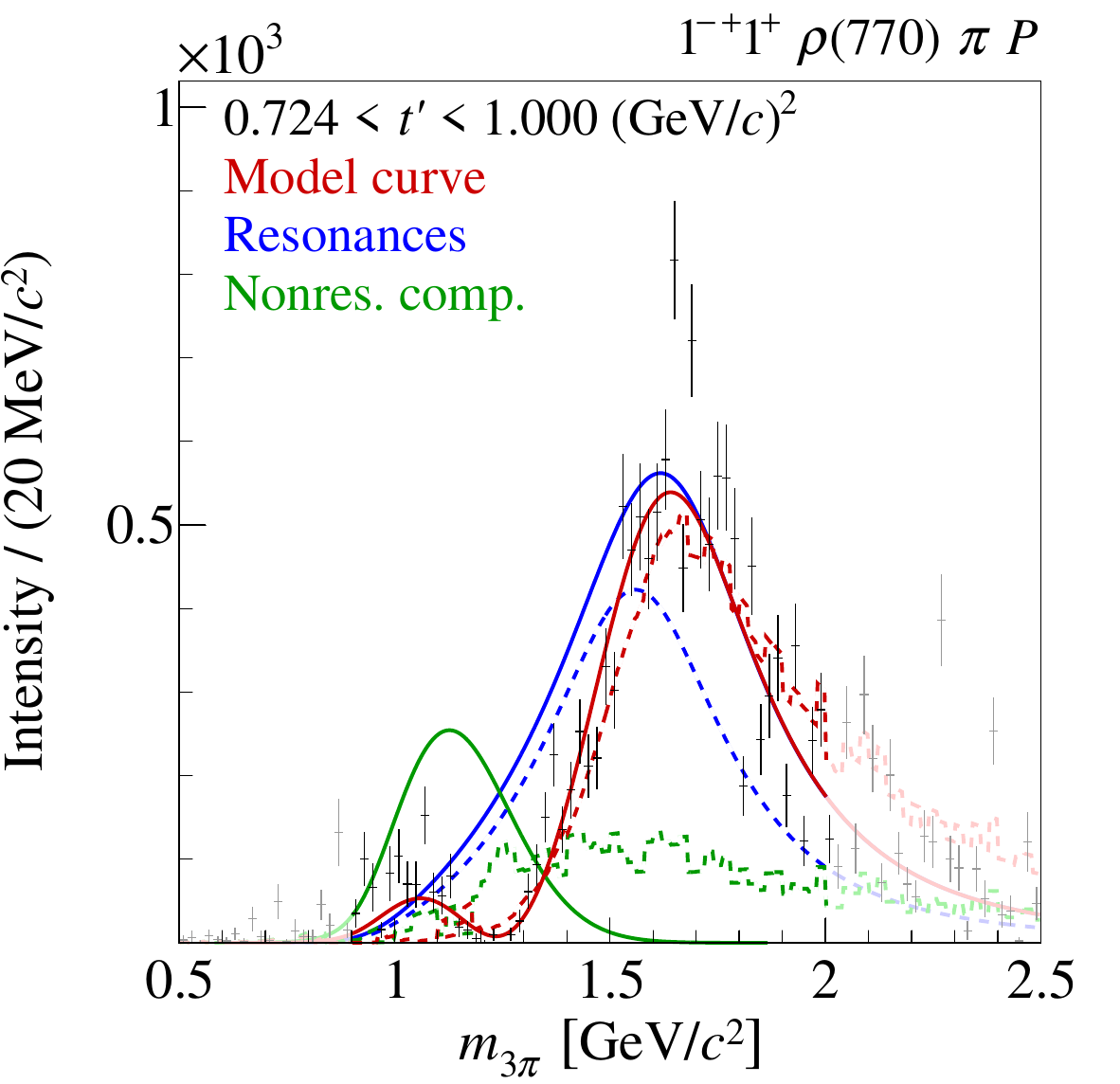}%
    \label{fig:intensity_1mp_deck_tbin11}%
  }%
  \hfill%
  \subfloat[][]{%
    \includegraphics[width=\threePlotWidth]{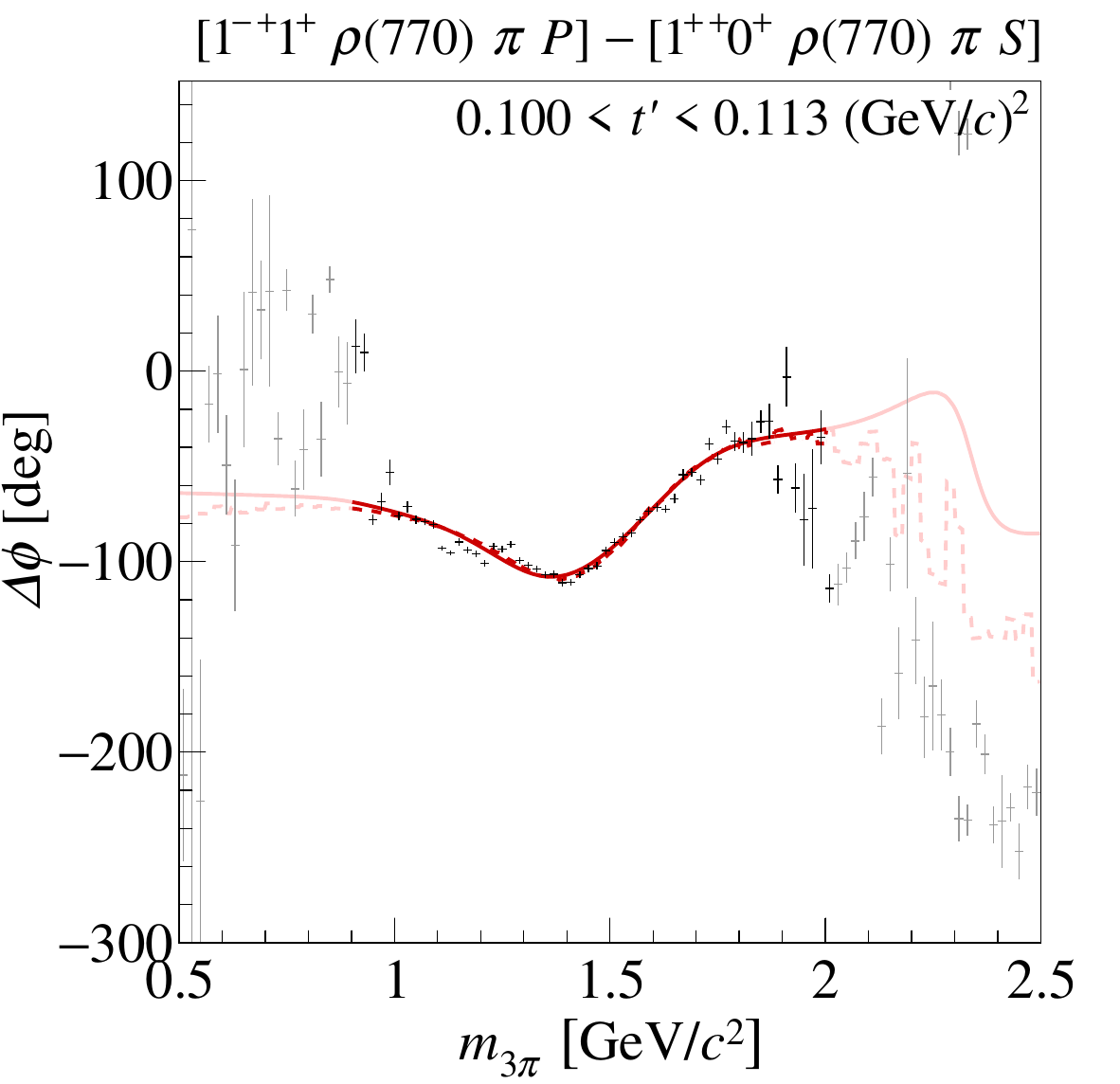}%
    \label{fig:phase_1mp_deck_tbin1}%
  }%
  \caption{Similar to \cref{fig:1mp}, except that the phase is shown
    for the lowest $t'$~bin~\cite{Akhunzyanov:2018lqa}.  Here, the
    dashed curves represent the result of a resonance-model fit, where
    the empirical parameterizations of the non-resonant components are
    replaced by the square root of the intensity distributions of the
    partial-wave decomposition of Deck Monte Carlo data.}
  \label{fig:1mp_deck}
\end{figure}

Due to the large contribution of the non-resonant component to the
$1^{-+}$ intensity, especially at low~$t'$, the fit result depends
strongly on the parameterization used for the non-resonant component.
This model dependence is studied by replacing the empirical
parameterizations of the non-resonant components (see
\cref{eq:dyn_amp_non_res_simple,eq:dyn_amp_non_res}) by the square
root of the intensity distributions obtained from the partial-wave
decomposition of Deck Monte Carlo data (see
\cref{sec:3pi_model:resonance}).  The result of this study is
represented by the dashed curves in \cref{fig:1mp_deck}.  Compared to
the main fit that employs the empirical parameterizations of the
non-resonant components, the description of the $1^{-+}$ amplitude is
only slightly worse.  The non-resonant component has a qualitatively
similar dependence on~$t'$ and has shapes similar to those in the main
fit.  Only in the highest $t'$~bin, the non-resonant component has a
different, actually more plausible shape than the one found in the
main fit.  In the Deck study, the yield of the non-resonant component
is significantly lower than in the main fit, whereas the yield of the
\PpiOne[1600] is significantly higher without affecting the resonance
parameters too much.  This shows that in our resonance model, the
\PpiOne[1600] yield has a large systematic uncertainty.

\begin{figure}[tbp]
  \centering
  \hfill%
  \subfloat[][]{%
    \includegraphics[width=\threePlotWidth]{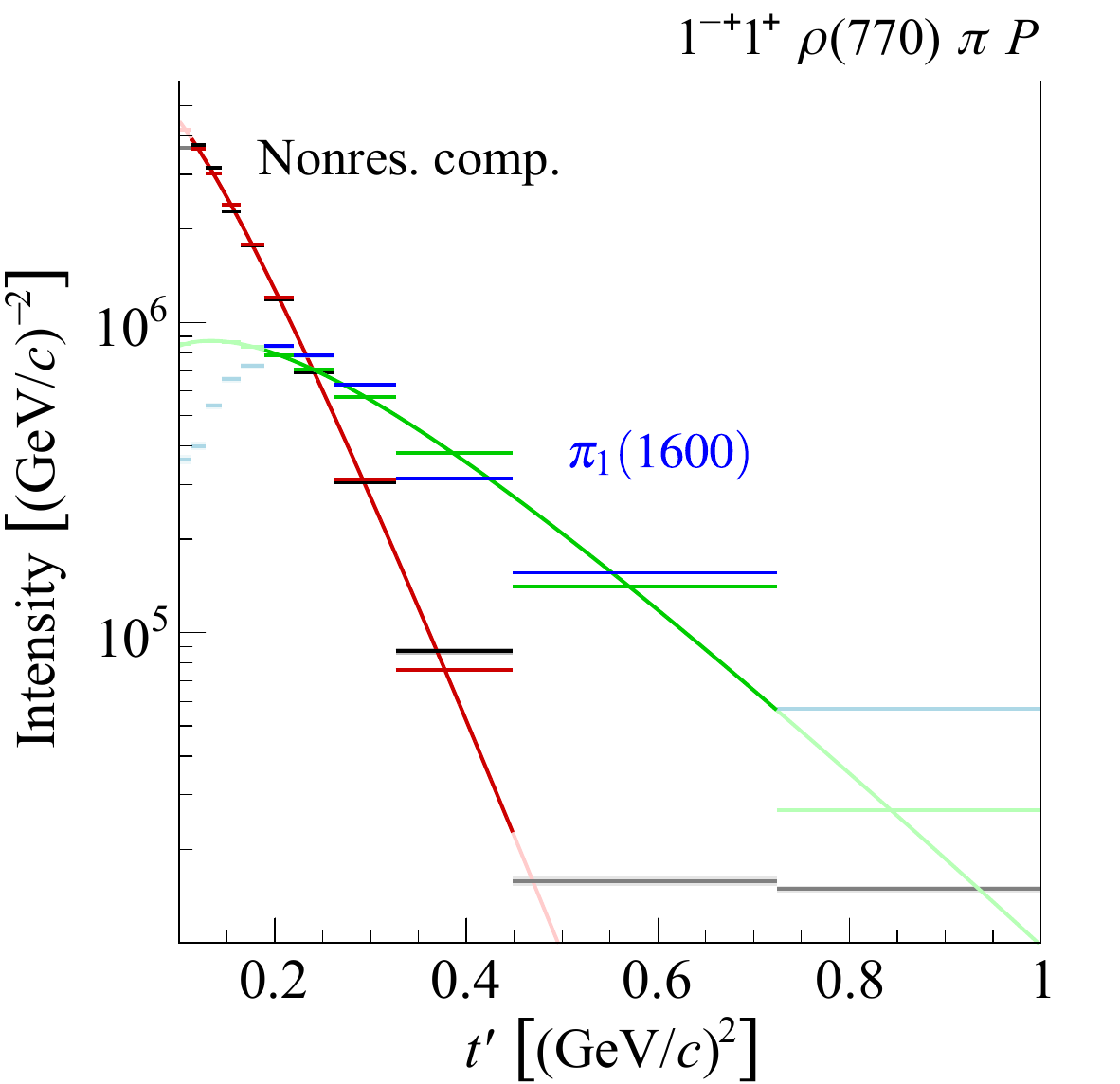}%
    \label{fig:1mp_t_spectrum_main}%
  }%
  \hfill%
  \subfloat[][]{%
    \includegraphics[width=\threePlotWidth]{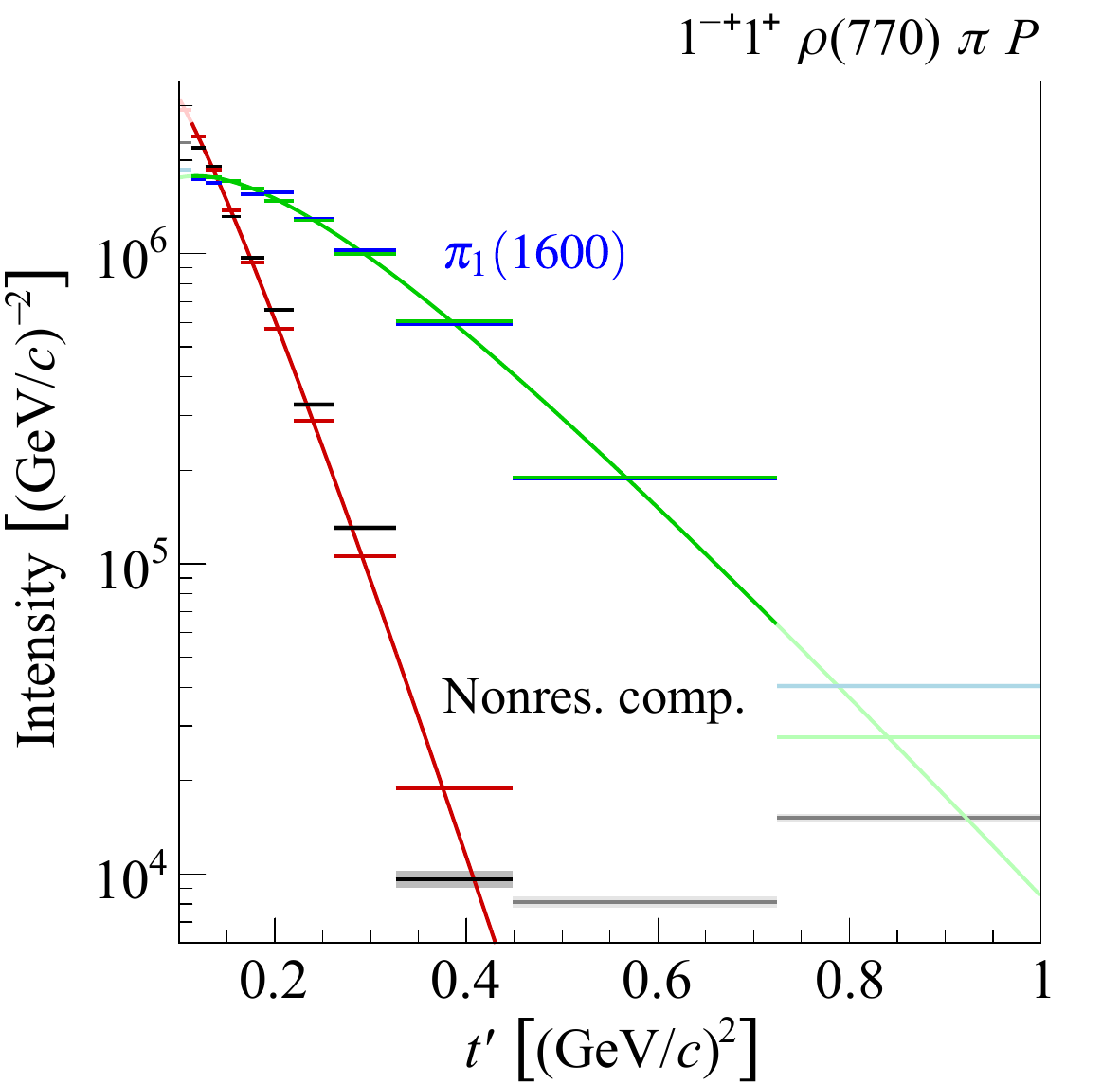}%
    \label{fig:1mp_t_spectrum_deck}%
  }%
  \hfill\null%
  \caption{Similar to \cref{fig:tspectrum_0mp_f0}, but showing the
    $t'$~spectra of the two components in the
    \wave{1}{-+}{1}{+}{\Pprho}{P} wave~\cite{Akhunzyanov:2018lqa}.
    The \PpiOne[1600] component is shown as blue lines and light blue
    boxes, and the non-resonant component as black lines and gray
    boxes.  \subfloatLabel{fig:1mp_t_spectrum_main}~Result of the main
    fit.  \subfloatLabel{fig:1mp_t_spectrum_deck}~Result of a fit, in
    which the empirical parameterizations of the non-resonant
    components are replaced by the square root of the intensity
    distributions of the partial-wave decomposition of Deck Monte
    Carlo data.}
    \label{fig:1mp_t_spectrum}%
\end{figure}

The $t'$~spectrum of the non-resonant component (black lines in
\cref{fig:1mp_t_spectrum_main}) falls steeply with~$t'$ and has an
exceptionally large slope parameter value of
$\SIaerr{19.1}{1.4}{4.7}{\perGeVcsq}$.  Hence the non-resonant
component dominates the $1^{-+}$ intensity at low~$t'$.  Only for
$t' \gtrsim \SI{0.3}{\GeVcsq}$, the intensity of the \PpiOne[1600]
component (blue lines in \cref{fig:1mp_t_spectrum_main}) becomes
larger than that of the non-resonant component.  The simple
exponential model in \cref{eq:t_spectrum_model} is not able to
reproduce the downturn of the \PpiOne[1600] $t'$~spectrum toward
low~$t'$.  However, this might be an artificial effect caused by an
inappropriate description of the non-resonant component by our
parameterizations.  It seems that the fit is not able to completely
separate the \PpiOne[1600] from the non-resonant component, which
dominates at low~$t'$.  This hypothesis is supported by the study
discussed above, where the shape of the non-resonant component is
determined from a model of the Deck process (see
\cref{fig:1mp_t_spectrum_deck}).  In this study, the \PpiOne[1600] has
a larger yield at low~$t'$ so that the resulting $t'$~spectrum of the
\PpiOne[1600] is well described by the exponential model in
\cref{eq:t_spectrum_model} and has a slope parameter of
\SI{7.3}{\perGeVcsq}, which lies in the range expected for resonances.

The results from our $t'$-resolved analysis of the \threePi
proton-target data explain the seemingly contradictory experimental
findings of the BNL E852, VES, and COMPASS experiments concerning the
existence of a \PpiOne resonance in the $\Pprho \pi$ $P$~wave.
\Cref{fig:1mp_compass_tsummed} shows that our intensity distribution
of the \wave{1}{-+}{1}{+}{\Pprho}{P} wave summed over the 11~$t'$~bins
is very similar to the one found in the VES
analysis~\cite{Zaitsev:2000rc} (\confer\ \cref{fig:1mp_ves}).  In
\cref{fig:1mp_e852_compass}, we show the intensity of the
\wave{1}{-+}{1}{+}{\Pprho}{P} wave in the range
\SIvalRange{0.18}{t'}{0.23}{\GeVcsq} that is obtained if we perform
the PWA using the same two wave sets as in \refCite{Dzierba:2005jg}.
The similarity of the results with \cref{fig:1mp_e852} confirms that
the \PpiOne[1600] intensity peak in the original BNL E852 analysis in
\refsCite{Adams:1998ff,Chung:2002pu} was mostly an artificial
structure caused by leakage of intensity from the
\wave{2}{-+}{1}{+}{\Pprho}{P}, \wave{2}{-+}{0}{+}{\Pprho}{F}, and
\wave{2}{-+}{1}{+}{\Pprho}{F} waves that were missing in the PWA
model.  We also confirm the finding of \refCite{Dzierba:2005jg} that
in the region $t' \lesssim \SI{0.5}{\GeVcsq}$ there is only weak
evidence for the \PpiOne[1600].  However, our data show that a
resonance-like signal is required to describe the data in the region
$t' \gtrsim \SI{0.5}{\GeVcsq}$ (see
\cref{fig:intensity_1mp_tbin11,fig:phase_1mp_tbin11}).  This
$t'$~region was excluded from the analysis in
\refCite{Dzierba:2005jg}.  In the COMPASS \threePi data taken with a
solid-lead target, the contribution of the non-resonant component is
much smaller than in the proton-target data.  The $t'$-integrated
lead-target data actually resemble the high-$t'$ region of the
proton-target data (compare
\cref{fig:1mp_compass_pb,fig:1mp_compass_tbin10}).  So far, no
explanation has been found for this effect.

\begin{figure}[tbp]
  \centering
  \hfill%
  \subfloat[]{%
    \includegraphics[width=\twoPlotWidth]{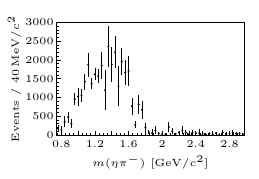}%
    \label{fig:int_eta_p_wave}%
  }%
  \hfill%
  \subfloat[]{%
    \includegraphics[width=\twoPlotWidth]{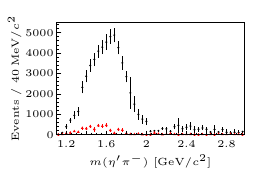}%
    \label{fig:int_eta_etaprime_p_wave}%
  }%
  \hfill\null%
  \\
  \null\hfill%
  \subfloat[]{%
    \includegraphics[width=\twoPlotWidth]{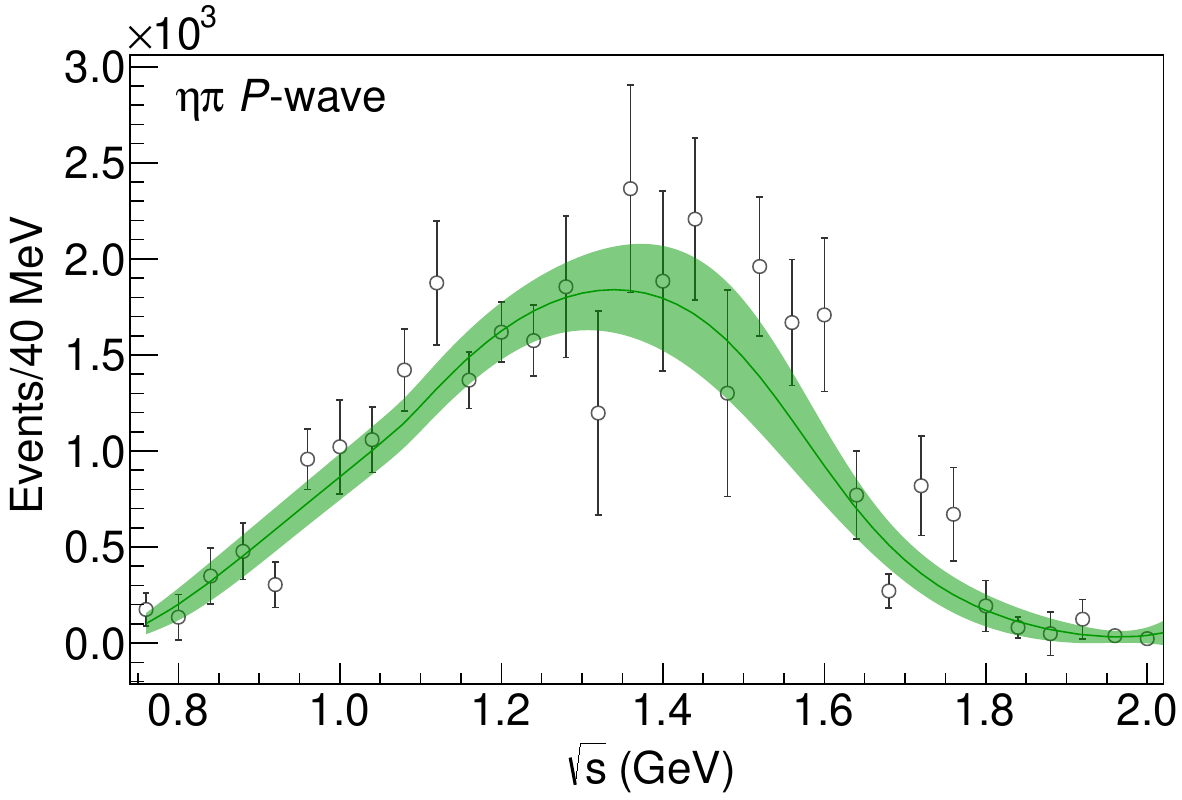}%
    \label{fig:int_eta_p_wave_fit}%
  }%
  \hfill%
  \subfloat[]{%
    \includegraphics[width=\twoPlotWidth]{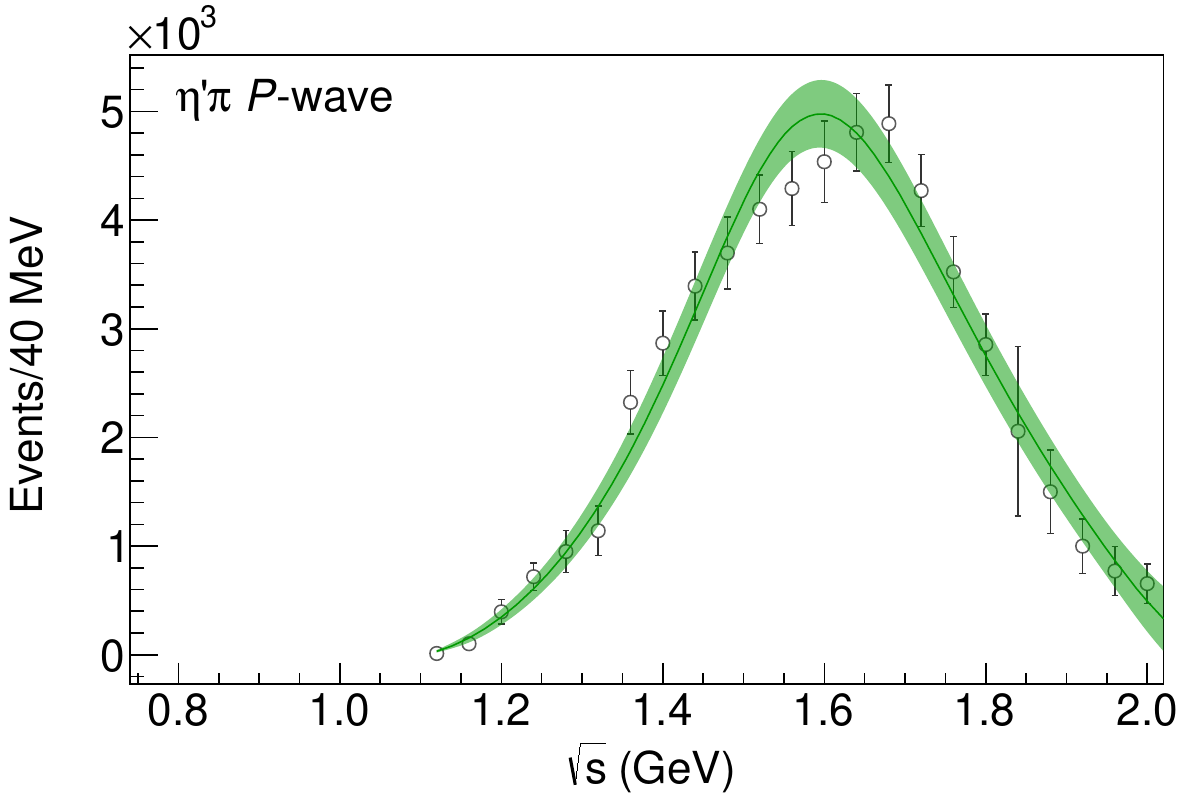}%
    \label{fig:int_etaprime_p_wave_fit}%
  }%
  \hfill\null%
  \\
  \null\hfill%
  \subfloat[]{%
    \includegraphics[width=\twoPlotWidth]{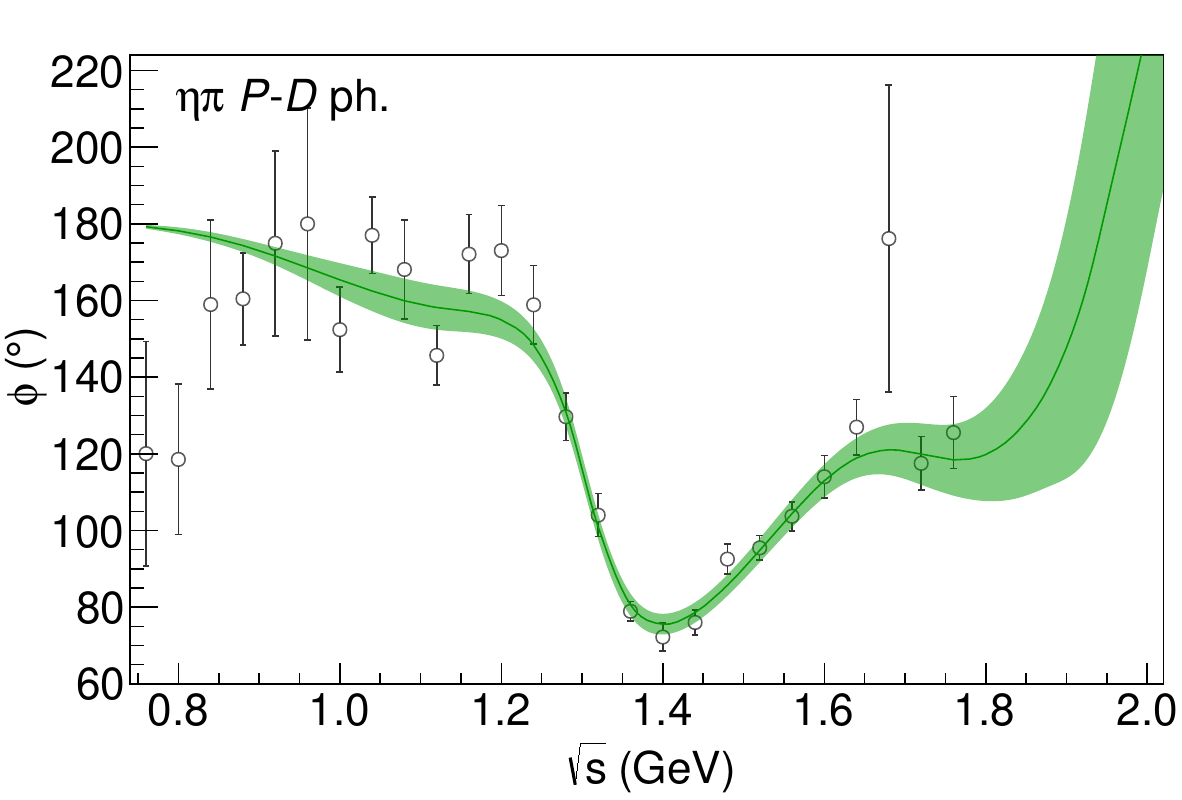}%
    \label{fig:phase_eta_p_d_wave_fit}%
  }%
  \hfill%
  \subfloat[]{%
    \includegraphics[width=\twoPlotWidth]{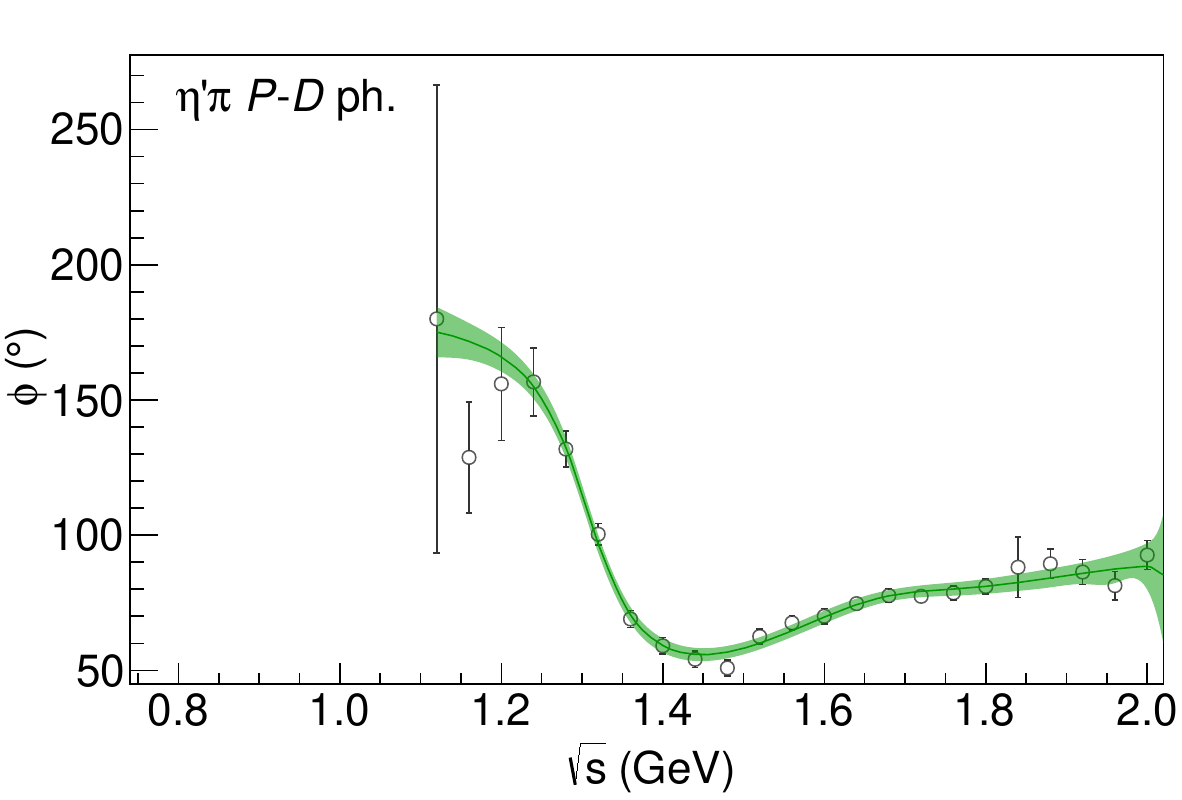}%
    \label{fig:phase_etaprime_p_d_wave_fit}%
  }%
  \hfill\null%
  \caption{Intensities of the $P$~wave with $\Mrefl = 1^+$
    \subfloatLabel{fig:int_eta_p_wave}~in the \etaPim and
    \subfloatLabel{fig:int_eta_etaprime_p_wave}~in the \etaPrPim
    COMPASS data (black dots)~\cite{Adolph:2014rpp}.  The red
    triangles in~\subfloatLabel{fig:int_eta_etaprime_p_wave} show the
    \etaPim intensity from~\subfloatLabel{fig:int_eta_p_wave} scaled
    by the kinematic factor in \cref{eq:etaprime_eta_r}.
    \subfloatLabel{fig:int_eta_p_wave_fit}~and~\subfloatLabel{fig:int_etaprime_p_wave_fit}:
    Same data as in
    \subfloatLabel{fig:int_eta_p_wave}~and~\subfloatLabel{fig:int_eta_etaprime_p_wave}
    overlaid with the result of a fit of the unitary model (green
    curves) described in
    \cref{sec:pwa.unitary_model}~\cite{Rodas:2018owy}.  The shaded
    bands represent the two-standard-deviation confidence interval.
    Note the different mass range.
    \subfloatLabel{fig:phase_eta_p_d_wave_fit}~and~\subfloatLabel{fig:phase_etaprime_p_d_wave_fit}:
    Phases of the $P$~wave relative to the $D$~wave both with
    $\Mrefl = 1^+$~\cite{Rodas:2018owy}.  The phase drop around
    \SI{1.3}{\GeVcc} is due to the \PaTwo (see
    \cref{sec:results_2pp}). Data and model curve as in
    \subfloatLabel{fig:int_eta_p_wave}~and~\subfloatLabel{fig:int_eta_etaprime_p_wave}.}
  \label{fig:int_phase_eta_etaprime_p_wave}
\end{figure}

The intensity distributions of the $P$~wave with $\Mrefl = 1^+$ in the
COMPASS \etaPim and \etaPrPim data (black dots in
\cref{fig:int_eta_p_wave,fig:int_eta_etaprime_p_wave}) are similar to
previous observations.  The \etaPim $P$-wave intensity shows a broad
peak of about \SI{400}{\MeVcc} width, centered at a mass of
\SI{1.4}{\GeVcc}.  For $m_{\etaPi} > \SI{1.8}{\GeVcc}$, the intensity
nearly vanishes.  The \etaPim $P$-wave intensity is about a factor
of~20 smaller than the dominant $D$-wave intensity.  Hence the
detector acceptance has to be simulated accurately in order to resolve
the small $P$-wave contribution.  In the \etaPrPim data, $P$ and
$D$~waves have comparable intensities.  The $P$-wave intensity
distribution also shows a broad peak but the peak position is shifted
to about \SI{1.65}{\GeVcc}.  Similar to \etaPim, the intensity is
small in the high-mass region $m_{\etaPrPi} > \SI{2}{\GeVcc}$.  The
differences in the shape of the $P$-wave intensity distributions in
\etaPim and \etaPrPim remain after scaling the \etaPim intensity
distribution by the kinematic factor in \cref{eq:etaprime_eta_r} (see
\cref{fig:int_eta_etaprime_p_wave}).  In addition, we observe an
enhancement of about a factor of~10 of the \etaPrPim $P$-wave
intensity \wrt the scaled \etaPi $P$-wave intensity.  This is in stark
contrast to the striking similarity of the \etaPrPim and the scaled
\etaPim intensity distributions observed for the $D$~and $G$~waves
(see \cref{fig:int_eta_etaprime_d_wave,fig:int_eta_etaprime_g_wave}).
The relative phases exhibit a pattern similar to the one of the
intensity distributions.  While the phases of the $P$~wave relative to
the $D$~wave are significantly different in \etaPim and \etaPrPim for
masses above \SI{1.4}{\GeVcc} (see
\cref{fig:phase_eta_p_d_wave_fit,fig:phase_etaprime_p_d_wave_fit}),
the phases between the $D$~and $G$~waves are nearly identical (see
\cref{fig:phase_eta_g_d_wave}).  Naively, one could interpret this as
different resonance content of the $P$~wave in the \etaPim and
\etaPrPim final states consistent with the observations of the
\PpiOne[1400] and the \PpiOne[1600] by previous experiments.

\begin{figure}[tbp]
  \centering
  \includegraphics[width=\textwidth]{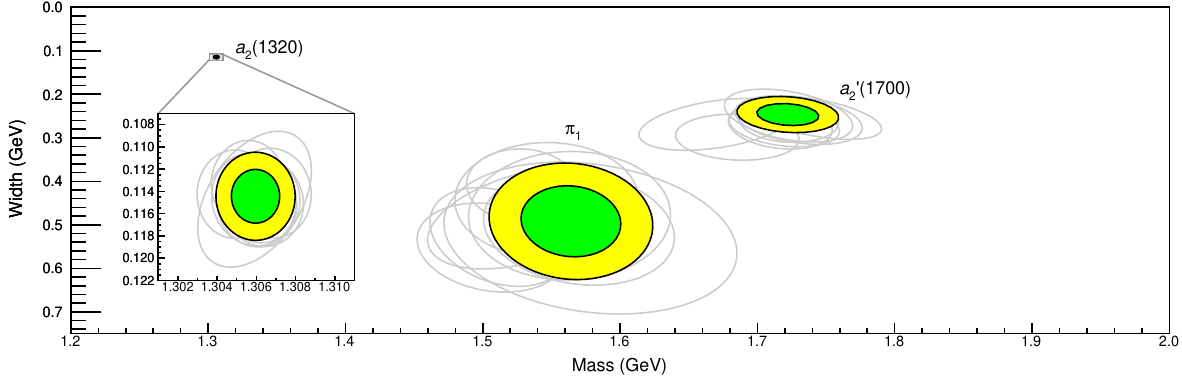}%
  \caption{Positions of poles extracted from the \etaPim and \etaPrPim
    coupled-channel analysis, corresponding to the \PaTwo,
    \PpiOne[1600], and \PaTwo[1700].  The inset shows an enlarged view
    of the \PaTwo region.  The green and yellow ellipses indicate the
    $1\sigma$ and $2\sigma$ confidence levels, respectively.  The
    light gray ellipses show the systematic variation of the $2\sigma$
    confidence contour upon changes to the
    model~\cite{Rodas:2018owy}.}
  \label{fig:poles_etapi_etaprpi}
\end{figure}

Fitting the COMPASS \etaOrPrPim data using simple Breit--Wigner
resonance-models, similar to the ones employed in previous analyses,
yields \PpiOne[1400] resonance parameters from the \etaPim data and
\PpiOne[1600] resonance parameters from the \etaPrPim data that are
consistent with previous
results~\cite{Chung:1999we,Ivanov:2001rv,Dorofeev:2001xu}.  However,
the results for the \PpiOne* resonances depend strongly on whether we
include a coherent non-resonant component in the $P$~waves and on how
we describe the high-mass shoulder in the $D$~waves (see
\cref{sec:results_2pp}).  A more stable result can be achieved by
performing an \etaPi-\etaPrPi coupled-channel fit to the $P$- and
$D$-wave data using the unitary analytic model described in
\cref{sec:pwa.unitary_model} (see green curves in
\cref{fig:int_phase_eta_etaprime_p_wave}).  Compared to
sum-of-Breit--Wigner models, this model improves, among other things,
the description of the interference of the various resonances and the
non-resonant components.  The data are described well by the model.
Since the model is analytical, the amplitude can be continued to
complex values of~$s$ in order to identify its poles.  In the mass
range between $1$ and $2\,\GeV/c^2$, two poles are identified in the
$D$~wave, corresponding to the well-known \PaTwo and the \PaTwo[1700]
resonances.  For the $P$~wave, only a single \PpiOne*~resonance pole
is extracted from the data, with parameters that are compatible with
those of the \PpiOne[1600] (see below).  This important and
astonishing result puts strong doubts on the existence of a separate
\PpiOne[1400] state in \etaPi and thereby resolves a long-standing
puzzle.  \cref{fig:poles_etapi_etaprpi} shows the positions of the
three poles extracted from the coupled-channel analysis of the \etaPim
and \etaPrPim final states.

Although the COMPASS \threePi proton-target data are the most precise
so far, the 14-wave resonance-model fit results in \PpiOne[1600]
Breit--Wigner parameters of
$m_{\PpiOne[1600]} = \SIaerr{1600}{110}{60}{\MeVcc}$ and
$\Gamma_{\PpiOne[1600]} = \SIaerr{580}{100}{230}{\MeVcc}$ that have
large systematic uncertainties~\cite{Akhunzyanov:2018lqa}.  This
mainly reflects the limitations of our current analysis model.  These
values are in good agreement with the mass and width values calculated
from the pole position~$s_\text{p}$ found in the \etaPi-\etaPrPi
coupled-channel analysis performed by JPAC~\cite{Rodas:2018owy}:
$m_{\PpiOne[1600]} \coloneqq \Re{\sqrt{s_\text{p}}} =
\SIerrs{1564}{24}{86}{\MeVcc}$ and
$\Gamma_{\PpiOne[1600]} \coloneqq -2\Im{\sqrt{s_\text{p}}} =
\SIerrs{492}{54}{102}{\MeVcc}$.

\begin{figure}[tbp]
  \centering
  \subfloat[]{%
    \includegraphics[width=0.5\textwidth]{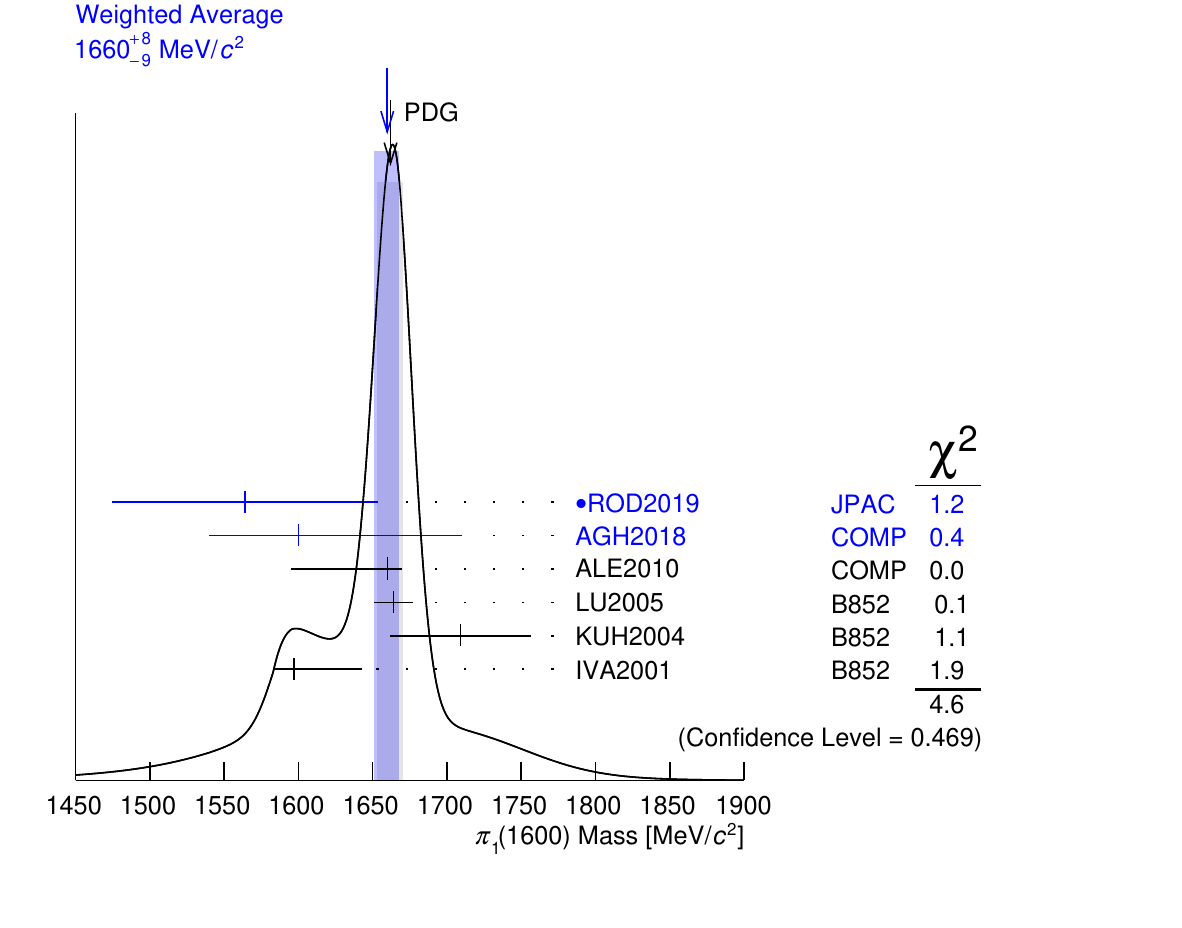}%
    \label{fig:ideogram_pi1_1600_mass}%
  }%
  \subfloat[]{%
    \includegraphics[width=0.5\textwidth]{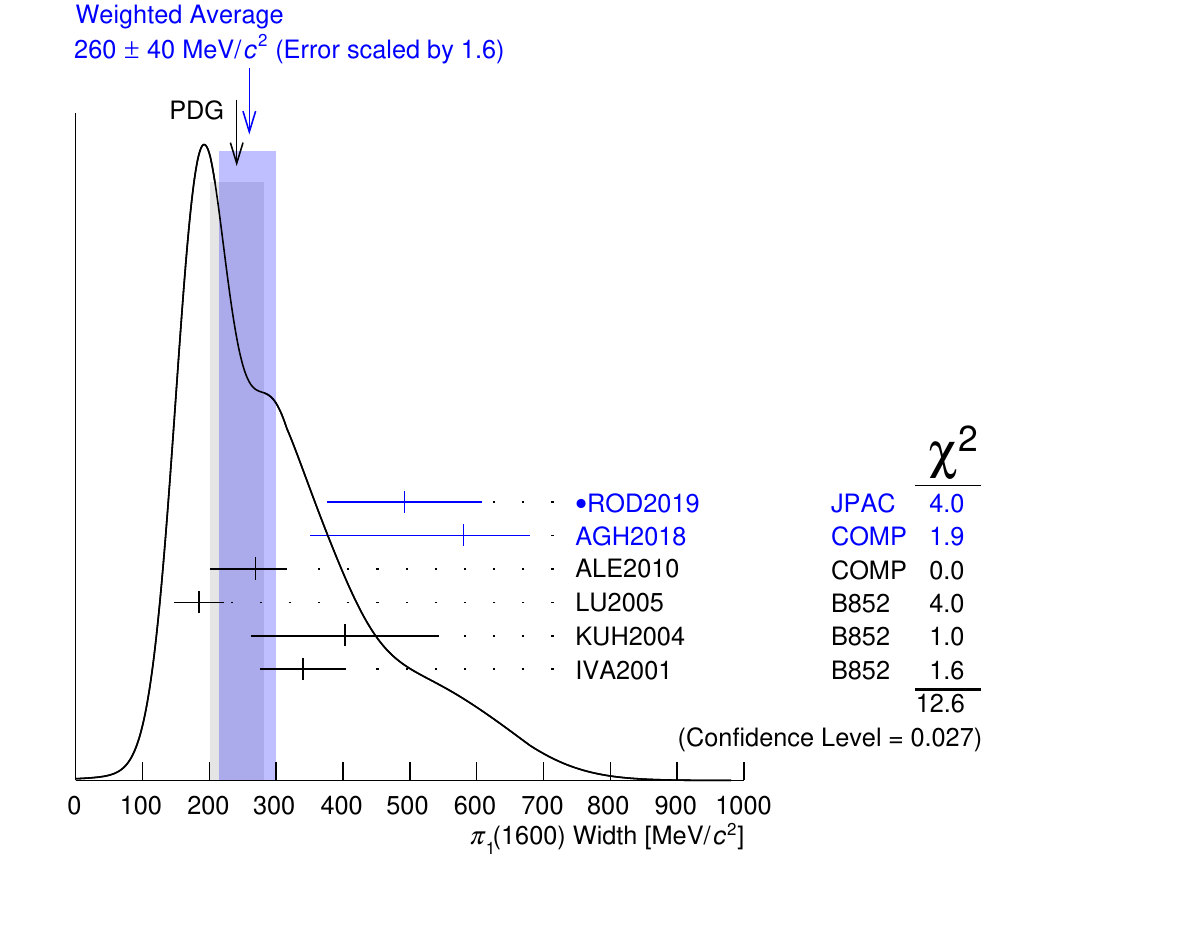}%
    \label{fig:ideogram_pi1_1600_width}%
  }%
  \caption{Ideograms similar to the ones in
    \cref{fig:ideogram_pi_1800} but for
    \subfloatLabel{fig:ideogram_pi1_1600_mass}~the mass and
    \subfloatLabel{fig:ideogram_pi1_1600_width}~the width of the
    \PpiOne[1600].  The pole parameters obtained from a
    coupled-channel fit of the COMPASS \etaOrPrPim data by the JPAC
    collaboration (ROD2019,~\cite{Rodas:2018owy}, marked by~$\bullet$)
    and the Breit--Wigner parameters obtained from fits of the COMPASS
    \threePi proton-target (AGH2018,~\cite{Akhunzyanov:2018lqa}) and
    lead-target data (ALE2010,~\cite{Alekseev:2009aa}) are compared to
    previous measurements~\cite{Tanabashi:2018zz}.}
  \label{fig:ideogram_pi1_1600}
\end{figure}

From the \threePi lead-target data, we obtain Breit--Wigner parameters
of $m_{\PpiOne[1600]} = \SIsaerrs{1660}{10}{0}{64}{\MeVcc}$ and
$\Gamma_{\PpiOne[1600]} =
\SIsaerrs{269}{21}{42}{64}{\MeVcc}$~\cite{Alekseev:2009aa}.  While the
mass values are compatible with previous measurements (see
\cref{fig:ideogram_pi1_1600_mass}), the widths from the \threePi
proton-target data and the \etaOrPrPim data are larger than the PDG
average of \SI{241(40)}{\MeVcc}~\cite{Tanabashi:2018zz}.  As shown in
\cref{fig:ideogram_pi1_1600_width}, the width values are mainly at
variance with the extremely small width value reported by the BNL E852
experiment in the $\PbOne \pi$ decay mode~\cite{Lu:2004yn}.

\begin{figure}[tbp]
  \centering
  \hfill%
  \subfloat[][]{%
    \includegraphics[width=\twoPlotWidth,valign=m]{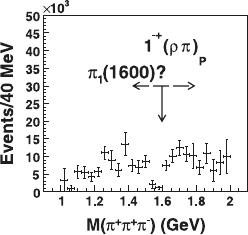}%
    \label{fig:exotic_clas}%
  }%
  \hfill%
  \subfloat[][]{%
    \includegraphics[width=\twoPlotWidth,valign=m]{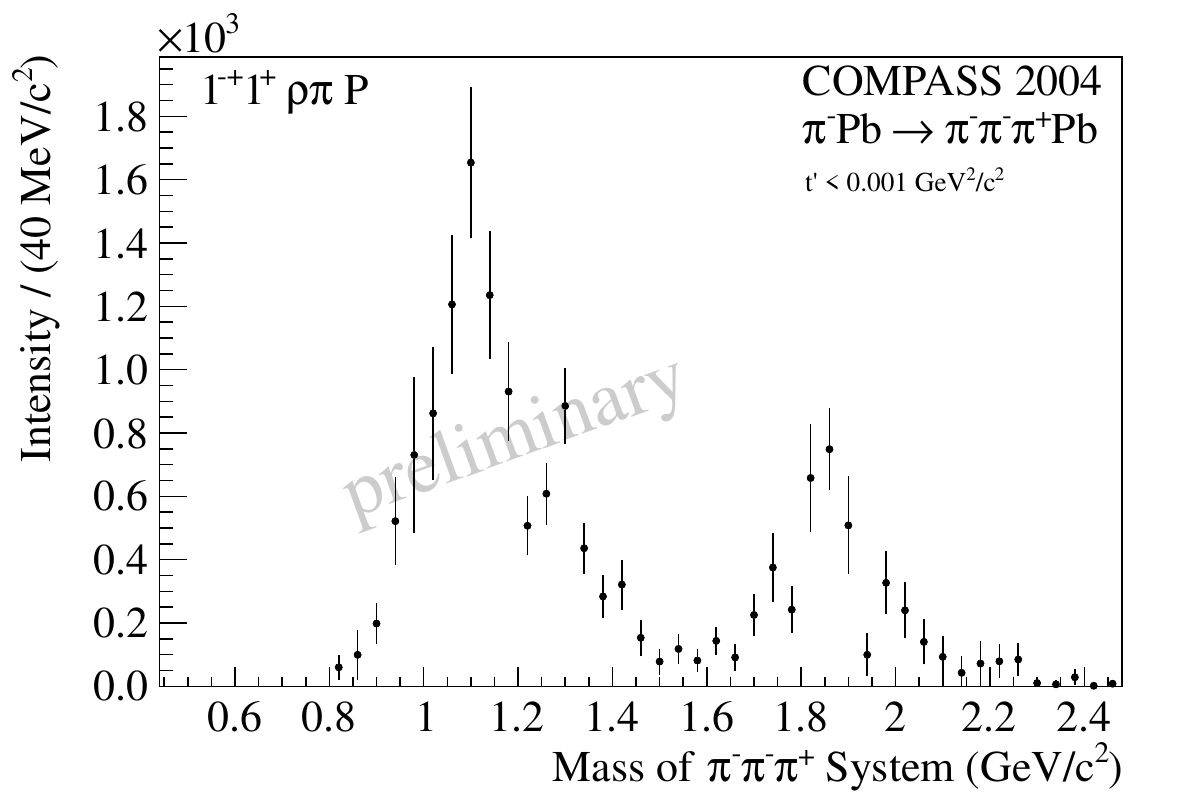}%
    \vphantom{\includegraphics[width=\twoPlotWidth,valign=m]{fig64a}}%
    \label{fig:exotic_primakoff}%
  }%
  \hfill\null%
  \caption{Intensities of the \wave{1}{-+}{1}{+}{\Pprho}{P} wave
    produced in $\gamma + \pi^\pm \to \pi^\pm \pi^- \pi^+$
    reactions. \subfloatLabel{fig:exotic_clas}~Result from the CLAS
    experiment~\cite{Nozar:2008aa}, where the process is embedded into
    $\gamma + p \to \pi^+ \pi^- \pi^+ + (n)_\text{miss}$.
    \subfloatLabel{fig:exotic_primakoff}~Result from the COMPASS
    experiment~\cite{Grabmuller:2012oja,Ketzer:2012vn}, where the
    process is embedded into
    $\pi^- + \text{Pb} \to \pi^- \pi^- \pi^+ +
    \text{Pb}$.\protect\footnotemark\ The PWA model includes a
    leading-order \chiPT amplitude for
    $\mThreePi < \SI{1.24}{\GeVcc}$, which is discussed in
    \cref{sec:pwa_chpt_amp}.}
  \label{fig:exotic_photo_prod}
\end{figure}

A remaining mystery is that the production of the \PpiOne[1600] in
photon-induced reactions seems to be much less prominent than
expected.  The CLAS experiment did not observe a \PpiOne[1600] signal
in the charge-exchange reaction
$\gamma + p \to \pi^+ \pi^- \pi^+ +
(n)_\text{miss}$~\cite{Nozar:2008aa,Eugenio:2013xua} (see
\cref{fig:exotic_clas}).  This finding is supported by an analysis of
COMPASS data of the reaction
$\pi^- + \text{Pb} \to \pi^- \pi^- \pi^+ + \text{Pb}$ in the range
$t' < \SI{e-3}{\GeVcsq}$, where photon exchange is
dominant~\cite{Grabmuller:2012oja} (see \cref{fig:exotic_primakoff};
more details of this analysis will be discussed in
\cref{sec:results_3pic_primakoff}).  In both processes, the dominant
underlying reaction is $\gamma + \pi^\pm \to \pi^\pm \pi^- \pi^+$.
Since we observe the \PpiOne[1600] to decay into $\Pprho \pi$, it
should couple to $\gamma \pi$ via vector-meson dominance.  This is
because the photon fluctuates into quark--antiquark pairs with
$\JPC_{\qqbar} = 1^{--}$ quantum numbers.  In addition, the
\PpiOne[1600] is a hybrid candidate and is compatible with the
lightest $1^{-+}$ hybrid state predicted by lattice QCD (see
\cref{sec:pheno.exotics,fig:lattice_spectrum_light_ns}).  Lattice QCD
and also many models predict that the lowest excitation of the gluon
field in a hybrid meson has $\JPC_g = 1^{+-}$ quantum numbers.  Hence
the \qqbar pair from the photon and the gluonic excitation can
directly couple to spin-exotic $\JPC = 1^{-+}$ quantum numbers, which
is not the case for pion-induced reactions.  Thus the production of
hybrids is expected to be enhanced in photon-induced
reactions~\cite{Afanasev:1997fp,Szczepaniak:2001qz}, which seems to be
contradicted by the nearly vanishing intensities that are observed in
the \SI{1.6}{\GeVcc} mass region.  This, however, could be due to
destructive interference of a \PpiOne[1600] with a non-resonant
component.  This hypothesis could be verified by performing
resonance-model fits on the existing data.  In the future, much more
precise photoproduction data from the GlueX experiment at JLab will
help to clarify the situation.

\footnotetext{\Cref{fig:exotic_primakoff} is the result of an analysis
  of COMPASS data taken 2004.  The analysis of a larger lead-target
  data sample taken 2009 is work in progress.}

\subsubsection{The $\JPC = 2^{++}$ Sector }
\label{sec:results_2pp}

Currently, the PDG lists seven isovector states with
$\JPC = 2^{++}$~\cite{Tanabashi:2018zz}: \PaTwo, \PaTwo[1700],
\PaTwo[1950], \PaTwo[1990], \PaTwo[2030], \PaTwo[2175], and
\PaTwo[2255] (see also \cref{fig:light_flavorless_spectrum}).  The
\PaTwo is the $2^{++}$ ground state, \ie the \termSym{1}{3}{P}{2}
quark-model state.  It is the best known $3\pi$~resonance and is the
dominant resonance in the $\eta \pi$ channel.  The \PaTwo[1700] is
omitted from the PDG summary table and hence not well known.  It is a
good candidate for the first radial excitation, \ie the
\termSym{2}{3}{P}{2} quark-model state.  The five higher-lying states
are listed as \textquote{further states} by the PDG and require
confirmation.

\begin{figure}[tbp]
  \centering
  \hfill%
  \subfloat[][]{%
    \includegraphics[width=\threePlotWidth]{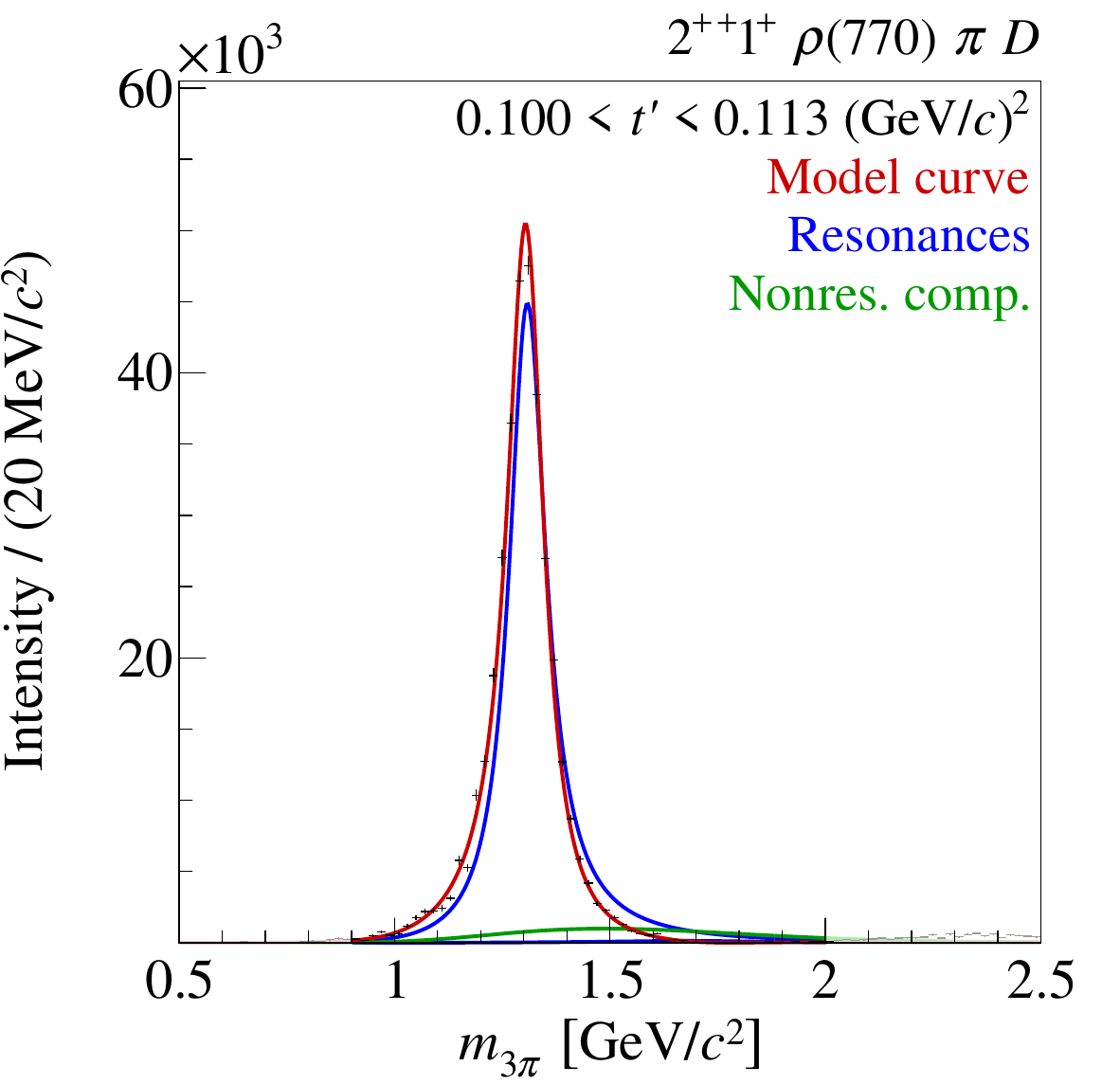}%
    \label{fig:intensity_2pp_tbin1}%
  }%
  \hfill%
  \subfloat[][]{%
    \includegraphics[width=\threePlotWidth]{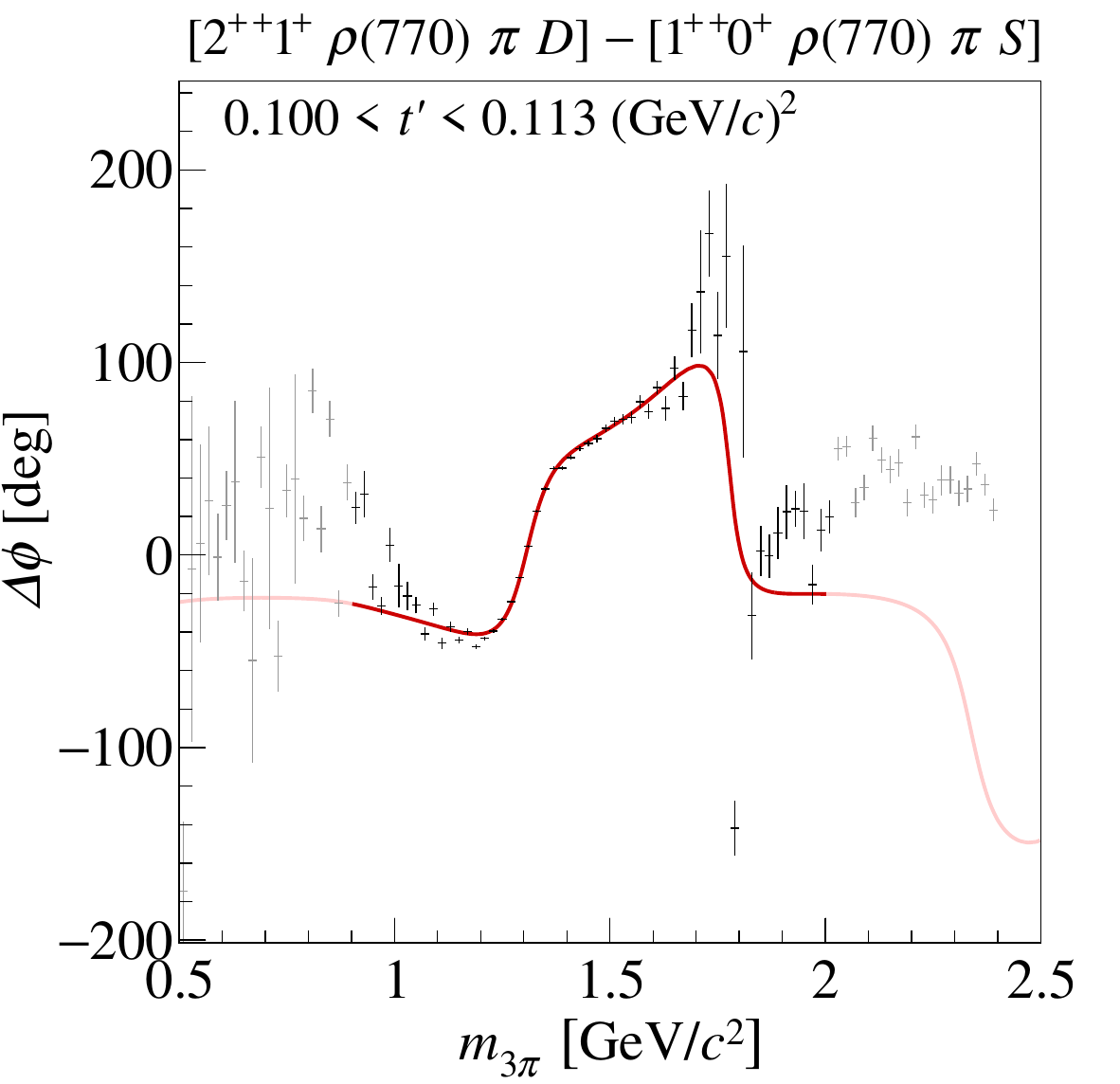}%
    \label{fig:phase_2pp_tbin1}%
  }%
  \hfill\null%
  \caption{\subfloatLabel{fig:intensity_2pp_tbin1}~Intensity
    distribution of the \wave{2}{++}{1}{+}{\Pprho}{D} wave and
    \subfloatLabel{fig:phase_2pp_tbin1}~phase of this wave \wrt the
    \wave{1}{++}{0}{+}{\Pprho}{S} wave, both in the lowest $t'$~bin of
    the \threePi proton-target data~\cite{Akhunzyanov:2018lqa}.  The
    curves represent the result of the resonance-model fit.  The model
    and the wave components are represented as in
    \cref{fig:intensity_phase_0mp}.  The dominant resonant component
    is the \PaTwo; the \PaTwo[1700] is barely visible (see
    \cref{fig:intensity_2pp_tbin1_log} for a log-scale version
    of~\subfloatLabel{fig:intensity_2pp_tbin1}).}
  \label{fig:2pp_a2_1320}
\end{figure}

\begin{figure}[tbp]
  \centering
  \hfill%
  \subfloat[]{%
    \includegraphics[width=\twoPlotWidth]{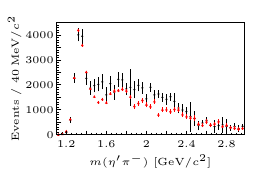}%
    \label{fig:int_eta_etaprime_d_wave}%
  }%
  \hfill%
  \subfloat[]{%
    \includegraphics[width=\twoPlotWidth]{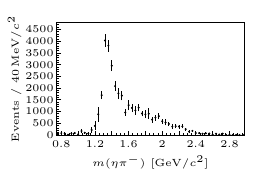}%
    \label{fig:int_eta_d_wave_m_2}%
  }%
  \hfill\null%
  \\
  \null\hfill%
  \subfloat[]{%
    \includegraphics[width=\twoPlotWidth]{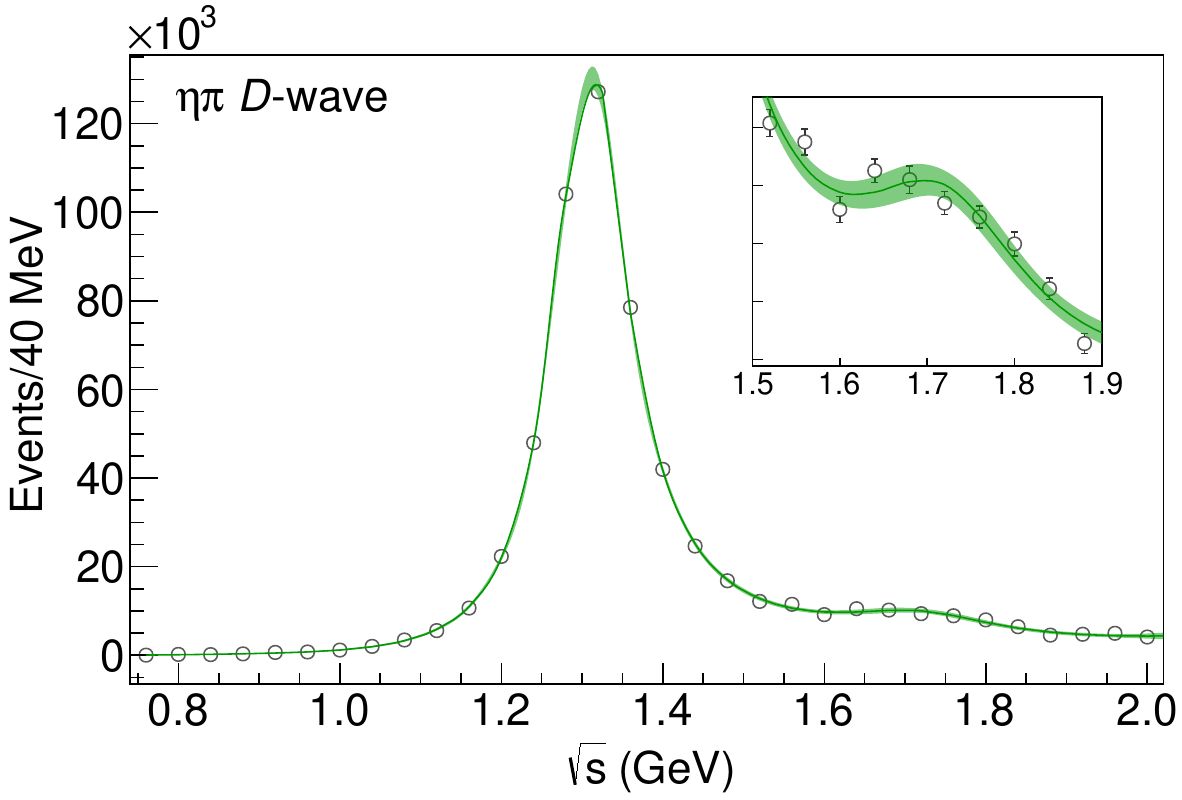}%
    \label{fig:int_eta_d_wave}%
  }%
  \hfill%
  \subfloat[]{%
    \includegraphics[width=\twoPlotWidth]{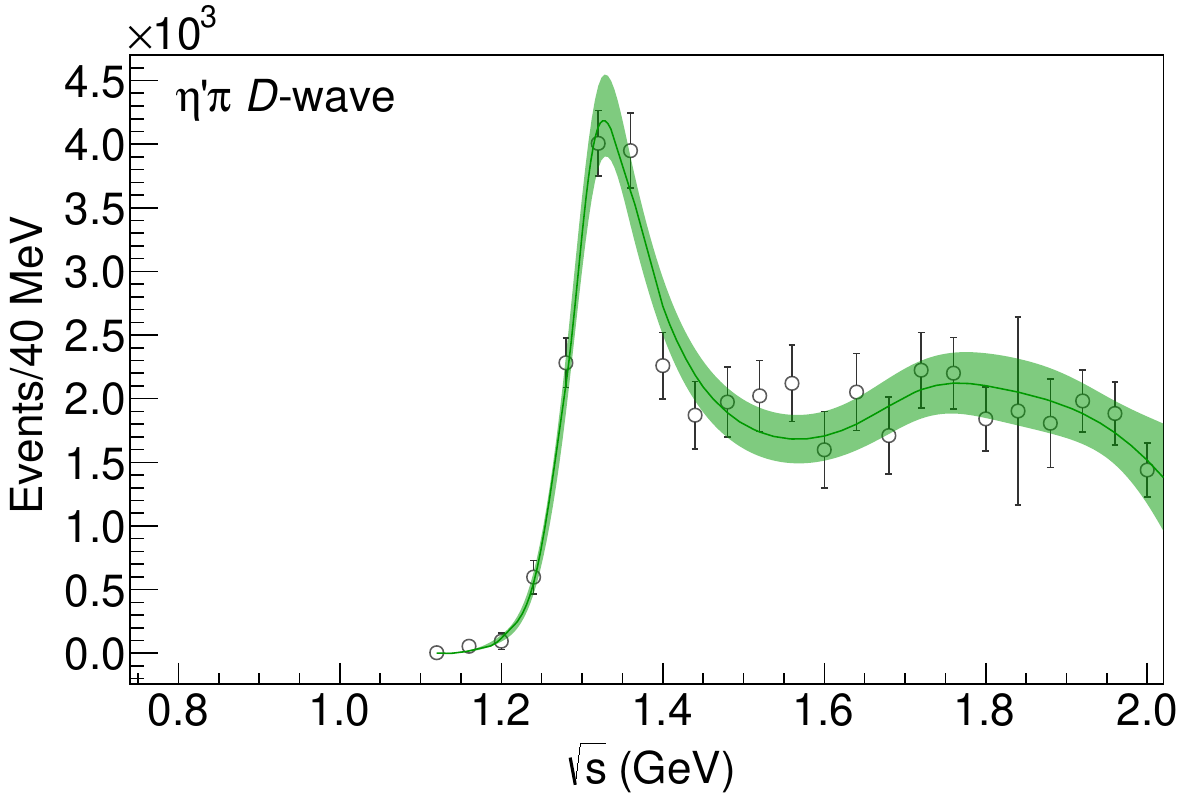}%
    \label{fig:int_etaprime_d_wave}%
  }%
  \hfill\null%
  \caption{\subfloatLabel{fig:int_eta_etaprime_d_wave}~Intensity
    distribution of the $D$~waves with $\Mrefl = 1^+$ in the COMPASS
    \etaPrPim data (black dots) overlaid with the intensity
    distribution of the same wave in the COMPASS \etaPim data (red
    triangles), scaled by the kinematic factor in
    \cref{eq:etaprime_eta_r}~\cite{Adolph:2014rpp}.
    \subfloatLabel{fig:int_eta_d_wave_m_2}~Intensity distribution of
    the \etaPim $D$~wave with $\Mrefl = 2^+$~\cite{Adolph:2014rpp}.
    \subfloatLabel{fig:int_eta_d_wave}~and~\subfloatLabel{fig:int_etaprime_d_wave}:
    Intensity distributions of the $D$~waves with $\Mrefl = 1^+$ in
    the COMPASS \etaPim and \etaPrPim data~\cite{Rodas:2018owy}. The
    points in~\subfloatLabel{fig:int_etaprime_d_wave} are the same as
    in~\subfloatLabel{fig:int_eta_etaprime_d_wave}.  Note the
    different mass range.  The green curves show the result of a fit
    of the unitary model from \refCite{Rodas:2018owy} that was
    described in \cref{sec:pwa.unitary_model}.  The shaded bands
    represent the two-standard-deviation confidence interval.}
  \label{fig:int_phase_eta_etaprime_d_wave}
\end{figure}

The \PaTwo has a large branching fraction into~$3\pi$ of
\SI{70.1(27)}{\percent}~\cite{Tanabashi:2018zz}.  The second most
probable decay mode is \etaPi with a branching fraction of
\SI{14.5(12)}{\percent}.  In contrast, the decay into \etaPrPi has
only a small branching fraction of \num{5.5(9)e-3} and was first
observed by the VES~experiment~\cite{Beladidze:1993km}.  The \PaTwo is
the narrowest $3\pi$~resonance and appears in the $3\pi$~channel
predominantly in the $\Pprho \pi$ $D$~waves with $M = 1$ and~2 with
very low background (see \eg\ \cref{fig:intensity_2pp_tbin1}).  The
\PaTwo peak is also associated with rapid phase motions \wrt other
waves (see \eg \cref{fig:phase_2pp_tbin1}).  This makes the \PaTwo the
cleanest resonance signal in our \threePi data.  The \etaPim data are
dominated by the $D$~wave with $\Mrefl = 1^+$ with a prominent \PaTwo
peak (see \cref{fig:int_eta_d_wave}).  In the \etaPim data, we observe
in addition a $D$~wave with $\Mrefl = 2^+$ and an intensity at the
\SI{5}{\percent} level \wrt the $D$~wave with $\Mrefl = 1^+$ (see
\cref{fig:int_eta_d_wave_m_2}).  Also the $D$~wave with $\Mrefl = 2^+$
contains a clear \PaTwo signal.  As expected, the \PaTwo signal is
much weaker in the \etaPrPim data (see
\cref{fig:int_etaprime_d_wave}).  As shown in
\cref{fig:int_eta_etaprime_d_wave}, the observed differences in the
intensity distributions of the \etaPim and \etaPrPim $D$~waves with
$\Mrefl = 1^+$ are nearly completely explained by the kinematic factor
in \cref{eq:etaprime_eta_r}.

\begin{figure}[tbp]
  \centering
  \subfloat[]{%
    \includegraphics[width=0.5\textwidth]{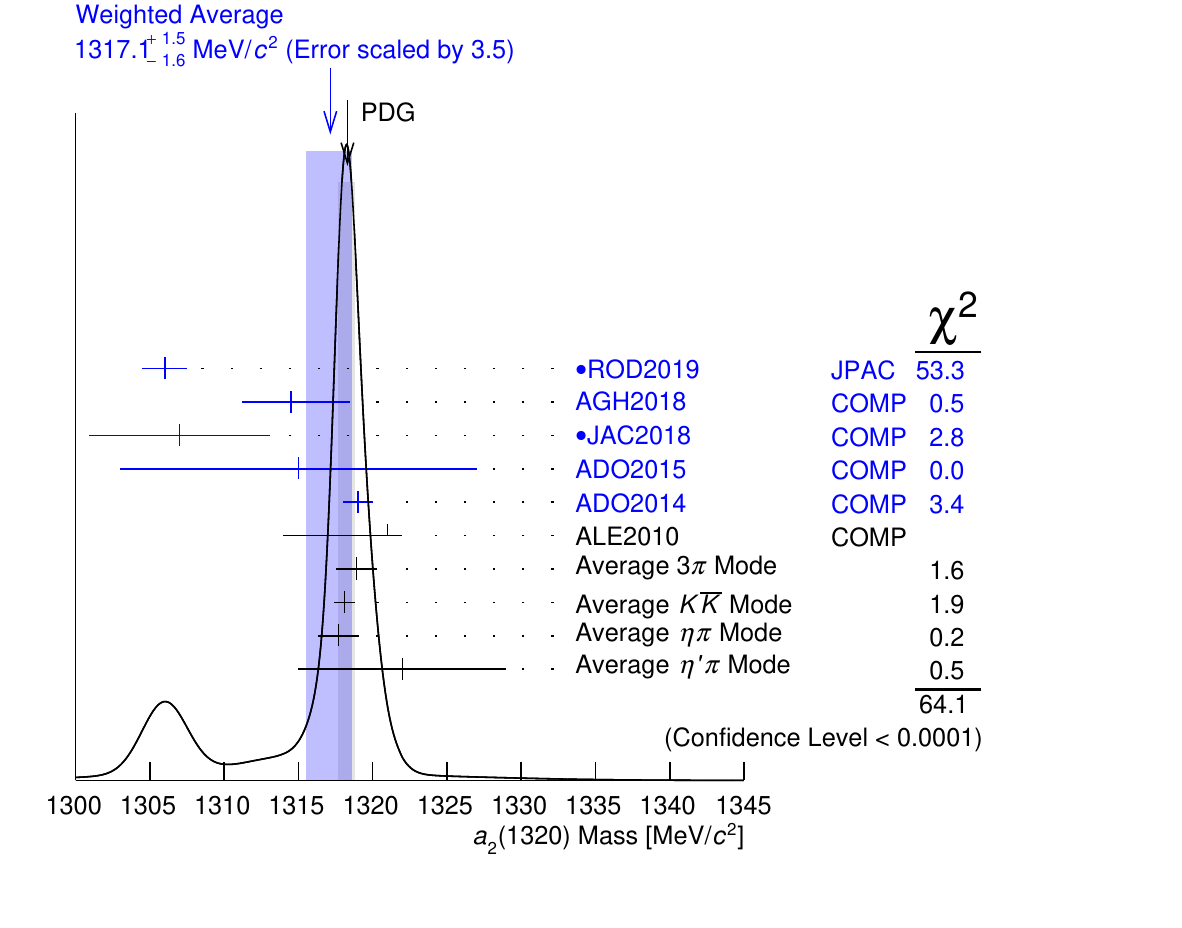}%
    \label{fig:ideogram_a2_1320_mass}%
  }%
  \subfloat[]{%
    \includegraphics[width=0.5\textwidth]{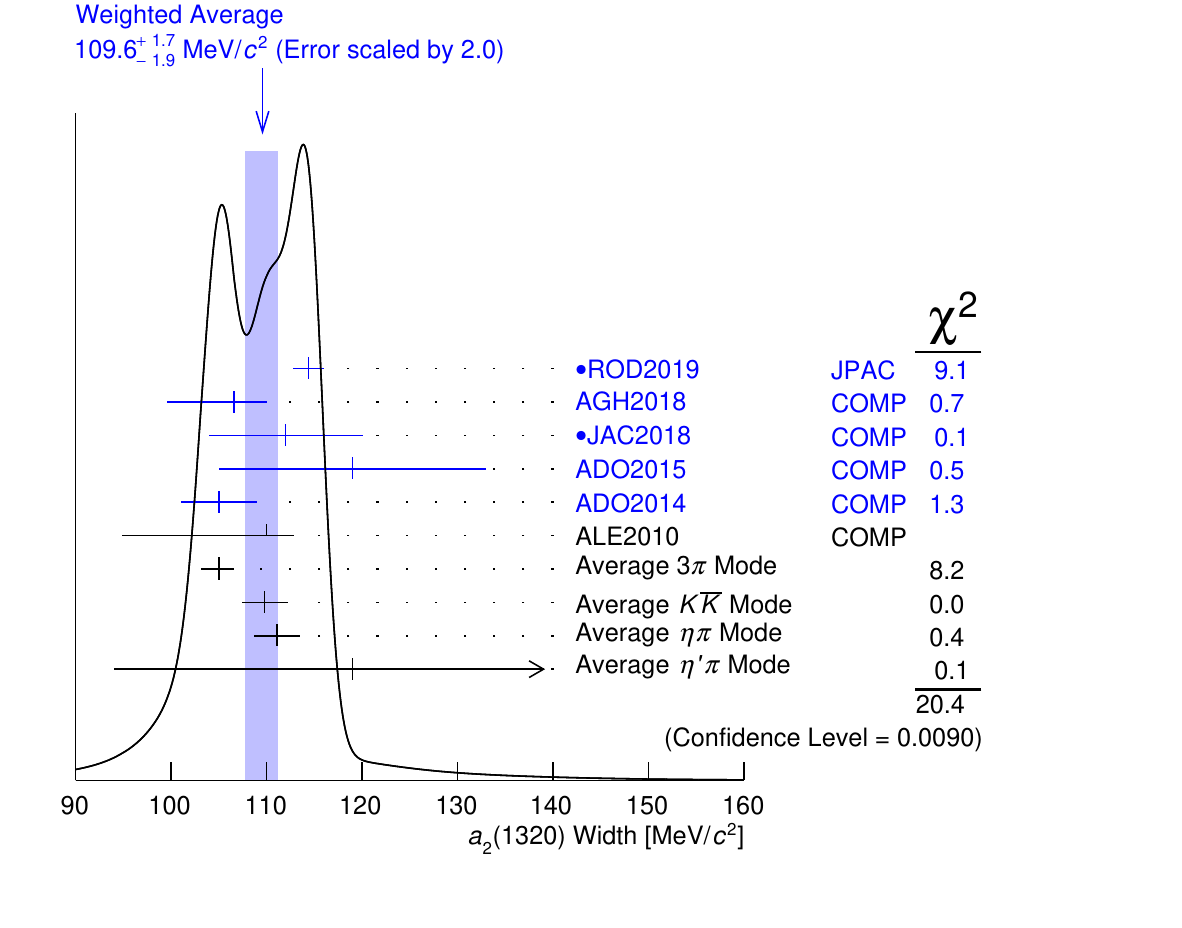}%
    \label{fig:ideogram_a2_1320_width}%
  }%
  \caption{Ideograms similar to the ones in
    \cref{fig:ideogram_pi_1800} but for
    \subfloatLabel{fig:ideogram_a2_1320_mass}~the mass and
    \subfloatLabel{fig:ideogram_a2_1320_width}~the width of the
    \PaTwo.  The pole parameters obtained from a coupled-channel fit
    of the COMPASS \etaOrPrPim data by the JPAC collaboration
    (ROD2019,~\cite{Rodas:2018owy}) and from a fit of the $D$-wave
    intensity distribution in the COMPASS \etaPim data
    (JAC2018,~\cite{Jackura:2017amb}) are marked by~$\bullet$.  These
    values and the Breit--Wigner parameters obtained from fits of the
    COMPASS diffractive \threePi proton-target data
    (AGH2018,~\cite{Akhunzyanov:2018lqa}), the diffractive \etaOrPrPim
    data (ADO2015,~\cite{Adolph:2014rpp}), the photo-production
    \threePi data (ADO2014,~\cite{Adolph:2014mup}) (uncertainties are
    statistical only; see \cref{sec:results_3pic_primakoff}), and of
    the diffractive \threePi lead-target data
    (ALE2010,~\cite{Alekseev:2009aa}) are compared with previous
    measurements.  Note that the values from the COMPASS \threePi
    lead-target data (ALE2010,~\cite{Alekseev:2009aa}) are already
    included in the present PDG averages for the $3\pi$~decay mode.
    For the \PaTwo width
    in~\subfloatLabel{fig:ideogram_a2_1320_width}, the PDG provides
    only an estimate, which is not shown.}
  \label{fig:ideogram_a2_1320}
\end{figure}

The \PaTwo peaks in the \etaOrPrPim and \threePi data are well
described by the employed Breit--Wigner amplitude.  In the 14-wave
resonance-model fit of the \threePi proton-target data, the \PaTwo
resonance parameters have the smallest uncertainties of all resonances
in this fit.  From this fit, we extract Breit--Wigner parameters of
$m_{\PaTwo} = \SIaerr{1314.5}{4}{3.3}{\MeVcc}$ and
$\Gamma_{\PaTwo} =
\SIaerr{106.6}{3.4}{7.0}{\MeVcc}$~\cite{Akhunzyanov:2018lqa}.  These
values are consistent with the Breit--Wigner parameters
$m_{\PaTwo} = \SIsaerrs{1321}{1}{0}{7}{\MeVcc}$ and
$\Gamma_{\PaTwo} = \SIsaerrs{110}{2}{2}{15}{\MeVcc}$ obtained from the
6-wave resonance-model fit of the \threePi lead-target
data~\cite{Alekseev:2009aa} and with the Breit--Wigner parameters
$m_{\PaTwo} = \SI{1315(12)}{\MeVcc}$ and
$\Gamma_{\PaTwo} = \SI{119(14)}{\MeVcc}$ obtained from the combined
fit of the \etaPim and \etaPrPim data~\cite{Adolph:2014rpp}.  Pole
positions of the \PaTwo are extracted from fits of more advanced
analytical models that were developed by the JPAC collaboration and
are based on the principles of the relativistic $S$-matrix (see
\cref{sec:pwa.unitary_model}).  From a fit to the intensity
distribution of the \etaPim $D$~wave with $\Mrefl = 1^+$, we obtain
the pole parameters $m_{\PaTwo} = \SIerrs{1307}{1}{6}{\MeVcc}$ and
$\Gamma_{\PaTwo} = \SIerrs{111}{1}{8}{\MeVcc}$~\cite{Jackura:2017amb}.
A coupled-channel fit of the \etaPim and \etaPrPim $P$- and $D$-wave
amplitudes with $\Mrefl = 1^+$ yields the pole parameters
$m_{\PaTwo} = \SIerrs{1306.0}{0.8}{1.3}{\MeVcc}$ and
$\Gamma_{\PaTwo} =
\SIerrs{114.4}{1.6}{0.0}{\MeVcc}$~\cite{Rodas:2018owy} (see curves in
\cref{fig:int_eta_d_wave,fig:int_etaprime_d_wave}).  As shown in
\cref{fig:ideogram_a2_1320}, the results of the Breit-Wigner fits are
in good agreement with the PDG averages.  The \PaTwo mass values from
the pole positions are smaller than the Breit--Wigner masses.  The
result from \refCite{Rodas:2018owy} also prefers a larger width
compared to the Breit--Wigner parameters.

\begin{figure}[tbp]
  \centering
  \subfloat[][]{%
    \includegraphics[width=\threePlotWidth]{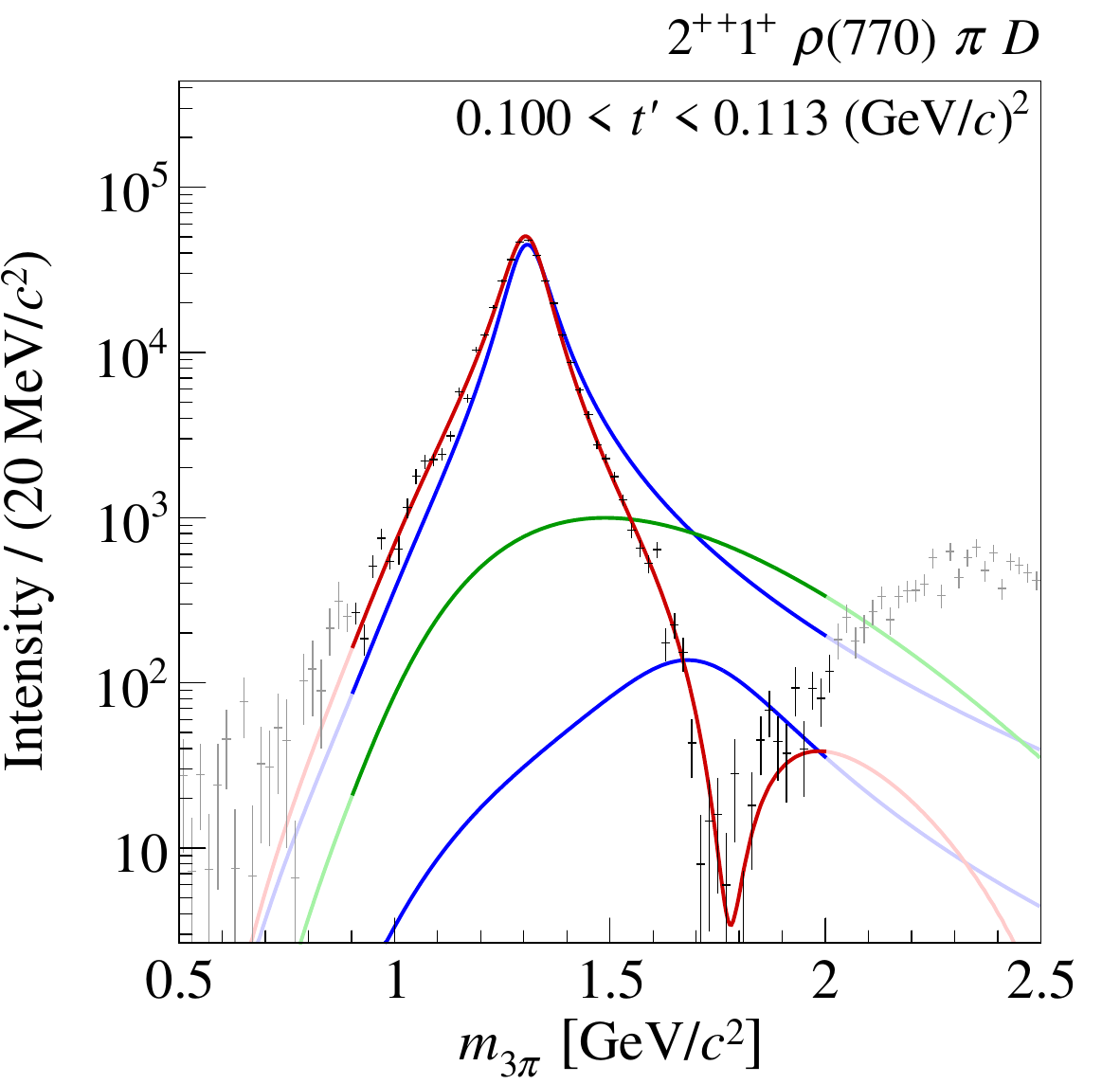}%
    \label{fig:intensity_2pp_tbin1_log}%
  }%
  \hfill%
  \subfloat[][]{%
    \includegraphics[width=\threePlotWidth]{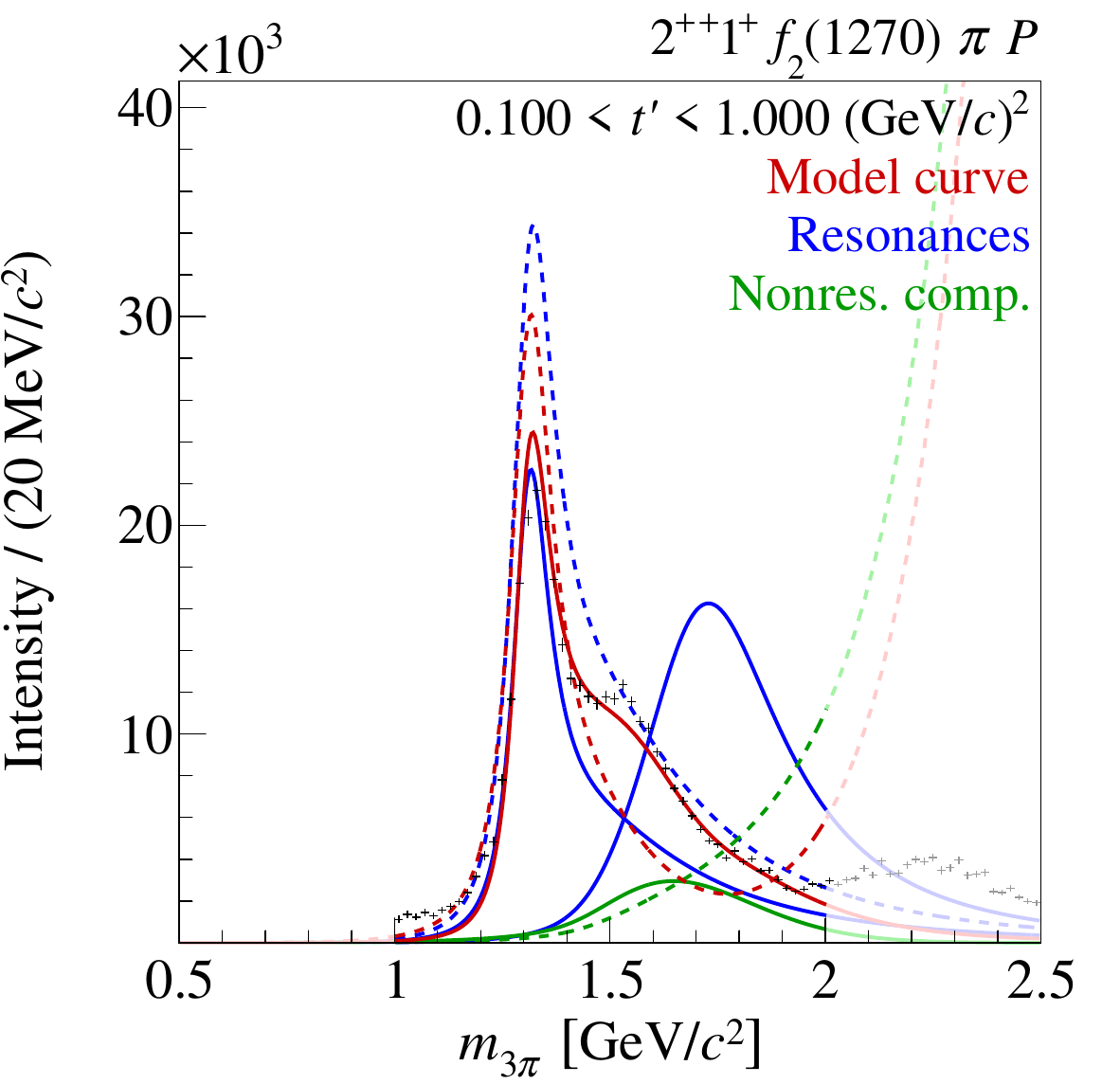}%
    \label{fig:intensity_2pp_f2_no_a2_1700}%
  }%
  \hfill%
  \subfloat[][]{%
    \includegraphics[width=\threePlotWidth]{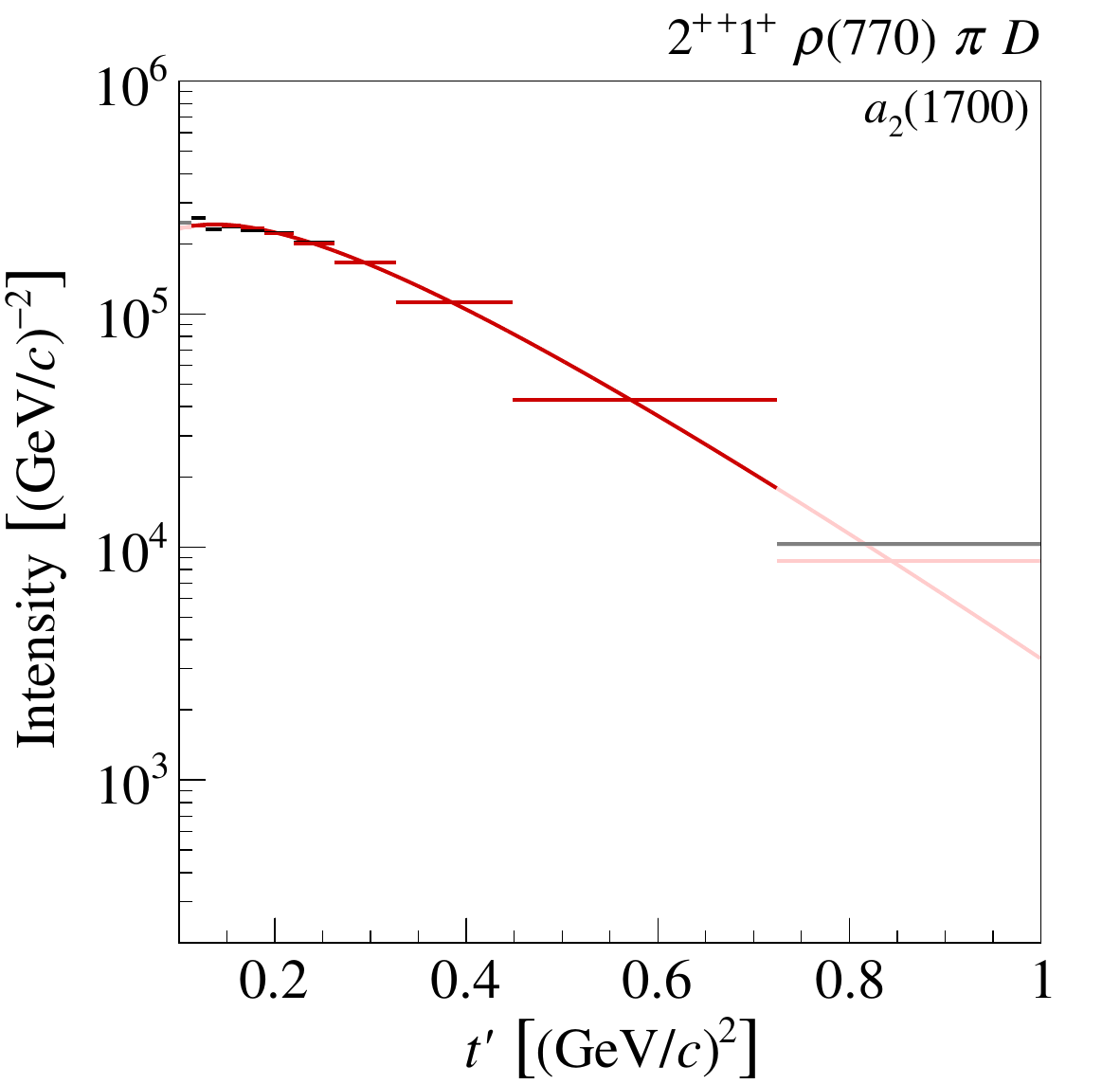}%
    \label{fig:a2_1700_t_spectrum}
  }%
  \caption{The \PaTwo[1700] signal in the \threePi proton-target
    data~\cite{Akhunzyanov:2018lqa}.
    \subfloatLabel{fig:intensity_2pp_tbin1_log}~Intensity distribution
    of the \wave{2}{++}{1}{+}{\Pprho}{D} wave in the lowest $t'$~bin
    (same as \cref{fig:intensity_2pp_tbin1} but in log scale).
    \subfloatLabel{fig:intensity_2pp_f2_no_a2_1700}~Intensity
    distribution of the \wave{2}{++}{1}{+}{\PfTwo}{P} wave summed over
    the 11~$t'$~bins.  The curves represent the result of the
    resonance-model fit.  The model and the wave components are
    represented as in \cref{fig:intensity_phase_0mp} except that the
    blue curves represent the \PaTwo and the \PaTwo[1700].  In
    addition to the continuous curves that represent the main
    resonance-model fit as
    in~\subfloatLabel{fig:intensity_2pp_tbin1_log}, the dashed curves
    in~\subfloatLabel{fig:intensity_2pp_f2_no_a2_1700} represent the
    result of a fit, where the \PaTwo[1700] component is removed from
    the resonance model.
    \subfloatLabel{fig:a2_1700_t_spectrum}~Similar to
    \cref{fig:tspectrum_0mp_f0}, but showing the $t'$~spectrum of the
    \PaTwo[1700] in the \wave{2}{++}{1}{+}{\Pprho}{D} wave.}
  \label{fig:2pp_a2_1700}
\end{figure}

Based on the Breit--Wigner resonance-model fit to the \etaOrPrPim data,
we measure the branching-fraction ratio of the \PaTwo decays into
\etaPi and \etaPrPi~\cite{Adolph:2014rpp}:
\begin{equation}
  \label{eq:branch_fract_ratio_a2_eta_etaprime_pi}
  B_{\etaPrPi, \etaPi}^{\PaTwo*}
  = \frac{\text{BF}\big[ \PaTwo \to \etaPrPi \big]}%
         {\text{BF}\big[ \PaTwo \to \etaPi \big]\hfill}
  = \num{0.05(2)}\eqPunctSpacing.
\end{equation}
This value is consistent with previous
measurements~\cite{Tanabashi:2018zz}.  As shown in
\cref{fig:intensity_2pp_f2_no_a2_1700}, we observe for the first time
the decay of the \PaTwo into $\PfTwo \pi$, which is a sub-threshold
decay and hence suppressed by phase space.  The branching-fraction
ratio for the decays of the \PaTwo into the $\Pprho \pi$~$D$ and
$\PfTwo \pi$ $P$~decay modes with $M = 1$ is
\begin{equation}
  \label{eq:branch_fract_ratio_a2_corr}
  B_{\Pprho* \pi D, \PfTwo* \pi P}^{\PaTwo*, \text{corr}}
  = \frac{\text{BF}\big[ \PaTwo^- \to \Pprho \pi \big]\hfill}%
         {\text{BF}\big[ \PaTwo^- \to \PfTwo \pi \big]}
  = \numaerr{16.5}{1.2}{2.4}\eqPunctSpacing.
\end{equation}
This number takes into account the unobserved decays into \threePiN
via isospin symmetry, the branching fraction of the \PfTwo to $2\pi$
and the effect of the different Bose symmetrizations in the \threePi
and \threePiN final states (see Section~VI.C.2 in
\refCite{Akhunzyanov:2018lqa} for details).

\begin{figure}[tbp]
  \centering
  \subfloat[]{%
    \includegraphics[width=0.5\textwidth]{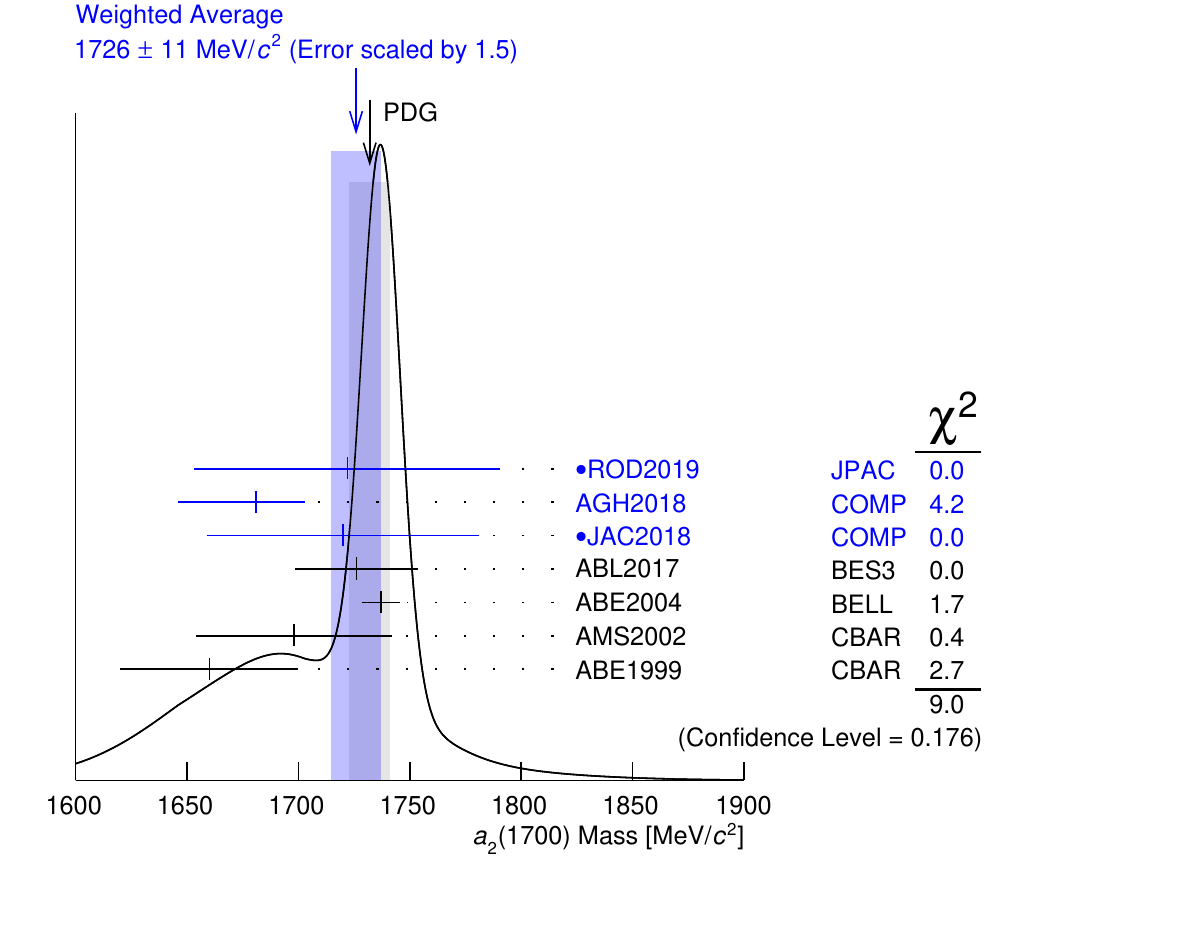}%
    \label{fig:ideogram_a2_1700_mass}%
  }%
  \subfloat[]{%
    \includegraphics[width=0.5\textwidth]{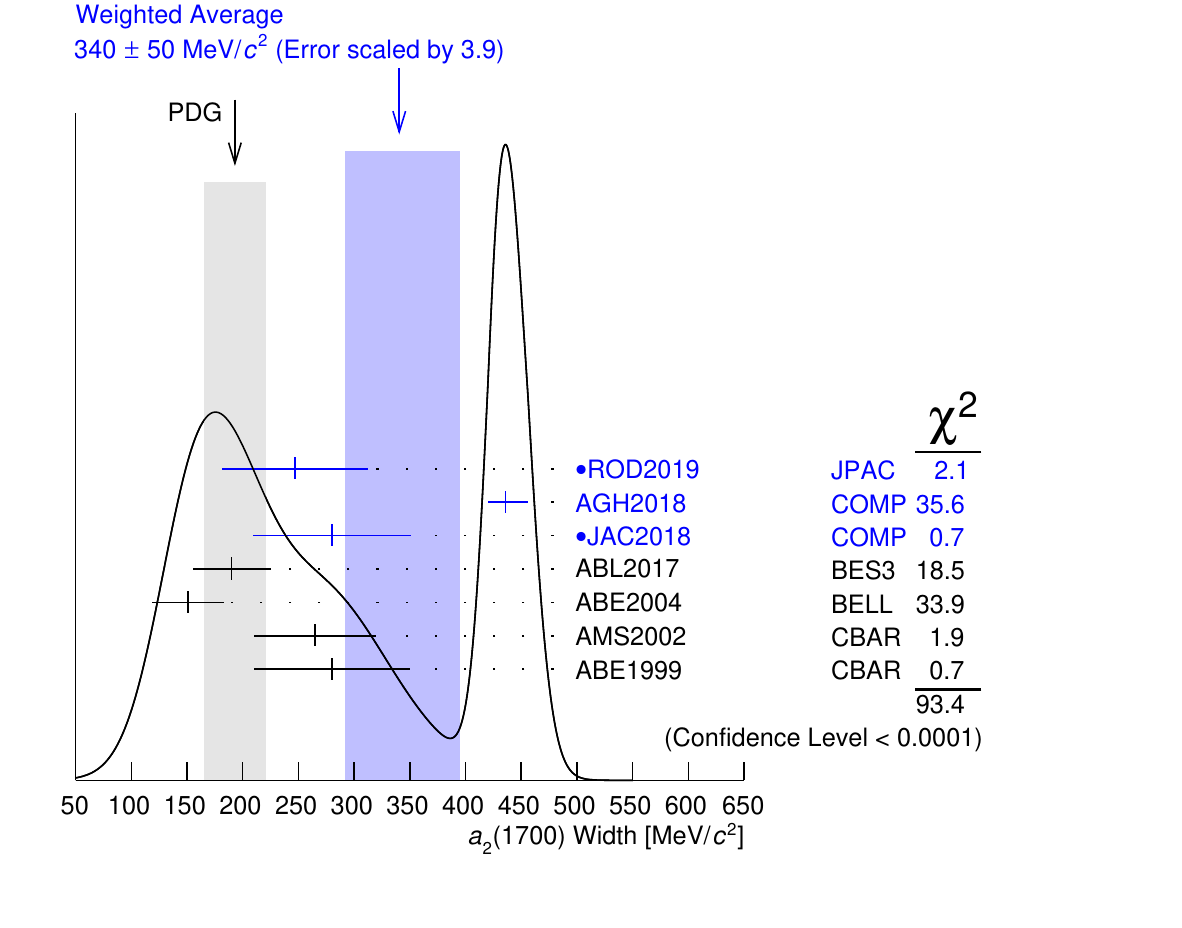}%
    \label{fig:ideogram_a2_1700_width}%
  }%
  \caption{Ideograms similar to the ones in
    \cref{fig:ideogram_pi_1800} but for
    \subfloatLabel{fig:ideogram_a2_1700_mass}~the mass and
    \subfloatLabel{fig:ideogram_a2_1700_width}~the width of the
    \PaTwo[1700].  The pole parameters obtained from a coupled-channel
    fit of the COMPASS \etaOrPrPim data by the JPAC collaboration
    (ROD2019,~\cite{Rodas:2018owy}) and from a fit of the $D$-wave
    intensity distribution in the COMPASS \etaPim data
    (JAC2018,~\cite{Jackura:2017amb}) are marked by~$\bullet$.  These
    values and the Breit--Wigner parameters obtained from a fit of the
    COMPASS \threePi proton-target data
    (AGH2018,~\cite{Akhunzyanov:2018lqa}) are compared with previous
    measurements~\cite{Tanabashi:2018zz}.}
  \label{fig:ideogram_a2_1700}
\end{figure}

In the \threePi data, the signal of the first radially excited
$2^{++}$ state, \ie the \PaTwo[1700], is approximately two orders of
magnitude smaller than the one of the ground state, \ie the \PaTwo.
This is similar to the situation in the $1^{++}$ sector (see
\cref{sec:results_1pp}).  As a consequence, the \PaTwo[1700] is only
included in the resonance-model fit of the much larger proton-target
data set.  In the \wave{2}{++}{1}{+}{\Pprho}{D} intensity
distribution, the \PaTwo[1700] appears in the lowest $t'$~bin as a dip
at about \SI{1.7}{\GeVcc} due to destructive interference (see
\cref{fig:intensity_2pp_tbin1_log} and \confer\
\cref{fig:intensity_2pp_tbin1}).  The interference pattern changes
with~$t'$ such that the dip disappears with increasing~$t'$ and turns
into a subtle shoulder.  As shown in
\cref{fig:intensity_2pp_f2_no_a2_1700}, the \PaTwo[1700] signal is
clearer in the \wave{2}{++}{1}{+}{\PfTwo}{P} intensity distribution,
where it appears as a high-mass shoulder of the suppressed \PaTwo
peak.  If the \PaTwo[1700] component is removed from the fit model,
the shoulder cannot be described (see dashed curves in
\cref{fig:intensity_2pp_f2_no_a2_1700}).
\Cref{fig:a2_1700_t_spectrum} shows that the $t'$~spectrum of the
\PaTwo[1700] is well described by the simple exponential model in
\cref{eq:t_spectrum_model} and has a slope parameter of
\SIaerr{7.3}{2.4}{0.9}{\perGeVcsq} in the $\Pprho \pi$ wave, which is
similar to the value of \SI{7.9(5)}{\perGeVcsq} for the
\PaTwo.\footnote{Since the $t'$~dependences of the $\Pprho \pi$ and
  $\PfTwo \pi$ amplitudes are constrained by \cref{eq:branching_amp},
  the slope parameters of \PaTwo and \PaTwo[1700] in the $\PfTwo \pi$
  wave have nearly identical values.}  This supports the resonance
interpretation of the \PaTwo[1700] signal.  From the 14-wave
resonance-model fit, we obtain Breit--Wigner parameters of
$m_{\PaTwo[1700]} = \SIaerr{1681}{22}{35}{\MeVcc}$ and
$\Gamma_{\PaTwo[1700]} =
\SIaerr{436}{20}{16}{\MeVcc}$~\cite{Akhunzyanov:2018lqa}.  Whereas our
mass value is consistent with the PDG average, our width value is
considerably larger and incompatible with the other measurements
included in the PDG average (see \cref{fig:ideogram_a2_1700}).  This
could in part be due to our simplifying model assumptions, which may
lead to an overestimation of the \PaTwo[1700] width.  This hypothesis
is supported by the results of the resonance-model fits of the
\etaOrPrPim data using the more advanced analytical models developed
by the JPAC collaboration (see \cref{sec:pwa.unitary_model}).  In the
\etaOrPrPim data, the \PaTwo[1700] appears as a broad high-mass
shoulder in the $D$-wave intensity distribution.  This shoulder is
more pronounced in \etaPrPim (see
\cref{fig:int_eta_d_wave,fig:int_etaprime_d_wave}).  From a fit to the
intensity distribution of the \etaPim $D$~wave with $\Mrefl = 1^+$, we
obtain the pole parameters
$m_{\PaTwo[1700]} = \SIerrs{1720}{10}{60}{\MeVcc}$ and
$\Gamma_{\PaTwo[1700]} =
\SIerrs{280}{10}{70}{\MeVcc}$~\cite{Jackura:2017amb}.  A
coupled-channel fit of the \etaPim and \etaPrPim $P$- and $D$-wave
amplitudes with $\Mrefl = 1^+$ yields the pole parameters
$m_{\PaTwo[1700]} = \SIerrs{1722}{15}{67}{\MeVcc}$ and
$\Gamma_{\PaTwo[1700]} =
\SIerrs{247}{17}{63}{\MeVcc}$~\cite{Rodas:2018owy} (see curves in
\cref{fig:int_eta_d_wave,fig:int_etaprime_d_wave}).  Both pole
positions are in agreement with the PDG average.

In the \threePi proton-target data, we observe that the \PaTwo[1700]
decays more often into $\PfTwo \pi$ than into $\Pprho \pi$.  This is
at odds with the result of the L3~experiment, which observed a
dominance of the $\Pprho \pi$ over the $\PfTwo \pi$ decay mode in an
analysis of the $\pi^+ \pi^- \pi^0$ final state produced in two-photon
collisions~\cite{Shchegelsky:2006es}.  At the moment, we do not have
an explanation for this discrepancy and more studies are needed.

\subsubsection{The $\JPC = 2^{-+}$ Sector }
\label{sec:results_2mp}

The PDG lists currently five isovector states with
$\JPC = 2^{-+}$~\cite{Tanabashi:2018zz}: \PpiTwo, \PpiTwo[1880],
\PpiTwo[2005], \PpiTwo[2100], and \PpiTwo[2285] (see also
\cref{fig:light_flavorless_spectrum}).  The latter three states
require confirmation: the \PpiTwo[2100] is omitted from the summary
table, the \PpiTwo[2005] and \PpiTwo[2285] are listed as
\textquote{further states}.  The \PpiTwo is the $2^{-+}$ ground state,
\ie the \termSym{1}{1}{D}{2} quark-model state.  The \PpiTwo[1880] is
an established state but its mass is too close to that of the \PpiTwo
ground state in order to be the radial excitation of the latter, \ie
the \termSym{2}{1}{D}{2} state.  The \PpiTwo[2005] or the
\PpiTwo[2100] would be more plausible candidates for the
\termSym{2}{1}{D}{2} quark-model state.

\begin{figure}[tbp]
  \centering
  \subfloat[][]{%
    \includegraphics[width=\threePlotWidth]{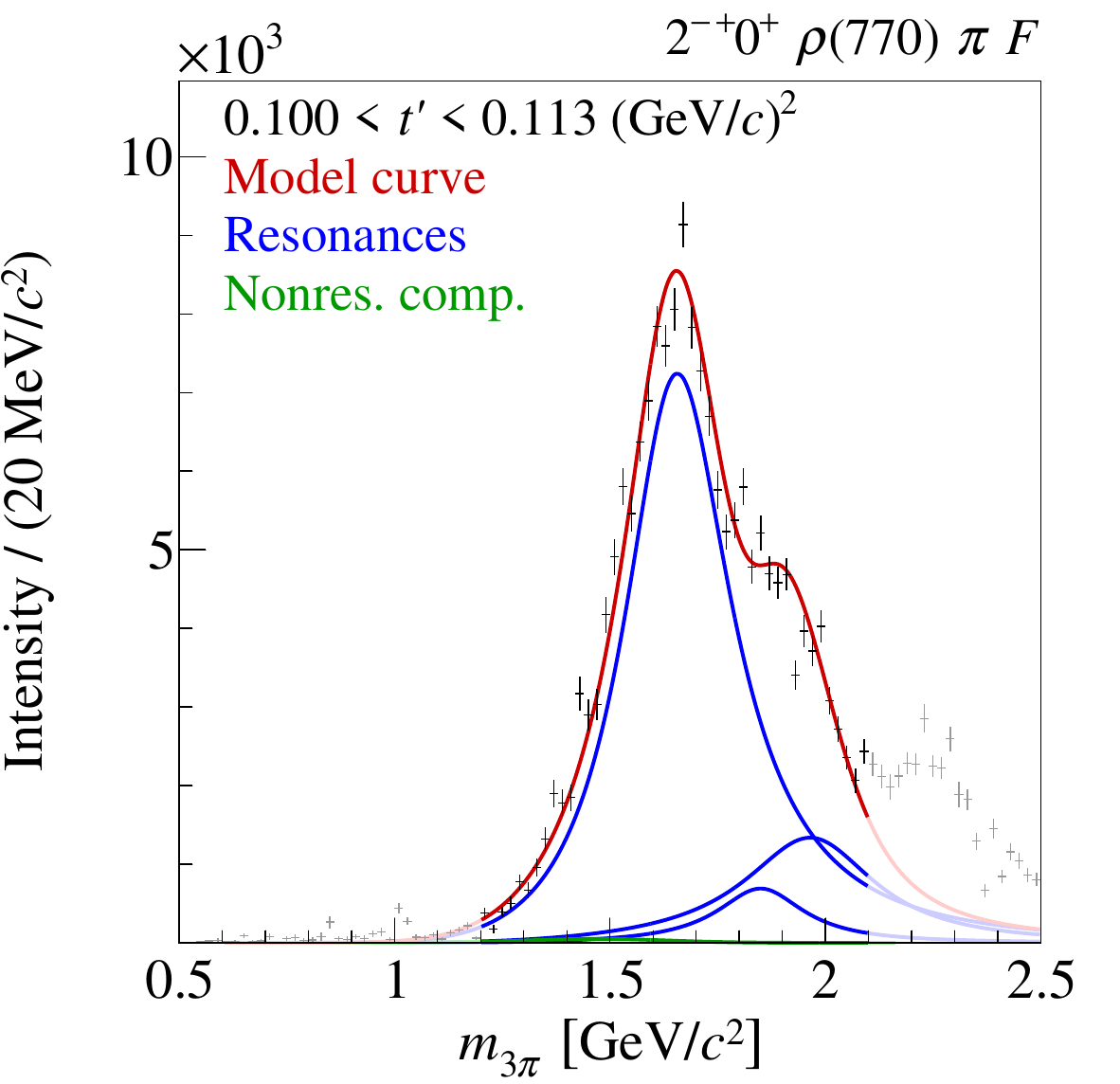}%
    \label{fig:intensity_2mp_rho_tbin1}%
  }%
  \hfill%
  \subfloat[][]{%
    \includegraphics[width=\threePlotWidth]{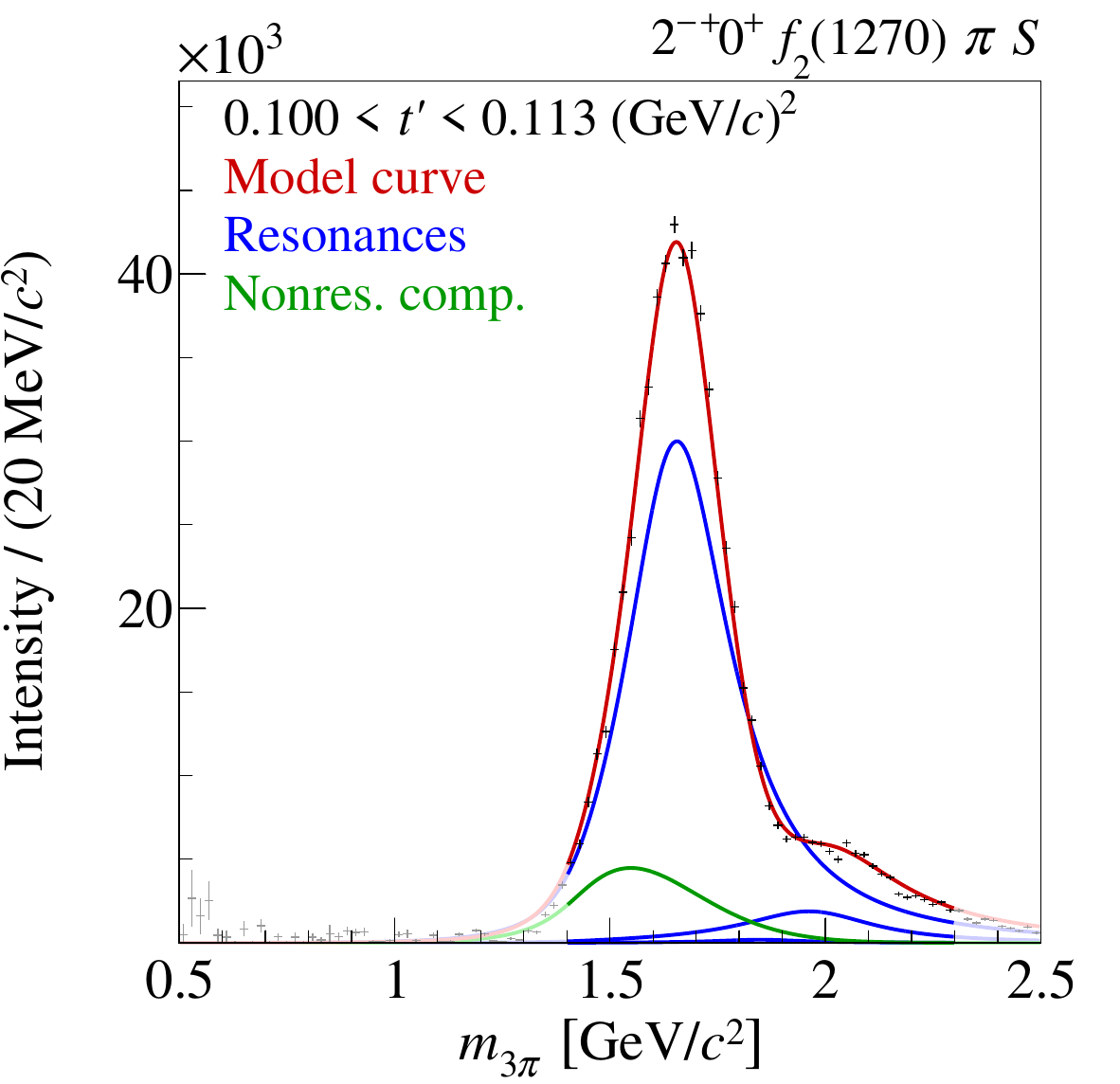}%
    \label{fig:intensity_2mp_f2_s_tbin1}%
  }%
  \hfill%
  \subfloat[][]{%
    \includegraphics[width=\threePlotWidth]{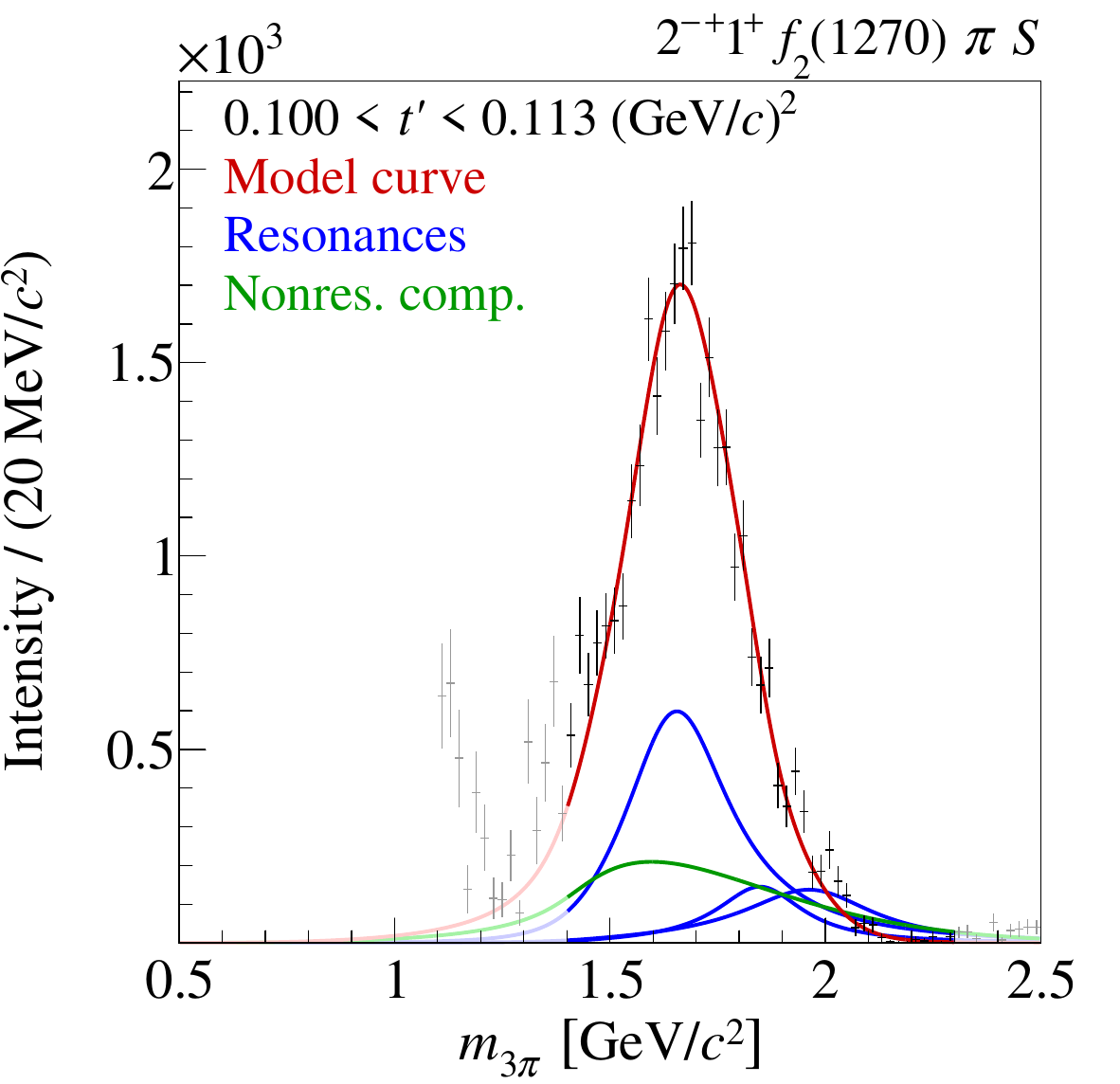}%
    \label{fig:intensity_2mp_f2_m1_tbin1}%
  }%
  \caption{Intensity distributions of
    \subfloatLabel{fig:intensity_2mp_rho_tbin1}~the
    \wave{2}{-+}{0}{+}{\Pprho}{F},
    \subfloatLabel{fig:intensity_2mp_f2_s_tbin1}~the
    \wave{2}{-+}{0}{+}{\PfTwo}{S}, and
    \subfloatLabel{fig:intensity_2mp_f2_m1_tbin1}~the
    \wave{2}{-+}{1}{+}{\PfTwo}{S} wave, all in the lowest $t'$~bin of
    the \threePi proton-target data~\cite{Akhunzyanov:2018lqa}.  The
    curves represent the result of the resonance-model fit.  The model
    and the wave components are represented as in
    \cref{fig:intensity_phase_0mp}, except that the blue curves
    represent the \PpiTwo, \PpiTwo[1880], and \PpiTwo[2005].  At
    low~$t'$, the \PpiTwo is the dominant resonant component in all
    three waves.}
  \label{fig:intensity_2mp_pi2_1670}
\end{figure}

\begin{figure}[tbp]
  \centering
  \subfloat[]{%
    \includegraphics[width=0.5\textwidth]{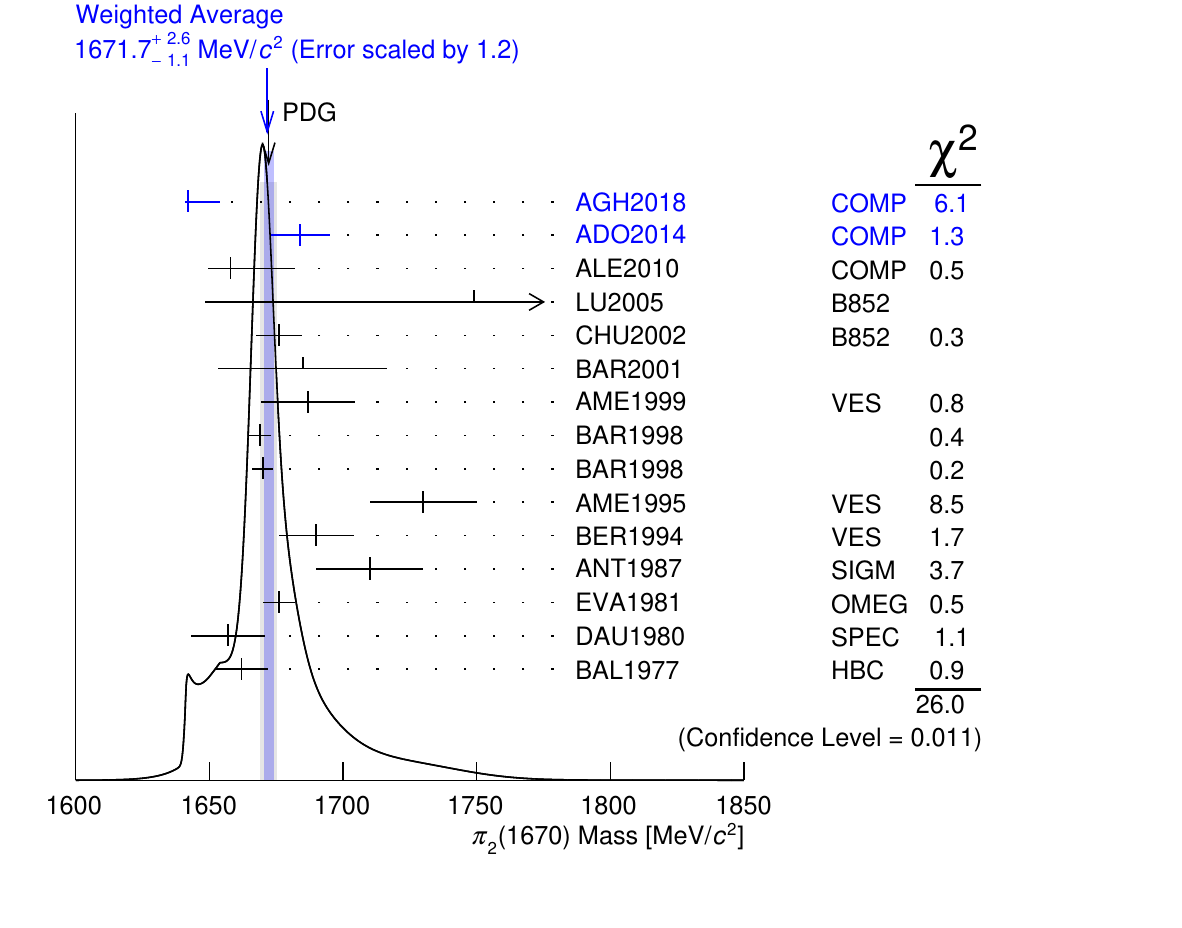}%
    \label{fig:ideogram_pi2_1670_mass}%
  }%
  \subfloat[]{%
    \includegraphics[width=0.5\textwidth]{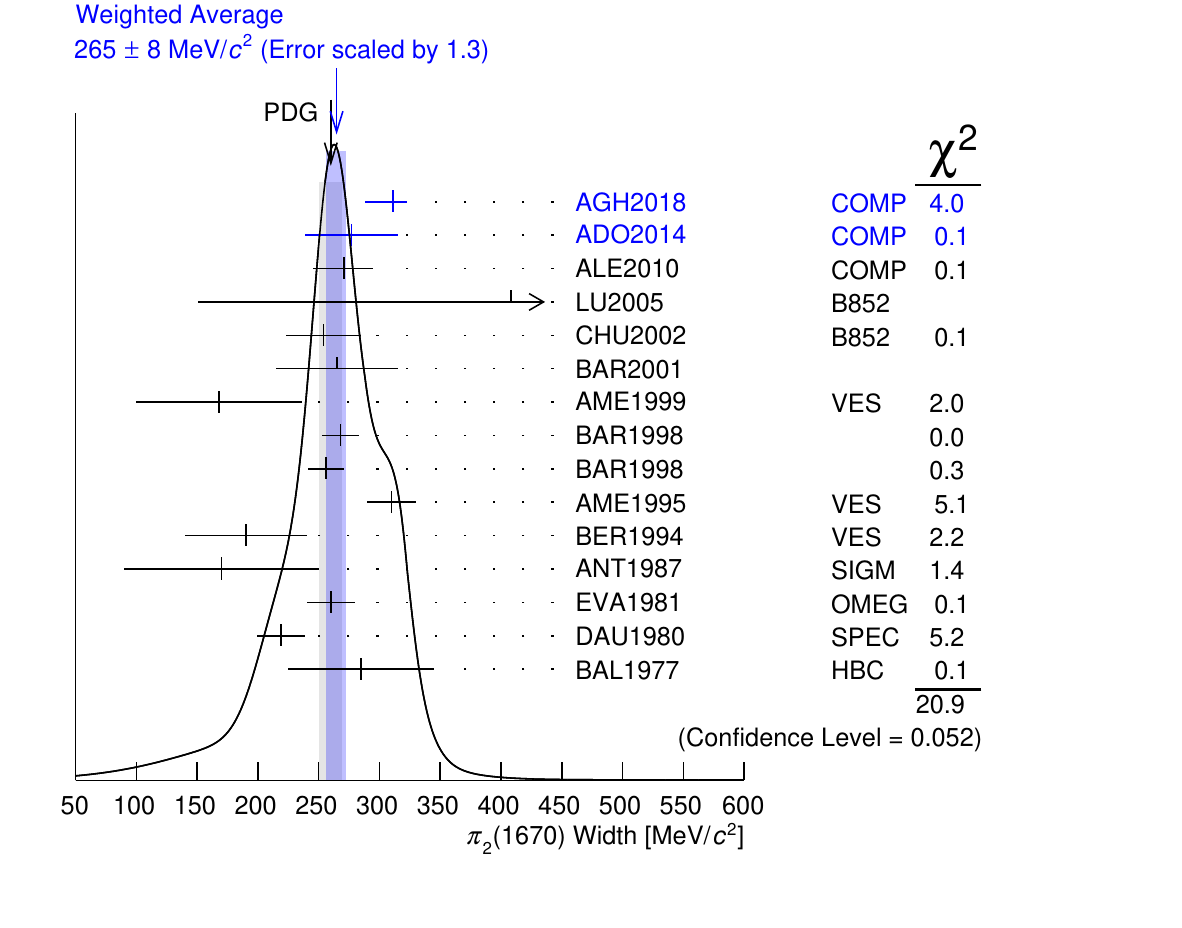}%
    \label{fig:ideogram_pi2_1670_width}%
  }%
  \caption{Ideograms similar to the ones in
    \cref{fig:ideogram_pi_1800} but for
    \subfloatLabel{fig:ideogram_pi2_1670_mass}~the mass and
    \subfloatLabel{fig:ideogram_pi2_1670_width}~the width of the
    \PpiTwo.  The Breit--Wigner parameters obtained from fits of the
    COMPASS diffractive \threePi proton-target
    (AGH2018,~\cite{Akhunzyanov:2018lqa}), the photo-production
    \threePi data (ADO2014,~\cite{Adolph:2014mup}) (uncertainties are
    statistical only; see \cref{sec:results_3pic_primakoff}), and
    diffractive lead-target data (ALE2010,~\cite{Alekseev:2009aa}) are
    compared to previous measurements~\cite{Tanabashi:2018zz}.}
  \label{fig:ideogram_pi2_1670}
\end{figure}

The \PpiTwo decays nearly exclusively into~$3\pi$ with a branching
fraction of \SI{95.8(14)}{\percent}~\cite{Tanabashi:2018zz}.  The
dominant $3\pi$~decay modes are $\PfTwo \pi$ with a branching fraction
of \SI{56.3(32)}{\percent} and $\Pprho \pi$ with \SI{31(4)}{\percent}.
Decays into $\pipiS \pi$ have also been observed using different
models for the
\pipiSW~\cite{Daum:1980ay,Baker:1999fc,Chung:2002pu}. In the \threePi
proton-target data, we see clear peaks of the \PpiTwo in three of the
four $2^{-+}$ waves that are included in the 14-wave resonance-model
fit, namely in the $\Pprho \pi$ $F$~wave and the $\PfTwo \pi$
$S$~waves with $M = 0$ and~1 (see \cref{fig:intensity_2mp_pi2_1670}).
The measured \PpiTwo Breit--Wigner parameters of
$m_{\PpiTwo} = \SIaerr{1642}{12}{1}{\MeVcc}$ and
$\Gamma_{\PpiTwo} = \SIaerr{311}{12}{23}{\MeVcc}$ have comparatively
small uncertainties~\cite{Akhunzyanov:2018lqa}.  These values are
consistent with the Breit--Wigner parameters
$m_{\PpiTwo} = \SIsaerrs{1658}{3}{24}{8}{\MeVcc}$ and
$\Gamma_{\PpiTwo} = \SIsaerrs{271}{9}{22}{24}{\MeVcc}$ obtained from
the \threePi lead-target data~\cite{Alekseev:2009aa}.  In the latter
analysis, only the \wave{2}{-+}{0}{+}{\PfTwo}{S} wave was included in
the 6-wave resonance-model fit.  As shown in
\cref{fig:ideogram_pi2_1670}, the \PpiTwo parameters from the
lead-target data are in good agreement with the PDG average.  Those
from the proton-target data are in fair agreement: the mass is
somewhat lower, while the width is somewhat larger.

\begin{figure}[tbp]
  \centering
  \subfloat[][]{%
    \includegraphics[width=\threePlotWidthTwoD]{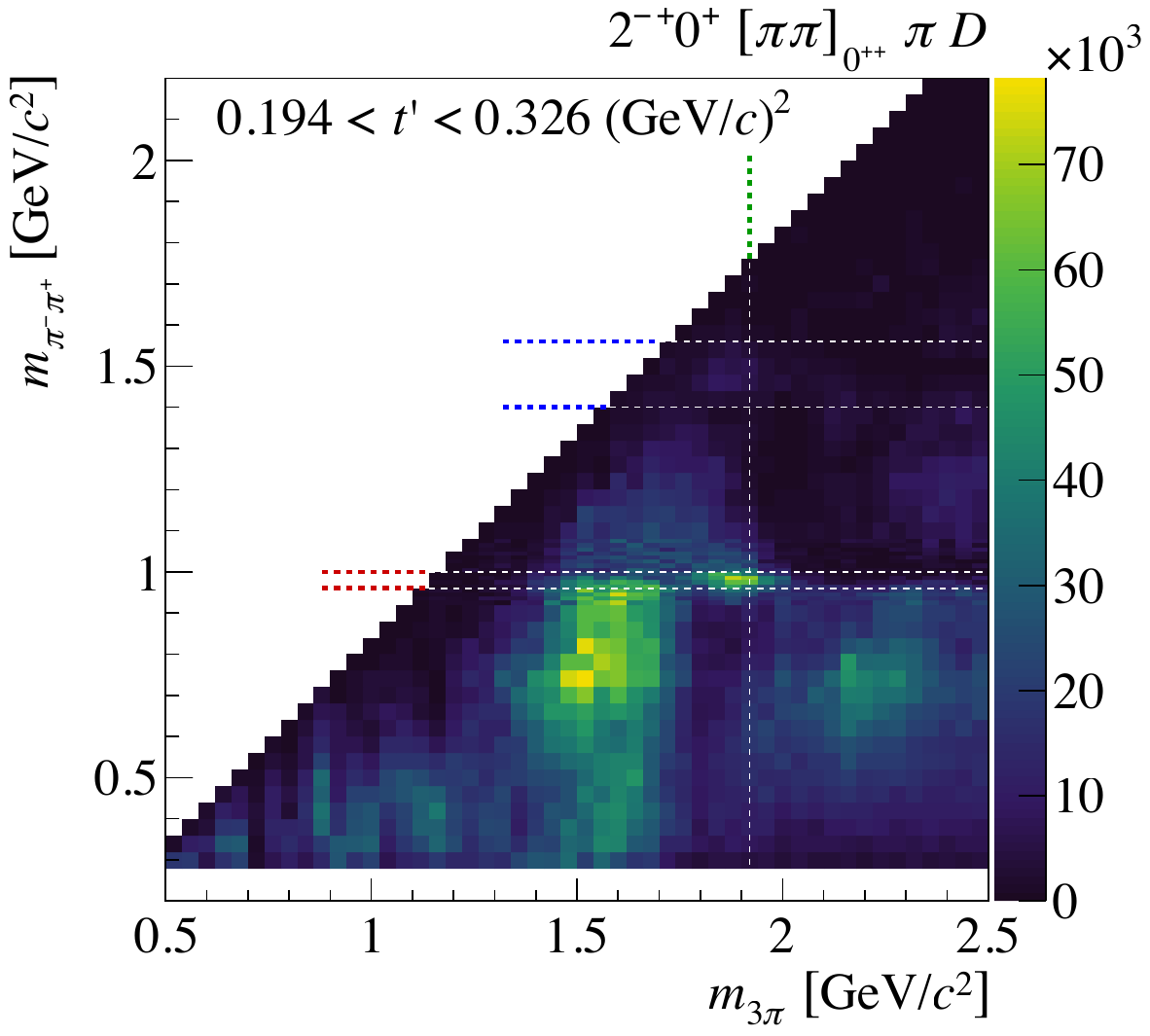}%
    \label{fig:2mp_pipiS_highT_2d}
  }%
  \hfill%
  \subfloat[][]{%
    \includegraphics[width=\threePlotWidth]{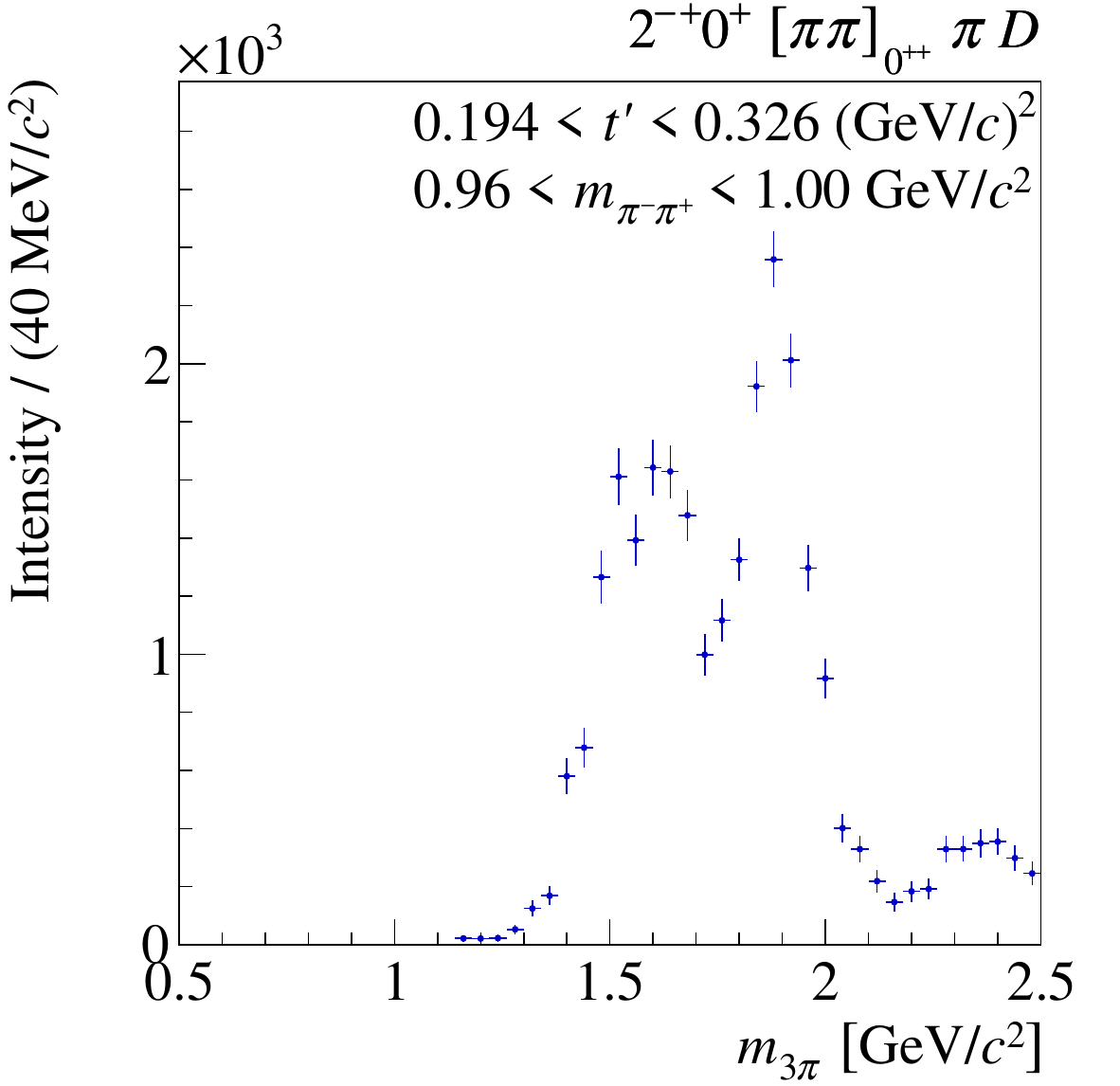}%
    \label{fig:2mp_pipiS_highT_f0980}
  }%
  \hfill%
  \subfloat[][]{%
    \includegraphics[width=\threePlotWidth]{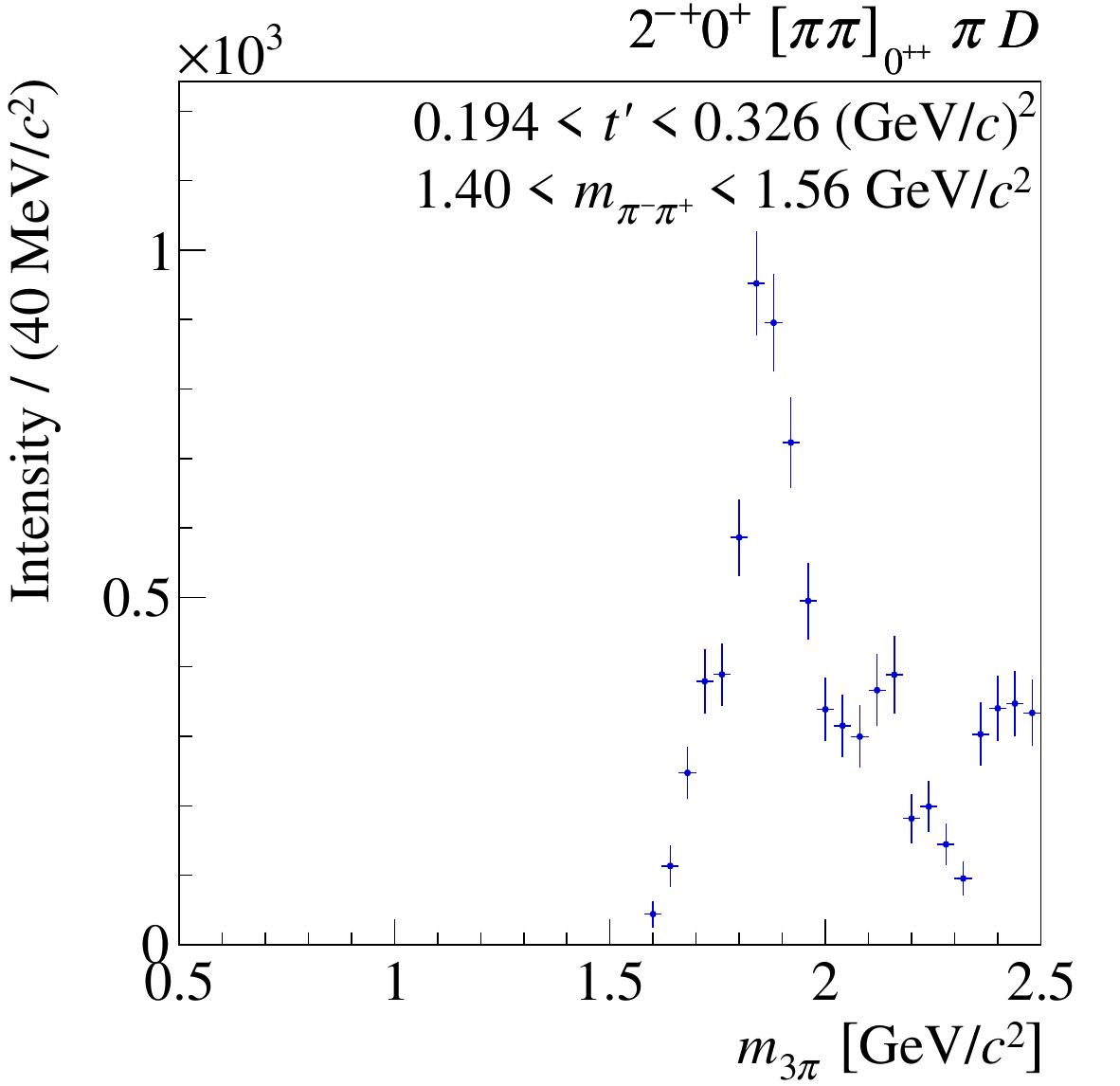}%
    \label{fig:2mp_pipiS_highT_f01500}
  }%
  \\
  \null\hfill%
  \subfloat[][]{%
    \includegraphics[width=\threePlotWidth]{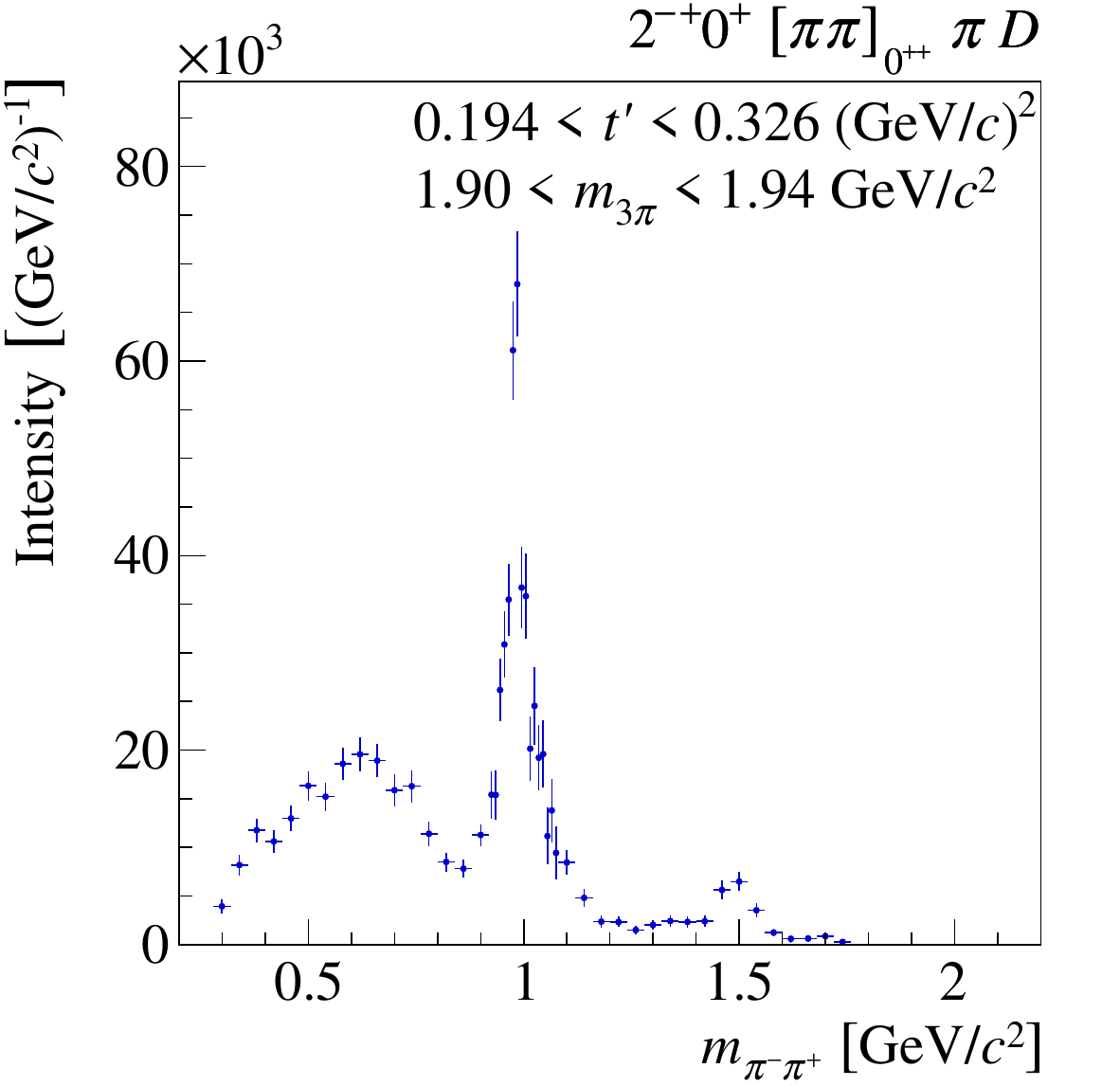}%
    \label{fig:PIPIS_2mp_1.9_m2pi}%
  }%
  \hfill%
  \subfloat[][]{%
    \includegraphics[width=\threePlotWidth]{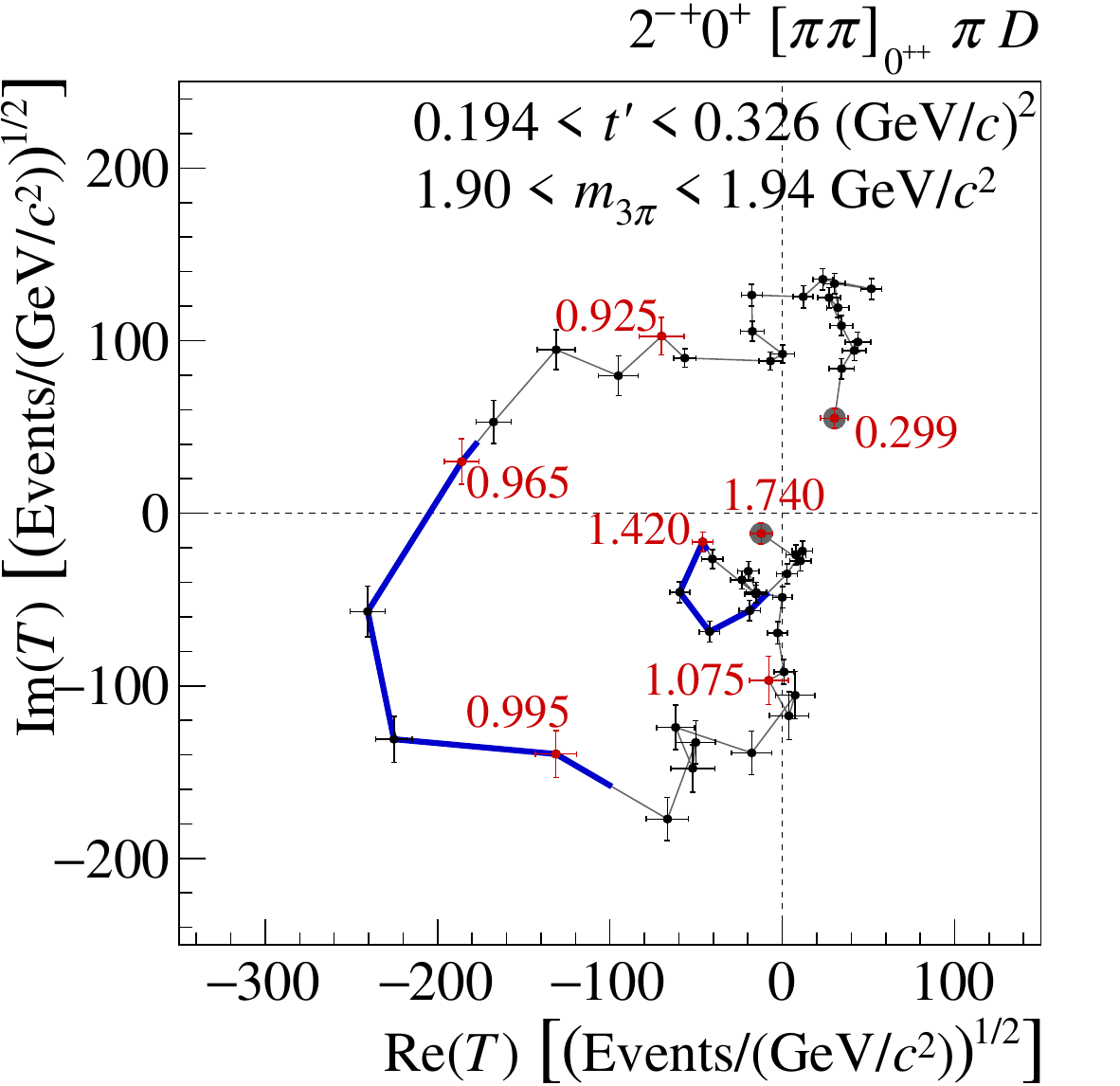}%
    \label{fig:PIPIS_2mp_1.9_argand}%
  }%
  \hfill\null%
  \caption{Similar to \cref{fig:pipi_s_wave_0mp_highT} but for the
    \wave{2}{-+}{0}{+}{\pipiSF}{D} wave with the freed-isobar
    amplitude \pipiSF in an intermediate $t'$~bin of the \threePi
    proton-target data~\cite{Adolph:2015tqa}.
    \subfloatLabel{fig:2mp_pipiS_highT_2d}~Intensity of the
    \wave{2}{-+}{0}{+}{\pipiSF}{D} wave as a function of~\mTwoPi
    and~\mThreePi.
    \subfloatLabel{fig:2mp_pipiS_highT_f0980}~and~\subfloatLabel{fig:2mp_pipiS_highT_f01500}:
    Intensity as a function of~\mThreePi summed over the selected
    \mTwoPi~intervals around
    \subfloatLabel{fig:2mp_pipiS_highT_f0980}~the \PfZero[980] and
    \subfloatLabel{fig:2mp_pipiS_highT_f01500}~the \PfZero[1500] as
    indicated by the pairs of horizontal dashed lines
    in~\subfloatLabel{fig:2mp_pipiS_highT_2d}.
    \subfloatLabel{fig:PIPIS_2mp_1.9_m2pi}~and~\subfloatLabel{fig:PIPIS_2mp_1.9_argand}:
    The \pipiSF freed-isobar amplitude for the \mThreePi~bin at the
    \PpiTwo[1880] mass as indicated by the vertical dashed line
    in~\subfloatLabel{fig:2mp_pipiS_highT_2d}.
    \subfloatLabel{fig:PIPIS_2mp_1.9_m2pi}~Intensity as a function
    of~\mTwoPi.  \subfloatLabel{fig:PIPIS_2mp_1.9_argand}~\Argand.}
  \label{fig:pipi_s_wave_2mp_highT}
\end{figure}

As already mentioned above, the \PpiTwo[1880] is peculiar.  Not only
is its mass too low in order to be the radial excitation of the
\PpiTwo ground state, but it also exhibits an unexpected decay
pattern.  The PDG does not list any $3\pi$~decay modes and in
particular no $\Pprho \pi$ decay mode~\cite{Tanabashi:2018zz}.
However, the \PpiTwo[1880] is expected to decay into~$3\pi$ because it
has been observed in
$\PfZero[1500]\pi \to \eta \eta \pi$~\cite{Anisovich:2001hj}.  We have
studied the \PpiTwo[1880] decay modes into \PfZero* isobars using the
freed-isobar PWA method (see
\cref{sec:pwa_cells:freed_isobar,sec:3pi_model:pwa}).
\Cref{fig:2mp_pipiS_highT_2d} shows the correlation of the
\mThreePi~intensity distribution of the \wave{2}{-+}{0}{+}{\pipiSF}{D}
wave with the \mTwoPi~intensity distribution of the freed-isobar
amplitude with $\JPC = 0^{++}$.  We observe a clear peak slightly
below $\mThreePi = \SI{1.9}{\GeVcc}$ and at
$\mTwoPi \approx \SI{1.0}{\GeVcc}$.  This \PpiTwo[1880] peak is also
clearly visible in the \mThreePi~intensity distribution in the
\PfZero[980] region that is shown in \cref{fig:2mp_pipiS_highT_f0980}.
The \mTwoPi~intensity distribution at the \PpiTwo[1880] peak position
is shown in \cref{fig:PIPIS_2mp_1.9_m2pi} and exhibits a narrow peak
of the \PfZero[980].  The resonant nature of this peak is confirmed by
the corresponding highlighted circular structure in the \Argand in
\cref{fig:PIPIS_2mp_1.9_argand}.  The \Argand exhibits an additional
smaller circular structure in the highlighted \PfZero[1500] region.
This structure corresponds to a small peak at about \SI{1.5}{\GeVcc}
in the \mTwoPi~intensity distribution shown in
\cref{fig:PIPIS_2mp_1.9_m2pi} and to a \PpiTwo[1880] peak in the
\mThreePi~intensity distribution shown in
\cref{fig:2mp_pipiS_highT_f01500}.  Our freed-isobar PWA result hence
establishes two $3\pi$~decay modes of the \PpiTwo[1880], namely
$\PfZero[980] \pi$ and $\PfZero[1500] \pi$.\footnote{Since in the
  conventional PWA the intensities of the $2^{-+}$ waves with
  $\JPC = 0^{++}$ isobars depend on the particular parameterizations
  chosen for the isobar amplitudes, these waves are not included in
  the 14-wave resonance-model fit (see also Section~VI.D.2 in
  \refCite{Akhunzyanov:2018lqa}).} The latter one is consistent with
the $\PfZero[1500] \pi$ decay mode seen in the $\eta \eta \pi$ final
state~\cite{Anisovich:2001hj}.

\begin{figure}[tbp]
  \centering
  \subfloat[][]{%
    \includegraphics[width=\threePlotWidth]{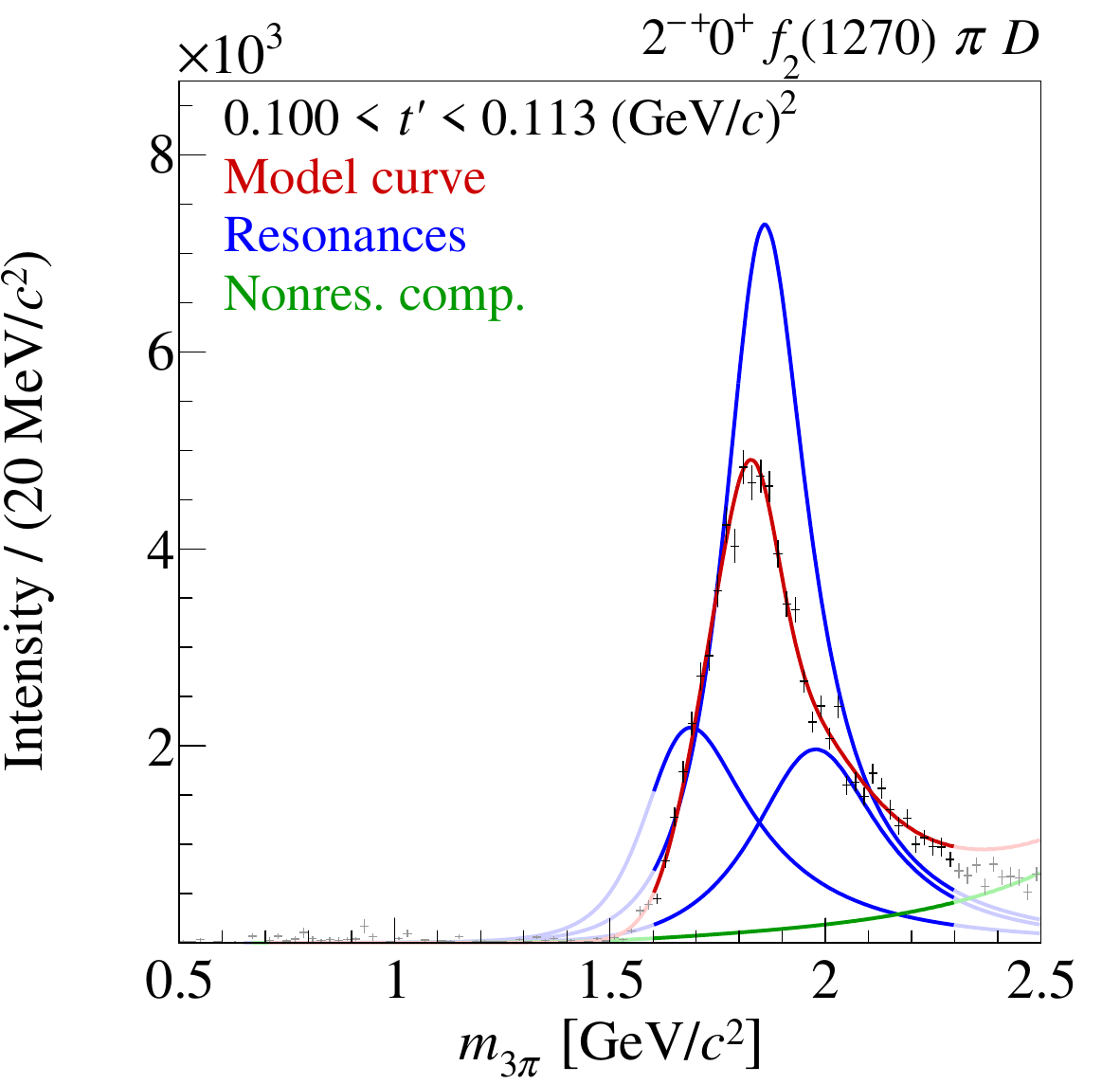}%
    \label{fig:intensity_2mp_f2_d_tbin1}%
  }%
  \hfill%
  \subfloat[][]{%
    \includegraphics[width=\threePlotWidth]{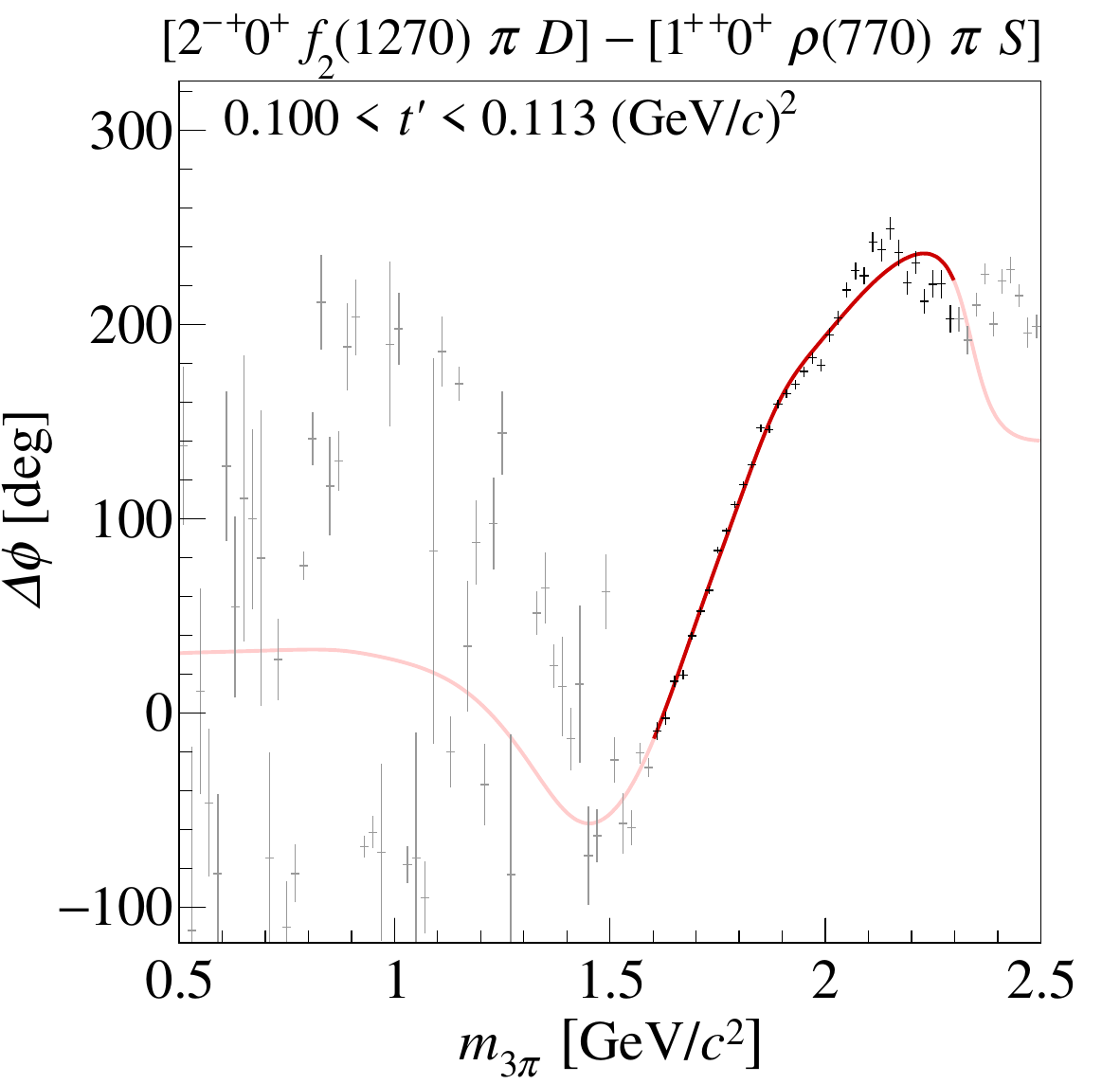}%
    \label{fig:phase_2mp_f2_d_1pp_rho_tbin1}%
  }%
  \hfill%
  \subfloat[][]{%
    \includegraphics[width=\threePlotWidth]{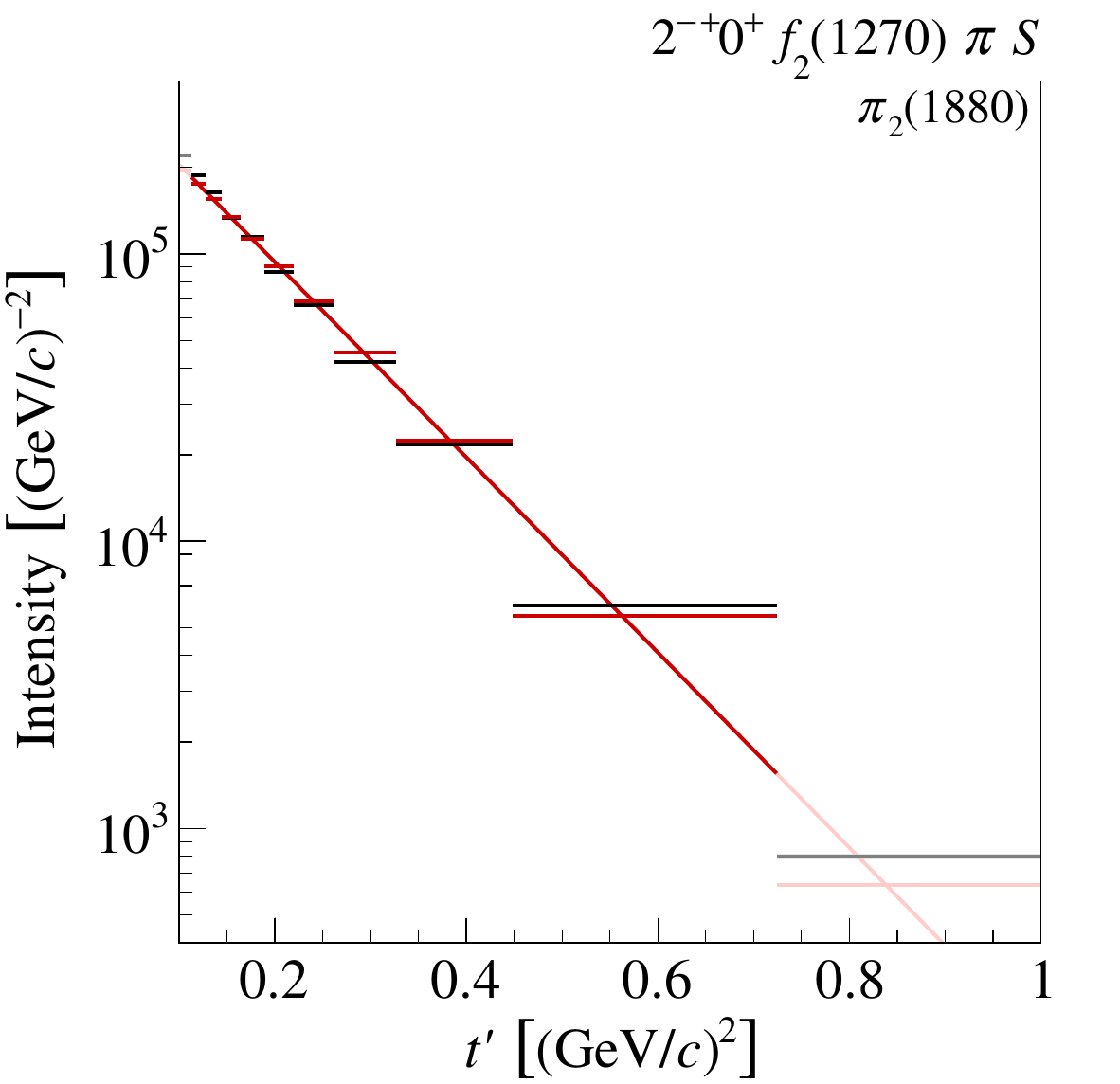}%
    \label{fig:tspectrum_2mp_f2}%
  }%
  \caption{\subfloatLabel{fig:intensity_2mp_f2_d_tbin1}~Intensity
    distribution of the \wave{2}{-+}{0}{+}{\PfTwo}{D} wave and
    \subfloatLabel{fig:phase_2mp_f2_d_1pp_rho_tbin1}~phase of this
    wave \wrt the \wave{1}{++}{0}{+}{\Pprho}{S} wave, both in the
    lowest $t'$~bin of the \threePi proton-target
    data~\cite{Akhunzyanov:2018lqa}.  The curves represent the result
    of the resonance-model fit.  The model and the wave components are
    represented as in \cref{fig:intensity_phase_0mp} except that the
    blue curves represent the \PpiTwo, \PpiTwo[1880], and
    \PpiTwo[2005].  \subfloatLabel{fig:tspectrum_2mp_f2}~Similar to
    \cref{fig:tspectrum_0mp_f0}, but showing the $t'$~spectrum of the
    \PpiTwo[1880] in the \wave{2}{-+}{0}{+}{\PfTwo}{S} wave.}
  \label{fig:intensity_phase_2mp_f2_d}
\end{figure}

\begin{figure}[tbp]
  \centering
  \subfloat[]{%
    \includegraphics[width=0.5\textwidth]{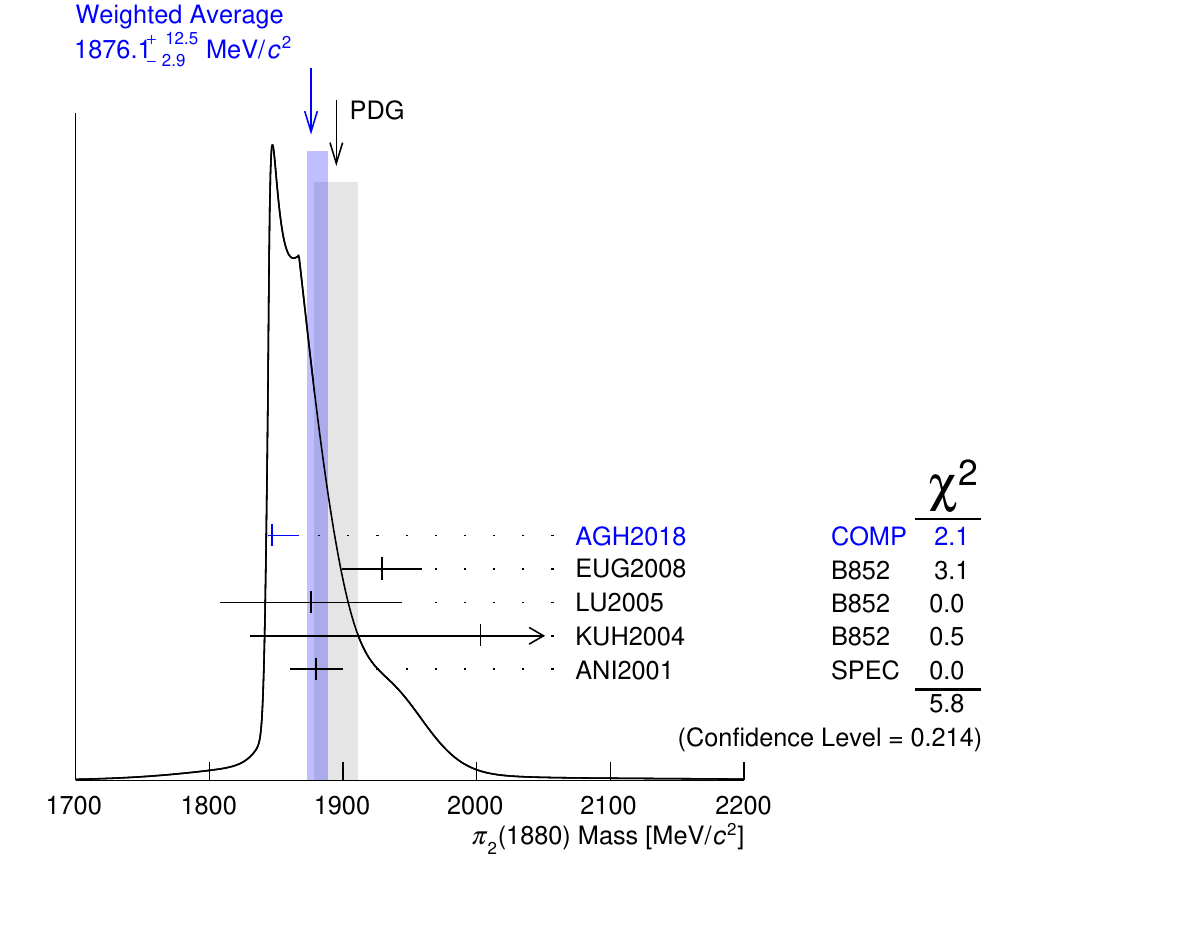}%
    \label{fig:ideogram_pi2_1880_mass}%
  }%
  \subfloat[]{%
    \includegraphics[width=0.5\textwidth]{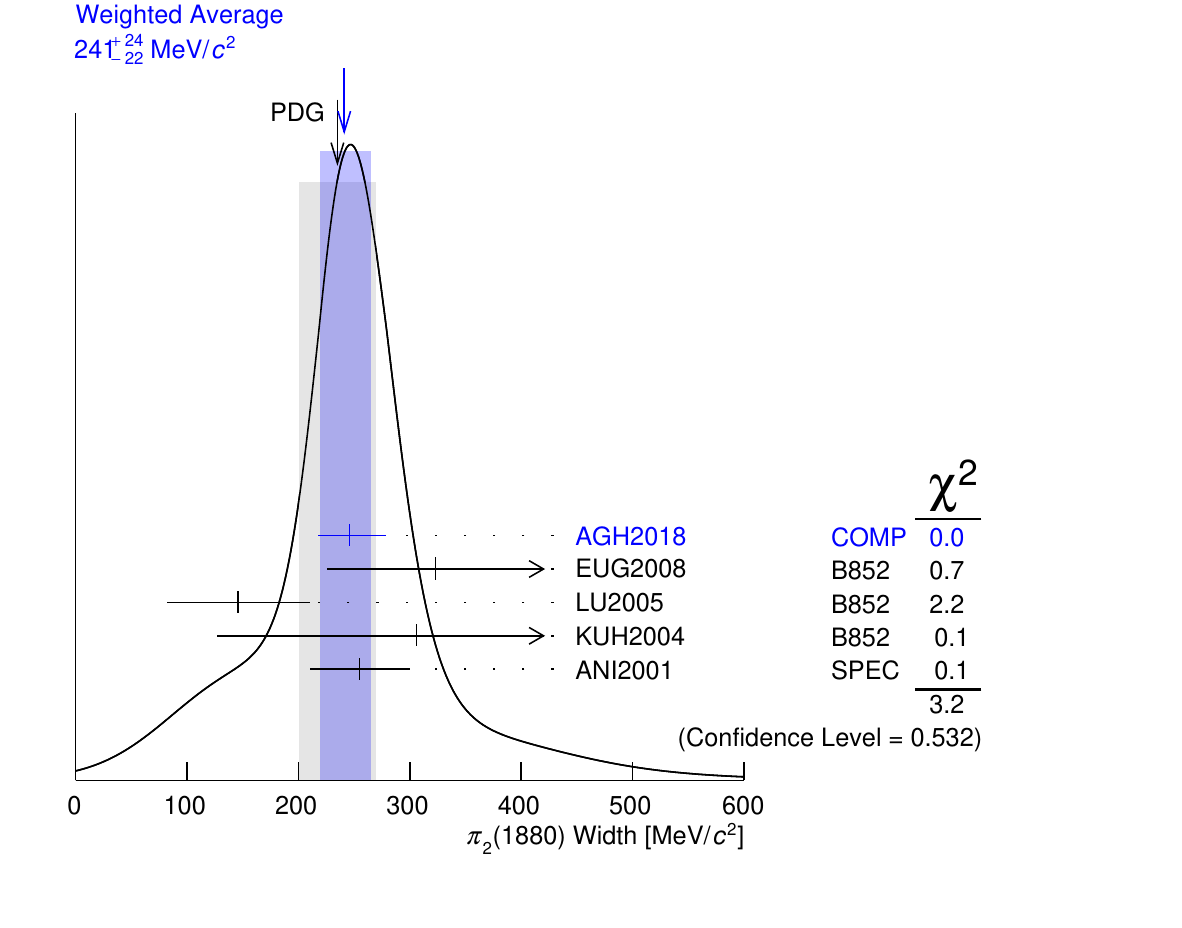}%
    \label{fig:ideogram_pi2_1880_width}%
  }%
  \caption{Ideograms similar to the ones in
    \cref{fig:ideogram_pi_1800} but for
    \subfloatLabel{fig:ideogram_pi2_1880_mass}~the mass and
    \subfloatLabel{fig:ideogram_pi2_1880_width}~the width of the
    \PpiTwo[1880].  The Breit--Wigner parameters obtained from a fit of
    the COMPASS \threePi proton-target data
    (AGH2018,~\cite{Akhunzyanov:2018lqa}) are compared to previous
    measurements~\cite{Tanabashi:2018zz}.}
  \label{fig:ideogram_pi2_1880}
\end{figure}

In the 14-wave resonance-model fit, we observe a strong \PpiTwo[1880]
signal only in the \wave{2}{-+}{0}{+}{\PfTwo}{D} wave.  In the
$\Pprho \pi$ $F$~wave and the two $\PfTwo \pi$ $S$~waves with $M = 0$
and~1, the \PpiTwo[1880] component is small.  The intensity
distribution of the $\PfTwo \pi$ $D$~wave as shown in
\cref{fig:intensity_2mp_f2_d_tbin1} exhibits a peak at about
\SI{1.8}{\GeVcc} that is described mostly by the \PpiTwo[1880]
component.  The non-resonant contribution is small.  There is,
however, considerable destructive interference of the \PpiTwo[1880]
with the \PpiTwo and the \PpiTwo[2005].  As is shown in
\cref{fig:phase_2mp_f2_d_1pp_rho_tbin1}, the peak is associated with a
rapid phase motion.  In our resonance model, the coupling amplitudes
of the \PpiTwo[1880] in the three $2^{-+}$ waves with $M = 0$ are
constrained by \cref{eq:branching_amp}.  Therefore, the $t'$~spectra
of the \PpiTwo[1880] are very similar in these three waves.  As an
example, \cref{fig:tspectrum_2mp_f2} shows the $t'$~spectrum of the
\PpiTwo[1880] in the \wave{2}{-+}{0}{+}{\PfTwo}{S} wave.  The
distribution is approximately exponential and has a slope parameter
value of \SIaerr{7.8}{0.5}{0.9}{\perGeVcsq}, which is typical for
resonances.  We find \PpiTwo[1880] Breit--Wigner parameters of
$m_{\PpiTwo[1880]} = \SIaerr{1847}{20}{3}{\MeVcc}$ and
$\Gamma_{\PpiTwo[1880]} =
\SIaerr{246}{33}{28}{\MeVcc}$~\cite{Akhunzyanov:2018lqa}.  As is shown
in \cref{fig:ideogram_pi2_1880}, our values are compatible with the
PDG world average~\cite{Tanabashi:2018zz}.

The peculiar properties of the \PpiTwo[1880] and the fact that it does
not appear in quark-model calculations (see \eg\
\cref{fig:light_flavorless_spectrum}) suggest that the \PpiTwo[1880]
could be a supernumerary state.  The authors of
\refsCite{Close:1994hc,Barnes:1996ff} propose it as a hybrid
candidate.  This would be also consistent with the predictions from
lattice QCD calculations (see \cref{fig:lattice_spectrum_light_ns}).
The authors of \refCite{Page:1998gz} predict for a hybrid meson that
the $\PfTwo \pi$ $D$-wave decay would be strongly suppressed \wrt the
$\PfTwo \pi$ $S$-wave decay.  It is therefore interesting to compare
the \PpiTwo[1880] signals in the \wave{2}{-+}{0}{+}{\PfTwo}{S} and
\wave{2}{-+}{0}{+}{\PfTwo}{D} waves.  Although the interference of the
model components in the $\PfTwo \pi$ $D$~wave is strongly
model-dependent and hence the \PpiTwo[1880] yield not well determined,
we can still conclude that the \PpiTwo[1880] signal in the
$\PfTwo \pi$ $D$~wave is enhanced \wrt the one in the $S$~wave by
about an order of magnitude (\confer\
\cref{fig:intensity_2mp_f2_s_tbin1,fig:intensity_2mp_f2_d_tbin1}).
The decay pattern that we observe in the \threePi final state is hence
exactly opposite to the one predicted in \refCite{Page:1998gz} for a
hybrid state and challenges the hybrid interpretation of the
\PpiTwo[1880].

The \PpiTwo[2005] appears as high-mass shoulders in the
\wave{2}{-+}{0}{+}{\Pprho}{F}, \wave{2}{-+}{0}{+}{\PfTwo}{S}, and
\wave{2}{-+}{0}{+}{\PfTwo}{D} waves, however, in the latter wave only
at high~$t'$ (see
\cref{fig:intensity_2mp_rho_tbin1,fig:intensity_2mp_f2_s_tbin1,fig:intensity_2mp_f2_d_tbin1}).
We also observe phase motions in the \SI{2}{\GeVcc} mass region (see
\eg\ \cref{fig:phase_2mp_f2_d_1pp_rho_tbin1}).  In the
$\Pprho \pi$~$F$ and the $\PfTwo \pi$ $S$~waves the \PpiTwo[2005]
component is significantly larger than the \PpiTwo[1880] component.

\begin{figure}[tbp]
  \centering
  \subfloat[][]{%
    \includegraphics[width=\threePlotWidth]{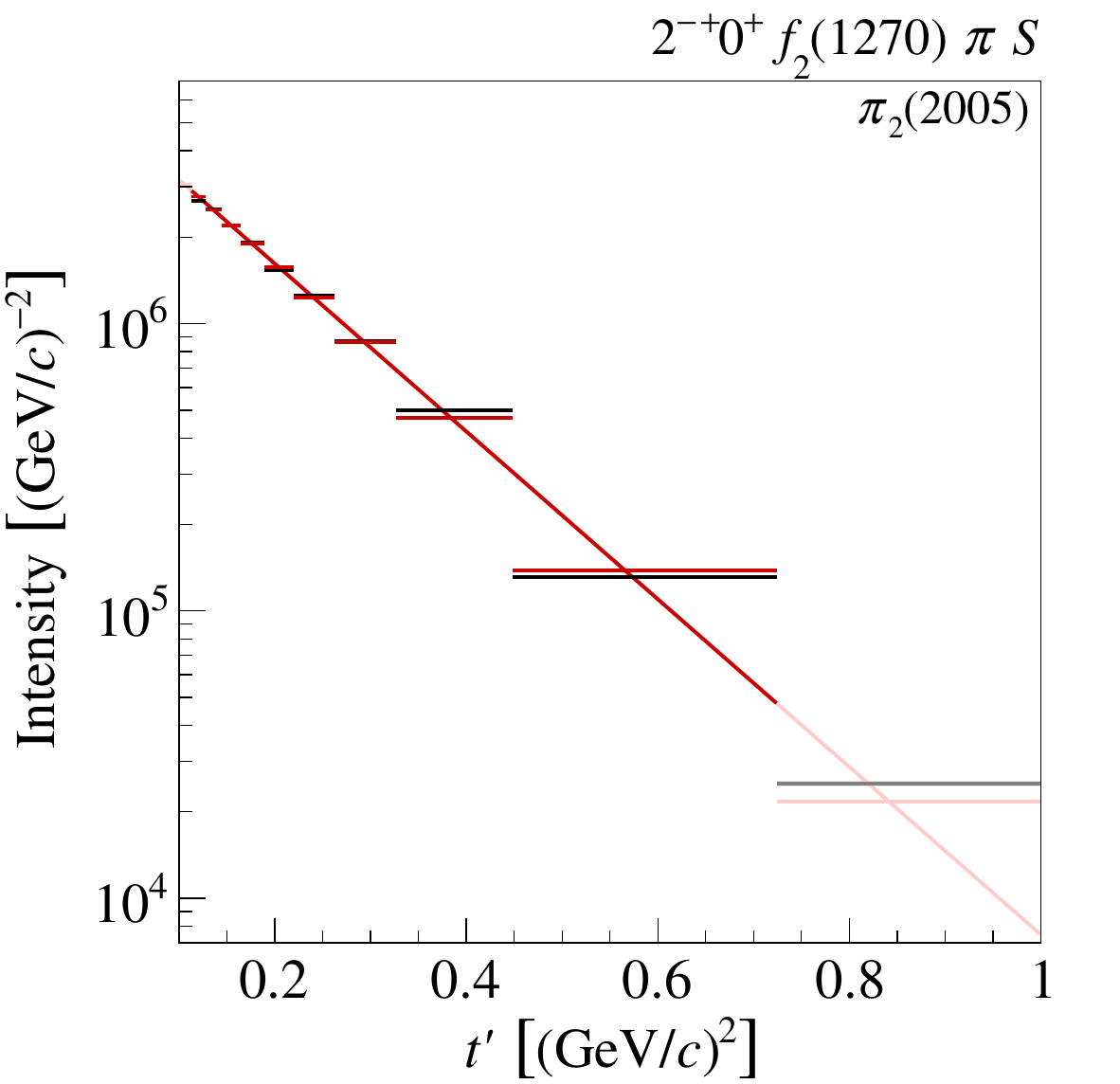}%
    \label{fig:tspectrum_2mp_f2_pi2_2005}%
  }%
  \hfill%
  \subfloat[][]{%
    \includegraphics[width=\threePlotWidth]{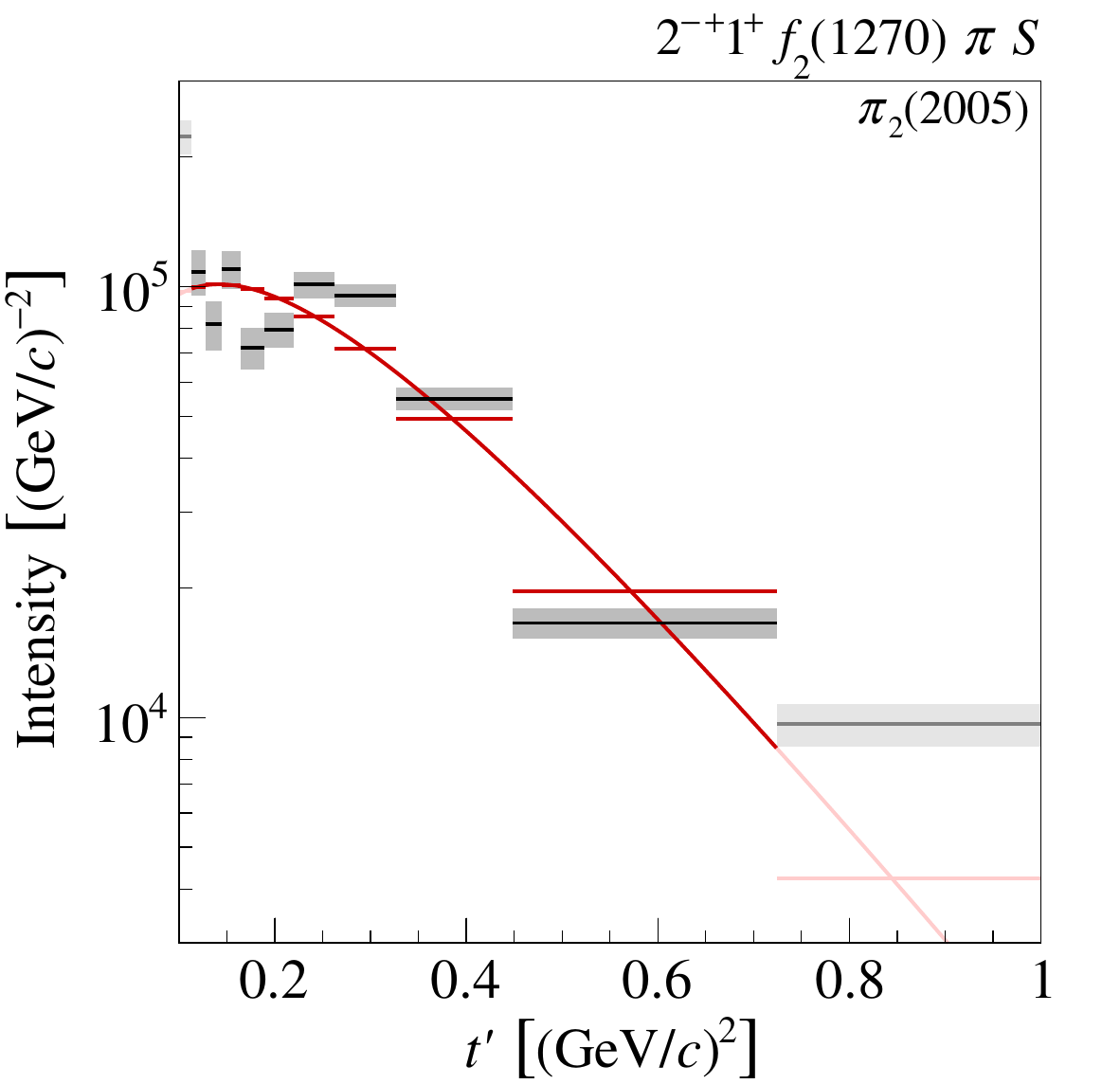}%
    \label{fig:tspectrum_2mp_f2_m1_pi2_2005}%
  }%
  \hfill%
  \subfloat[][]{%
    \includegraphics[width=\threePlotWidth]{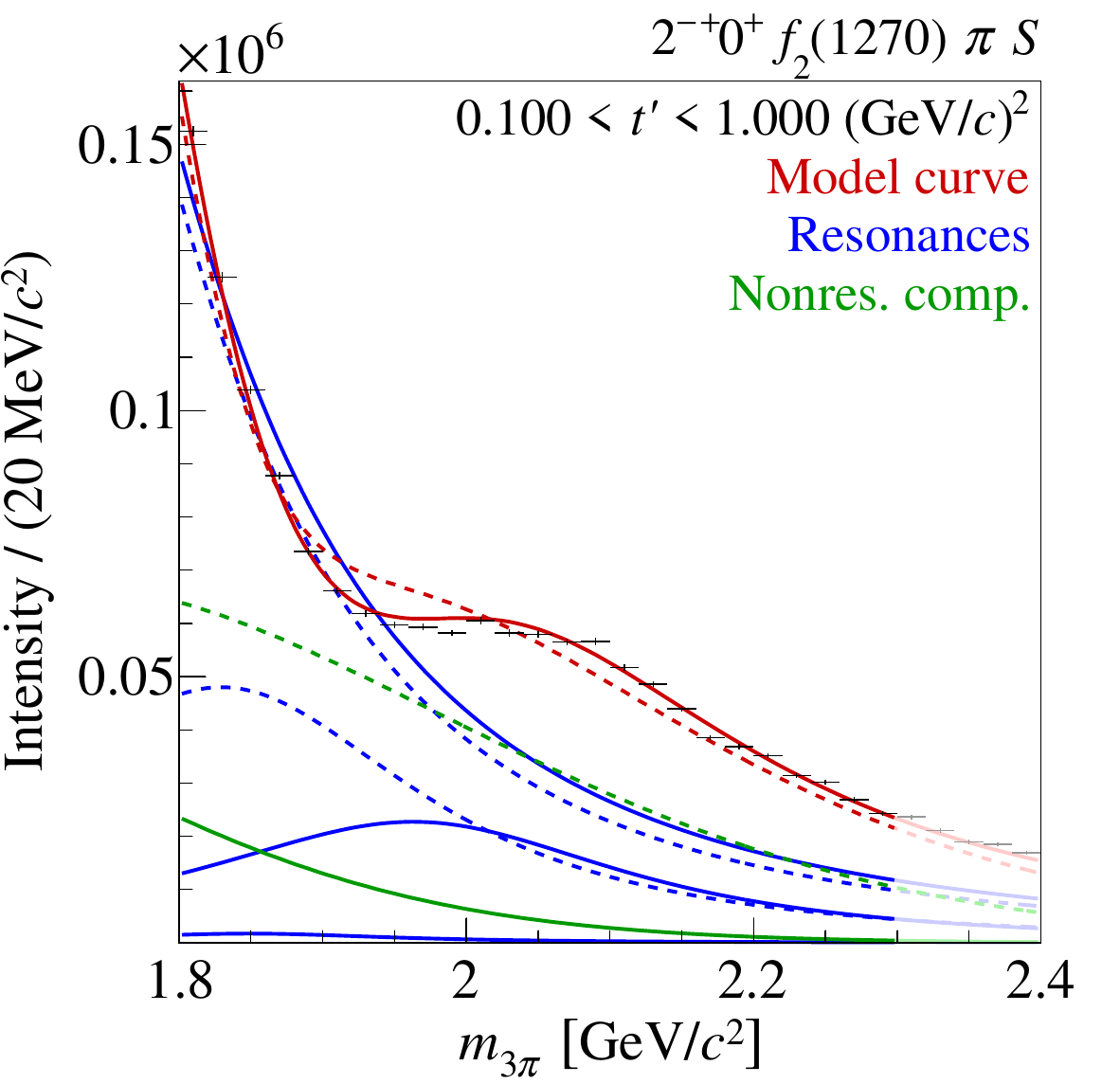}%
    \label{fig:intensity_2mp_f2_zoom}%
  }%
  \caption{\subfloatLabel{fig:tspectrum_2mp_f2_pi2_2005}~and~\subfloatLabel{fig:tspectrum_2mp_f2_m1_pi2_2005}:
    Similar to \cref{fig:tspectrum_0mp_f0}, but showing the
    $t'$~spectra of the \PpiTwo[2005] in the
    \wave{2}{-+}{0}{+}{\PfTwo}{S} and \wave{2}{-+}{1}{+}{\PfTwo}{S}
    waves.  \subfloatLabel{fig:intensity_2mp_f2_zoom}~Zoomed view of
    the intensity distribution of the \wave{2}{-+}{0}{+}{\PfTwo}{S}
    wave summed over the analyzed $t'$~range.  The curves represent
    the result of two resonance-model fits.  The model and the wave
    components are represented as in \cref{fig:intensity_phase_0mp}
    except that the blue curves represent the \PpiTwo, \PpiTwo[1880],
    and \PpiTwo[2005].  The result of the main resonance-model fit is
    represented by the continuous curves.  The dashed curves represent
    the result of a fit, where the \PpiTwo[2005] component is removed
    from the resonance model.  From \refCite{Akhunzyanov:2018lqa}.}
  \label{fig:pi2_2005}
\end{figure}

The $t'$~spectra of the \PpiTwo[2005] in the three $2^{-+}$ waves with
$M = 0$ are constrained by \cref{eq:branching_amp} and hence very
similar.  We show, as an example, in
\cref{fig:tspectrum_2mp_f2_pi2_2005} the $t'$~spectrum of the
\PpiTwo[2005] in the \wave{2}{-+}{0}{+}{\PfTwo}{S} wave.  It is
described well by the exponential model in \cref{eq:t_spectrum} with a
slope parameter value of \SIaerr{6.7}{0.4}{1.3}{\perGeVcsq} that is in
the range typical for resonances.  Due to the slightly shallower
$t'$~slope of the \PpiTwo[2005] in comparison of the other $2^{-+}$
wave components, its signal is more pronounced at high~$t'$.  The
\wave{2}{-+}{1}{+}{\PfTwo}{S} wave is the only $2^{-+}$ wave in the
14-wave resonance-model fit with $M = 1$.  Hence in this wave, the
\PpiTwo[2005] amplitude is not constrained by \cref{eq:branching_amp}.
Nevertheless, the corresponding $t'$~spectrum is well described by
\cref{eq:t_spectrum} with a slope parameter value of
\SIaerr{7.1}{3.5}{2.6}{\perGeVcsq} that is consistent with the ones in
the other $2^{-+}$ waves.  The measured $t'$~spectra hence support the
resonance interpretation of the \PpiTwo[2005] signal.  That the data
require a resonance in the \SI{2}{\GeVcc} mass region is verified by
comparing the main resonance-model fit with a fit without the
\PpiTwo[2005] component.  In the latter fit, the model describes the
intensity distributions and interference terms of the $2^{-+}$ waves
less well, in particular for the \wave{2}{-+}{0}{+}{\PfTwo}{S} and
\wave{2}{-+}{0}{+}{\PfTwo}{D} waves.  \Cref{fig:intensity_2mp_f2_zoom}
shows, for example, that the high mass-shoulder in the
\wave{2}{-+}{0}{+}{\PfTwo}{S} intensity distributions is not
reproduced well.  In addition the \PpiTwo[1880] becomes
\SI{20}{\MeVcc} lighter and \SI{100}{\MeVcc} wider, which would be in
tension with previous measurements.

We find \PpiTwo[2005] Breit--Wigner parameters of
$m_{\PpiTwo[2005]} = \SIaerr{1962}{17}{29}{\MeVcc}$ and
$\Gamma_{\PpiTwo[2005]} =
\SIaerr{371}{16}{120}{\MeVcc}$~\cite{Akhunzyanov:2018lqa} that are
consistent with the results of previous
measurements~\cite{Anisovich:2001pp,Anisovich:2001pn,Lu:2004yn}.  Our
\PpiTwo[2005] signal could be related to the \PpiTwo[2100] that has a
similar mass but a larger width of
\SI{625(50)}{\MeVcc}~\cite{Tanabashi:2018zz}.  The \PpiTwo[2100] PDG
entry is based on two observations in diffractively produced \threePi
reported by the ACCMOR~\cite{Daum:1980ay} and the VES
experiments~\cite{Amelin:1995gu}.  For both analyses, the set of
$2^{-+}$ waves that were included in the resonance-model fit differs
from our choice.  In addition, the resonance model was based on a
$K$-matrix approach.\footnote{A more detailed comparison of the ACCMOR
  and VES analyses with our approach can be found in
  \refCite{Akhunzyanov:2018lqa}.}  A potential caveat of our resonance
model that might affect our measurement of the \PpiTwo* resonance
parameters is that the three \PpiTwo* resonances have considerable
overlap.  As a consequence, our sum-of-Breit--Wigner ansatz might not
be a good approximation anymore.  Applying more advanced models is the
topic of future
research~\cite{Jackura:2016llm,Mikhasenko:2017jtg,Mikhasenko:2019}.
However, first studies also find three \PpiTwo* resonance poles in the
analyzed mass region and seem to confirm the findings from our
Breit--Wigner analysis at least qualitatively.

\subsubsection{The $\JPC = 4^{++}$ Sector }
\label{sec:results_4pp}

The \PaFour and the \PaFour[2255] are the only isovector states with
$J^{PC} = 4^{++}$ that are listed by the PDG~\cite{Tanabashi:2018zz}
(see also \cref{fig:light_flavorless_spectrum}).  The latter is listed
as a \textquote{further state} and requires confirmation.  The \PaFour
is known to decay into all final states considered here, \ie into
\etaOrPrPi and $3\pi$.  However, the branching fractions for these
decay modes are still unknown.

\begin{figure}[tbp]
  \centering
  \subfloat[][]{%
    \includegraphics[width=\threePlotWidth]{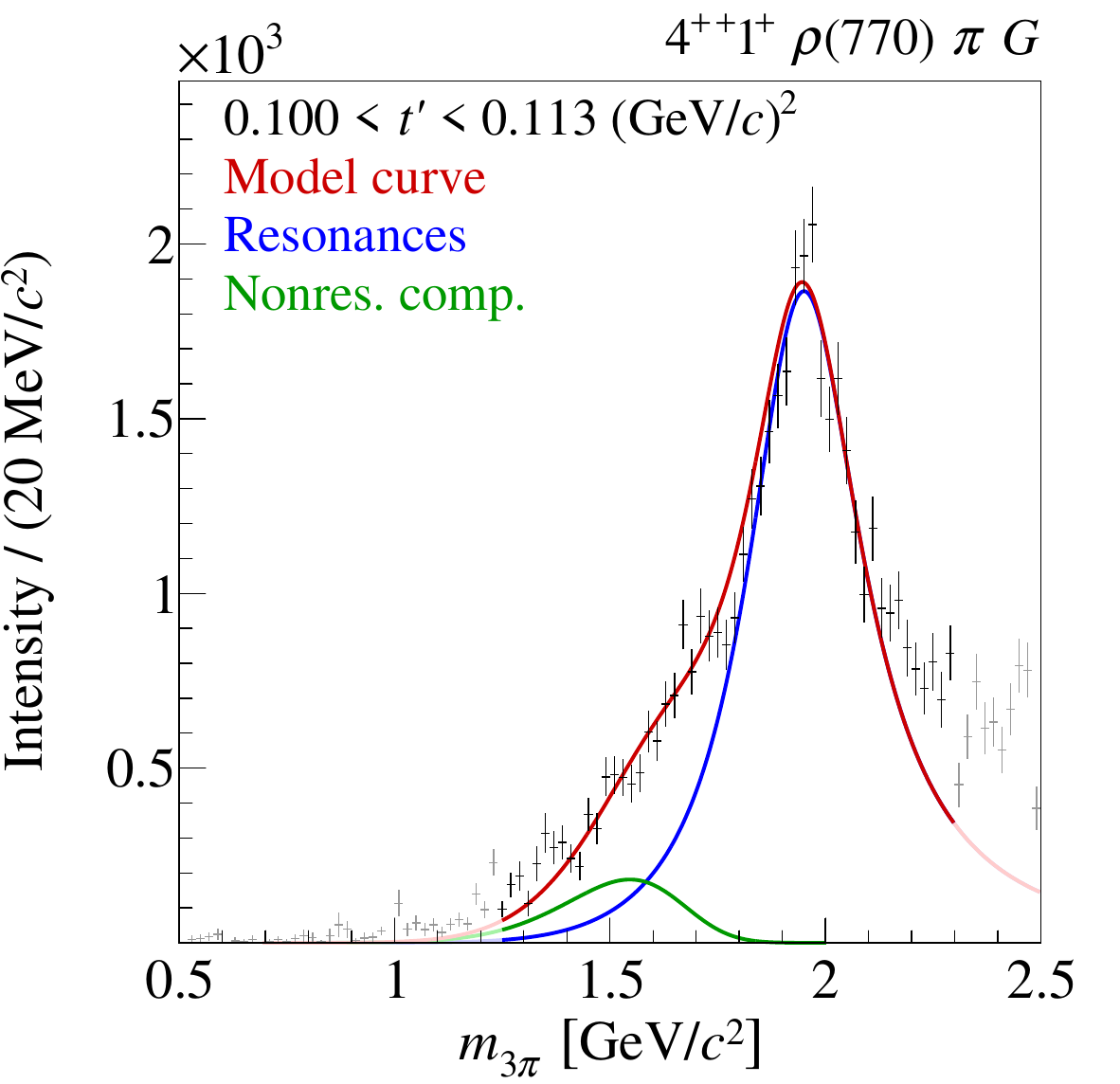}%
    \label{fig:intensity_4pp_rho_tbin1}%
  }%
  \hfill%
  \subfloat[][]{%
    \includegraphics[width=\threePlotWidth]{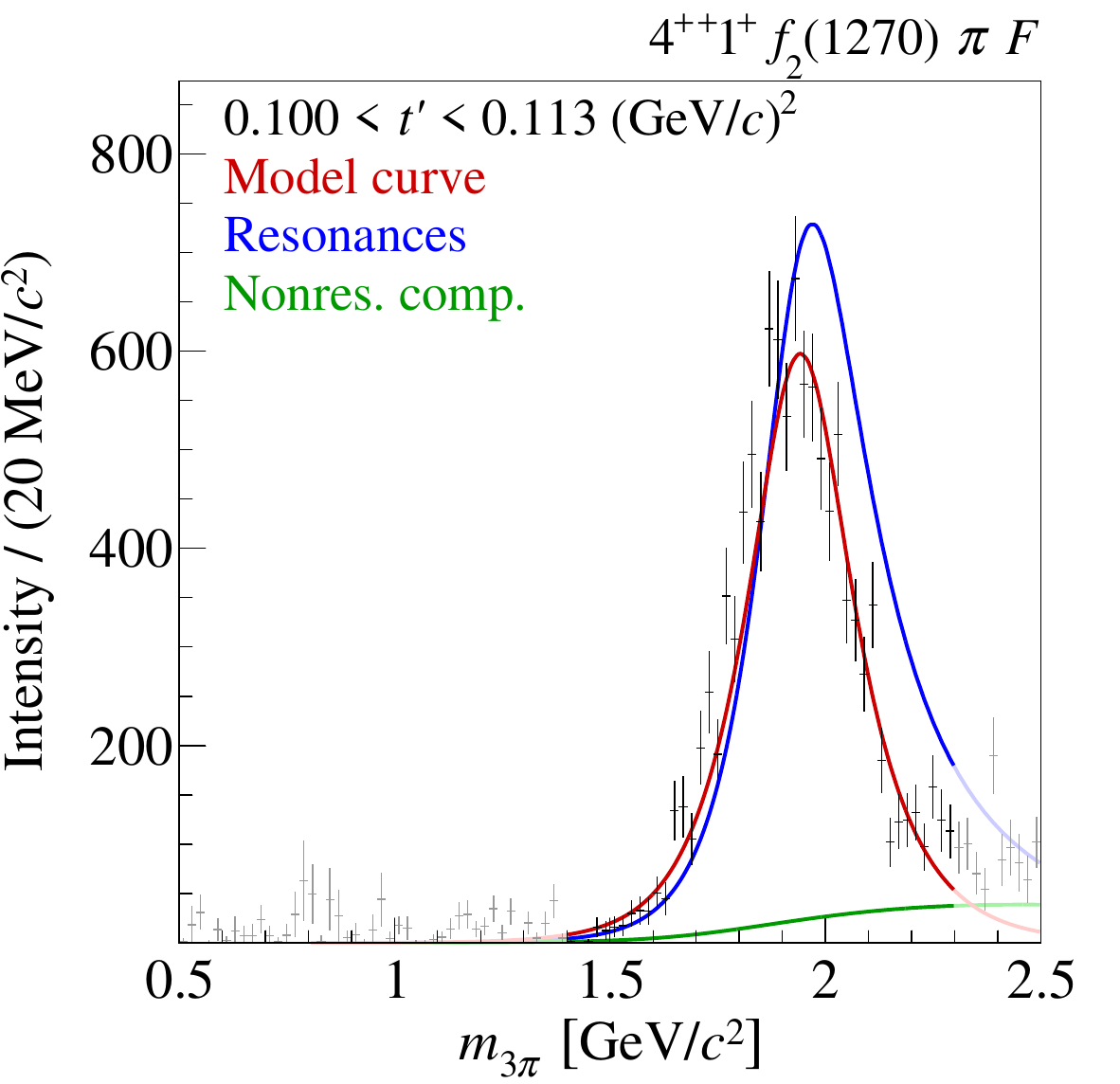}%
    \label{fig:intensity_4pp_f2_tbin1}%
  }%
  \hfill%
  \subfloat[][]{%
    \includegraphics[width=\threePlotWidth]{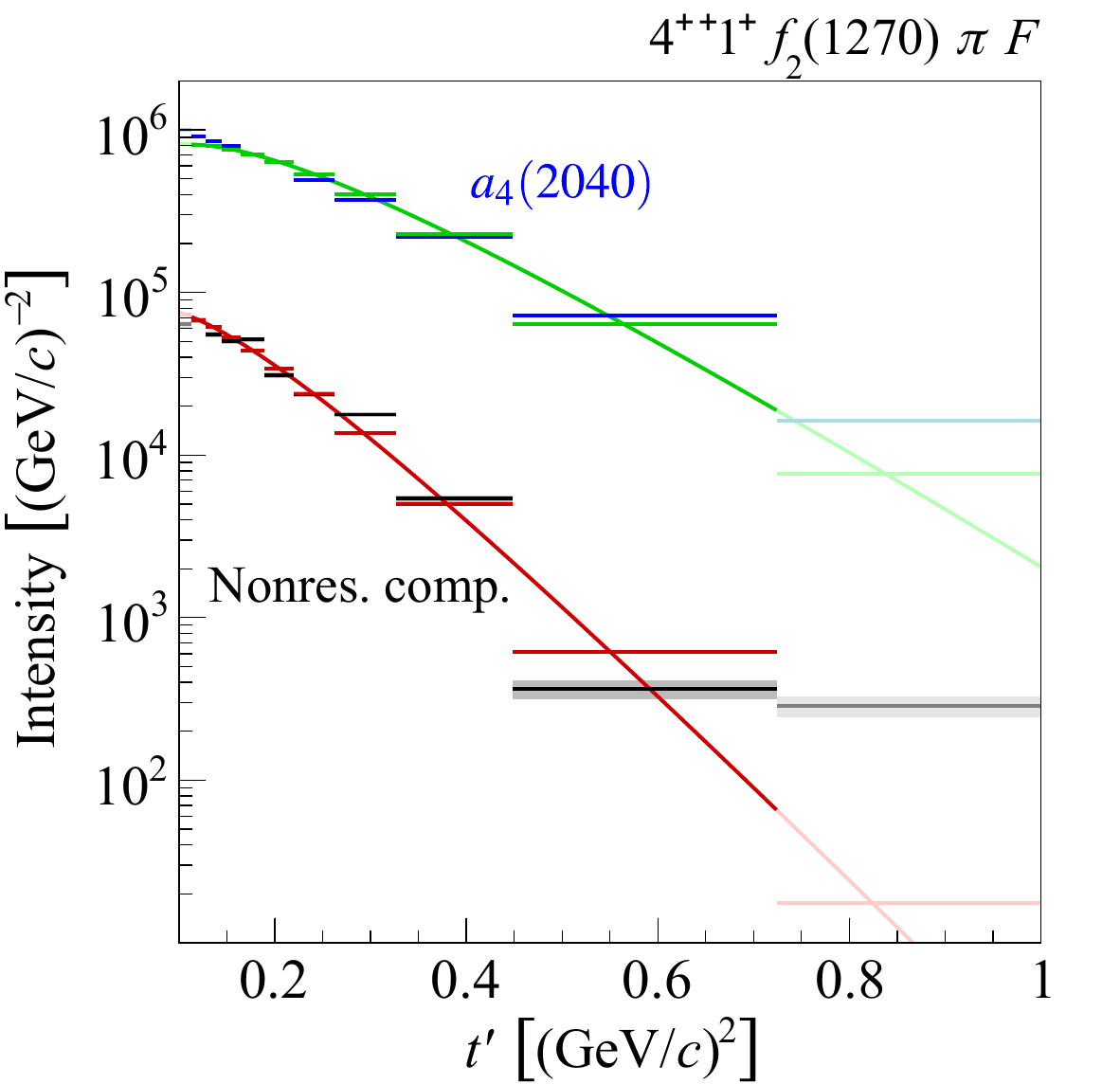}%
    \label{fig:a4_2040_t_spectrum}
  }%
  \caption{Intensity distributions of
    \subfloatLabel{fig:intensity_4pp_rho_tbin1}~the
    \wave{4}{++}{1}{+}{\Pprho}{G} and
    \subfloatLabel{fig:intensity_4pp_f2_tbin1}~the
    \wave{4}{++}{1}{+}{\PfTwo}{F} waves in the lowest $t'$~bin of the
    \threePi proton-target data~\cite{Akhunzyanov:2018lqa}.  The
    curves represent the result of the resonance-model fit.  The model
    and the wave components are represented as in
    \cref{fig:intensity_phase_0mp} except that the blue curve
    represents the \PaFour.
    \subfloatLabel{fig:a4_2040_t_spectrum}~Similar to
    \cref{fig:tspectrum_0mp_f0}, but showing the $t'$~spectrum of the
    \PaFour (blue lines and light blue boxes) and the non-resonant
    component (black lines and gray boxes) in the
    \wave{4}{++}{1}{+}{\PfTwo}{F} wave.}
  \label{fig:4pp}
\end{figure}

\begin{figure}[tbp]
  \centering
  \subfloat[]{%
    \includegraphics[width=0.5\textwidth]{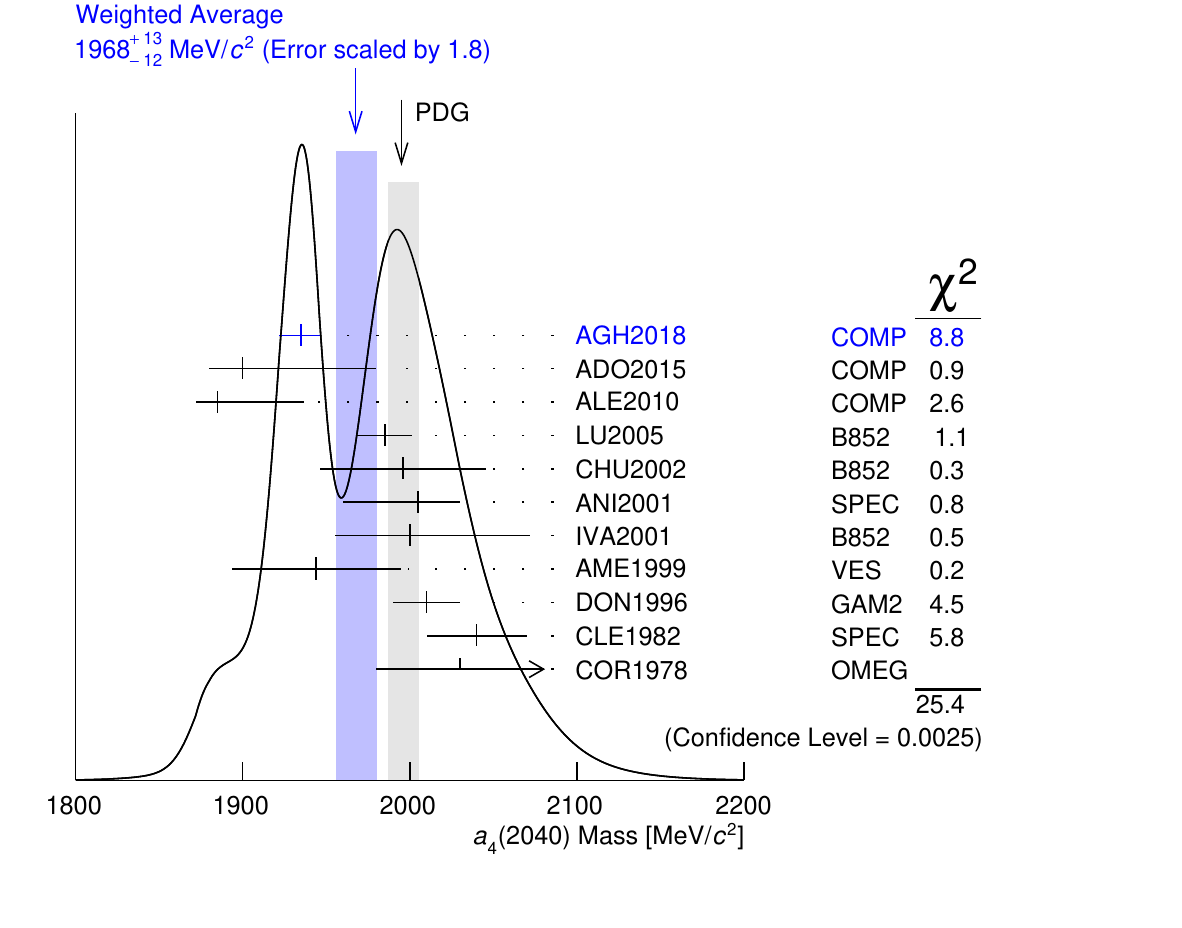}%
    \label{fig:ideogram_a4_2040_mass}%
  }%
  \subfloat[]{%
    \includegraphics[width=0.5\textwidth]{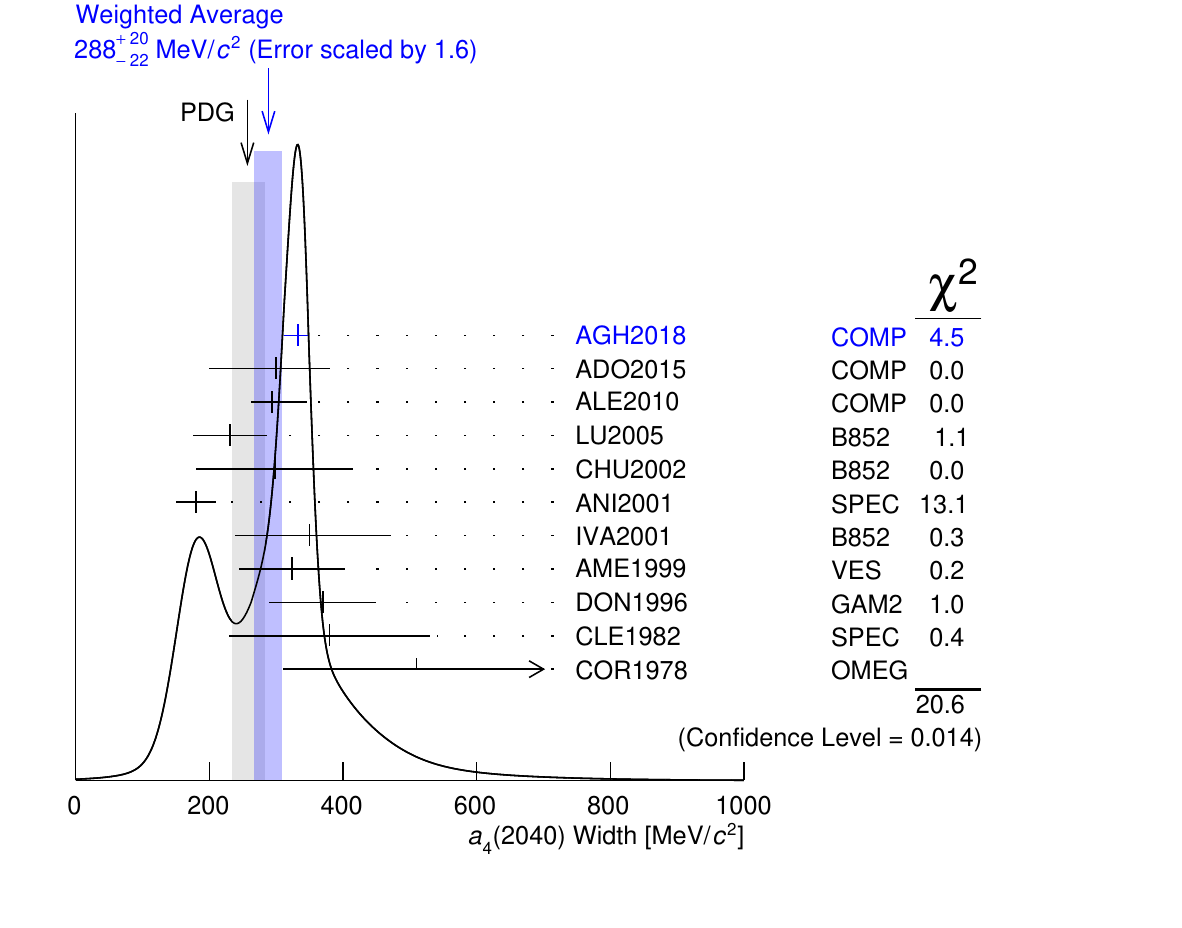}%
    \label{fig:ideogram_a4_2040_width}%
  }%
  \caption{Ideograms similar to the ones in
    \cref{fig:ideogram_pi_1800} but for
    \subfloatLabel{fig:ideogram_a4_2040_mass}~the mass and
    \subfloatLabel{fig:ideogram_a4_2040_width}~the width of the
    \PaFour.  The Breit--Wigner parameters obtained from fits of the
    COMPASS \threePi proton-target data
    (AGH2018,~\cite{Akhunzyanov:2018lqa}), the \etaOrPrPim data
    (ADO2015,~\cite{Adolph:2014rpp}), and the \threePi lead-target
    data (ALE2010,~\cite{Alekseev:2009aa}) are compared to previous
    measurements~\cite{Tanabashi:2018zz}.}
  \label{fig:ideogram_a4_2040}
\end{figure}

In the \threePi proton-target data, we observe clear signals of the
\PaFour in the \wave{4}{++}{1}{+}{\Pprho}{G} and
\wave{4}{++}{1}{+}{\PfTwo}{F} waves, which are well described by the
resonance model (see
\cref{fig:intensity_4pp_rho_tbin1,fig:intensity_4pp_f2_tbin1}).
\Cref{fig:a4_2040_t_spectrum} shows the $t'$~spectrum of the \PaFour
in the $\PfTwo \pi$ wave.  The simple exponential model in
\cref{eq:t_spectrum_model} is in fair agreement with the data and
yields a slope parameter value of \SIaerr{9.2}{0.8}{0.5}{\perGeVcsq},
which is larger than for most of the other resonances in the fit.  Due
to the constraint in \cref{eq:branching_amp}, the $t'$~spectrum of the
\PaFour in the $\Pprho \pi$ wave is practically identical.  We obtain
Breit--Wigner parameters of
$m_{\PaFour} = \SIaerr{1935}{11}{13}{\MeVcc}$ and
$\Gamma_{\PaFour} =
\SIaerr{333}{16}{21}{\MeVcc}$~\cite{Akhunzyanov:2018lqa}.  This is the
most precise measurement of the \PaFour parameters so far.  As shown
in \cref{fig:ideogram_a4_2040}, this result is in agreement with the
Breit--Wigner parameters
$m_{\PaFour} = \SIsaerrs{1885}{11}{50}{2}{\MeVcc}$ and
$\Gamma_{\PaFour} = \SIsaerrs{294}{25}{46}{19}{\MeVcc}$ from our
6-wave resonance-model fit of the \threePi lead-target data, where
only the $\Pprho \pi$ wave was included in the fit.

\begin{figure}[tbp]
  \centering
  \hfill%
  \subfloat[]{%
    \includegraphics[width=\twoPlotWidth]{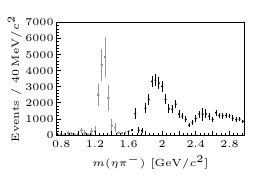}%
    \label{fig:int_eta_g_wave}%
  }%
  \hfill%
  \subfloat[]{%
    \includegraphics[width=\twoPlotWidth]{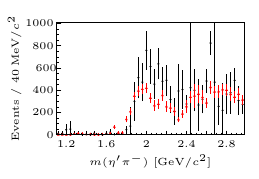}%
    \label{fig:int_eta_etaprime_g_wave}%
  }%
  \hfill\null%
  \\
  \null\hfill%
  \subfloat[]{%
    \includegraphics[width=\twoPlotWidth]{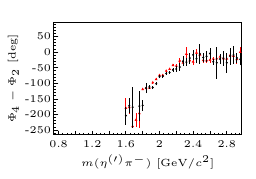}%
    \label{fig:phase_eta_g_d_wave}%
  }%
  \hfill\null%
  \caption{Intensities of the $G$~wave with $\Mrefl = 1^+$
    \subfloatLabel{fig:int_eta_g_wave}~in the \etaPim and
    \subfloatLabel{fig:int_eta_etaprime_g_wave}~in the \etaPrPim
    COMPASS data (black dots)~\cite{Adolph:2014rpp}.  The gray data
    points for $m_{\etaPi} \leq \SI{1.6}{\GeVcc}$
    in~\subfloatLabel{fig:int_eta_g_wave} indicate leakage from the
    \etaPim $D$~wave with $\Mrefl = 1^+$ (\confer\
    \cref{fig:int_eta_d_wave}).  The red triangles
    in~\subfloatLabel{fig:int_eta_etaprime_g_wave} show the \etaPim
    intensity from~\subfloatLabel{fig:int_eta_g_wave} scaled by the
    kinematic factor in \cref{eq:etaprime_eta_r}.
    \subfloatLabel{fig:phase_eta_g_d_wave}~Phases of the $G$~wave
    relative to the $\Mrefl = 1^+$ $D$~wave in \etaPim (red triangles)
    and \etaPrPim (black dots).}
  \label{fig:int_phase_eta_etaprime_g_wave}
\end{figure}

In the \etaPim data, the \PaFour appears in the intensity distribution
of the $G$~wave with $\Mrefl = 1^+$ as a broad peak below
\SI{2}{\GeVcc} (see \cref{fig:int_eta_g_wave}).  In addition, the data
exhibit a broad high-mass shoulder.  The $G$-wave intensity
distribution is contaminated by leakage of the order of
\SI{3}{\percent} from the dominant \PaTwo in the $D$~wave (gray data
points in \cref{fig:int_eta_g_wave}).  This contamination is, however,
well separated from the \PaFour region.  The intensity distribution of
the \etaPrPim $G$~wave with $\Mrefl = 1^+$ is qualitatively similar to
that of the \etaPim $G$~wave (see black dots in
\cref{fig:int_eta_etaprime_g_wave}).  It exhibits a peak consistent
with the \PaFour that is, however, a bit broader and shifted toward
higher masses.  The high-mass shoulder is more pronounced.
\Cref{fig:int_eta_etaprime_g_wave} shows that after scaling of the
\etaPim intensity distribution with the kinematic factor in
\cref{eq:etaprime_eta_r} (red triangles), the two $G$-wave intensity
distributions become nearly identical.  In particular the leakage from
the \PaTwo is completely suppressed by the angular-momentum barrier
factor in \cref{eq:etaprime_eta_r}.  Also the phases between the $D$-
and $G$-wave amplitudes exhibit a striking similarity in \etaPim and
\etaPrPim, as shown in \cref{fig:phase_eta_g_d_wave}.  For both final
states, the \PaFour resonance produces a slow rise of the phase around
\SI{2.0}{\GeVcc}.  A combined resonance-model fit of the \etaPim and
\etaPrPim data yields Breit--Wigner parameters of
$m_{\PaFour} = \SIaerr{1900}{80}{20}{\MeVcc}$ and
$\Gamma_{\PaFour} =
\SIaerr{300}{80}{100}{\MeVcc}$~\cite{Adolph:2014rpp}, which are
consistent with the parameters obtained from the COMPASS \threePi
data.

\Cref{fig:ideogram_a4_2040} shows that, compared to the PDG average,
the COMPASS data prefer a lighter and wider \PaFour.  This is
particularly apparent for the parameter values from the \threePi
proton-target data, where we find the \PaFour mass to be
\SI{60}{\MeVcc} smaller and the width \SI{76}{\MeVcc} larger.  The
latter is in tension only with the extremely low width value of
\SI{180(30)}{\MeVcc} reported by the authors of
\refCite{Anisovich:2001pn}, our mass values, however, are lower than
any of the other mass values included in the PDG average.

Based on the Breit--Wigner resonance-model fit to the \etaOrPrPim data,
we measure the branching-fraction ratio of the \PaFour decays into
\etaPi and \etaPrPi~\cite{Adolph:2014rpp}:
\begin{equation}
  \label{eq:branch_fract_ratio_a4_eta_etaprime_pi}
  B_{\etaPrPi, \etaPi}^{\PaFour*}
  = \frac{\text{BF}\big[ \PaFour \to \etaPrPi \big]}%
         {\text{BF}\big[ \PaFour \to \etaPi \big]\hfill}
  = \num{0.23(7)}\eqPunctSpacing.
\end{equation}
This constitutes the first measurement of this quantity.  From the
\threePi proton-target data, we obtain a branching-fraction ratio for
the decays of the \PaFour into the $\Pprho \pi$~$G$ and $\PfTwo \pi$
$F$~modes with $M = 1$:
\begin{equation}
  \label{eq:branch_fract_ratio_a4_corr}
    B_{\Pprho* \pi G, \PfTwo* \pi F}^{\PaFour*, \text{corr}}
    = \frac{\text{BF}\big[ \PaFour^- \to \Pprho \pi \big]\hfill}%
           {\text{BF}\big[ \PaFour^- \to \PfTwo \pi \big]}
    = \numaerr{2.9}{0.6}{0.4}\eqPunctSpacing.
\end{equation}
This number takes into account the unobserved decays into \threePiN
via isospin symmetry, the branching fraction of the \PfTwo to $2\pi$
and the effect of the different Bose symmetrizations in the \threePi
and \threePiN final states (see Section~VI.B.2 in
\refCite{Akhunzyanov:2018lqa} for details).  Our value is in good
agreement with the value of~3.3 predicted by the ${}^3P_0$ decay
model~\cite{Barnes:1996ff}.  This model describes the decay of \qqbar
states into $(q\bar{q}')(q'\bar{q})$ meson pairs via production of a
$q'\bar{q}'$ pair with vacuum quantum numbers, \ie $\JPC = 0^{++}$.
\clearpage{}%
\clearpage{}%
\section{Primakoff Production of \threePi}
\label{sec:results_3pic_primakoff}

As was discussed in \cref{sec:exp.prod_reactions}, high-energy
scattering reactions such as
\begin{equation}
  \label{eq:primakoff_reaction}
  \pi + A \to 3\pi + A\eqPunctSpacing,
\end{equation}
where a beam pion interacts coherently with a target nucleus with mass
number~$A$ and dissociates into $3\pi$, are dominated by diffraction,
\ie Pomeron exchange, at low reduced four-momentum transfer
squared~$t'$.  The differential cross section
$\dif{\sigma} / \dif{t'}$ follows approximately an exponential
distribution $e^{-b\, t'}$ (see \cref{eq:dsigma_dt.regge.2}), which
corresponds to a dominant contribution from spin projection $M = 0$.
For heavy nuclei, the slope parameter~$b$ is related to the
geometrical size of the nucleus, \ie $b \propto R^2$, where
$R \propto A^{1/3}$ is the nuclear radius.  The cross section
integrated over~$t'$ from~0 to the first diffractive minimum is
approximately proportional to~$A^{2/3}$, \ie to the surface of the
nucleus.  The features of diffractive-dissociation reactions described
above were, for example, studied by the FNAL E272 experiment with a
\SI{200}{\GeVc} pion beam impinging on~C, Al, Cu, and
Pb~targets~\cite{Zielinski:1983ty}.

\begin{figure}[bp]
  \centering
  \includegraphics[width=\threePlotWidth]{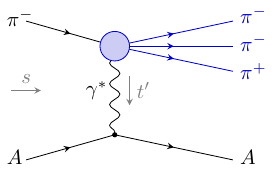}%
  \caption{Diagram for pion-induced $3\pi$ production on a
      target nucleus via photon exchange.}
  \label{fig:primakoff_feyn}
\end{figure}

In addition to diffractive production, \ie Pomeron exchange between
the incoming beam particle and the target nucleus, also photon
exchange is possible (see \cref{fig:primakoff_feyn} in
\cref{sec:exp.prod_reactions}).  Such a reaction is called Coulomb or
Primakoff production.  Primakoff production is described theoretically
using the Weizs\"{a}cker--Williams equivalent-photon approximation, which
assumes that ultra-relativistic, \ie quasi-stable, beam pions scatter
off quasi-real photons of the electromagnetic field of the heavy
target nucleus~\cite{vonWeizsacker:1934nji,Williams:1934ad}.  The
equivalent-photon approximation relates the cross
section~$\sigma_{\pi A}$ for the experimentally observed
process~\eqref{eq:primakoff_reaction} to the cross
section~$\sigma_{\pi \gamma}$ for the process
$\pi + \gamma \to
3\pi$~\cite{Pomeranchuk:1961,Halprin:1966zz,Zielinski:1986mg}:
\begin{equation}
  \label{eq:primakoff_xsec}
  \frac{\dif{\sigma_{\pi A}}}{\dif{\mThreePi^2}\, \dif{t'}\, \dif{\Phi_3}}
  = \frac{\alpha_\text{em}\, Z^2}{\pi}\, \frac{1}{\mThreePi^2 - m_\pi^2}\,
  \frac{t'}{\big( t' + \tMin \big)^2}\, \big| F_\text{eff}(t', \tMin) \big|^2\,
  \frac{\dif{\sigma_{\pi \gamma}}(\mThreePi)}{\dif{\Phi_3}}\eqPunctSpacing.
\end{equation}
Here, \mThreePi~is the invariant mass of the $3\pi$~system, which is
equal to the center-of-momentum energy of the pion--photon system,
$\dif{\Phi_3}$ is the differential phase-space element of the $3\pi$
final state as defined in \cref{eq:dPhi_N}, $t'$~and~\tMin are defined
by \cref{eq:tprime,eq:t.minmax}.  The fine-structure constant is
denoted by~$\alpha_\text{em}$, $Z$~is the charge number of the
nucleus, and $F_\text{eff}(t', \tMin)$ is the electromagnetic nuclear
form factor specific to $\pi^-$--$A$ scattering.  This form factor
accounts not only for the charge distribution in the nucleus but also
for the distortion of the pion wave function in the Coulomb field and
depends on~$t'$
and~\tMin~\cite{Faldt:2008yw,Faldt:2010ss,Adolph:2014mup}.

The cross section for Primakoff processes grows with the charge number
as~$Z^2$, while the diffractive cross section grows with the mass
number as~$A^{2/3}$.  As a consequence, Primakoff processes are
enhanced on heavy target nuclei.  The different $A$~and $Z$~dependence
of the Primakoff and the diffractive cross sections also provides a
handle to separate the two mechanisms.

The FNAL E272 experiment first observed Primakoff production of $3\pi$
final states off heavy nuclear targets~\cite{Zielinski:1983ty}.  The
signature was a sharp peak in the $t'$~spectrum toward $t' = 0$ on top
of an exponential distribution from coherent diffraction events.
Later, a partial-wave analysis was performed to study the Primakoff
production of \PaTwo and \PaOne~\cite{Zielinski:1984mt}.  A much
larger data sample of $3\pi$ events was acquired by the FNAL E781
(SELEX) experiment using a \SI{600}{\GeVc} pion beam on~Cu and
Pb~targets~\cite{Molchanov:2001qk}.  They measured the radiative width
$\Gamma[\PaTwo \to \pi \gamma]$.  To separate Primakoff production of
\PaTwo from diffractive production, a statistical subtraction method
was applied.

\begin{figure}[tbp]
  \centering
  \includegraphics[width=\twoPlotWidth]{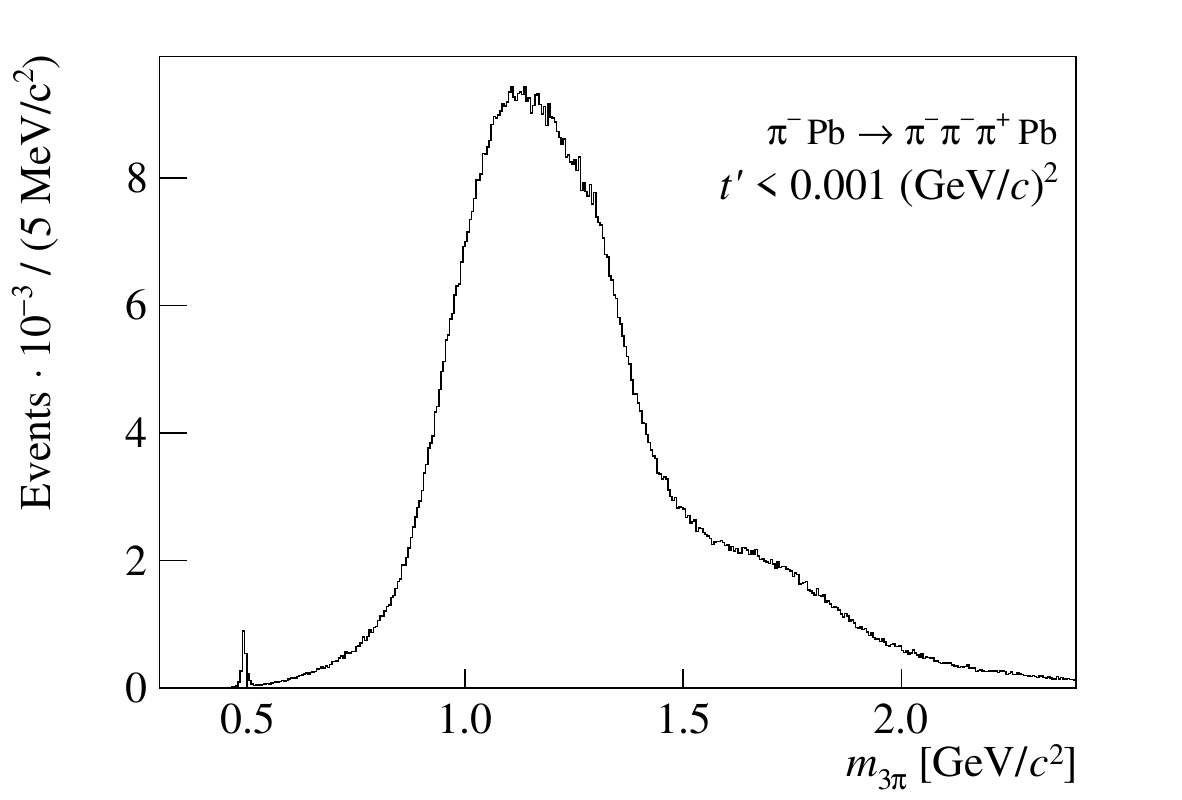}
  \caption{Invariant mass spectrum (not acceptance corrected) of
    \threePi events produced in the reaction
    $\pi^- + \text{Pb} \to \threePi + \text{Pb}$ with
    $t' < \SI{e-3}{\GeVcsq}$~\cite{Adolph:2014mup}.  The sharp peak
    below \SI{0.5}{\GeVcc} originates from in-flight decays of beam
    kaons into \threePi.}
  \label{fig:mass_3pic_prim}
\end{figure}

The COMPASS experiment has collected a data sample of about
\num{e6}~events for the reaction
$\pi^- + \text{Pb} \to \threePi + \text{Pb}$ in the kinematic region
$t' < \SI{e-3}{\GeVcsq}$, \ie in the Primakoff
region~\cite{Adolph:2014mup}.  The \threePi invariant mass spectrum of
these events is shown in \cref{fig:mass_3pic_prim}, where the main
contributions from diffractive production of the \PaOne and \PpiTwo
resonances are visible.

\subsection{Extraction of Primakoff Process via $t'$~Spectrum}
\label{sec:primakoff_t_spectrum}

The $t'$~spectrum of diffractively produced $3\pi$ states with spin
projection~$M$ is given by (\confer\
\cref{eq:dsigma_dt.regge.2,eq:t_spectrum_model})
\begin{equation}
  \label{eq:tspectrum_diff}
  \frac{\dif{N_\text{diff}}}{\dif{t'}}
  \propto (t')^{\abs{M}}\, e^{-b_\text{diff}(\mThreePi)\, t'}\eqPunctSpacing.
\end{equation}
Due to the $(t')^{\abs{M}}$ factor, diffractive production of states
with $\abs{M} > 0$ becomes negligible at sufficiently low~$t'$ so that
only states with $M = 0$ remain.  This is
in contrast to quasi-real photoproduction, where only states with
$\abs{M} = 1$ are produced because for quasi-real photons the state
with helicity~0 is suppressed due to the very small virtuality (see
\eg\ \refCite{Jurisic:1970ra}).

\begin{figure}
  \centering
  \hfill%
  \subfloat[][]{%
    \includegraphics[width=\twoPlotWidth]{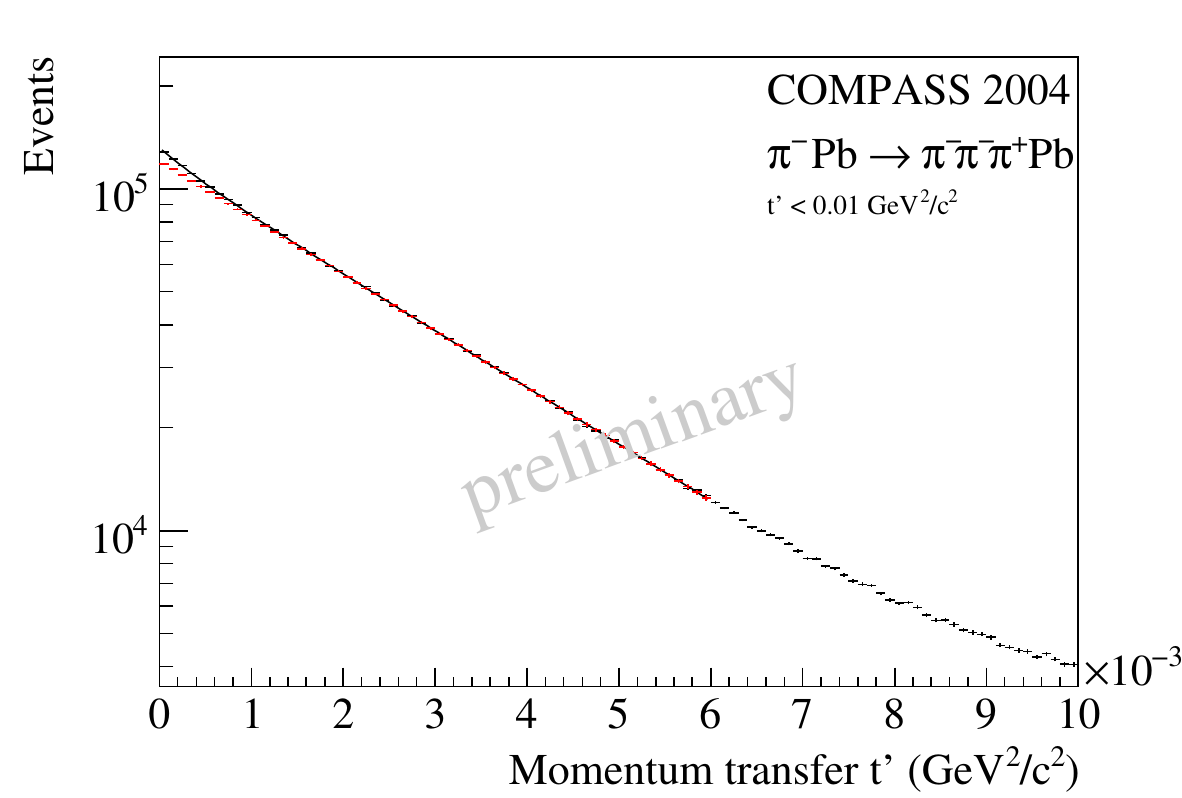}%
    \label{fig:stat_sub_tspect}%
  }%
  \hfill%
  \subfloat[][]{%
    \includegraphics[width=\twoPlotWidth]{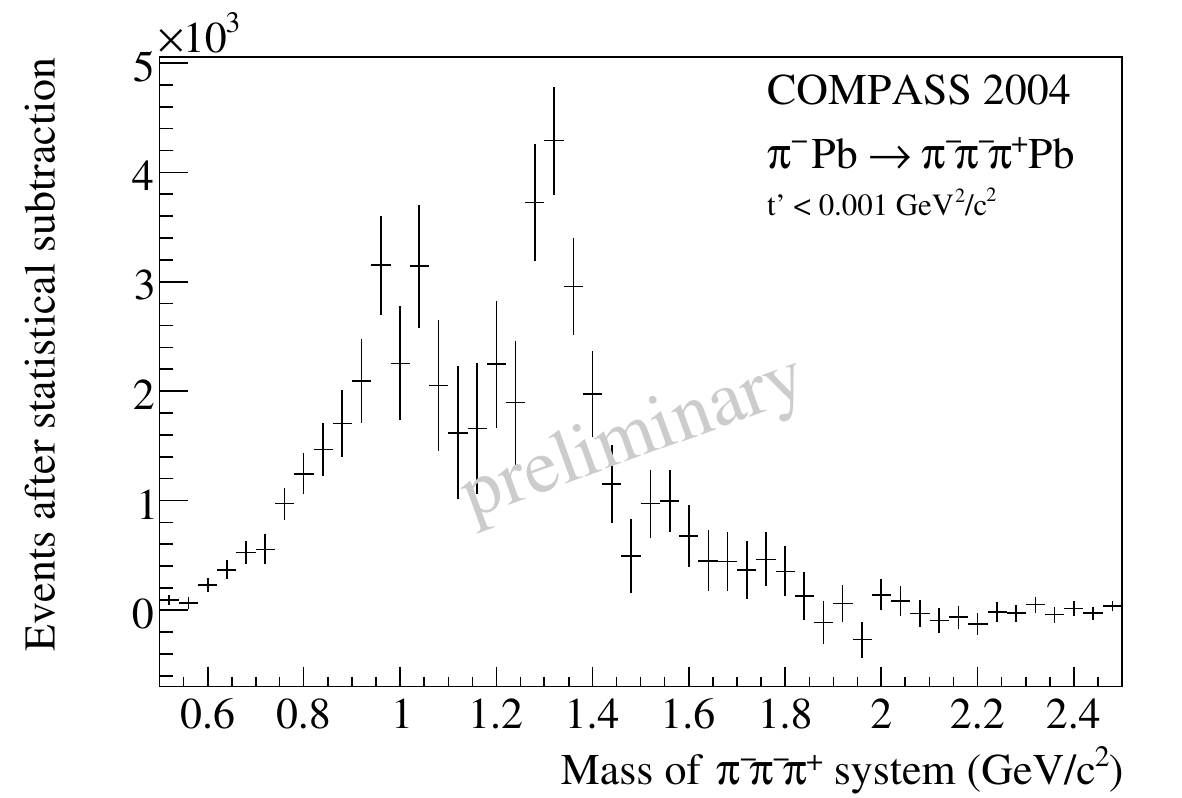}%
    \label{fig:stat_sub_mass_spect}%
  }%
  \hfill\null%
  \caption{\subfloatLabel{fig:stat_sub_tspect}~Measured $t'$~spectrum
    for the reaction $\pi^- + \text{Pb} \to \threePi + \text{Pb}$ in
    the analyzed range \SIvalRange{0.42}{\mThreePi}{2.50}{\GeVcc}
    (black points)~\cite{Grabmuller:2012oja}.  The black curve
    represents the result of a fit of \cref{eq:tspectrum_total}.  The
    red points represent the diffractive component in the fit model.
    \subfloatLabel{fig:stat_sub_mass_spect}~\mThreePi distribution of
    the Primakoff component (not acceptance corrected) obtained by
    fitting the $t'$~spectrum in each \SI{40}{\MeVcc} wide
    \mThreePi~bin with
    \cref{eq:tspectrum_total,eq:tspectrum_mc_prim_model} (see
    text)~\cite{Grabmuller:2011zz,Friedrich:2010zz}.\protect\footnotemark}
  \label{fig:stat_sub}
\end{figure}

\Cref{fig:stat_sub_tspect} shows the $t'$~spectrum in the analyzed
range \SIvalRange{0.42}{\mThreePi}{2.50}{\GeVcc} and provides first
insights into the Primakoff production of \threePi states.  Two
production mechanisms, diffraction and Primakoff production, are
assumed to be present at low~$t'$.  In the region
$t' < \SI{6e-3}{\GeVcsq}$, the $t'$~spectrum can be fitted well by a
sum of two exponentials, \ie
\begin{equation}
  \label{eq:tspectrum_total}
  \frac{\dif{N_{\pi A}}}{\dif{t'}}
  = \frac{\dif{N_\text{diff}}}{\dif{t'}} + \frac{\dif{N_\text{Prim}}}{\dif{t'}} \\
  \quad\text{with}\quad
  \frac{\dif{N_\text{diff}}}{\dif{t'}}
  \propto e^{-b_\text{diff}(\mThreePi)\, t'}
  \quad\text{and}\quad
  \frac{\dif{N_\text{Prim}}}{\dif{t'}}
  \propto e^{-b_\text{Prim}(\mThreePi)\, t'}\eqPunctSpacing,
\end{equation}
which is shown as the black curve in \cref{fig:stat_sub_tspect}.  The
shallower slope of $b_\text{diff} \approx \SI{400}{\perGeVcsq}$
corresponds to coherent diffractive production (red points in
\cref{fig:stat_sub_tspect}), while the much steeper slope of
$b_\text{Prim} \approx \SI{1500}{\perGeVcsq}$ corresponds to Primakoff
production.

\footnotetext{\Cref{fig:stat_sub_tspect,fig:stat_sub_mass_spect} are
  unpublished auxiliary plots from the analysis presented in
  \refCite{Adolph:2014mup}.}

According to \cref{eq:primakoff_xsec} the theoretical $t'$~spectrum
for Primakoff production peaks at \tMin and vanishes at $t' = 0$.  The
measured $t'$~spectrum is, however, modified significantly by
experimental resolution. To obtain parameterizations of the smeared
Primakoff $t'$~spectrum, we performed dedicated Monte Carlo (MC)
simulations.  In the Primakoff region $t' < \SI{e-3}{\GeVcsq}$, the
reconstructed Primakoff MC events approximately follow an exponential
distribution, \ie
$\dif{N_\text{Prim, MC}} / \dif{t'} \propto e^{-b_\text{Prim,
    MC}(\mThreePi)\, t'}$.  This indicates that the dip at $t' = 0$ is
fully smeared out due to resolution effects.  The slope parameter
changes with~\mThreePi from
$b_\text{Prim, MC}(\mThreePi = \SI{0.5}{\GeVcc}) \approx
\SI{1500}{\perGeVcsq}$ to
$b_\text{Prim, MC}(\mThreePi = \SI{2.5}{\GeVcc}) \approx
\SI{700}{\perGeVcsq}$.  The flattening of the reconstructed MC
$t'$~spectrum with increasing~\mThreePi is mainly due to the increase
of~\tMin with~\mThreePi.  The \mThreePi~dependence of the Primakoff
slope parameter is parameterized by
\begin{equation}
  \label{eq:bprim_polyn}
  b_\text{Prim, MC}(\mThreePi)
  = \SI{2108.0}{\perGeVcsq} - \SI{812.1}{GeV^{-3}/\clight^{-4}}\, \mThreePi
  + \SI{46.4}{GeV^{-4}/\clight^{-6}}\, \mThreePi^2\eqPunctSpacing.
\end{equation}

In order to separate the small Primakoff contribution from the
dominant diffractive background, a fitting range of
$t' < \SI{6e-3}{\GeVcsq}$ is needed, which extends beyond the
Primakoff region.  It turns out that in this extended $t'$~range the
reconstructed Primakoff MC $t'$~spectrum cannot be well described
anymore by an exponential distribution.  Instead, the empirical
formula
\begin{equation}
  \label{eq:tspectrum_mc_prim_model}
  \frac{\dif{N_\text{Prim, MC}}}{\dif{t'}}
  \propto \frac{A(\mThreePi)}{\big[ t' + P(\mThreePi) \big]^{3.9}}
\end{equation}
is used, where $A(\mThreePi)$ is a normalization parameter and
$P(\mThreePi)$ is a fifth-order polynomial.

We fit a modified version of \cref{eq:tspectrum_total} to the measured
$t'$~spectra in \SI{40}{\MeVcc} wide \mThreePi~bins, where we replace
the exponential distribution for the Primakoff component with
\cref{eq:tspectrum_mc_prim_model} with the parameters fixed to the
values from the Primakoff MC.  These fits of the measured data yield
the contributions of diffractive and Primakoff production and the
slope parameter $b_\text{diff}$ of the diffractive component as
functions of~\mThreePi.  A significant variation of the diffractive
slope parameter is observed.  It falls monotonically from
$b_\text{diff}(\mThreePi = \SI{0.5}{\GeVcc}) \approx
\SI{400}{\perGeVcsq}$ to
$b_\text{diff}(\mThreePi = \SI{2.5}{\GeVcc}) \approx
\SI{300}{\perGeVcsq}$.  The explicit \mThreePi~dependence is
parameterized by
\begin{equation}
  \label{eq:bdiff_polyn}
  b_\text{diff}(\mThreePi)
  = \SI{456}{\perGeVcsq} - \SI{72}{GeV^{-3}/\clight^{-4}}\, \mThreePi
  + \SI{6}{GeV^{-4}/\clight^{-6}}\, \mThreePi^2\eqPunctSpacing.
\end{equation}
The extracted Primakoff intensity integrated over
$t' < \SI{e-3}{\GeVcsq}$ is shown as a function of~\mThreePi in
\cref{fig:stat_sub_mass_spect}.  The intensity distribution shows a
clear peak at \SI{1.3}{\GeVcc}, stemming from Primakoff-produced
\PaTwo.  We do not observe clear signals of \PaOne or \PpiTwo.  The
intensity from $3\pi$ threshold to about \SI{1.1}{\GeVcc} is dominated
by non-resonant low-energy processes.  These processes are calculable
in the framework of Chiral Perturbation Theory (see
\cref{sec:theory.chiPT}) and will be discussed in
\cref{sec:pwa_chpt_amp}.

\subsection{Model for Partial-Wave Decomposition in the Primakoff Kinematical Region}
\label{sec:pwa_prim_specifics}

Compared to the statistical subtraction method based on the
$t'$~spectrum (see \cref{sec:primakoff_t_spectrum}), partial-wave
analysis techniques are a much more powerful tool\ to separate
Primakoff and diffractive contributions because they exploit the full
information contained in the multi-dimensional distribution of the
$3\pi$ phase-space variables.  A number of special features of
photoproduction and of the analyzed Primakoff region
$t' < \SI{e-3}{\GeVcsq}$ require modifications of the PWA model that
was derived in
\cref{sec:pwa.analysis_model,sec:pwa_cells,sec:3pi_model:pwa}.  Both
Pomeron and photon have positive naturality (see
\cref{sec:pheno.qm.quantum-numbers}).  A PWA of diffractive \threePi
production on a lead target in the range
\SIvalRange{e-3}{t'}{e-2}{\GeVcsq}~\cite{Weitzel:2008qia}, showed that
contributions from unnatural exchanges have negligibly small
contributions, which is consistent with the expected Pomeron dominance
at high energies (see \cref{sec:exp.prod_reactions}).  We therefore
include only waves with positive reflectivity (see
\cref{sec:pwa_cells.reflectivity}) in the PWA model.  As was discussed
in \cref{sec:primakoff_t_spectrum}, in the analyzed $t'$~range the
spin-projection quantum number of diffractively produced states is
limited to $M = 0$, while states produced by quasi-real photon
exchange have $M = 1$.\footnote{Note that we use the reflectivity
  basis, where $M \geq 0$ (see \cref{sec:pwa_cells.reflectivity}).}
For both production processes, the scattering is coherent, \ie the
beam pion interacts with the target nucleus as a whole.  Therefore,
the total amplitude for reaction~\eqref{eq:primakoff_reaction} is the
coherent sum of the amplitudes for the two production processes and
hence a rank-1 spin-density matrix (see \cref{sec:pwa_cells.rank}) is
used.  Based on the model for the intensity distribution in
\cref{eq:intensity_model_final}, our ansatz for the Primakoff PWA
model is\footnote{We omit the sums over reflectivity and rank index
  because $\refl = +1$ and $N_r = 1$.}
\begin{equation}
  \label{eq:primakoff_coh_formula}
  \mathcal{I}(\tau_3; \mThreePi, t')
  = \bigg| \sum_i^{N_\text{waves}} \mathcal{T}_i(\mThreePi, t') \, \Psi_i(\tau_3; \mThreePi) \bigg|^2
  + |\mathcal{T}_\text{flat}(\mThreePi, t')|^2\eqPunctSpacing.
\end{equation}

At low~$t'$, detector resolution effects play an important role.  They
significantly affect the measured $t'$~spectra (see
\cref{sec:primakoff_t_spectrum}) and also the production plane defined
by the incoming pion and the intermediate state~$X$ is only poorly
reconstructed because of the very small scattering angle of~$X$.  As
was explained in \cref{sec:pwa_cells.likelihood_fit}, we construct
from the model in \cref{eq:primakoff_coh_formula} the probability
density function in \cref{eq:data_pdf} that describes the
$\tau_3$~distribution for given values of~\mThreePi and~$t'$.  This
function takes into account the acceptance of the detector but
neglects smearing of the phase-space variables~$\tau_3$ due to
detector resolution.  Taking resolution effects into account would
require to convolve the theoretical model with a smearing function as
in \cref{eq:data_pdf_smeared}, where this function describes the
experimental resolution in the whole phase space.  Such a convolution
approach is computationally very expensive.  We thus use a different
approach, where we modify the PWA model such that we take into account
resolution effects in an effective way.

Due to the significant smearing of~$t'$ and the limited size of the
data sample, we do not perform the PWA in narrow~\mThreePi and
$t'$~bins as was described in \cref{sec:pwa_cells}.  Instead, we model
the explicit $t'$~dependence of the transition amplitudes
$\mathcal{T}_i(\mThreePi, t')$ in \cref{eq:primakoff_coh_formula} in
the analyzed $t'$~range using the same approach as for the analysis of
\threePi diffractive production on a lead target that was discussed in
\cref{sec:3pi_model:pwa}.  Following \cref{eq:intensity_model_lead},
we factorize the~\mThreePi and the $t'$~dependence of the transition
amplitudes by introducing real-valued functions $f_i(t')$.  For
diffractively produced waves, which all have $M = 0$, the
$t'$~dependence of the transition amplitudes is parameterized by
\begin{equation}
  \label{eq:primakoff_diff_amp_tdep}
  f_\text{diff}^{M = 0}(t')
  = e^{-\frac{1}{2}\, b_\text{diff}(\mThreePi)\, t'}\eqPunctSpacing.
\end{equation}
Here, we use the \mThreePi~dependence of the slope parameter in
\cref{eq:bdiff_polyn} that was extracted from the experimental data.
For Primakoff-produced waves, which all have $M = 1$, the
theoretically expected $t'$~dependence as given by
\cref{eq:primakoff_xsec} is
\begin{equation}
  \label{eq:primakoff_prim_amp_tdep_theo}
  f_\text{Prim, theo}^{M = 1}(t')
  =  F_\text{eff}(t', \tMin)\, \frac{\sqrt{t'}}{t' + \tMin}\eqPunctSpacing.
\end{equation}
As was discussed in \cref{sec:primakoff_t_spectrum}, this
$t'$~dependence is significantly modified by resolution effects. Thus
\cref{eq:primakoff_prim_amp_tdep_theo,eq:intensity_model_lead} are used
in the PWA model in \cref{eq:primakoff_coh_formula} only for Monte
Carlo simulation.  To analyze real data, we employ the same
parameterization as in \cref{eq:tspectrum_total} for the Primakoff
component, \ie
\begin{equation}
  \label{eq:primakoff_prim_amp_tdep}
  f_\text{Prim}^{M = 1}(t')
  = e^{-\frac{1}{2}\, b_\text{Prim}(\mThreePi)\, t'}\eqPunctSpacing,
\end{equation}
where we use the \mThreePi~dependence of the slope parameter in
\cref{eq:bprim_polyn} that was extracted from simulated Primakoff
events.

Smearing of the $3\pi$ phase-space variables, in particular of the
azimuthal angle \phiGJ in the Gottfried--Jackson frame (see
\cref{sec:pwa.analysis_model.coordsys}), causes two specific features
that we observe in the PWA results: \one~contrary to expectation,
negative-reflectivity waves with $M = 1$ are significant and \two~the
amplitudes of waves with $\Mrefl =0^+$ and~$1^+$ quantum numbers are
not fully coherent.  We hence have to modify the PWA model in
\cref{eq:primakoff_coh_formula} in order to take into account these
effects.  The most economical PWA model in terms of fit parameters,
that provides a satisfactory description of the data, is obtained by
\one~including negative-reflectivity waves with $M = 1$ in the PWA
model and \two~introducing an additional real-valued parameter
$r_{M = 0, M = 1}$ that quantifies the degree of coherence between the
$\Mrefl = 0^+$ and~$1^+$ waves:
\begin{multline}
  \label{eq:primakoff_partcoh_formula}
  \mathcal{I}(\tau_3; \mThreePi, t')
  = \bigg| \sum_k^{N_\text{waves}^{M = 0}} \mathcal{T}_k^{(\refl = +1)}\, \Psi_k^{(\refl = +1)} \bigg|^2
  + \bigg| \sum_l^{N_\text{waves}^{M = 1}} \mathcal{T}_l^{(\refl = +1)}\, \Psi_l^{(\refl = +1)} \bigg|^2
  + \bigg| \sum_l^{N_\text{waves}^{M = 1}} \mathcal{T}_l^{(\refl = -1)}\, \Psi_l^{(\refl = -1)} \bigg|^2 \\
  + r_{M = 0, M = 1}\, 2 \Re\!\bigg[ \sum _k^{N_\text{waves}^{M = 0}} \mathcal{T}_k^{(\refl = +1)}\, \Psi_k^{(\refl=+1)}
  \sum_l^{N_\text{waves}^{M = 1}} \mathcal{T}_l^{(\refl = +1) \text{*}}\, \Psi_l^{(\refl = +1) \text{*}} \bigg]
  + |\mathcal{T}_\text{flat}(\mThreePi, t')|^2\eqPunctSpacing.
\end{multline}
Here, index~$k$ runs over all $M = 0$ waves, whereas index~$l$ runs
over all $M = 1$ waves.  For the latter, we use the same wave set for
$\refl = +1$ and~$-1$.  The PWA model contains 17~waves with
$J \leq 3$ and $\Mrefl = 0^+$ attributed to diffractive dissociation
and 8~waves with $J \leq 2$ and $M = 1$ attributed to Primakoff
production.  Each of the $M = 1$ waves appears with both
reflectivities.  For $\mThreePi < \SI{1.56}{\GeVcc}$, the model
contains in addition a \chiPT amplitude with $\refl = \pm 1$ (see
\cref{sec:pwa_chpt_amp}).  We also include an amplitude that takes
into account the $K^- \to \threePi$ events close to threshold.  In
total, the wave set consists of 37~waves including the flat wave (see
Tab.~4 in \refCite{Adolph:2014mup} for a complete list).  In order to
define waves, we use the same short-hand notation
\wave{J}{PC}{M}{\refl}{r}{L} that we already introduced for
diffractively produced \threePi in \cref{sec:3pi_model:pwa}.

It is important to note that the modifications in
\cref{eq:primakoff_partcoh_formula} compared to
\cref{eq:primakoff_coh_formula} do not reflect any physical feature of
the scattering process.  The modifications are applied only to
effectively take into account effects of experimental resolution.  To
demonstrate this, we performed a dedicated Monte Carlo study, which is
based on the result from a fit of the PWA model in
\cref{eq:primakoff_partcoh_formula} to the measured
data.\footnote{Note that in PWA fits to measured data, we use the
  smeared $t'$~dependence for the $M = 1$ waves, \ie\
  \cref{eq:primakoff_prim_amp_tdep} with \cref{eq:bprim_polyn}.}  From
the set of transition amplitudes that we obtain in each \mThreePi~bin,
we then construct a PWA model according to
\cref{eq:primakoff_coh_formula} that we use to generate MC
data.\footnote{Note that for generating MC data, we use the unsmeared
  $t'$~dependence from \cref{eq:primakoff_prim_amp_tdep_theo} for the
  $\Mrefl = 1^+$ waves.}  In order to construct this MC PWA model, we
scale each $\Mrefl = 1^+$ transition amplitude in
\cref{eq:primakoff_coh_formula} such that the corresponding intensity
is equal to the sum of the $\Mrefl = 1^+$ and~$1^-$ intensities of
this wave in the PWA model in \cref{eq:primakoff_partcoh_formula}.
The full Monte Carlo and reconstruction chain, which includes a
detailed simulation of the detector resolution effects, is applied to
the simulated events.  The reconstructed Monte Carlo events are then
fit using the same PWA model in~\cref{eq:primakoff_partcoh_formula}
that we use for the measured data. The fit result exhibits the same
two features that we already observed in the PWA of the measured data:
\one~for each $\Mrefl = 1^+$ wave in the simulated PWA model in
\cref{eq:primakoff_coh_formula}, we find significant $\Mrefl = 1^+$
and~$1^-$ components of this wave in the PWA result of the
reconstructed MC data, with their intensity sum being close to the
intensity of the simulated $\Mrefl = 1^+$ wave, and \two~we observe
reduced coherence between the $\Mrefl = 0^+$ and~$1^+$ amplitudes.
This study hence points to the necessity of summing up the intensities
of the $\Mrefl = 1^+$ and~$1^-$ components of each wave in each
\mThreePi~bin to get the proper intensities of the Primakoff-produced
waves.  In \cref{sec:primakoff_phi_smearing}, we show that the effects
discussed above can be explained by a substantial smearing of the
\phiGJ~angle.

In order to study the production mechanisms for states with spin
projections $M = 0$ and~1 in detail, we perform a dedicated PWA in
$t'$~bins.  This analysis is performed in the region
$t' < \SI{e-2}{\GeVcsq}$ using 31~narrow non-equidistant $t'$~bins in
order to resolve the $t'$~spectra of Primakoff and diffractive
production.  Due to the limited size of the experimental data sample,
we perform this PWA in two broad mass ranges:
\one~\SIvalRange{1.26}{\mThreePi}{1.38}{\GeVcc}, which we use to study
the \PaTwo signal, and
\two~\SIvalRange{1.50}{\mThreePi}{1.80}{\GeVcc}, which we use to study
the \PpiTwo signal.  Since the PWA is performed in narrow $t'$~bins,
the transition amplitudes are assumed to be independent of~$t'$ in
each bin, \ie the functions $f_i^\refl(t')$ in the PWA model in
\cref{eq:primakoff_partcoh_formula} that model these $t'$~dependences
are set to~1.

As the transition amplitudes vary significantly over the analyzed
broad \mThreePi~ranges, we have to model this \mThreePi~dependence.
We use the same ansatz in \cref{eq:res_model_amp_wave} that we use for
the resonance-model fits.\footnote{Here, the amplitude
  $\mathcal{P}(\mThreePi, t')$ that models the average production
  probability is set to~1.}  This means that we model the transition
amplitudes as a sum of dynamical amplitudes $\mathcal{D}_k(\mThreePi)$
for the wave components~$k$, which are, in the case of resonances,
Breit--Wigner amplitudes.  The main difference \wrt a resonance-model
fit is that here the dynamical amplitudes have no free parameters.
Due to this restriction, the standard PWA fit procedure (see
\cref{sec:pwa_cells.likelihood_fit}), including the pre-calculation of
the integral matrices, can be applied.  The coupling amplitudes
$\mathcal{C}_{k i}^\refl(t')$, which are free parameters that are
determined independently in each $t'$~bin, contain information about
the $t'$~dependence of production strength and phase of component~$k$
in wave~$i$ with reflectivity~\refl (see
\cref{sec:pwa.analysis_model.extension}).  Since we perform the PWA
fits in broad \mThreePi~ranges that fully contain the peak of the
studied $3\pi$ resonance~$k$, we effectively integrate over the
analyzed \mThreePi~range so that $|\mathcal{C}_{k i}^\refl(t')|^2$
corresponds to the total intensity of resonance~$k$ in the given
$t'$~bin.  Dividing by the $t'$~bin width, we hence obtain the
$t'$~spectrum $\dif{N_{k i}^\refl} / \dif{t'}$ of the resonance
(\confer\ \cref{eq:t_spectrum}).  We also measure the coupling
phases~$\phase_{ki, lj}^\refl$ defined in \cref{eq:phase_comp_rank1}
as a function of~$t'$.  These are the phase differences between
component~$k$ in wave~$i$ and component~$l$ in wave~$j$, which contain
information about the production phases of the wave components.  The
$t'$~dependence of these phases allows us to study the production
mechanism of resonances.

For the \PaTwo mass range, the \mThreePi~dependence of the transition
amplitudes of the \wave{1}{++}{0}{+}{\Pprho}{S} and
\wave{2}{++}{1}{\pm}{\Pprho}{D} waves are modeled by Breit--Wigner
amplitudes for \PaOne and \PaTwo, respectively.  The \PaTwo parameters
are fixed to the PDG values~\cite{Beringer:2012zz}; for the \PaOne, we
use the parameters $m_{\PaOne} = \SI{1220}{\MeVcc}$ and
$\Gamma_{\PaOne} = \SI{370}{\MeVcc}$.  For the dynamic widths of the
two resonances, we use the same parameterizations as for the
resonance-model fit of the diffractive \threePi proton-target data
(see \cref{sec:3pi_model:resonance} and Section~IV.A.1 in
\refCite{Akhunzyanov:2018lqa} for details).  All the other transition
amplitudes in the PWA model are assumed to be independent
of~\mThreePi.  For the broader \PpiTwo mass range, all transition
amplitudes depend on~\mThreePi.  The \mThreePi~dependence of the
$\JPC = 2^{-+}$ transition amplitudes of the $\PfTwo \pi$~$S$ and the
$\Pprho \pi$~$P$ and $F$~waves with $M = 0$ and~1 is modeled by a
constant-width Breit--Wigner amplitude
(\cref{eq:bw.rel,eq:BW_const_width}) with mass and width fixed to the
PDG parameters of the \PpiTwo~\cite{Beringer:2012zz}.  In addition,
the transition amplitudes of the \wave{2}{-+}{0}{+}{\PfTwo}{D} wave
and of selected $M = 0$ waves with $\JPC = 0^{-+}$ and $1^{++}$ are
modeled by a coherent sum of a Breit--Wigner amplitude and a
non-resonant component (see Appendix~A.5 and Table~5 in
\refCite{Adolph:2014mup} for details).  The \mThreePi~dependence of
the remaining transition amplitudes is modeled by the square root of
polynomials that were obtained by fitting the intensity distributions
of the respective waves, as obtained from the PWA in \mThreePi~bins,
in the analyzed \mThreePi~range.

\subsection{Results for \PaTwo and \PpiTwo from Partial-Wave Decomposition}
\label{sec:pwa_results_a2_pi2}

\begin{figure}[tbp]
  \subfloat[][]{%
    \includegraphics[width=\twoPlotWidth]{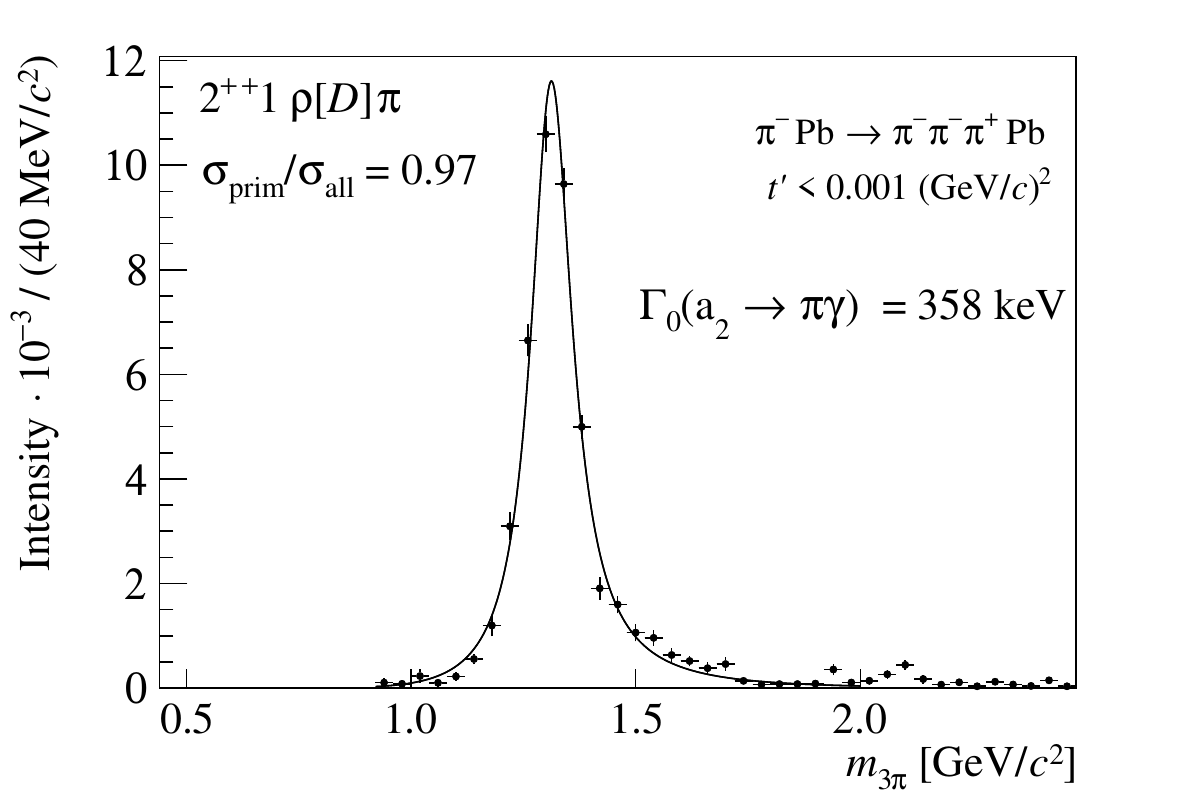}%
    \label{fig:pwa_prim_2pp1_mspect_fitted}%
  }%
  \hfill%
  \subfloat[][]{%
    \includegraphics[width=\twoPlotWidth]{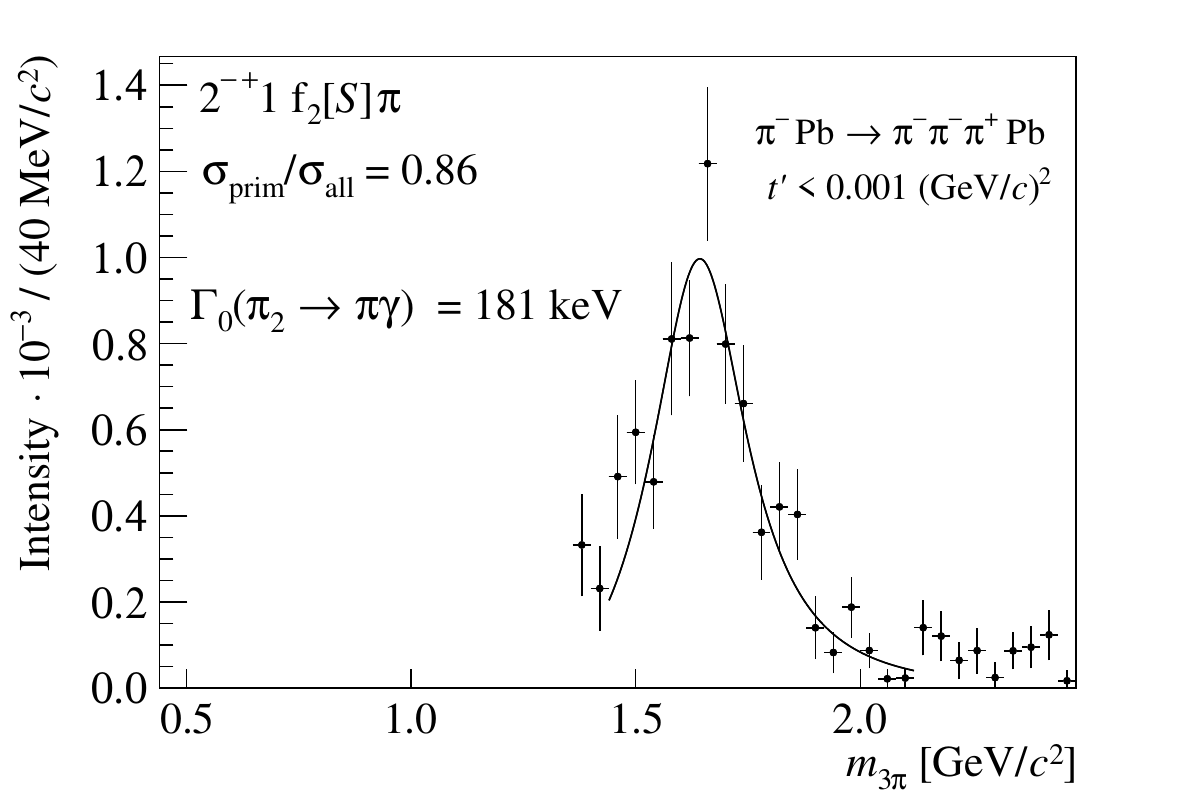}%
    \label{fig:pwa_prim_2mp1_mspect_fitted}%
  }%
  \\
  \subfloat[][]{%
    \includegraphics[width=\twoPlotWidth]{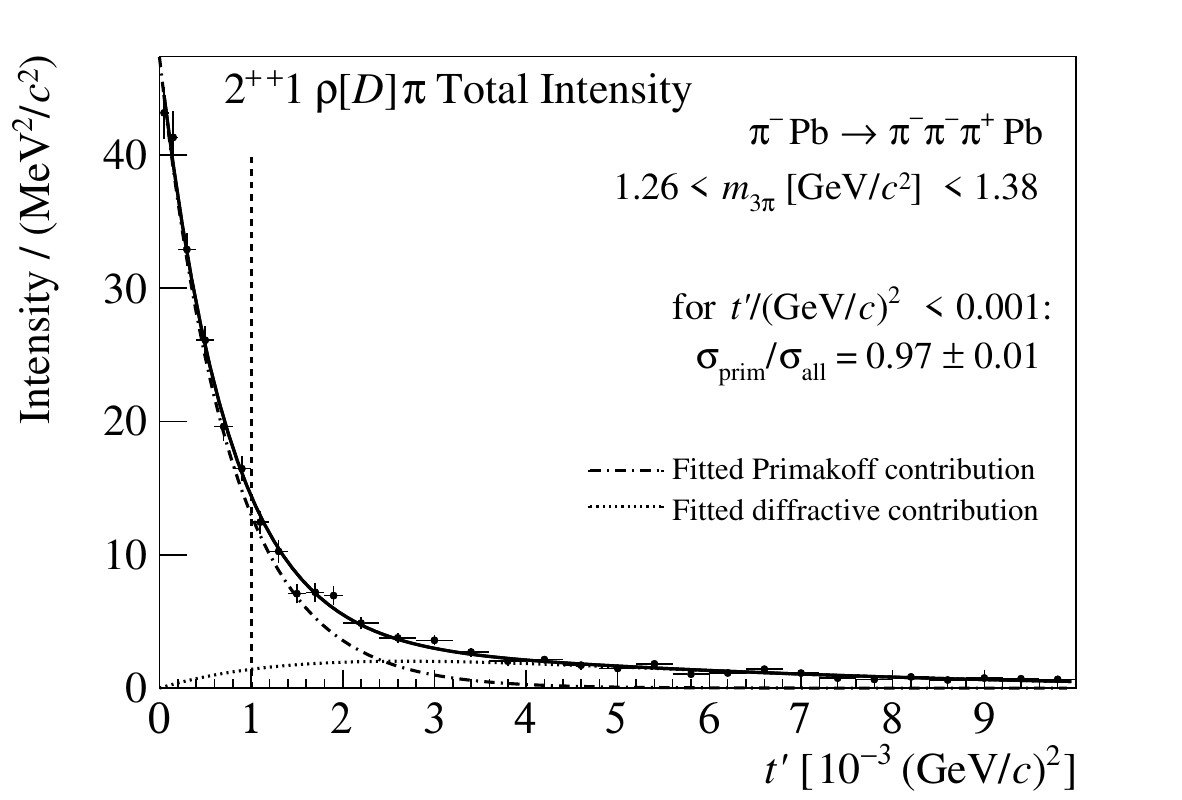}%
    \label{fig:pwa_prim_2pp1_tspect_fitted}%
  }%
  \hfill%
  \subfloat[][]{%
    \includegraphics[width=\twoPlotWidth]{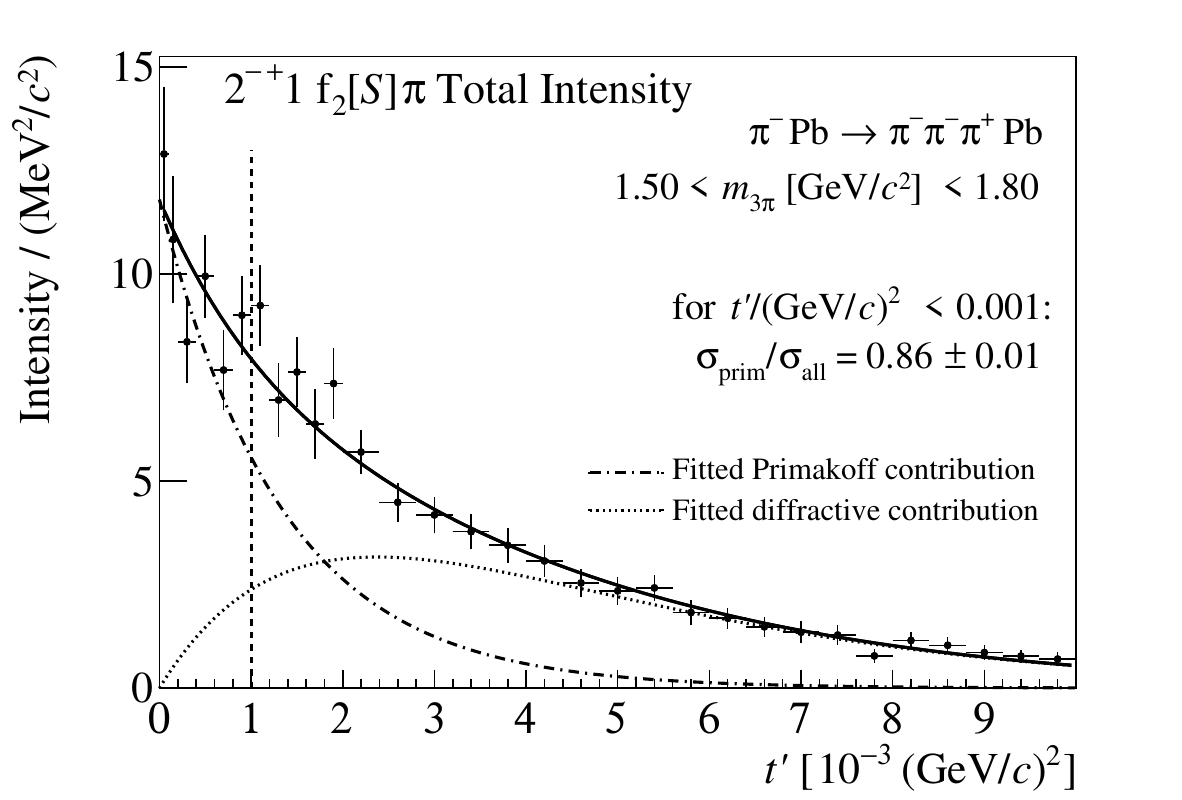}%
    \label{fig:pwa_prim_2mp1_tspect_fitted}%
  }%
  \\
  \subfloat[][]{%
    \includegraphics[width=\twoPlotWidth]{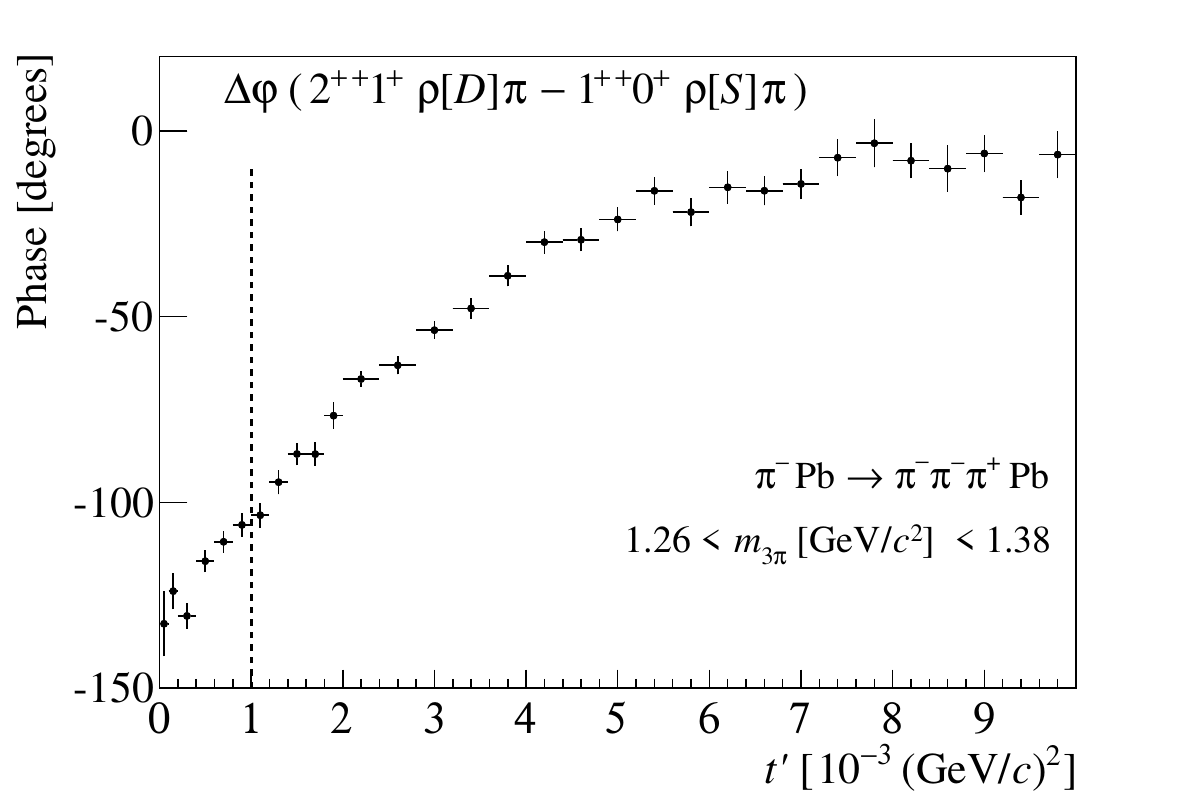}%
    \label{fig:pwa_prim_2pp1p_tphase}%
  }%
  \hfill%
  \subfloat[][]{%
    \includegraphics[width=\twoPlotWidth]{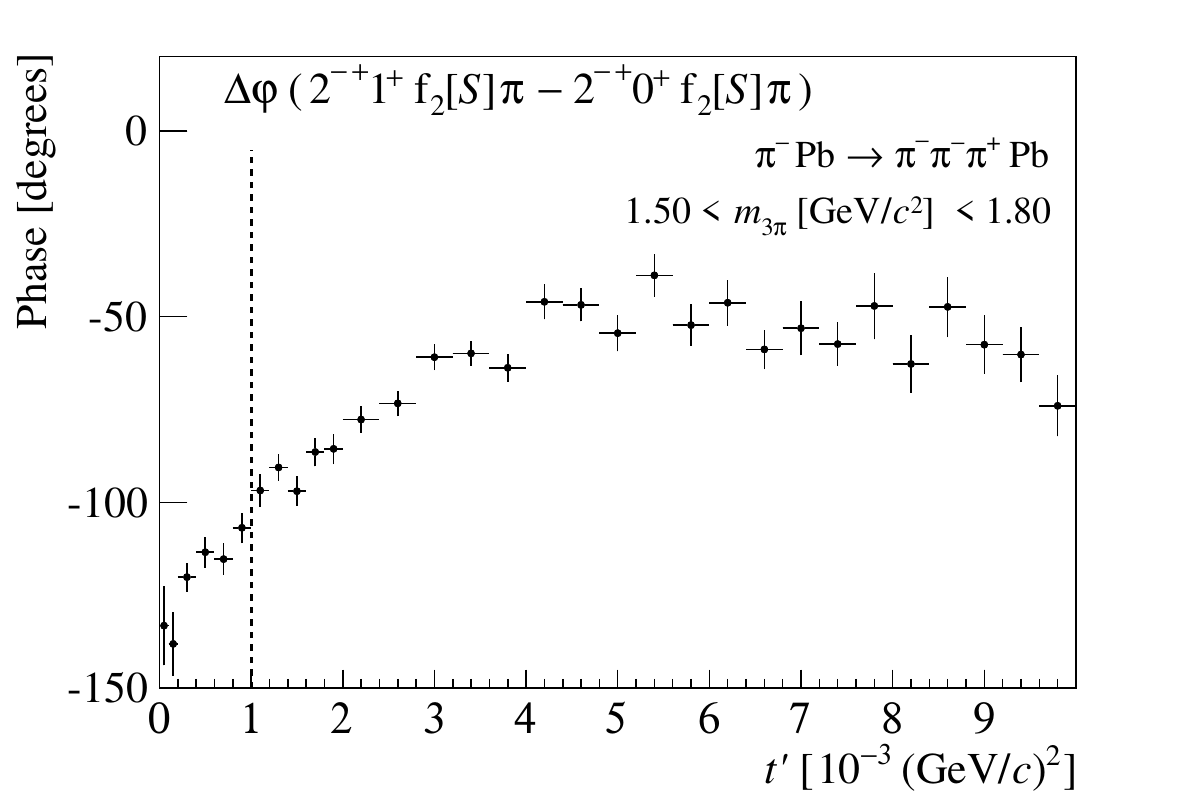}%
    \label{fig:pwa_prim_2mp1p_tphase}%
  }%
  \caption{\subfloatLabel{fig:pwa_prim_2pp1_mspect_fitted}~and~\subfloatLabel{fig:pwa_prim_2pp1_tspect_fitted}:
    Intensity sum of the \wave{2}{++}{1}{}{\Pprho}{D} waves with
    $\refl = +1$ and~$-1$,
    \subfloatLabel{fig:pwa_prim_2pp1_mspect_fitted}~as a function
    of~\mThreePi for $t' < \SI{e-3}{\GeVcsq}$ and
    \subfloatLabel{fig:pwa_prim_2pp1_tspect_fitted}~as a function
    of~$t'$ for a broad \mThreePi~range around the \PaTwo.  The curve
    in~\subfloatLabel{fig:pwa_prim_2pp1_mspect_fitted} represents the
    result of a fit with \cref{eq:ww_reson_xsec} integrated over
    $t' < \SI{e-3}{\GeVcsq}$.  The continuous curve
    in~\subfloatLabel{fig:pwa_prim_2pp1_tspect_fitted} represents the
    result of a fit with \cref{eq:prim_res_t_spectrum_model}.  The
    dotted curve represents the contribution from diffractive
    production, the dashed--dotted curve the one from Primakoff
    production.
    \subfloatLabel{fig:pwa_prim_2mp1_mspect_fitted}~and~\subfloatLabel{fig:pwa_prim_2mp1_tspect_fitted}:
    Similar to
    \subfloatLabel{fig:pwa_prim_2pp1_mspect_fitted}~and~\subfloatLabel{fig:pwa_prim_2pp1_tspect_fitted}
    but for the \wave{2}{-+}{1}{\pm}{\PfTwo}{S} waves and the
    \mThreePi~range around the \PpiTwo.
    \subfloatLabel{fig:pwa_prim_2pp1p_tphase}~Relative phase between
    the \PaTwo in the \wave{2}{++}{1}{+}{\Pprho}{D} wave and the
    \PaOne in the \wave{1}{++}{0}{+}{\Pprho}{S} wave as a function
    of~$t'$.  \subfloatLabel{fig:pwa_prim_2mp1p_tphase}~Relative phase
    between the \PpiTwo in the \wave{2}{-+}{\!\!}{}{\PfTwo}{S} waves
    with $\Mrefl = 1^+$ and~$0^+$ as a function of~$t'$.
    In~\subfloatLabel{fig:pwa_prim_2pp1_tspect_fitted}
    and~\subfloatLabel{fig:pwa_prim_2pp1p_tphase}, the
    \mThreePi~dependence of the \wave{2}{++}{1}{\pm}{\Pprho}{D}
    amplitudes is modeled by a Breit--Wigner for the \PaTwo.
    In~\subfloatLabel{fig:pwa_prim_2mp1_tspect_fitted}
    and~\subfloatLabel{fig:pwa_prim_2mp1p_tphase}, the
    \mThreePi~dependence of the \wave{2}{-+}{1}{\pm}{\PfTwo}{S}
    amplitudes is modeled by a Breit-Wigner for the \PpiTwo.}
  \label{fig:pwa_prim}
\end{figure}

In \cref{fig:pwa_prim}, we summarize the PWA results for Primakoff
production of \PaTwo and \PpiTwo.  The points with error bars
represent the result of the partial-wave decomposition in
\mThreePi~and $t'$~bins, respectively, using the PWA models that were
discussed in \cref{sec:pwa_prim_specifics}.  The PWA in the Primakoff
region $t' < \SI{e-3}{\GeVcsq}$ is performed in \SI{40}{\MeVcc} wide
\mThreePi~bins.  In addition, we perform partial-wave analyses in
31~narrow non-equidistant $t'$~bins in the extended range
$t' < \SI{e-2}{\GeVcsq}$ for two broad mass regions around \PaTwo and
\PpiTwo.

\Cref{fig:pwa_prim_2pp1_mspect_fitted} shows the intensity sum of the
\wave{2}{++}{1}{}{\Pprho}{D} waves with $\refl = +1$ and~$-1$ with a
clear \PaTwo signal.  The curve represents the result of a fit with a
resonance model that will be discussed in \cref{sec:rad_widths}.
\Cref{fig:pwa_prim_2mp1_mspect_fitted} shows the corresponding
intensity sum of the \wave{2}{-+}{1}{}{\PfTwo}{S} waves with a clear
\PpiTwo peak.

In
\crefrange{fig:pwa_prim_2pp1_tspect_fitted}{fig:pwa_prim_2mp1p_tphase},
we show the results of the PWA performed in 31~non-equidistant
$t'$~bins in broad mass ranges around the \PaTwo and \PpiTwo.  As was
discussed in \cref{sec:pwa_prim_specifics}, in the employed PWA model
the \wave{2}{++}{1}{}{\Pprho}{D} amplitudes are described by a
Breit--Wigner amplitude for the \PaTwo.  The $t'$~spectrum in
\cref{fig:pwa_prim_2pp1_tspect_fitted} hence represents the \PaTwo
intensity, where we again sum the intensities of the $\refl = +1$
and~$-1$ contributions.  The $t'$~spectrum is fitted well by an
incoherent sum of two exponentials similar to
\cref{eq:tspectrum_total}:
\begin{multline}
  \label{eq:prim_res_t_spectrum_model}
  \frac{\dif{N_{\PaTwo}}}{\dif{t'}}
  = \frac{\dif{N_{\PaTwo}^\text{diff}}}{\dif{t'}} + \frac{\dif{N_{\PaTwo}^\text{Prim}}}{\dif{t'}} \\
  \quad\text{with}\quad
  \frac{\dif{N_{\PaTwo}^\text{diff}}}{\dif{t'}}
  \propto t'\, e^{-b_{\PaTwo}^\text{diff}\, t'}
  \quad\text{and}\quad
  \frac{\dif{N_{\PaTwo}^\text{Prim}}}{\dif{t'}}
  \propto e^{-b_{\PaTwo}^\text{Prim}\, t'}\eqPunctSpacing.
\end{multline}
We obtain slope parameters of
$b_{\PaTwo}^\text{diff} = \SI{374(25)}{\perGeVcsq}$ and
$b_{\PaTwo}^\text{Prim} = \SI{1292(53)}{\perGeVcsq}$.  The latter
value is in fair agreement with the one obtained in the Primakoff
MC~study that was described in \cref{sec:primakoff_t_spectrum}
(\confer\ \cref{eq:bprim_polyn}).  The fit of the $t'$~spectrum also
provides an estimate of the diffractive background.  For the \PaTwo,
it amounts to approximately \SI{3}{\percent} of the total intensity in
the Primakoff region.  Applying the model in
\cref{eq:prim_res_t_spectrum_model} to the $t'$~spectrum of the
\PpiTwo (see \cref{fig:pwa_prim_2mp1_tspect_fitted}), which is
obtained by summing the intensities of the $\refl = +1$ and~$-1$
contributions, we find that the relative contribution of the
diffractive background is significantly larger than for the \PaTwo.
In the Primakoff region, it amounts to about \SI{14}{\percent}.  This
makes the separation of the contributions from Primakoff and
diffractive production more difficult.  As a consequence, additional
constraints are needed to stabilize the fit (see Section~3.2 in
\refCite{Adolph:2014mup} for details).

As explained in \cref{sec:pwa_results_a2_pi2}, the PWA in $t'$~bins
also allows us to extract the relative phases between resonances, \ie
the coupling phases $\phase_{ki, lj}^\refl$, as a function of~$t'$.
\Cref{fig:pwa_prim_2pp1p_tphase} shows the relative phase of the
\PaTwo with $\Mrefl = 1^+$ and the \PaOne with $\Mrefl = 0^+$.  For
$t' < \SI{6e-3}{\GeVcsq}$, this phase rises monotonically with~$t'$
and levels off for larger~$t'$ reaching a plateau at about
\SI{0}{\degree}.  This reflects the continuous evolution from nearly
pure Primakoff production of the \PaTwo at low~$t'$ to nearly pure
diffractive production at higher~$t'$.  In the Primakoff region
$t' < \SI{e-3}{\GeVcsq}$, the relative phase is given by the phase
difference of the Primakoff production amplitude for the \PaTwo, which
is approximately purely real, and the diffractive production amplitude
of the \PaOne, which is approximately purely imaginary.  This results
in $\phase_{\PaTwo,~\PaOne}^{(\refl = +1)} \approx \SI{-90}{\degree}$.
Assuming that the phase of the diffractive production amplitude of the
\PaTwo has a $t'$~dependence similar to that of the \PaOne,
interference effects between diffractive and Primakoff production of
the \PaTwo are expected to be small.  Hence the sum of two
non-interfering contributions in \cref{eq:prim_res_t_spectrum_model}
is a good approximation to describe the $t'$~spectrum.  At higher
values of~$t'$, both \PaTwo and \PaOne are diffractively produced so
that $\phase_{\PaTwo,~\PaOne}^{(\refl = +1)}(t')$ flattens and reaches
approximately \SI{0}{\degree}.  Like for the \PaTwo, we observe for
the \PpiTwo a phase difference of roughly \SI{-90}{\degree} between
the $\Mrefl = 1^+$ (Primakoff-produced) and~$0^+$ (diffractively
produced) states in the $\PfTwo \pi$ $S$~decay mode (see
\cref{fig:pwa_prim_2mp1p_tphase}).  However, at larger values of~$t'$
the phase difference does not approach \SI{0}{\degree} but levels off
already at approximately \SI{-50}{\degree}.  This behavior is not yet
understood.

\subsection{Radiative Widths of \PaTwo and \PpiTwo}
\label{sec:rad_widths}

Electromagnetic decays of mesons are important to understand their
internal structure.  Like in atomic and nuclear physics, the
multipolarities of the electromagnetic transitions, probe different
aspects of the meson wave function.  Several phenomenological models
have been developed to describe electromagnetic decays of mesons, for
example, the relativized quark model~\cite{Godfrey:1985xj}, the
covariant oscillator quark
model~\cite{Ishida:1984tq,Ishida:1988uw,Maeda:2013dka}, and the
vector-meson dominance model (see \eg\ \refCite{Feynman:1973xc}).  The
latter model assumes that any soft interaction of photons with hadrons
proceeds via intermediate vector-meson states, predominantly via the
lightest ones, \ie\ \Pprho, \Ppomega, and \Ppphi.  Some
electromagnetic decays such as $\pi^0 \to \gamma \gamma$ are also
calculable in the framework of \chiPT.  The latest experimental result
for $\Gamma[\pi^0 \to \gamma \gamma]$ in \refCite{Larin:2010kq} is in
excellent agreement with the theoretical prediction from
\refCite{Goity:2002nn}.

For most light mesons, the direct observation of their radiative
decays is impossible because the corresponding partial decay widths
are usually only \num{e-4}--\num{e-3} of the total decay width.
Hence backgrounds originating from $\pi^0 \to \gamma \gamma$, where
one photon is not detected, obscure the signal.  However, there has
been a lot of progress in the development of experimental approaches
to measure these radiative decays nevertheless.  The Primakoff
approach is widely employed to measure the $\pi^\pm \gamma$ and
$K^\pm \gamma$ partial decay widths of various pseudoscalar,
(pseudo)vector or (pseudo)tensor mesons~$X$ by measuring the cross
section of reactions of the form $h^\pm + A \to X^\pm + A$ on a target
nucleus~$A$, where $h^\pm$ stands for $\pi^\pm$ or $K^\pm$~beam
particles.  The Primakoff approach was used to measure
$\Gamma[\Pprho \to \pi
\gamma]$~\cite{Capraro:1987rp,Huston:1986wi,Jensen:1982nf},
$\Gamma[\PpKstar \to K
\gamma]$~\cite{Carlsmith:1985ep,Chandlee:1983hf},
$\Gamma[\PaOne \to \pi \gamma]$~\cite{Zielinski:1984au},
$\Gamma[\PbOne \to \pi \gamma]$~\cite{Collick:1985yi},
$\Gamma[\PaTwo \to \pi
\gamma]$~\cite{Adolph:2014mup,Molchanov:2001qk,Cihangir:1982ti,May:1977ra},
and $\Gamma[\PpiTwo \to \pi \gamma]$~\cite{Adolph:2014mup}.

The cross section for pion-induced Primakoff production of a broad
resonance~$X$ with spin~$J$, nominal mass~$m_0$, and nominal total
width~$\Gamma_0$ that is measured in the \threePi partial wave with
index~$i$ is given by integrating \cref{eq:primakoff_xsec} over the
\threePi phase space~$\Phi_3$~\cite{Halprin:1966zz,Zielinski:1986mg}.
This yields
\begin{equation}
  \label{eq:ww_reson_xsec}
  \frac{\dif{\sigma_{\pi + A \to X + A}}}{\dif{\mThreePi}\, \dif{t'}}
  = \Underbrace{\frac{\alpha_\text{em}\, Z^2}{\pi}\, \frac{2\mThreePi}{\mThreePi^2 - m_\pi^2}\,
  \frac{t'}{\big( t' + \tMin \big)^2}\, \big| F_\text{eff}(t', \tMin) \big|^2}%
  {\displaystyle{\eqqcolon \mathcal{F}(\mThreePi, t')}}\,
  \sigma_{\pi + \gamma \to X \to i}(\mThreePi)\eqPunctSpacing,
\end{equation}
where
\begin{equation}
  \label{eq:ww_reson_xsec_bw}
  \sigma_{\pi + \gamma \to X \to i}(\mThreePi)
  = 8\pi\, (2J + 1) \left[ \frac{\mThreePi}{ \mThreePi^2 - m_\pi^2} \right]^2
  \frac{m_0\, \Gamma_{\pi \gamma}(\mThreePi)\; m_0\, \Gamma_i(\mThreePi)}%
  {\big( \mThreePi^2 - m_0^2 \big)^2 + m_0^2\, \Gamma^2(\mThreePi)}\eqPunctSpacing.
\end{equation}
Here, we have used $\dif{\mThreePi^2} = 2\mThreePi\, \dif{\mThreePi}$
and we have parameterized the resonance propagator by a relativistic
Breit--Wigner amplitude with the dynamic total width
$\Gamma(\mThreePi)$ (see \cref{eq:BW_mass-dep_width}).  The dynamic
partial width of the decay $X \to \pi + \gamma$, \ie the radiative
width of~$X$, is denoted by $\Gamma_{\pi \gamma}(\mThreePi)$; the
dynamic partial width for the decay of~$X$ into \threePi via the
partial wave~$i$ by $\Gamma_i(\mThreePi)$.

We parameterize the radiative width $\Gamma_{\pi \gamma}(\mThreePi)$
following \cref{eq:BW_mass-dep_width} but omitting the term
$m_0 / \mThreePi$~\cite{Pisut:1968zza}, \ie
\begin{equation}
  \label{eq:extract:dynwidth_pigamma}
  \Gamma_{\pi \gamma}(\mThreePi)
  = \Gamma_{\pi \gamma, 0}\, \frac{q}{q_0}\, \frac{F^2_L(q)}{F^2_L(q_0)}\eqPunctSpacing,
\end{equation}
where $\Gamma_{\pi \gamma, 0}$ is the nominal radiative width that we
want to measure.  For the $\pi \gamma$ decay of the \PaTwo we use
$L = 2$, \ie $D$~wave, and for that of the \PpiTwo we use $L = 1$, \ie
$P$~wave.

For an accurate description of the resonance line shape, the dynamic
partial decay width $\Gamma_i(\mThreePi)$ must take into account the
width of the $2\pi$ isobar and self-interference effects due to Bose
symmetrization (see \cref{sec:pwa.analysis_model.symmetrization}).
This is achieved by parameterizing the \mThreePi~dependence of the
partial width for the decay chain represented by the wave index~$i$ by
$\int \dif{\Phi_3}(\tau_3)\, \big\vert \Psi_i(\mThreePi, \tau_3)
\big\vert^2$, where $\Psi_i$ is Bose-symmetrized according to
\cref{eq:decay_amp_bose}.  Normalizing to the nominal partial decay
width $\Gamma_{i, 0}$, we obtain:
\begin{equation}
  \label{eq:dynwidth_dalplot}
  \Gamma_i(\mThreePi)
  = \Gamma_{i, 0}\,
  \frac{\dint\! \dif{\Phi_3}(\tau_3)\, \big\vert \Psi_i(\mThreePi, \tau_3) \big\vert^2}%
       {\dint\! \dif{\Phi_3}(\tau_3)\, \big\vert \Psi_i(m_0, \tau_3) \big\vert^2}\eqPunctSpacing.
\end{equation}
We measure the \PaTwo and \PpiTwo in their dominant $3\pi$ decay
channels, \ie $\PaTwo \to \Pprho \pi$~$D$ and
$\PpiTwo \to \PfTwo \pi$~$S$.  The values of the nominal partial
widths are taken from the PDG~\cite{Beringer:2012zz}.  For
$\PaTwo \to \Pprho \pi$~$D$, we assume that this decay mode saturates
the branching fraction
$\text{BF}\big[ \PaTwo \to 3\pi \big] = \Gamma_{\PaTwo \to 3\pi, 0} /
\Gamma_{\PaTwo, 0} = \SI{70.1(27)}{\percent}$ and we apply an isospin
factor of $1 / 2$ to correct for the unobserved
$\PaTwo^- \to \Pprho^- \pi^0 \to \threePiN$ decay mode.  For
$\PpiTwo \to \PfTwo \pi$~$S$, we assume that this decay saturates the
branching fraction
$\text{BF}\big[ \PpiTwo \to \PfTwo \pi \big] = \Gamma_{\PpiTwo \to
  \PfTwo \pi, 0} / \Gamma_{\PpiTwo, 0} = \SI{56.3(32)}{\percent}$ and
we apply an isospin factor of $2 / 3$ to correct for the unobserved
$\PpiTwo^- \to \PfTwo \pi^- \to \threePiN$ decay mode.

The employed parameterization for the total width of the \PaTwo is
similar to the one used in the resonance-model fit of the diffractive
\threePi proton-target data (see \cref{sec:3pi_model:resonance} and
Section~IV.A.1 in \refCite{Akhunzyanov:2018lqa} for details).  We also
assume that the \PaTwo decays only into $\Pprho \pi$~$D$
(\SI{82}{\percent}) and $\eta \pi$~$D$ (\SI{18}{\percent}), \ie we
neglect all other decay modes.  We use \cref{eq:dynwidth_dalplot} for
$\Gamma_{\PaTwo \to \Pprho \pi D}(\mThreePi)$ and
\cref{eq:BW_mass-dep_width} for
$\Gamma_{\PaTwo \to \eta \pi D}(\mThreePi)$.  For the \PpiTwo, the
situation is more complicated.  The \PpiTwo decays nearly exclusively,
\ie with a branching fraction of \SI{95.8(14)}{\percent}, into $3\pi$,
which includes the decays into $\PfTwo \pi$ (\SI{56.3(32)}{\percent}),
$\Pprho \pi$ (\SI{31(4)}{\percent}), $\sigma \pi$
(\SI{10.9(34)}{\percent}), and $\pipiS \pi$
(\SI{8.7(34)}{\percent})~\cite{Tanabashi:2018zz}.
In the analysis of the diffractive \threePi data (see
\cref{sec:results_2mp}), we observe sizable interference of different
$3\pi$ decay chains of the \PpiTwo.  Therefore, the branching fraction
information in the PDG is insufficient to construct a realistic $3\pi$
decay model that takes into account these interferences.  However, for
our analysis the details of the parameterization of
$\Gamma(\mThreePi)$ are not very important.  They have only to a
subtle influence on the peak shape, which mostly affects the estimates
for~$m_0$ and~$\Gamma_0$, whereas the resonance yield that we are
mainly interested in (see below) is basically unaffected by the choice
of parameterization.  Considering in addition the limited precision of
our data in \cref{fig:pwa_prim_2mp1_mspect_fitted}, we choose a simple
parameterization for the total width of the \PpiTwo, where we assume
that the $\PfTwo \pi$~$S$ decay mode saturates the total width and use
\cref{eq:dynwidth_dalplot}.

With the above definitions, only three unknowns remain in
\cref{eq:ww_reson_xsec,eq:ww_reson_xsec_bw}: the nominal radiative
width $\Gamma_{\pi \gamma, 0}$, which defines the height of the
resonance peak, the nominal mass~$m_0$, and the nominal total
width~$\Gamma_0$ of the resonance, which both determine the shape of
the resonance peak.  In order to measure $\Gamma_{\pi \gamma, 0}$, we
use the fact that
$\dif{\sigma_{\pi + A \to X + A}} / (\dif{\mThreePi}\, \dif{t'})
\propto \Gamma_{\pi \gamma, 0}$ and perform the analysis in three
steps.

In the first step, we integrate \cref{eq:ww_reson_xsec} over
$t' < \SI{e-3}{\GeVcsq}$ and fit the resulting function of~\mThreePi
to the \wave{2}{++}{1}{}{\Pprho}{D} and \wave{2}{-+}{1}{}{\PfTwo}{S}
intensity distributions.  The results are represented by the curves in
\cref{fig:pwa_prim_2pp1_mspect_fitted,fig:pwa_prim_2mp1_mspect_fitted}.
In the $\chi^2$~function that is used in the fit, we convolve the
model with the \mThreePi~resolution obtained from MC simulations and
compare the integral of the model function over each \mThreePi~bin
with the measured intensity in this bin.  As a result, we
determine~$m_0$, $\Gamma_0$, and an overall normalization factor that
contains $\Gamma_{\pi \gamma, 0}$.  For the \PaTwo, we obtain the
Breit--Wigner parameters $m_{\PaTwo} = \SI{1319(1)}{\MeVcc}$ and
$\Gamma_{\PaTwo} = \SI{105(4)}{\MeVcc}$, and for the \PpiTwo,
$m_{\PpiTwo} = \SI{1684(11)}{\MeVcc}$ and
$\Gamma_{\PpiTwo} = \SI{277(38)}{\MeVcc}$.  Here, the uncertainties
are statistical only.  For both resonances, the measured parameter
values are in good agreement with the PDG averages and with the PWA
results from the COMPASS diffractive data (see
\cref{fig:ideogram_a2_1320,fig:ideogram_pi2_1670}).

In the second step, we obtain the acceptance-corrected yields of
\PaTwo and \PpiTwo in terms of number of events by integrating the
curves in
\cref{fig:pwa_prim_2pp1_mspect_fitted,fig:pwa_prim_2mp1_mspect_fitted}
over the \mThreePi~fit range.  The yields are corrected for the
contributions from diffractive background, which are estimated by
fitting \cref{eq:prim_res_t_spectrum_model} to the $t'$~spectra in
\cref{fig:pwa_prim_2pp1_tspect_fitted,fig:pwa_prim_2mp1_tspect_fitted}
as obtained from the PWA in $t'$~bins (see
\cref{sec:pwa_results_a2_pi2}).  A second correction factor is applied
to account for the migration of events from above and below the
Primakoff limit at $t' = \SI{e-3}{\GeVcsq}$, which is caused by the
limited $t'$~resolution.  The correction factor is estimated by
simulating Primakoff events and calculating the ratio of the number of
generated events in the region $t'_\text{gen} < \SI{e-3}{\GeVcsq}$ and
the number of reconstructed events in the region
$t'_\text{rec} < \SI{e-3}{\GeVcsq}$.\footnote{Since the contribution
  of diffractive events with $M = 1$ is negligible (see below), the
  migration of these events is not considered.}  For the \PaTwo, this
ratio is \num{1.348}; for the \PpiTwo it is \num{1.359}.

In the third step, we use the corrected yield~$N_X$ from above to
calculate the Primakoff production cross
section~$\sigma_{\text{Prim}, X}$ for resonance~$X$.  For this, we
need the integrated luminosity~$\mathscr{L}$, for which we need in
turn the total integrated $\pi^-$~beam flux.  The effective beam flux
is determined using \threePi decays of beam kaons that are observed
outside the target.  Based on the known $K^-$~fraction of
\SI{2.4}{\percent} in the beam, the branching fraction
$\text{BF}\big[ K^- \to \threePi \big] =
\SI{5.54}{\percent}$~\cite{Beringer:2012zz}, the acceptance for
$K^- \to \threePi$ of \SI{45.9}{\percent}, which was estimated by a
dedicated Monte Carlo simulation, the limited decay volume, and the
thickness of the Pb~target of \SI{3}{mm}, we obtain an integrated
effective luminosity of $\mathscr{L} = \SI{208}{\per\micro\barn}$.
Using the numbers from \refCite{Adolph:2014mup}, we calculate the
Primakoff cross sections
$\sigma_{\text{Prim}, \PaTwo} = \SI{280}{\micro\barn}$ and
$\sigma_{\text{Prim}, \PpiTwo} = \SI{39.2}{\micro\barn}$.  These
values correspond to the integral of \cref{eq:ww_reson_xsec} over
$t' < \SI{e-3}{\GeVcsq}$ and the \mThreePi~fit range.\footnote{For
  \PaTwo, the fit range is \SIvalRange{0.92}{\mThreePi}{2.00}{\GeVcc};
  for \PpiTwo, it is \SIvalRange{1.44}{\mThreePi}{2.12}{\GeVcc}.}
Since the radiative width is a multiplicative parameter in
\cref{eq:ww_reson_xsec} that is independent of~$t'$ and~\mThreePi, we
can write this integral as
\begin{equation}
  \label{eq:xsec_radwidth}
  \sigma_{\text{Prim}, X}
  = \int_0^{t'_\text{max}}\! \dif{t'} \int_{m_{3\pi, \text{min}}}^{m_{3\pi, \text{max}}}\! \dif{\mThreePi}\,
  \frac{\dif{\sigma_{\pi + A \to X + A}}}{\dif{\mThreePi}\, \dif{t'}}
  = \Gamma_{\pi \gamma, 0}\, I_X\eqPunctSpacing,
\end{equation}
where $I_X$~represents the integral of \cref{eq:ww_reson_xsec} with
$\Gamma_{\pi \gamma, 0}$ taken out.  The constant~$I_X$ is calculable
using the resonance parameters of~$X$ and the parameterizations of the
partial and total widths discussed above.

The results for the radiative widths of \PaTwo and \PpiTwo are listed
in \cref{tab:table_rad_wid} together with the results of previous
measurements and theoretical calculations using various models.  The
values for the \PaTwo are compatible with previous measurements and
with calculations using a vector-meson dominance
model~\cite{Rosner:1980ek} and a relativistic quark
model~\cite{Aznauryan:1988aa}.  The values obtained using a covariant
oscillator quark model~\cite{Ishida:1988uw,Maeda:2013dka} lie below
the experimental values.  For the \PpiTwo, we performed the first
measurement of its radiative width.  The predictions from the
covariant oscillator quark model from \refCite{Maeda:2013dka} are both
significantly higher than our value.

\begin{table}[tbp]
  \centering
  \renewcommand{\arraystretch}{1.2}
  \caption{Comparison of the COMPASS values for the radiative widths
    of \PaTwo and \PpiTwo with the ones from previous measurements and
    theoretical calculations.}
  \label{tab:table_rad_wid}
  \begin{small}
    \begin{tabular}{lll}
      \toprule
      &
      $\Gamma\big[ \PaTwo \to \pi \gamma \big]$ &
      $\Gamma\big[ \PpiTwo \to \pi \gamma \big]$ \\
      \midrule

      \textbf{Experiments} & & \\
      COMPASS~\cite{Adolph:2014mup} & \SIerrsnol{358}{6}{42}{\keVcc} & \SIerrsnol{181}{11}{27}{\keVcc} \\
      FNAL E781 (SELEX)~\cite{Molchanov:2001qk} & \SIerrsnol{284}{25}{25}{\keVcc} &  \\
      FNAL E272~\cite{Cihangir:1982ti} & \SI{295(60)}{\keVcc} & \\
      E.~N. May \etal~\cite{May:1977ra} & \SI{0.46(11)}{\MeVcc} & \\
      \midrule

      \textbf{Theory} & & \\
      Vector-meson dominance model~\cite{Rosner:1980ek} & \SI{375(50)}{\keVcc} & \\
      Relativistic quark model~\cite{Aznauryan:1988aa}  & \SI{324}{\keVcc} & \\
      Covariant oscillator quark model~\cite{Ishida:1988uw} & \SI{235}{\keVcc} & \\
      Covariant oscillator quark model~\cite{Maeda:2013dka} & \SI{237}{\keVcc} & 2~values: \SI{335}{\keVcc} and \SI{521}{\keVcc} \\
      \bottomrule
    \end{tabular}
  \end{small}
\end{table}

When comparing the experimental values in \cref{tab:table_rad_wid}, it
is important to note that the value of $\Gamma_{\pi \gamma, 0}$
depends on the choice of the parameterization of the form factor
$F_\text{eff}$ in \cref{eq:ww_reson_xsec}.  We use the form factor
from \refCite{Faldt:2008yw} that includes corrections, which take into
account the distortion of the pion wave function in the Coulomb field.
These Coulomb corrections were neglected in the previous measurements
of the \PaTwo radiative width that used the Primakoff technique.
However, as noted by the authors of \refCite{Faldt:2008yw}, the
Coulomb corrections can significantly reduce the Primakoff cross
section and hence increase the estimate for the radiative width.  This
may explain that the COMPASS value for the \PaTwo radiative width lies
above the values from the FNAL SELEX~\cite{Molchanov:2001qk} and E272
experiments~\cite{Cihangir:1982ti}.  For COMPASS kinematics,
accounting for the Coulomb corrections increases both radiative widths
by \SI{24}{\percent}.  For SELEX kinematics, the increase is
\SI{15}{\percent}.  The latter value is at variance with the
conclusion of the authors of \refCite{Molchanov:2001qk}, who report
that taking into account the Coulomb corrections had only a
\textquote{minor impact} on their result.

\subsection{Test of Chiral Perturbation Theory}
\label{sec:pwa_chpt_amp}

In addition to Primakoff production of resonances also non-resonant
processes contribute to the reaction $\pi^- \gamma \to \threePi$ and
may interfere with the resonant production.  In the case of quasi-real
photon exchange and in the region of low~\mThreePi, the non-resonant
processes can be calculated using \chiPT (see
\cref{sec:theory.chiPT}).  The leading-order (LO) calculation was
performed by Kaiser and Friedrich in \refCite{Kaiser:2008ss}.  The
next-to-leading-order (NLO) \chiPT calculations include contributions
from loops~\cite{Kaiser:2010zg} and from the
\Pprho~\cite{Kaiser:2013rho}.  They are expected to describe the
non-resonant physics also at higher~\mThreePi.

We take into account the non-resonant processes described above in the
PWA by including a dedicated non-isobaric decay amplitude, called
\chiPT amplitude.  Since we perform the analysis in the range
$\mThreePi < \SI{0.72}{\GeVcc}$ below the $\Pprho \pi$ threshold,
\Pprho contribution are expected to be negligible.  It was shown in
\refCite{Kaiser:2010zg} that at low~\mThreePi also the contributions
from NLO diagrams with loops are negligible for the \threePi final
state.\footnote{The loop corrections are significantly larger for the
  \threePiN final state.}  We thus use for the PWA in the
low-\mThreePi region the LO \chiPT amplitude.  This amplitude
decomposes into $3\pi$ partial waves with $\Mrefl = 1^+$ and various
\JPC quantum numbers.  Due to the resolution effects that were
discussed in \cref{sec:pwa_prim_specifics}, the \chiPT amplitude has
to be included with $\refl = \pm 1$ into the PWA model in
\cref{eq:primakoff_partcoh_formula}~\cite{Kaiser:2010zg,Halprin:1966zz}.

We performed a detailed study of the process
$\pi^- + A \to \threePi + A$ in the Primakoff region
$t' < \SI{e-3}{\GeVcsq}$ and in the low-mass range
\SIvalRange{0.44}{\mThreePi}{0.72}{\GeVcc}~\cite{Adolph:2011it}, where
\chiPT is applicable.  The PWA model in
\cref{eq:primakoff_partcoh_formula} is used with a wave set that
consists of the LO~\chiPT amplitude with $\Mrefl = 1^\pm$ that
describes the quasi-real photoproduction plus a set of 5~isobaric
amplitudes with $\Mrefl = 0^+$ and $J \leq 2$ that describe the
diffractive production.  We also include a wave that represents the
$K^- \to \threePi$ events close to threshold.  We verified that the
\chiPT amplitude can indeed replace the isobaric amplitudes with
$M = 1$ by performing a PWA, where we included 6~isobaric waves with
$M = 1$ instead of the \chiPT amplitude.  The intensity of the \chiPT
amplitude is found to be similar to the intensity of the coherent sum
of the $M = 1$ isobaric waves~\cite{Adolph:2011it} (see also
\refCite{Grabmuller:2012oja} for details).

\begin{figure}
  \centering
  \hfill%
  \subfloat[][]{%
    \includegraphics[width=\twoPlotWidth]{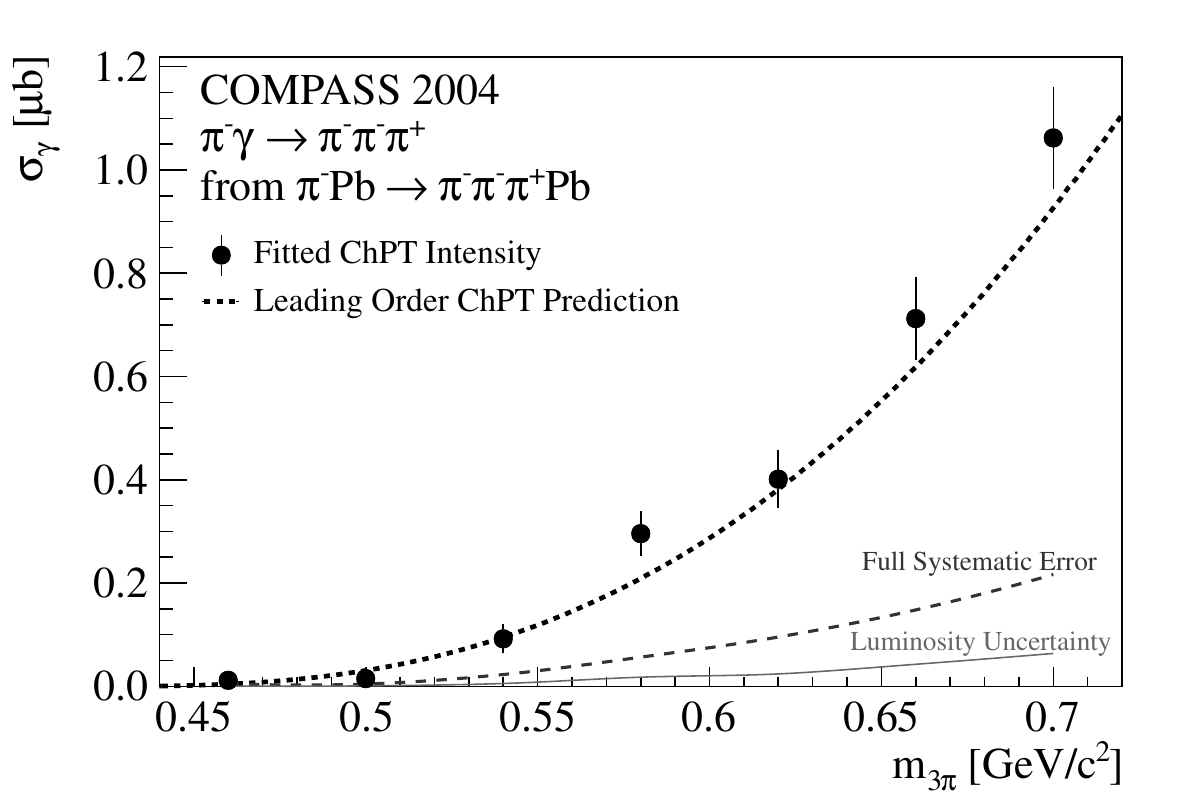}%
    \label{fig:chi_pt_massbins}%
  }%
  \hfill%
  \subfloat[][]{%
    \includegraphics[width=\twoPlotWidth]{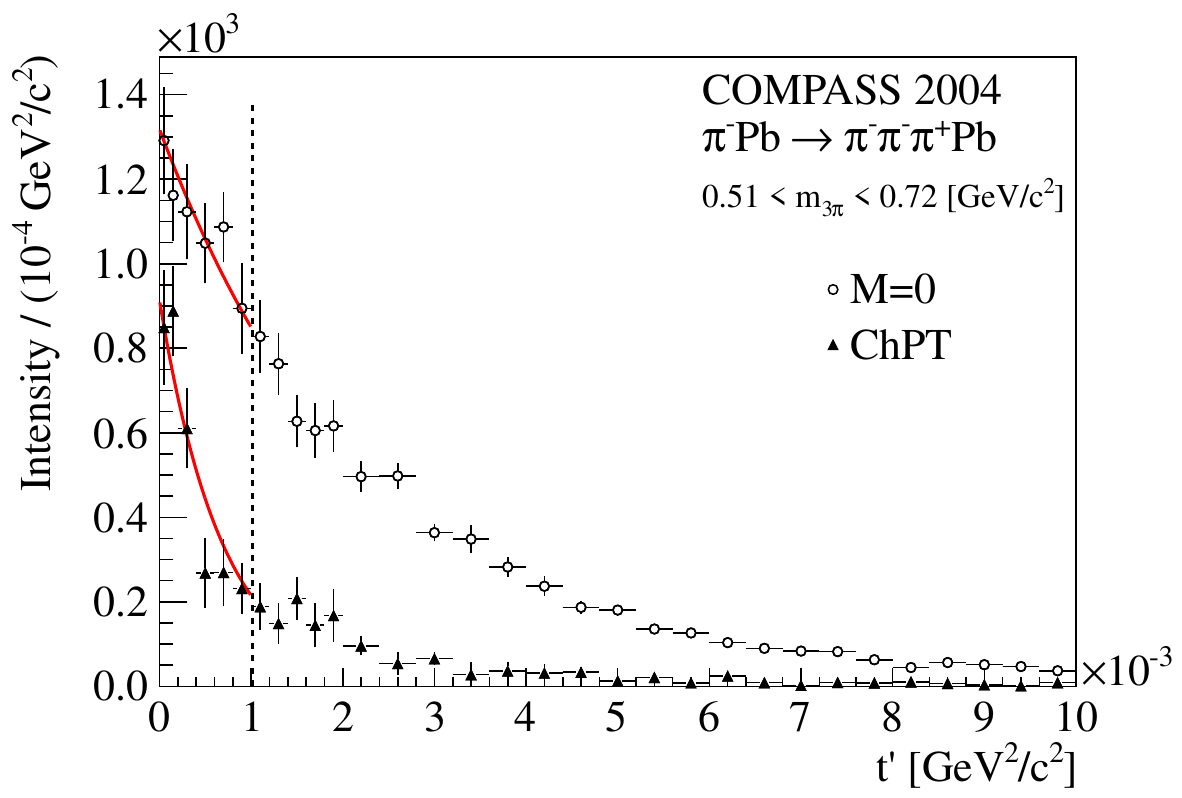}%
    \label{fig:chi_pt_tbins}%
  }%
  \hfill\null%
  \caption{\subfloatLabel{fig:chi_pt_massbins}~Cross section of the
    \chiPT process $\pi + \gamma \to \threePi$ as a function
    of~\mThreePi.  The points with error bars are calculated from the
    total intensity of the \chiPT amplitude as obtained from the PWA;
    error bars are statistical only.  The short-dashed curve shows the
    leading-order \chiPT prediction from \refCite{Kaiser:2008ss}.  The
    long-dashed curve represents the total systematic uncertainty, the
    continuous curve the contribution from the luminosity uncertainty.
    \subfloatLabel{fig:chi_pt_tbins}~The $t'$~spectra of the total
    $M = 0$ intensity (open circles) and the total \chiPT intensity
    (triangles).  The curves represent the results of
    single-exponential fits of the form $e^{-b\, t'}$.}
  \label{fig:chi_pt}
\end{figure}

The total \chiPT intensity is obtained by performing the PWA in seven
\SI{40}{\MeVcc} wide \mThreePi~bins and by summing up the intensities
of the two \chiPT amplitudes with $\refl = \pm 1$.  The such obtained
numbers of events are corrected for migration of events into and out
of the Primakoff region using the same MC technique as for the
radiative-width measurement (see \cref{sec:rad_widths}).  We do not
correct for the diffractive background since it was found to be
negligibly small (see discussion of $t'$~spectra below).  For each
\mThreePi~bin, the corrected number of events is converted into a
cross-section value for the process $\pi + A \to \threePi + A$ using
the same value for the integrated luminosity as for the
radiative-width measurements (see \cref{sec:rad_widths}).  Finally, by
integrating \cref{eq:ww_reson_xsec} over each \mThreePi~bin and over
$t' < \SI{e-3}{\GeVcsq}$ and by dividing out the quasi-real photon
flux factor $\mathcal{F}(\mThreePi, t')$, we obtain
$\sigma_{\pi + \gamma \to \threePi}(\mThreePi)$ at each \mThreePi~bin
center.\footnote{Here, we assume
  $\sigma_{\pi + \gamma \to \threePi}(\mThreePi)$ to be a constant in
  each \mThreePi~bin.}  The resulting cross-section values are shown
in \cref{fig:chi_pt_massbins}.  The total systematic uncertainty is
about \SI{20}{\percent}, where the dominant contributions come from
varying the fitting model and from the determination of the
luminosity.  The short-dashed curve shows the LO \chiPT prediction
from \refCite{Kaiser:2008ss}, which is in good agreement with the
measured cross-section values.

To study the details of the production mechanisms, we performed
another PWA in 31~narrow non-equidistant $t'$~bins, using now the
broad mass region \SIvalRange{0.51}{\mThreePi}{0.72}{\GeVcc}, where
the lower mass limit is chosen to exclude $K^- \to \threePi$ events.
In the PWA model, all transition amplitudes have \mThreePi~dependences
that are modeled using polynomials.  The parameters of these
polynomials are determined by fitting the \mThreePi~intensity
distribution of each partial wave as determined by the PWA in
\mThreePi~bins described above.  \Cref{fig:chi_pt_tbins} shows the
$t'$~dependence of the intensity of the coherent sum of all $M = 0$
waves and of the total \chiPT intensity.  The dominant $M = 0$
intensity shows the typical shape for diffraction on a Pb~nucleus,
while the \chiPT intensity forms a sharper peak toward $t' = 0$.  A
single-exponential fit to the latter results in a slope parameter of
$b_{\chiPT} = \SI{1447(196)}{\perGeVcsq}$.  This value is consistent
with $b_\text{\chiPT, MC} = \SI{1600}{\perGeVcsq}$ that is obtained
from MC simulation of quasi-real photon exchange.  Since no isobaric
waves with $M = 1$ are included in the PWA model, the \chiPT amplitude
may wrongly absorb diffractive $M = 1$ contributions at higher~$t'$.
However, at $t' \approx \SI{3e-3}{\GeVcsq}$, where for~Pb we expect
the maximum of the diffractive $M = 1$ contributions, the \chiPT
intensity nearly vanishes.  Fitting the \chiPT $t'$~spectrum over a
broader $t'$~range and adding a diffractive contribution of the form
$t'\, e^{-b_{\text{diff}}\, t'}$, we estimate that the diffractive
background is well below \SI{5}{\percent} for
$t' < \SI{e-3}{\GeVcsq}$.

The results of our analysis confirm the LO \chiPT prediction for the
process $\pi + \gamma \to \threePi$ for
$\mThreePi < \SI{0.72}{\GeVcc}$ and $t' < \SI{e-3}{\GeVcsq}$.  The
\chiPT prediction is confirmed on two levels of the analysis: \one~We
demonstrated that the \chiPT amplitude describes the data similarly
well than a PWA model with 6~isobaric $M = 1$ waves.  This confirms
that at low~\mThreePi and low~$t'$ the $3\pi$ phase-space distribution
of the physical process is consistent with the \chiPT prediction.
\two~The cross section $\sigma_{\pi + \gamma \to \threePi}(\mThreePi)$
that is calculated from the total intensity of the \chiPT amplitude as
obtained by the PWA fit, agrees well with the LO \chiPT prediction.
In addition, we find that the $t'$~spectrum of the \chiPT amplitude
shows a sharp peak at $t' = 0$ with a shape that is consistent with
Primakoff production, convolved with the experimental resolution.  The
\chiPT intensity nearly vanishes for $t' > \SI{3e-3}{\GeVcsq}$, which
demonstrates that possible diffractive $M = 1$ background is
negligibly small.
\clearpage{}%
\clearpage{}%
\section{Conclusions and Outlook}
\label{sec:conclusions_outlook}

The COMPASS experiment has collected world-leading data samples on the
scattering of a \SI{190}{\GeVc} pion beam on proton and nuclear
targets to perform a high-precision measurement of the excitation
spectrum of isovector mesons.  Based on these data, COMPASS has
performed the so far most comprehensive and most detailed partial-wave
analyses of \etaPi, \etaPrPi, and \threePi final states from
diffractive production in the squared four-momentum transfer range
\SIvalRange{0.1}{t'}{1.0}{\GeVcsq} and of the \threePi final state
from quasi-real photoproduction in the range $t' < \SI{e-3}{\GeVcsq}$.
At the used beam energy, diffractive production is dominated by
Pomeron exchange.

The PWA of \num{116000}~\etaPi and \num{39000}~\etaPrPi diffractive
events of the reaction $\pi^- + p \to \etaOrPrPi + p$ was performed
using a PWA model, which includes waves with relative orbital angular
momentum~$L$ between the two final-state particles ranging from~1
to~6.  This analysis revealed a striking pattern in the partial-wave
amplitudes.  On the one hand, even partial waves with $L = 2$, 4,
and~6 exhibit a close similarity of their intensities and phases as a
function of the \etaOrPrPi mass after the intensities have been scaled
by a kinematic factor that accounts for the different phase-space and
angular-momentum barrier factors in the two final states.  On the
other hand, applying the same scaling, odd partial waves with $L = 1$,
3, and~5, which carry spin-exotic, \ie non-\qqbar, \JPC quantum
numbers, are suppressed in \etaPi\ \wrt\ \etaPrPi.

COMPASS also studied diffractive production of \threePi using proton
and Pb~targets.  The Pb~target data sample consists of
\num{420000}~events, whereas the proton-target data sample is much
larger and consists of \num{46e6}~events.  For the latter data, a PWA
model with 88~partial waves is used, which is the largest wave set
used so far in a PWA of $3\pi$ final states.  It consists of 80~waves
with positive reflectivity, 7~with negative reflectivity, and one
incoherent wave that is isotropic in the phase space and represents
uncorrelated three pions.  Due to the large size of the \threePi
proton-target data sample, a $t'$-resolved PWA is performed by
subdividing the analyzed range \SIvalRange{0.1}{t'}{1.0}{\GeVcsq} into
11~non-equidistant bins.

For the \threePi proton-target data, we developed the novel
freed-isobar PWA method to extract the dynamical amplitudes of \twoPi
subsystems with well-defined \IGJPC quantum numbers in selected
\threePi partial waves as a function of two-pion mass, three-pion
mass, and squared four-momentum transfer.  This approach not only
reduces the model dependence of the PWA but in addition allows us to
study \twoPi isobar resonances and possible distortions of their
amplitudes due to final-state interactions.  We applied this method to
the \threePi proton-target data in order to study the isoscalar
$\JPC = 0^{++}$ \twoPi isobars, \ie \PfZero* states.  Comparing the
result of the freed-isobar PWA with the one of the conventional PWA,
we find that the parameterizations employed in the conventional PWA do
not cause artificial resonance-like structures in the extracted
partial-wave amplitudes and that partial waves with \pipiS and
\PfZero[980] isobars can be well separated.  Using the freed-isobar
PWA method, we establish, for the first time, the $3\pi$ decay modes
$\Pppi[1800] \to \PfZero[1500] \pi$,
$\PpiTwo[1880] \to \PfZero[980] \pi$, and
$\PpiTwo[1880] \to \PfZero[1500] \pi$ in a model-independent way.

We performed resonance-model fits for all analyzed data samples using
different models.  Due to the high precision of the \threePi
proton-target data, the resonance model employed for this data sample
is the most comprehensive one.  Out of the 88~waves, a subset of
14~partial waves with $\JPC = 0^{-+}$, $1^{++}$, $2^{++}$, $2^{-+}$,
$4^{++}$, and spin-exotic $1^{-+}$ quantum numbers was selected for
the resonance-model fit.  Compared to previous analyses of the $3\pi$
final state, this is by far the largest wave set that is consistently
described in a single resonance-model fit.  The amplitudes of the
14~waves including all of their mutual interference terms are
described simultaneously using 11~resonances.  The resonance models
for the other data samples contain only subsets of these resonances.
For the \threePi proton-target data, we performed, for the first time,
a simultaneous resonance-model fit in all 11~$t'$~bins.  Using this
novel $t'$-resolved approach, we extracted the $t'$~dependences of the
amplitudes of the resonant and non-resonant wave components in
individual partial waves from the data in a model-independent way.
This allows us to study the production mechanism of resonances in
terms of their $t'$~spectra and of the $t'$~dependences of their
relative phases \wrt other wave components.  Most resonances are
produced with a phase that is approximately independent of~$t'$, which
is expected if a single production mechanism contributes over the
analyzed $t'$~range.  With the $t'$-resolved approach, we also take
into account the change of the shape of the \mThreePi intensity
distributions of the partial waves with~$t'$, which for some waves is
very pronounced.  Hence our approach avoids a potential broadening of
resonance peaks and artificial incoherences between waves and wave
components that may have led to dilutions and distortions of resonance
signals in previous analyses.  In addition, the $t'$-resolved approach
exploits the generally different $t'$~dependences of the resonant and
non-resonant wave components in order to better disentangle the two.

We have measured the Breit--Wigner masses and widths of the $a_J$-like
resonances: \PaOne, \PaOne[1640], \PaTwo, \PaTwo[1700], \PaFour, and
the novel resonance-like \PaOne[1420]; and those of the $\pi_J$-like
resonances: \Pppi[1800], \PpiTwo, \PpiTwo[1880], \PpiTwo[2005], and
the spin-exotic \PpiOne[1600].  The parameters of \PaTwo, \PaFour,
\Pppi[1800], and \PpiTwo are reliably extracted with comparatively
small uncertainties.  The values from the \threePi proton- and
Pb-target data are in good agreement.  The \PaTwo and \PaFour
parameters from the \etaOrPrPi data also agree with the \threePi
results.  The \PaTwo and \Pppi[1800] parameter values are consistent
with the ones from previous experiments.  Our values for the \PaFour
mass and width are the most accurate so far.  We find a lower \PaFour
mass and a larger width than some of the previous experiments.  The
\PpiTwo[1880] is found to decay into $\PfTwo \pi D$,
$\PfZero[980] \pi D$, and $\PfZero[1500] \pi D$.  Its coupling to the
$\Pprho \pi F$ and $\PfTwo \pi S$ decay modes is found to be small.
This decay pattern contradicts model predictions that assume a hybrid
interpretation of the \PpiTwo[1880].  The measured \PpiTwo[1880] width
is consistent with the world average, its mass is found to be smaller.
The \threePi proton-target data require a third $\JPC = 2^{-+}$
resonance, the \PpiTwo[2005], which has so far been reported by only
two previous experiments.  The \PaOne parameters have large systematic
uncertainties mainly because of large backgrounds from non-resonant
components.  Due to the dominant \PaOne and \PaTwo signals, the
parameters of their radial excitations, \PaOne[1640] and \PaTwo[1700],
are difficult to measure, which leads to large systematic
uncertainties of the respective resonance parameters in the analysis
of the \threePi proton-target data.  Resonance-model fits of the
\etaOrPrPi data using Breit--Wigner amplitudes were inconclusive
concerning the \PaTwo[1700] parameter values.  This shows the
limitations of our simple sum-of-Breit--Wigner approach.  Together with
the JPAC collaboration, we performed a resonance-model fit of the
\etaPi $D$-wave intensity distribution using a more advanced model
that adheres to the principles of the relativistic $S$-matrix, in
particular analyticity and unitarity.  This analysis yielded pole
positions for \PaTwo and \PaTwo[1700] and showcases the superiority of
this approach.

The \threePi proton- and Pb-target data require a spin-exotic
resonance, the \PpiOne[1600], in the \wave{1}{-+}{1}{+}{\Pprho}{P}
wave.  The $t'$-resolved analysis allows us to establish, for the
first time, that for a proton target a significant \PpiOne[1600]
signal appears only for $t' \gtrsim \SI{0.5}{\GeVcsq}$, whereas at
low~$t'$ the intensity of the spin-exotic wave is saturated by
Deck-like non-resonant contributions.  This finding reconciles the
seemingly contradictory results from previous experiments that led to
a controversy about the existence of the \PpiOne[1600].  Our values
for the \PpiOne[1600] Breit--Wigner parameters have large
uncertainties, which are mainly due to the large background
contributions from non-resonant processes.  The width measured in the
\threePi proton-target data is larger than the values reported by
previous experiments including our own result from the \threePi
Pb-target data.  Resonance-model fits of the \etaOrPrPi data using
Breit--Wigner amplitudes were inconclusive concerning the parameters of
the \PpiOne* in the spin-exotic $P$-waves.  However, an
\etaPi-\etaPrPi coupled-channel analysis performed by the JPAC
collaboration using an analytic and unitary model based on $S$-matrix
principles shows that the COMPASS data can be described consistently
by only a single \PpiOne[1600] resonance pole.  Breit--Wigner-based
analyses of previous experiments required an additional \PpiOne[1400]
resonance in order to explain the \etaPi data.  However, compared to
predictions from models and lattice QCD a \PpiOne[1400] would be too
light and too close to the \PpiOne[1600].  The JPAC result shows that
no extra \PpiOne[1400] is required and that the \PpiOne[1400] signals
found in previous analyses are probably due to the \PpiOne[1600].

The highly precise \threePi proton-target data also revealed an
unexpected novel resonance-like signal, the \PaOne[1420], which is
observed only in the $\PfZero[980] \pi$ decay channel and is
well-described by a Breit--Wigner amplitude.  The
$\PaOne[1420] \to \PfZero[980] \pi$ signal was confirmed by the
freed-isobar PWA, without separating the data into non-orthogonal
waves with $\IGJPC = 0^+\, 0^{++}$ isobar resonances.  If interpreted
as a genuine resonance, the \PaOne[1420] would be a supernumerary
state and a candidate for a four-quark state.  However, the peculiar
properties of the \PaOne[1420] signal indicate that it might not be
due to a resonance but a triangle singularity, which is related to the
\PaOne ground state.  Preliminary studies show that the latter
hypothesis fits the data equally well as the Breit--Wigner amplitude.

In addition to diffractive production, COMPASS has studied the
pion-induced production of \threePi in the Coulomb field of a
Pb~target nucleus, \ie by quasi-real photon exchange.  This process is
strongly enhanced at very low squared four-momentum transfer
$t' < \SI{e-3}{\GeVcsq}$.  In this kinematic region, photoproduced
states have spin-projection quantum numbers of~$M = \pm 1$ \wrt the
beam axis because of the transverse nature of the photon, whereas
diffractively produced states have $M = 0$.  We separated these
contributions by means of a partial-wave analysis.  Using the
equivalent-photon approach, we determined the partial widths for
$\PaTwo \to \pi \gamma$ and, for the first time, the one for
$\PpiTwo \to \pi \gamma$ from the measured production cross sections
of the resonances.  In this analysis, we take into account sizable
corrections from the distortion of the pion wave function in the
Coulomb field of the nucleus, which have been neglected in previous
measurements.  The subprocess $\pi^- + \gamma \to \threePi$ that is
embedded in the measured scattering reaction also allows us to test
predictions from chiral perturbation theory at low three-pion mass.
COMPASS has measured the cross section for \threePi Coulomb production
in the range $m_{3\pi} < 5 m_\pi$ using PWA techniques and confirms
the leading-order predictions from chiral perturbation theory.

A focus of current research is the improvement of the analysis models,
in particular for the analysis of the high-precision \threePi
proton-target data, which is dominated by systematic uncertainties.
We pursue two strategies in order to reduce these uncertainties: \one
by reducing the model dependence, \eg by extending the freed-isobar
PWA to more freed waves, and \two by employing more advanced models
that incorporate more physical constraints, similar to what was done
for \etaOrPrPi.  For the latter, in particular a better understanding
of the non-resonant contributions from double-Regge exchange
processes, such as the Deck effect, is essential.

COMPASS data could also settle the claims by previous experiments of
\PpiOne* signals in \threePiN, $\pi^-\pi^0\Pomega$, and $\pi^-\PfOne$
from pion diffraction, which is another focus of future work.  An
additional line of research is the study of kaon diffraction into
various final states by using the $K^-$~component in the beam.  Since
kaons are not eigenstates of $G$~parity, nearly all kaon states can be
produced in kaon diffraction.  A first analysis of the diffractive
process $K^- + p \to K^- \pi^- \pi^+ + p$ on a subset of the available
data showed promising results~\cite{Jasinski:2012dia,Wallner:2020}.

The exploration of the light-meson spectrum is a global effort with
several experiments focusing on different parts of the spectrum.  In
addition to COMPASS, the VES experiment at IHEP (Russia) studies
diffractive-dissociation reactions.  Using a \SI{29}{\GeVc} pion beam
on a Be target, the VES experiment collected data samples of
\num{87e6} \threePi and \num{32e6} \threePiN events that are even
larger than the COMPASS data samples of \num{46e6} \threePi and
\num{3.5e6} \threePiN
events~\cite{Khokhlov:2012ck,Kachaev:2016yph,Ryabchikov:2016rji,Kachaev:2016ddk,Riabchikov:2017pfm,Ryabchikov:2019rgx}.
Despite the lower beam energy and the different target, the PWA
results from VES are in good agreement with the COMPASS
results~\cite{Ryabchikov:2019rgx}.  In particular, the VES experiment
also observes a narrow peak in the \wave{1}{++}{0}{+}{\PfZero[980]}{P}
wave of the $3\pi$ data that is consistent with the \PaOne[1420]
signal observed by COMPASS.  Two experiments at JLab are dedicated to
the study of the light-meson spectrum: GlueX and MesonEx.  The GlueX
experiment~\cite{GlueX,Ghoul:2015ifw,Dobbs:2016mal,Dobbs:2017vjw,Austregesilo:2018mno,Britton:2019cda}
uses a high-intensity beam of linearly polarized photons that is
produced from the primary \SI{12}{GeV} electron beam of the CEBAF
accelerator using coherent Bremsstrahlung from a thin diamond
radiator.  The produced photons have a most probable energy of about
\SI{9}{GeV} and a polarization of about \SI{40}{\percent}.  The
MesonEx experiment at CLAS12~\cite{MesonEx,Celentano:2013ata} studies
electroproduction of mesons using an \SI{11}{GeV} high-intensity
electron beam from CEBAF.  The quasi-real photons exchanged with the
target have energies from \SIrange{6.5}{10.5}{GeV}.  Both experiments
study photoproduction of mesons on proton targets at intermediate
energies.  Due to the---compared to COMPASS---lower center-of-momentum
energies, the photons may interact via various exchange processes with
the target proton and also target excitations can be studied.  This
has the advantage that in contrast to high-energy pion diffraction,
where the produced states are limited to isovector states with
negative $G$~parity, states with a wide variety of \IGJPC quantum
numbers are accessible.  In particular, both experiments will search
for the spin-exotic $\eta_1$, which would be the isoscalar partner of
the \PpiOne[1600].  The $\eta_1$ is expected to decay, for example,
into $\eta\PfTwo$ and $\PaTwo\pi$, which can be measured \eg in the
$\eta\pi\pi$ final state.  But also spin-exotic states with
$\JPC = 0^{+-}$ and $2^{+-}$ quantum numbers (\ie $b_0$, $h_0$,
$h_0'$, and $b_2$, $h_2$, $h_2'$) that cannot be produced in pion
diffraction are in principle accessible.  As was discussed in
\cref{sec:results_1mp}, photoproduction of hybrid mesons is expected
to be enhanced \wrt pion-induced reactions.  In contrast, currently
available data on $3\pi$ photoproduction seem to suggest that the
production rate of the hybrid candidate \PpiOne[1600] is suppressed
although it is observed to decay into $\Pprho\pi$.  The data from the
JLab experiments will help to solve this puzzle.  A complication
caused by the lower beam energies is the fact that particles emitted
from the beam and the target vertex of the scattering reaction, \ie
beam and target fragmentation, are less clearly kinematically
separated than in high-energy diffraction, making it necessary to take
into account baryon excitations in the analysis of excited mesons.
Light mesons are also studied at $e^+e^-$ machines using two-photon
processes, processes with initial-state radiation, and multi-body
decays of tau leptons or heavy-quark mesons.  The
CMD-3~\cite{Khazin:2008zz} and
SND~\cite{Achasov:1999ju,Achasov:2012nw} experiments at BINP perform
high-precision exclusive measurements of $e^+e^- \to \text{hadrons}$
cross sections up to center-of-momentum energies of \SI{2}{GeV}.
These data allow detailed studies of the properties of the light
vector mesons $\rho$, $\omega$, $\phi$ and of their excited states in
a large variety of final states.  The BESIII
experiment~\cite{Asner:2008nq} at IHEP (China) also studies the
light-meson spectrum, for example, by high-precision measurements of
radiative $J/\psi$ decays.  These decays are a rich source of light
mesons and allow \eg the study of $f_0$~and $f_2$~states up to high
masses.  Also the $B$-factories, \ie the
\textit{BABAR}~\cite{Bevan:2014iga}, Belle~\cite{Bevan:2014iga},
Belle~II~\cite{Kou:2018nap}, and LHCb
experiments~\cite{Bediaga:2012py}, have studied and are studying light
mesons, although this research is not their main goal.

In the future, the PANDA experiment at GSI~\cite{Lutz:2009ff} will
measure proton--antiproton annihilations in flight and will continue
the successful meson-spectroscopy programs carried out by the Crystal
Barrel experiment~\cite{Aker:1992ny} at LEAR and by the
E760~\cite{Armstrong:1991yk,Armstrong:1992wu,Armstrong:1995nn} and
E835 experiments~\cite{Garzoglio:2004kw} at Fermilab.  In addition to
a broad heavy-meson spectroscopy program, which will in particular
focus on the illusive $X$, $Y$, $Z$ states, PANDA will also study
light mesons.  Proton--antiproton annihilations provide a gluon-rich
environment, which should enhance the production of glueballs or
hybrid states~\cite{Wiedner:2011mf}.  For example, isoscalar scalar
mesons, \ie $f_0$~states, can be studied in reactions like
$\ppbar \to 3\pi^0$, $\eta\eta\pi^0$, $\eta\eta'\pi^0$, $\KKbar\pi^0$,
and
$5\pi^0$~\cite{Amsler:1995gf,Amsler:1995bf,Abele:1996fr,Abele:1996nn},
for which PANDA is expected to record high-precision data samples.  At
CERN, there are plans for a new fixed-target experiment covering a
wide range of QCD-related research topics~\cite{Denisov:2018unj}.  One
of these topics is the high-precision study of the kaon excitation
spectrum by measuring high-energy kaon-diffraction reactions.
Compared to the non-strange light mesons, the kaon spectrum is not
very well known~\cite{Tanabashi:2018zz}.  From the spectroscopy point
of view, it is of most importance to complete the light-meson
SU(3)$_\text{flavor}$ nonets and to search for the
SU(3)$_\text{flavor}$ partners of exotic states.  A better knowledge
of the kaon spectrum will also help to improve the analyses that
search for $CP$~violation in multi-body hadronic decays of~$D$ and
$B$~mesons, where kaon resonances appear in the subsystems of various
final states.  Using RF-separation
techniques~\cite{Bernard:1968gg,Panofsky:1956jg,Citron:1978sr}, the
kaon fraction of the high-intensity hadron beam in the SPS M2 beam
line could be dramatically enhanced so that data samples of much
higher precision than any previous experiment could be obtained.  For
example, for the $K^- \pi^- \pi^+$ final state, data samples of the
order of \num{e7}~events could be collected.
\clearpage{}%
\addcontentsline{toc}{section}{Acknowledgments}
\clearpage{}%
\section*{Acknowledgments}

We are greatly indepted to our COMPASS collaborators and colleagues,
especially to S.~U.~Chung, W.~D\"unnweber, M.~Faessler, J.~M.~Friedrich,
and S.~Paul for many enlightening discussions.
We thank in particular the students
A.~Austregesilo,
J.~Bernhard
K.~A.~Bicker,
S.~Grabm\"uller,
F.~Haas,
S.~Huber,
A.~Jackura,
P.~K.~Jasinski,
F.~M.~Kaspar,
M.~C.~Kr\"amer,
F.~M.~Krinner,
M.~Mikhasenko,
T.~Nagel,
S.~Neubert,
A.~Rodas,
T.~Schl\"uter,
S.~Schmeing,
S.~Uhl,
M.~Wagner,
S.~Wallner, and
Q.~Weitzel,
who played an important role in collecting and analyzing the data
discussed here.
We also gratefully acknowledge the support of the European Organization for Nuclear Research (CERN) management and
staff as well as the skills and efforts of the technicians of the
collaborating institutions.

We are especially indebted to the members of the JPAC collaboration,
in particular to
M.~Pennington and
A.~Szczepaniak,
for many useful discussions and for a fruitful collaboration on the
analysis of COMPASS spectroscopy data.
We also would like to thank
E.~L. Berger,
C.~Hanhart,
N.~Kaiser,
E.~Klempt,
B.~Kubis,
W.~Ochs,
J.~Pelaez,
A.~Sarantsev,
A.~M.~Zaitsev, and
Q.~Zhao
for important theory input and discussions.
We have in addition profited greatly from conversations and
discussions during a series of PWA workshops: a joint COMPASS-JLab-GSI
Workshop on Physics and Methods in Meson Spectroscopy (Garching/2008),
Workshops on Spectroscopy at COMPASS held 2009 and 2011 in Garching,
and in the context of the PWA/ATHOS workshop series (Camogli/2012,
Kloster Seeon/2013, Ashburn/2015, Bad Honnef/2017, and Beijing/2018).
We thank J.~Beringer for allowing us to use the PDG computer code to
create the ideograms.
D.~R. would like to thank the Excellence Cluster \enquote{Universe}
for supporting many visits to Munich during the past years.

We acknowledge the support by the German Bundesministerium für Bildung
und Forschung (BMBF), the DFG cluster of excellence \textquote{Origin
  and Structure of the Universe}, the DFG Collaborative Research
Centre/Transregio~110, and the computing facilities of the
Computational Center for Particle and Astrophysics (C2PAP).
\clearpage{}%
\appendix
\gdef\thesection{\Alph{section}}  %
\makeatletter
\renewcommand\@seccntformat[1]{Appendix \csname the#1\endcsname\@seccntDot\hskip 0.5em}
\makeatother
\titlecontents{section}%
[1.5em]%
{\addvspace{2.3ex}\bfseries}%
{\contentslabel[\hspace*{0.19em}\hyperlink{appendix.\thecontentslabel}{\thecontentslabel}]{1.5em}}%
{\hspace*{-1.5em}\hspace*{0.19em}}%
{\hfill\contentspage\hspace*{0.16em}}%
\clearpage{}%
\addcontentsline{toc}{section}{Appendices}

\section{$n$-Body Phase-Space Element}
\label{sec:nbody_phase_space}

A convenient parameterization for the $n$-body differential
phase-space element for $n > 2$ is obtained by applying the
phase-space recurrence
relation~\cite{pdg_kinematics:2018,James:1968gu} such that the
$n$-body phase space is expressed as a product of $(n - 1)$ two-body
differential phase-space elements.

For the decay of a parent state with mass~$m$ into two daughter
particles with masses~$m_1$ and~$m_2$, the two-body differential
phase-space element is given by \cref{eq:dPhi_2.cms}:
\begin{equation}
  \label{eq:two-body_dLIPS}
  \dif{\Phi_2}(m,
  \Underbrace{\vartheta, \phi}{\displaystyle{\eqqcolon \Omega}}; m_1, m_2)
  = \frac{1}{(4\pi)^2}\, \frac{q(m; m_1, m_2)}{m}\,
  \Underbrace{\dif{\cos \vartheta}\, \dif{\phi}}{\displaystyle{= \dif{\Omega}}}\eqPunctSpacing.
\end{equation}
Here, $q(m; m_1, m_2)$ is the magnitude of the two-body breakup
momentum in the parent rest frame as given by
\cref{eq:breakup_mom.cms.f,eq:kaellen}.  The polar angle~$\vartheta$
and the azimuthal angle~$\phi$ in \cref{eq:two-body_dLIPS} describe
the direction of daughter~1 in the parent rest frame.\footnote{Since
  in the parent rest frame the two daughter particles are emitted back
  to back, the choice of daughter particle~1 is a matter of
  convention.}

The decomposition of the $n$-body decay $X \to 1 + 2 + \ldots + n$
into a chain of $(n - 1)$ successive two-body decays is not unique,
but all decompositions are mathematically equivalent.  Using
\cref{eq:two-body_dLIPS}, the $n$-body phase-space element can be
written as~\cite{James:1968gu,Chung:1971ri}
\begin{multline}
  \label{eq:n-body_dLIPS}
  \dif{\Phi_n}(m_{s_1},
  \Underbrace{m_{s_2}, \ldots, m_{s_{n - 1}}, \Omega_{s_1}, \Omega_{s_2}, \ldots, \Omega_{s_{n - 1}}}{\displaystyle{= \tau_n}}) \\
  = \frac{1}{(4\pi)^2}\, \frac{q_{s_1}}{m_{s_1}}\, \dif{\Omega_{s_1}}
  \prod_{i = 2}^{n - 1} \frac{2 m_{s_i}\, \dif{m_{s_i}}}{2\pi}\, \frac{1}{(4\pi)^2}\,
  \frac{q_{s_i}}{m_{s_i}}\, \dif{\Omega_{s_i}}\eqPunctSpacing.
\end{multline}
The~$m_{s_i}$ are the parent masses of the $n - 1$ two-body
systems~$s_i$, where $m_{s_1} = m_X$.  The~$\Omega_{s_i}$ are the
angles and the~$q_{s_i}$ the breakup momenta of the two-body systems.
The $(3n - 4)$ phase-space variables that define the kinematics of the
$n$~final-state particles are represented by~$\tau_n$.

\section{Wigner $D$-Function}
\label{sec:wigner_D_function}

The Wigner $D$-function
$D_{M'\; M}^{J}(\alpha, \beta, \gamma)$~\cite{Wigner:1931,Wigner:1959}
represents the transformation property of a spin state $\ket{J, M}$
under an arbitrary active rotation $\hat{\mathcal{R}}$ defined by the
three Euler angles $\alpha$, $\beta$, and $\gamma$.  Since the
$\ket{J, M}$ basis is complete, the rotated state can be expressed as
a linear combination of the basis states:
\begin{equation}
  \label{eq:rot_property}
  \begin{split}
    \hat{\mathcal{R}}(\alpha, \beta, \gamma)\, \ket{J, M}
    &= \overunderbraces{&&\br{2}{\displaystyle{\eqqcolon D_{M'\; M}^{J}(\alpha, \beta, \gamma)}}}%
    {&\sum_{M' = -J}^{+J} \ket{J, M'}\, &\bra{J, M'}&\, \hat{\mathcal{R}}(\alpha, \beta, \gamma)\, \ket{J, M}&}%
    {&\br{2}{\displaystyle{= \mathds{1}}}} \\
    &= \sum_{M' = -J}^{+J}
    D_{M'\; M}^{J}(\alpha, \beta, \gamma)\, \ket{J, M'}\eqPunctSpacing.
  \end{split}
\end{equation}
We use the $y$--$z$--$y$ convention from \refCite{Rose:1957}, where
\begin{equation}
  \hat{\mathcal{R}}(\alpha, \beta, \gamma)
  = e^{-i\, \alpha\, \hat{J}_z}\, e^{-i\, \beta\, \hat{J}_y}\, e^{-i\, \gamma\, \hat{J}_z}\eqPunctSpacing.
\end{equation}
Here,~$\hat{J}_y$ and~$\hat{J}_z$ are the $y$~and $z$~components of
the angular momentum operator $\hat{J}$, respectively.  Hence
\begin{align}
  D_{M'\; M}^{J}(\alpha, \beta, \gamma)
  &= \bra{J, M'}\, e^{-i\, \alpha\, \hat{J}_z}\, e^{-i\, \beta\, \hat{J}_y}\, e^{-i\, \gamma\, \hat{J}_z}\, \ket{J, M} \nonumber \\
  \label{eq:D_func}
  &= e^{-i\, M'\, \alpha}\, \Underbrace{\bra{J, M'}\, e^{-i\, \beta\, \hat{J}_y}\, \ket{J, M}}%
  {\displaystyle{\eqqcolon d_{M'\; M}^{J}(\beta)}}\, e^{-i\, M\, \gamma}
\end{align}
with the Wigner (small) $d$-function (see \eg\ \refCite{Koelbig:1989}
and Eq.~(3) in Section~4.3.1 of \refCite{Varshalovich:1988ye})
\begin{multline}
  \label{eq:d_func}
  d_{M'\; M}^{J}(\beta)
  = (-1)^{J + M'}\, \sqrt{(J + M')!\, (J - M')!\, (J + M)!\, (J - M)!} \\
  \times \sum_k \left[ \frac{(-1)^k}{k!\, (J + M' - k)!\, (J + M - k)!\, (k - M' - M)!}\,
  \left( \cos \frac{\beta}{2} \right)^{2k - M' - M}\,
  \left( \sin \frac{\beta}{2} \right)^{2J + M' + M - 2k} \right]\eqPunctSpacing.
\end{multline}
The sum in \cref{eq:d_func} runs over all values of~$k$, for which the
factorials are non-negative.

The Wigner $D$-functions are orthogonal, \ie
\begin{equation}
  \label{eq:wigner_d_orthogonality}
  \int_0^{2\pi}\!\!\!\!\dif{\alpha} \int_{-1}^{1}\!\!\!\!\dif{\cos \beta} \int_0^{2\pi}\!\!\!\!\dif{\gamma}\,
  D_{M\; K}^{J}(\alpha, \beta, \gamma)\, D_{M'\; K'}^{J' \text{*}}(\alpha, \beta, \gamma)
  = \frac{8 \pi^2}{2 J + 1}\, \delta_{J J'}\, \delta_{M M'}\, \delta_{K K'}\eqPunctSpacing.
\end{equation}
Since the decay amplitudes $\Psi_i^\refl(\tau_n; m_X)$ in our PWA
model in \cref{eq:intensity_model_final} are constructed by recursive
application of \cref{eq:2_body_decay_amp}, they are linear
combinations of products of Wigner $D$-functions and hence inherit the
orthogonality property in \cref{eq:wigner_d_orthogonality}.  As a
consequence, the off-diagonal elements of the phase-space integral
matrix $I_{ij}^\refl$ for waves~$i$ and~$j$ with reflectivity~\refl as
defined in \cref{eq:int_matrix_def} vanish, unless the quantum numbers
that determine the $D$-functions in the decay amplitudes are the same
for both waves, \ie the two waves have the same $\JPMrefl$ quantum
numbers and the same isobar and orbital-angular-momentum quantum
numbers in the decay chain.\footnote{Note that the total intrinsic
  spins~$S_r$ of the daughter particles of the isobars~$r$ in the
  decay chain appear in \cref{eq:2_body_decay_amp} only in
  Clebsch--Gordan coefficients.  Hence the decay amplitudes of two
  partial waves that differ only in the~$S_r$ are linear dependent,
  \ie not orthogonal.}
For orthogonal decay amplitudes, also the
overlap $\ovl_{ij}^\refl(m_X, t')$ as defined in \cref{eq:overlap_def}
vanishes.  Depending on the analyzed final state, the orthogonality of
the decay amplitudes is broken by the Bose symmetrization in
\cref{eq:decay_amp_bose}.  In practice, the corresponding off-diagonal
elements of $I_{ij}^\refl$ are often still close to zero.  It is
important to note that decay amplitudes that correspond to decay
chains with different radially excited states of an isobar resonance
are in general not orthogonal and the corresponding elements of the
integral matrix may have large magnitudes.  In this case, the two
decay amplitudes correspond to similar phase-space distributions,
which may lead to ambiguity and distinguishability issues at the
partial-wave decomposition stage (see also
\cref{sec:pwa_cells.discussion}).

\section{Angular Distribution for Two-Body Decay}
\label{sec:ang_dist}

In order to derive the angular distribution for the two-body decay
$r \to 1 + 2$ in \cref{eq:ang_dist}, we consider the special case
where $\vartheta_r = \phi_r = 0$, \ie where particles~1 and~2 move in
opposite directions along the quantization axis.  Hence the
corresponding two-particle plane-wave center-of-momentum helicity
state $\ket{0, 0; \lambda_1, \lambda_2}$ is an eigenstate of the
$z$~component~$\hat{J}_z$ of the angular-momentum operator~$\hat{J}$.
The total spin projection of this state is
$\lambda = \lambda_1 - \lambda_2$, if we assume without loss of
generality that particle~1 moves in the $+z$~direction.  We express
the two-particle plane-wave state in terms of angular-momentum
helicity states:
\begin{equation}
  \ket{0, 0; \lambda_1, \lambda_2}
  = \sum_{J_r = 0}^\infty \ket{J_r, \lambda; \lambda_1, \lambda_2}\,
  \Underbrace{\braket{J_r, \lambda; \lambda_1, \lambda_2| 0, 0; \lambda_1, \lambda_2}}%
  {\displaystyle{\eqqcolon C_{J_r}}}\eqPunctSpacing.
\end{equation}
From this state, we can construct an arbitrary two-particle plane-wave
center-of-momentum helicity state by applying the active rotation
$\hat{\mathcal{R}}(\phi_r, \vartheta_r, 0)$, where~$\vartheta_r$
and~$\phi_r$ describe the direction of particle~1:\footnote{Since a
  direction is defined by only two angles, we use the convention of
  \refCite{Chung:1971ri} and set the third Euler angle $\gamma = 0$.
  Another commonly used convention is
  $\gamma = -\phi_r$~\cite{Jacob:1959at}.}
\begin{equation}
  \label{eq:2-body_hel_state}
  \ket{\vartheta_r, \phi_r; \lambda_1, \lambda_2}
  = \hat{\mathcal{R}}(\phi_r, \vartheta_r, 0)\, \ket{0, 0; \lambda_1, \lambda_2}
  = \sum_{J_r = 0}^\infty C_{J_r}\, \hat{\mathcal{R}}(\phi_r, \vartheta_r, 0)\,
  \ket{J_r, \lambda; \lambda_1, \lambda_2}\eqPunctSpacing.
\end{equation}
The coefficients $C_{J_r}$ are fixed by the normalization of the
Wigner $D$-function and the two-particle states.  Together with the
transformation property of a spin state under rotations as given in
\cref{eq:rot_property}, this yields
\begin{equation}
  \label{eq:2-body_hel_state_2}
  \ket{\vartheta_r, \phi_r; \lambda_1, \lambda_2}
  = \sum_{J_r = 0}^\infty \sum_{M_r = -J_r}^{+J_r} \sqrt\frac{2 J_r + 1}{4 \pi}\,
  D_{M_r\; \lambda}^{J_r}(\phi_r, \vartheta_r, 0)\, \ket{J_r, M_r; \lambda_1, \lambda_2}\eqPunctSpacing.
\end{equation}
From the above equation, \cref{eq:ang_dist} follows directly by using
the orthonormality of the two-particle states.

\section{List of Angular-Momentum Barrier Factors}
\label{sec:barrier_factors}

Using the parameterization for the angular-momentum barrier
factor~$F_{L}(z)$ in \cref{eq:bw_factor} from
\refCite{VonHippel:1972fg} and the normalization $F_L(z = 1) = 1$, the
barrier factors for the lowest values of~$L$ read
\begin{align}
  \label{eq:bw_factor_0}
  F_0^2(z) &= 1\eqPunctSpacing, \\
  \label{eq:bw_factor_1}
  F_1^2(z) &= \frac{2 z}{z + 1}\eqPunctSpacing, \\
  \label{eq:bw_factor_2}
  F_2^2(z) &= \frac{13 z^2}{z^2 + 3 z + 9}\eqPunctSpacing, \\
  \label{eq:bw_factor_3}
  F_3^2(z) &= \frac{277 z^3}{z^3 + 6 z^2 + 45 z + 225}\eqPunctSpacing, \\
  \label{eq:bw_factor_4}
  F_4^2(z) &= \frac{\num{12746} z^4}{z^4 + 10 z^3 + 135 z^2 + 1575 z + 11025}\eqPunctSpacing, \\
  \label{eq:bw_factor_5}
  F_5^2(z) &= \frac{\num{998881} z^5}{z^5 + 15 z^4 + 315 z^3 + \num{6300} z^2 + \num{99225} z + \num{893025}}\eqPunctSpacing,~\text{and} \\
  \label{eq:bw_factor_6}
  F_6^2(z) &= \frac{\num{118394977} z^6}{z^6 + 21 z^5 + 630 z^4 + \num{18900} z^3 + \num{496125} z^2 + \num{9823275} z + \num{108056025}}\eqPunctSpacing.
\end{align}

\section{Chung--Trueman Parameterization of the Spin-Density Matrix}
\label{sec:chung_trueman}

The Chung--Trueman parameterization of the spin-density
matrix~\cite{Chung:1974fq} exploits the fact that any Hermitian
positive-semidefinite matrix~$\bm{A}$ can be written as
\begin{equation}
  \label{eq:spin-dens_cholesky}
  \bm{A} = \bm{L}\, \bm{L}^\dagger\eqPunctSpacing,
\end{equation}
where $\bm{L}$ is a lower triangular matrix with real-valued and
non-negative diagonal entries.  This so-called Cholesky decomposition
is unique in case $\bm{A}$~is positive-definite.  If we reinterpret
the wave index~$i$ defined in \cref{eq:wave_index} as an index that
enumerates the various partial waves, we can treat the transition
amplitudes $\overline{\mathcal{T}_i}^r$ in
\cref{eq:intensity_model_bin_rank_rho} as elements of an
$N_\text{waves} \times N_r$ matrix~$\bm{\overline{\mathcal{T}}}$,
where $N_\text{waves}$~is the dimension of the spin-density matrix and
$N_r$~its rank, so that
\begin{equation}
  \label{eq:trans_amp_matrix}
  \bm{\overline{\mathcal{T}}}
  = \begin{pmatrix*}[c]
    \overline{\mathcal{T}_1}^1 & 0 & 0 & \ldots & 0\\
    \overline{\mathcal{T}_2}^1 & \overline{\mathcal{T}_2}^2 & 0 & \ldots & 0 \\
    \overline{\mathcal{T}_3}^1 & \overline{\mathcal{T}_3}^2 & \overline{\mathcal{T}_3}^3 & \ldots & 0 \\
    \vdots & \vdots & \vdots & \ddots & \vdots \\
    \overline{\mathcal{T}}_{\!\!\!N_r}^1 & \overline{\mathcal{T}}_{\!\!\!N_r}^2 & \overline{\mathcal{T}}_{\!\!\!N_r}^3 & \ldots & \overline{\mathcal{T}}_{\!\!\!N_r}^{N_r} \\
    \vdots & \vdots & \vdots &  & \vdots \\
    \overline{\mathcal{T}}_{\!\!\!N_\text{waves}}^1 & \overline{\mathcal{T}}_{\!\!\!N_\text{waves}}^2 & \overline{\mathcal{T}}_{\!\!\!N_\text{waves}}^3 & \ldots & \overline{\mathcal{T}}_{\!\!\!N_\text{waves}}^{N_r} \\
  \end{pmatrix*}
  \quad\text{and}\quad
  \bm{\overline{\rho}} = \bm{\overline{\mathcal{T}}}\, \bm{\overline{\mathcal{T}}}^\dagger\eqPunctSpacing.
\end{equation}
The diagonal elements $\overline{\mathcal{T}_i}^i$ of
$\bm{\overline{\mathcal{T}}}$ are real-valued and
positive.\footnote{The corresponding waves are called
  \textquote{anchor waves}, because they define the overall phase(s).
  Note that apart from differences due to numerical effects, the
  values of the spin-density matrix elements do not depend on the
  choice of the anchor waves.}  In addition to the $N_r$~diagonal
elements, the matrix $\bm{\overline{\mathcal{T}}}$ contains
$N_r\, (2 N_\text{waves} - N_r - 1) / 2$ complex-valued off-diagonal
elements, which yields a total of $N_r\, (2 N_\text{waves} - N_r)$
real-valued free parameters that need to be determined from data.

\section{Maximization of the Likelihood Function and Uncertainty Estimation}
\label{sec:max_likelihood_procedure}

In order to determine the maximum likelihood estimate for the
transition amplitudes, \ie
\begin{equation}
  \{\hat{\mathcal{T}}_i^{r \refl}\}
  = \argmax_{\{\mathcal{T}_i^{r \refl}\}}
  \Big[ \ln \mathcal{L}_\text{ext}(\{\mathcal{T}_i^{r \refl}\}; \{\tau_{n, k}\}, N) \Big]\eqPunctSpacing,
\end{equation}
using \cref{eq:likelihood_ext_final} one has to solve the coupled
system of
$\big( N_r^{\refl = +1}\, [2 N_\text{waves}^{\refl = +1} - N_r^{\refl
  = +1}] + N_r^{\refl = -1}\, [2 N_\text{waves}^{\refl = -1}
-N_r^{\refl = -1} ] \big)$ equations of the form
\begin{equation}
  \label{eq:mle_trans_amp_2}
  0 = \frac{\pdif{\ln \mathcal{L}_\text{ext}(\{\mathcal{T}_i^{r \refl}\}; \{\tau_{n, k}\}, N)}}{\pdif{\Re\!\big[ \mathcal{T}_j^{r \refl} \big]}}\bigg|_{\{\hat{\mathcal{T}}_i^{r \refl}\}}
  \quad\text{and}\quad
  0 = \frac{\pdif{\ln \mathcal{L}_\text{ext}(\{\mathcal{T}_i^{r \refl}\}; \{\tau_{n, k}\}, N)}}{\pdif{\Im\!\big[ \mathcal{T}_j^{r \refl} \big]}}\bigg|_{\{\hat{\mathcal{T}}_i^{r \refl}\}}
  \quad \forall~\text{$j$, $r$, and \refl}
\end{equation}
for $\{\hat{\mathcal{T}}_i^{r \refl}\}$.  However, in practice usually
numerical approaches are used to determine the maximum of
$\mathcal{L}_\text{ext}$ \wrt the $\{\mathcal{T}_i^{r \refl}\}$.
Traditionally, the numerical methods find the minimum of a given
function.  Hence we minimize the negative log-likelihood function
$-\ln \mathcal{L}_\text{ext}$.

The minimization programs usually employ iterative methods to find the
minimum.  The most widely used minimization algorithm for PWA is the
MIGRAD algorithm from the MINUIT or MINUIT2
packages~\cite{James:1975dr,minuit}.  The algorithm uses an iterative
decent method that belongs to the class of quasi-Newton methods, which
are generalizations of the secant method to find the root of the first
derivative for multi-dimensional functions.  Since for these kind of
problems, the secant equation does not specify a unique solution, the
various methods differ in how they constrain the solution.  The MIGRAD
algorithm is based on an improved version of the
Davidon--Fletcher--Powell (DFP) variable-metric
algorithm~\cite{Davidon:1966mm,Davidon:1991,Fletcher:1963,Fletcher:1970,Nocedal:2006},
Since the method depends heavily on the knowledge of the first
derivatives, it performs best if they are known precisely.  Luckily,
the log-likelihood function in \cref{eq:likelihood_ext_final} is
simple enough to calculate the Jacobian analytically (see \eg\
\refCite{msc_thesis_drotleff}) or using automatic
differentiation~\cite{Griewank:2008}.  In PWA fits, which may easily
have hundreds of free parameters, the MIGRAD algorithm is known to
find the minimum reliably.

Due to the central limit theorem, the likelihood function in
\cref{eq:likelihood_def} approaches in the asymptotic limit a
multivariate Gaussian in the parameters $\vec{\theta}$, with the
maximum at $\hat{\vec{\theta}}$ and the covariance matrix
$\bm{V}_{\hat{\vec{\theta}}}$ of $\hat{\vec{\theta}}$ given by
\begin{equation}
  \label{eq:mle_cov}
  \big( V_{\hat{\vec{\theta}}}^{-1} \big)_{ij}
  = E \bigg[ -\Underbrace{\frac{\pdif[2]{\ln \mathcal{L}(\vec{\theta}; \vec{x})}}{\pdif{\theta_i}\, \pdif{\theta_j}}}%
  {\displaystyle{= H_{ij}(\vec{\theta}; \vec{x})}} \bigg]\eqPunctSpacing,
\end{equation}
where $E[~]$ is the expectation value \wrt $\vec{x}$ and $\bm{H}$ is
the Hessian matrix of the log-likelihood function.  In practice, the
analytic calculation of the expectation value is often impractical.
For sufficiently large~$N$, a good estimate
for~$\bm{V}_{\hat{\vec{\theta}}}$ is obtained by calculating the
inverse of the Hessian matrix with the measured data at the maximum
likelihood estimate, \ie
\begin{equation}
  \label{eq:mle_cov_est}
  \hat{\bm{V}}_{\hat{\vec{\theta}}}
  = - \bm{H}^{-1}(\hat{\vec{\theta}}; \vec{x})\eqPunctSpacing.
\end{equation}
MIGRAD provides a numerical estimate for the covariance matrix.
However, the estimate of the HESSE routine of MINUIT, which calculates
the covariance matrix according to \cref{eq:mle_cov_est} through
inversion of the numerically estimated Hessian matrix of the minimized
function, is usually better.

The main caveat of the MIGRAD algorithm is that it requires a
comparatively large number of calls of the likelihood function to
converge to the minimum.  Depending on the number of events in the
$(m_X, t')$ cell and the number of waves in the model, the computation
of the log-likelihood function can be computationally expensive.  In
such cases, the Limited-memory Broyden--Fletcher--Goldfarb--Shanno
(L-BFGS) algorithm~\cite{Nocedal:1980,Liu:1989,Nocedal:2006} is a
better alternative.  It also belongs to the class of quasi-Newton
methods and requires an analytic Jacobian of the minimized function.
L-BFGS uses a sparse approximation to the inverse Hessian matrix to
find the minimum, so that the memory requirements grow only linearly
with the number of free parameters.  Therefore, it is well suited for
PWA fits, which often have a large number of parameters.  Another
alternative minimization algorithm that is successfully used for PWA
fits is called FUMILI~\cite{Dymov:1998zu,Sitnik:2014zwa}.  It is based
on the conjugate gradient method~\cite{Hestenes:1952,Nocedal:2006}.
Both L-BFGS and FUMILI estimate only the function minimum.  The
covariance matrix of the parameters at the minimum has to be
calculated separately.  Due to the simple structure of the
log-likelihood function in \cref{eq:likelihood_ext_final} the Hessian
matrix can be calculated analytically (see \eg\
\refCite{msc_thesis_drotleff}) or using automatic
differentiation~\cite{Griewank:2008}.  Using \cref{eq:mle_cov_est},
this yields a more precise estimate for the covariance matrix compared
to numerical methods.

\section{Numerical Calculation of Integral Matrices}
\label{sec:integral_matrices}

The integral matrices~$I_{ij}^\refl$ and
$\prescript{\text{acc}\!}{}{I}_{ij}^\refl$ that are defined in
\cref{eq:int_matrix_def,eq:likelihood_ext_final}, respectively, are
calculated using Monte Carlo integration techniques.  To this end, we
generate $N_\text{MC}$~Monte Carlo events that are uniformly
distributed in the $n$-body phase space of the final-state particles
using \eg\ \cref{eq:n-body_dLIPS} (see
\refsCite{James:1968gu,Block:1991} for more details).\footnote{These
  events are hence distributed according to the phase-space density
  $\rho_n(\tau_n)$ that is defined in \cref{eq:dLIPS_dens}.}  The
integral matrix~$I_{ij}^\refl$ is approximated by
\begin{equation}
  \label{eq:int_matrix_mc}
  I_{ij}^\refl
  = \int\! \dif{\Phi_n}(\tau_n)\,
  \overline{\Psi}_i^\refl(\tau_n)\, \overline{\Psi}_j^{\refl \text{*}}(\tau_n)
  \approx \frac{V_n}{N_\text{MC}}\,
  \sum_{k = 1}^{N_\text{MC}} \overline{\Psi}_i^\refl(\tau_{n, k})\,
  \overline{\Psi}_j^{\refl \text{*}}(\tau_{n, k})\eqPunctSpacing,
\end{equation}
where $V_n$~is the volume of the phase space as defined in
\cref{eq:phase-space_vol}.  To calculate the integral
matrix~$\prescript{\text{acc}\!}{}{I}_{ij}^\refl$, the
$N_\text{MC}$~phase-space Monte Carlo events are processed through the
detector simulation and are then subjected to the event reconstruction
and event selection procedure like real data.  This yields a sample of
$N_\text{MC}^\text{acc}$~accepted phase-space Monte Carlo events,
which are used to calculate an approximation to the integral matrix
\begin{align}
  \prescript{\text{acc}\!}{}{I}_{ij}^\refl
  &= \int\! \dif{\Phi_n}(\tau_n)\, \acc(\tau_n)\,
    \Psi_i^\refl(\tau_{n, k})\, \Psi_i^{\refl \text{*}}(\tau_{n, k}) \nonumber \\
  \label{eq:int_matrix_acc_mc}
  &\approx \frac{V_n}{N_\text{MC}}\, \frac{1}{\sqrt{I_{ii}^\refl\, I_{jj}^\refl}}
  \sum_{k = 1}^{N_\text{MC}} \acc(\tau_{n, k})\,
    \overline{\Psi}_i^\refl(\tau_{n, k})\, \overline{\Psi}_j^{\refl \text{*}}(\tau_{n, k})
    = \frac{V_n}{N_\text{MC}}\, \frac{1}{\sqrt{I_{ii}^\refl\, I_{jj}^\refl}}
  \sum_{k = 1}^{N_\text{MC}^\text{acc}}
    \overline{\Psi}_i^\refl(\tau_{n, k})\, \overline{\Psi}_j^{\refl \text{*}}(\tau_{n, k})\eqPunctSpacing.
\end{align}
The acceptance weight $\acc(\tau_{n, k})$ for an individual Monte
Carlo event is either~1, if the event was detected and selected for
the analysis, or~0 otherwise.\footnote{Note that the~$\tau_{n, k}$ in
  \cref{eq:int_matrix_acc_mc} represent the phase-space variables of
  the generated Monte Carlo sample, \ie the Monte Carlo truth, whereas
  the calculation of $\acc(\tau_{n, k})$ uses the reconstructed Monte
  Carlo information obtained from the detector simulation.  Resolution
  effects in the phase-space variables are therefore neglected by this
  approach.}  Hence the acceptance is taken into account by summing
only over the accepted phase-space Monte Carlo events.  Due to the
finite size of the Monte Carlo data samples, the integral matrices
have statistical uncertainties.  Their size depends on the number of
Monte Carlo events, the phase-space volume, and the shape of the
amplitudes.  The Monte Carlo samples are chosen to be large enough so
that these uncertainties are negligible compared to the statistical
uncertainties of the real data.

Note that the Monte Carlo approximation for~$I_{ij}^\refl$ in
\cref{eq:int_matrix_mc} is proportional to the phase-space
volume~$V_n$ and that consequently the Monte Carlo approximation
for~$\prescript{\text{acc}\!}{}{I}_{ij}^\refl$ in
\cref{eq:int_matrix_acc_mc} is independent of~$V_n$.  As was discussed
in \cref{sec:pwa_cells.normalization,sec:pwa_cells.likelihood_fit},
the normalization of the decay and transition amplitudes in
\cref{eq:decay_amp_norm,eq:decay_amp_flat_norm,eq:trans_amp_norm,eq:trans_amp_flat_norm}
fixes only the relative normalization of the transition amplitudes up
to an arbitrary common normalization factor that is fixed via
\cref{eq:abs_norm_cond,eq:events_exp} by maximizing the extended
likelihood function in \cref{eq:likelihood_ext_final}.  We hence
obtain the same maximum likelihood estimate for the transition
amplitudes when we use the integral matrix
$\tilde{I}_{ij}^\refl \coloneqq I_{ij}^\refl / V_n$ instead
of~$I_{ij}^\refl$ in \cref{eq:decay_amp_norm,eq:decay_amp_flat_norm}
to normalize the decay amplitudes.  This way we do not need to
know~$V_n$ in order to perform the partial-wave decomposition, \ie the
first analysis stage.

\section{Additional Observables}
\label{sec:additional_observables}

In order to interpret the relative phase $\phase_{ij}^\refl(m_X, t')$
between two waves~$i$ and~$j$ with reflectivity~\refl, it is important
to take into account the \emph{degree of coherence}
$\coh_{ij}^\refl(m_X, t')$ between the two partial-wave amplitudes.
This quantity is given by
\begin{equation}
  \label{eq:coherence_def}
  \coh_{ij}^\refl(m_X, t')
  \coloneqq \frac{\big| \varrho_{ij}^\refl(m_X, t') \big|}%
  {\sqrt{\varrho_{ii}^\refl(m_X, t')\, \varrho_{jj}^\refl(m_X, t')}}
  = \sqrt{\frac{\Re^2\!\big[ \varrho_{ij}^\refl(m_X, t') \big] + \Im^2\!\big[ \varrho_{ij}^\refl(m_X, t') \big]}%
    {\intens_i^\refl(m_X, t')\, \intens_j^\refl(m_X, t')}}\eqPunctSpacing.
\end{equation}
Note that due to the Cauchy--Schwartz inequality,
$\coh_{ij}^\refl \leq 1$.  Also, $\coh_{ii}^\refl = 1$ by definition.
For PWA models with rank-1 spin-density matrix, \ie $N_r^\refl = 1$,
all partial-wave amplitudes are fully coherent, \ie
\begin{equation}
  \label{eq:coherence_rank_1}
  \coh_{ij}^\refl(m_X, t')
  = \frac{\big| \mathcal{T}_i^\refl(m_X, t')\, \mathcal{T}_j^{\refl \text{*}}(m_X, t') \big|}%
  {\sqrt{\big| \mathcal{T}_i^\refl(m_X, t') \big|^2\, \big| \mathcal{T}_j^\refl(m_X, t') \big|^2}}
  = 1\eqPunctSpacing.
\end{equation}
If the rank of the spin-density matrix is larger than~1, the coherence
is in general reduced.  In order to interpret resonance signals in the
relative phases, the degree of coherence of the respective waves
should be close to unity.

Similar to \cref{eq:events_pred_int_ovl}, we can calculate the number
of events $N_\text{pred}^\text{acc}(m_X, t')$ that it is predicted by
the model to be measured by a detector with acceptance $\acc(m_X, t')$
in a given $(m_X, t')$ cell:
\begin{multline}
  \label{eq:events_pred_acc}
  N_\text{pred}^\text{acc}(m_X, t') \\
  = \sum_{\refl = \pm 1} \Bigg\{ \sum_i^{N_\text{waves}^\refl} \intens_i^\refl(m_X, t')\, \prescript{\text{acc}\!}{}{I}_{ii}^\refl
  + \sum_{i, j; i < j}^{N_\text{waves}^\refl} 2 \Re\!\big[ \varrho_{ij}^\refl(m_X, t')\, \prescript{\text{acc}\!}{}{I}_{ij}^\refl \big] \Bigg\}
  + \intens_\text{flat}(m_X, t')\, \prescript{\text{acc}\!}{}{I}_\text{flat}\eqPunctSpacing.
\end{multline}
Limiting the sums in \cref{eq:events_pred_acc} to a subset of partial
wave or even a single wave and comparing the
$N_\text{pred}^\text{acc}(m_X, t')$ value with the corresponding
$N_\text{pred}(m_X, t')$ value from \cref{eq:events_pred_int_ovl}
allows us to study the effect of the detector acceptance on the
partial-wave intensities.

\section{Resonance Parameters of~$a_J$ and $\pi_J$~Mesons}
\label{sec:res_par_aJ_piJ}

In \cref{tab:PDG_mesons_2018}, we list the known~$a_J$ and
$\pi_J$~mesons with spin~$J$ in the mass region below \SI{2.2}{\GeVcc}
according to the PDG~\cite{Tanabashi:2018zz}.

\begin{table}[htbp]
  \centering
  \renewcommand{\arraystretch}{1.2}
  \caption{Resonance parameters of~$a_J$ and $\pi_J$~mesons in the
    mass region below \SI{2.2}{\GeVcc} as given by the
    PDG~\cite{Tanabashi:2018zz}.}
  \label{tab:PDG_mesons_2018}
  \begin{small}
    \begin{tabular}[t]{lcll}
      \toprule
      \textbf{Particle} &
      \textbf{\JPC} &
      \textbf{Mass [\si{\MeVcc}]} &
      \textbf{Width [\si{\MeVcc}]} \\
      \midrule

      \addlinespace[1mm]
      \multicolumn{4}{c}{\textbf{Established states}} \\
      \addlinespace[1mm]

      \PaZero       & $0^{++}$ & $\phantom{0}980 \pm 20$ & $\phantom{0}50~\text{to}~100$ \\
      \PaOne        & $1^{++}$ & $1230 \pm 40$        & $250~\text{to}~600$ \\
      \PaTwo        & $2^{++}$ & $1318.3^{+0.5}_{-0.6}$ & $107 \pm 5$ \\
      \PaZero[1450] & $0^{++}$ & $1474 \pm 19$        & $265 \pm 13$ \\
      \PaFour       & $4^{++}$ & $1995^{+10}_{-8}$     & $257^{+25}_{-23}$ \\

      \midrule

      \Pppi[1300]   & $0^{-+}$ & $1300 \pm 100$   & $200~\text{to}~600$ \\
      \PpiOne[1400] & $1^{-+}$ & $1354 \pm 25$    & $330 \pm 35 $ \\
      \PpiOne[1600] & $1^{-+}$ & $1662^{+8}_{-9}$   & $241 \pm 40$ \\
      \PpiTwo       & $2^{-+}$ & $1672.2 \pm 3.0$ & $260 \pm 9$ \\
      \Pppi[1800]   & $0^{-+}$ & $1812 \pm 12$    & $208 \pm 12$ \\
      \PpiTwo[1880] & $2^{-+}$ & $1895 \pm 16$    & $235 \pm 34$ \\

      \midrule

      \addlinespace[1mm]
      \multicolumn{4}{c}{\textbf{States omitted from summary table}} \\
      \addlinespace[1mm]

      \PaOne[1420]  & $1^{++}$ & $1414^{+15}_{-13}$ & $153^{+8}_{-23}$ \\
      \PaOne[1640]  & $1^{++}$ & $1654 \pm 19$     & $240 \pm 27$ \\
      \PaTwo[1700]  & $2^{++}$ & $1732 \pm 9$      & $193 \pm 27$ \\
      \PaZero[1950] & $0^{++}$ & $1931 \pm 26$     & $271 \pm 40$ \\

      \midrule

      \PpiTwo[2100] & $2^{-+}$ & $2090 \pm 29$ & $625 \pm 50$ \\

      \midrule

      \addlinespace[1mm]
      \multicolumn{4}{c}{\textbf{Further states}} \\
      \addlinespace[1mm]

      \PaThree[1875] & $3^{++}$ & $1874 \pm 43 \pm 96$   & $385 \pm 121 \pm 114$ \\
      \PaOne[1930]   & $1^{++}$ & $1930^{+30}_{-70}$       & $155 \pm 45$ \\
      \PaTwo[1950]   & $2^{++}$ & $1950^{+30}_{-70}$       & $180^{+30}_{-70}$ \\
      \PaTwo[1990]   & $2^{++}$ & $2050 \pm 10 \pm 40$   & $190 \pm 22 \pm 100$ \\[-1ex]
                     &         & $2003 \pm 10 \pm 19$   & $249 \pm 23 \pm 32$ \\
      \PaZero[2020]  & $0^{++}$ & $2025 \pm 30$          & $330 \pm 75$ \\
      \PaTwo[2030]   & $2^{++}$ & $2030 \pm 20$          & $205 \pm 30$ \\
      \PaThree[2030] & $3^{++}$ & $2031 \pm 12$          & $150 \pm 18$ \\
      \PaOne[2095]   & $1^{++}$ & $2096 \pm 17 \pm 121 $ & $451 \pm 41 \pm 81$ \\
      \PaTwo[2175]   & $2^{++}$ & $2175 \pm 40$          & $310^{+90}_{-45}$ \\

      \midrule

      \PpiTwo[2005] & $2^{-+}$ & $1974 \pm 14 \pm 83$ & $341 \pm 61 \pm 139$ \\[-1ex]
                    &         & $2005 \pm 15$        & $200 \pm 40$ \\
      \PpiOne[2015] & $1^{-+}$ & $2014 \pm 20 \pm 16$ & $230 \pm 32 \pm 73$ \\[-1ex]
                    &         & $2001 \pm 30 \pm 92$ & $333 \pm 52 \pm 49$ \\
      \Pppi[2070]   & $0^{-+}$ & $2070 \pm 35$        & $310^{+100}_{-50}$ \\

      \midrule

      \PX[1775] & $\text{?}^{-+}$       & $1763 \pm 20$ & $192 \pm 60$ \\[-1ex]
                &                      & $1787 \pm 18$ & $118 \pm 60$ \\
      \PX[2000] & $\text{?}^{\text{?}+}$ & $1964 \pm 35$ & $225 \pm 50$ \\[-1ex]
                &                      & $\sim 2100$   & $\sim 500$ \\[-1ex]
                &                      & $2214 \pm 15$ & $355 \pm 21$ \\[-1ex]
                &                      & $2080 \pm 40$ & $340 \pm 80$ \\

      \bottomrule
    \end{tabular}
  \end{small}
\end{table}

\section{Effect of \phiGJ Smearing on the Partial-Wave Decomposition of \threePi}
\label{sec:primakoff_phi_smearing}

As discussed in \cref{sec:pwa_prim_specifics}, dedicated Monte Carlo
simulations show that at low~$t'$ detector resolution leads to a
splitting of $\Mrefl = 1^+$ waves in the physics process into
$\Mrefl = 1^+$ and~$1^-$ components in the PWA result and to a reduced
coherence between $\Mrefl = 0^+$ and~$1^+$ waves.  Here we show that
these two effects can be explained by a substantial smearing of the
\phiGJ~angle that is defined in
\cref{sec:pwa.analysis_model.coordsys}.

The limiting case of maximal smearing of \phiGJ, \ie when \phiGJ
cannot be measured, is equivalent to an integration over~\phiGJ.  We
study such integrals of products
$\Psi_i^\refl\, \Psi_j^{\refl \text{*}}$ of decay amplitudes.  For the
diagonal term of an $\Mrefl = 1^+$ wave, we obtain:
\begin{equation}
  \label{eq:phi_gj_integration_intens}
  \frac{1}{2\pi} \int_{0}^{2\pi}\!\! \dif{\phiGJ}\, \big| \Psi_{(M = 1)}^{(\refl = +1)}(\phiGJ) \big|^2
  =  \frac{1}{2}\, \Big[ \big| \Psi_{(M = 1)}^{(\refl = +1)}(\phiGJ) \big|^2
  + \big| \Psi_{(M = 1)}^{(\refl = -1)}(\phiGJ) \big|^2 \Big]\eqPunctSpacing.
\end{equation}
Here, both sides of the equation are independent of~\phiGJ.  This
means that adding the two intensities with opposite~\refl cancels the
\phiGJ~dependence of the right-hand side while the dependence on the
remaining phase-space variables remains unchanged.  For the
interference term between an $\Mrefl = 0^+$ and a~$1^+$ wave, the
integration gives
\begin{equation}
  \label{eq:phi_gj_integration_interf}
  \int_{0}^{2\pi}\!\! \dif{\phiGJ}\, \Psi_{(M = 0)}^{(\refl = +1)}(\phiGJ)\,
  \Psi_{(M = 1)}^{(\refl = +1) \text{*}}(\phiGJ)
  = 0\eqPunctSpacing.
\end{equation}
It is important to note that the above equation holds for any value of
the remaining phase-space variables.

In a PWA, we fit the model in \cref{eq:primakoff_partcoh_formula} to
the smeared data.  The model is differential in~\phiGJ, \ie in the PWA
we do not integrate over~\phiGJ.  In order to effectively satisfy
\cref{eq:phi_gj_integration_intens,eq:phi_gj_integration_interf},
which are a property of the data for the extreme case of maximal
smearing, the transition amplitudes must hence fulfill the following
conditions: \one~the initial intensity of a pure $\Mrefl = 1^+$ wave
splits into equal amounts of
$\big| \mathcal{T}_{(M = 1)}^{(\refl = +1)} \big|^2$ and
$\big| \mathcal{T}_{(M = 1)}^{(\refl = -1)} \big|^2$ and \two~the
interference terms
$\mathcal{T}_{(M = 0)}^{(\refl = +1)}\, \mathcal{T}_{(M = 1)}^{(\refl
  = +1) \text{*}}$ vanish.

In a realistic case with low \phiGJ~resolution, the intensity ratio of
an $\Mrefl = 1^+$ wave and the corresponding $1^-$~wave is always
larger than one and increases toward higher~$t'$, where the
measurement of the orientation of the production plane and hence
of~\phiGJ is recovered.  The small interference terms between the
decay amplitudes of waves with $\Mrefl = 0^+$ and~$1^+$ lead to a
reduced coherence that is effectively described by $r_{M = 0, M = 1}$
in \cref{eq:primakoff_partcoh_formula}.  In the Primakoff region
$t' < \SI{e-3}{\GeVcsq}$, $r_{M = 0, M = 1}$ has a value of about 0.5
in the peak region of the \PaTwo and it is found to increase with
increasing~$t'$.
\clearpage{}%
\addcontentsline{toc}{section}{References}
\bibliographystyle{./bib/model1a-num-names}

\end{document}